\documentclass{dune} 
\pdfoutput=1            
\graphicspath{ {figures/}{executive-summary/figures/}{generated/} }

\def\expshort{DUNE\xspace}
\def\explong{The Deep Underground Neutrino Experiment\xspace}

\def\thedocsubtitle{Deep Underground Neutrino Experiment (DUNE)} 

\def\voltitledpfd{Volume 3: Dual-Phase Module\xspace}

\def\volspfd{\textbf{Single-Phase Module\xspace}  
}

\newcommand{\kamland}{KamLAND\xspace}

\newcommand{\microboone}{MicroBooNE\xspace}
\newcommand{\minerva}{MINER$\nu$A\xspace}
\newcommand{\nova}{NO$\nu$A\xspace}

\newcommand{\lariat}{LArIAT\xspace}

\newcommand{\lartpc}{LArTPC\xspace}

\newcommand{\larsoft}{LArSoft\xspace}

\def\argon40{$^{40}$Ar}  
\def\Ar39{$^{39}$Ar}
\def\Cl40{$^{40}$Cl}
\def\K40{$^{40}$K}
\def\B8{$^{8}$B}

\def\fdfiducialmass{\SI{40}{\kt}\xspace}

\def\larmass{\SI{17.5}{\kt}\xspace} 
\def\tpcheight{\SI{12}{\meter}\xspace} 
\def\nominalmodsize{\SI{10}{kt}\xspace} 

\def\dunelifetime{\SI{20}{year}\xspace} 

\def\spmaxfield{\SI{500}{\volt/\centi\meter}} 
\def\spmaxdrift{\SI{3.53}{\m}\xspace}
\def\sptpclen{\SI{58}{\meter}\xspace} 
\def\spfcmodlen{\SI{3.5}{\m}} 
\def\spnumch{\num{384000}\xspace} 
\def\spnumpdch{\num{6000}\xspace} 

\def\dpmaxdrift{\SI{12}{\m}\xspace} 

\def\spreadout{\SI{5.4}{\ms}\xspace}
\def\dpreadout{\SI{16.4}{\ms}\xspace}
\def\snbtime{\SI{30}{\s}\xspace}
\def\snbpretime{\SI{10}{\s}\xspace}
\def\spsnbsize{\SI{45}{\PB}\xspace}

\def\offsitepbpy{\SI{30}{\PB/\year}\xspace}
\def\offsitegbyteps{\SI{1}{\GB/\s}\xspace}
\def\offsitegbps{\SI{8}{\Gbps}\xspace}

\def\surffnalbw{\SI{100}{\Gbps}\xspace}
\newcommand{\fnal}{Fermilab\xspace}
\newcommand{\surf}{SURF\xspace}

\newcommand{\efield}{E field\xspace}

\newcommand{\rms}{RMS\xspace} 
\newcommand{\threed}{3D\xspace}
\newcommand{\twod}{2D\xspace}

\newcommand{\dual}{DP\xspace}

\newcommand{\single}{SP\xspace}

\newcommand{\lar}{LAr\xspace}

\newcommand{\fdth}{feedthrough\xspace}

\newcommand{\phel}{photoelectron\xspace}

\newcommand{\pwrsupps}{power supplies\xspace}

\DeclareSIUnit \s {\second}
\DeclareSIUnit \MB {\mega\byte}
\DeclareSIUnit \GB {\giga\byte}
\DeclareSIUnit \TB {\tera\byte}
\DeclareSIUnit \PB {\peta\byte}
\DeclareSIUnit \Mbps {\mega\bit/\s}
\DeclareSIUnit \Gbps {\giga\bit/\s}
\DeclareSIUnit \Tbps {\tera\bit/\s}
\DeclareSIUnit \Pbps {\peta\bit/\s}
\DeclareSIUnit \kton {\kilo\tonne} 
\DeclareSIUnit \kt {\kilo\tonne}
\DeclareSIUnit \Mt {\mega\tonne}
\DeclareSIUnit \eV {\electronvolt}
\DeclareSIUnit \keV {\kilo\electronvolt}
\DeclareSIUnit \MeV {\mega\electronvolt}
\DeclareSIUnit \GeV {\giga\electronvolt}
\DeclareSIUnit \m {\meter}
\DeclareSIUnit \cm {\centi\meter}
\DeclareSIUnit \in {\inchcommand}
\DeclareSIUnit \km {\kilo\meter}
\DeclareSIUnit \kV {\kilo\volt}
\DeclareSIUnit \kW {\kilo\watt}
\DeclareSIUnit \MW {\mega\watt}
\DeclareSIUnit \MHz {\mega\hertz}
\DeclareSIUnit \mrad {\milli\radian}
\DeclareSIUnit \year {year}
\DeclareSIUnit \POT {POT}
\DeclareSIUnit \sig {$\sigma$}
\DeclareSIUnit\parsec{pc}
\DeclareSIUnit\lightyear{ly}
\DeclareSIUnit\foot{ft}
\DeclareSIUnit\ft{ft}
\DeclareSIUnit \ppb{ppb}
\DeclareSIUnit \ppt{ppt}
\DeclareSIUnit \samples{S}

\usepackage[toc]{glossaries}
\makeglossaries

\newcommand{\dshort}[1]{\glsentrytext{#1}}

\newcommand{\dlong}[1]{\glsentrylong{#1}}

\newcommand{\dfirst}[1]{\glsfirst{#1}}
\newcommand{\dfirsts}[1]{\glsfirstplural{#1}}

\newcommand{\dword}[1]{\gls{#1}}
\newcommand{\dwords}[1]{\glspl{#1}}
\newcommand{\Dword}[1]{\Gls{#1}}
\newcommand{\Dwords}[1]{\Glspl{#1}}

\newcommand{\newduneword}[3]{
    \newglossaryentry{#1}{
        text={#2},
        long={#2},
        name={\glsentrylong{#1}},
        first={\glsentryname{#1}},
        firstplural={\glsentrylong{#1}\glspluralsuffix},
        description={#3}
    }
}

\newcommand{\newduneabbrev}[4]{
  \newglossaryentry{#1}{
    text={#2},
    long={#3},
    shortplural={{#2}\glspluralsuffix},
    longplural={{#3}\glspluralsuffix{}},
    name={\glsentrylong{#1}{} (\glsentrytext{#1}{})},
    first={\glsentryname{#1}},
    firstplural={\glsentrylong{#1}\glspluralsuffix{} (\glsentrytext{#1}\glspluralsuffix{})},
    description={#4}
  }
}

\newcommand{\newduneabbrevs}[5]{
  \newglossaryentry{#1}{
    text={#2},
    long={#3},
    plural={#4},
    shortplural={{#2}\glspluralsuffix},
    longplural={#4},
    name={\glsentrylong{#1}{} (\glsentrytext{#1}{})},
    first={#3 (#2)},
    firstplural={#4 (\glsentrytext{#1}\glspluralsuffix{})},
    description={#5}
  }
}

\newduneword{dword}{DUNE Word}{A term in the DUNE lexicon}

\newduneabbrev{nd}{ND}{near detector}{Refers to the detector(s) 
 installed close to the neutrino source at \fnal }

\newduneabbrev{fd}{FD}{far detector}{Refers to the \fdfiducialmass fiducial mass DUNE detector 
  to be installed at the far site at \surf in
  Lead, SD, to be composed of four \nominalmodsize modules}

\newduneabbrev{sp}{SP}{single-phase}{Distinguishes one of the DUNE far detector technologies by the fact that it operates using argon in its liquid phase only
  }

\newduneabbrev{dp}{DP}{dual-phase}{Distinguishes one of the DUNE far detector technologies by the fact that it operates using argon 
 in both gas and liquid phases}

\newduneabbrev{pds}{PDS}{photon detection system}{The detector 
  subsystem sensitive to light produced in the \lar }

\newduneabbrev{tpc}{TPC}{time projection chamber}{The portion of each
  DUNE \gls{detmodule} that records ionization electrons
  after they drift away from a cathode through the \lar, and also
  through gaseous argon in a \dual module. 
  The activity is recorded by digitizing the waveforms of current
  induced on the anode as the distribution of ionization charge passes by
  or is collected on the electrode}

\newduneabbrevs{apa}{APA}{anode plane assembly}{anode plane assemblies}{A unit of the \single
  detector module containing the elements sensitive to ionization in the \lar. 
  It contains two faces each of three planes of wires, and interfaces to the cold
  electronics and photon detection system} 

\newduneabbrev{cro}{CRO}{charge readout}{The system for detecting
  ionization charge distributions in a \dual detector module}

\newduneabbrev{lro}{LRO}{light readout}{The system for detecting
  scintillation photons in a \dual  detector module}

\newduneabbrev{fe}{FE}{front-end}{The front-end refers a point that is
  ``upstream'' of the data flow for a particular subsystem. 
  For example the front-end electronics is where the cold electronics
  meet the sense wires of the TPC and the front-end DAQ is where the
  DAQ meets the output of the electronics}

\newduneabbrev{adc}{ADC}{analog-to-digital converter}{A sampling of a voltage
  resulting in a discrete integer count corresponding in some way to
  the input}

\newduneabbrev{daq}{DAQ}{data acquisition}{The data acquisition system
  accepts data from the detector FE electronics, buffers
  the data, performs a \gls{trigdecision}, builds events from the selected
  data and delivers the result to the offline \gls{diskbuffer}}

\newduneword{detmodule}{detector module}{The entire DUNE far detector is
  segmented into four modules, each with a nominal \SI{10}{\kton}
  fiducial mass}

\newduneword{detunit}{detector unit}{A \gls{submodule} may be partitioned
  into a number of similar parts. 
  For example the single-phase TPC \gls{submodule} is made up of APA
  units}

\newduneword{diskbuffer}{secondary DAQ buffer}{A secondary
  \dshort{daq} buffer holds a small subset of the full rate as
  selected by a \gls{trigcommand}. 
  This buffer also marks the interface with the DUNE Offline}

\newduneabbrev{om}{OM}{online monitoring}{Processes that run inside
  the DAQ on data ``in flight,'' specifically before landing on the
  offline disk buffer, and that provide feedback on the operation of
  the DAQ itself and the general health of the data it is marshalling}

\newduneabbrev{dqm}{DQM}{data quality monitoring}{Analysis of the raw
  data to monitor the integrity of the data and the performance of the
  detectors and their electronics. This type of monitoring may be
  performed in real time, within the \gls{daq} system, or in later
  stages of processing, using disk files as input}

\newduneword{dumpbuffer}{DAQ dump buffer}{This \dshort{daq} buffer
  accepts a high-rate data stream, in aggregate, from an associated
  \gls{submodule} sufficient to collect all data likely relevant to
  a potential \gls{snb}}

\newduneabbrev{etl}{ETL}{external trigger logic}{Trigger processing
  that consumes \gls{detmodule} level \gls{trignote} information
  and other global sources of trigger input and emits
  \gls{trigcommand} information back to the \glspl{mtl}}

\newduneword{trignote}{trigger notification}{Information provided by
  \gls{mtl} to \gls{etl} about \gls{trigdecision} its processing}

\newduneword{trigprimitive}{trigger primitive}{Information derived by
  the DAQ \gls{fe} hardware that describes a region of space (e.g.,
  one or several neighboring channels) and time (e.g., a contiguous set
  of ADC sample ticks) associated with some activity}

\newduneword{externtrigger}{external trigger candidate}{Information
  provided to the \gls{mtl} about events external to a
  \gls{detmodule} so that it may be considered in forming
  \glspl{trigcommand}}

\newduneabbrev{daqoob}{OOB dispatcher}{out-of-band trigger command
  dispatcher}{This component is responsible for dispatching a \gls{snb} dump
  command to all \glspl{daqfer} in the \gls{detmodule}}

\newduneabbrev{mtl}{MTL}{module trigger logic}{Trigger processing
  that consumes \gls{detunit} level \gls{trigcommand} information
  and emits \glspl{trigcommand}. 
  It provides the \gls{etl} with \glspl{trignote} and receives back any
  \glspl{externtrigger}}

\newduneword{octant}{octant}{Any of the eight parts into which 4$\pi$
  is divided by three mutually perpendicular axes. 
  In particular in referencing the value for the mixing angle
  $\theta_{23}$}

\newduneword{submodule}{subdetector}{A detector unit of granularity less
  than one \gls{detmodule} such as the TPC of either a \single or \dual 
  module}

\newduneword{trigcandidate}{trigger candidate}{Summary information derived
  from the full data stream and representing a contribution toward
  forming a \gls{trigdecision}}

\newduneword{trigcommand}{trigger command}{Information derived from
  one or more \glspl{trigcandidate}  that directs elements of the
  \gls{detmodule} to read out a portion of the data stream}

\newduneabbrev{tcm}{TCM}{trigger command message}{A message flowing
  down the trigger hierarchy from global to local context.  Also see \gls{tpm}}

\newduneword{trigdecision}{trigger decision}{The process by which
  \glspl{trigcandidate} are converted into \glspl{trigcommand}}

\newduneabbrev{tpm}{TPM}{trigger primitive message}{A message flowing
  up the trigger hierarchy from local to global context.  Also see \gls{tcm}}

\newduneabbrev{eb}{EB}{event builder}{A software agent servicing one
  \gls{detmodule} by executing \glspl{trigcommand} by reading out
  the requested data}

\newduneabbrev{cob}{COB}{Cluster On Board}{An ATCA motherboard housing four RCEs}

\newduneabbrev{rce}{RCE}{reconfigurable computing element}{Data processor located outside of the cryostat on a \gls{cob} which contains \gls{fpga}, RAM and SSD resources, responsible for buffering data, producing trigger primitives, responding to triggered requests for data and sinking \gls{snb} dumps}

\newduneabbrev{bow}{BOW}{Bump On Wire}{A working name for the front-end readout computing elements used in the nominal DAQ design to interface the \dual  crates to the DAQ front-end computers}

\newduneabbrev{atca}{ATCA}{Advanced Telecommunications Computing
  Architecture}{An advanced computer architecture specification developed for the telecommunications, military, and aerospace industries that incorporates the latest trends high-speed interconnect technologies, next-generation processors, and improved reliability, availability and serviceability} 

\newduneabbrev{utca}{$\mu$TCA}{Micro Telecommunications Computing Architecture}{The computer architecture specification followed by the crates that house charge and light readout electronics in the dual-phase module} 

\newduneabbrev{udp}{UDP}{user datagram protocol}{A simple,
  connectionless Internet protocol that supports data integrity
  checksums, requires no handshaking, and does not guarantee packet delivery}

\newduneabbrev{amc}{AMC}{advanced mezzanine card}{Holds digitizing
  electronics and lives in \gls{utca} crates}

\newduneabbrev{rf}{RF}{radio frequency}{Electromagnetic emissions
  that are within the (radio) frequency band of sensitivity of the detector
  electronics}

\newduneabbrev{fpga}{FPGA}{field programmable gate array}{An
integrated circuit technology that allows the hardware to be reconfigured to
execute different algorithms after its manufacture and deployment}

\newduneabbrev{felix}{FELIX}{Front-End Link eXchange}{A
  high-throughput interface between front-end and trigger electronics
  and the standard PCIe computer bus}

\newduneword{daqpart}{DAQ partition}{A cohesive and
 coherent collection of DAQ hardware and software working together to trigger and read out some portion of one detector module; it consists of an integral number of \glspl{daqfrag}. 
 Multiple DAQ partitions may operate simultaneously, but each instance operates independently}

\newduneabbrev{fec}{FEC}{front-end computer}{The portion of one
  \gls{daqpart} that hosts the \gls{daqdr}, \gls{daqbuf} and
  \gls{daqds}.  It is connected to the \gls{daqfer} via fiber optic. 
Each \gls{detunit} of a  certain granularity, such as two \single \glspl{apa}, has one front-end computer
  that receives data from the readout hardware, hosts the primary DAQ
  memory buffer for that data, emits trigger candidates derived from
  that data, and satisfies requests for producing subsets of that data
  for egress}

\newduneword{daqfrag}{DAQ front-end fragment}{The portion of one
  \gls{daqpart} relating to a single \gls{fec} and corresponding to an
  integral number of \glspl{detunit}.  See also \gls{datafrag}}

\newduneword{datafrag}{data fragment}{A block of data read out from a single \gls{daqfrag} that
span a contiguous period of time as requested by a \gls{trigcommand}}

\newduneabbrev{daqfer}{FER}{DAQ front-end readout}{The portion of a
  \gls{daqfrag} that accepts data from the detector electronics and
  provides it to the \gls{fec}. 
  In the nominal design it is also responsible for generating channel
  level \glspl{trigprimitive}}

\newduneabbrev{daqdr}{DDR}{DAQ data receiver}{The portion of the
  \gls{daqfrag} that accepts data from the \gls{daqfer}, emits
  trigger candidates produced from the input trigger primitives, and
  forwards the full data stream to the \gls{daqbuf}}

\newduneabbrev{daqbuf}{primary buffer}{DAQ primary buffer}{The portion
  of the \gls{daqfrag} that accepts full data stream from the
  corresponding \gls{detunit} and retains it sufficiently long for it
  to be available to produce a \gls{datafrag}}

\newduneword{daqds}{data selector}{The portion of the \gls{daqfrag}
  that accepts \glspl{trigcommand} and returns the corresponding
  \gls{datafrag}}

\newduneabbrev{femb}{FEMB}{front-end mother board}{Refers a unit of
  the \gls{sp} cold electronics that contains the front-end amplifier
  and ADC ASICs covering 128 channels}

\newduneword{asic}{ASIC}{application-specific integrated circuit}

\newduneword{lv}{LV}{low voltage}

\newduneword{coldata}{COLDATA}{a 64-channel control and communications ASIC}

\newduneword{larasic}{LArASIC}{A 16-channel front-end ASIC that provides signal amplification and pulse shaping}

\newduneword{cmos}{CMOS}{Complementary metal-oxide-semiconductor}

\newduneword{enc}{ENC}{equivalent noise charge}

\newduneword{dre}{DRE}{dynamic range enhancement}

\newduneword{sar}{SAR}{successive approximation register}

\newduneword{protodune}{ProtoDUNE}{Either of the two DUNE prototype detectors constructed at CERN and operated in
  a CERN test beam (expected fall 2018). 
  One prototype implements \single and the other \dual technology}

\newduneword{pdsp}{ProtoDUNE-SP}{The \single ProtoDUNE detector}

\newduneword{pddp}{ProtoDUNE-DP}{The \dual ProtoDUNE detector}

\newduneword{wa105}{WA105 DP demonstrator}{The \SI[product-units=power]{3x1x1}{m} WA105 dual-phase prototype detector at CERN}

\newduneword{rawevent}{DAQ event block}{The unit of data output by the
  DAQ. 
  It contains trigger and detector data spanning a unique, contiguous
  time period and a subset of the detector channels.}

\newduneabbrev{ssd}{SSD}{solid-state disk}{Any storage device that
  may provide sufficient write throughput to receive, both collectively and
  distributed, the sustained full rate of data from a \gls{detmodule}
  for many seconds}

\newduneabbrev{hlt}{HLT}{high-level trigger}{A source of triggering at
  the module level}

\newduneabbrev{pid}{PID}{particle ID}{Particle identification}

\newduneword{readout window}{readout window}{A fixed, atomic and
  continuous period of time over which data from a \gls{detmodule}, in
  whole or in part, is recorded. 
  This period may differ based on the trigger that initiated the
  readout}

\newduneabbrev{zs}{ZS}{zero-suppression}{Used to delete some portion of a
  data stream that does not significantly deviate from zero or
  intrinsic noise levels. 
  It may be applied at different granularity from per-channel to per
  \dword{detunit}}

\newduneabbrev{rc}{RC}{run control}{The system for configuring,
  starting and terminating the DAQ}

\newduneabbrev{snb}{SNB}{supernova neutrino burst}{A prompt 
  increase in the flux of low-energy neutrinos emitted in the first few seconds of a core-collapse supernova.  It can also refer to a trigger command type that may be due to an SNB,
  or detector conditions that mimic its interaction signature}

\newduneabbrev{snble}{SNB/LE}{supernova neutrino burst and low
  energy}{Supernova neutrino burst and low-energy physics program}

\newduneabbrev{snews}{SNEWS}{SuperNova Early Warning System}{A global
  supernova neutrino burst trigger formed by a coincidence of SNB
  triggers collected from participating experiments}

\newduneabbrev{pps}{1PPS signal}{one-pulse-per-second signal}{An
  electrical signal with a fast rise time and that arrives in real
  time with a precise period of one second}

\newduneabbrev{sls}{SLS}{spill location system}{A system residing at
  the DUNE far detector site that provides information, possibly
  predictive, indicating periods of time when neutrinos are being
  produced by the \fnal Main Injector beam spills}

\newduneabbrev{wib}{WIB}{warm interface board}{Digital electronics
  situated just outside the \single cryostat that receives digital data
  from the FEMBs over cold copper connections and sends it to the RCE
  FE readout hardware}

\newduneabbrev{nic}{NIC}{network interface controller}{Hardware for controlling the interface to a communication network.  Typically, one that obeys the Ethernet protocol}

\newduneabbrev{wiec}{WIEC}{warm interface electronics crate}{Crates mounted on the signal flanges that contain the warm interface boards}

\newduneabbrev{ptc}{PTC}{power and timing cards}{Cards that provide further processing and distribution of the signals entering and exiting the \single cryostat}

\newduneabbrev{sipm}{SiPM}{silicon photomultiplier}{A solid-state
  avalanche photodiode sensitive to single \phel signals}

\newduneabbrev{cisc}{CISC}{cryogenic instrumentation and slow controls}{A DUNE
  consortium responsible for the cryogenic instrumentation and slow controls components}

\newduneword{consortium}{consortium}{A unit of organization in the
  DUNE project focused on one major component of the far detector}

\newduneword{art}{\textit{art}}{A software framework implementing an
  event-based execution paradigm} 

\newduneabbrev{sam}{SAM}{sequential
  access via metadata}{A data-handling system to store and retrieve
  files and associated metadata, including a complete record of the
  processing that has used the files}

\newduneword{artdaq}{\textit{art}DAQ}{A general-purpose \fnal data acquisition framework for distributed data combination and processing. 
It is designed to exploit the parallelism that is possible with
modern multi-core and networked computers}

\newduneword{order}{$\mathcal{O}(n)$}{of order $n$}

\newduneword{beamline}{beamline}{A sequence of control and monitoring devices used for the formation of a directed collection of particles}
\newduneabbrev{cdr}{CDR}{conceptual design report}{A formal project
  document 
   that describes the experiment
  at a conceptual level}

\newduneabbrev{cf}{CF}{conventional facilities}{Pertaining to
  construction and operation of buildings or caverns and conventional infrastructure}

\newduneabbrev{cp}{CP}{charge parity}{Product of charge and parity
  transformations}

\newduneword{cpt}{CPT}{product of charge, parity
  and time-reversal transformations}

\newduneabbrev{cpv}{CPV}{charge-parity symmetry violation}{Lack of
  symmetry in a system before and after charge and parity
  transformations are applied}

\newduneword{doe}{DOE}{U.S. Department of Energy}

\newduneword{dune}{DUNE}{Deep Underground Neutrino Experiment}

\newduneword{esh}{ES\&H}{Environment, Safety and Health}

\newduneabbrev{fgt}{FGT}{fine-grained tracker}{A near detector module ??}

\newduneabbrev{fscf}{FSCF}{far site conventional facilities}{The
  \gslpl{cf} at the DUNE far detector site, \surf}
  
\newduneabbrev{nscf}{NSCF}{near site conventional facilities}{The
  \gslpl{cf} at the DUNE near detector site, \fnal}

\newduneabbrev{gut}{GUT}{grand unified theory}{A class of theories that
  unifies the electro-weak and strong forces}

\newduneabbrev{lar}{LAr}{liquid argon}{The liquid phase of argon}

\newduneabbrev{lartpc}{LArTPC}{liquid argon time-projection chamber}{A
  class of detector technology that forms the basis for the DUNE far
  detector modules. 
  It typically entails observation of ionization activity by
  electrical signals and of scintillation by optical signals}

\newduneabbrev{lbl}{LBL}{long-baseline}{Refers to the distance between the 
  neutrino source  and the far detector.  It can also refer to the distance between the near and far detectors. 
  The ``long'' designation is an approximate and relative distinction. For DUNE, this distance  (between \fnal and \surf) is approximately \SI{1300}{km}}

\newduneabbrev{lbnf}{LBNF}{Long-Baseline Neutrino Facility}{The
  organizational entity responsible for developing the neutrino beam, the cryostats
  and cryogenics systems, and the conventional facilities for DUNE}

\newduneabbrev{mh}{MH}{mass hierarchy}{Describes the separation
  between the mass squared differences related to the solar and
  atmospheric neutrino problems}

\newduneabbrev{mi}{MI}{\fnal Main Injector}{An accelerator at
  \fnal that provides a beam of high-energy protons that upon
  striking a target produce secondaries that decay to provide the
  neutrinos directed toward the DUNE far detector}

\newduneabbrev{pot}{POT}{protons on target}{Typically used as a unit
  of normalization for the number of protons striking the neutrino
  production target}

\newduneabbrev{qa}{QA}{quality assurance}{The process by which quality is maintained so as to preserve
high availability and precise function}

\newduneabbrev{qc}{QC}{quality control}{A system of maintaining
  quality through testing products against a specification}

\newduneabbrev{sm}{SM}{Standard Model}{Refers to a theory describing
  the interaction of elementary particles}

\newduneabbrev{tdr}{TDR}{technical design report}{A formal project
  document 
  that describes the experiment at a technical level}

\newduneabbrev{tp}{IDR}{interim design report}{An intermediate
milestone on the path to a full \gls{tdr}} 

\newduneabbrev{ckm}{CKM matrix}{Cabibbo-Kobayashi-Maskawa
  matrix}{Refers to the matrix describing the mixing between mass and
  weak eigenstates of quarks}

\newduneabbrev{cl}{CL}{confidence level}{Refers to a probability
  used to determine the value of a random variable given its
  distribution}

\newduneabbrev{pmns}{PMNS matrix}{Pontecorvo-Maki-Nakagawa-Sakata
  matrix}{Describes the mixing between mass and weak eigenstates of
  the neutrino}

\newduneabbrevs{cpa}{CPA}{cathode plane assembly}{cathode plane assemblies}{The component of the \single detector module that provides the drift HV cathode}

\newduneabbrev{fc}{FC}{field cage}{The component of a LArTPC that contains and shapes the applied \efield}

\newduneabbrev{topfc}{top FC}{top field cage}{The horizontal portions of the \single FC on the top}

\newduneabbrev{botfc}{bottom FC}{bottom field cage}{The horizontal portions of the \single FC on the bottom}

\newduneabbrev{ewfc}{endwall FC}{endwall field cage}{The vertical portions of the \single FC near the wall}

\newduneabbrev{gp}{GP}{ground plane}{An electrode that is held to be
  electrically neutral relative to Earth ground voltage}

  \newduneword{gg}{ground grid}{An electrode that is held to be
  electrically neutral relative to Earth ground voltage  ??}

\newduneabbrev{alara}{ALARA}{as low as reasonably
  achievable}{Typically used with regard management of radiation
  exposure but may be used more generally. It means making every
  reasonable effort to maintain e.g., exposures, to as far below the
  limits as practical, consistent with the purpose for that the
  activity is undertaken}

\newduneabbrev{ecal}{ECAL}{electromagnetic calorimeter}{A detector
  component that measures energy deposition of traversing particles}

\newduneabbrev{hv}{HV}{high voltage}{Generally describes a voltage
  applied to drive the motion of free electrons through some media}

\newduneword{spmod}{SP module}{single-phase detector module}
\newduneword{dpmod}{DP module}{dual-phase detector module}

\newduneabbrev{tc}{TC}{technical coordination}{A dedicated DUNE
  project effort responsible for assuring proper interfaces between
  the collaboration consortia}

\newduneabbrev{cc}{CC}{charged current}{Refers to an interaction
  between elementary particles where a charged weak force carrier
  ($W^+$ or $W^-$) is exchanged}

\newduneabbrev{dis}{DIS}{deep inelastic scattering}{Refers to
  interaction of an elementary charged particle with a nucleus in an
  energy range where the interaction can be modeled as being with
  individual nucleons}

\newduneabbrev{fsi}{FSI}{final-state interactions}{Refers to
  interactions between elementary or composite particles subsequent to
  the initial, fundamental particle interaction, such as may occur as
  the products exit a nucleus}

\newduneabbrev{geant4}{Geant4}{GEometry ANd Tracking, version 4}{A
  software toolkit for the simulation of the passage of particles
  through matter using Monte Carlo methods}

\newduneabbrev{genie}{GENIE}{Generates Events for Neutrino Interaction
  Experiments}{Software providing an object-oriented neutrino
  interaction simulation resulting in kinematics of the products of
  the interaction}

\newduneabbrev{mc}{MC}{Monte Carlo}{Refers to a method of numerical
  integration that entails the statistical sampling of the integrand
  function. 
  Forms the basis for some types of detector and physics simulations}

\newduneabbrev{qe}{QE}{quasi-elastic}{Refers to interaction between
  elementary particles and a nucleus in an energy range where the
  interaction can be modeled as occurring between constituent quarks
  of one nucleon and resulting in no bulk recoil of the resulting
  nucleus}

\newduneabbrev{mou}{MOU}{memorandum of understanding}{A document
  summarizing an agreement between two or more parties}

\newduneabbrev{pip2}{PIP-II}{Proton Improvement Plan II}{A plan for
  improving the protons on target delivered for the DUNE neutrino
  production beam. 
  This is revision two of this plan and is planned to be followed by a PIP-III}

\newduneabbrev{sdsta}{SDSTA}{South Dakota Science and Technology
  Authority}{The legal entity that manages the Sanford Underground
  Research Facility, \surf, in Lead, S.D}

\newduneabbrev{wbs}{WBS}{work breakdown structure}{An organizational
  project management tool by which the tasks to be performed are
  partitioned in a hierarchical manner}

\newduneabbrev{br}{BR}{branching ratio}{A fractional probability for a
  decay of a composite particle to occur into some specified set or
  sets of products}

\newduneabbrev{dm}{DM}{dark matter}{The term given to the unknown
  matter or force that explains measurements of motion of galaxies
  that are otherwise inconsistent with the amount of mass associated
  with observed amount of photon production}

\newduneabbrev{dsnb}{DSNB}{Diffuse Supernova Neutrino Background}{The
  term describing the pervasive, constant flux of neutrinos due to all
  past supernova neutrino bursts}

\newduneabbrev{globes}{GLoBES}{General Long-Baseline Experiment
  Simulator}{A software package for simulating energy spectra of
  neutrino flux, interaction and measured (after application of some
  model of a detector response) energy spectra}

\newduneabbrev{nc}{NC}{neutral current}{Refers to an interaction
  between elementary particles where a neutrally charged weak force carrier
  ($Z^0$) is exchanged}

\newduneabbrev{nh}{NH}{normal hierarchy}{Refers to the neutrino mass
  eigenstate ordering whereby the sign of the mass squared difference
  associated with the atmospheric neutrino problem is positive}

\newduneabbrev{ih}{IH}{inverted hierarchy}{Refers to the neutrino mass
  eigenstate ordering whereby the sign of the mass squared difference
  associated with the atmospheric neutrino problem is negative}

\newduneabbrev{nsi}{NSI}{nonstandard interactions}{A general class of
  theory of elementary particles other than the Standard Model}

\newduneabbrev{msw}{MSW}{Mikheyev-Smirnov-Wolfenstein effect}{Explains
  the oscillatory behavior of neutrinos produced inside the sun as
  they traverse the solar matter}

\newduneabbrev{susy}{SUSY}{supersymmetry}{Theoretical symmetry between a fermion and a boson}

\newduneabbrev{wimp}{WIMP}{weakly-interacting massive particle}{A
  hypothesized particle that may be a component of dark matter}

\newduneabbrev{ce}{CE}{cold electronics}{Refers to readout electronics that operate at cryogenic temperatures}

\newduneabbrev{crp}{CRP}{charge-readout plane}{In the \dual technology, a  collection of
  electrodes in a planar arrangement placed at a particular voltage
  relative to some applied \efield such that drifting electrons
  may be collected and their number and time may be measured}

\newduneabbrev{dram}{DRAM}{dynamic random access memory}{A computer memory technology}

\newduneword{fermilab}{\fnal}{Fermi National Accelerator Laboratory (in Batavia, IL, the Near Site)}
\newduneword{fnal}{FNAL}{see \fnal}

\newduneabbrev{fs}{FS}{full stream}{Relates to a data stream that has not undergone selection, compression or other form of reduction}

\newduneabbrev{lem}{LEM}{large electron multiplier}{A micro-pattern detector suitable for use in ultra-pure argon vapor; LEMs consist of copper-clad PCB boards with sub-millimeter-size holes through which electrons undergo amplification}
\newduneabbrev{lng}{LNG}{liquefied natural gas}{Pertaining to natural gas in its liquid phase}

\newduneabbrev{mip}{MIP}{minimum ionizing particle}{Refers to a
  momentum traversing some medium such that the particle is losing
  near the minimum amount of energy per distance traversed} 

\newduneabbrev{pd}{PD}{photon detector}{Refers to the detector
  elements involved in measurement of number and arrival times of
  optical photons produced in a detector module} 

\newduneabbrev{pmt}{PMT}{photomultiplier tube}{A device that makes use
  of the photoelectric effect to produce an electrical signal from the
  arrival of optical photons}

\newduneabbrev{ppm}{ppm}{parts per million}{A number equal to $10^{-6}$}
\newduneabbrev{ppb}{ppb}{parts per billion}{A number equal to $10^{-9}$}
\newduneabbrev{ppt}{ppt}{parts per trillion}{A number equal to $10^{-12}$}

\newduneword{rio}{RIO}{reconfigurable input output}

\newduneword{rpc}{RPC}{resistive plate chamber}
\newduneword{s/n}{S/N}{signal-to-noise (ratio)}
\newduneword{ssp}{SSP}{SiPM signal processor}
\newduneword{sbn}{SBN}{Short-Baseline Neutrino program (at \fnal)}
\newduneword{stt}{STT}{straw tube tracker}
\newduneword{tr}{TR}{transition radiation}
\newduneword{wire board}{wire board}{At the head end of the APA in the \single TPC, stacks of electronics boards referred to as ``wire boards'' are arrayed to anchor the wires.  They also provide the connection between the wires and the cold electronics} 

\newduneabbrev{wls}{WLS}{wavelength shifting}{A material or process by
  which incident photons are absorbed by a material and photons are
  emitted at a different, typically longer, wavelength}

\newduneabbrev{tpb}{TPB}{tetra-phenyl butadiene}{A type of wavelength shifting material}

\newduneabbrev{sft}{SFT}{signal feedthrough}{A cryostat penetration allowing for the passage of cables or other extended parts}
\newduneabbrev{sftchimney}{SFT chimney}{signal feedthrough chimney}{In the \dual technology, a volume above the cryostat penetration used for a signal feedthrough}

\newduneword{catiroc}{CATIROC}{charge and time integrated readout chip}

\newduneabbrev{wr}{WR}{White Rabbit}{A component of the timing system that forwards clock signal and time-of-day reference data to the master timing unit}

\newduneabbrev{mch}{MCH}{MicroTCA Carrier Hub}{An network switching device}

\newduneabbrev{wrmch}{WR-MCH}{White Rabbit \gls{utca} Carrier Hub}{add def} 

\newduneword{cmp}{CMP}{configuration management plan}

\newduneword{qap}{QAP}{quality assurance plan} 

\newduneword{ieshp}{IESHP}{integrated environmental, safety and health plan}

\newduneword{dmp}{DMP}{data management plan} 

\newduneword{qam}{QAM}{quality assurance manager} 

\newduneabbrev{dss}{DSS}{detector support system}{The system used to support the \single detector within the cryostat}

\newduneabbrev{itf}{ITF}{integration and test facility}{A facility where various detector components will be tested prior to installation}

\newduneabbrev{tco}{TCO}{temporary construction opening}{An opening in the side of a cryostat through which detector elements are brought into the cryostat; utilized during construction and installation}

\newduneabbrev{uit}{UIT}{underground installation team}{An organizational unit responsible for installation in the underground area at the \surf site}

\newduneabbrev{pdr}{PDR}{preliminary design review}{A project management device by which an early design is reviewed}
\newduneabbrev{fdr}{FDR}{final design review}{A project management device by which a final design is reviewed}
\newduneabbrev{prr}{PRR}{production readiness review}{A project management device by which the production readiness is reviewed}
\newduneabbrev{orr}{ORR}{operational readiness review}{A project management device by which the operational readiness is reviewed}
\newduneabbrev{ppr}{PPR}{production progress review}{A project management device by which the progress of production is reviewed}
\newduneabbrev{edms}{EDMS}{electronic document management system}{A computerized system used at CERN by which documents are managed}

\newduneword{wrgm}{WR grandmaster}{White Rabbit grandmaster}

\newduneword{larsoft}{\larsoft}{Liquid Argon Software (\larsoft),  a shared base of physics software across \lartpc experiments}
\newduneword{nova}{\nova}{The \nova off-axis neutrino oscillation experiment at \fnal }
\newduneword{minerva}{\minerva}{The \minerva neutrino cross sections experiment at \fnal }
\newduneword{microboone}{\microboone}{The \lartpc-based \microboone neutrino oscillation experiment at \fnal }
\newduneword{wirecell}{wire-cell}{A tomographic automated \threed neutrino event reconstruction method for \lartpc{}s}
\newduneword{ftslite}{F-FTS-lite}{Light-weight version of the \fnal File Transfer system used for rapid data transfers out of the online systems}
\newduneword{fts}{FTS}{File Transfer System developed at \fnal to catalog and move data to permanent storage}

\newduneword{35t}{35 ton prototype}{The 35 ton prototype cryostat and \gls{sp} detector built at \fnal before the \gls{protodune} detectors}

\newduneabbrev{cuc}{CUC}{central utility cavern}{The central underground cavern  containing utilities such as central cryogenics and other systems, and the underground data center and control room}
\newduneabbrev{cfd}{CFD}{computational fluid dynamics}{High performance computer-assisted modeling of fluid dynamical systems}
\newduneword{vuv}{VUV}{vacuum ultra-violet}
\newduneword{tallbo}{TallBo}{A cylindrical cryostat at \fnal primarily used for developing scintillation light collection technologies for \lartpc detectors}

\newduneabbrev{eos}{EOS}{EOS}{The XRootD based distributed file system developed by CERN}
\newduneabbrev{ehn1}{EHN1}{Experiment Hall North One}{Location at CERN of the ProtoDUNE experiments}
\newduneword{led}{LED}{Light-emitting diode}
\newduneword{rtd}{RTD}{Cryostat level and temperature monitoring devices}
\newduneword{swc}{SWC}{Software \& Computing}
\newduneabbrev{las}{LAS}{LEM-anode Sandwich}{In the \dual technology, a \gls{lem} and its corresponding anode are mounted together in a module called a LEM-anode sandwich}

\newduneword{roi}{ROI}{region of interest}
\newduneword{hpc}{HPC}{high-performance computing facilities; generally computing facilities emphasizing parallel computing with aggregate power of more than a teraflop}

\newduneword{comfund}{common fund}{The shared resources of the collaboration}
\newduneabbrev{ims}{IMS}{integrated master schedule}{A project management device consisting of linked tasks and milestones}

\newduneword{hvdb}{HVDB}{dual-phase HV divider board}
\newduneword{sas}{SAS}{A pass-through chamber used to ensure safe transfer of materials, avoiding contamination in both directions}

\newduneword{fea}{FEA}{finite element analysis}

\newduneword{fss}{FSS}{field shaping strips}
\newduneword{lvds}{LVDS}{low-voltage differential signaling}

\hypersetup{
    pdftitle={\expshort TDR \thedocsubtitle},
    pdfauthor={\expshort Collaboration},
    final=true,
    colorlinks=false,
    linktocpage=true,
    linkbordercolor=blue,
    citebordercolor=green,
    urlbordercolor=magenta,
    filecolor=black,
    pdfpagemode=UseOutlines,
    pdfborderstyle={/S/U},  
}

\renewcommand\thedoctitle{\volspfd} 
\begin{document}

\pagestyle{titlepage}
\includepdf[pages={-}]{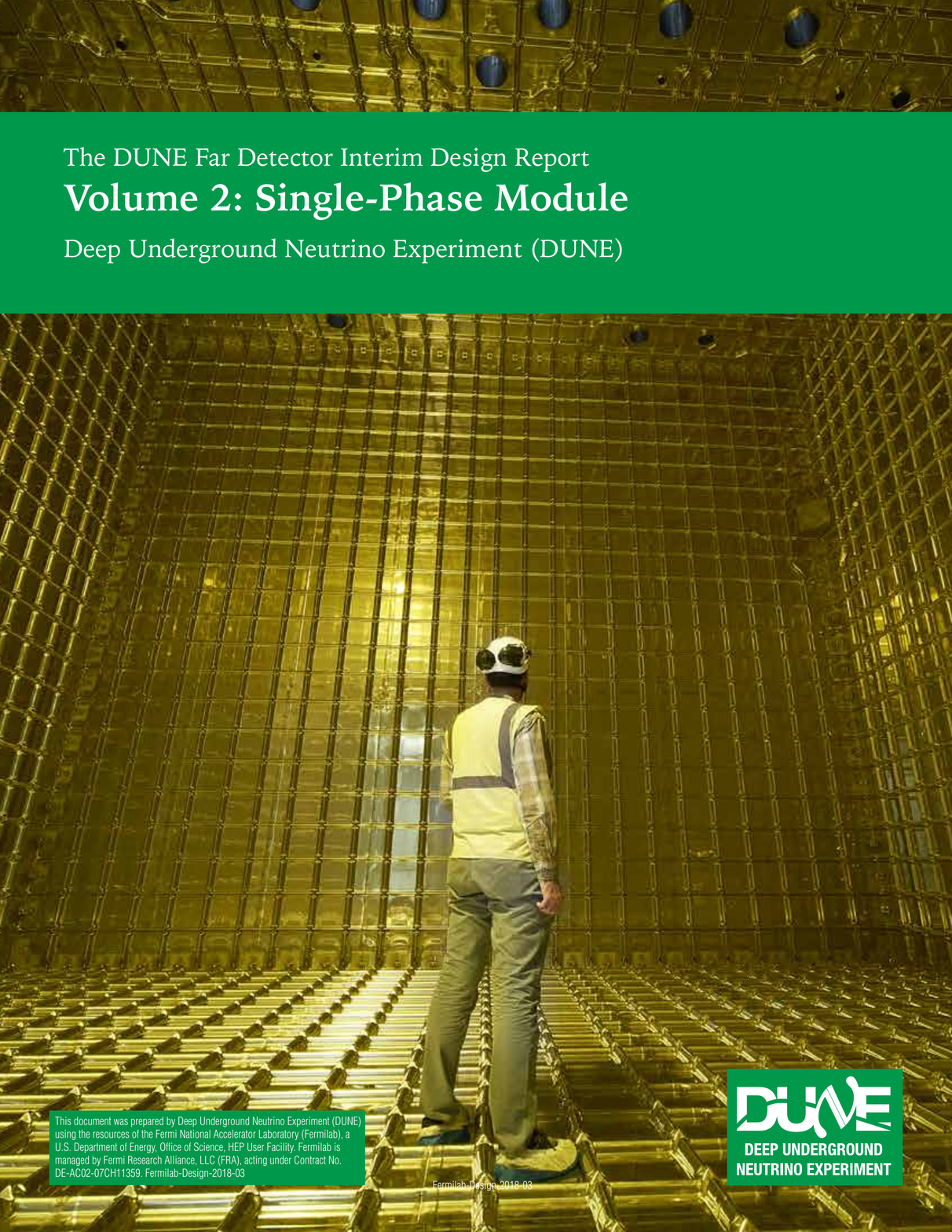}
\cleardoublepage



\cleardoublepage

\includepdf[pages={-}]{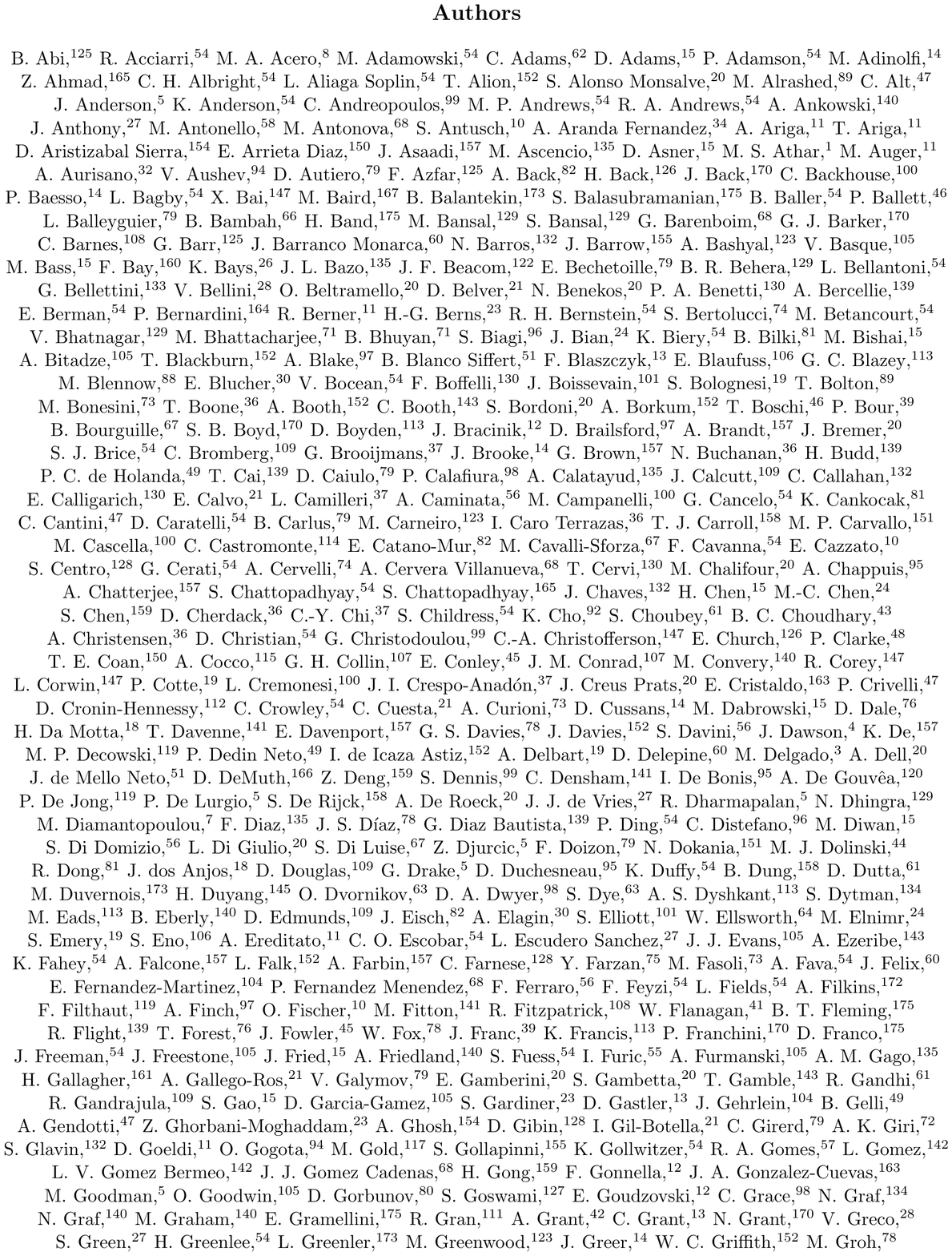}

\renewcommand{\familydefault}{\sfdefault}
\renewcommand{\thepage}{\roman{page}}
\setcounter{page}{0}

\pagestyle{plain} 


\textsf{\tableofcontents}

\textsf{\listoffigures}

\textsf{\listoftables}



\renewcommand{\thepage}{\arabic{page}}
\setcounter{page}{1}

\pagestyle{fancy}

\renewcommand{\chaptermark}[1]{%
\markboth{Chapter \thechapter:\ #1}{}}
\fancyhead{}
\fancyhead[RO,LE]{\textsf{\footnotesize \thechapter--\thepage}}
\fancyhead[LO,RE]{\textsf{\footnotesize \leftmark}}

\fancyfoot{}
\fancyfoot[RO]{\textsf{\footnotesize The DUNE Far Detector Interim Design Report}}
\fancyfoot[LO]{\textsf{\footnotesize \thedoctitle}}
\fancypagestyle{plain}{}

\renewcommand{\headrule}{\vspace{-4mm}\color[gray]{0.5}{\rule{\headwidth}{0.5pt}}}

\nocite{CD0}


\chapter{Design Motivation and Overview}
\label{ch:fdsp-apa-design}

\section{Introduction to the DUNE Single-Phase Far Detector Design}
\label{sec:fdsp-design-intro}

The DUNE single-phase (SP) \lartpc detector module will be the culmination of several decades
of \lartpc technology development, and once operational, it will open new windows of opportunity in the study of neutrinos.  DUNE's rich physics program, with discovery
potential for \dword{cpv} in the neutrino sector, and capability to make
significant observations of nucleon decay and astrophysical events, is enabled
by the exquisite resolution of the \lartpc detector technique.

Experience with design, construction, operation, and data
analysis with numerous single-phase \lartpc experiments and prototypes has informed the approach to
realizing the massive DUNE \dwords{spmod}. Each far \dword{detmodule} will feature the largest \lartpc{}s ever
constructed, at approximately \nominalmodsize active volume each.  Aside from the
challenges inherent in such a large undertaking, DUNE presents the added complication of construction and operation in a location
that is \SI{1.5}{km} (one mile) underground with limited access.

The design of the DUNE \dword{spmod} presented in this document
reflects an approach to achieving the science goals of the experiment, and
addresses the challenges of constructing and operating a massive detector in a deep
underground environment.

\section{Single-Phase \lartpc Operational Principle}
\label{sec:fdsp-design-op}

The precision tracking and calorimetry offered by the single-phase \lartpc
technology provides excellent capabilities for identifying interactions of interest
while mitigating sources of background.  The operational principle of a
single-phase \lartpc is summarized here for reference.

Charged particles traversing the active volume of the \lartpc ionize the medium,
while also producing scintillation light.  The ionization drifts along
an \efield that is present throughout the volume, towards a series of
anode layers.  Each anode layer is composed of finely spaced wires arranged at
characteristic angles, and appropriate biasing of these wires allows the
ionization to drift through the successive layers before terminating on a wire
in the collection layer.  The individual wires in the anode layers can be
instrumented with low-noise electronics that record the current in the wire as
a function of time.  The argon scintillation light, which at \SI{127}{nm} wavelength
is deep in the UV spectrum, can be recorded by \dwords{pd} that shift the
wavelength closer to the visible spectrum and subsequently record the time and
pulse characteristics of the incident light.



The performance of the \lartpc hinges on several key factors.  First, the
purity of the \lar must be extremely high in order to allow ionization to 
drift over several meters towards the anode planes.  The levels of
electronegative contaminants (e.g., oxygen, water), must be reduced and
maintained to \dword{ppt} levels in order to achieve minimum charge attenuation
over the longest drift lengths in the \lartpc.   Second, the electronic readout
of the \lartpc requires very low noise levels so that the signal of drifting
ionization is clearly discernible over the baseline of the electronics.  
Third, a uniform \efield must be established over the detector volume, requiring a robust and stable high voltage system.  Finally, the sheer size of the \dword{spmod} means that once it is filled with \lar, all components within the cryostat are inaccessible for decades.  All internal devices must have long operating lifetimes at \lar temperatures.

\section{Motivation of Single-Phase \lartpc Design at DUNE}
\label{sec:fdsp-design-impl}

The DUNE Single-Phase \dword{fd} design builds on several decades of experience in designing, constructing, and operating \lartpc{}s.  It implements unique design features to maximize the capability of the experiment, as well as new features motivated by the unprecedented scale of the \dword{fd} modules and the deep underground location where construction will occur.

Among the features driven by the underground location of the experiment, all detector components are sized to fit within the constraints of the \surf shafts and access pathways.

A drift time of several milliseconds is typical for ionization to arrive at the anode wires after drifting several meters.  This lengthy duration of time, as well as aspects of the DUNE physics program looking for rare and low-energy processes, makes the deep underground location essential for the \dword{spmod}.  The  $\sim$\SI{1.5}{km} overburden of earth greatly reduces the rate of cosmic rays reaching the active volume of the \dword{detmodule}, greatly enhancing the ability to search for rare and low-energy signatures without the influence of cosmic-induced backgrounds.

\section{Overview of the Single-Phase Design}
\label{sec:fdsp-ov-model}

The DUNE \dword{spmod} features a \nominalmodsize active mass \lartpc, with all associated cryogenic, electronic readout, computing, and safety systems.  The \dword{spmod} is designed to maximize the active volume within the confines of the membrane cryostat while minimizing dead regions.  The detector elements have been modularized such that their production can proceed in parallel with the construction of the DUNE caverns and cryostats, and sized so that they conform to the access restrictions for transport underground.  Table~\ref{tab:dune-sp-parameters} summarizes some of the high-level parameters of the \dword{spmod}.

\begin{dunetable}[\dword{spmod} parameters]{lll}{tab:dune-sp-parameters}{\dword{spmod} parameters}
Parameter & Value & Note \\ \toprowrule
Cryostat \lar mass & \larmass & \\ \colhline 
Active \lar mass & \nominalmodsize & \\ \colhline 
Active Height & \tpcheight & \\ \colhline 
Active Length & \sptpclen & \\ \colhline 
Maximum Drift & \spmaxdrift & \\ \colhline 
Number of \dword{apa} channels & \spnumch & \\\colhline 
Number of \dword{pds} channels & \spnumpdch & \\ 
\end{dunetable}


The cryostat is constructed such that its long axis is aligned with the beam arriving from \fnal.  The TPC  inside the cryostat is composed of two rows of \dwords{cpa} oriented along the long axis of the cryostat, flanked on either side by rows of \dwords{apa}.  A \dword{fc} completely surrounds the four
open sides of the four drift regions to ensure that the \efield within is uniform and unaffected by the presence of the cryostat walls and other nearby conductive structures.  Integrated within each \dword{apa} are elements of the \dword{pds} as well as electronics to process the \dword{apa} signals.  Around the periphery of the TPC various instrumentation for monitoring the cryogenic environment is present.  Outside of the cryostat, additional electronic readout and data acquisition equipment is present to transfer information from the \dword{detmodule}.  Figure~\ref{fig:dune-sp-overview} illustrates the basic arrangement of the TPC elements within the \dword{spmod}.

\begin{dunefigure}[DUNE \dword{spmod} diagram]{fig:dune-sp-overview}{A diagram showing the arrangement of the main TPC elements in the \dword{spmod}.  Two rows of \dwords{cpa} are interleaved with three rows of \dwords{apa}.  The \dword{fc} structure (only partially depicted to enable visibility of other elements) surrounds the outer area of the \dword{apa} and \dword{cpa} rows.  Elements of the \dword{pds} are integrated within the \dword{apa} structure.}
\includegraphics[width=0.6\linewidth]{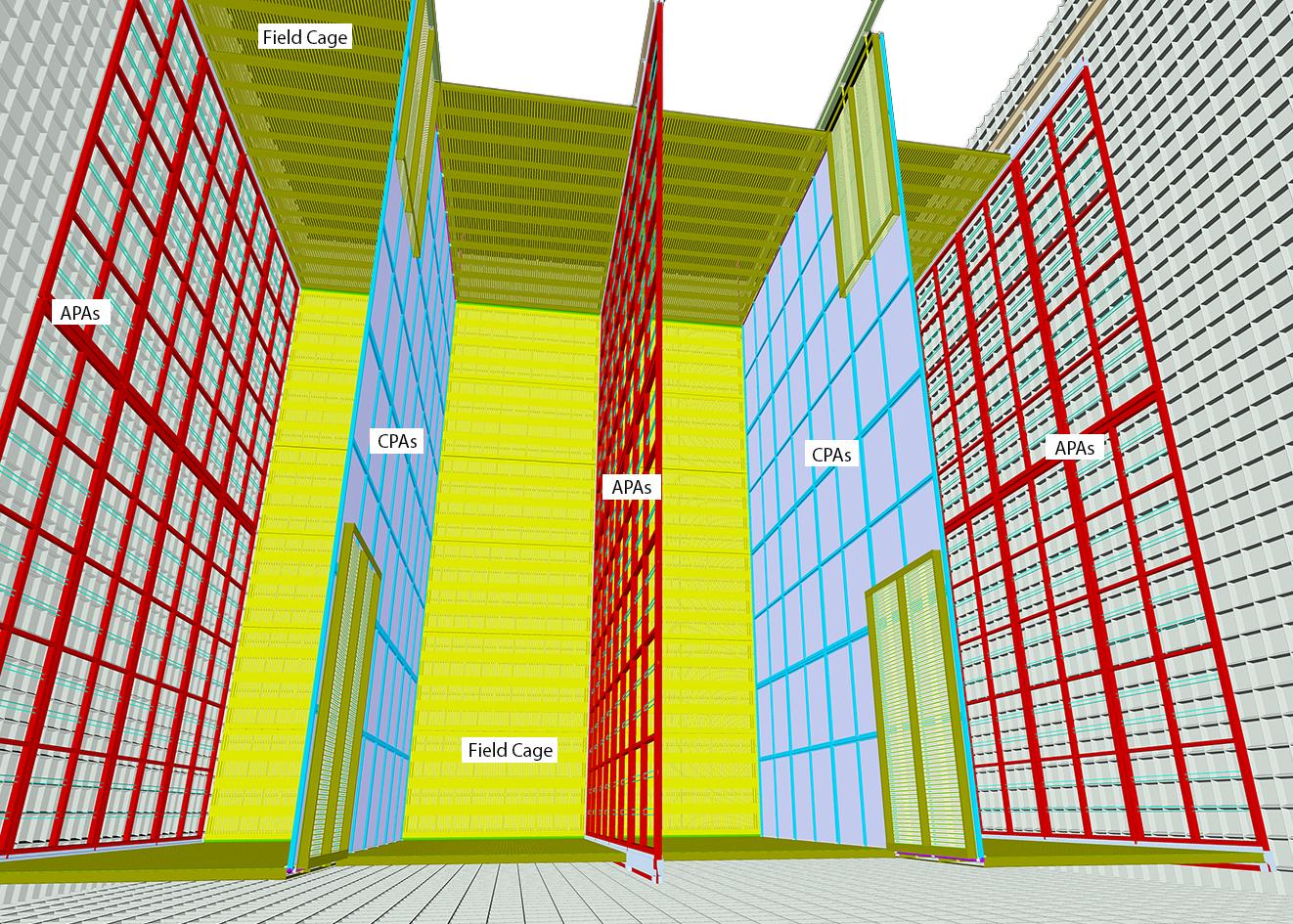}
\end{dunefigure}

\section{Detector Systems}
\label{sec:fdsp-ov-sys}

 

Table~\ref{tab:tpc-systems} lists the principal detection systems of the \dword{spmod} along with the primary purpose of each system.  In this section, the primary detector systems are introduced briefly.  The subsequent chapters of this document provide extensive descriptions of each of these systems. 

\begin{dunetable}[\dword{spmod} systems]{lll}{tab:tpc-systems}{\dword{spmod} systems.}
System & Name  & Purpose   \\  \toprowrule
\hyperref[ch:fdsp-apa]{\dshort{apa}}  & anode plane assemblies & ionization signal development \\ \colhline
\hyperref[ch:fdsp-hv]{\dshort{hv}} & high voltage & establish uniform drift field \\ \colhline
\hyperref[ch:fdsp-tpc-elec]{\dshort{ce}} & cold electronics & process \dword{apa} signals  \\ \colhline
\hyperref[ch:fdsp-pd]{\dshort{pd}} & photon detection & light collection and triggering\\ \colhline
\hyperref[ch:fdsp-daq]{\dshort{daq}} & data acquisition & record and handle digital data \\ \colhline
\hyperref[ch:fdsp-slow-cryo]{\dshort{cisc}} & cryogenics instrumentation and slow controls & maintain and monitor \lar volume\\ 
\end{dunetable}


\subsection{Anode Plane Assemblies}
\label{sec:fdsp-ov-apa}

The \dword{apa} system, described in full detail in Chapter~\ref{ch:fdsp-apa}, is used to capture the signals created by ionization drifting in the TPC volume.  Each \dword{apa} features a metal frame, on each side of which there are three instrumented and two uninstrumented anode layers.  The design of the anode layers is arranged to provide three complementary views of the ionization present in the TPC that can be combined to form \threed representations of the distribution of the charge. 

Among the novel features of the \single \lartpc is the presence of wrapped anode wires that follow a helical trajectory around the height of the \dword{apa}.  This design choice was made to minimize the need to tile electronic readout around the perimeter of the \dword{apa}, which would lead to dead space between neighboring \dwords{apa}.  This choice also was driven by reconstruction performance, with the angle of the wrap chosen such that a given induction plane wire does not intersect a given collection plane wire more than once, which greatly reduces pathologies in pattern recognition.

\subsection{TPC Electronics}
\label{sec:fdsp-ov-elec}

The electronics system, described in full detail in Chapter~\ref{ch:fdsp-tpc-elec}, is responsible for manipulating the signals present on the \dword{apa} wires and ultimately transferring them out of the cryostat and on to the \dword{daq} system.  
Several stages of signal processing occur within the cryostat, including \dword{fe} amplification and pulse shaping, analog-to-digital conversion, and control and communication functions.

\subsection{CPA, Field Cage and High Voltage}
\label{sec:fdsp-ov-hv}

The \dword{hv} system, described in full detail in Chapter~\ref{ch:fdsp-hv}, creates the uniform electric field in the TPC volume that causes ionization to drift towards the \dwords{apa}.  The \dword{hv} system contains both the \dwords{cpa}, which are operated at a 
voltage of \SI{-180}{kV}, as well as the \dword{fc} elements which progressively step the \dword{cpa} voltage down in magnitude.  

A novel feature of the 
\dword{hv} system is the use of resistive panels for the \dwords{cpa}, which serves to control the flow of stored energy in this system in the event of an unexpected electrical discharge.  This feature provides protection to the \dword{spmod} elements and guards against damage that would negatively impact detector performance.

\subsection{Photon Detection}
\label{sec:fdsp-ov-pds}

The \dword{pds}, described in full detail in Chapter~\ref{ch:fdsp-pd}, is used to capture scintillation light produced by interactions in the TPC.  The scintillation light of argon is very deep in the ultraviolet, so the \dword{pd} elements are designed to shift this wavelength closer to the visible spectrum where the \dword{spmod} has high efficiency.   
The \dword{pds} light detectors are geometrically arranged as approximately \SI{15}{cm} wide vertical strips mounted in the \dwords{apa}, with ten strips per \dword{apa}.  The light collection implementation continues to be optimized.  Electronic signals are generated via \dwords{sipm} immersed in the \lar and passed on to readout modules outside the cryostat.

\subsection{Data Acquisition}
\label{sec:fdsp-ov-daq}

The \dword{daq} system is described in full detail in Chapter~\ref{ch:fdsp-daq}.
DUNE physics requires that the \dword{daq} system record \dword{apa} and \dword{pds} signals with high efficiency both from relatively high-energy (>\SI{100}{MeV}) single interactions from beam and atmospheric neutrinos, interactions from proton decay (that are localized in both space and time), and from multiple low-energy (<\SI{100}{MeV}) interactions distributed throughout the detector over tens of seconds from \dwords{snb}.
\subsection{Cryogenic Instrumentation and Slow Controls}
\label{sec:fdsp-ov-instr}

The \dword{cisc} system, described in full detail in Chapter~\ref{ch:fdsp-slow-cryo}.
This system provides comprehensive monitoring for all \dword{detmodule} components as well as for the \lar quality and behavior.  Beyond passive monitoring, \dword{cisc} also provides a control system for some of the detector components.

\section{Technical Coordination}
\label{sec:fdsp-ov-tc}
Chapter~\ref{ch:fdsp-coord} provides an overview of detector integration and installation functions. These include project support, integration, infrastructure, the DUNE \dword{itf}, and installation.

\cleardoublepage

\chapter{Anode Plane Assemblies}
\label{ch:fdsp-apa}

\section{Anode Plane Assembly (APA) Overview}
\label{sec:fdsp-apa-intro}

Anode planes (or wire planes) are the DUNE \dword{spmod} elements used to sense, through both signal induction and direct collection, the ionization electrons created when charged particles traverse the \lar volume inside the \dword{spmod}. To facilitate fabrication and installation underground, the anode design is modular, with \dwords{apa} tiled together to form the readout system for a \SI{10}{kt} \dword{detmodule}. A single \dword{apa} is \SI{6}{m} high by \SI{2.3}{m} wide, but two of them are connected vertically, and twenty-five of these vertical stacks are linked together to define a \SI{12}{m} tall by \SI{58}{m} long mostly-active readout plane.  As described below, the planes are active on both sides, so three such wire readout planes are interleaved with two high voltage surfaces to define four \SI{3.6}{m} wide drift regions inside each \dword{spmod}, as shown in the detector schematic views in Figure~\ref{fig:FarDet-interior}.  Each single-phase \SI{10}{kt} module, therefore, will contain 150 \dwords{apa}.

\begin{dunefigure}[Schematic view of a DUNE \SI{10}{kt} \dword{spmod} 
]{fig:FarDet-interior}
{Left: End-on schematic view of the active argon volume showing the four drift regions and anode-cathode plane ordering of the TPC inside the detector. Right: View of the partially installed DUNE TPC inside the membrane cryostat. The \dwords{apa} are shown in red, \dwords{cpa} are in cyan, \dword{fc} modules in yellow/green.  Some of the \dword{fc} modules are in their folded position against the cathode to providing aisle access during installation.}
\setlength{\fboxsep}{0pt}
\setlength{\fboxrule}{0.5pt}
\includegraphics[width=0.445\textwidth, trim=0mm 3.7mm 0mm 0mm, clip]{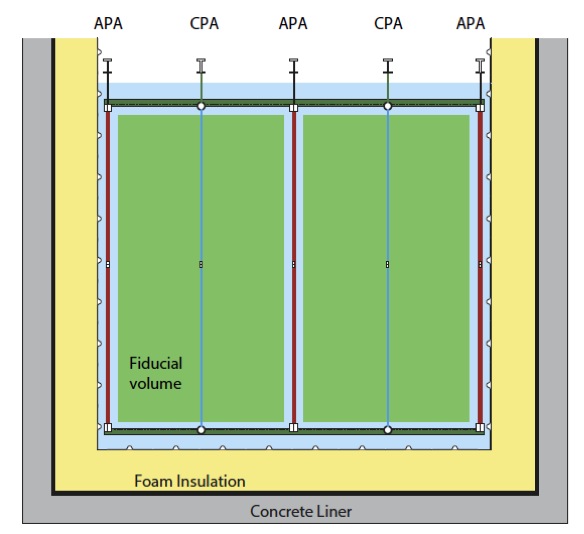}\hspace{0.01\textwidth}
\fbox{\includegraphics[width=0.52\textwidth]{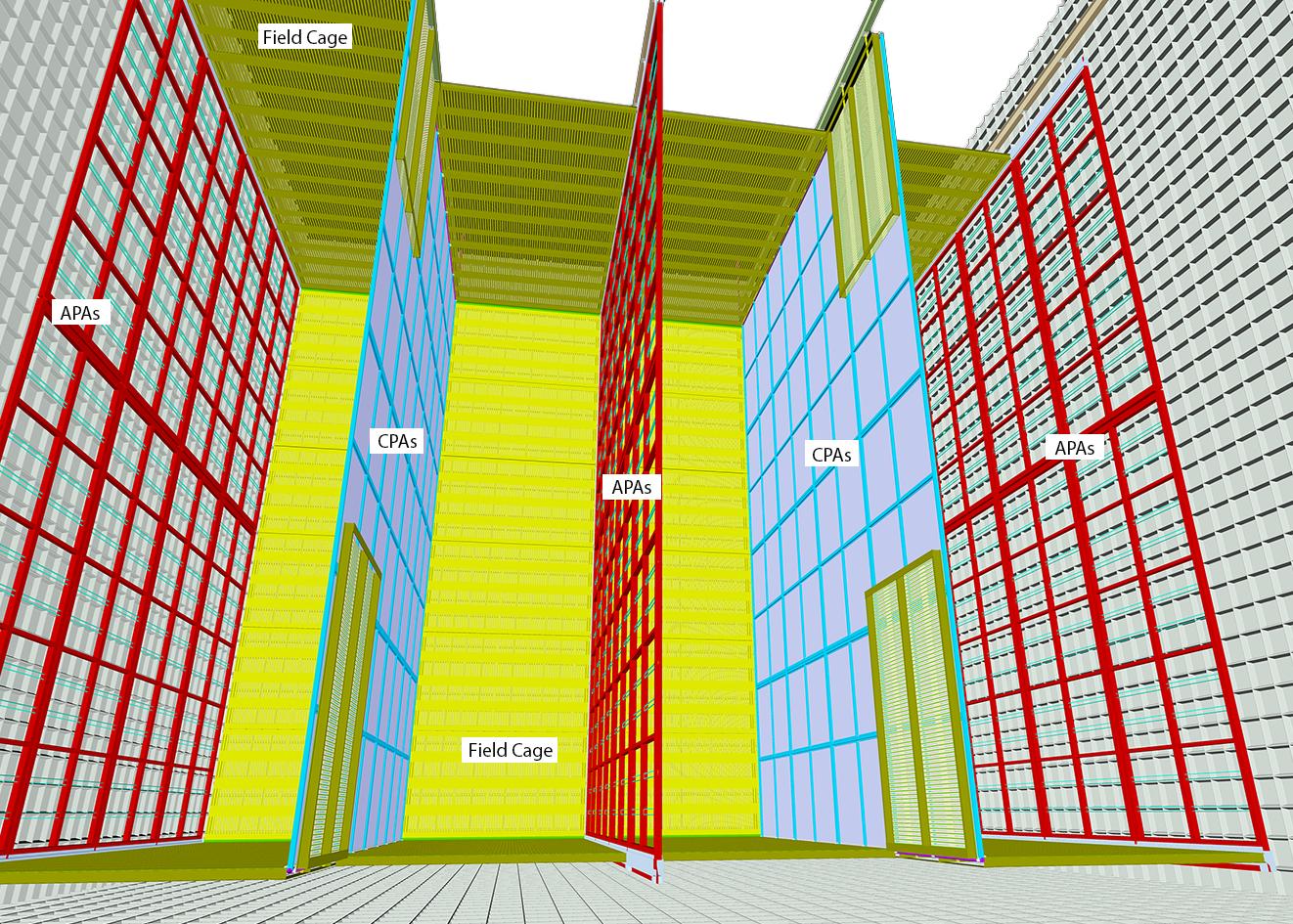}}
\end{dunefigure}

Each \dword{apa} frame is covered by over \num{2500} sense wires laid in three planes  oriented at angles to each other: a vertical collection plane, $X$, and two induction planes at $\pm35.7^\circ$ to the vertical, $U$ and $V$. These enable multi-dimensional reconstruction of particle tracks.  An additional \num{960} wires that are not read out make up an outer shielding plane, $G$, to improve signal shapes on the $U$ induction channels.  The angled wires are wrapped around the frame from one side to the other, allowing all channels to be read out from one end of the \dword{apa} only (the top or bottom), and thereby minimizing the dead regions between neighboring \dwords{apa}. Signals induced or collected on the wires are transferred through soldered connections to wire termination boards mounted at the end of the \dword{apa} frame that in turn connect to \dword{fe} readout electronics sitting in the \lar.  Figures~\ref{fig:tpc_apa1} and \ref{fig:tpc_apa2} illustrate the layout of the wires on an \dword{apa}, showing how they wrap around the frame and terminate on wire boards at the head end where readout electronics are mounted.

The \dwords{apa} represent a critical interface point between the various detector subsystems within the \dword{spmod}.  As already mentioned, the TPC readout electronics mount directly to the \dword{apa} frames.  The \dwords{pd} for detecting scintillation light produced in the \lar are also housed inside the frames, sandwiched between the wires on the two sides, requiring careful coordination in the frame design as well as placing a requirement on the transparency of the \dword{apa} structures.  In addition, the electric \dfirst{fc} panels connect directly to the edges of the \dword{apa} frames.  Finally, the \dwords{apa} must support the routing of cables for both the TPC electronics and the photon detector systems. All of these considerations have important impacts on the design, fabrication, and installation planning for the \dwords{apa}.   

\begin{dunefigure}[Illustration of the \dword{apa} wire layout]{fig:tpc_apa1}
{Illustration of the DUNE \dword{apa} wire wrapping scheme showing small portions of the wires from the three signal planes ($U,V,X$). The fourth wire plane ($G$) above these three, and parallel to $X$, is present to improve the pulse shape on the $U$ plane signals. The TPC electronics boxes, shown in blue on the right, mount directly to the frame and process signals from both the collection and induction channels. The \dword{apa} is shown turned on its side in a horizontal orientation.} 
\includegraphics[width=\textwidth]{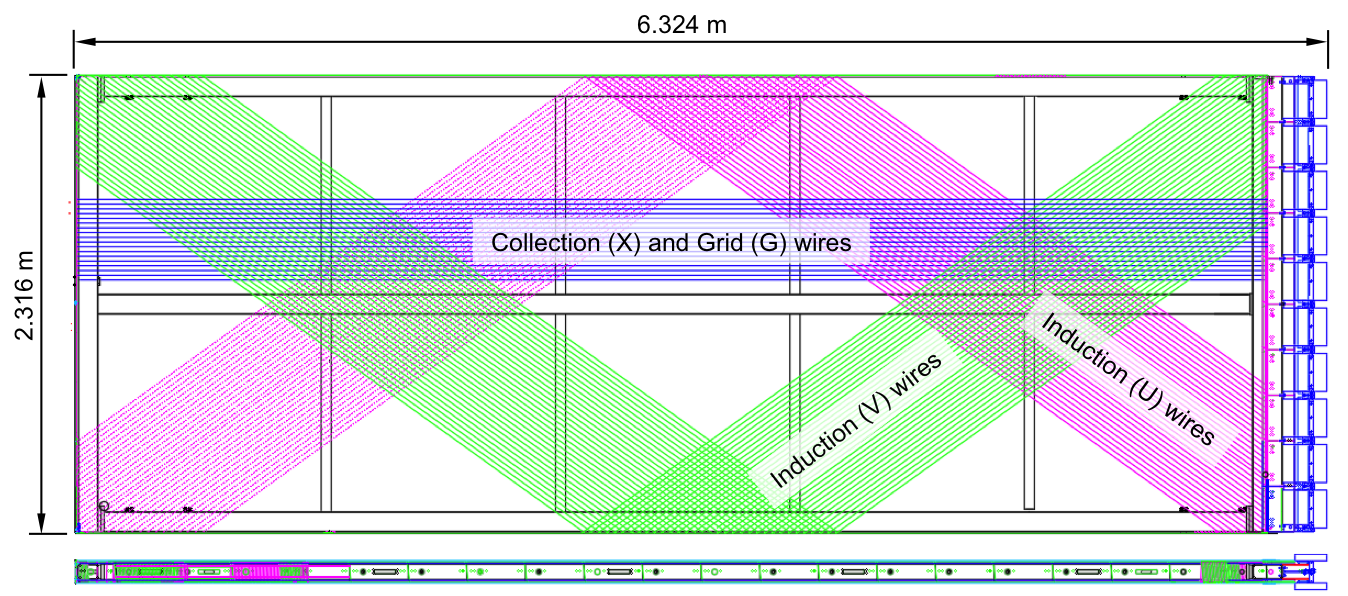} 
\end{dunefigure} 

\begin{dunefigure}[Cross section view of the head end and wire layers of an \dword{apa}]{fig:tpc_apa2}
{Cross section view of an \dword{apa} frame near the head end showing the layers of wires ($X,V,U,G$ inside to out) on both sides of the frame and terminating on wire boards at the head end of the frame, which connect directly to TPC readout electronics.} 
\includegraphics[width=0.95\textwidth]{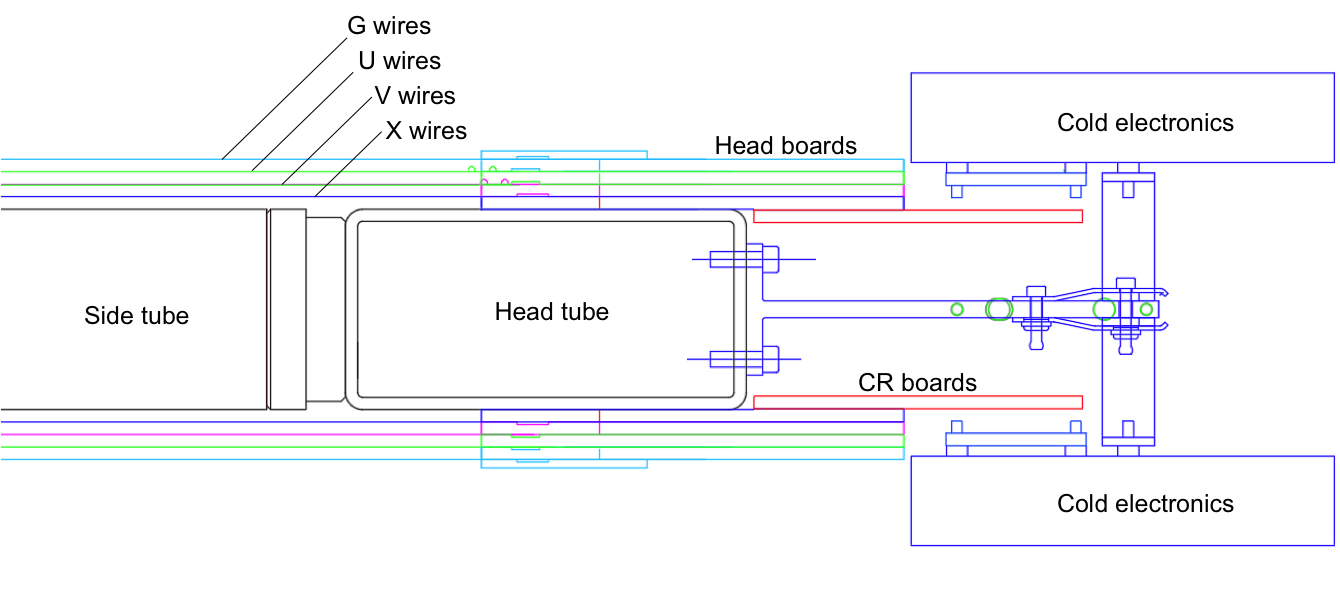} 
\end{dunefigure} 

Full-scale \dwords{apa} have recently been produced at the Physical Sciences Laboratory (PSL) at the University of Wisconsin and at the Daresbury Laboratory in the UK for the \dword{pdsp} project at CERN. Figure~\ref{fig:apa-photo} shows a completed \dword{apa} produced at PSL just before shipment to CERN for use in \dword{pdsp}. This effort has greatly informed the design and production planning for the DUNE \dwords{detmodule}, and future \dword{pdsp} running is expected to provide valuable validation information for many fundamental aspects of the  \dword{apa} design. 

The design, construction, testing, and installation of the \dwords{apa} is overseen by the \dword{apa} consortium within the DUNE collaboration. Multiple \dword{apa} production sites will be set up in the USA and the UK, with each nation producing approximately half of the \dwords{apa} needed for the 
\dwords{spmod}.  Factory setup is anticipated to begin in 2020, with \dword{apa} fabrication for the first \SI{10}{kt} far detector module running from 2021--2023.  

\begin{dunefigure}[Photo of a completed \dword{pdsp} \dword{apa}.]{fig:apa-photo}
{Completed \dword{pdsp} \dword{apa} ready for shipment to CERN.}
\setlength{\fboxsep}{0pt}
\setlength{\fboxrule}{0.5pt}
\fbox{\includegraphics[width=0.9\textwidth,trim=20mm 80mm 0mm 60mm,clip]{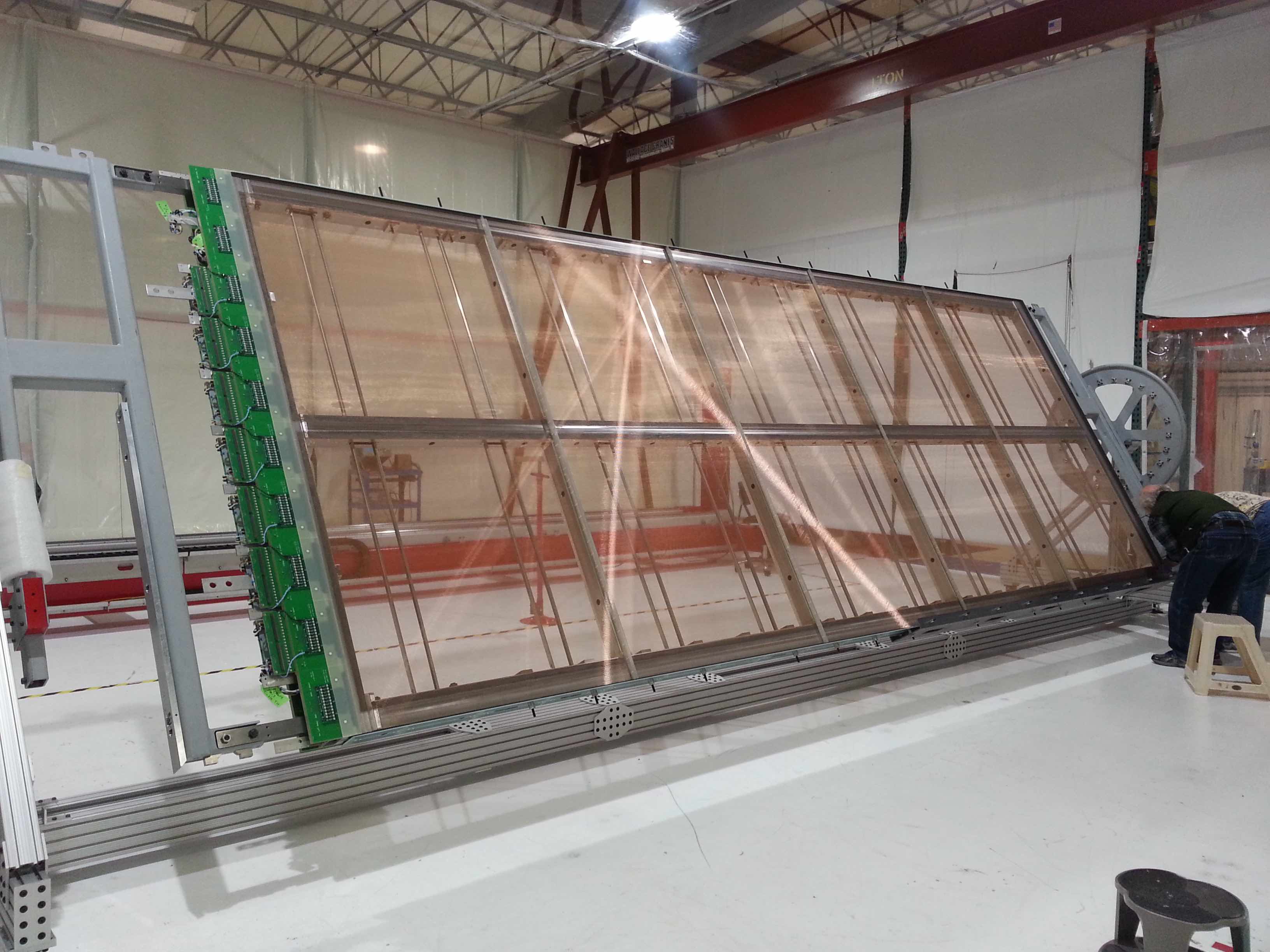}} 
\end{dunefigure}

\section{Design}
\label{sec:fdsp-apa-design}

The physics performance of the \dword{spmod} is a function of many intertwined detector parameters including argon purity, drift distance, electric field, wire pitch, wire length, and noise levels in the readout electronics.  Energy deposits from \dwords{mip}
originating anywhere inside the active volume of the detector should be identifiable with near \num{100}\% efficiency.  This requirement sets constraints on the \dword{apa} design, such as limits on the wire pitch, wire length, and choice of wire material.  This section details the current 
\dword{apa} design and discuss some areas where design enhancements are being considered based on experience from prototypes.  We begin with an overview of the key fundamental parameters of the \dwords{apa} and their connection to the physics requirements of the experiment.

\subsection{APA Overview and Key Design Parameters}
\label{sec:fdsp-apa-design-overview}

Each \dword{apa} is \SI{6}{m} high, \SI{2.3}{m} wide, and \SI{12}{cm} thick.  The underlying support frame is made from stainless steel hollow tube sections that are precisely machined and bolted together. A fine, conducting mesh covers the rectangular openings in the frame on both sides, to define a uniform electrical \dword{gp} behind the wires. The four layers of sense and shielding wires at varying angles relative to each other completely cover the frame. The wires are terminated on boards that anchor them and also provide the connection to the TPC readout electronics. Starting from the outermost wire layer, there is first an uninstrumented shielding plane (vertical, $G$), followed by two induction planes ($\pm 35.7^{\circ}$ to the vertical, $U,V$), and finally the collection plane (vertical, $X$). All wire layers span the entire height of the \dword{apa} frame. The two planes of induction wires wrap in a helical fashion around the long edges and over both sides of the \dword{apa}. The layout of the wire layers is illustrated above in Figures~\ref{fig:tpc_apa1} and \ref{fig:tpc_apa2}.  Below we summarize the key design parameters and the considerations driving the main design choices for the \dwords{apa}.  

\begin{itemize}
\item \textbf{\dword{apa} size.} The size of the \dwords{apa} is chosen for fabrication purposes, compatibility with over-the-road shipping, and for eventual transport to the \SI{4850}{ft}. level at SURF and installation into the membrane cryostat of a detector module. The dimensions are also chosen such that an integral number of electronic readout channels and boards fill in the full area of the \dword{apa}. 

\item \textbf{Detector active area.} \dwords{apa} should be sensitive over most of the full area of an \dword{apa} frame, with dead regions between \dwords{apa} due to frame elements, wire boards, electronics, or cabling kept to a minimum.  The wrapped style shown in Figure~\ref{fig:tpc_apa1} allows all channels to be read out at the top of the \dword{apa}, eliminating the dead space between \dwords{apa} that would otherwise be created by electronics and associated cabling. In addition, in the design of the \dword{spmod}, a central row of \dwords{apa} is flanked by drift-field regions on either side (see Figure~\ref{fig:FarDet-interior}), and the wrapped design allows the induction plane wires to sense drifted ionization that originates from either side of the \dword{apa}.  This double-sided feature is also effective for the \dwords{apa} located against the cryostat walls where there is a drift field on only one side, since the grid layer facing the wall effectively blocks any ionization generated outside the TPC from drifting in to the wires on that side of the \dword{apa}.        

\item \textbf{Wire angles.} The $X$ wires run vertical to provide optimal reconstruction of beam-induced particle tracks, which are predominantly forward (in the beam direction). The angle of the induction planes on the \dword{apa}, $\pm35.7^{\circ}$, was chosen to ensure that each induction wire only crosses a given collection wire once, reducing the ambiguities that the reconstruction must address.  Simulation studies (see next item) show that this configuration performs similarly to an optimal 45$^\circ$ wire angle for the primary DUNE physics channels.  The design angle of the induction wires, coupled with their pitch, was also chosen such that an integer multiple of electronics boards are needed to read out one \dword{apa}.

\item \textbf{Wire pitch.} The choice of wire pitch, \SI{4.7}{mm}, combined with key parameters for other TPC systems (described in their respective sections of the \dword{tdr}), can achieve the required performance for energy deposits by \dwords{mip} while providing good tracking resolution and good granularity for particle identification. The \single requirement that it be possible to determine the fiducial volume to \num{1}\% implies a vertex resolution of \SI{1.5}{cm} along each coordinate direction. The \SI{4.7}{mm} wire pitch achieves this for the $y$ and $z$ coordinates.  The resolution on $x$, the drift-coordinate, will be better than in the $y$--$z$ plane, due to the combination of drift-velocity and electronics sampling-rate.  Finally, as already mentioned, the total number of wires on an \dword{apa} should match the granularity of the electronics boards (each \dword{fe} motherboard can read out \num{128} wires, mixed between the $U,V,X$ planes). This determines the exact wire spacings of \SI{4.790}{mm} and \SI{4.669}{mm} on the collection and induction planes, respectively.  To achieve the reconstruction precision required (e.g., for $dE/dx$ reconstruction accuracy and multiple Coulomb scattering determination), the tolerance on the wire pitch is $\pm$\SI{0.5}{mm}.

In 2017, the DUNE Far Detector Task Force, utilizing a full \dword{fd} simulation and reconstruction chain, performed many detector optimization studies to quantify the impact of design choices, including wire pitch and wire angle, on DUNE physics performance.  The results indicated that a reduction in wire spacing (to \SI{3}{mm}) or a change in wire angle (to \num{45}$^\circ$) would not significantly impact the performance for the main physics goals of DUNE, including $\nu_\mu $ to $\nu_e$ oscillations and \dword{cpv} sensitivity.  Figure~\ref{fig:e-gamma} reproduces two plots from the Task Force report showing the impact of wire pitch and orientation on distinguishing electrons versus photons in the detector.  This is a key low-level metric for oscillation physics since photon induced showers can fake electron showers and create neutral-current generated backgrounds in the $\nu_e$ charged-current event sample. Two important handles for reducing this contamination are the energy density at the start of the shower and the visible gap between a photon shower and the vertex of the neutrino interaction due to the non-zero photon interaction length.  Figure~\ref{fig:e-gamma}(a) shows the reconstructed ionization energy loss density ($dE/dx$) in the first centimeters of electron and photon showers, illustrating the separation between the single \dword{mip} signal from electrons and the double \dword{mip} signal when photons pair-produce an $e^+e^-$.  The final electron signal selection efficiency is shown as a function of the background rejection rate for different wire configurations in Figure~\ref{fig:e-gamma}(b). At a signal efficiency of \num{90}\,\%, for example, the background rejection can be improved by about \num{1}\,\% using either \SI{3}{mm} spacing or 45$^\circ$ wire angles for the induction planes.  This slight improvement in background rejection with more dense hit information or more optimal wire angles is not surprising, but the impact on high-level physics sensitivities from these changes is very small. The conclusions of the Far Detector Task Force, therefore, are that the substantial cost impacts of making such changes to the \dword{spmod} design are not justified.    

\begin{dunefigure}[Electron-photon separation dependence on wire pitch and angle]{fig:e-gamma}
{Summary of electron--photon separation performance studies from the DUNE Far Detector Task Force. (a) $e$--$\gamma$ separation by $dE/dx$ for the nominal wire spacing and angle (\SI{4.7}{mm}/$37.5^\circ$) compared to \SI{3}{mm} spacing or 45$^\circ$ induction wire angles. (b) Electron signal selection efficiency versus photon (background) rejection for the different detector configurations. The \SI{3}{mm} wire pitch and 45$^\circ$ wire angle have similar impacts, so the 45$^\circ$ curve is partially obscured by the \SI{3}{mm} curve.}
(a)
\includegraphics[height=0.24\textheight]{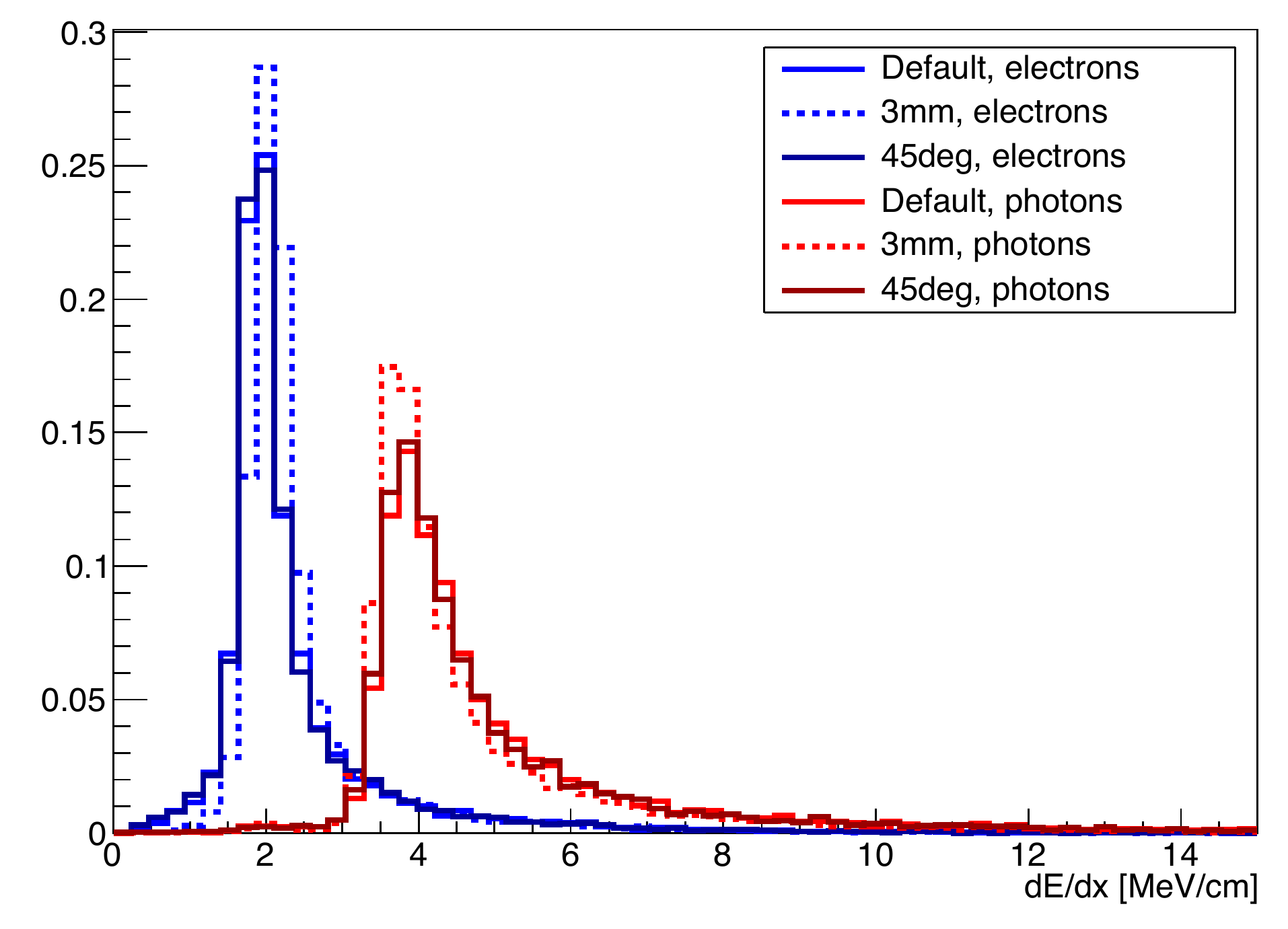} \qquad
(b)
\includegraphics[height=0.24\textheight]{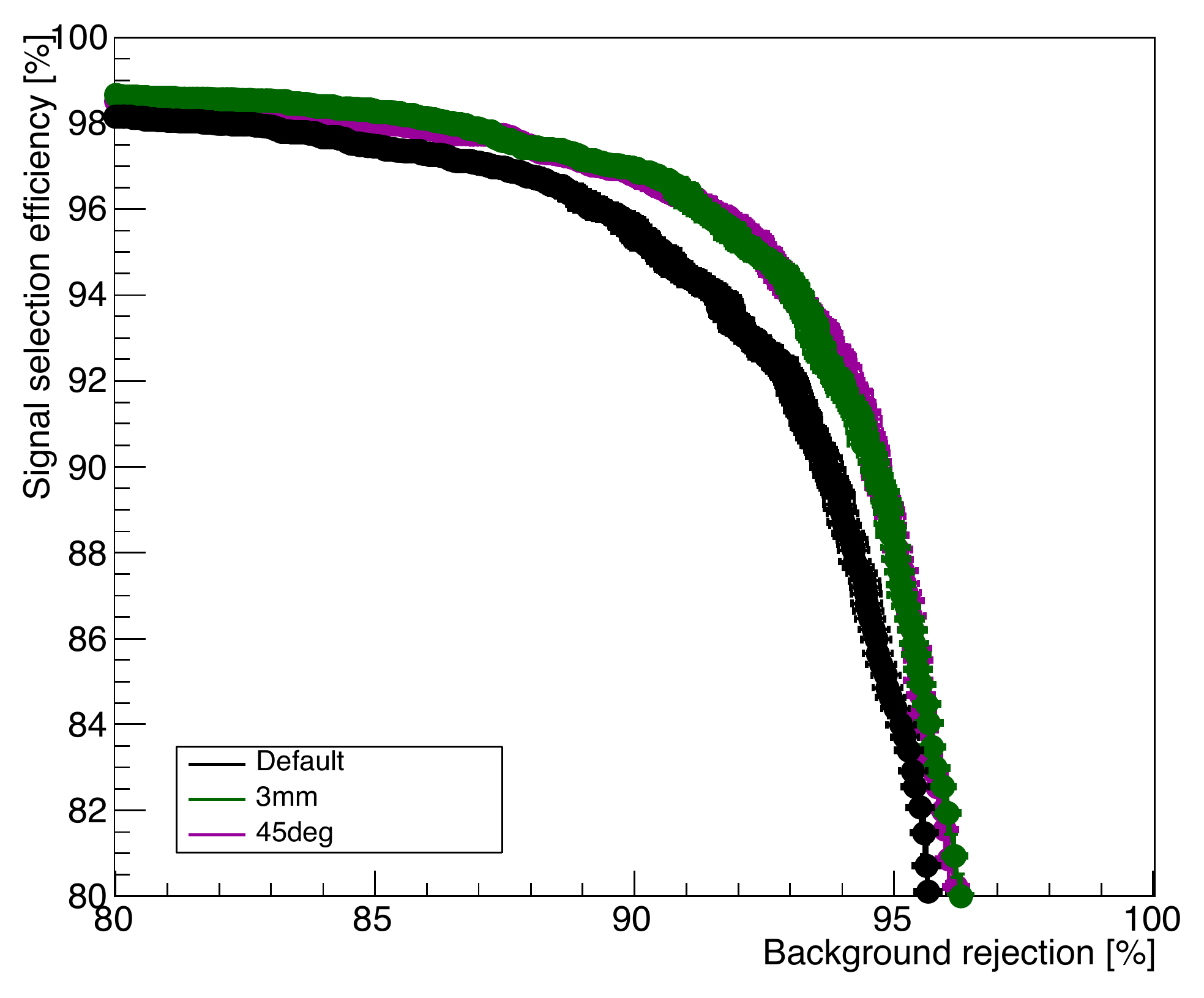} 
\end{dunefigure}

\item \textbf{Wire plane transparency and signal shapes.}  The ordering of the layers, from the outside in, is $G$-$U$-$V$-$X$, followed by the grounding mesh. The operating voltages of the \dword{apa} layers are listed in Table~\ref{tab:bias}.  Figure~\ref{fig:apa-fields} shows the field simulation and expected signal shapes for the bias voltages listed in the table.  When operated at these voltages, the drifting ionization follows trajectories around the grid and induction wires, ultimately terminating on a collection plane wire. The grid and induction layers are completely transparent to drifting ionization, and the collection plane is completely opaque.  The grid layer is present for pulse-shaping purposes and not connected to the electronics readout; it effectively shields the first induction plane from the drifting charge and removes the long leading edge from the signals on that layer.

\begin{table}[ht]
\begin{minipage}[b]{0.46\linewidth}
\centering
\begin{tabular}{ l  r }
    \hline
    \textbf{Anode Plane} & \textbf{Bias Voltage} \\ \toprowrule
	$G$ - Grid & \SI{-665}{V} \\ \colhline
	$U$ - Induction & \SI{-370}{V{}} \\ \colhline
	$V$ - Induction & \SI{0}{V} \\ \colhline
	$X$ - Collection & \SI{820}{V} \\ \colhline
	Grounding Mesh & \SI{0}{V} \\ \colhline
    \end{tabular}
    \caption{Baseline bias voltages for \dword{apa} wire layers.}
    \label{tab:bias}
\end{minipage}\hfill
\begin{minipage}[t]{0.5\linewidth}
\centering
\includegraphics[width=0.9\linewidth]{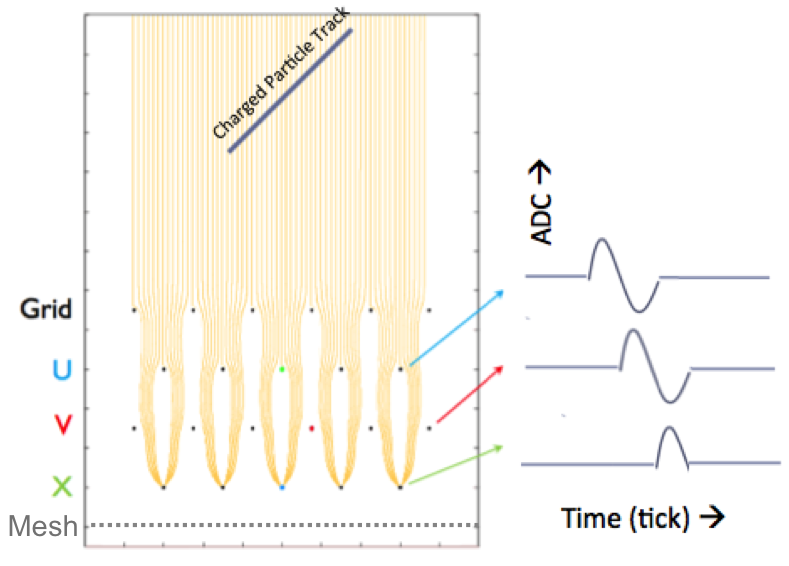}
\captionof{figure}{Field lines and resulting signal shapes on the APA induction and collection wires.}
\label{fig:apa-fields}
\end{minipage}
\end{table}

\item \textbf{Wire type and tension.}  The wire selected for the \dwords{apa} is \SI{150}{$\mu$m} beryllium (\num{1.9}\%) copper wire, 
chosen for its mechanical and electrical properties, ease of soldering, and cost.  The tension on the wires, combined with intermediate support combs (described in Section~\ref{sec:combs}) on the \dword{apa} frame cross beams, ensure that the wires are held taut in place with minimal sag.  Wire sag can impact the precision of reconstruction, as well as the transparency of the TPC wire planes.  The tension must be low enough that when the wires are cooled, which increases their tension due to thermal contraction, they stay safely below the break load of the beryllium copper wire.  To further mitigate wire slippage and its impact on detector performance, each wire in the \dword{apa} is anchored twice at all end points, with both solder and epoxy.  See Section~\ref{sec:fdsp-apa-wires} for more details about the wires.
\end{itemize}

Some of the principal design parameters for the DUNE 
\dwords{apa} are summarized in Table~\ref{tab:apaparameters}.

\begin{dunetable}[\dword{apa} design parameters]{lr}{tab:apaparameters}
{\dword{apa} design parameters}   
Parameter & Value  \\ \toprowrule
Active height & \SI{5.984}{m} \\ \colhline
Active width & \SI{2.300}{m} \\ \colhline
Wire pitch ($U,V$) & \SI{4.669}{mm} \\ \colhline
Wire pitch ($X,G$) & \SI{4.790}{mm} \\ \colhline
Wire pitch tolerance & $\pm$\SI{0.5}{mm} \\ \colhline
Wire plane spacing & \SI{4.75}{mm} \\ \colhline
Wire plane spacing tolerance & $\pm$\SI{0.5}{mm} \\ \colhline
Wire Angle (w.r.t. vertical) ($U,V$) & 35.7$^{\circ}$\\ \colhline
Wire Angle (w.r.t. vertical) ($X,G$) & 0$^{\circ}$\\ \colhline
Number of wires / \dword{apa} & 960 ($X$), 960 ($G$), 800 ($U$), 800 ($V$) \\ \colhline
Number of electronic channels / \dword{apa} & 2560 \\ \colhline
Wire material & beryllium copper \\ \colhline
Wire diameter & 150 $\mu$m \\ \colhline
\end{dunetable}

\subsection{APA Frames}
\label{sec:fdsp-apa-frames}

The \dword{apa} frames are an assembly of rectangular hollow section (RHS) stainless steel tubes.  As seen in Figure~\ref{fig:apa-frame-2}, there are three long tubes, a foot tube, a head tube, and eight cross-piece ribs that bolt together to create the \SI{6.0}{m} tall by \SI{2.3}{m} wide frame. All hollow sections are \SI{3}{in}. deep with varying widths depending on their role, see Figure~\ref{fig:apa-frame-2}.

\begin{dunefigure}[Bare \dword{apa} frame drawing]{fig:apa-frame-2}
{A \dword{pdsp} \dword{apa} frame showing overall dimensions and the \num{13} separate stainless steel tube sections that bolt together to form a complete frame.  The long tubes and foot tube are 3$\times$\SI{4}{in} cross section, the head tube is 3$\times$\SI{6}{in} and the ribs are \num{3}$\times$\SI{2}{in}. Also shown are the slots and guide rails used to house the light guide bar \dwords{pd} in \dword{pdsp}.}
\includegraphics[width=0.9\textwidth]{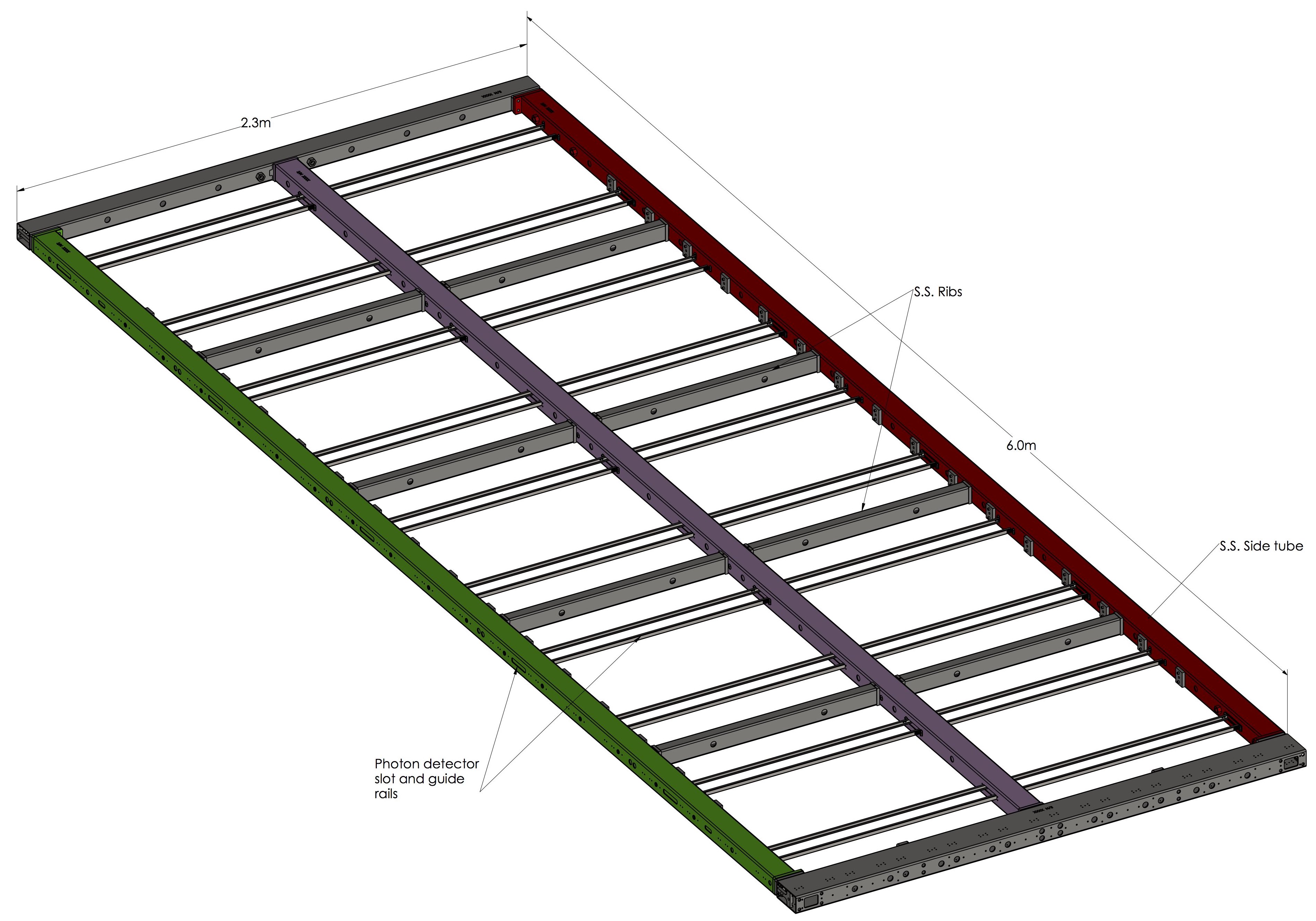} 
\end{dunefigure}

The head and foot tubes are bolted to the side and center pieces via abutment flanges welded to the tubes. In production, the pieces can be individually machined and cleaned prior to assembly, which gives flexibility both in the production process and helps achieve the flatness and shape tolerances.  During final assembly, shims are used to create a flat, rectangular frame of the specified dimensions.  The central cross pieces are attached to the side pieces in a similar manner.  Figure~\ref{fig:tpc_apa_boltedjointdrawing} shows models of the different joints.   

\begin{dunefigure}[Details of \dword{apa} bolted joints]{fig:tpc_apa_boltedjointdrawing}
{The bolted joints in the \dword{apa} frame. Left: Connection between the head tube and a side tube. Right: Connections between the center tube and the rib pieces on either side.  These bolted connections can be shimmed during assembly to ensure the frame meets dimensional and flatness specifications.}
\includegraphics[width=0.45\textwidth]{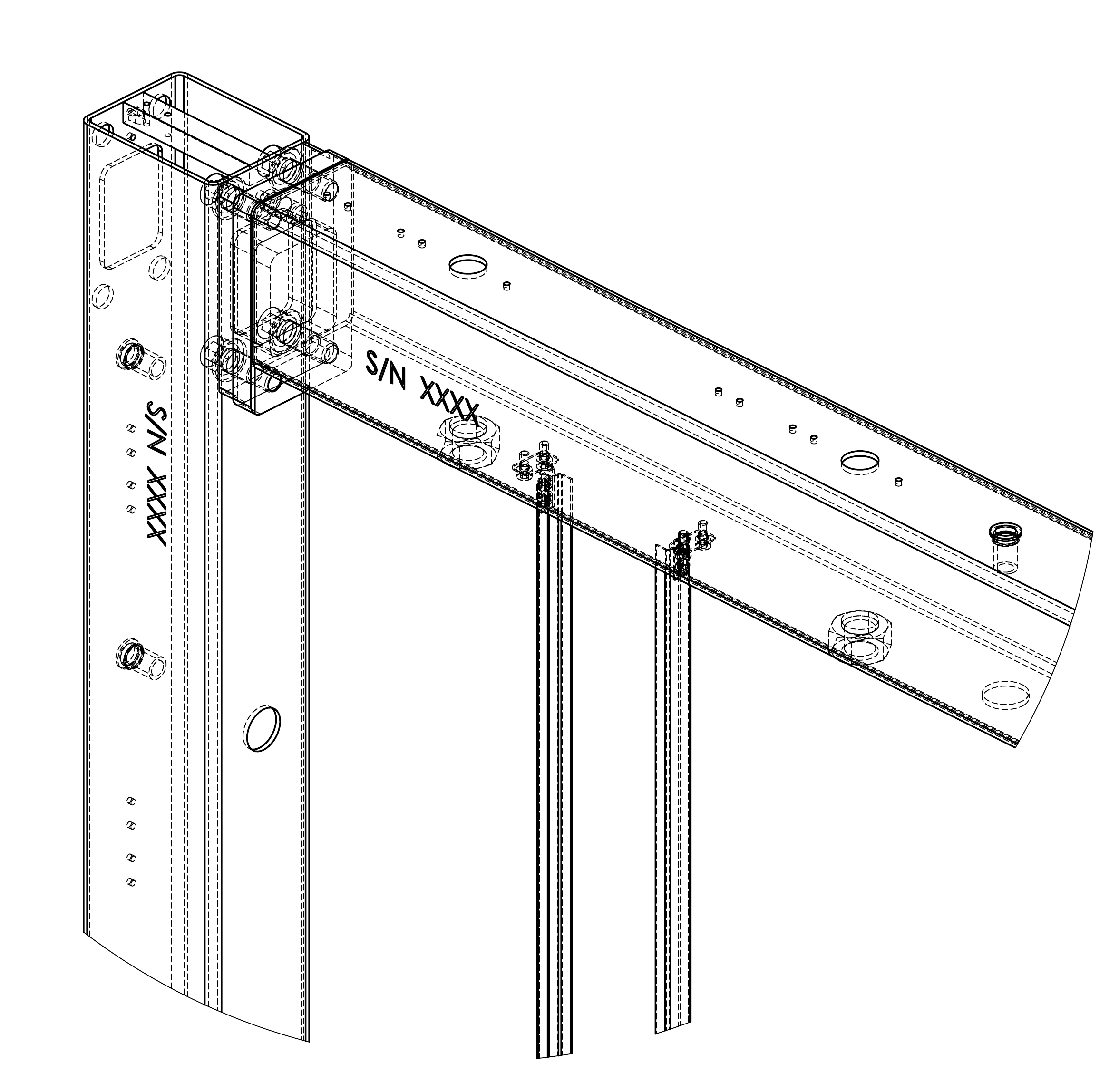} \quad
\includegraphics[width=0.45\textwidth]{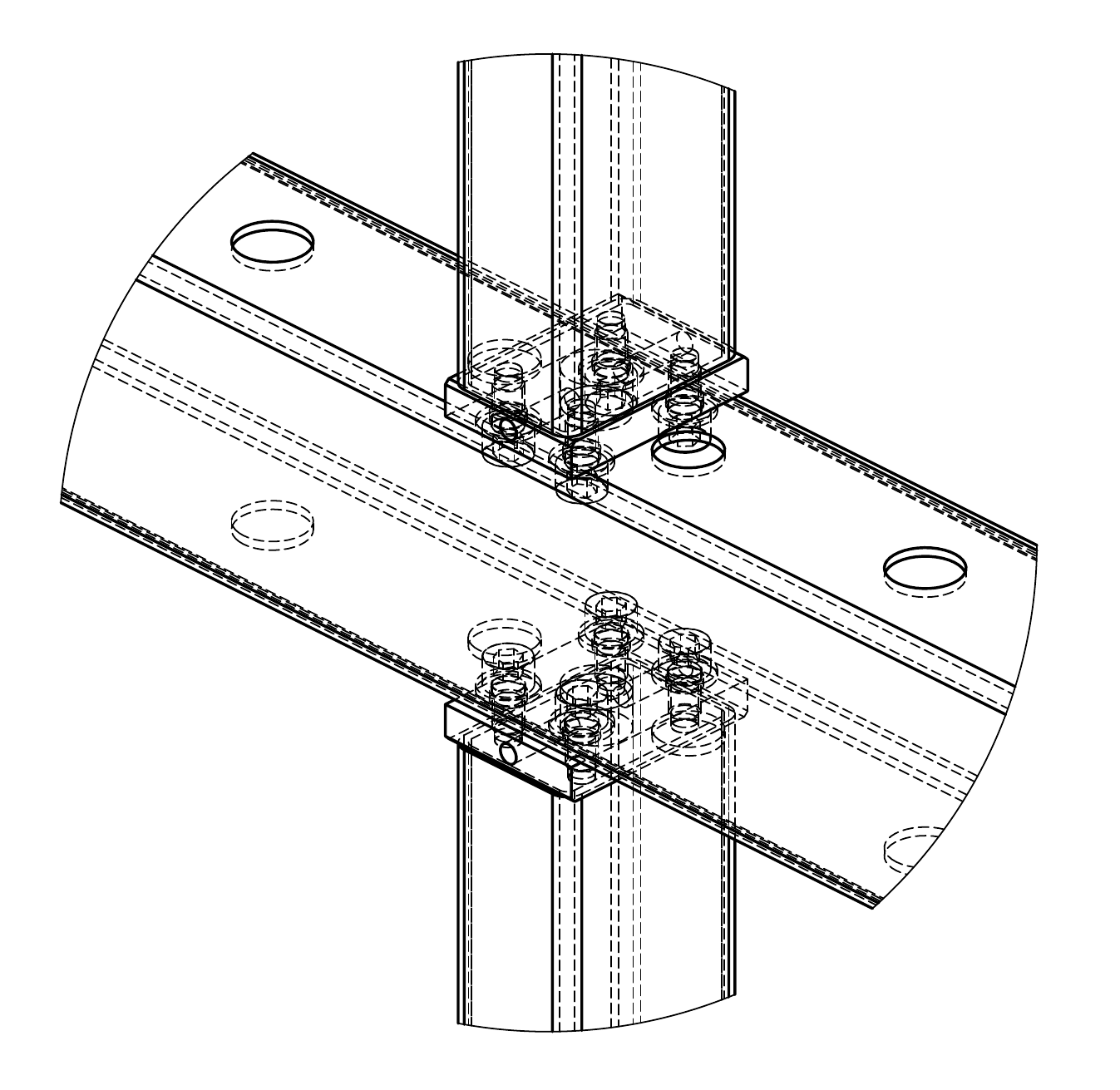} 
\end{dunefigure}

The \dword{apa} frames also house the \dword{pds}.  In the \dword{pdsp} design, rectangular slots are machined in the outer frame tubes and guide rails are used to secure a light guide bar detector, though alternative \dword{pd} designs are being considered for the \dwords{spmod}. 
(See Section~\ref{sec:fdsp-apa-intfc} for more details on interfacing with the \dword{pds}.)   Also, in a \dword{spmod}, pairs of \dword{apa} frames will be mechanically connected to form a \SI{12}{m} tall structure, as shown in Figure~\ref{fig:tpc_apa_dual}, with electronics for TPC readout located at both the top and bottom of this two-frame assembly and \dwords{pd} installed throughout.  The \dword{apa} frame design, therefore, must support the routing of cables to the top of the detector from both the bottom \dword{apa} readout electronics and the \dwords{pd} mounted throughout both \dwords{apa}.  The dimensions of the stainless steel tube sections used in the frame are currently being revisited from that used in \dword{pdsp} to ensure sufficient space is available to accommodate all detector cables.  See again Section~\ref{sec:fdsp-apa-intfc} on interfaces or Chapter~\ref{ch:fdsp-pd}.

\begin{dunefigure}[Dimensioned diagram of a pair of \dwords{apa} hanging vertically]{fig:tpc_apa_dual}{Diagram of an \dword{apa} pair, with bottom \dword{apa} hung from the top \dword{apa}. The dimensions of the \dword{apa} pair, including the accompanying \dword{ce} (\dword{ce}) and mechanical supports (the yoke), are indicated.}
\includegraphics[width=0.9\textwidth]{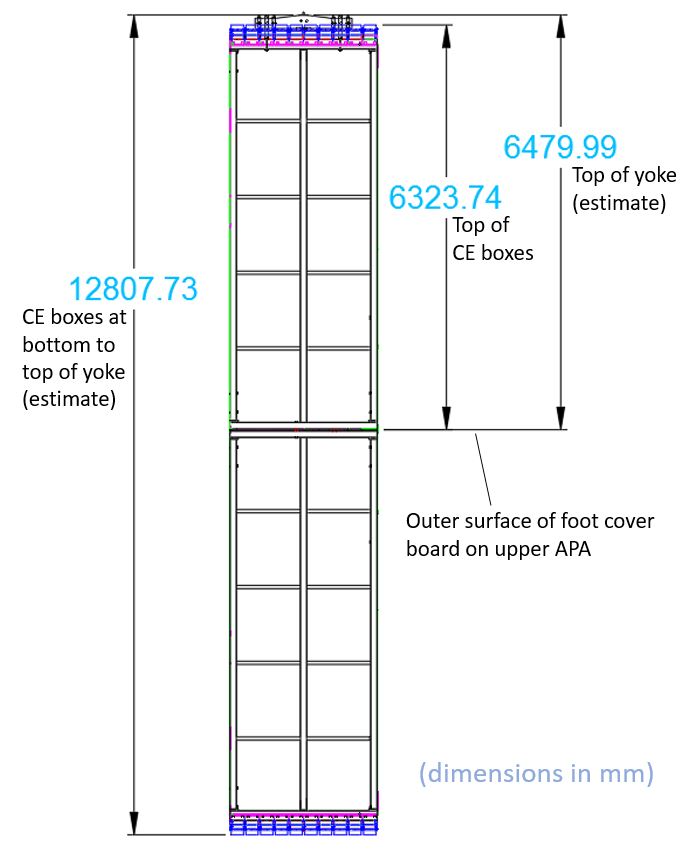} 
\end{dunefigure}

\subsection{Grounding Mesh}
\label{sec:fdsp-apa-mesh}

A fine mesh screen is mounted on both sides of the frames beneath the sense wires to provide a ground plane that evenly terminates the \efield and improves the uniformity of field lines around the wire planes.  This mesh also shields the wires to minimize any potential signal pickup from other detectors.  The mesh used is formed from a 
woven conducting wire (\SI{80}{$\mu$m} bronze) and is \num{85}\,\% transparent to allow scintillation photons to pass through to the \dwords{pd} mounted inside the frame. 

In the \dword{pdsp} \dwords{apa}, the mesh was installed in four parts, along the length of the left- and right-hand halves of each side of the \dword{apa}. The mesh was clamped around the perimeter of the opening and then pulled tight (by opening and closing clamps, as needed, during the process).  Once the mesh was taut, a \SI{25}{mm} wide strip was masked off around the opening and epoxy was applied through the mesh to attach it directly to the steel frame.  Although measurements have shown that this gives good electrical contact between the mesh and the frame, a deliberate electrical connection was also made.  Figure~\ref{fig:tpc-apa-mesh-application} depicts the mesh application setup for a full-size \dword{pdsp} \dword{apa}.

\begin{dunefigure}[Photos of \dword{apa} grounding mesh application in \dword{pdsp}]{fig:tpc-apa-mesh-application}
{Grounding mesh being clamped to the \dword{apa} and taped off, ready for gluing to a ProtoDUNE-SP frame.}
\setlength{\fboxsep}{0pt}
\setlength{\fboxrule}{0.5pt}
\fbox{\includegraphics[height=0.6\textwidth]{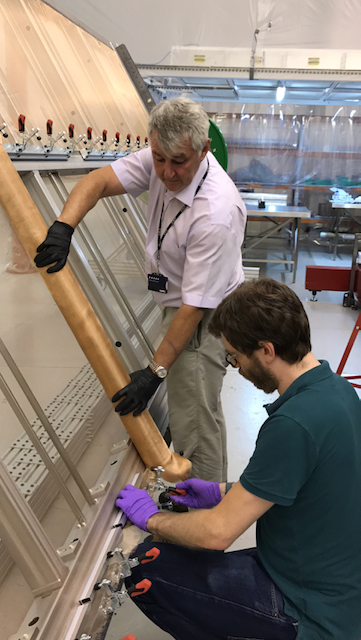}
} \quad
\fbox{\includegraphics[height=0.6\textwidth]{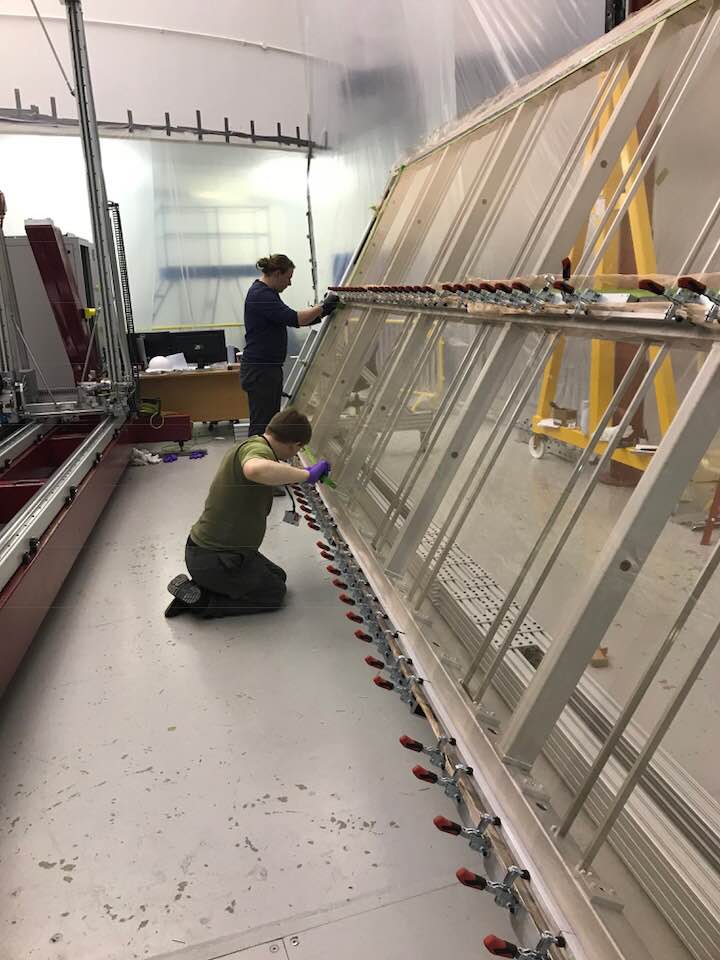}
}
\end{dunefigure}
\fixme{Replacing ProtoDUNE-SP in the caption with the dword generated an error. Don't know why. Anne}

The mesh installation procedure described above is difficult and prone to wrinkles remaining 
in the mesh that can affect the \efield uniformity and the transparency of the wire planes. For the DUNE mass production, a window-frame design is being considered, where mesh is pre-stretched over smaller sub-frames that can be clipped into each gap between cross beams in the full \dword{apa} frame.  See Section~\ref{sec:fdsp-apa-prod-assy} for more information.

\subsection{Wires}
\label{sec:fdsp-apa-wires}

The \SI{150}{$\mu$m} (\SI{.006}{in}) diameter beryllium copper (CuBe) wire chosen for use in the \dwords{apa} is known for its high durability and yield strength. It is composed of \num{98}\,\% copper, \num{1.9}\,\% beryllium, and a negligible amount of other elements. Each \dword{apa} contains a total of \SI{23.4}{km} of wire.  

The key properties for its use in the \dwords{apa} are low resistivity, high tensile or yield strength, and a coefficient of thermal expansion suitable for use with the \dword{apa}'s stainless steel frame (see Table~\ref{tab:wire} for a summary of properties).  Tensile strength of the wire describes the wire-breaking stress.  The yield strength is the stress at which the wire starts to take a permanent (inelastic) deformation, and is the important limit stress for this case.  The wire purchased from Little Falls Alloys~\footnote{Little Falls Alloys\texttrademark, \url{http://www.lfa-wire.com/}} for use on \dword{pdsp} had tensile strength over \SI{1380}{MPa} and yield strength more than \SI{1100}{MPa} (\SI{19.4}{N} for \SI{150}{$\mu$m} diameter wire).  The stress while in use is around \SI{280}{MPa} (\SI{5}{N}), leaving a comfortable margin.

The coefficient of thermal expansion (CTE) describes how a material expands or contracts with changes in temperature.  The CTEs of CuBe alloy and \num{304} stainless steel are very similar.  Integrated down to \SI{87}{K}, they are \SI{2.7}{mm/m} for stainless steel and \SI{2.9}{mm/m} for CuBe. Since the wire contracts slightly more than the frame, for a wire starting at \SI{5}{N} at room temperature, for example, the tension increases to around \SI{5.5}{N} when everything reaches \lar temperature.  

The change in wire tension during cool-down is also important to consider.  In the worst case, the wire cools quickly to \SI{87}{K} before any significant cooling of the much larger frame.  In the limiting case with complete contraction of the wire and none in the frame, the tension would peak around \SI{11.7}{N}, which is still well under the \SI{20}{N} yield tension. In practice, however, the cooling will be done gradually to avoid this tension spike as well as other thermal shocks to the detectors.

\begin{dunetable}[Beryllium copper (CuBe) wire properties]{lr}{tab:wire}{Summary of properties of the beryllium copper wire used on the \dwords{apa}.}
Parameter & Value \\ \toprowrule
Resistivity & 7.68 $\mu\Omega$-cm $@$ 20$^{\circ}$ C \\ \colhline
Resistance & 4.4 $\Omega$/m $@$ 20$^{\circ}$ C \\ \colhline
Tensile strength (from property sheets)  & \SI{1436}{MPa} / \SI{25.8}{N} for \SI{150}{$\mu$m} wire \\ \colhline
CTE of beryllium copper integrated to \SI{87}{K}  & \SI{2.9e-3}{m/m} \\ \colhline
CTE of stainless steel integrated to \SI{87}{K}  & \SI{2.7e-3}{m/m} \\
\end{dunetable}

\subsection{Wire Boards and Anchoring Elements}
\label{sec:fdsp-apa-boards}

To guide and secure the \num{3520} wires on an \dword{apa}, stacks of custom FR4 circuit boards attach to the outside edges of the frame, as shown in the engineering drawings in Figure~\ref{fig:apa-wire-boards}.  There are \num{204} \textit{wire boards} on each \dword{apa} and \num{337} total circuit boards, where this number includes the wire boards, cover boards, capacitive-resistive (CR) boards, $G$-layer bias boards, adapter boards, and one SHV board.

\begin{dunefigure}[Wire carrier board layout on the \dword{apa} frames]{fig:apa-wire-boards}
{Engineering drawings that illustrate the layering of the wire carrier boards that are secured along the perimeter of the \dword{apa} steel frames. Left: The full set of $V$-layer boards.  Right: Detail showing the full stack of four boards at the head end of the \dword{apa}.}
\includegraphics[width=0.48\textwidth]{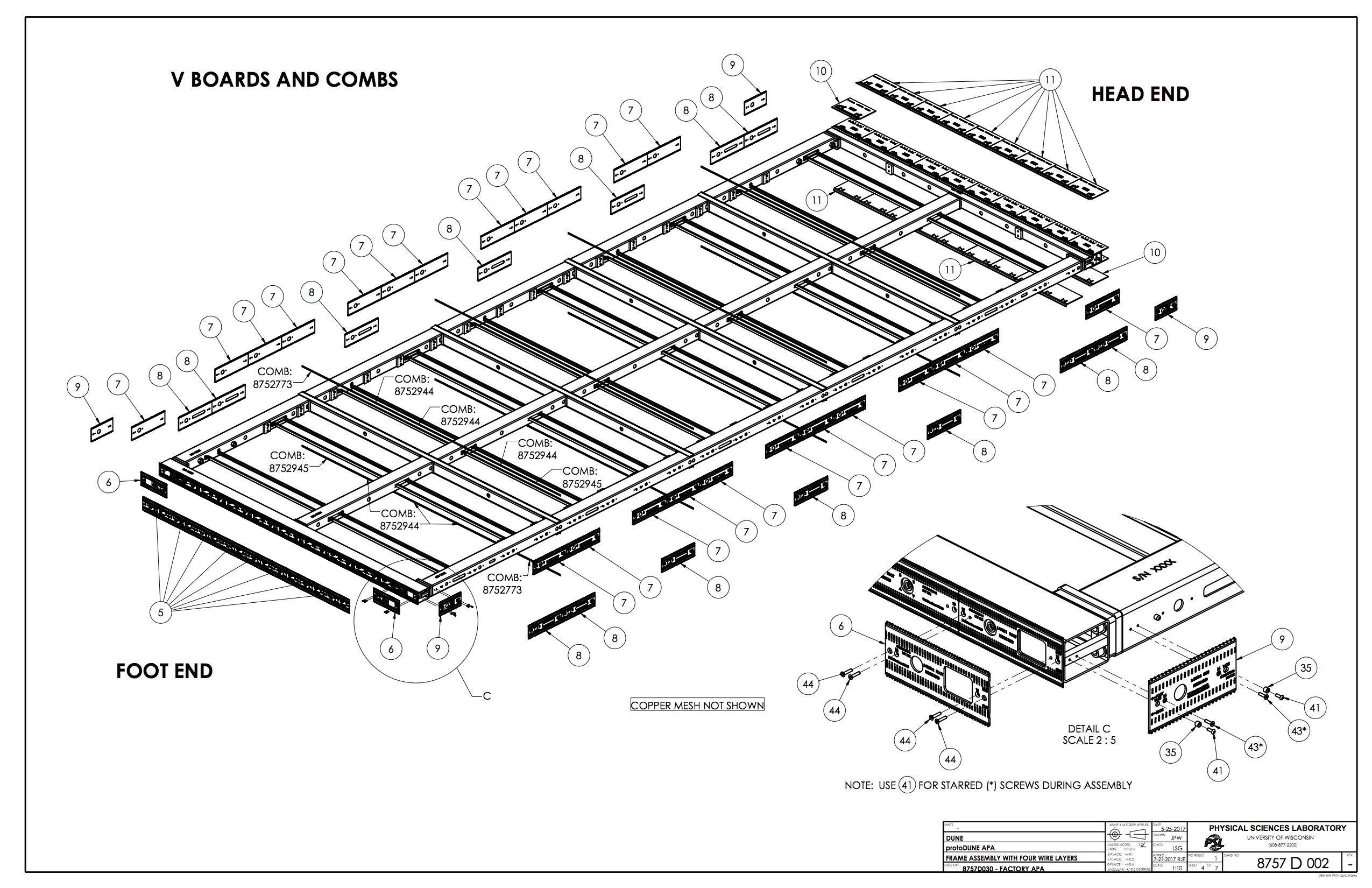}
\includegraphics[width=0.48\textwidth]{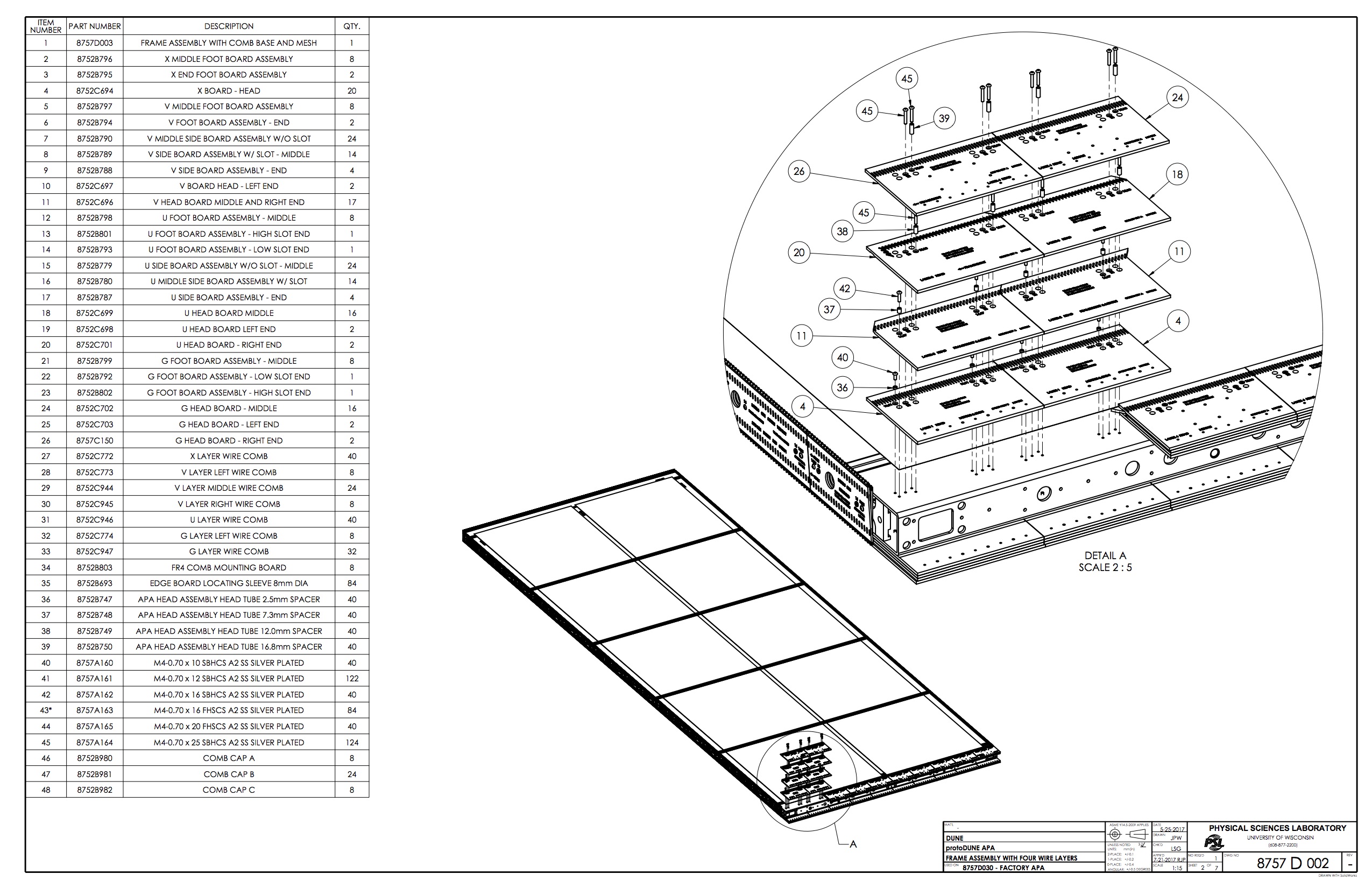}
\end{dunefigure}

\subsubsection{Head Electronics Boards}

All \dword{apa} wires are terminated on wire boards that are stacked along the electronics end of the \dword{apa} frame.  The board stack at the head end is shown in Figure~\ref{fig:apa-wire-boards}. Attachment of the wire boards begins with the $X$-plane (lowest). Once the $X$-plane wires are strung on both sides of the \dword{apa} frame, they are soldered and epoxied to their wire boards and trimmed. The remaining wire board layers are attached as each previous layer of wires are placed.  The wire plane spacing of \SI{4.75}{mm} is set by the thickness of these wire boards.   

Mill-Max~\footnote{Mill-Max\texttrademark{}, \url{https://www.mill-max.com/}} pins and sockets provide electrical connections between circuit boards within a stack. They are pressed into the circuit boards and are not repairable if damaged. To minimize the possibility of damaged pins, the boards are designed so that the first wire board attached to the frame has only sockets. All boards attached subsequently contain pins that plug into previously mounted boards. This process eliminates exposure of any pins to possible damage during winding, soldering, or trimming processes.

The $X$, $U$ and $V$ layers of wires are connected to the \dword{ce} (housed in boxes mounted on the \dword{apa}) either directly or through DC-blocking capacitors.  Ten stacks of wire boards are installed across the width of each side along the head of the \dword{apa}.  The $X$-layer board in each stack has room for \num{48} wires, the $V$-layer has 40 wires, the $U$-layer \num{40} wires and the $G$-layer \num{48} wires.  Each board stack, therefore, has 176 wires but only \num{128} signal channels since the $G$ wires are not read out. With a total of \num{20} stacks per \dword{apa}, this results in \SI{2560} signal channels per \dword{apa} and a total of \SI{3520} wires starting at the top of the \dword{apa} and ending at the bottom. Many of the capacitors and resistors that in principle could be on these wire boards are instead placed on the attached CR boards (see next section) to improve their accessibility in case of component failure. Figure~\ref{fig:tpc_apa_electronics_connectiondiagram} depicts the connections between the different elements of the \dword{apa} electrical circuit at the head end of the frame. 

\begin{dunefigure}[\dword{apa} wire board connection to electronics]{fig:tpc_apa_electronics_connectiondiagram}{The wire board stack at the head end of an \dword{apa} and the connection to the \dword{ce}. The set of wire boards within a stack can be seen on both sides of the \dword{apa}, with the CR board extending further to the right to provide a connection to the \dword{ce}.}
\includegraphics[width=0.60\textwidth, trim=0mm 0mm 5mm 0mm, clip]{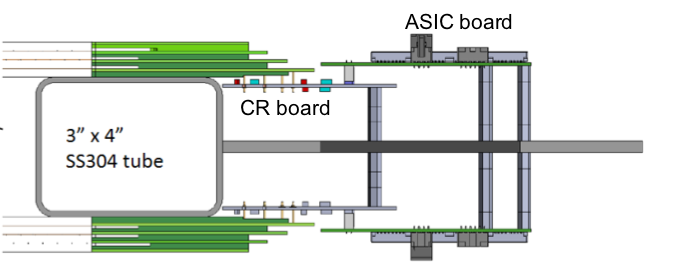}
\includegraphics[width=0.35\textwidth]{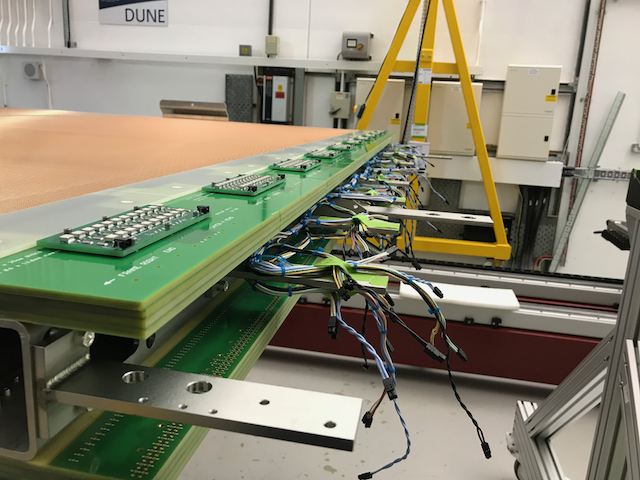}

\end{dunefigure}

\subsubsection{CR Boards}
\label{sec:crboards}

The capacitive-resistive (CR) boards carry a bias resistor and a DC-blocking capacitor for each wire in the $X$ and $U$-planes. These boards are attached to the board stacks after fabrication of all wire planes.   Electrical connections to the board stack are made though Mill-Max pins that plug into the wire boards. Connections from the CR boards to the \dword{ce} are made through a pair of \num{96}-pin Samtec~\footnote{Samtec\texttrademark \url{https://www.samtec.com/}} connectors.

Surface-mount bias resistors on the CR boards have resistance of \SI{50}{\mega\ohm} and are constructed with a thick film on a ceramic substrate. Rated for \SI{2.0}{kV} operation, the resistors measure \num{3.0}$\times$\SI{6.1}{mm} (\num{0.12}$\times$\SI{0.24}{in}). The selected DC-blocking capacitors have capacitance of \SI{3.9}{nF} and are rated for \SI{2.0}{kV} operation. Measuring \num{5.6}$\times$\SI{6.4}{mm} (\num{0.22}$\times$\SI{0.25}{in}) across and \SI{2.5}{mm} (\SI{0.10}{in}) high, the capacitors feature flexible terminals to comply with PC board expansion and contraction. They are designed to withstand \num{1000} thermal cycles between the extremes of the operating temperature range. Tolerance is also \num{5}\,\%.

In addition to the bias and DC-blocking capacitors for all $X$ and $U$-plane wires, the CR boards include two R-C filters for the bias voltages. The resistors are of the same type used for wire biasing except with a resistance of \SI{2}{M$\Omega$}. Wire plane bias filter capacitors are \SI{39}{nF}, consisting of ten \SI{3.9}{nF} surface-mount capacitors connected in parallel. They are the same capacitors as those used for DC blocking.

The selected capacitors were designed by the manufacturer to withstand repeated temperature excursions over a wide range. Their mechanically compliant terminal structure accommodates CTE mismatches. The resistors employ a thick-film technology that is also tolerant of wide temperature excursions.  Capacitors and resistors were qualified for \dword{pdsp} by subjecting samples to repeated testing at room temperature and at \num{-190}\,$^\circ$C.  Performance criteria were measured across five thermal cycles, and no measurable changes were observed. During the production of \num{140} CR boards, more than \num{10000} units of each component were tested at room temperature, at \lar temperature, and again at room temperature. No failures or measurable changes in performance were observed.

\subsubsection{Side and Foot Boards}

The boards along the sides and foot of the \dword{apa} have notches, pins, and other location features to hold the wires in the correct position as they wrap around the edge from one side of the \dword{apa} to the other.  

\begin{dunefigure}[Photos of \dword{apa} side boards showing traces that connect wires around openings]{fig:tpc_apa_sideboardmodel}
{Side boards with traces that connect wires around openings.  The wires are wound straight over the openings, then soldered to pads at the ends of the traces, then the wire sections between the pads are trimmed away.}
\setlength{\fboxsep}{0pt}
\setlength{\fboxrule}{0.5pt}
\fbox{\includegraphics[height=0.28\textheight]{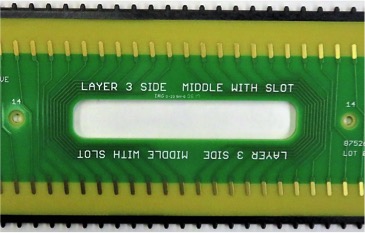}
}\quad
\fbox{\includegraphics[height=0.28\textheight,trim=0mm 0mm 0mm 25mm,clip]{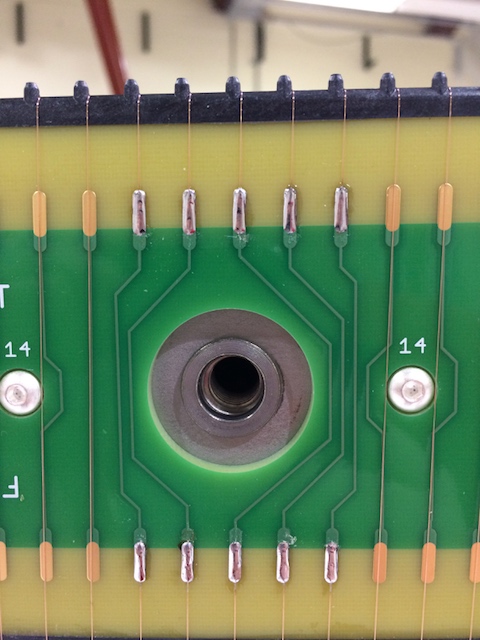}
}
\end{dunefigure}

A number of hole or slot features are needed in the edge boards to provide access to the underlying frame (see Figure~\ref{fig:tpc_apa_sideboardmodel} for examples).  In order that these openings not be covered by wires, the sections of wire that would go over the openings are replaced by traces on the boards.  After the wires are wrapped, the wires over the opening are soldered to pads at the ends of the traces and the section of wire between the pads is snipped out.  These traces are easily and economically added to the boards by the many commercial fabricators who make circuit boards. 

The placement of the angled wires are fixed by teeth that are part of an injected molded strip that is glued to the edge of the FR4 boards.  The polymer used for the strips is Vectra e130i (a trade name for 30$\%$ glass filled liquid crystal polymer, or LCP). It retains its strength at cryogenic temperature and has a CTE similar enough to FR4 that differential contraction is not a problem.  The wires make a partial wrap around the pin as they change direction from the face of the \dword{apa} to the edge.

\subsubsection{Support Combs}
\label{sec:combs}

Support combs are glued at four points along each side of the \dword{apa}, along the four cross beams. These combs maintain the wire and plane spacing along the length of the \dword{apa}. A dedicated jig is used to install the combs and provides the alignment and the pressure to allow the glue to dry. The glue used is the Gray epoxy \num{2216} described below. An eight-hour cure time is required after comb installation on each side of the \dword{apa} before the jig can be removed and production can continue.  Figure~\ref{fig:tpc_apa_sideboardphoto} shows a detail of the wire support combs on a \dword{pdsp} \dword{apa}.

\begin{dunefigure}[Photos of \dword{apa} side boards on the frame]{fig:tpc_apa_sideboardphoto}
{Left: \dword{apa} corner where end boards meet side boards.  The injection molded teeth that guide the $U$ and $V$ wires around the edge are visible at the bottom. Right: The wire support combs.}
\setlength{\fboxsep}{0pt}
\setlength{\fboxrule}{0.5pt}
\fbox{\includegraphics[height=0.3\textheight]{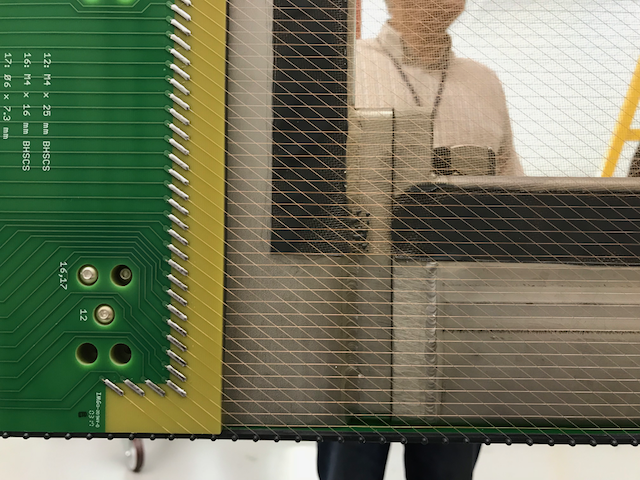}
}\quad
\fbox{\includegraphics[height=0.3\textheight]{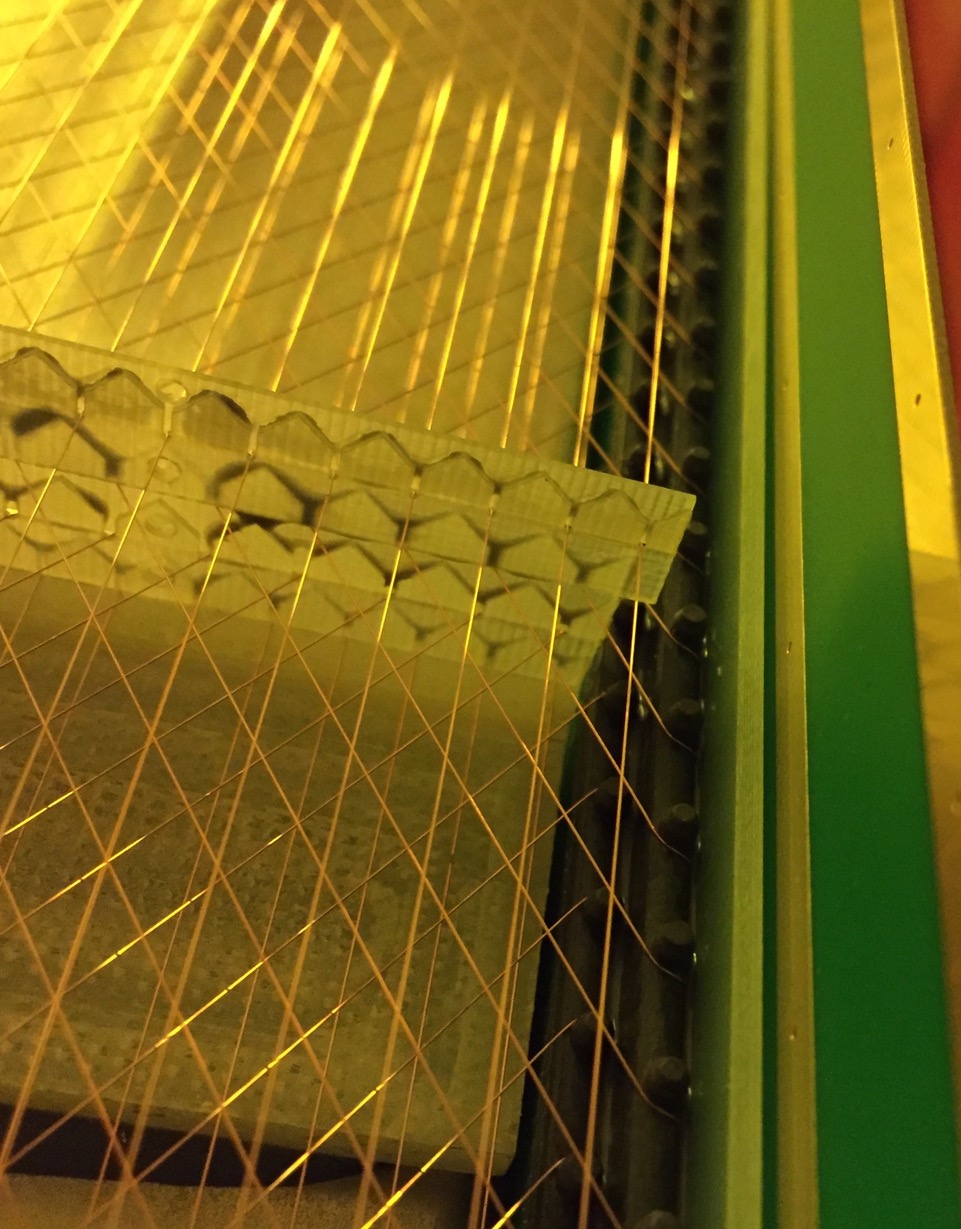}
}
\end{dunefigure}

\subsubsection{Solder and Epoxy}
\label{sec:glue-solder}

The ends of the wires are soldered to pads on the edges of the wire boards.  Solder provides both an electrical connection and a physical anchor to the wire pads. A 62$\%$ tin, 36$\%$ lead, and 2$\%$ silver solder was chosen.  A eutectic mix (63/37) is the best of the straight tin-lead 
 solders but the 2$\%$ added silver gives better creep resistance. 

Once a wire layer is complete, the next layer of boards is glued on, this glue providing an additional physical anchor. Gray epoxy \num{2216} by 3M~\footnote{3M\texttrademark \url{https://www.3m.com/}} is used for the glue.  It is strong, widely used (therefore much data is available), and it retains good properties at cryogenic temperatures.

\section{Interfaces} 
\label{sec:fdsp-apa-intfc}

The interface between the \dword{apa} consortium and other detector consortia, facilities, and working groups covers a wide range of activities. Table~\ref{tab:apa_interface_docdb} summarizes the interface control documents under development. In the following, we elaborate slightly on the interfaces with the TPC readout electronics and the \dword{pds}, as well as the connections between neighboring \dwords{apa} in the \dword{spmod} and cable routing.  Other important interfaces are to the TPC \dword{hv} system (the \dword{fc}) and the \dword{dss} inside the DUNE cryostats.  

\begin{dunetable}[\dword{apa} interface control documents]{lr}{tab:apa_interface_docdb}
{Summary of interface control documents being developed.}  
  Interface Document & DUNE doc-db number \\\colhline 
  Interface to TPC electronics & 6670 \\ \colhline 
  Interface to photon detector system & 6667 \\ \colhline
  Interface to drift high voltage system & 6673 \\ \colhline
  Interface to DAQ & 6676 \\ \colhline
  Interface to slow controls and cryogenics infrastructure & 6679 \\\specialrule{1.5pt}{1pt}{1pt}
  Integration facility interface & 7021 \\ \colhline
  Facility interfaces (Detector Hall, Cryostat, and Cryogenics) & 6967 \\ \colhline
  Installation interface & 6994 \\ \colhline
  Calibration interface & 7048 \\\specialrule{1.5pt}{1pt}{1pt}
  Software computing interface & 7102 \\ \colhline
  Physics interface & 7075 \\ \colhline
\end{dunetable}

\subsection{TPC Cold Electronics}
\label{sec:fdsp-apa-intfc-elec}

The TPC readout electronics is directly mounted to the \dword{apa} immersed in \lar in order to reduce the input capacitance and thus the inherent electronics noise.  With the wire-wrapped design, all \num{2560} wires to be read out (recall \num{960} are $G$-plane wires used for charge shielding only and so not read out) are terminated on wire boards that stack along one end (the head) of the \dword{apa} frame.  The \num{2560} channels are read out by \num{20} \dword{fe} motherboards (\num{128} channels per board), each of which includes eight \num{16}-channel \dword{fe} ASICs, eight \num{16}-channel ADC ASICs, low-voltage regulators, and input signal protection circuits.  A schematic view of the head end of an \dword{apa} with electronics installed and a cable tray mounted above is shown in Figure~\ref{fig:apa_ce}. 

\begin{dunefigure}[\dword{apa} interface with TPC electronics]{fig:apa_ce}
{The head region of an \dword{apa} frame showing the 10 wire board stacks on each side, \num{20} \dword{fe} motherboard boxes, and the cable tray mounted above.}
\includegraphics[width=0.85\textwidth,trim=10mm 0mm 10mm 20mm, clip]{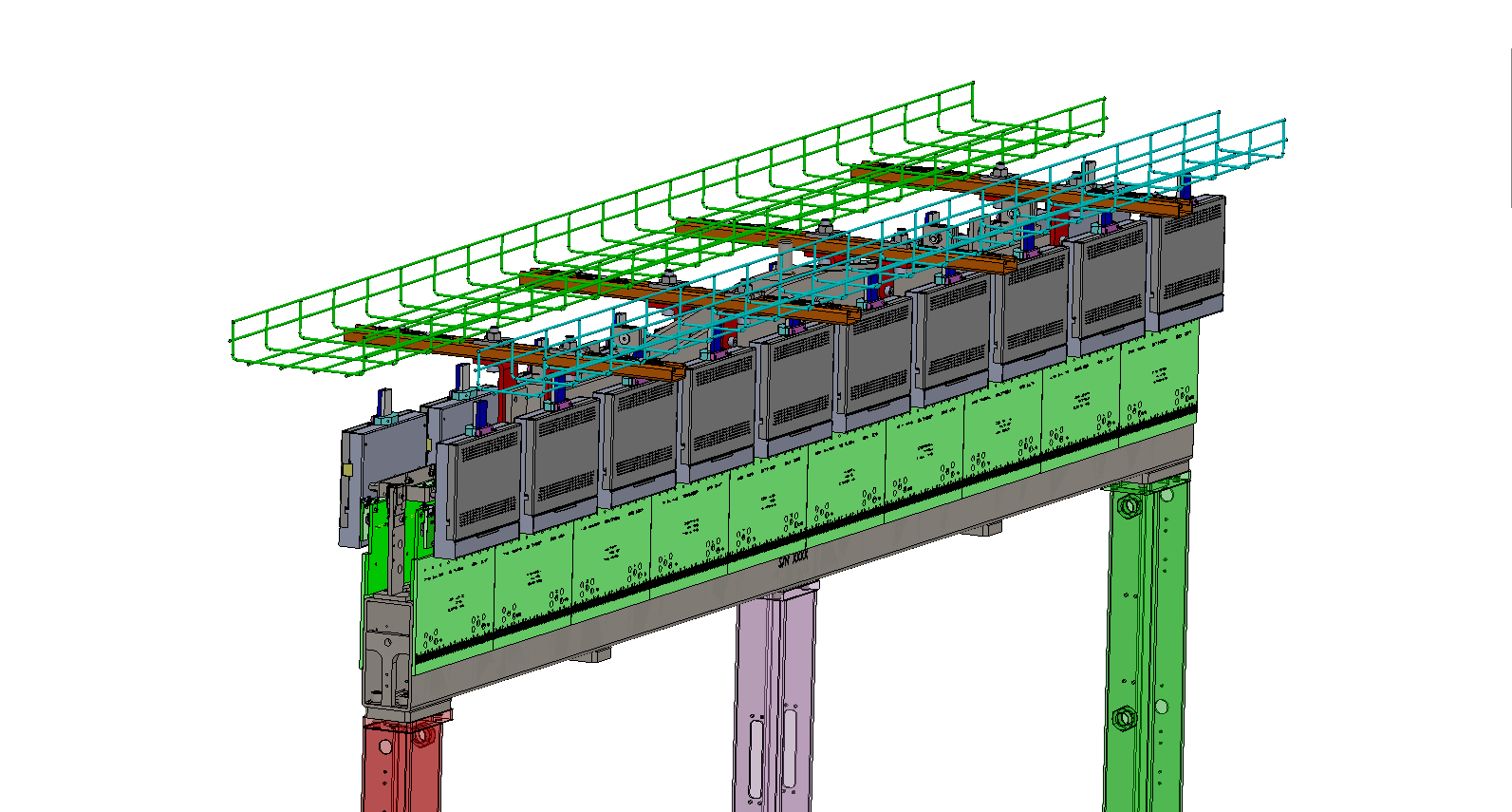}
\end{dunefigure}

The interface between the \dword{apa} and TPC \dword{ce} covers a wide range of topics, including the hardware design and production, testing, integration, installation, and commissioning. The hardware interface has two basic components, mechanical and electrical. The mechanical interface includes the support of the \num{20} \dword{ce} boxes, with each housing a \num{128} channel \dword{fe} motherboard.  These are the gray colored, vertically oriented boxes shown in Figure~\ref{fig:apa_ce}. 

The electrical interface covers the choice of wire-bias voltages to the four wire planes so that \num{100}\% transparency can be achieved for drifting ionization electrons, cable connection for the wire bias voltages from the cryostat feedthroughs to the CR boards, interface boards providing connection between CR boards and \dword{ce} boxes, filtering of the wire-bias voltages through CR boards in order to suppress potential introduction of electronics noise, and an overall grounding scheme and electrical isolation scheme for each \dword{apa}. The last item is particularly important in order to reach the low electronics noise levels required.  See Chapter~\ref{ch:fdsp-tpc-elec} for information on all of these aspects of the \dword{fe} electronics system.

\subsection{Photon Detection System}
\label{sec:fdsp-apa-intfc-pds}

While the design of the \dword{pds} is still under development, it is expected that it is integrated into the \dword{apa} frame to form a single unit for the detection of both ionization charge and scintillation light.  Cables for the \dwords{pd} must also be accommodated in the \dword{apa} frame design.  Figure~\ref{fig:apa-pd} shows the interface for a light-guide bar based \dword{pds} as has been deployed in \dword{pdsp}. Individual bars were inserted through \num{10} slots left on the side steel tubes of the frame. Rails mounted in the \dword{apa} frame, as shown in Figure~\ref{fig:apa-frame-2}, support the bars in their final positions. 

\begin{dunefigure}[\dword{apa} interface with \dwords{pd} in \dword{pdsp}]{fig:apa-pd}
{Installation of a light-guide bar photon detector module into the available slots in the \dword{apa} frame. Also shown is a concept for routing \dword{pds} cables through the rib tubes of the \dword{apa} frame and up the central vertical tube section.}
\includegraphics[height=0.3\textheight]{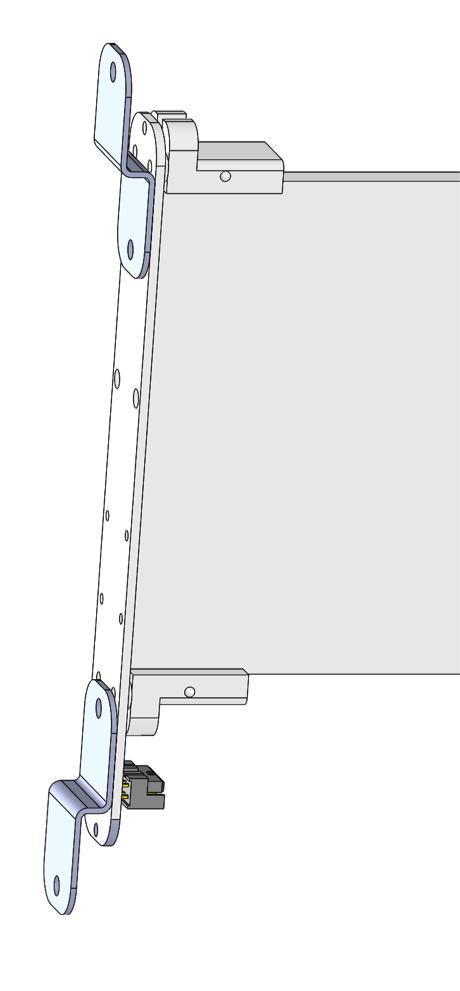}\qquad\qquad
\includegraphics[height=0.3\textheight]{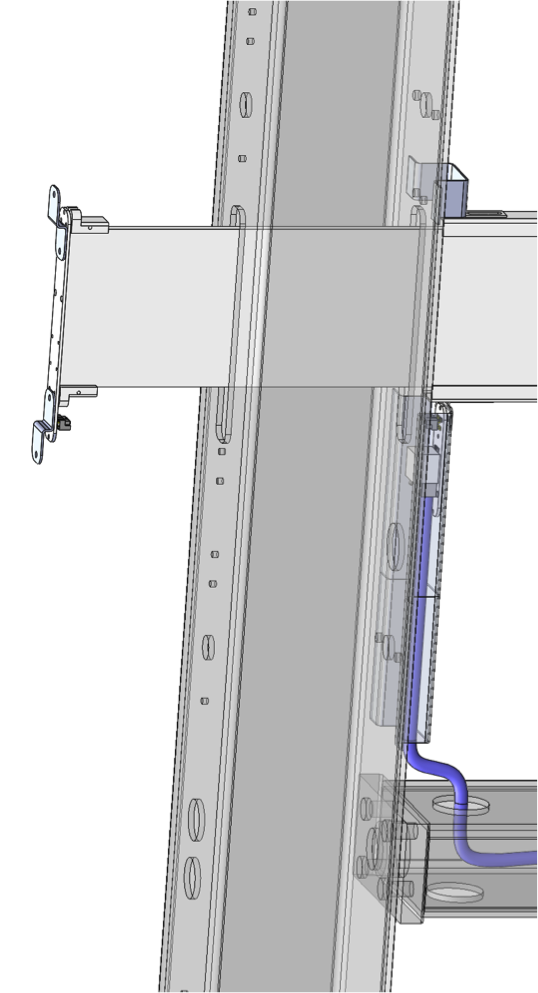}\qquad\qquad
\includegraphics[height=0.3\textheight]{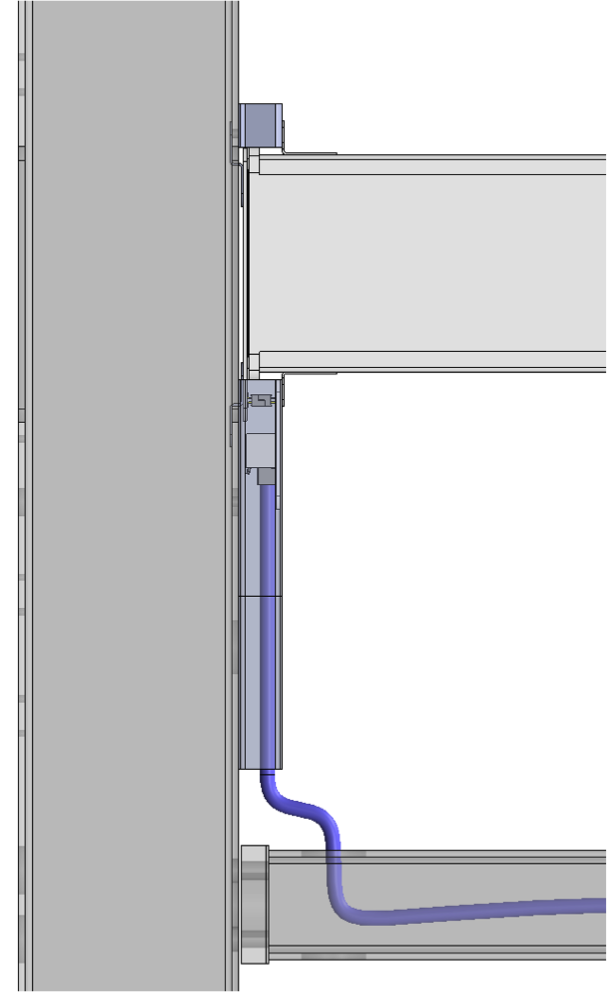}
\end{dunefigure}

Similar to that of the \dword{ce}, the interface between the \dword{pds} and \dwords{apa} covers a wide range of topics, including the hardware design and production, testing, integration, installation, and commissioning. Depending on the final design of the \dword{pds}, the geometry of the \dword{apa}, including access slot dimensions, locations, and number, may require modification. Any proposed changes by the \dword{pds} consortium must be evaluated by the \dword{apa} consortium to understand structural impacts or interferences with other components.  The electrical interface includes a grounding scheme and electrical insulation. Due to the strict requirements on the noise from the \dword{ce}, the electrical interface must be defined together with the \single electronics consortium. 

For more information on the photon system, see Chapter~\ref{ch:fdsp-pd} 

\subsection{APA-to-APA Connections and Cable Routing}
\label{sec:fdsp-apa-intfc-apa}

The TPC readout electronics require that the \dword{apa} frames must be electrically isolated.  The left panel of Figure~\ref{fig:apa-cabling} shows the current conceptual design for mechanically connecting the two \dwords{apa} in a vertical stack while maintaining electrical isolation.  The green elements are an insulating panel and bolt sleeve made from G10. 

Cable routing schemes for both the TPC electronics and \dword{pds} are actively being developed.  A concept currently under evaluation is to run the cables of the \dword{pds} inside the crossing rib tubes to the central beam tube of the \dword{apa} frames to get to the top.  The \dword{ce} signal and power cables also need to be routed so that the head end of the lower \dword{apa} in the two-\dword{apa} assembly can be reached. The current concept is to route the electronics cables inside the two side beams of the \dword{apa} frames. The right panel of Figure~\ref{fig:apa-cabling} depicts such a cable routing scheme. To fully accommodate the cables from two \dwords{apa}, 
using larger hollow tube sections is under consideration. The final design is in progress, and prototyping is planned for later this year to verify a cabling and installation solution.     

\begin{dunefigure}[\dword{apa}-to-\dword{apa} connection and cable routing]{fig:apa-cabling}
{Left: Conceptual design for the \dword{apa}-to-\dword{apa} connection.  The green insulator pieces act to electrically isolate the two frames, as required by the \dword{fe} electronics.  Right: A concept for TPC electronics and \dword{pds} cable routing. Photon detector cables would go through the central beam and be distributed inside the supporting tubes of the \dword{apa} frame.  \Dword{ce} cables (both data and power) from the bottom \dword{apa} electronics would go through the outside tubes to reach the top of the stack.}
\setlength{\fboxsep}{0pt}
\setlength{\fboxrule}{0.5pt}
\fbox{\includegraphics[height=0.26\textheight]{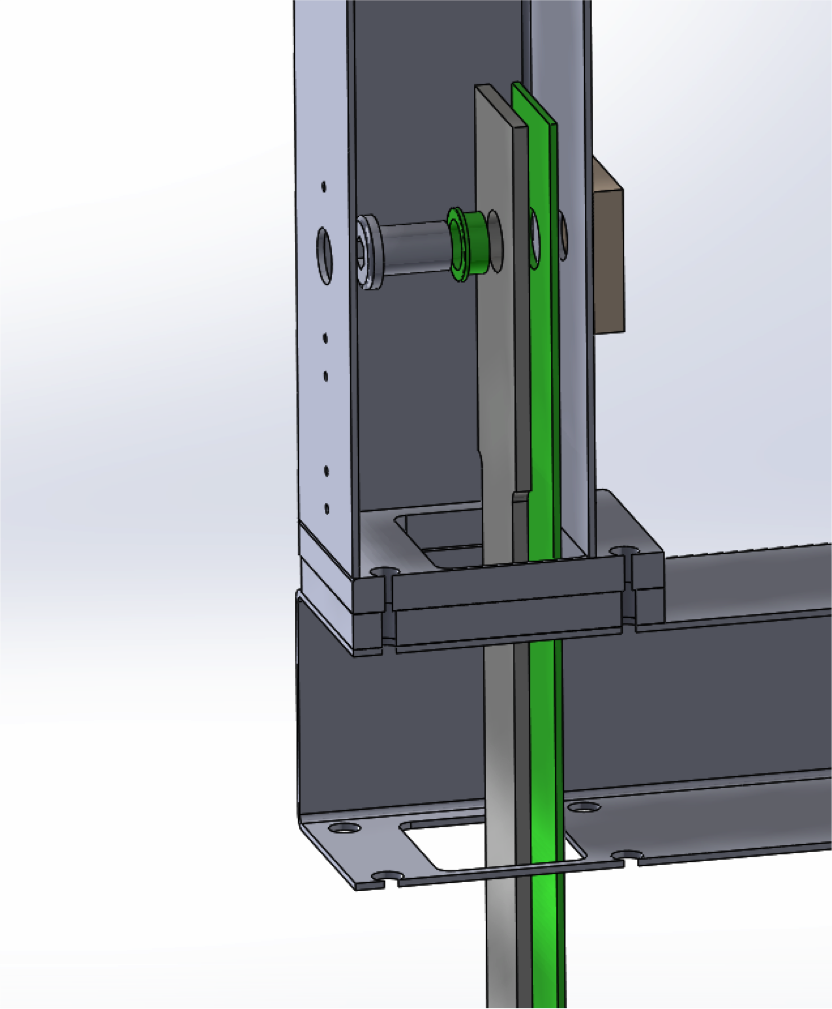}}\qquad \quad
\includegraphics[height=0.26\textheight]{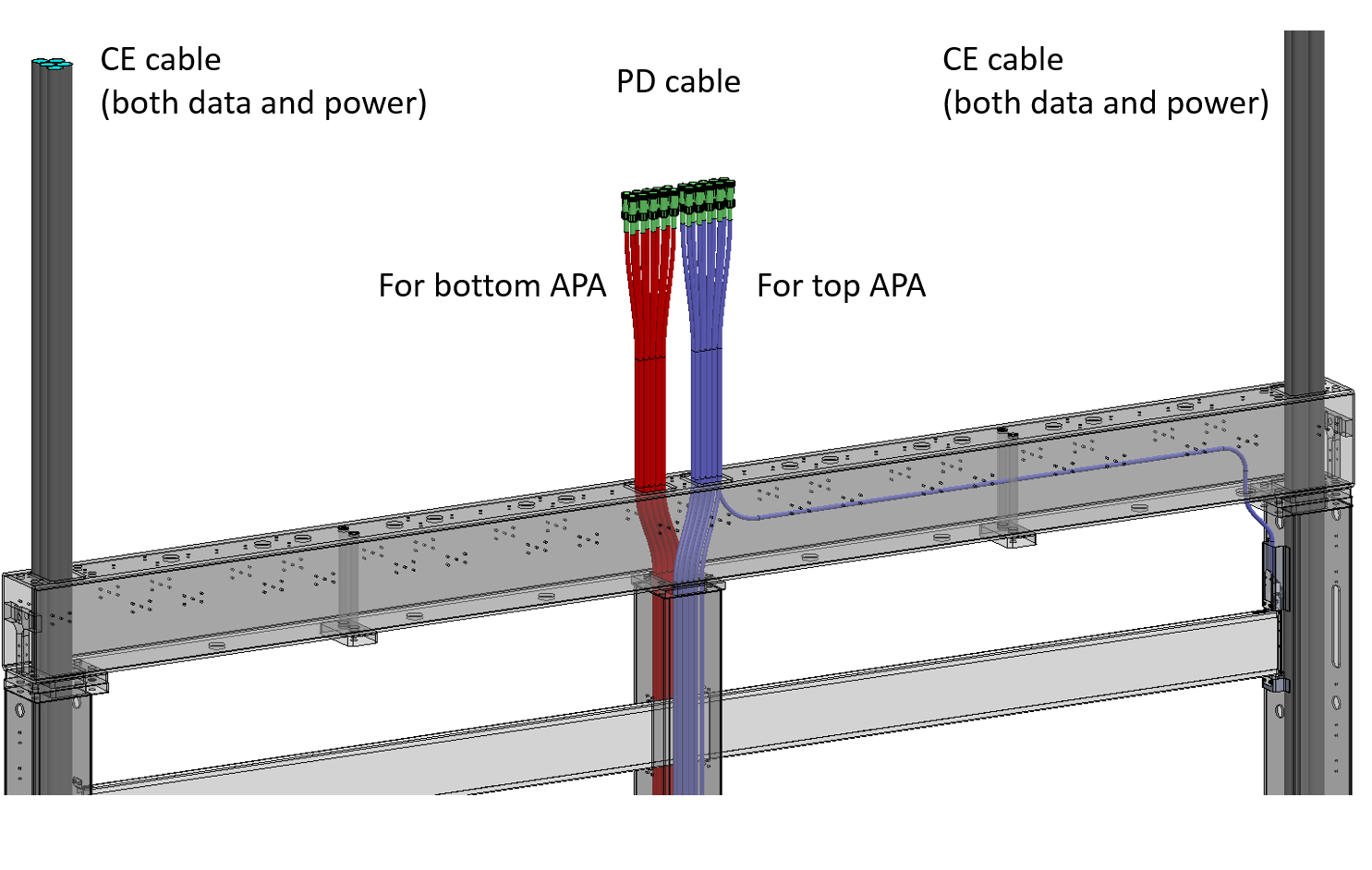}
\end{dunefigure}

\section{Production and Assembly}
\label{sec:fdsp-apa-prod-assy}

Design, construction, and testing of the DUNE \dword{spmod} \dwords{apa} is overseen by the \dword{apa} consortium. 
The \dword{apa} consortium takes a \textit{factory style} approach to the construction with multiple factories being planned in the USA and UK. This approach allows the consortium to produce \dwords{apa} at the rate required to meet overall construction milestones and at the same time reduce risk to the project if any location encounters problems that slow the pace of production.

The starting point for the \dword{apa} production plan for the \dwords{spmod} is the experience and lessons learned from \dword{pdsp} construction. For \dword{pdsp}, \dwords{apa} have been constructed both at the Physical Sciences Laboratory (PSL) at the University of Wisconsin in the USA and at Daresbury Laboratory in the UK.  \dword{apa} construction for DUNE is also envisaged to be done at USA and UK collaborating institutions, and assuming construction begins in 2021, a minimum of six production lines is required to build \num{150} \dwords{apa} within \num{2.5} years for the first 
\SI{10}{kt} \dword{spmod}.   

Based on the \dword{pdsp} experience, we estimate that each \dword{apa} requires approximately \num{50} shifts (eight-hour intervals) of effort to construct, with a mix of engineering, technical, and scientific personnel. This estimate involves only the wiring stages of production, and assumes that completed frames and all other hardware necessary for construction are ready to go at the factories. Currently an \dword{apa} can be completed in \num{64} shifts. Several improvements to the process and tooling are planned that will bring this down to the required \num{50} shifts. The production model assumes that factories run two shifts per day and that two weeks per year are devoted to maintenance of equipment. 

Each production line is centered around a wire winding robot, or \textit{winder}, that enables the continuous wrapping of wire on a \SI{6}{m} long frame. The winder can also be used to make wire tension measurements by replacing the winding head with a laser photodiode system that then can determine an individual wire's natural frequency and hence its tension. A production line also requires two process carts. These carts support the \dword{apa} and are used during various steps in the construction process, e.g., continuity testing, board epoxy installation, etc. A production line, therefore, requires a means of lifting the \dword{apa} in and out of the winder. A gantry-style crane has been used for \dword{pdsp} construction.

Having multiple \dword{apa} production sites in two different countries presents quality assurance and quality control (\dshort{qa}/\dshort{qc}) challenges. Key among the requirements of production is that every \dword{apa} be the same, regardless of where it was constructed. To achieve this goal we are building on \dword{pdsp} experience where six identical \dwords{apa} were built, four in the USA and two in the UK. This was achieved by using the same tooling, fabrication drawings, assembly and test procedures, and identical acceptance criteria at both sites.  This uniform approach to construction for DUNE is necessary, and the \dword{apa} consortium is developing the necessary management structure to ensure that each factory and production line follows the agreed upon approach to achieve \dword{apa} performance requirements.

\subsection{Facility Plans}
\label{sec:fdsp-apa-facility}

Construction of \dword{spmod} \dwords{apa} is planned to take place in both the USA and the UK. Daresbury Lab in the UK will house multiple production lines, one of which already exists from \dword{pdsp}. In the USA, it is anticipated that production lines will be set up at the University of Chicago, Yale University, and the already existing production facility at the University of Wisconsin, PSL. At least eight \dword{apa} production lines spread over multiple facilities will provide some margin on the production schedule and provide backup in the event that technical problems occur at any particular site. 
\fixme{specified future partner sites}
The space requirements for each production line are driven by the large size of the \dword{apa} frames and the winding robot used to build them. The approximate dimensions of a class \num{100000} clean space needed to house winder operations and associated tooling is \SI{175}{m$^2$}. The estimated requirement for inventory, work in progress, and completed \dwords{apa} is about \SI{600}{m$^2$}. Each facility  also needs temporary access to shipping and crating space of about \SI{200}{m$^2$}. Possible floor layouts at each institution are currently in development. Adequate space is available at each site and commitments have been expressed by the institutions for its use on DUNE. 


The University of Wisconsin has space available within the Physical Sciences Lab Rowe Technology Center. A portion of the facility has been used for the past two years for the \dword{pdsp} project. There is approximately \SI{20000}{ft$^2$} (\SI{1850}{m$^2$}) available for DUNE and the possibility exists to expand the current clean tent to house another production line. 

\dword{pdsp} construction has also taken place at Daresbury Lab.  The current facility cannot accommodate multiple production lines, but the ``Inner Hall'' on the Daresbury site has been identified as an area that is sufficiently large to be used for DUNE \dword{apa} construction. It has good access and crane coverage throughout. Daresbury Laboratory management have agreed that the area is available, but investment is needed to establish a safe working environment. Preparation work for the construction area is underway to clear the current area of existing facilities, obsolete cranes, and ancillary equipment. Also planned is the renovation of a plant room to be used for storage and as a shipping area. This work is ongoing. The production factory is being designed to hold four winding machines and associated process equipment and tooling. 

The Enrico Fermi Institute at the University of Chicago and the Wright Laboratory at Yale University each have the needed infrastructure to house up to two \dword{apa} production lines. Development work that is relevant for local planning at each site has begun at those institutions, as well.

\subsection{Assembly Procedures and Tooling}
\label{sec:fdsp-apa-assy}

The central piece of equipment used in \dword{apa} production is the custom-designed wire winder machine, shown in use in Figure~\ref{fig:winder-photos}.  An important centerpiece of the winder machine is the wiring head.  The head releases wire as motors move it up and down and across the frame, controlling the tension in the wire as it gets laid. Currently, the head then positions the wire at solder connection points for soldering by hand. The fully automated motion of the winder head is controlled by software, which is written in the widely used numerical control G programming language.  The winder also includes a built-in vision system to assist operators during winding, which is currently used at winding start-up to find a locator pin on the wire boards.  In the current scheme used for \dword{pdsp}, during the winding process an \dword{apa} moves on and off the winder machine multiple times for wiring, soldering, testing, etc.  

\begin{dunefigure}[Photos of the \dword{apa} wire winding machine]{fig:winder-photos}
{Left: Partially wired \dword{pdsp} \dword{apa} on the winding machine at Daresbury Lab, UK. Right: Partially wired \dword{pdsp} \dword{apa} on the winding machine during wire tension measurements at University of Wisconsin, PSL.}
\setlength{\fboxsep}{0pt}
\setlength{\fboxrule}{0.5pt}
\fbox{\includegraphics[height=0.3\textheight,trim=25mm 0mm 4mm 0mm,clip]{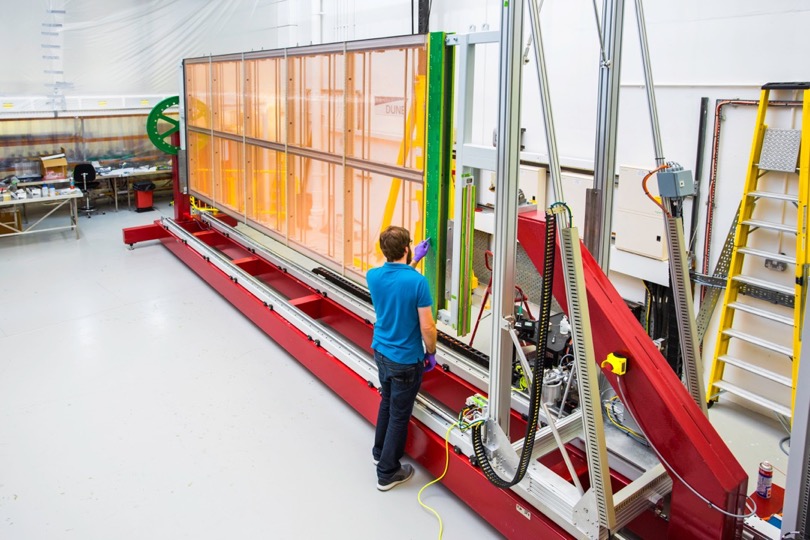}}
\fbox{\includegraphics[height=0.3\textheight,trim=200mm 0mm 30mm 0mm,clip]{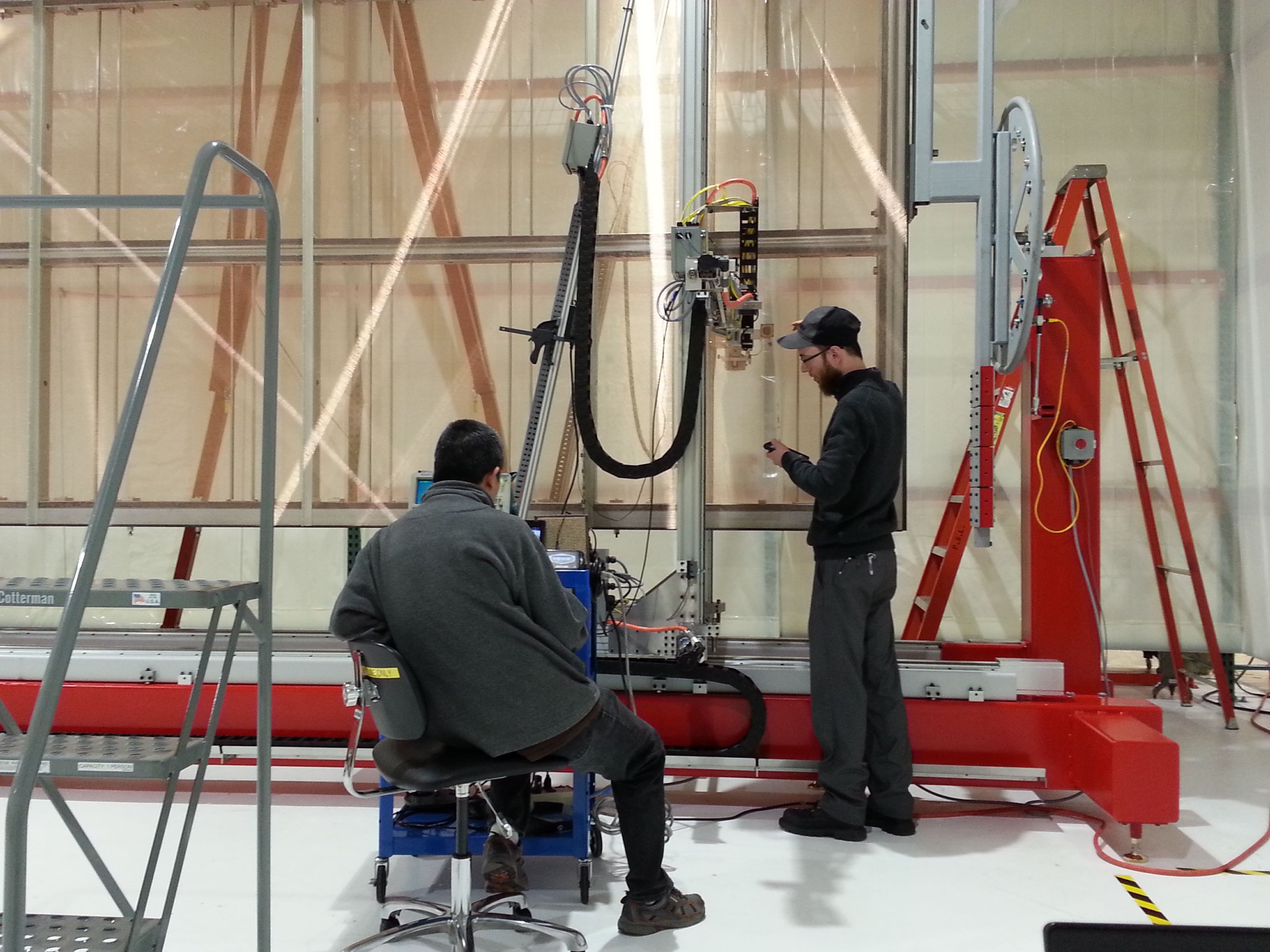}}
\end{dunefigure}

Two large process carts, shown in Figure~\ref{fig:apa-process-cart}, are used to move \dwords{apa} around the assembly facility. 
One process cart with regular casters remains in the assembly area and is maintained at a particular height that coordinates with other construction tooling such as jack stands and platform ladders. A second process cart has been fitted with specialized 360$^\circ$ rotating casters that allow the process cart loaded with a fully assembled \dword{apa} to maneuver corners during the journey from the assembly area to the shipping/packing location.

Before wiring can begin, the first operation with a bare frame is to install the grounding mesh. In the \dword{pdsp} design, a large jig is needed to hold the mesh in place for gluing. Once the jig is leveled sufficiently to the frame, a mesh sheet is laid into place and hold down bars are iteratively moved and repositioned until the mesh is flat and tight.  The outside edge of the mesh panel then gets epoxied, and the jig and hold down bars remain in place for a 12 hour epoxy cure cycle.  This process is then repeated for the next three shifts until all four panels of mesh have been attached to the bare \dword{apa} frame.  As described below, changes to this lengthy procedure are being considered for DUNE \dwords{apa}.  

Another custom construction jig is needed for installing the wire combs that hold the wires at intermediate points above the four cross beams of the \dword{apa}.  Currently there are two jigs that can be loaded and installed at a time, and each installation requires a six-hour epoxy cure cycle. 

\begin{dunefigure}[Photos of an \dword{apa} on a process cart during construction]{fig:apa-process-cart}{(Left) \dword{apa} being moved around a production facility on the process cart. (Right) \dword{apa} frame with the grounding mesh already installed is shown sitting on a process cart.  Two technicians are using a custom jig to place the wire combs above a horizontal cross beam on the \dword{apa}.}
\setlength{\fboxsep}{0pt}
\setlength{\fboxrule}{0.5pt}
\fbox{\includegraphics[height=0.24\textheight,trim=50mm 0mm 70mm 0mm,clip]{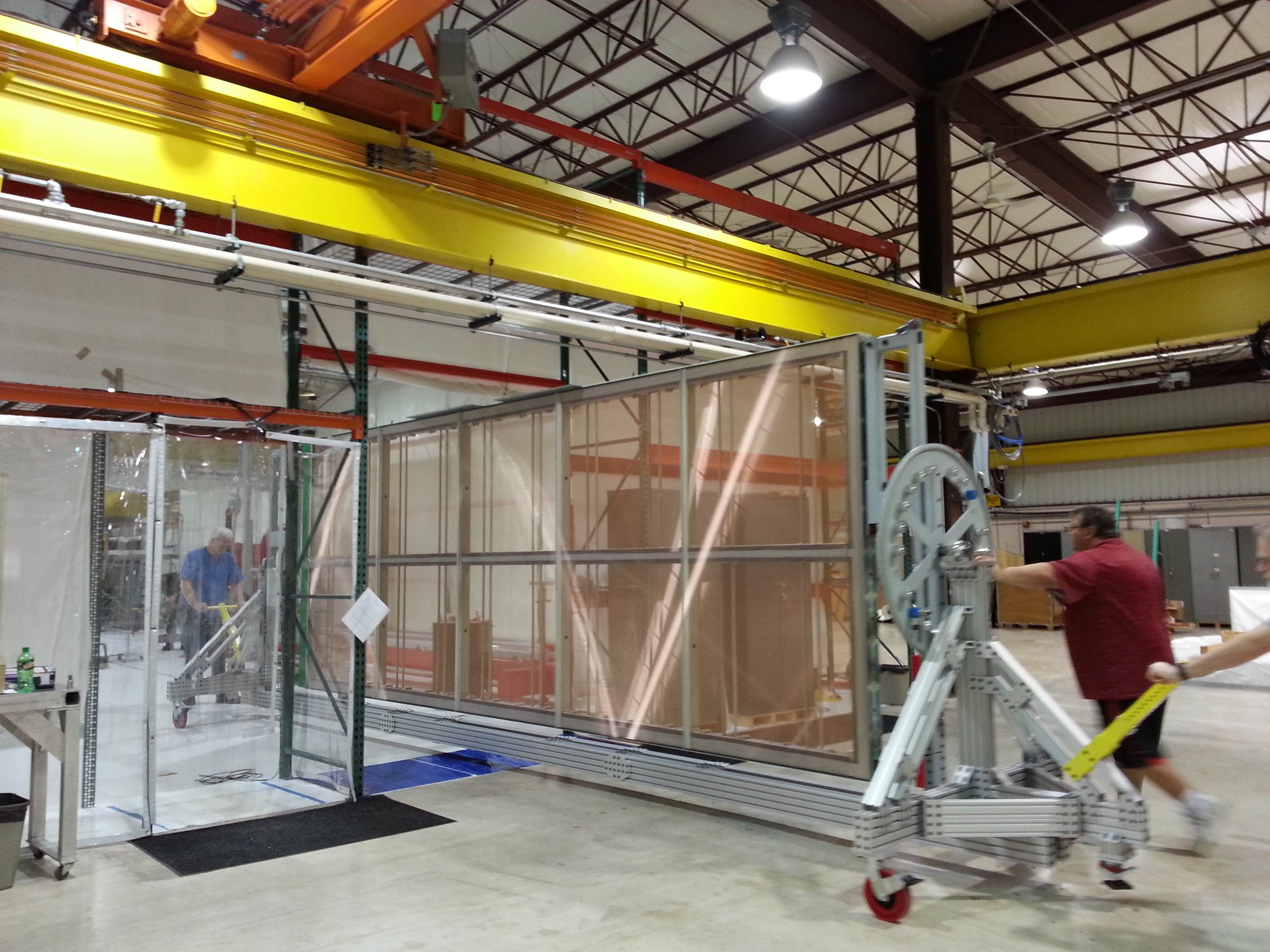}}
\fbox{\includegraphics[height=0.24\textheight,trim=15mm 0mm 35mm 0mm,clip]{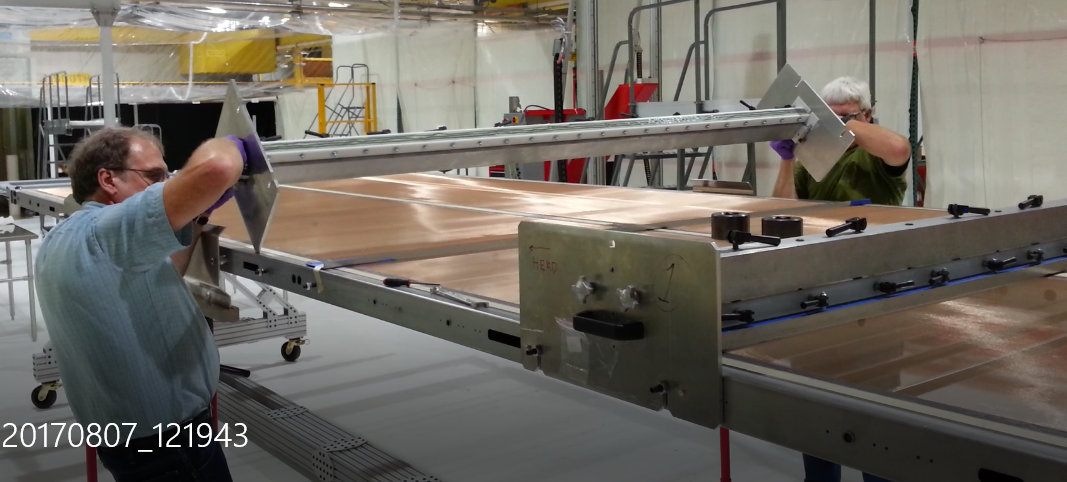}}
\end{dunefigure}

\subsection{Material Supply}  

Ensuring the reliable supply of raw materials and parts to each of the factories is critical to keeping \dword{apa} production on schedule through multiple years of construction. Here the consortium institutions play a pivotal role taking on the responsibility for the delivery of \dword{apa} sub-elements to each of the factories. Supplier institutions have responsibility for the sourcing, inspection, cleaning, testing, quality assurance, and delivery of hardware to each of the factories. 

\begin{itemize}

\item Frame construction: We envision two sources of frames, one in the USA and one in the UK. The institutions responsible will rely on many lessons learned from \dword{pdsp}. The effort requires specialized resources and skills including a large assembly area, certified welding capability, large scale metrology tools and experience, and large scale tooling and crane support. Two approaches are under consideration for sourcing; one is a total outsource strategy with an industrial supplier, the other is to procure all of the major machined and welded components and then assemble and survey in-house. Material suppliers have been identified and used with good results on \dword{pdsp}.

\item Mesh supply and construction: Elsewhere in this proposal we describe the current mesh installation procedure. However, our \dword{pdsp} experience leads us to believe that moving to smaller self-supporting \textit{window screen} panels may save assembly time and improve overall \dword{apa} quality. An excellent source of mesh exists and was used on \dword{pdsp}.

\item Wire procurement: Wire is a significant element in the assembly of an \dword{apa}. There is approximately \SI{24}{km}| of wire wound on each unit. Through \dword{pdsp} 
an excellent supplier 
has worked with us to provide wire that is of high quality and wound on spools that we provide. These spools are then used directly on the winder head with no additional handling or re-spooling required. Wire samples from each spool are strength tested prior to use.

\item Comb procurement: An institution will work with either our existing comb supplier or find additional suppliers that can meet our requirements. The \dword{pdsp} supplier has been very reliable.

\item Wire wrapping board procurement: One or more consortium institutions will take on the responsibility of wire-wrapping board supply. The side and foot boards are unique to suppliers as they have electrical traces and provide wire placement support through a separately bonded tooth strip. There are \num{276} boards per \dword{apa}, or \num{41400} needed for \num{150} \dwords{apa}. The institutions that have responsibility for boards will spend time working with multiple vendors to reduce risk and ensure quality. 

\item Capacitor resistor boards: These boards are unique given their thickness, 
\dword{hv} components, and leakage current requirements. A reliable source of bare boards was found for \dword{pdsp}. Assembly and testing was performed at PSL. We will conduct a more exhaustive search of vendors that are willing to take on assembly and testing for the \num{3000} plus boards needed for DUNE.

\item Winders and tooling: We propose that PSL and Daresbury work together to supply tooling and winding machines for additional production lines at new locations and for additional lines in-house. This is a natural collaboration that has been in place for nearly two years on \dword{pdsp}.

\end{itemize}

\subsection{Planned Improvements to Production Process}

Based on our \dword{pdsp} experience, we have identified several potential improvements to tooling and process that allow the \dwords{apa} to be constructed in a more efficient and reliable manner, including:

\begin{itemize}

\item Wiring head design: Efforts to improve winder head performance are already underway. We envision improved tension control, continuous tension feedback, improved clutch, and an improvement to the compensator mechanism all leading to better, more consistent, and more reliable winder performance.  The current winding head uses a magnetic clutch mechanism that is manually adjusted to increase or decrease the tension of the wire as it is wound around the \dword{apa}. The clutch regularly needs adjustment as the diameter of the wire on the spool reduces during the winding process. In addition, if the mechanism is run from a cold start, it has been observed that the tension changes after $\sim$10 minutes of running. Experience winding the \dword{pdsp} \dwords{apa} has shown that it is difficult to maintain the target tension within tolerances (5$\pm$\SI{1}{N} for \dword{pdsp}).

A solution to this issue is to design a winder head with active tension control. This can be achieved by replacing the magnetic clutch with a servo motor and introducing a potentiometer on a dancer arm for the feedback loop (see Figure~\ref{fig:winding-head}). This only works if there are no signal losses when transferring the winding head to the compensator latching mechanism and back. The system can be driven in torque mode and compensates for any wire spool changes. It must be able to operate from a cold start. This development is well underway and tests are currently being carried out.

\begin{dunefigure}[Exploded view of the winding machine head]{fig:winding-head}
{Exploded view of winder head with active tension control.}
\includegraphics[width=0.7\textwidth]{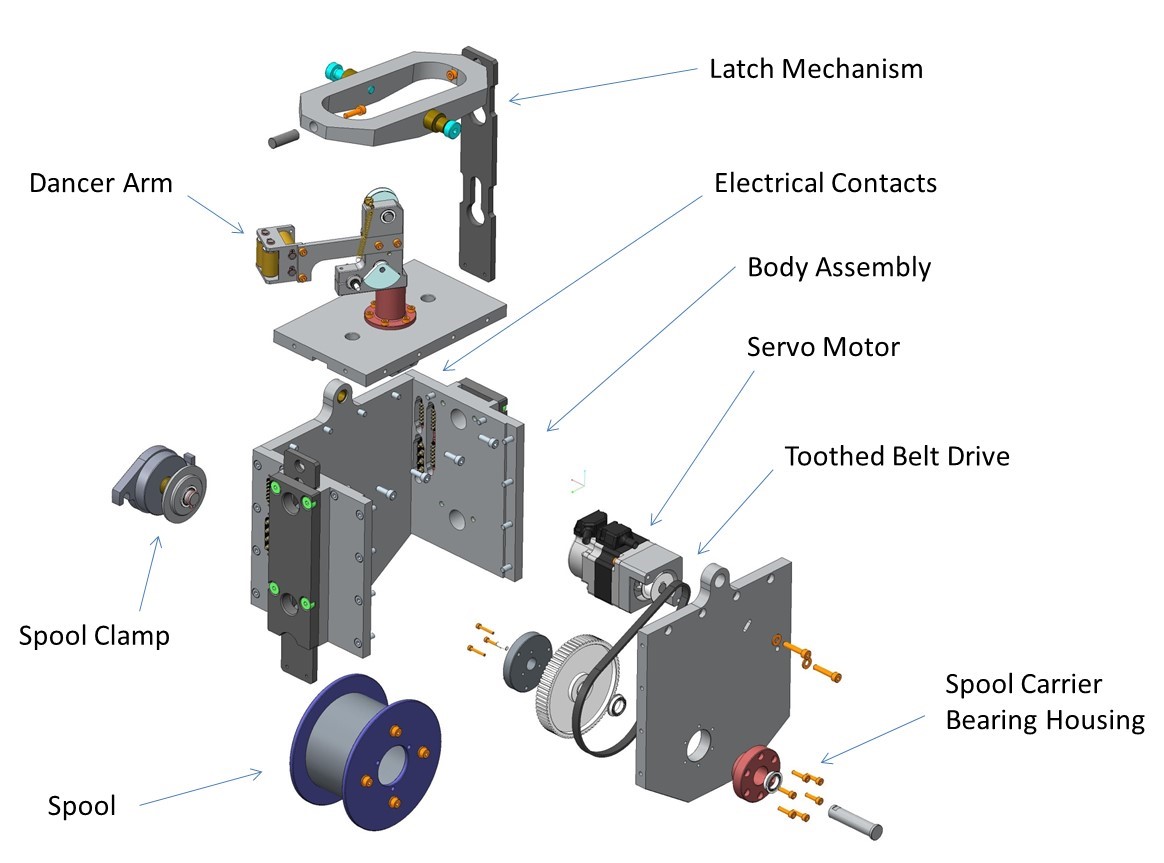}
\end{dunefigure}

\item Winder interface arm design: The current winder interface only allows one-half of a wire plane to be wired at a time. The \dword{apa} frame must be moved to the process cart where the interface arms are flipped 180$^\circ$ to wind the second half of the wire plane.  A new design concept, illustrated in Figure~\ref{fig:winding-dev}, allows the winder head to pass from one side to the other in a nearly continuous fashion without removal from the winding machine.  The interface frames are replaced at either end by retractable linear guided shafts. These can be withdrawn to allow passing of the winding head around the frame over the full height of the frame. These shafts have conical ends and locate in shafts that are fixed to the internal frame tube to provide guided location. This design change does not alter the design of the frame. The design also allows for rotation in the winding machine, so that it should also be possible to carry out board installation and gluing \& soldering in the winding machine. This eliminates the need to transfer the \dword{apa} to the process cart for the whole of the production operation, which is inherently a safer and faster production method as it cuts down the amount of handling of the \dword{apa}.

\begin{dunefigure}[Winding machine schematic showing ongoing development]{fig:winding-dev}
{Work is ongoing to modify the wiring machine design to allow an \dword{apa} frame to be rotated without removing it from the frame.  This would reduce the required handling of the frame during fabrication and speed production substantially.}
\includegraphics[width=0.95\textwidth]{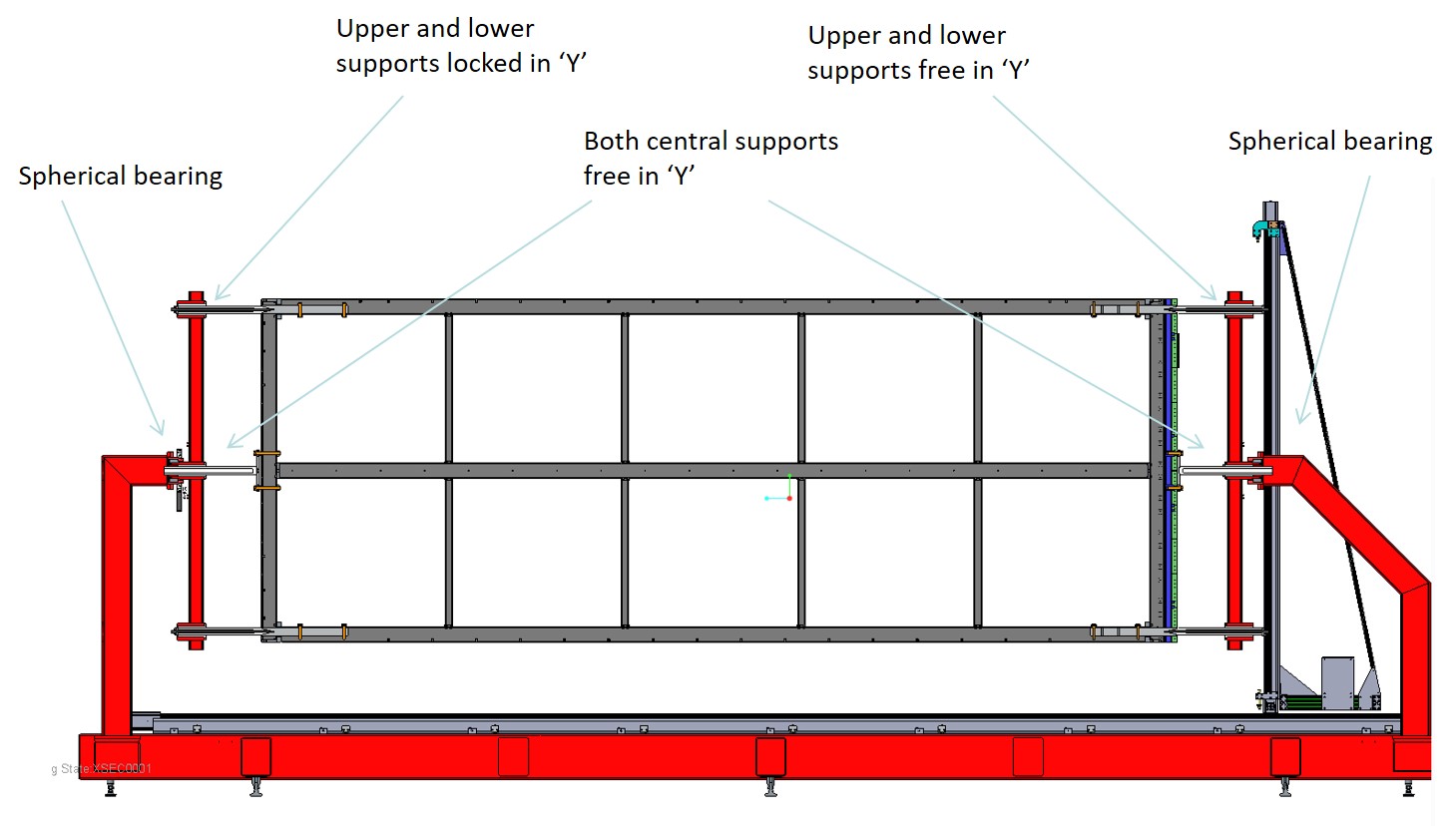} 
\end{dunefigure}

\item Modular mesh panels: The current approach to mesh installation is slow and cumbersome. We will improve this aspect of construction by moving toward a modular window screen design that improves the reliability of the installed mesh (more uniform tension across the mesh), and allows much easier installation on the \dword{apa} frame.

\item Epoxy process improvements: There are many epoxy application steps during the construction process. These steps require careful work that takes many hours between winding each successive wire plane. We already have concepts for improved epoxy application jigs from \dword{pdsp}, and we will investigate whether epoxy pre-forms or accelerated heat curing can yield time or reliability improvements.

\item Automated soldering: Every solder joint on the six \dword{pdsp} \dwords{apa} was done by hand. We will investigate automated soldering techniques to improve process and reduce the amount of manual effort required. 

\item Wire tension measurement techniques: Verifying wire tension is an important, but time consuming process during construction. The current technique utilizes a laser photodiode tool mounted on the winder to measure tension one wire at a time. This takes many hours for each wire plane. Techniques are under development by collaborators at the University of Manchester  to electronically measure groups of \num{20} or more wires at one time. This technique  provides much faster tension measurements and shorter turnaround between wire planes. 

\item Winder maintenance plan: The approach to winder maintenance used during \dword{pdsp} construction was not well formed. As a result, winding machine problems that can be traced back to lack of routine maintenance occurred from time to time, which shut the production line down until a repair or maintenance was performed. We will formulate a routine and preventive maintenance plan that minimizes winder downtime during \dword{apa} production.
\end{itemize}

\subsection{Quality Assurance and Quality Control in APA Production}
\label{sec:fdsp-apa-qa}

A key input to \dword{qa} for the \dword{apa} design and manufacturing procedures is the experience with \dword{pdsp}, including upcoming operations and data analysis results from the detector.  Much has already been learned regarding design, component testing, and fabrication procedures that will go into formulating the detailed design and plans for the 
\dword{apa} construction project over the next year.  The set of final design drawings and detailed procedures documentation generated over the next year leading to the \dword{tdr} represent an important element of the \dword{qa} plan for the fabrication of the \dwords{apa}.  

Summaries of all \dword{qa} testing performed for elements used in the final design of the \dwords{apa} will also be prepared for the \dword{tdr}.  Much data already exists, and again, \dword{pdsp} will provide valuable additional information regarding the robustness of the detector components and construction.  

\subsubsection{Incoming Inspections}

Some components require inspection and \dword{qc} checks prior to use on an \dword{apa}, including:

\begin{itemize}

\item Frame components: If the \dword{apa} steel frames are produced in-house, then upon receipt of the rectangular hollow section steel for the frames, a selection procedure is followed to choose the sections of the steel most suited to achieving the geometrical tolerances. 

\item Wire testing: The CuBe wire is provided on spools from the supplier. Samples from each spool are strength tested prior to use on an \dword{apa}.

\item Circuit boards: All circuit boards that get installed on an \dword{apa} are inspected for dimensional accuracy prior to being routed through various epoxy and cleaning processes as they are prepped for assembly. Inspection results are documented, and if anomalies are found, an electronic non-conformance report is written.  

\item CR and $G$-plane bias board testing: Acceptance tests of these boards include leakage current measurements ($<$\SI{0.5}{nA}) and continuity tests on each channel.  This test is performed at room temperature. \dword{pdsp} was used to perform design validation on over \num{100} boards that were cycled and tested at LN temperature. No failures were seen during these tests. 

\end{itemize}

\subsubsection{\dword{apa} Acceptance Tests} 

The following are examples of quality control data to be collected for each \dword{apa} during production:  

\begin{itemize}

\item Frame flatness: A laser survey is performed to measure the flatness of the assembled bare frame. Three sets of data are compiled into a map that shows the amount of bow, twist, and fold in the frame. Each of these parameters is compared to an allowable amount that does not cause wire plane-to-plane spacing to be out of tolerance ($\pm$\SI{0.5}{mm}).  A visual file is created for each \dword{apa} from measured data. A final frame survey is completed after all electrical components have been installed, and the as-built plane-to-plane separations are measured to verify the distance between adjacent wire planes.

\item Mesh to frame connection: To confirm sufficient electrical contact between these two components a resistance measurement is taken in each of \num{20} zones of mesh bounded by the outside frame perimeter and the four cross beam ribs. This measurement is completed immediately after mesh install, prior to any winding.

\item Wire tension: The tension of each wire is measured after each new plane of wires is installed on an \dword{apa}. The optimal target tension is still under discusion (was \SI{5}{N} in \dword{pdsp}), as are the necessary tolerances.  \Dword{pdsp} data, where the tensions have substantial variation, will provide important data for quantifying the impacts of varying tensions.  

\end{itemize}

\subsubsection{Documentation} 

Each \dword{apa} is delivered with a traveler document in which specific assembly information is gathered, initially by hand on a paper copy, then entered into an electronic version for longer term storage.  The traveler database contains a detailed log of the production of each \dword{apa}, including where and when the \dword{apa} was built and the origin of all parts used in its construction. 

As assembly issues arise during the construction of an \dword{apa}, they are gathered in an issue log for each \dword{apa}, and separate short reports are created to provide details of what caused the occurrence, how the issue was immediately resolved, and what measures should be taken in the future to ensure the specific issue has a reduced risk of occurring.  

\section{Integration and Installation}
\label{sec:fdsp-apa-install}

Completed \dwords{apa} are shipped from the \dword{apa} production sites to an \dword{itf} for integration with the TPC \dword{fe} electronics and \dwords{pd}.  The \dword{itf} location is not decided, but facilities near the \surf site are being considered.   Activities at the \dword{itf}  include extensive \dword{qc} testing to ensure the functioning of the fully integrated \dwords{apa}.  Once checked, the \dwords{apa} are repackaged for final transport to \surf. Each \dword{apa}, still in its transport crate, are hung from the Ross Shaft cage by a sling and transported underground where it is stored in a waiting area.  Pairs of \dwords{apa} must be linked in their vertical configuration and cables ran from both the lower and upper \dwords{apa} in an area just outside the cryostat.  Once completed, the pair enters through the \dword{tco} onto the \dword{dss} and is moved into its final position.  Final checkout tests are performed once the \dwords{apa} are in place.

The integration with the \dwords{pd} is expected to be done at the Integration Facility. 
An alternative plan entailing \dword{pds} installation at the \dword{apa} production sites is also under consideration. The TPC \dword{fe} electronics are installed  at the \dword{itf} and the exact installation sequence will be developed with the electronics consortium.

A conceptual layout of the space required at the Integration Facility is being developed. An overhead crane is needed to lift \dwords{apa} out of their shipping crate and maneuver them through the facility.  Most of the handling areas need to be embedded in a class \num{100000} clean tent. Finally, a cold box will be available for \dword{qc} testing of the electronics once installed on the \dword{apa} (see Section~\ref{sec:fdsp-apa-install-qc_if}).  


\subsection{Transport and Handling}
\label{sec:fdsp-apa-install-transport}

Custom designed crates are used for transport between the production sites and the \dword{itf}, and between the \dword{itf} and \surf{}. The design of the crates is still being finalized, but there are currently two possible approaches. The first is to use less expensive, disposable crates for transport to the \dword{itf} and fewer, more expensive crates for transport underground, which are reused between the \dword{itf} and underground. The second option is a single crate that is used for all transport stages. The transport underground requires a design that allows a 180$^{\circ}$ rotation of the crate. 


The handling of the \dwords{apa} at the \dword{itf} and underground is done with overhead cranes. Once the \dwords{apa} are repackaged in the crates, they are loaded on a truck, driven to the mine, transported to the cage, secured on the sling under the cage (see figure \ref{fig:cagetransport}), lowered down and moved to the underground storage area.

\begin{dunefigure}[\dword{apa} suspended beneath the mine shaft cage]{fig:cagetransport}{The \dword{apa} crate (in blue) is brought underground with a sling under the elevator cage (the green box at the top of the figure). The insertion into the crate at the surface is done from the back of the cage, but the extraction underground must be done from the front of the cage. The \dword{apa} crate must, therefore, be able to be rotated  by 180$^\circ$ in the sling.}
\setlength{\fboxsep}{0pt}
\setlength{\fboxrule}{0.5pt}
\fbox{\includegraphics[width=0.4\textwidth, trim=0mm 20mm 0mm 20mm,clip]{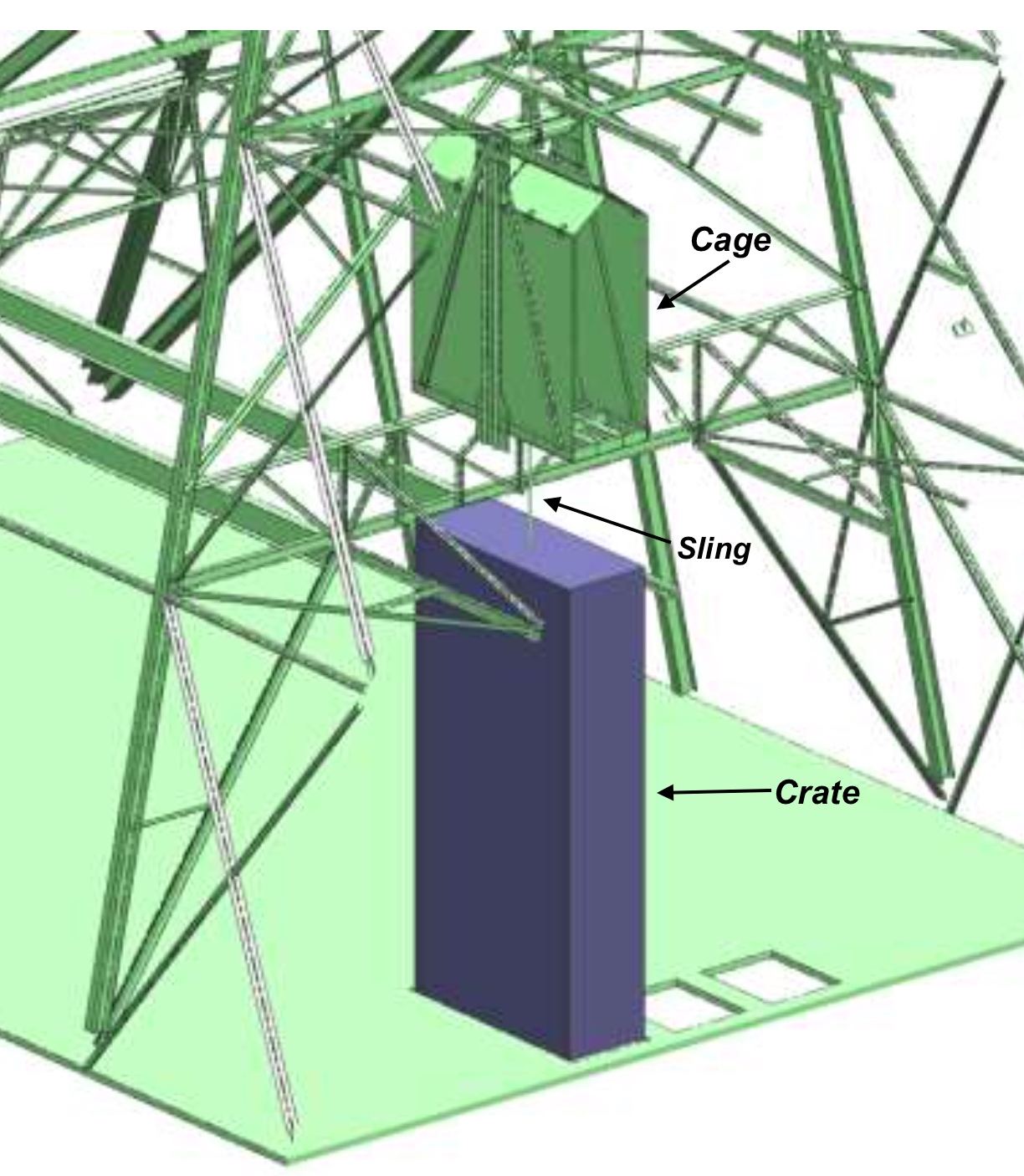}} 
\end{dunefigure}

\subsection{APA-to-CPA Assembly and Installation in the Cryostat}
\label{sec:fdsp-apa-install-cryostat}

Once underground, there will be a small storage area for stockpiling \dwords{apa} (see Figure~\ref{fig:handling}). When ready for installation, each \dword{apa} is extracted from its crate, inspected and rotated to be lowered into the area just outside of the \dword{tco} in the cryostat. Two \dwords{apa} are lowered in front of the \dword{tco} where they are linked and cabled. The details of the cabling are still being finalized, but the main option is currently to pass all the cables inside the \dword{apa} frame tubes (see Section~\ref{sec:fdsp-apa-intfc-apa}).



Finally, when the two \dwords{apa} are fully cabled, they are placed onto the DSS inside the cryostat (see bottom right of Figure~\ref{fig:handling}) and moved to their location in the cryostat where final integration tests are performed.  For more information on the detector support structure and installation into the cryostat, see 
Chapter~\ref{ch:fdsp-coord}. 

\begin{dunefigure}[Underground handling of the \dwords{apa}]{fig:handling}{(Top row) Handling of an \dword{apa} in the underground storage area where the \dwords{apa} are extracted from the crates, inspected, and readied for installation in the cryostat. (Bottom row) A pair of \dwords{apa} are brought into the space just outside the \dword{tco} to be linked and cabled, then connected to the \dword{dss} and moved into their final position inside the cryostat.}
\setlength{\fboxsep}{0pt}
\setlength{\fboxrule}{0.5pt}
\centering
\fbox{\includegraphics[height=0.195\textheight,trim=8mm 8mm 20mm 4mm,clip]{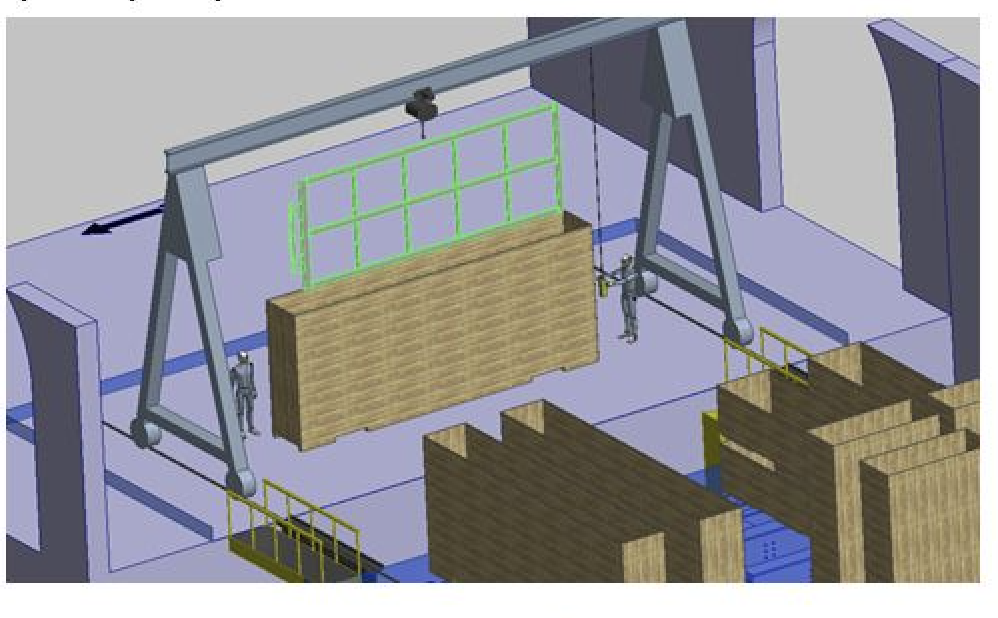}} 
\fbox{\includegraphics[height=0.195\textheight,trim=8mm 4mm 20mm 4mm,clip]{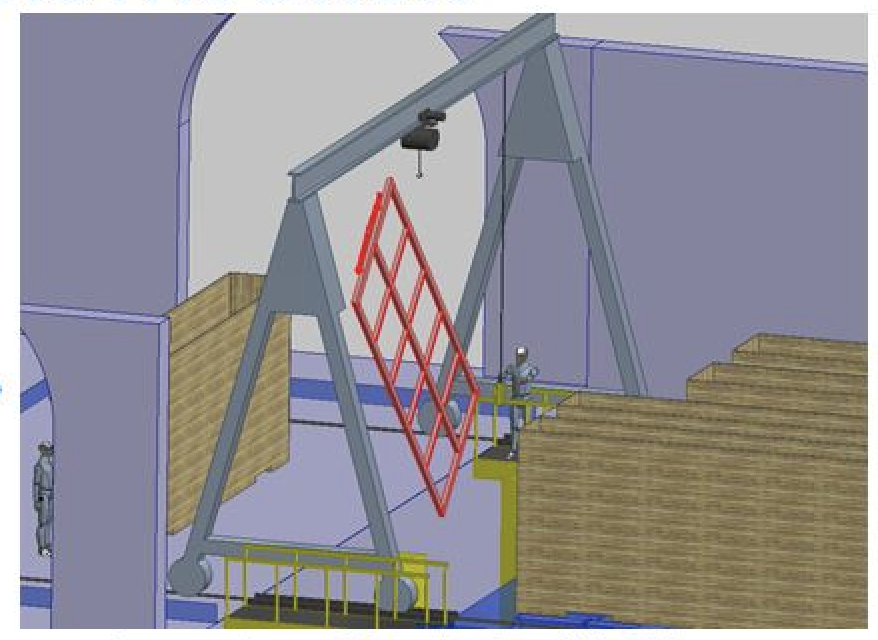}} 
\fbox{\includegraphics[height=0.195\textheight,trim=4mm 4mm 4mm 4mm,clip]{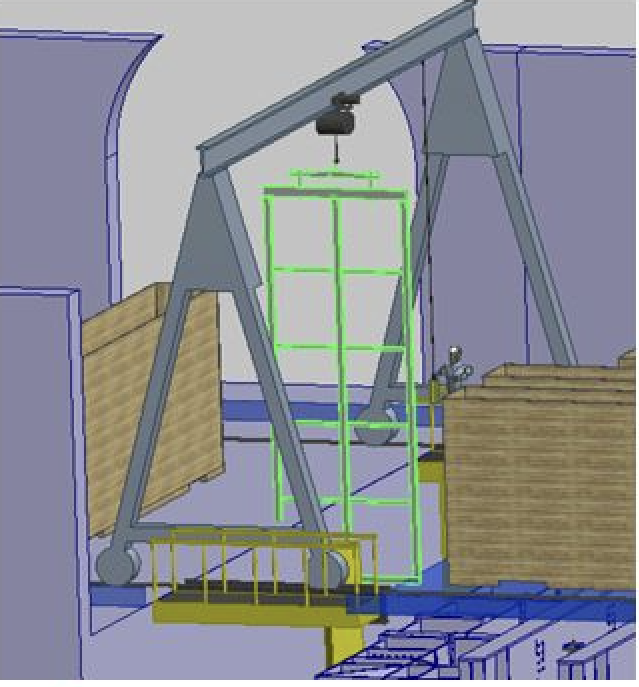}} 
\\ \vspace*{1.5mm}
\hspace*{-.25mm}
\fbox{\includegraphics[height=0.37\textheight,trim=4mm 4mm 4mm 4mm,clip]{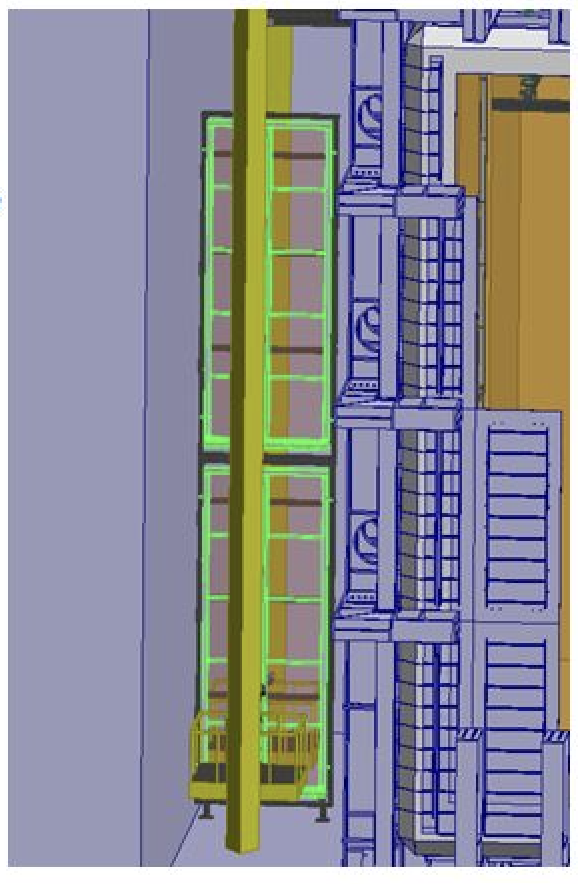}}
\hspace*{1.mm}
\fbox{\includegraphics[height=0.37\textheight,trim=4mm 4mm 4mm 4mm,clip]{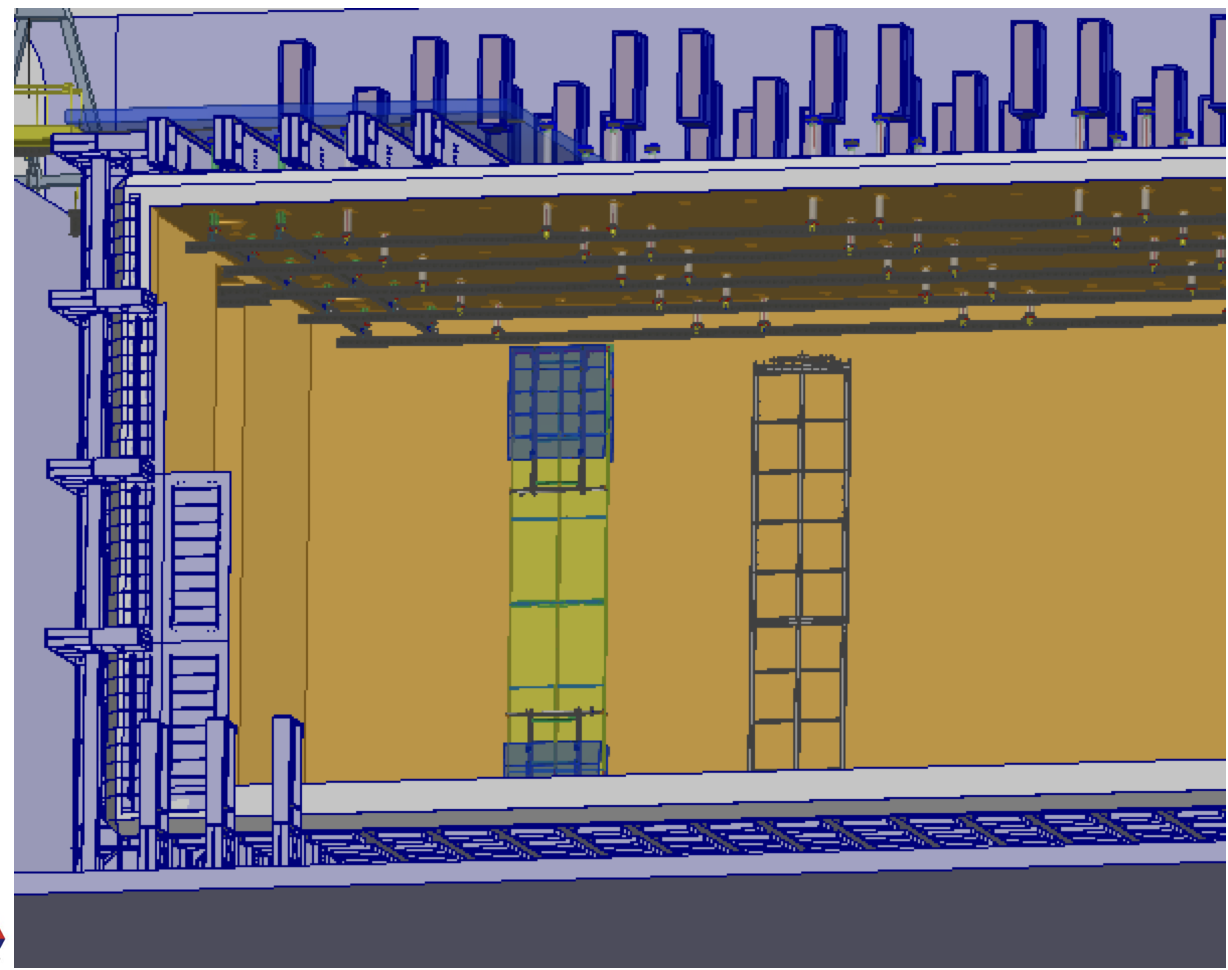}}
\end{dunefigure}

\subsection{Quality Assurance and Quality Control in Integration and Installation}
\label{sec:fdsp-apa-install-calib}

The \dword{qc} related to integration and installation has two main testing campaigns, one at the Integration Facility (\dword{itf}) and one once the \dwords{apa} are installed into their location in the cryostat. Some details are still under development by the installation and integration team within the \dword{apa} consortium. 

A dedicated database for \dword{qc} is required to keep track of all the components for all the \dwords{apa} at the different stages of the integration and installation. A simple and practical method of tagging critical parts in the \dword{apa} is also under development for efficient integration.

The \dword{qa} related to integration and installation is heavily based on the \dword{pdsp} experience and at this point no dedicated \dword{qa} protocol is developed. The full development is done by the installation and integration team in the \dword{apa} consortium. 

\subsubsection{Quality Control at the Integration Facility}
\label{sec:fdsp-apa-install-qc_if}

All the active detector components are shipped to the \dword{itf} for integration and for testing, where more time is available 
to perform tests. This step is critical for ensuring high performance of the integrated \dwords{apa}. The exact time scale of \dword{apa} testing needs to be finalized based on information from the production sites and on the installation schedule.

After unpacking an \dword{apa} at the \dword{itf}, a thorough visual inspection is performed. Tension measurements are made for a sample of around \num{350} wires (representing $\sim$10\% of the wires). The default technique is the laser method that has been used for \dword{pdsp}.  The method works well, but is time consuming, so alternative methods that use voltage measurements are also being pursued to reduce the measuring time. Such improved methods could allow a larger number of wires (even the full \dword{apa}) to be measured. 

Tension values are recorded in the database and compared with the original tension measurements performed at the production sites. Definite guidance for the acceptable tension values will be available to inform decisions on the quality of the \dword{apa}. Clear pass/fail criteria 
will be provided as well as clear procedures to deal with individual wires laying outside the acceptable values. 
This guidance will be based on the \dword{pdsp} experience, where the tension of some wires have changed during the production to installation process. In addition, a continuity test and a leakage current test is performed on all the wires and the data is also recorded in the database. 

Once the electronics are installed by the electronics consortium, dedicated testing of the \dword{apa} readout is performed. The integrated \dword{apa} is inserted in the cold box so that the electronics performance can be tested adequately. Strict guidance is provided 
for assessing the pass/fail criteria for each \dword{apa} during these tests. Here too, guidance from \dword{pdsp} and development tests will guide the exact criteria. Close collaboration with the electronics consortium is necessary.

When all the tests have been successfully performed and more than \num{99}\,\% of the channels are confirmed functional, the \dword{apa} is  tagged as `'good'' and prepared for shipment to \surf{}.

\subsubsection{Quality Control Underground}
\label{sec:fdsp-apa-install-qc_underground}

There are three opportunities to test the \dwords{apa} underground: in the storage area, once secured in front of the \dword{tco}, or once positioned at their final location in the cryostat. The latter is the most important and it may save time to perform the final tests once the full \textit{APA-CPA-APA-CPA-APA} wall is installed (\dwords{cpa} are described in Chapter~\ref{ch:fdsp-hv}). This brings the risk that if serious problems are found, \dwords{apa} are harder (more time-consuming) to move out.

The \dwords{apa} are unpacked in the storage area underground (see Figure~\ref{fig:storage}). Space in this area is very limited and only visual inspection is performed during unpacking. If clear defects are visible, the \dword{apa} is returned to the \dword{itf} for further investigation.

\begin{dunefigure}[Schematics of the underground storage area; full A-C-A-C-A wall in the cryostat]{fig:storage}{Left: A schematic of the layout for the storage and unpacking area underground. Right: A schematic of the layout of a full APA-CPA-APA-CPA-APA wall installed in the cryostat.}
\setlength{\fboxsep}{0pt}
\setlength{\fboxrule}{0.5pt}
\fbox{\includegraphics[height=0.26\textheight]{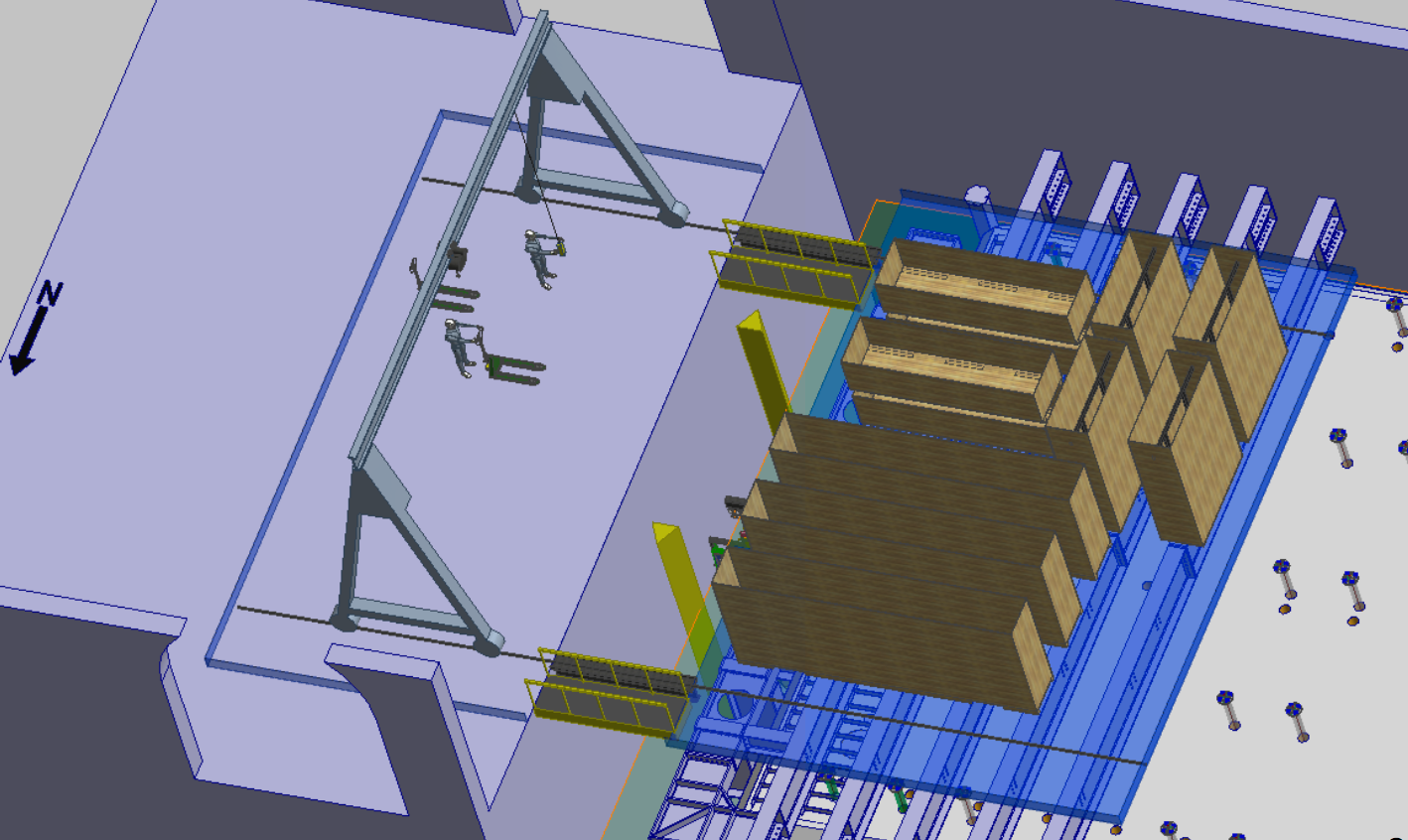}} 
\fbox{\includegraphics[height=0.26\textheight,trim=0mm 2mm 2mm 0mm,clip]{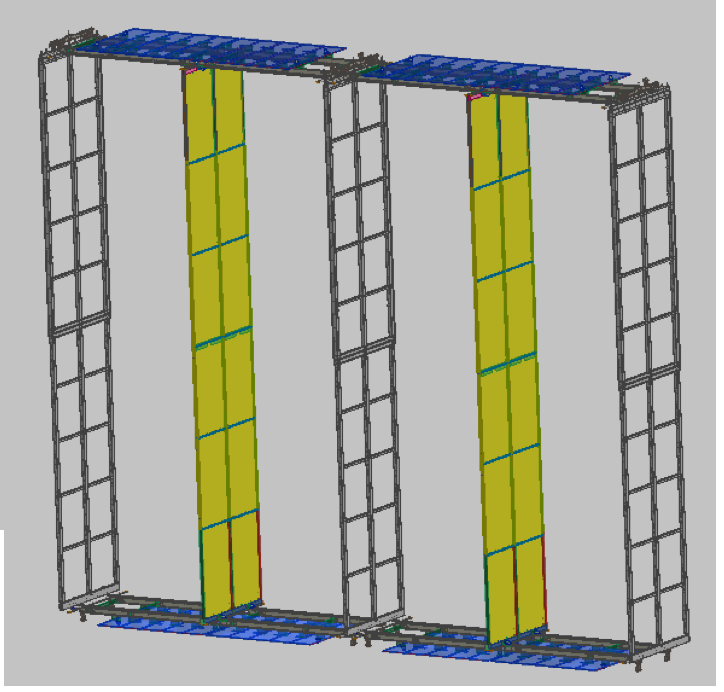}}
\end{dunefigure}

Pairs of \dwords{apa} (top and bottom) are lowered in front of the \dword{tco} to be linked and cabled. Once the cabling is finished a connection test is performed to ensure adequate cabling. Due to the very restrictive space near the \dword{tco} (see Figure~\ref{fig:handling}), no additional tests other than visual inspection are performed at that time, and the cabled and linked \dwords{apa} are positioned in their final location in the cryostat.


The current goal is to install a full APA-CPA-APA-CPA-APA wall every week (see Figure~\ref{fig:storage}, right). After each wall is installed, the night crew has time for final testing of the installed \dword{apa}. There are currently two testing models, one where the night crew tests \dword{apa} pairs as they are installed (every two days), and the other model where the night crew tests the full wall at once. The decision between the two models will be made when accurate estimates of the time needed for the testing become available.


The tests are 
the same described above at the \dword{itf}. Tension on a smaller set of wires is measured ($\sim$5\%, potentially more if a quicker tension method is developed) to ensure that the installation operations did not alter the \dwords{apa}. Since the complete integration is now done, a full readout test can be performed. Short runs are taken with the \dword{daq} system to ensure that the readout is fully operational. The details of these tests still need to be developed to provide efficient assessment of the integrated \dwords{apa}. If an \dword{apa} appears to have more than \num{1}\,\% of the channels not functioning, the \dword{apa} is sent back to the \dword{itf}.

\subsubsection{Quality Assurance}

We will rely on the \dword{pdsp} experience to assess most of the \dword{qa} protocols. The dedicated \dword{qa} plan during production should ensure that the \dwords{apa} meet the requirements and the installation steps should not modify them. The control of the quality of each wire along the installation steps will ensure fully functioning \dwords{apa}. The detailed \dword{qa} program is currently under development by the installation and integration group in the \dword{apa} consortium.

\section{Safety}
\label{sec:fdsp-apa-safety}

Building on the experience of \dword{pdsp}, a full safety analysis will be performed and a set of safe work procedures developed for all stages of the fabrication process before the start of DUNE \dword{apa} production.  In the final design of the winding machine, central to the production process, safety must be taken into account right from the design stage and must be kept in mind at all stages in the life of the machine: design, manufacture, installation, adjustment, operation, and maintenance.  Handling of the large, but delicate frames is a major challenge and safe procedures will be developed for all phases of construction, including frame assembly, wiring, transport, and integration and installation in the cryostat.         

At the factory sites, safety is ultimately the responsibility of the host institutions, and all local rules and regulations must be followed.  However, common job hazard analyses can be performed and documents prepared for many shared aspects of the tooling and activities.  In addition, safety will be an important element of production readiness reviews that are conducted for the project overall and for the factory sites individually.   

\section{Organization and Management}
\label{sec:fdsp-apa-org}

\subsection{APA Consortium Organization}
\label{sec:fdsp-apa-org-consortium}

The \dword{apa} consortium comprises \num{21} institutions, of which \num{13} are from the USA, seven from the UK, and one from the Czech Republic. The consortium is organized along the main deliverables, which are the final design of the \dword{apa} and the \dword{apa} production and installation procedures. Since the two main centers for \dword{apa} construction are expected to be located in the USA and the UK, there are usually two leaders of each working group, representing the main stakeholders (Figure~\ref{fig:apa-consortium-structure}). This is particularly important to ensure that common procedures and tooling are developed. 

\begin{dunefigure}[\dword{apa} Consortium organizational chart]{fig:apa-consortium-structure}
{\dword{apa} Consortium organizational chart}
\includegraphics[width=0.80\textwidth]{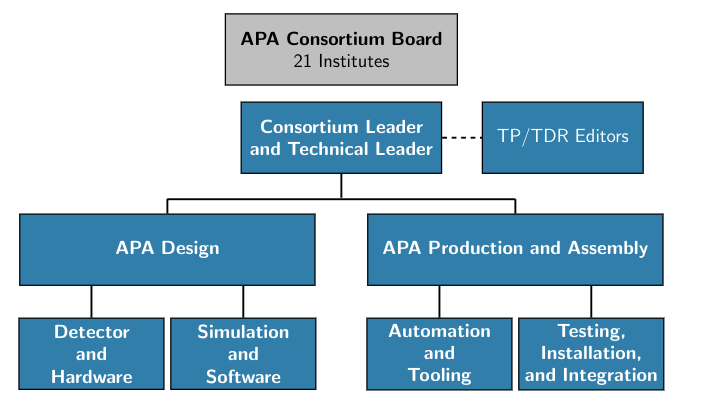}  
\end{dunefigure}

\subsection{Planning Assumptions}
\label{sec:fdsp-apa-org-assmp}

The planning assumptions are based on having eight to nine \dword{apa} assembly lines, at different locations in the UK and the USA. 
We assume about one year of setup time for the factories.
It will take of the order \num{50} shifts to construct a single \dword{apa}. Assuming a multi-shift system, we will be able to construct the \num{150} \dwords{apa} required for one \dword{spmod} 
within about two years.

\subsection{WBS and Responsibilities}
\label{sec:fdsp-apa-org-wbs}

Here, we only discuss the top-level WBS elements, which are (1)~design, engineering and R\&D, (2)~production setup, (3)~production, (4)~integration, and (5)~installation.

The validation of the design is mainly a responsibility of the university groups and BNL, while engineering and the production setup will be developed at PSL in Madison (USA) and Daresbury Laboratory (UK), where the \dwords{apa} for \dword{pdsp} have been built, with contributions from university groups. In addition to PSL and Daresbury Laboratory, the University of Chicago and Yale University have been identified as candidate sites for the production. The production sites will require significant contributions from university groups during the production process. 

In total, we expect half of the \dwords{apa} to be produced in the USA and half in the UK. The steel for the frames is most likely to be bought from a single vendor. The assembly of the frames will be performed in the USA and the UK separately. The options to assemble the frames in house or in collaboration with industrial partners are still being explored. 

Other significant components include on-\dword{apa} electronics boards. Design modifications relative to \dword{pdsp} are the responsibility of BNL. The boards will be produced by industry, while the testing will be distributed among consortium institutions. The shipping of the \dwords{apa} is the responsibility of the production factories in the USA and the UK.  The integration and installation are a joint responsibility of the Consortium, with ANL providing the interface with the technical coordination group.

\subsection{High-level Milestones and Schedule}
\label{sec:fdsp-apa-org-cs}

The high-level milestones for the period 2018 to 2024 are given in Table~\ref{tab:milestones} for the periods before and after the \dword{tdr}. The final design of the \dwords{apa} to be proposed in the \dword{tdr} will be informed by the experience of the \dword{pdsp} \dword{apa} production and performance, which will be reviewed in early 2019. Additional design considerations that cannot be directly tested through \dword{pdsp}, such as the two-\dword{apa} assembly and the related cabling issues, will require a full test with cabling of a two-\dword{apa} assembly also in early 2019. The production schedule, the required number of assembly lines, and the location of the production factories will depend on the improvements of the wire winding procedures, which will formally be reviewed in January 2019. The post-\dword{tdr} milestones are driven by the high-level international project milestones and are based on a schedule with one year factory preparation and about two years of \dword{apa} construction time.

\begin{dunetable}[\dword{apa} design and construction milestones]{ll}{tab:milestones}{\dword{apa} design and construction milestones}
Date &  Milestone   \\ \toprowrule
\multicolumn{2}{c}{Pre-\dword{tdr}}\\ \specialrule{1.5pt}{1pt}{1pt}
December 2018 & Test two-\dword{apa} assembly   \\ \colhline
January 2019  & Formal review of complete modifications to the winder design\\ \colhline
February 2019 & Formal review of \dword{pdsp} \dword{apa} performance \\ \colhline
February 2019 & Complete assembly test of \dword{fd} prototype \dword{apa}\\ \colhline
March 2019 & Decision on location of factories and required number of assembly lines \\ \colhline
March 2019 & \dword{apa} cost estimate for \dword{spmod} \\ \colhline
March 2019 & \dword{apa} schedule for \dword{spmod} \\ \colhline
April 2019 & \dword{apa} section of \dword{tdr} delivered \\ \colhline
\multicolumn{2}{c}{Post-\dword{tdr}}\\ \specialrule{1.5pt}{1pt}{1pt}
2020 & Preparation of \dword{apa} factories \\ \colhline
2021 -- 2023 & Construction of \dwords{apa} \\ \colhline
2022/3 & Installation of \dwords{apa} in \dword{spmod} 1\\ \colhline
2024 & Commissioning of \dword{spmod} 1 \\ \colhline
\end{dunetable}

\cleardoublepage

\chapter{TPC Electronics}
\label{ch:fdsp-tpc-elec}

\section{TPC Electronics (CE) System Overview}
\label{sec:fdsp-tpc-elec-ov}

\subsection{Introduction}
\label{sec:fdsp-tpc-elec-ov-intro}
DUNE single-phase time projection chamber TPC electronics hardware signal processing takes place inside the \lar, in boards that are directly mounted on the \dword{apa}; accordingly, the TPC readout electronics are referred to as the \dword{ce}.  The electronics are mounted inside the \lar to exploit the fact that charge carrier mobility in silicon is higher and that thermal fluctuations are lower at \lar temperature than at room temperature.  For \dword{cmos} (complementary metal-oxide-semiconductor) electronics, this results in substantially higher gain and lower noise at \lar temperature than at room temperature~\cite{larCMOS}.  Mounting the front-end electronics on the \dword{apa} frames also minimizes the input capacitance.  Furthermore, placing the digitizing and multiplexing (MUX) electronics inside the cryostat reduces the total number of penetrations into the cryostat and minimizes the number of cables coming out of the cryostat.  As the full TPC electronics chain for the \dword{spmod} includes many components on the warm side of the cryostat as well, the DUNE consortium designated to organize development of this system is called the DUNE \textit{Single-Phase TPC Electronics} consortium. It is sometimes referred to as the \dword{ce} consortium for short.

The overall noise requirement drives the choice of architecture of the TPC electronics. This requirement is difficult to establish precisely, but it is clear that the lower the electronic noise is, the greater the physics reach of the DUNE experiment will be.  An equivalent noise charge (\dword{enc}) of less than approximately 1000$e^-$ is required for satisfactory reconstruction of accelerator neutrino interactions, but a lower noise level will yield significantly better two-track separation and primary vertex resolution, and thus higher efficiency and/or lower background for identifying electron neutrino interactions.  Setting the noise level requirement for the DUNE \dword{spmod} more precisely is an ongoing effort.

The noise level enabled by having the front-end electronics in the cold (roughly half as much noise at \lar temperature than at room temperature) greatly extends the reach of the DUNE physics program.  Decreasing the noise level allows for smaller charge deposits to be measurable, which acts as a source of risk mitigation in the case that the desired drift field can not be reached or the electron lifetime in the detector is lower than desired (due to the electronegative impurities in the detector), and also increases the reach of low-energy physics measurements such as those associated with stellar core-collapse supernova burst neutrinos.  Finally, the low noise level allows the experiment to utilize low-energy $\mathrm{{}^{39}Ar}$ beta decays for the purpose of calibration in the DUNE \dword{spmod}.  The noise level requirement of \dword{enc}\,$<\num{1000}\,e^-$ will allow for the use of $\mathrm{{}^{39}Ar}$ beta decays in calibrations at the DUNE \dword{spmod}.

In order to retain maximum flexibility to optimize reconstruction algorithms after the DUNE data is collected, the \dword{spmod} electronics are designed to produce a digital record that is a representation of the waveform of the current produced by charge collection/induction on the anode wires.  Each anode wire signal is input to a charge sensitive amplifier, followed by a pulse shaping circuit and an \dword{adc}.  In order to minimize the number of cables and cryostat penetrations, the \dwords{adc} as well as the amplifier/shapers are located in the \lar, and digitized data from many wires are merged onto a much smaller set of high speed serial links.  


\subsection{System Description, Scope and Current Status}
\label{sec:fdsp-tpc-elec-ov-scope}
The \dword{ce} signal processing is implemented in application-specific integrated circuits (\dwords{asic})
using \dword{cmos} technology.  The \dword{ce} is continuously read out, resulting in a digitized \dword{adc}
sample from each \dword{apa} channel (wire) up to every \SI{500}{ns} (\SI{2}{MHz} sampling rate).

Each individual \dword{apa} has \num{2560} channels that are read out by \num{20} 
\dfirsts{femb}, with
each \dshort{femb} enabling digitized wire readout from \num{128} channels.  One cable bundle connects each \dshort{femb} to
the outside of the cryostat via a \dword{ce} signal cable flange located at the \dword{ce} \fdth at the
top of the cryostat, where a single flange services each \dword{apa}, as shown in Figure~\ref{fig:connections}.  Two \dword{ce} signal flanges are located on each \fdth, together accounting for all electronics channels associated with a pair of \dwords{apa} (upper and lower, vertically arranged).
Each cable bundle contains wires for low-voltage (\dword{lv}) power, high-speed data readout, and
clock or digital-control signal distribution.  Eight separate cables carry the TPC wire bias voltages
from the signal flange to the \dword{apa} wire bias boards, in addition to the bias voltages for the field
cage termination electrodes and for the electron diverters.  An additional flange on the top of each \fdth services the \dword{pds} cables associated with the \dword{apa} pair.

\begin{dunefigure}
[Connections between the signal flanges and \dword{apa}]
{fig:connections}
{Connections between the signal flanges and \dword{apa}. Only the upper \dword{apa} of the hanging \dword{apa} pair, described in Section~\ref{sec:fdsp-apa-frames}, and its connection paths are shown. The lower \dword{apa} shares the \dword{pd} flange with the upper \dword{apa} but has a separate TPC readout flange. A \textit{\dword{ce} module} consists of all \dword{ce} associated with \num{128} channels of digitized readout.}
\includegraphics[width=0.9\textwidth]{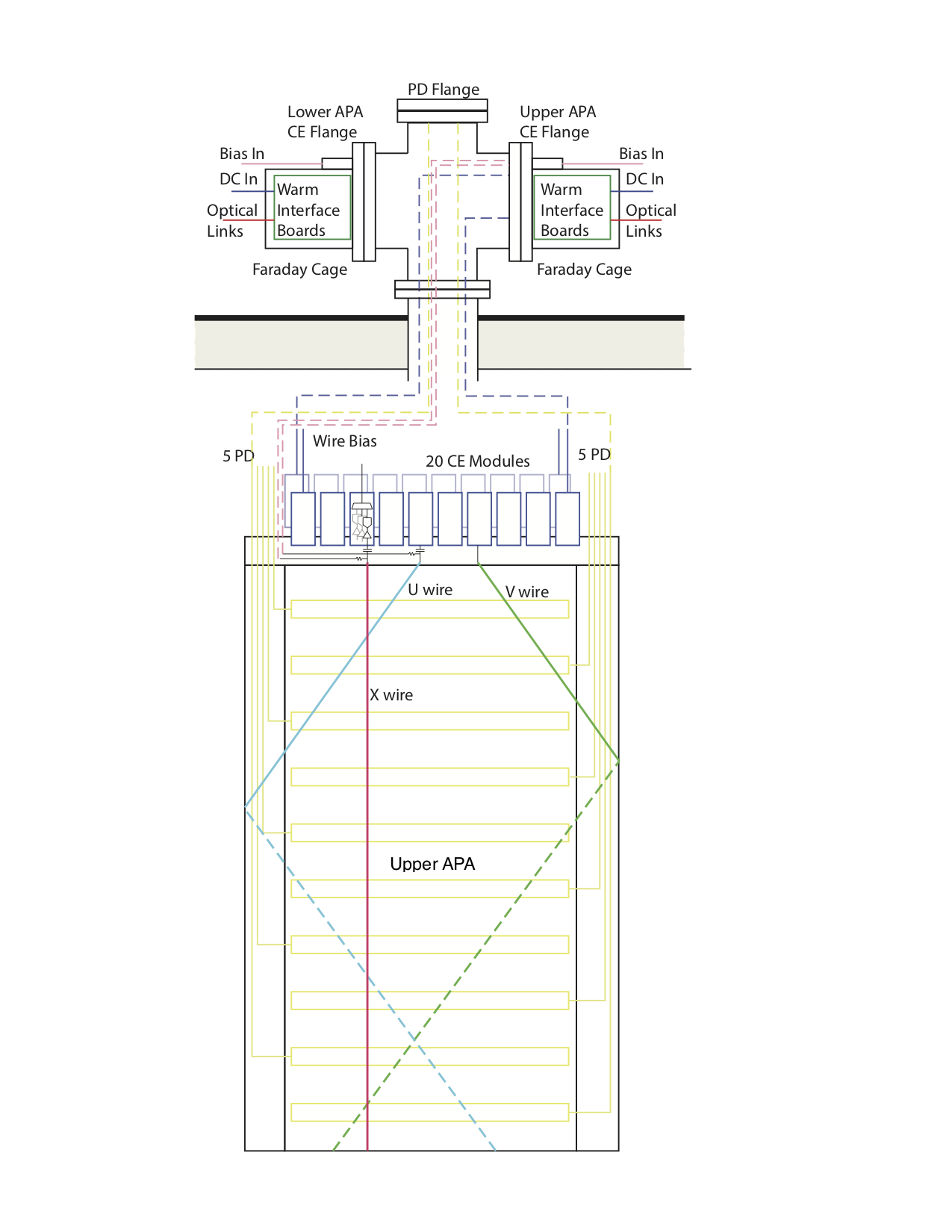}
\end{dunefigure}

The components of the \dword{ce} system are the following:
\begin{itemize}
\item{\dwords{femb}, on which the \dwords{asic} are mounted, which are installed on the \dwords{apa};}
\item{cables for the data, clock and control signals, \dword{lv} power, and wire bias voltages between the \dword{apa} and the signal flanges (cold cables);}
\item{signal flanges with a \dword{ce} \fdth to pass the data, clock and control signals, \dword{lv} power, and \dword{apa} wire-bias voltages between the inside and outside of the cryostat, in addition to the corresponding cryostat penetrations and spool pieces;}
\item{
\dwords{wiec} that are mounted on the signal flanges and contain
the 
\dwords{wib} and 
\dwords{ptc} for further processing
and distribution of the signals entering and exiting the cryostat;}
\item{cables for \dword{lv} power and wire bias voltages between the signal flange and external power
supplies (warm cables); and}
\item{\dword{lv} power supplies for the \dword{ce} and bias-voltage power supplies for the \dwords{apa}.}
\end{itemize}

Table~\ref{tab:elecNums} lists the component type, the quantity required for each type  and the number of channels per component of each type.

\begin{dunetable}
[TPC electronics components and quantities for a single \dword{apa} of a 
\dword{spmod}.]
{llr}
{tab:elecNums}
{TPC electronics components and quantities for a single \dword{apa} of the DUNE \dword{spmod}.}
\textbf{Element} &\textbf{Quantity} & \textbf{Channels per element}\\ \toprowrule
Front-end mother board (\dword{femb}) & \num{20} per \dword{apa} & \num{128} \\ \colhline
FE \dword{asic} chip & \num{8} per \dword{femb} & \num{16} \\ \colhline
\dword{adc} \dword{asic} chip & \num{8} per \dword{femb} & \num{16} \\ \colhline
\dword{coldata} \dword{asic} chip & \num{2} per \dword{femb} & \num{64} \\ \colhline
Cold cable bundle & \num{1} per \dword{femb} & \num{128} \\ \colhline
Signal flange & \num{1} per \dword{apa} & \num{2560} \\ \colhline
\dword{ce} \fdth & \num{1} per \dword{apa} & \num{2560} \\ \colhline
Warm interface board (\dword{wib}) & \num{5} per \dword{apa} & \num{512} \\ \colhline
Warm interface electronics crate (\dword{wiec}) & \num{1} per \dword{apa} & \num{2560} \\ \colhline
Power and timing card (\dword{ptc}) & \num{1} per \dword{apa} & \num{2560} \\ \colhline
Power and timing backplane (PTB) & \num{1} per \dword{apa} & \num{2560} \\
\end{dunetable}

The baseline design for the \dword{spmod} TPC electronics calls for three types of \dwords{asic} to be located inside of the \lar:
\begin{itemize}
\item{a \num{16}-channel \dword{fe} \dword{asic} for amplification and pulse shaping, referred to as \dword{larasic} in the following;}
\item{a \num{16}-channel \num{12}-bit \dword{adc} \dword{asic} operating at \SI{2}{MHz}; and}
\item{a \num{64}-channel control and communications \dword{asic}, referred to as \dword{coldata} in the following.}
\end{itemize}

The \dword{fe} \dword{asic} has been prototyped and is close to meeting requirements (discussed in Section~\ref{sec:fdsp-tpc-elec-ov-req}). Another prototype to address issues in the version deployed in \dword{pdsp} is expected in the spring of 2018. Key portions of the control and communications \dword{asic} (also referred to as the \dword{coldata} \dword{asic}) have been prototyped and meet requirements.  However, it has been determined that the BNL-designed P1-\dword{adc} \dword{asic} now being used in \dword{pdsp} does not meet requirements, and accordingly, its development has been terminated.  A new \dword{adc} \dword{asic} (referred to as the cold \dword{adc} \dword{asic}) is being developed by an LBNL-\fnal-BNL collaboration and first prototypes are expected by the end of summer 2018.  The first full prototype of the controls and communication \dword{asic} is also expected to be available for testing by the end of summer 2018. 

In order to maximize the probability of developing a complete design for cold TPC \dword{fe} electronics in a timely fashion, an alternative solution is also being investigated, a single \num{64}-channel \dword{asic} that will consolidate all three functions described above.  This design is being done at SLAC and first prototypes are expected in summer 2018.  An \dword{adc} solution in the form of a developmental \dword{adc} chip for an upgrade of the ATLAS detector provides an additional backup option; this option will be explored further if the performance of the other two \dword{adc} solutions being considered do not meet the requirements for DUNE.

While the higher charge carrier mobility at \lar temperature than at room temperature is central to the improved performance of \dword{ce}, it also leads to the \textit{hot carrier effect}.  In n-type MOS transistors, the carriers (electrons) can acquire enough kinetic energy to ionize silicon in the active channel.  This charge can become trapped and lead to effects (including threshold shifts) similar to those caused by radiation damage.  This effect can cause MOS circuits to age much more quickly at \lar temperature than at room temperature, reducing performance and potentially causing failure.  In order to mitigate this effect, the maximum \efield in transistor channels must be lower than the field that can be reliably used at room temperature.  
This is accomplished by using transistors that are fabricated with longer than typical length and operated at reduced bias voltage.  Any commercial circuits that are used in the \lar must be carefully tested to ensure that they will perform well for the expected \num{20}-year lifetime of DUNE.

A series of tests are planned to demonstrate that the \dword{ce} system design will meet DUNE requirements. These include two system tests: one using the \dword{pdsp} \textit{cold box} at CERN, and one using a new small \lartpc at \fnal. The latter will also accommodate one half-length DUNE \dword{pd}, and will provide a low-noise environment that will allow one to make detailed comparisons of the performance of the new \dwords{asic}. It will also enable the study of interactions between the TPC readout and other systems, including the \dword{pd} readout and the \dword{hv} distribution. These test facilities are discussed in more detail in Section~\ref{sec:fdsp-tpc-elec-qa-facilities}. Plans are also being made for a second period of data taking for the \dword{pdsp} detector, with final \dwords{apa} including the final \dwords{asic} and \dwords{femb} replacing the current prototypes. This second run of \dword{pdsp} is planned for 2021-2022.

\subsection{System Requirements}
\label{sec:fdsp-tpc-elec-ov-req}
In addition to the noise requirement (less than \num{1000}\,$e^{-}$), several additional requirements determine most of the other important TPC electronics specifications.  These are:

\begin{itemize}
\item{The \dword{fe} peaking time must  
be  in the range \numrange{1}{3}\,\si{\micro\second}.  This requirement is derived primarily from the time required for drifting charges to travel from one plane of anode wires to the next.}
\item{The \dword{fe} must 
 have an adjustable baseline.  This requirement reflects the fact that the signal from induction wires is bipolar while the signal from the collection wires is mostly unipolar.}
\item{The \dword{adc} sampling frequency must 
be \SI{2}{MHz}.  This value is chosen to match a \dword{fe} shaping time of \SI{1}{\micro\second} (approximate Nyquist condition) while minimizing the data rate.}
\item{The system must 
have a linear response up to an impulse input of at least \num{500000}\,$e^{-}$.  This roughly corresponds to the charge collected on a single wire from one stopping proton and two more highly ionizing protons, all assumed to have trajectories at \num{45}\,$^{\circ}$ with respect to the beam axis.  This number was chosen so that saturation will occur in less than \SI{5}{\%} of beam related events.  Studies are ongoing (including an evaluation of \lariat~\cite{Cavanna:2014iqa} data and simulation studies) to better understand this requirement.}
\item{The dynamic range of the system must be at least \num{3000}:\num{1}. This number is given by the ratio between the maximum signal for no saturation and 50\% of the lowest possible noise level.  It implies a \num{12}-bit \dword{adc}.}
\item{The \dword{adc} must not contribute significantly to overall \dword{fe} noise. This requirement is dependent on the gain of the \dword{fe}, but for each gain setting translates into requirements on \dword{adc} parameters including non-linearity and noise.}
\item{The power dissipated by the electronics located in the \lar must 
be less than \SI{50}{mW/channel}.  Lower power dissipation is desirable because the mass of the power cables scales with the power.  Studies are ongoing to understand if the amount of power dissipated by the electronics should be minimized further due to potential complications from argon boiling; in principle this should not be a problem because the \dword{ce} boxes housing the \dwords{femb} are designed to channel bubbles to the \dword{apa} frames.}
\end{itemize}

Finally, all electronics located in the \lar must 
be highly reliable because it will not be possible to access the \dword{ce} for repair once the cryostat is filled with \lar. Studies are ongoing to quantify the impact of failures in the TPC and electronics, including single wire failures, and failures of groups of \num{16}, \num{64}, or \num{128} channels.

\section{System Design}
\label{sec:fdsp-tpc-elec-design}

\subsection{Grounding and Shielding}
\label{sec:fdsp-tpc-elec-design-ground}
In order to minimize system noise, the \dword{ce} cables for each \dword{apa} enter 
the cryostat through a single \dword{ce} flange, as shown in Figure~\ref{fig:connections}, creating, for grounding purposes, and integrated unit consisting of an \dword{apa} frame, \dword{femb} ground for all \num{20} \dword{ce} modules, TPC flange, and warm interface
electronics. To accomplish this,
the input amplifiers on the \dword{fe} \dwords{asic} have their ground terminals connected to the \dword{apa} frame. 
All power-return leads and cable shields are connected to both the ground plane of the \dword{femb} and to the TPC signal flange.

The only location where this integrated unit makes electrical contact with the 
cryostat, which defines \textit{detector ground}, is at a single point on the \dword{ce} \fdth board in the TPC signal flange where the 
cables exit the cryostat. Mechanical suspension of the \dwords{apa} is accomplished using insulated supports. 
To avoid structural ground loops, the \dword{apa} frames described in Chapter~\ref{ch:fdsp-apa} are 
insulated from each other.

Filtering circuits for the \dword{apa} wire-bias voltages are locally referenced to the ground plane of the \dwords{femb} through low-impedance electrical connections. This approach ensures a ground-return path in close proximity to the bias-voltage and signal paths. The close proximity of the current paths minimizes the size of potential loops to further suppress noise pickup.

Signals associated with the \dword{pds}, described in Chapter~\ref{ch:fdsp-pd}, are carried directly on shielded, 
twisted-pair cables to the signal \fdth. The cable shields are connected to the cryostat 
at the \dword{pd} flange shown in Figure~\ref{fig:connections}, and to the PCB shield layer on the \dwords{pd}. There is no electrical connection between the cable shields and the \dword{apa} frame.


\subsection{Connections from Wire to Front-End}
\label{sec:fdsp-tpc-elec-design-bias}
Each side of an \dword{apa} includes four wire layers as described in Section~\ref{sec:fdsp-apa-design}. 
Electrons passing through the wire grid must drift unimpeded until they reach the $X$-plane 
collection layer. The nominal bias voltages are predicted to result in this electrically 
transparent configuration, and are given in Section~\ref{sec:fdsp-apa-design}. 

The filtering of wire bias voltages and AC coupling of wire signals passing
onto the charge amplifier circuits is done on capacitance-resistance (CR) boards that plug in between the \dword{apa} wire-board stacks and \dwords{femb}.
Each CR board includes single RC filters for the $X$- and $U$-plane wire bias voltages. In addition, each board has \num{48} 
pairs of bias resistors and AC coupling capacitors for $X$-plane wires, and \num{40} pairs for the $U$-plane wires. The coupling capacitors block DC while passing AC 
signals to the \dword{ce} motherboards.  A schematic diagram of the \dword{pdsp} \dword{apa} wire bias subsystem is illustrated in Figure~\ref{fig:CR-board}.

\begin{dunefigure}
[\dword{pdsp} \dword{apa} wire bias schematic diagram, including the CR board.]
{fig:CR-board}
{\dword{pdsp} \dword{apa} wire bias schematic diagram, including the CR board.}
\includegraphics[width=0.8\linewidth]{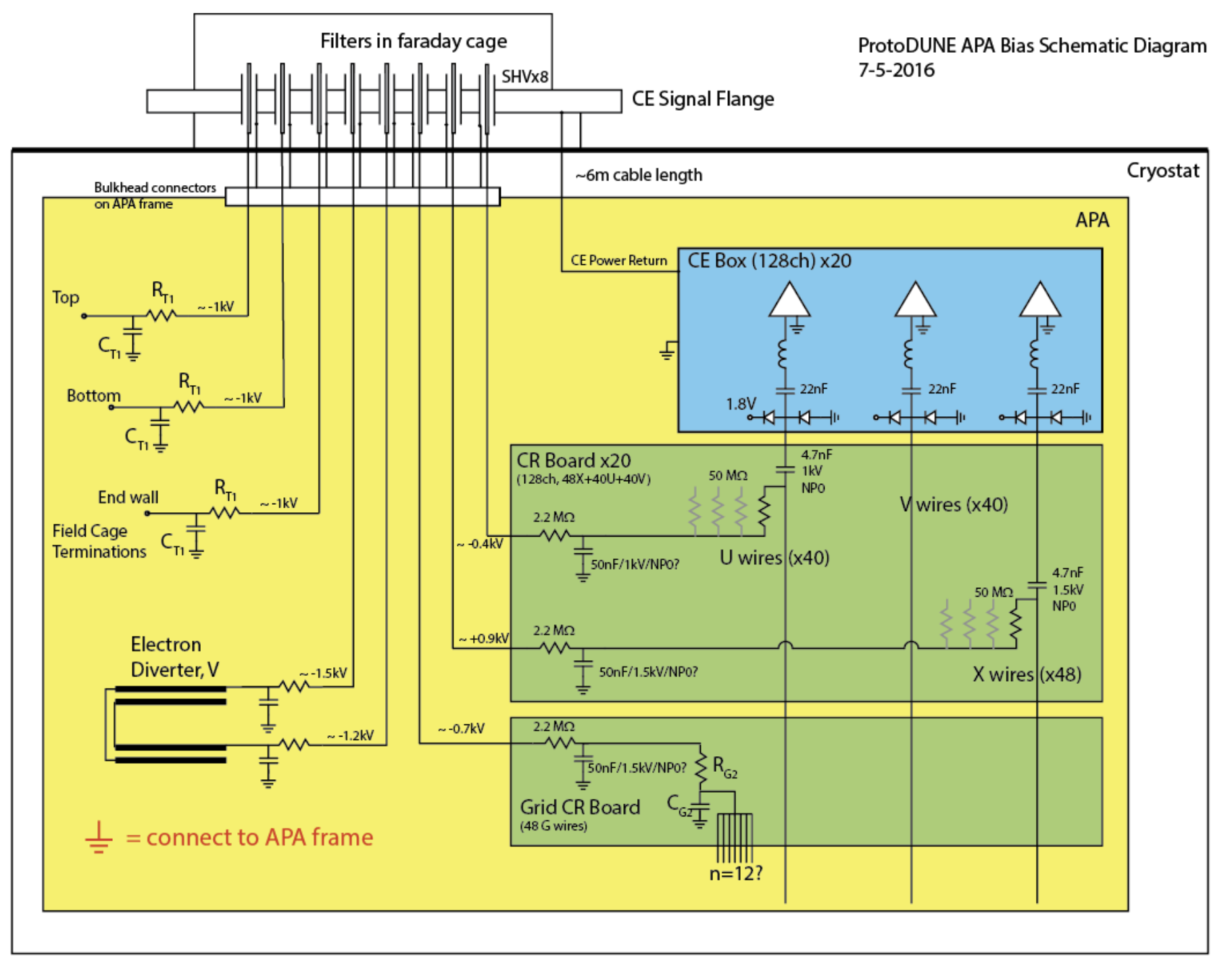}
\end{dunefigure}


Clamping diodes limit the input voltage received at the amplifier circuits to between \SI{1.8}{V}\,$\pm$\,U$_D$, where U$_D$
is the breakdown voltage of the diode, $\sim$\SI{0.7}{V}.
The amplifier circuit has a \SI{22}{nF} coupling capacitor at input to avoid leakage current from the protection clamping diodes. 


Bias resistance values should be at least \SI{20}{\mega\ohm} to maintain negligible noise contributions.
The higher value helps to achieve a longer time constant for the high-pass coupling networks.
Time constants should be at least \num{25} times the electron drift time so that the undershoot in the digitized waveform
is small and easily correctable.
However, leakage currents can develop on PC boards that are exposed to high voltages over extended periods.
If the bias resistors are much greater than \SI{50}{\mega\ohm}, leakage currents may affect the bias voltages applied to the wires. The target value of \SI{50}{\mega\ohm} was used in \dword{pdsp}.

The bias-voltage filters are RC low-pass networks.
Resistance values should be much smaller than the bias resistances to control crosstalk between wires
and limit the voltage drop if any of the wires becomes shorted to the \dword{apa} frame.
The value of \SI{2.2}{\mega\ohm} was used in \dword{pdsp}.
Smaller values may be considered for 
the \dword{spmod} although a larger filter capacitor would be required to maintain a given level of noise reduction.
The target value of \SI{47}{nF} was used in \dword{pdsp} for the filter capacitors.


\subsection{Front-End Mother Board (FEMB)}
\label{sec:fdsp-tpc-elec-design-femb}

\subsubsection{Overview}
\label{sec:fdsp-tpc-elec-design-femb-ov}
Each \dword{apa} is instrumented with \num{20} 
\dwords{femb}.
The \dwords{femb} plug into the \dword{apa} CR boards, making the connections from the wires to the charge amplifier circuits as short as possible.
Each \dword{femb} receives signals from \num{40} $U$ wires, \num{40} $V$ wires, and \num{48} $X$ wires.
The baseline \dword{femb} design contains eight \num{16}-channel \dword{fe} (\dword{larasic}) \dwords{asic}, eight \num{16}-channel Cold \dword{adc} \dwords{asic}, and two \dword{coldata} control and communication \dwords{asic} (see Figure~\ref{fig:ce-scheme}).
The \dword{femb} also contains regulators that produce the voltages required by the \dwords{asic} and 
filter those voltages.
The \dword{larasic} inputs are protected by diodes and a series inductor.

\begin{dunefigure}
[The baseline \dword{ce} architecture.]
{fig:ce-scheme}
{The baseline \dword{ce} architecture. The basic unit is the \num{128}-channel \dword{femb}. Note that only one \dword{ce} flange is shown to simplify the illustration. Note that \dword{ssp} stands for \textit{SiPM Signal Processor} (see Chapter~\ref{ch:fdsp-pd}).}
\includegraphics[width=0.9\linewidth]{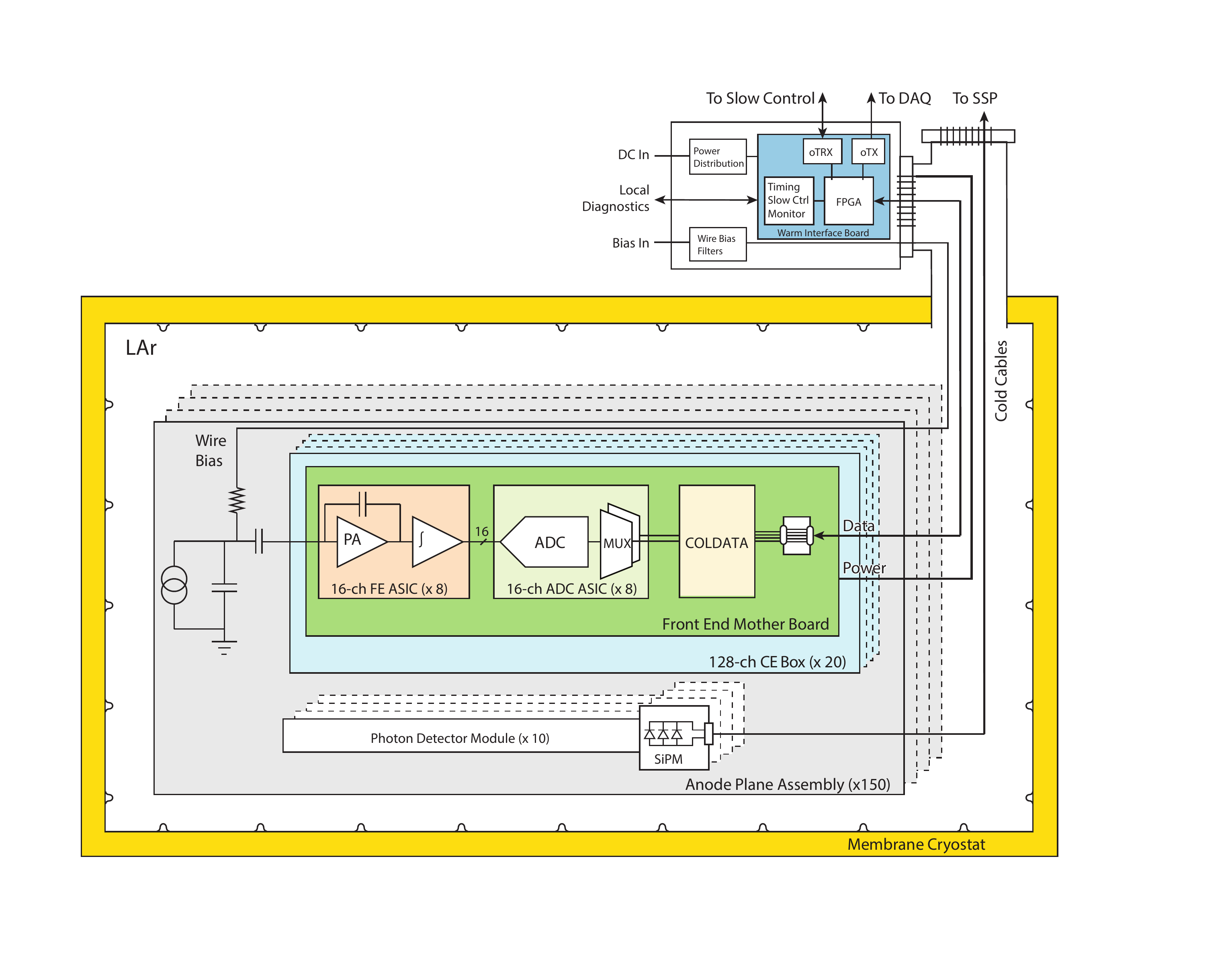}
\end{dunefigure}

The \dword{pdsp} version of the \dword{femb} (which uses a single \dword{fpga} on a mezzanine card instead of two \dword{coldata} \dwords{asic}) is shown in Figure~\ref{fig:femb}.

\begin{dunefigure}
[The complete \dword{femb} assembly as used in \dword{pdsp}.]
{fig:femb}
{The complete \dword{femb} assembly as used in the \dword{pdsp} detector. The cable shown is the high-speed data, clock, and control cable.}
\includegraphics[width=0.6\linewidth]{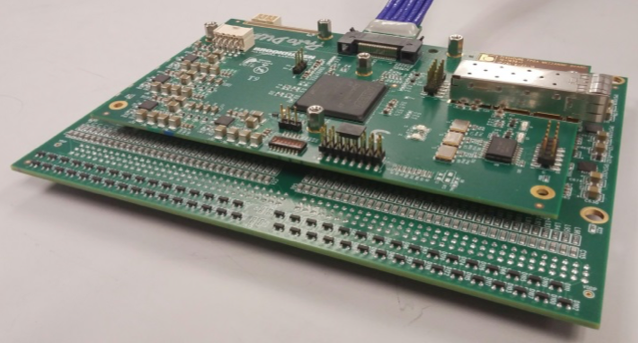}
\end{dunefigure}

\subsubsection{Front-End ASIC}
\label{sec:fdsp-tpc-elec-design-femb-fe}
The \dword{larasic} receives signals from the CR board and
provides a means to amplify and shape the current signals originally coming from the TPC wires; the
shaping serves as an anti-aliasing filter for the TPC signals.
Each \dword{larasic} channel has a charge amplifier circuit with a programmable
gain selectable from one of \num{4.7}, \num{7.8}, \num{14} or \SI{25}{mV/fC}
(corresponding to full-scale charge of \num{300}, \num{180}, \num{100} and \SI{55}{fC}),
a high-order anti-aliasing filter with programmable time
constant (semi-Gaussian with peaking time \num{0.5}, \num{1}, \num{2}, and \SI{3}{\micro\second}),
an option to enable AC coupling,
and a baseline adjustment for operation with either the collecting (\SI{200}{mV} nominal) or the non-collecting (\SI{900}{mV} nominal) wires.

Figure~\ref{fig:fe-output} (left) shows the simulated pulse response for all gains and peaking times and both baselines.
Note that the gain is independent of the peaking time;  the same amount of charge produces the same peak voltage signal regardless of the peaking time.  

Shared among the \num{16} channels in the \dword{larasic} are the bias circuits, programming registers,
a temperature monitor, an analog buffer for signal monitoring, and the digital interface.
The power dissipation of \dword{larasic} is about \SI{6}{mW} per channel at \SI{1.8}{V} supply voltage.

The \dword{larasic} is implemented using the TSMC \SI{180}{nm} \dword{cmos} process.\footnote{TSMC 0.18-micron Technology\texttrademark{}, Taiwan Semiconductor Manufacturing Company Ltd., \url{http://www.tsmc.com/english/dedicatedFoundry/technology/0.18um.htm}.}  The charge sensitive amplifier uses a very large p-channel field effect transistor (PFET) with a width of \SI{20}{mm} and a length of \SI{270}{nm} followed by a dual cascode stage, a pulse shaping network, and a baseline restoration circuit.  

Each channel also implements a high-performance output driver 
that can be used to drive a long cable, but which is disabled when interfaced to an \dword{adc} \dword{asic} to reduce the power consumption.
The \dword{asic} integrates a band-gap reference (BGR) to generate all the internal bias voltages and currents.
This guarantees a high stability of the operating point over a wide range of
temperatures, including cryogenic temperatures.
The \dword{asic} is packaged in a commercial, fully encapsulated plastic QFP~80 package.

\begin{dunefigure}
[Simulated \dword{fe} response to an instantaneous injected charge]
{fig:fe-output}
{Simulated \dword{fe} response to an instantaneous injected charge for all gains and peaking times and both baselines (left); also shown are measured calibration pulse response overlays for \num{2560} electronics channels (baseline subtracted) attached to a \dword{pdsp} \dword{apa} (right).  Note that the truncated negative pulses are due to effects of saturation associated with the collection plane threshold being close to the lower \dword{adc} boundary.}
\includegraphics[width=0.47\linewidth]{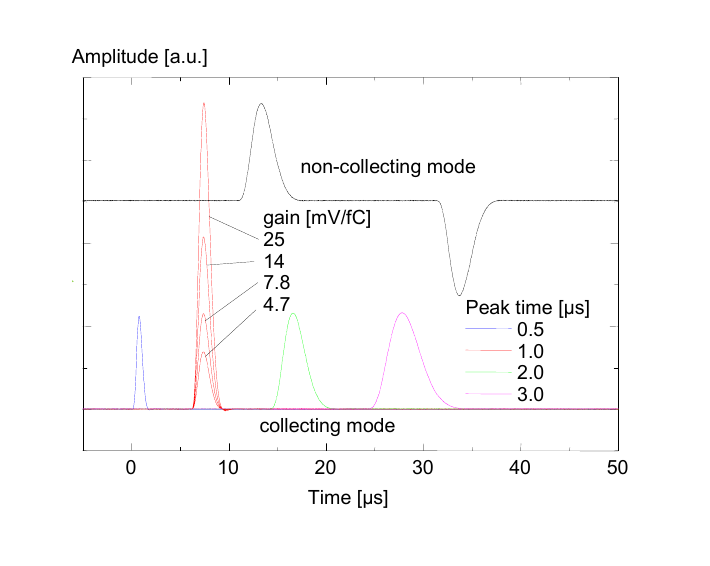}
\includegraphics[width=0.5\linewidth]{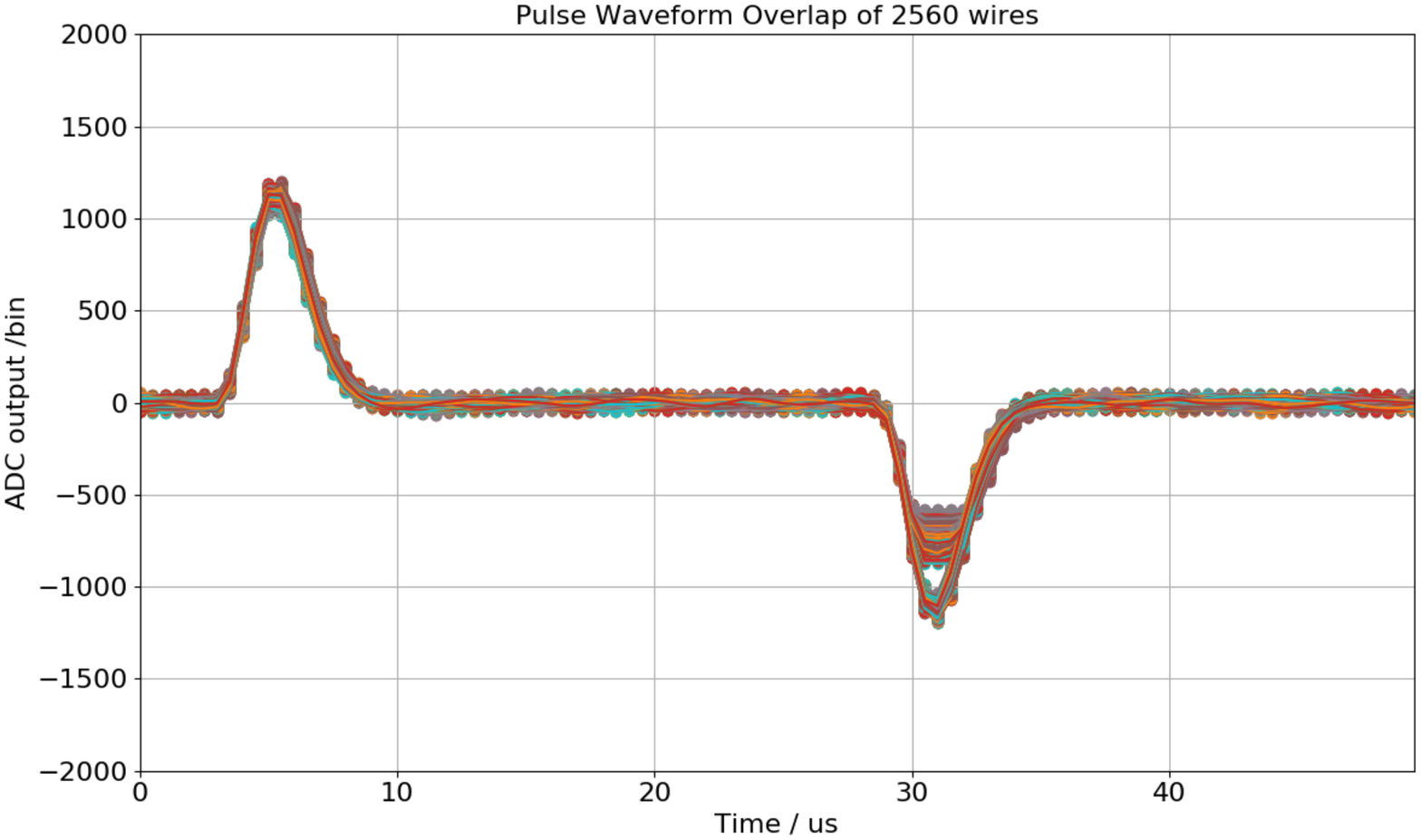}
\end{dunefigure}

Each \dword{fe} \dword{larasic} channel is equipped with an injection capacitor which can be used
for test and calibration and can be enabled or disabled through a
dedicated register. The injection capacitance has been measured to \num{0.5}\% using 
a calibrated external capacitor. The measurements show
that the calibration capacitance is extremely stable, changing from
\SI{184}{fF} at room temperature to \SI{183}{fF} at \SI{77}{K}. This result and the measured
stability of the peaking time demonstrate the high stability of the
passive components as a function of temperature. Channel-to-channel and chip-to-chip
variation in the calibration capacitor are typically less than \num{1}\%. 

Prototype \dwords{larasic} have been evaluated and characterized at room temperature and LN 
(\SI{77}{K}) temperature.
During testing the circuits have been cycled multiple times
between the two temperatures and operated without any change in performance.
Figure~\ref{fig:fe-output} (right) shows the measured injection pulse response overlaid with the baseline subtracted for one full \dword{apa} 
(\num{2560}) electronics channels from \dword{pdsp} \dword{femb} attached to a \dword{pdsp} \dword{apa} in a 
shielded environment at approximately \SI{180}{K}. This contains \num{1600} induction (high-baseline)
and \num{960} collection (low-baseline) channels, the latter of which saturate the negative pulse at the low 
end of the \dword{fe} output. The spread in saturation values between \num{-500} and \num{-750} \dword{adc} bins is due to the
variation in relative position of the \dword{fe} baseline to the low end of the \dword{fe} output 
in the \dword{pdsp} version of the \dword{larasic}.

\subsubsection{Cold ADC}
\label{sec:fdsp-tpc-elec-design-femb-adc}
The baseline option for the DUNE cold \dword{adc} is a new \num{16}-channel low-noise \dword{adc} \dword{asic} intended to read out the \dword{larasic} preamps in the \dword{spmod} \dword{ce}. The \dword{adc} is \num{12} bits and digitizes each channel at a rate of \SI{2}{MHz}. The \dword{adc} accepts single-ended or differential inputs, and outputs a serial data stream to \dword{coldata}, the \dword{spmod} 
digital serializer chip. The \dword{adc} \dword{asic} is implemented using \SI{65}{nm} \dword{cmos} technology. The \dword{asic} uses a conservative, industry standard design along with digital calibration. A block diagram of the \dword{adc} \dword{asic} is shown in Figure~\ref{fig:adc-blockdiagram}. The design and testing of the baseline \dword{adc} \dword{asic} is being carried out by a collaboration of scientists and engineers at BNL, \fnal, and LBNL.

\begin{dunefigure}
[Baseline cold \dword{adc} \dword{asic} block diagram.]
{fig:adc-blockdiagram}
{Baseline cold \dword{adc} \dword{asic} block diagram.}
\includegraphics[width=0.75\textwidth]{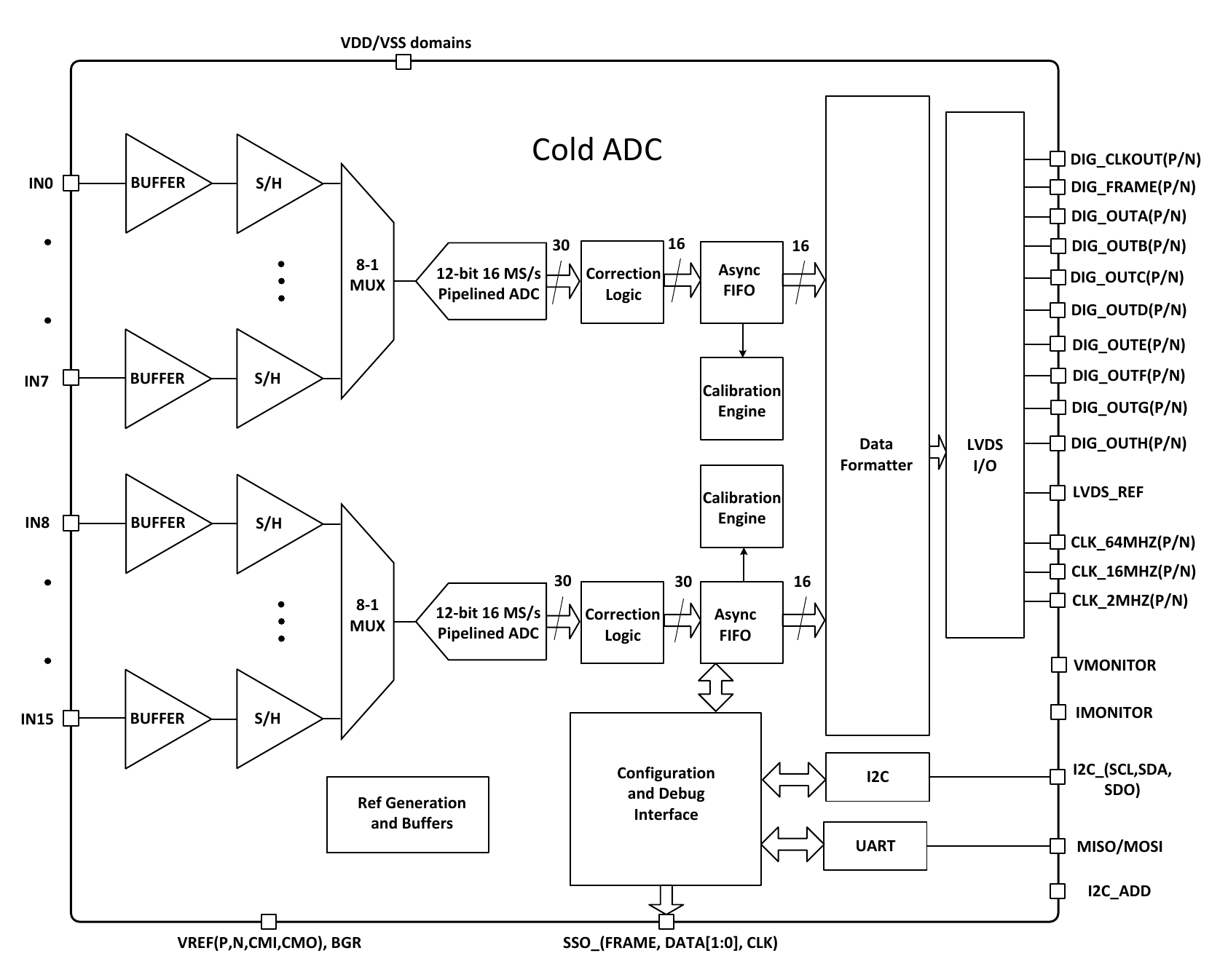}
\end{dunefigure}

Each cold \dword{adc} receives \num{16} single-ended voltage outputs from a single \dword{larasic} chip. The voltage is buffered and then sampled at a rate of \SI{2}{MHz}. The analog samples are multiplexed by eight and digitized by calibrated \num{12}\,bit pipelined \dwords{adc} operating at \SI{16}{MHz}. The \dword{adc} uses the well-known pipelined architecture with redundancy to reduce the impact of component non-idealities on the linearity of the \dword{adc}~\cite{121557}. The linearity of the raw output samples from the \dwords{adc} is improved using an on-chip calibration. The corrected \dword{adc} output is then multiplexed onto eight \dword{lvds} channels and sent to \dword{coldata} for further aggregation and transmission via a copper link to the warm electronics sitting outside the cryostat.

The \dword{adc} \dword{asic} is designed for low-noise operation, with a noise specification of  \SI{175}{\micro\volt} \rms. This noise specification was chosen to ensure that \dword{larasic} will dominate the overall noise performance of the channel.

The \dword{adc} is digitally calibrated using the proven Soenen-Karanicolas algorithm~\cite{280084,372864}. The algorithm exploits the observation that in a pipelined \dword{adc} with redundancy, the \dword{adc} nonlinearity is caused almost entirely by errors in the closed-loop interstage gain~\cite{121557}. Traditionally, the \dword{adc} output bits are assumed to be in radix two and are simply combined to generate the \dword{adc} output. However, due to unavoidable non-idealities such as finite op-amp gain and capacitor mismatch, the true radix of each stage is slightly different from two. The extent to which the true radix is different from two leads to DNL and INL in the \dword{adc} transfer characteristic. The Soenen-Karanicolas algorithm provides a way to measure the radix of a given stage by forcing events at the decision boundaries and using the following stages of the \dword{adc} to record the stage's response. The radix is then decomposed into a set of weights and during normal operation the \dword{adc} output is converted from the true radix to radix two using pipelined digital adders. This way, static linearity can be greatly improved without any post-processing required. To provide additional ease-of-use, all calibration hardware (including test signal generation) is included on the \dword{adc} \dword{asic}. To control power dissipation, the stages of the \dword{adc} are scaled in area to take advantage of the fact that the accuracy requirements of the stages decline down the pipeline~\cite{494191}.

To reduce the number of pads and to improve performance, all required reference voltages and currents are generated internally by a resistor-programmed reference generator on the \dword{asic}.

The cold \dword{adc} is highly configurable (see Table~\ref{tab:adcconfig}) and includes two redundant slow control interfaces for configuration (either UART or I2C). The configurability of the chip is included primarily to reduce risk by providing a high degree of flexibility and observability. First, many of the components on the \dword{asic} can be bypassed and their functions assumed at the board level if desired. For example, the \dword{adc} reference voltages can be supplied externally and the input buffers can be bypassed. Second, the \dword{adc} digital calibration algorithm can be implemented externally with the calculated stage weights loaded back into the chip using the configuration interface. Third, various internal voltages and currents can be monitored and test data can be introduced at various parts of the digital processing to observe the function of the \dword{asic}. Lastly, the bias point of the analog circuits in the \dword{asic} can be adjusted to compensate for expected component variations between room temperature and \lar temperature.

\begin{dunetable}
[Baseline cold \dword{adc} \dword{asic} configurability.]
{p{3cm} p{9cm} p{4cm}}
{tab:adcconfig}
{Baseline cold \dword{adc} \dword{asic} configurability.}
\textbf{BLOCK} &\textbf{Configurability} & \textbf{Comment}\\ \toprowrule
Input Buffer & Single-ended/differential, bypass, bias current adjust & Reduces design risk \\ \colhline
Sample-and-hold Amplifiers & Multiplexer freeze, bias current adjust & Simplifies evaluation of prototype \\ \colhline
\dword{adc} & Bias currents, clock edge fine adjustment, sync and test modes & Simplifies evaluation of prototype and reduces risk \\ \colhline
References & All reference voltages can be adjusted in \SI{8}{mV} increments; all references can be powered down and external voltages used & Reduces design risk \\ \colhline
Calibration & Number of stages and amount of digital filtering; all calibration commands can be implemented through configuration interface for offline calibration; known data can be injected at various points for testing & Simplifies evaluation of prototype and reduces risk \\ \colhline
Output Monitor & Various internal bias voltages and currents can be sent off-chip for evaluation & Simplifies evaluation of prototype \\
\end{dunetable}


\subsubsection{COLDATA ASIC}
\label{sec:fdsp-tpc-elec-design-femb-coldata}
The \dword{coldata} \dword{asic} is responsible for all communication between the 
\dword{ce} on \dwords{femb} and electronics located outside the cryostat.  The \dword{coldata} \dword{asic} is being designed by engineers from \fnal and Southern Methodist University.  Each \dword{femb} contains two \dword{coldata} \dwords{asic}.  \dword{coldata} receives command and control information; it provides clocks to the cold \dword{adc} \dwords{asic} and relays commands to the \dword{larasic} front-end and to the cold \dword{adc} \dwords{asic} to set operating modes and initiate calibration procedures.  \dword{coldata} receives data from the \dword{adc} \dwords{asic}, reformats these data, merges data streams, formats data packets, and sends these data packets to the warm electronics using \SI{1.28}{Gbps} links.  These links include line drivers with pulse pre-emphasis.  All the components of \dword{coldata}, with the exception of the line drivers and of the interface to the \dword{adc}, have been implemented in the CDP1 prototype \dword{asic} and demonstrated to work as designed both at room temperature and at \SI{77}{K}.  A block diagram of \dword{coldata} is shown in Figure~\ref{fig:coldata}.  

\begin{dunefigure}
[Block diagram of \dword{coldata} \dword{asic} design.]
{fig:coldata}
{Block diagram of \dword{coldata} \dword{asic} design.}
\includegraphics[width=0.85\linewidth]{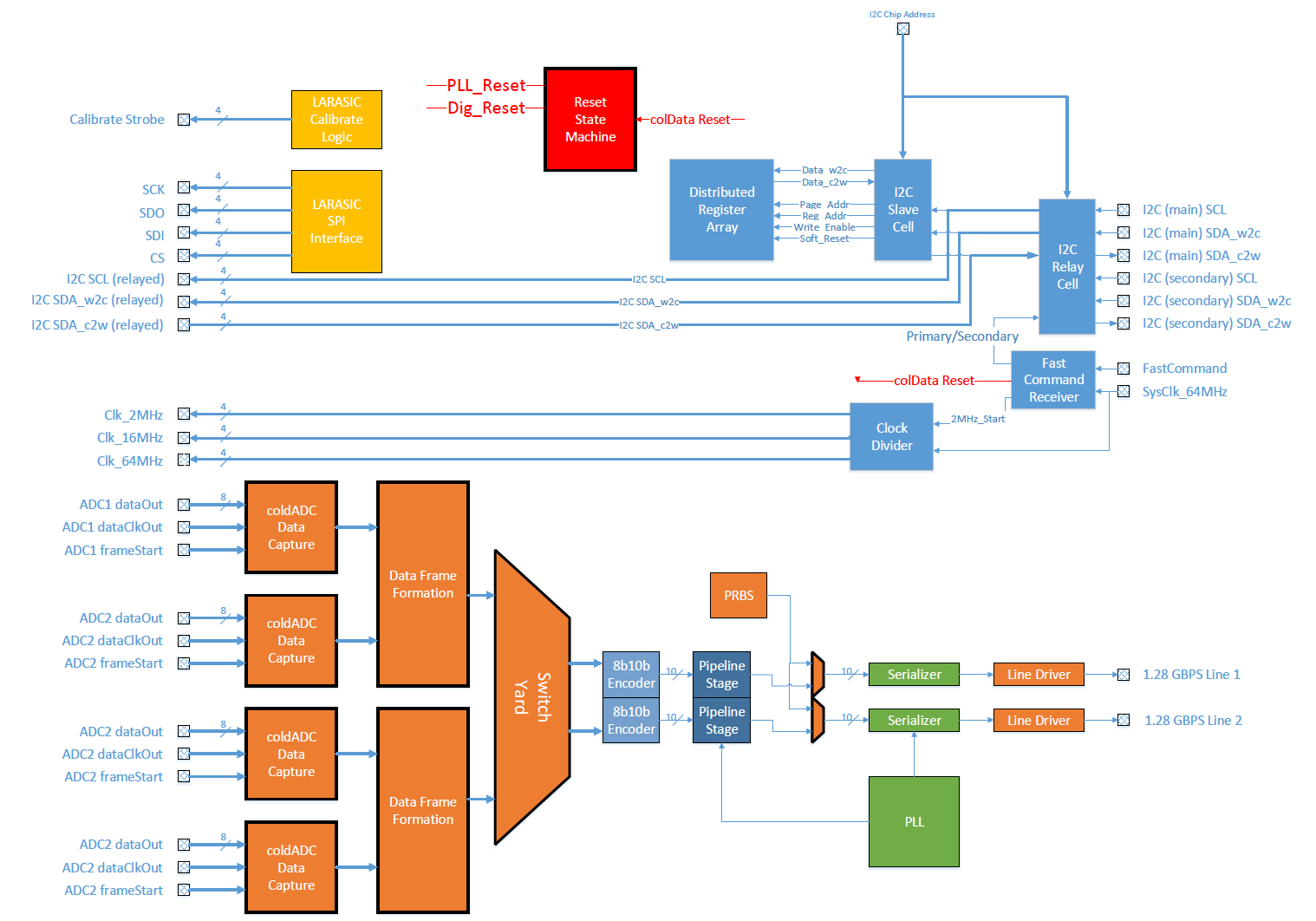}
\end{dunefigure}

Both \dword{coldata} and cold \dword{adc} are implemented in TSMC \SI{65}{nm} \dword{cmos}\footnote{TSMC 65 Nanometer Technology\texttrademark{}, Taiwan Semiconductor Manufacturing Company Ltd., \url{http://www.tsmc.com/english/dedicatedFoundry/technology/65nm.htm}.} 
using cold transistor models produced by Logix Consulting~\footnote{Logix Consulting\texttrademark{}, \url{http://www.lgx.com/}}.  Logix made measurements of \fnal-supplied TSMC \SI{65}{nm} transistors at a variety of temperatures (including room temperature and LN$_2$ temperature).  They extracted and provided to \fnal SPICE\footnote{SPICE\texttrademark{}, is a general-purpose circuit simulation program for nonlinear DC, nonlinear transient, and linear AC analyses. \url{https://bwrcs.eecs.berkeley.edu/Classes/IcBook/SPICE/}.} models as a function of temperature.  A special library of standard cells, based on these SPICE models and using a minimum channel length of \SI{90}{nm}, was developed by members of the University of Pennsylvania and \fnal groups.  This library was designed to eliminate the risk posed by the hot carrier effect.  The digital sections of \dword{coldata} and cold \dword{adc} use these standard cells and were synthesized from RTL (register-transfer level) using automatic place and route tools.

\subsubsection{Cold Electronics Box}
\label{sec:fdsp-tpc-elec-design-femb-box}
Each \dword{femb} is enclosed in a mechanical \dword{ce} box to provide support, cable strain
relief, and control of gas argon bubbles in the \lar from the \dword{femb} attached to the lower \dword{apa}
(which could in principle lead to discharge of the \dword{hv} system).
The \dword{ce} box, illustrated in Figure~\ref{fig:ce-box}, is designed to make the electrical connection 
between the \dword{femb} and the \dword{apa} frame, as defined in Section~\ref{sec:fdsp-tpc-elec-design-ground}.
Mounting hardware inside the \dword{ce} box connects the ground plane of the \dword{femb} to the box casing. The
box casing is electrically connected to the \dword{apa} frame via twisted conducting wire (not 
shown in Figure~\ref{fig:ce-box}). This is the only point of contact between the \dword{femb} and
\dword{apa}, except for the input amplifier circuits connected to the CR board, which also terminate to
ground at the \dword{apa} frame, as shown in Figure~\ref{fig:CR-board}.

\begin{dunefigure}
[Prototype \dword{ce} box used in \dword{pdsp}.]
{fig:ce-box}
{Prototype \dword{ce} box used in \dword{pdsp}.}
\includegraphics[width=0.45\linewidth]{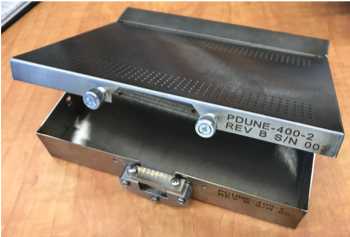}
\end{dunefigure}

\subsection{Additional FEMB/ASIC Designs}
\label{sec:fdsp-tpc-elec-design-alt}
In addition to the baseline \dword{femb} and \dword{asic} designs discussed in Section~\ref{sec:fdsp-tpc-elec-design-femb}, 
two other \dword{femb} and \dword{asic} options are currently under consideration. 
There is one official alternative design, the SLAC nEXO three-chip \textit{CRYO} \dword{asic}, and one fallback option for the \dword{adc} \dword{asic}, the Columbia University ATLAS-style \dword{adc} \dword{asic}.  These options are described in Section~\ref{sec:fdsp-tpc-elec-design-alt-cryo} and Section~\ref{sec:fdsp-tpc-elec-design-alt-atlas}, respectively.

\subsubsection{nEXO CRYO ASIC}
\label{sec:fdsp-tpc-elec-design-alt-cryo}
The SLAC CRYO \dword{asic} differs from the baseline three-chip design in that it combines the functions of an analog preamplifier, \dword{adc}, and data serialization and transmission for \num{64}~wire channels, into a single chip.
It is based on a design developed for the nEXO experiment\footnote{Enriched Xenon Observatory, \url{https://www-project.slac.stanford.edu/exo/about.html}.} and differs from it only in the design of the preamplifier, which is modified to account for the higher capacitance of the DUNE \dword{spmod} wires compared to the small pads of nEXO.
The \dwords{femb} constructed using this chip would use only two \dwords{asic}, compared to the \num{18} (eight~FE, eight~\dword{adc} and two~COLDATA) needed in the baseline design.
This drastic reduction in part count may significantly improve \dword{femb} reliablity, reduce power, and reduce costs related to production and testing. 

Figure~\ref{fig:cryo-architecture} shows the overall architecture of the CRYO \dword{asic}, which will be implemented in \SI{130}{nm} \dword{cmos}.
It comprises two identical, \num{32}-channel blocks. 
The current signal from each wire is amplified using a preamplifier with pole zero cancellation and an anti-alias fifth-order Bessel filter applied. 
Provisions are also made for injection of test pulses. 
Gain and peaking time are adjustable to values similar to those of the baseline design.

\begin{dunefigure}
[Overall architecture of the CRYO \dword{asic}.]
{fig:cryo-architecture}
{Overall architecture of the CRYO \dword{asic}.}
\includegraphics[width=0.8\textwidth]{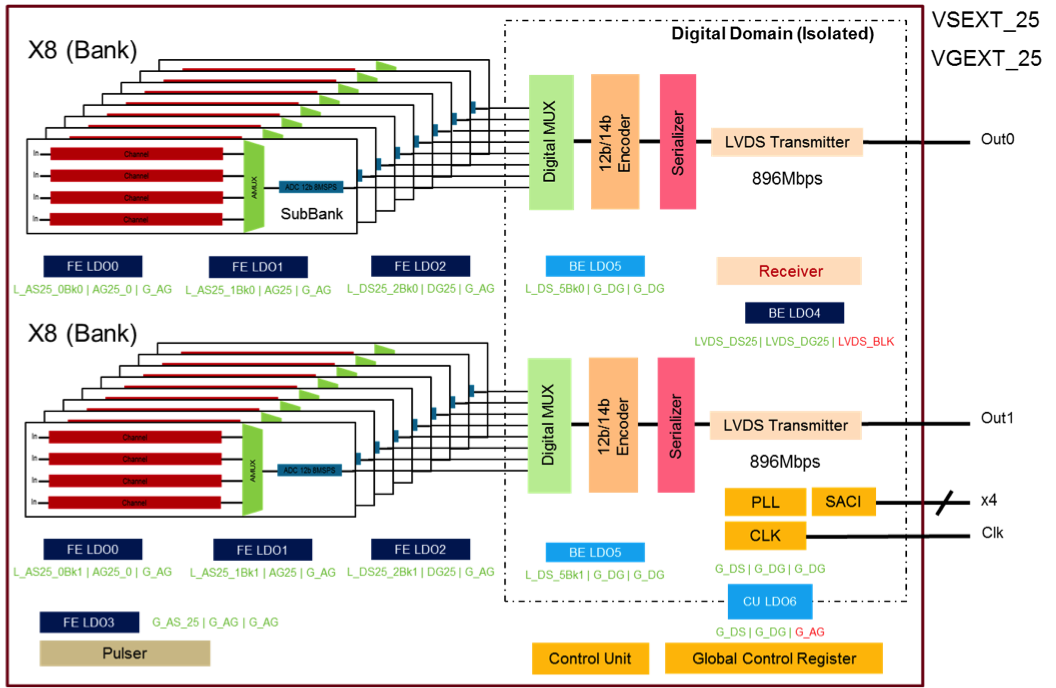}
\end{dunefigure}

The \dword{adc} uses \SI{8}{MHz} successive approximation registration (SAR), so that four input channels are multiplexed onto a single \dword{adc}. The data serialization and transmission block employs a custom 12b/14b encoder, so that \num{32} channels of \num{12}-bit, \SI{2}{MHz} data can be transmitted with a digital bandwidth of only \SI{896}{Mbps}, which is significantly less than the required bandwidth of the baseline, which is \SI{1.28}{Gbps}.

One key concern with mixed signal \dwords{asic} is the possibility of interference from the digital side causing noise on the very sensitive preamplifier. 
Fortunately, there are well established techniques for substrate isolation described in the literature~\cite{yeh}, which have been successfully employed in previous \dwords{asic} produced by the SLAC group.


The infrastructure requirements for a CYRO \dword{asic}-based system are similar to those of the baseline option. However, in most cases, somewhat fewer resources are needed:
\begin{itemize}
\item{A single voltage is needed for the power supply. This is used to generate two supply voltages using internal voltage regulators.}
\item{The output digital bandwidth on each of the four lines in an \dword{femb} is \SI{896}{Mbps}. This is lower than the baseline option due to the custom 12b/14b encoder of the CRYO chip. }
\item{The warm interface is different. Only a single clock is needed (\SI{56}{MHz}) and the configuration protocol is the SLAC \dword{asic} Control Interface (SACI)~\cite{SACI} rather than I2C.}
\end{itemize}

The first prototype of the CRYO \dword{asic} is in the final design and simulation stages. Simulation-based studies have already been performed; at \SI{0.8}{\micro\second} peaking time and an input capacitance of \SI{200}{pF} (similar to that expected in the DUNE \dword{spmod}), the \dword{enc} is approximately \num{500}\,e$^-$.  This noise level is similar to that expected with the baseline \dword{fe} and \dword{adc} \dword{asic} design in \lar with the same input capacitance.  Submission to the \dword{asic} foundry is imminent and the first prototypes should be received by summer 2018. They will first be tested in an existing test stand at SLAC. Subsequent tests are planned for a small test TPC at \fnal and on an \dword{apa} in the \dword{pdsp} cold box; these test facilities are described in Section~\ref{sec:fdsp-tpc-elec-qa-facilities}.

\subsubsection{ATLAS ADC ASIC}
\label{sec:fdsp-tpc-elec-design-alt-atlas}
An alternative \dword{adc} solution is to adapt the \dword{adc} chip under development for the 
ATLAS \lar calorimeter readout upgrade for the high luminosity LHC.  The main ATLAS 
requirements are given in Table~\ref{tab:ATLAS-adc-reqs}.  Adapting the chip to 
the \dword{spmod} needs 
would require doubling the number of channels per chip as well as adapting the output 
architecture.  These are both relatively simple changes compared to the overall complexity of the chip.

\begin{dunetable}
[Performance requirements for the ATLAS-style \dword{adc} \dword{asic}.]
{ll}
{tab:ATLAS-adc-reqs}
{Performance requirements for the ATLAS-style \dword{adc} \dword{asic}.}
\textbf{Parameter} &\textbf{Specification}\\ \toprowrule
Channels/chip & eight preferred, four minimum \\ \colhline
Sampling Frequency & \SI{40}{MHz} \\ \colhline
Dynamic Range & \SI{14}{bits}  \\ \colhline
Precision & \SI{11}{ENOB}\\ \colhline
Power & $<$~\SI{100}{mW}/channel at \SI{40}{MHz}\\ \colhline
Input & \SI{2}{V} differential\\ \colhline
Output & E-link interface operating at \SI{640}{Mbps}\\
\end{dunetable}

To achieve a \SI{14}{bit} dynamic range, each analog channel is comprised
of two main sections: a dynamic range enhancement (\dword{dre}) block that determines the
most significant two bits of the \SI{14}{bit} digital code, followed by a \SI{12}{bit} \dword{sar} block. 
The input signal to the \dword{dre} block is sampled on two
paths, one with unity gain and the other of gain four. A comparator determines which gain to use.
The signal from the selected \dword{dre} gain is presented at the \dword{dre} output,
which is connected to the input of the \SI{12}{bit} \dword{sar} \dword{adc} block. The \dword{dre} design has been carefully
optimized so that its output preserves the required \SI{12}{bit} performance.


Following current state-of-the-art \dword{adc} development techniques, a two-stage 
\dword{sar} architecture is used, exploiting the high speed of the technology while maintaining the \dword{sar} input 
capacitance at a reasonable value. Since capacitor matching in this technology might not meet the 
precision required, the \dword{adc} will use bit redundancy, i.e., determine more bits than its actual output, 
and the redundant bits will be used to both calibrate the \dword{adc} and produce correct output codes. 
Such procedures are well understood and applied to both pipeline~\cite{Kuppambatti:2013nfa} and 
\dword{sar}~\cite{5999734} \dwords{adc} using foreground or background calibration techniques. Details of the \dword{sar} design are shown in Figure~\ref{fig:65nmadcarchitecture_sar}. 

\begin{dunefigure}
[Block diagram of the two-stage \dword{sar} design of the ATLAS \dword{adc} \dword{asic}.]
{fig:65nmadcarchitecture_sar}
{Block diagram depicting the two-stage \dword{sar} design of the ATLAS \dword{adc} \dword{asic}.}
\includegraphics[width=.9\textwidth]{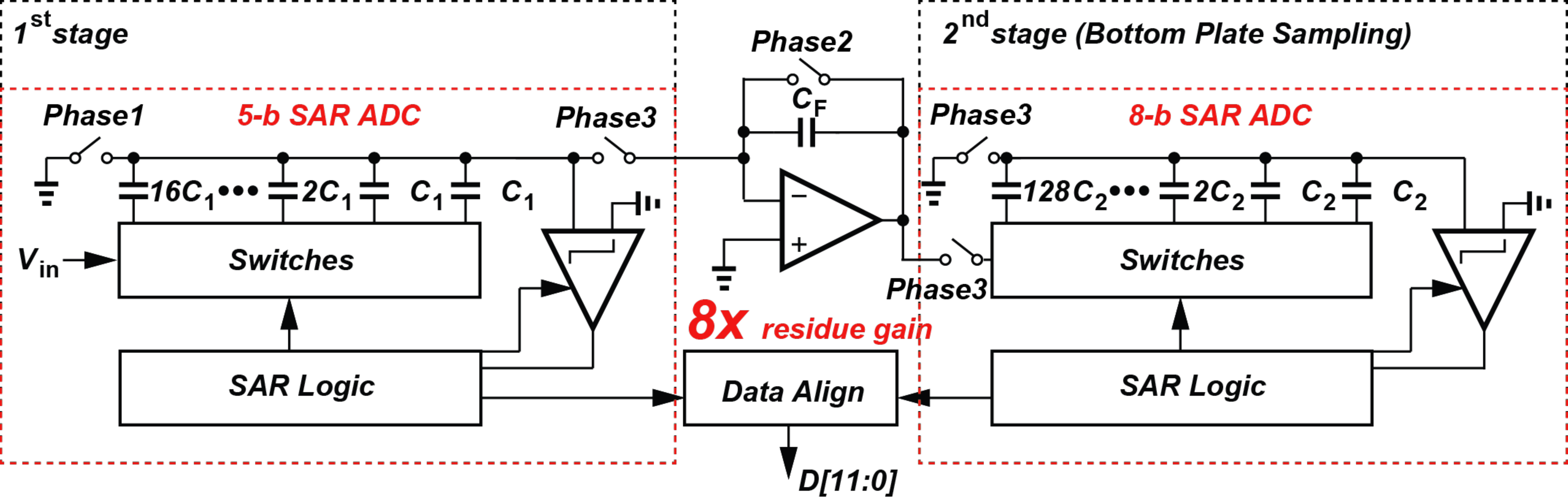}
\end{dunefigure}

An \dword{adc} test chip, dubbed COLUTA65V1, was designed and submitted for fabrication in May 2017 and received
in September of 2017.  The \dword{dre} and \dword{sar} blocks of the COLUTA65V1 were first tested independently. Measurements were made of the \dword{sar} precision using the sine-wave fast Fourier transform method. An effective number of bits
(ENOB) of \SI{11.6}{bits} at \SI{20}{MHz} (after calibration) was obtained.
Both \dword{dre} and \dword{sar} were successfully integrated with negligible degradation in performance. The COLUTA65V1 chip was also tested at \SI{2}{MHz} and shown to work as designed, meeting the requirement for the \dword{spmod}. 
Tests of an updated design in liquid nitrogen are planned for spring 2018. Additional tests associated with meeting power requirements will be carried out if this \dword{adc} option is further pursued.

\subsection{Cold Electronics Feedthroughs and Cold Cables}
\label{sec:fdsp-tpc-elec-design-ft}
All cold cables originating from inside the cryostat connect to the outside warm electronics through PCB board \fdth{}s
installed in the signal flanges that are distributed along the cryostat roof.
The TPC data rate per \dword{apa}, with an overall \num{32}:\num{321} MUX and eighty $\sim$1~Gbps data channels per \dword{apa},
is sufficiently low that the \dword{lvds} signals can be driven over copper twin-axial transmission lines.
Additional transmission lines are available for the distribution of \dword{lvds} clock signals and I$^2$C control information,
which are transmitted at a lower bit rate.
Optical fiber is employed externally from the \dwords{wib} on the signal flange to the \dword{daq} and slow control systems described in Chapter~\ref{ch:fdsp-daq} and Chapter~\ref{ch:fdsp-slow-cryo}, respectively.

\begin{dunefigure}
[TPC \dword{ce} \fdth.]
{fig:tpcelec-signal_FT}
{TPC \dword{ce} \fdth. The \dwords{wib} are seen edge-on in the left panel, and in an oblique side-view in the right panel, which also shows the warm crate for a \dword{spmod} 
in a cutaway view.}
\includegraphics[width=0.9\linewidth]{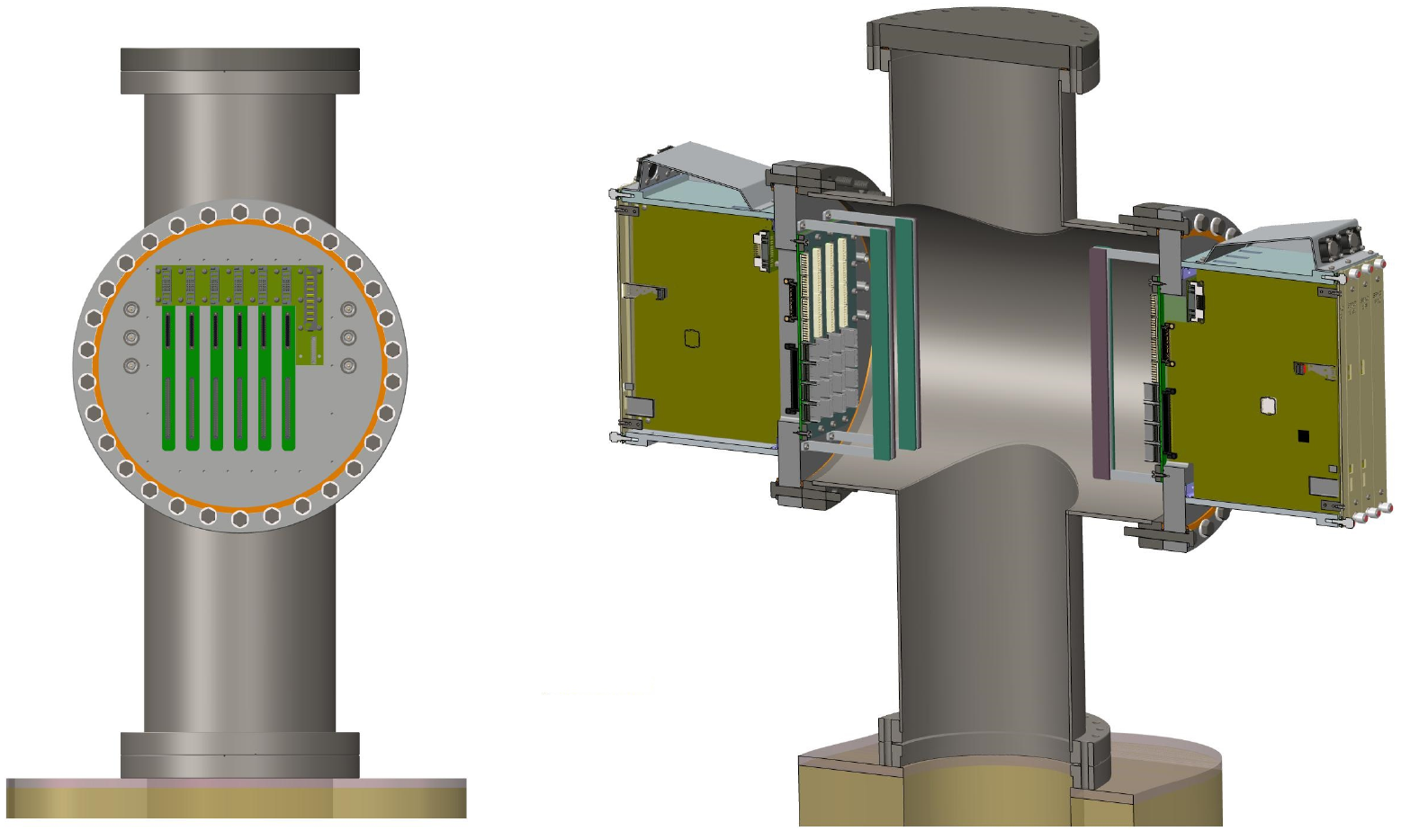}
\end{dunefigure}

The design of the signal flange includes a four-way cross spool piece, separate PCB \fdth{}s for the \dword{ce} and \dword{pds} cables, and
an attached crate for the TPC warm electronics, as shown in Figure~\ref{fig:tpcelec-signal_FT}.
The wire bias voltage cables connect to standard SHV (safe high voltage) connectors machined directly into the \dword{ce} \fdth,
ensuring no electrical connection between the wire bias voltages and other signals passing through the signal flange.
Each \dword{ce} \fdth serves the bias, power, and digital I/O needs of one \dword{apa}.  

Data and control cable bundles are used to send system clock and control signals from the 
signal flange to the \dword{femb}, stream the $\sim$\SI{1}{Gbps} high-speed data from the \dword{femb} to the signal flange.  Each \dword{femb} 
connects to a signal flange via one data cable bundle, leading to 20 bundles between one \dword{apa} and one flange.  Each data bundle contains 12 low-skew twin-axial cables with a drain wire, 
to transmit the following differential signals:

\begin{itemize}
    \item four \SI{1.28}{Gbps} data (two from each \dword{coldata});
    \item two \SI{64}{MHz} clocks (one input to each \dword{coldata});
    \item two fast command lines (one input to each \dword{coldata});
    \item three I$^2$C-like control lines (clock, data-in, and data-out); and
    \item one multipurpose \dword{larasic} output (temperature, reference voltage, or analog test output).
\end{itemize}

The \dword{lv} power is passed from the signal flange to the \dword{femb} by bundles of
20AWG twisted-pair wires. Half of the wires are power feeds; the others
are attached to the grounds of the input amplifier circuits, as described in Section~\ref{sec:fdsp-tpc-elec-design-bias}.
For a single \dword{femb}, the resistance is measured to be  <\SI{30}{\milli\ohm} at room temperature or $<10$~m$\Omega$ at 
\lar temperature. Each \dword{apa} has a copper cross section of approximately 
\SI{80}{mm$^2$}, with a 
resistance <\SI{1.5}{\milli\ohm} at room temperature or $<0.5$~m$\Omega$ at \lar temperature.

The bias voltages are applied to the $X$-, $U$-, and $G$-plane wire layers, three \dword{fc} terminations, 
and an electron diverter, as shown in Figure~\ref{fig:CR-board}. The voltages are supplied 
through eight SHV connectors mounted on the signal flange. RG-316 coaxial cables carry the voltages 
from the signal flange to a patch panel PCB which includes noise filtering mounted on the top 
end of the \dword{apa}. 

From there, wire bias voltages are carried by single wires to 
various points on the \dword{apa} frame, including the CR boards, a small PCB mounted on or near 
the patch panel that houses a noise filter and termination circuits for the field cage voltages, and 
a small mounted board near the electron diverter that also houses wire bias voltage filters.

\subsection{Warm Interface Electronics}
\label{sec:fdsp-tpc-elec-design-warm}
The warm interface electronics 
provide an interface between the \dword{ce}, \dword{daq}, timing, and slow control systems, including local power control at the flange and a real-time diagnostic readout. They are housed in the \dwords{wiec} attached directly to the \dword{ce} flange.  The \dword{wiec} shown in Figure~\ref{fig:tpcelec-flange} 
contains one power and timing card (\dword{ptc}), five warm interface boards (\dwords{wib}) and a passive
power and timing backplane (PTB), which fans out signals and \dword{lv} power from the \dword{ptc} to the \dwords{wib}. The \dword{wiec} must provide a Faraday-shielded housing, robust ground connections from the \dwords{wib} to the detector ground described in Section~\ref{sec:fdsp-tpc-elec-design-ground}, and only optical fiber links to the \dword{daq} and slow control in order to mitigate noise introduced at the \dword{ce} \fdth.

\begin{dunefigure}
[Exploded view of the \dword{ce} signal flange for \dword{pdsp}.]
{fig:tpcelec-flange}
{Exploded view of the \dword{ce} signal flange for \dword{pdsp}.  The design will be very similar for the \dword{spmod} \dword{ce} signal flange (with two \dword{ce} signal flanges per \fdth).}
\includegraphics[width=0.9\linewidth]{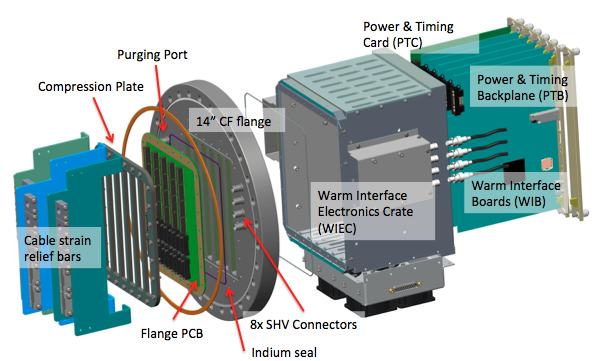}
\end{dunefigure}

The \dword{wib} is the interface between the \dword{daq} system and four
\dwords{femb}. It receives the system clock and control signals from the
timing system and provides for processing and fan-out of those signals to the four
\dwords{femb}. The \dword{wib} also receives the high-speed data signals from the four 
\dwords{femb} and transmits them to the \dword{daq} system over optical
fibers. The data signals are recovered onboard the \dword{wib} with commercial equalizers.
The \dwords{wib} are attached directly to the TPC
\dword{ce} \fdth on the signal flange. The \fdth
board is a PCB with connectors to the cold signal and \dword{lv} power cables fitted
between the compression plate on the cold side, and sockets for
the \dword{wib} on the warm side. Cable strain relief for the cold cables is 
supported from the back end of the \fdth.

\begin{dunefigure}
[\dword{pdsp} PTC and timing distribution to the WIB and FEMBs]
{fig:tpcelec-wib_timing}
{Power and timing card (\dword{ptc}) and timing distribution to the \dword{wib} and \dwords{femb} used in \dword{pdsp}.}
\includegraphics[width=0.75\linewidth]{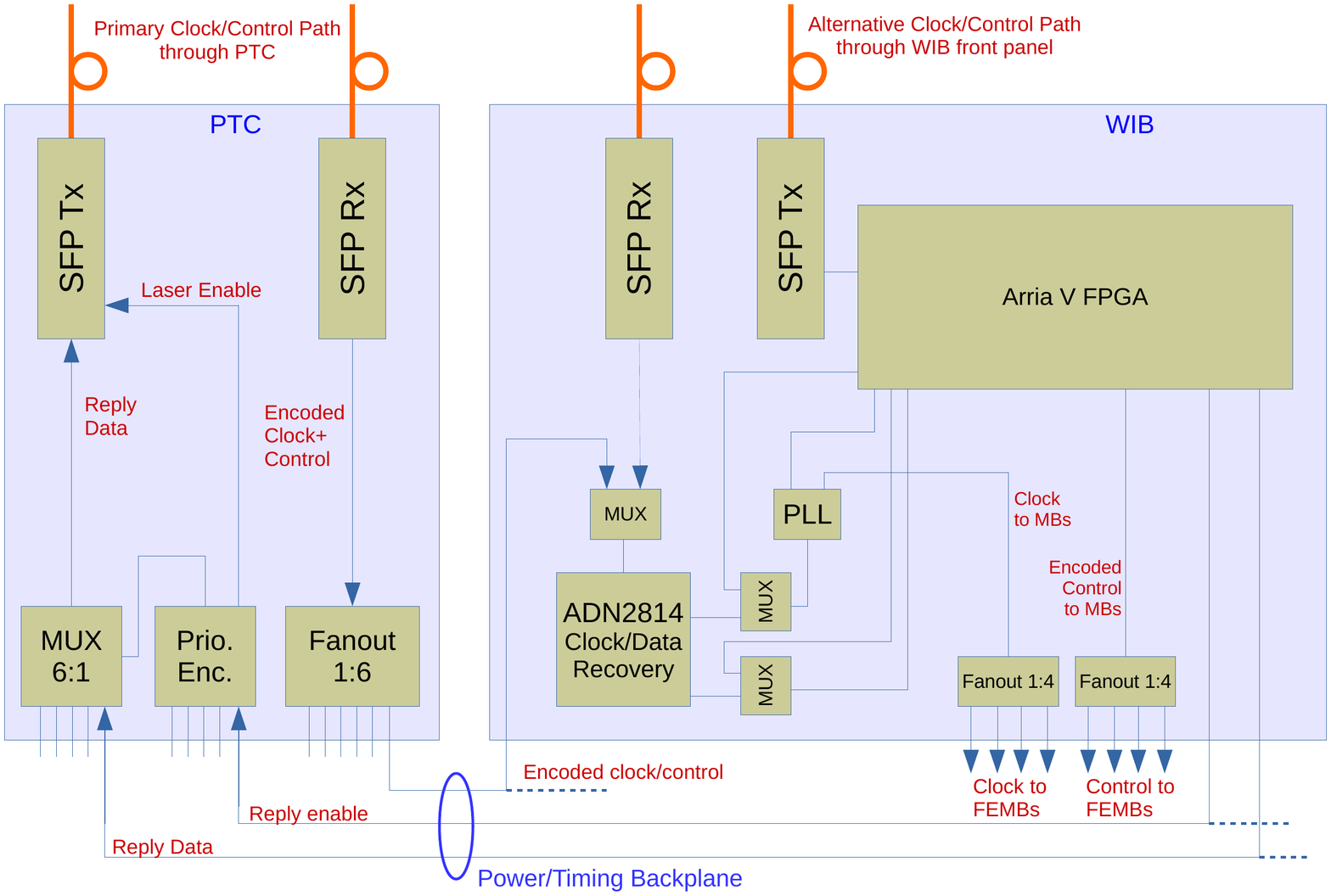}
\end{dunefigure}

The \dword{pdsp} \dword{ptc} provides a bidirectional fiber interface to the
timing system. The clock and data
streams are separately fanned out to the five \dwords{wib} as shown in
Figure~\ref{fig:tpcelec-wib_timing}. The \dword{ptc} fans the clocks out to the \dword{wib} over the
PTB, which is a passive backplane attached directly to the \dword{ptc} and
\dwords{wib}.  The received clock on the \dword{wib} is separated into clock and
data using a clock-data separator. Timing endpoint firmware to receive and transmit
the clock is integrated in the \dword{wib} \dword{fpga} (the Altera Arria V\footnote{Altera Arria\texttrademark{}, V FPGA family, \url{https://www.altera.com/products/fpga/arria-series/arria-v/overview.html}.} was used for \dword{pdsp}).
The \dword{spmod} timing system, described in Section~\ref{sec:fd-daq-timing}, is a development of the \dword{pdsp} system, and expected to require the nearly identical functionality at the \dword{wib} endpoint.

\begin{dunefigure}
[\dword{pdsp} \dword{lv} power distribution to the \dword{wib} and \dwords{femb}]
{fig:tpcelec-wib_power}
{\dword{lv} power distribution to the \dword{wib} and \dwords{femb} implemented for \dword{pdsp}. This will be modified for the \dword{spmod} to provide the required voltage or voltages depending on which \dwords{asic} are used on the \dwords{femb}. In particular the voltages to the \dword{femb} \numrange{0}{3} will change as the \dword{pdsp} \dword{fpga} is replaced by \dword{coldata}. }
\includegraphics[width=0.65\linewidth]{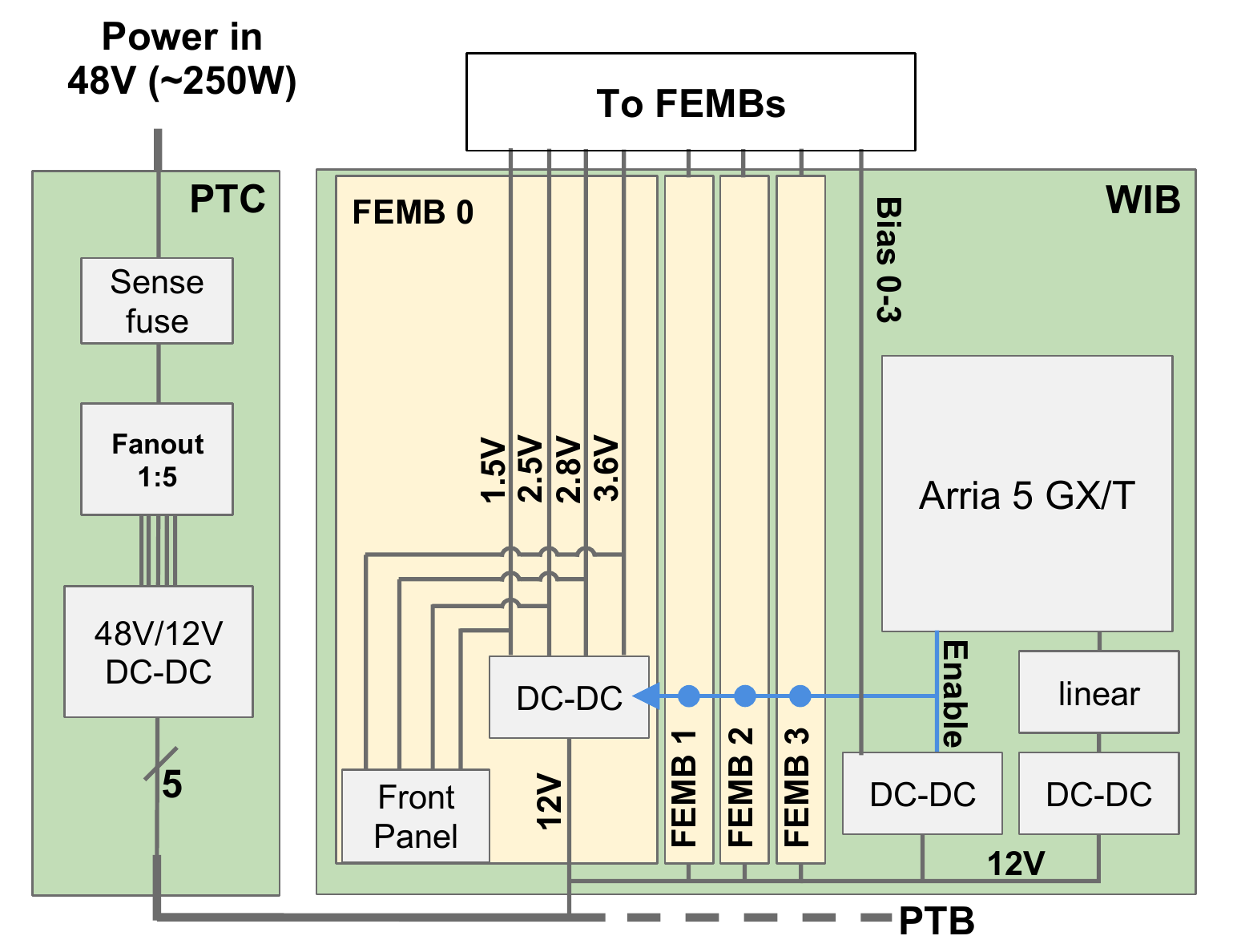}
\end{dunefigure}

The \dword{ptc} also receives \SI{48}{V} \dword{lv} power for all cold
electronics connected through the TPC signal flange: one \dword{ptc}, five \dword{wib}, and \num{20}~\dword{femb}. The \dword{lv} power is then stepped down
to \SI{12}{V} via a DC-DC converter onboard the \dword{ptc}. The output of the \dword{ptc} converters is filtered with a common-mode choke and fanned out
on the PTB to each \dword{wib}, which provides the necessary \SI{12}{V} DC-DC conversions and fans
the \dword{lv} power out to each of the cold \dwords{femb} supplied by that \dword{wib}, 
as shown in Figure~\ref{fig:tpcelec-wib_power}. The output of the \dword{wib} converters is further filtered by a common-mode choke. The 
majority of the power drawn by a full flange is dissipated in the \lar by the cold \dword{femb}.


As shown in Figure~\ref{fig:tpcelec-dune_wib}, the \dword{wib} is capable of receiving \dword{lv} power in the front panel and distributing it directly to the \dword{femb}, bypassing all DC/DC converters.
It can also receive the encoded system timing signals over bi-directional optical
fibers on the front panel, and process these using either
the on-board \dword{fpga} or clock synthesizer chip to provide the clock required by the \dword{ce}.
The baseline \dword{asic} design currently uses 8b/10b encoding; if the SLAC CRYO \dword{asic} is selected for
the DUNE \dword{spmod}, 12b/14b encoding will be used instead of 8b/10b.

\begin{dunefigure}
[Warm interface board (\dword{wib})]
{fig:tpcelec-dune_wib}
{Warm interface board (\dword{wib}). Note that front panel inputs include a LEMO connector and alternate inputs for \dword{lv} power and timing.}
\includegraphics[width=0.8\linewidth]{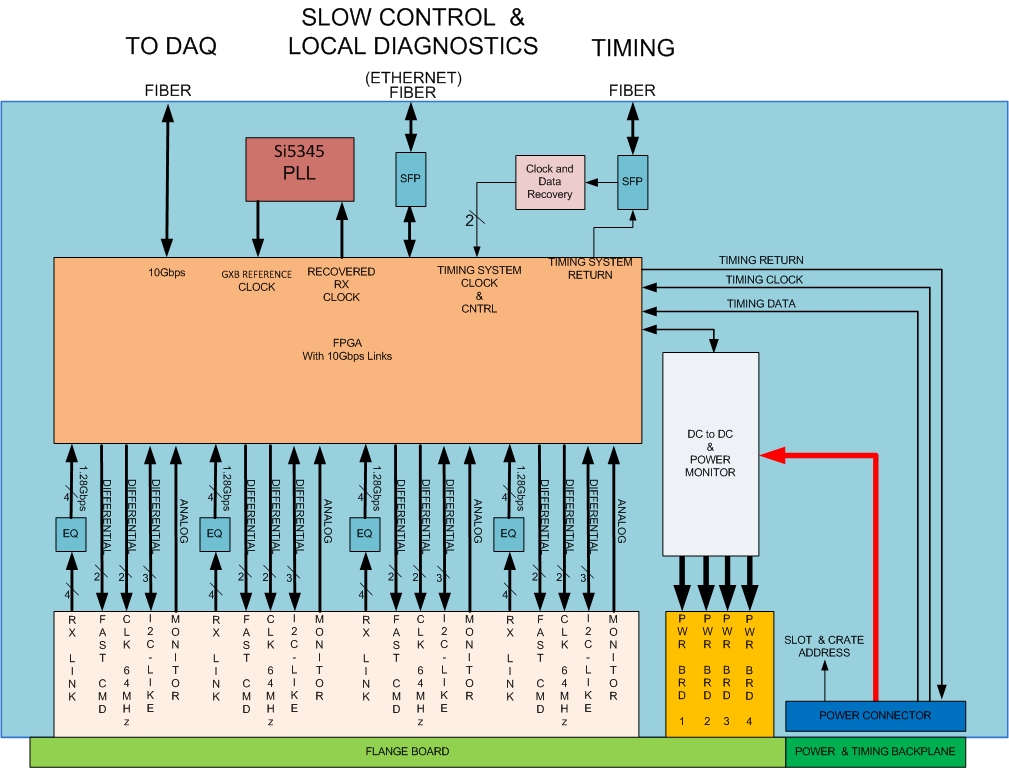}
\end{dunefigure}

The \dword{fpga} on the \dword{pdsp} \dword{wib} is an Altera Arria V GT variant, which has
transceivers that can drive the high-speed data to the \dword{daq} system up to
10.3125~Gbps per link, implying that all data from
two \dword{femb} (2$\times$5~Gbps) could be transmitted on a single link.
The \dword{fpga} has an additional Gbps Ethernet transceiver I/O based on the \SI{125}{MHz} clock, which 
provides real-time digital data readout to the slow control system.

\subsection{External Power and Supplies}
\label{sec:fdsp-tpc-elec-design-external}
As implemented for \dword{pdsp}, a fully loaded \dword{wib} (one \dword{wib} plus four \dwords{femb}) requires
\SI{12}{V} and draws up to approximately \SI{4}{A}. The full electronics for one \dword{apa} (one \dword{ptc}, five \dwords{wib}, and \num{20} \dwords{femb}) 
requires \SI{12}{V} and draws approximately \SI{20}{A}, for a total power of approximately \SI{240}{W}, as 
described in Section~\ref{sec:fdsp-tpc-elec-design-warm}. The \dword{spmod} implementation should require much 
less power as the \dword{fpga} will be replaced by the \dword{coldata} chips.

As the \dword{lv} power is delivered at \SI{48}{V} to the \dword{ptc}, each \dword{lv} power mainframe is chosen to bracket that value; each has  
roughly \numrange{30}{60}{V}, \SI{13.5}{A}, \SI{650}{W} maximum capacity per \dword{apa}. Using 10AWG cable, a \SI{0.8}{V} drop is 
expected along the cable with a required power of \SI{306}{W} out of \SI{650}{W} available.  
This leaves a significant margin that allows for larger distances between the power supplies and 
the warm interface crates than the \SI{20}{\meter} in \dword{pdsp}.

Four wires are used for each module; two 10AWG, shielded, twisted-pair cables for the power and return; and two 20AWG, shielded, twisted-pair cables for the sense.
The primary protection is the over-current protection circuit in the \dword{lv} supply modules, 
which is set above the \SI{20}{A} current draw of the \dword{wiec}.  Secondary sense line fusing is 
provided on the \dword{ptc}.  The \dword{lv} power cable uses FCi micro TCA\footnote{MicroTCA\texttrademark{} (\dword{utca}) vertical card-edge connectors, Amphenol ICC,  \url{https://www.amphenol-icc.com/product-series/micro-tca-card-edge.html}.} connectors, shown in
Figure~\ref{fig:tpcelec-power_conn}.

\begin{dunefigure}
[FCi microTCA power connector at the \dword{ptc} end of the cable.]
{fig:tpcelec-power_conn}
{FCi microTCA power connector at the \dword{ptc} end of the cable.}
\includegraphics[width=0.9\linewidth]{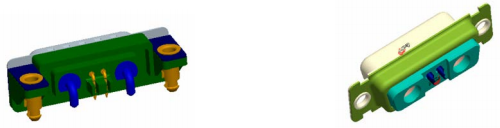}
\end{dunefigure}

Bias voltages for the \dword{apa} wire planes, the electron diverters, and the last \dword{fc} electrodes are generated by supplies which are the responsibility of the TPC Electronics consortium.  The current from each of these supplies is expected to be very close to zero in normal operation.  However, the ripple voltage must be carefully controlled to avoid injecting noise into the front-end electronics.  RG-58 coaxial cables connect the wire bias voltages from the mini-crate to the standard SHV
connectors machined directly into the \dword{ce} \fdth, so there is no electrical connection between 
the \dword{lv} power and data connectors and wire bias voltages.

Optical fibers are used for all connections between the \dwords{wiec}, which act as
Faraday-shielded boxes, and the \dword{daq} and slow control systems.  
The \dword{wib} reports
its onboard temperature and the current draw from each \dword{femb} to the slow control system, while the
current draw for each \dword{apa} is monitored at the mainframe itself.

\section{Production and Assembly}
\label{sec:fdsp-tpc-elec-prod}
A single \dword{spmod} requires \num{3000} \dwords{femb}, \num{750} \dwords{wib}, and \num{50} \dwords{ptc}.  A total of \num{3300} \dwords{femb}, \num{900} \dwords{wib}, and \num{60} \dwords{ptc} will be built.

If the three-\dword{asic} \dword{femb} solution is chosen (the baseline option), then two \dword{asic} production contracts will be required, one for \dword{larasic} (\SI{180}{nm} \dword{cmos}) and one for the cold \dword{adc} and \dword{coldata} (\SI{65}{nm} \dword{cmos}).  If the one-\dword{asic} \dword{femb} solution is chosen, only one \dword{asic} production contract will be required.

In either case, all \dwords{asic} will be packaged in plastic quad flat-pack (PQFP) or thin quad flat-pack (TQFP) surface mount packages.  No wafer probing will be done before the chips are packaged.  Rather, the wafers will be diced and all chips located more than $\sim$\SI{10}{mm} from the outside of the wafer will be selected for packaging.  The packaged parts will be tested by DUNE collaborators (see Section~\ref{sec:fdsp-tpc-elec-qc}) before being assembled onto printed circuit boards.

All printed circuit boards will be fabricated and tested by qualified vendors. Circuit boards will also be assembled by qualified vendors.  The completed boards will be acceptance tested by DUNE collaborators promptly after assembly.

All cable assemblies (including terminations) will be fabricated and tested by qualified vendors.  At least a fraction of the cable assemblies will be retested by DUNE collaborators promptly after purchase.


\section{Interfaces}
\label{sec:fdsp-tpc-elec-intfc}

\subsection{Overview}
\label{sec:fdsp-tpc-elec-intfc-ov}
Some of the components designed and built by the \dword{ce} consortium are
mounted on or need to work together with detector components provided by other DUNE
consortia. Interface documents have been developed to ensure that the boundaries
between systems are fully understood and that no detector components are missed or
not properly defined when organizing the detector construction project into
consortia. These interface documents are a work-in-progress. With time they will
evolve into a very detailed definition of mechanical and electrical interfaces,
including in some cases the description of data transmission protocols. These
interface documents include a list of the responsibilities of each consortium during the
R\&D, design, and prototyping phases, and discuss all of the procedures to be
followed during the integration of detector components and the following testing
and commissioning process. In some cases multiple consortia (\dword{apa}, \dword{pds}, \dword{hv}, \dword{cisc},
and \dword{daq}) 
can be involved, depending on the complexity and maturity of the test setup.

The most important interfaces for the \dword{ce} consortium are with the \dword{apa} and \dword{daq}
consortia, followed by those with \dword{pds}, \dword{hv}, and \dword{cisc}. One of the most
important aspects of all these interfaces is the enforcement of appropriate grounding rules
and of the physical separation between electrical circuits to minimize the noise and
crosstalk between different detector components. Different \dwords{apa} should be electrically
insulated, and the same should apply for the \dword{pds} and the
\dword{ce} readout of the \dwords{apa}. The flanges on the \fdth{}s are used to provide,
separately for the two \dwords{apa} and for the \dword{pds}, the reference
voltage for all the detector elements. The flanges are electrically connected
to the cryostat structure that acts as the detector ground. The same approach
has to be implemented for the \dword{cisc} instrumentation in the \lar.

The two most complex interfaces to the TPC electronics subsystem, the
interfaces to the \dwords{apa} and \dword{daq}, are discussed below.


\subsection{APAs}
\label{sec:fdsp-tpc-elec-intfc-apa}
The \dword{ce} consortium provides all of the electronics used for reading out the charge
collection and induction signals produced on the \dword{apa} wires by particles that ionize
the \lar. The \dword{apa} consortium
is responsible for all of the printed circuit boards mounted on the \dword{apa} frame
holding the anode wires. These boards provide filtering of the wire bias voltages,
connection of the bias voltages to the wires and AC or direct coupling of the
wire signals to the \dwords{femb} built by the \dword{ce} consortium.
The cold boxes containing the \dwords{femb} are also mounted on the \dwords{apa} and are connected
electrically to the CR boards. The \dword{ce} consortium is responsible for providing the bias
voltage to the \dwords{apa}, and also the bias voltage for the electron diverters and the \dword{fe} termination electrodes (these last two items are part of the interface with
the \dword{hv} consortium).

A crucial aspect of the interface between the \dword{ce} and \dword{apa} consortia
is the choice of routing for the cables that provide power, control, and readout
for the bottom \dword{apa}. Studies are ongoing to understand whether it is feasible to
route these cables inside the \dword{apa} frames of the two stacked \dwords{apa}, and whether it
is necessary to modify the size of the \dword{apa} frame in order to achieve this goal.
Mockups of the \dword{apa} frames will be used for routing tests together with \dword{pdsp}
cables, under the assumption that there will be minimal changes in the cross
section of cables between \dword{pdsp} and DUNE (we expect to be able to remove one
of the seven pairs of power lines used in \dword{pdsp}, when the \dword{fpga} on the
\dword{femb} is replaced by the \dword{coldata} \dword{asic}). Before spring 2019 we expect to perform
a realistic test of the cable insertion in a pair of stacked \dwords{apa} using
mechanical prototypes (i.e.,~without wires) of the \dword{apa}. We are also investigating
the possibility of routing the cables for the bottom \dword{apa} outside the field
cage, which increases significantly the length of the cables for the central
\dword{apa}; this also complicates the installation procedure and the engineering of the
support structures inside the cryostat.

\subsection{DAQ}
\label{sec:fdsp-tpc-elec-intfc-daq}
There are two components in the definition of the interface between the \dword{ce}
and \dword{daq} consortia. The first one is a decision on whether to implement any
firmware and buffering related to the trigger decision inside the \dword{wib}, or
instead transmit the data as they are produced from the \dwords{femb}, possibly with
some serialization taking place in the \dword{wib}. For the \dword{spmod} 
we have
chosen to adopt the latter option, which minimizes the requirements on the
\dword{fpga} inside the \dword{wib}, and also reduces the power and cooling requirements for the \dword{wiec}.
Based on this decision, the interface between the \dword{ce} and \dword{daq} consortia is
defined by the fiber plant
used to transmit the data from the \dwords{wib} to the \dword{daq} components housed in the
\dword{cuc}, and to broadcast the clock and controls in the
opposite direction. Only optical links are used in the connection with the \dword{daq},
which guarantees that the \dword{daq} electronics will not induce any noise on the
\dword{apa} wires.

The interface is fully defined with the selection of the number
and type of optical fiber links, their speed, and the type of connectors.
The \dword{fpga} inside the \dword{wib} can be used to reformat the data with changes to
the headers and trailers that include time stamps and geographical addresses
of the \dwords{femb}. It can also be used to serialize the data from multiple
\dword{coldata} \dwords{asic} into a single stream. In the simplest scheme each electrical
link from the \dword{femb} is routed to a single optical fiber, transmitting data
at \SI{1.28}{Gbps}. Depending on the availability and cost of transmitters
capable of sending data at higher speeds data from multiple electrical
links (four, eight, or more) could be serialized onto a single link. Using higher transmission
speeds reduces the number of links that are needed, possibly reducing the
cost of the \dword{daq} part of the detector. The use of links with speeds
of \SI{5}{Gbps} or larger may present the
drawback that the data has to be deserialized on the \dword{cuc} side, depending
on the availability of resources for the extraction of trigger primitives
on the \dword{daq} boards. The final choice of the number of fibers, link speed, and
matching between \dwords{femb} and \dword{daq} processing units in the \dword{cuc} should be
the result of a cost optimization process that can be delayed until
the \dword{tdr}.

Another aspect of the interface between the \dword{daq} and the \dword{ce} consortia is the
transmission of clock and command signals. In \dword{pdsp}, a single fiber
carries this information to the \dword{ptc} card at each \dword{ce} flange, which then re-broadcasts
the information to the five \dwords{wib}, using the \dword{wiec} backplane. For the \dword{spmod}, we
foresee the possibility of transmitting the information directly to each
\dword{wib}, the functionality for which is already implemented in the \dword{pdsp} \dword{wib}, as
shown in Figure~\ref{fig:tpcelec-wib_timing}.

The final aspect of the interface between the \dword{daq} and the \dword{ce} is
the definition of the format of the data transmitted by the \dword{ce}
to the \dword{daq}.  This format has been defined previously for \dword{pdsp}.
Some changes are needed for DUNE to accommodate a larger number of \dwords{apa}.


\section{Quality Assurance}
\label{sec:fdsp-tpc-elec-qa}

\subsection{Initial Design Validation}
\label{sec:fdsp-tpc-elec-qa-initial}
The \dword{qa} program for the DUNE \dword{spmod} electronics has been ongoing for several years.  The \dword{qa} program started with the appropriate design choices for operation
in \lar, including the measurement of transistor properties, and later
continued with tests of all of the cold components, cables, \fdth and flange mechanicals, and warm electronics for suitability 
with respect to the \dword{spmod} requirements.  The current focus of the \dword{qa} program is the instrumenting of the \dword{pdsp} detector with \num{120} \dword{femb} (\num{960} of each of the current \dword{fe} and \dword{adc} \dword{asic} designs) and six full \dword{apa} readout chains as described in Section~\ref{sec:fdsp-tpc-elec-design}. There are aspects of the DUNE design that are not going to be fully validated in the 2018 \dword{pdsp} data taking and that will require additional confirmation. Some aspects of the detector, e.g., the use of stacked \dwords{apa} with long cables routed possibly through the \dword{apa} frames, and the integration and installation procedure, will have to be demonstrated in independent tests. The \dwords{asic} and \dwords{femb} used for the \dword{spmod} will be an evolution of the current \dword{pdsp} ones. Below we focus mostly on the electronics and on system tests to demonstrate that the final design meets the \dword{spmod} specifications, but plans are being made to test all the detector components and their assembly and installation prior to the submission of the \dword{tdr}.

The existing \dword{larasic} design will be revised from the \dword{pdsp} version. Several issues, 
including pedestal uniformity and baseline restoration in the existing \dword{larasic}, have been 
addressed by BNL in a spring 2018 submission. The new design will be tested at BNL and other sites to 
verify that the issues have been resolved. All new \dword{asic} designs, including the new cold \dword{adc}, \dword{coldata}, and CRYO will be tested first at the component level, both at room temperature and at \lar (or LN$_2$) temperature. The next step will be to modify the \dword{femb} to accommodate the new \dwords{asic}.  Tests of the new \dwords{femb} (at least two versions) will be followed by small-scale system tests as described below, and plans are being made to test three full-scale \dwords{apa} with the final \dword{asic}(s) and \dwords{femb} in a second run in the \dword{pdsp} cryostat.  The cold data and \dword{lv} cables used in \dword{pdsp} have been selected to be candidates for the \dword{spmod} baseline design, and tests are in progress to demonstrate that they can successfully transmit the high-speed data over the longest possible cable length in the \dword{spmod}. It should be noted that the current schedule for the \dword{spmod} construction, which is discussed in Section~\ref{sec:fdsp-tpc-elec-org-timeline}, foresees the possibility of two more iterations in the design of the \dwords{asic} and the \dwords{femb}, including the time for performing system tests.

The updates to the \dword{spmod} cryostat penetrations and spool pieces, as well as the warm interface electronics, are expected to be small iterations on the already existing system for \dword{pdsp}.  Prototypes will be ordered in small batches and tested at the responsible institutions for the different components.  After individual testing, integrated system tests including other \dword{spmod} components will be critical to validate that the performance of the \dword{ce} meets the DUNE \dword{fd} requirements.  Issues identified in the integrated system tests will be fed back into the design requirements of the individual components.

\subsection{Integrated Test Facilities}
\label{sec:fdsp-tpc-elec-qa-facilities}
DUNE will plan for system tests of the baseline and alternative option in both the CERN cold box and a small test TPC at \fnal.  The cold box tests establish performance of the electronics coupled to a full-scale \dword{apa} in a correctly grounded environment, but in a gaseous environment no colder than \SI{150}{K}, and without TPC drift or \dword{hv}.  The small test TPC will provide tests in an operational \lartpc but at much smaller scale, with quick turn around (two to four weeks) for changing components and refilling.  The \num{40}\,\% \dword{apa} at BNL employs liquid nitrogen instead of \lar, does not integrate the \dword{ce} with the \dword{pds}, and has no TPC drift.  However, it allows for quick turn-around and is located very close to where the development is happening (BNL and \fnal), and is thus invaluable for initial board and component testing.

Generic board and component testing often includes measurement of (1) the baseline noise level and frequency spectrum, (2) the response to the calibration input signal provided on the \dword{femb}, and (3) the response to cosmic rays in a TPC.  Measurements are often done at both room temperature and liquid nitrogen or argon temperature in three configurations: nothing attached to the inputs, dummy capacitive load attached to the inputs, and an \dword{apa} attached to the inputs as the capacitive coupling of long wires is subtly different than a dummy capacitor with the equivalent capacitance of a single wire.  Measurements specific to the characterization of the \dword{adc} performance, such as integral nonlinearity (INL) and differential nonlinearity (DNL) determination, are done before the \dword{adc} is mounted on the \dword{femb}.

\subsubsection{Cold Box at CERN}
\label{sec:fdsp-tpc-elec-qa-facilities-coldbox}
A cold box at CERN used in electronics tests for \dword{pdsp} is available for electronics testing in cold nitrogen gas. The cold box is designed to cycle one full-size \dword{apa} with the full set of \num{20} \dword{femb} through gaseous nitrogen temperatures around \SI{150}{K} to check out the \dword{apa} performance prior to installing the \dword{apa} into the \dword{pdsp} cryostat. It is designed to be a Faraday cage, using the same grounding and shielding scheme that is implemented for the \dword{pdsp} cryostat. It is read out by a complete \dword{ce} system for a single \dword{apa}, including a \dword{ce} flange and fully-loaded \dword{wiec} with five \dwords{wib} and one \dword{ptc}.

Preliminary results from the \dword{pdsp} cold box indicate that
the noise performance of the TPC readout will satisfy the DUNE \dword{fd} noise requirements of
\dword{enc}\,<\,\num{1000}\,e$^-$ on \dword{spmod}-length wires. These measurements suggest that the noise level in
\lar would be around \num{500}\,e$^-$ and \num{600}\,e$^-$ for the collection and induction plane channels,
respectively; this is well below the requirement.  The \dword{enc} and temperature of the second \dword{pdsp}
\dword{apa} delivered to CERN in the cold box are shown in Figure~\ref{fig:cb_results}.

\begin{dunefigure}
[\dword{enc} in electrons for several cases]
{fig:cb_results}
{\dword{enc} in electrons (left axis) for the wrapped induction wires (red and blue curves) and 
straight collection wires (green curve) as well as the temperature in degrees Kelvin (right axis) for the temperature
sensors in the CERN cold box (orange curves) as a function of cold cycle time in gaseous nitrogen for \dword{pdsp} \dword{apa}2. 
At the lowest temperature of \SI{160}{K}, the wrapped wires measured 480e$^-$ noise and the straight 
wires \num{400}\,e$^-$. This noise level is consistent with all other \dword{pdsp} \dwords{apa} tested in the cold box.}
\includegraphics[width=0.73\linewidth]{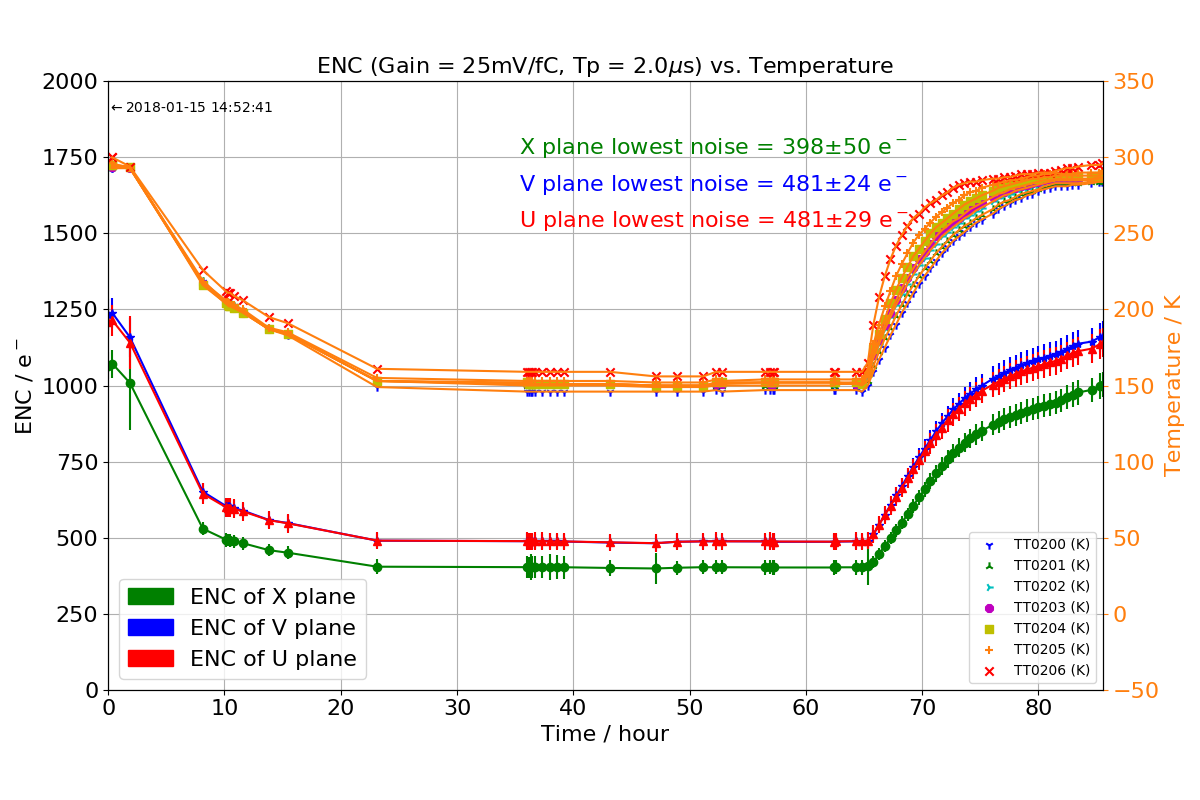}
\end{dunefigure}

\subsubsection{ProtoDUNE-SP}
\label{sec:fdsp-tpc-elec-qa-facilities-pdune}
\dword{pdsp} is intended to be a full slice of the \dword{spmod} as close as possible to the final DUNE \single design. It contains six full-size \dwords{apa} instrumented with \num{20} \dwords{femb} each for a total readout channel count of \num{15360} digitized sense wires. Critically, the \dword{ce} on each \dword{apa} is read out via a full \dword{ce} readout system, including a \dword{ce} flange and \dword{wiec} with five \dwords{wib} and one \dword{ptc}. Each \dword{apa} also has a full \dword{pd} readout system installed. Five of the six \dword{pdsp} \dwords{apa} have been validated in the cold box at CERN, and then installed in the \dword{pdsp} cryostat, while the last \dword{apa} was installed after passing only room temperature tests. Any issues that are discovered either during the cold box tests or the \dword{pdsp} commissioning and data-taking will be incorporated into the next iteration of the system design for the \dword{spmod}.

In addition to the tests described in Section~\ref{sec:fdsp-tpc-elec-qa-facilities-coldbox}, tests have also been done on the \dword{pdsp} \dword{apa} to check for any additional noise introduced on the TPC wire readout by operating the \dword{pds} or enabling the wire bias \dword{hv} system. So far, no significant increase in the noise on the \dword{apa} wire readout has been observed when operating these other systems.

The \dwords{apa} and the readout electronics will be different from the ones used in \dword{pdsp}; for this reason, plans are being made for re-opening the \dword{pdsp} cryostat and replacing three of the six \dwords{apa} with final DUNE prototypes that will also include the final versions of the \dwords{asic} and \dwords{femb}. A second period of data-taking with this new configuration of \dword{pdsp} is being planned for 2021-2022. This will also allow for another opportunity to check for interference between the readout of the \dword{apa} wires and the \dword{pds}.

\subsubsection{Small Test TPC at Fermilab}
\label{sec:fdsp-tpc-elec-qa-facilities-small}
A small test TPC is essential to qualify the different prototype \dwords{asic}, offering quick turn around for changing components and refilling; this allows one to study the response of the electronics to signals from cosmic-ray muons.  A new reduced-size \dword{apa} will be constructed with many similarities to the DUNE \dword{apa} design.  The \dword{apa} will be half the width of a DUNE \dword{apa} (along the beam direction) or \SI{1.3}{m}, with half the number of readout channels.  This amounts to \num{10} \dwords{femb} with a total of \num{1280} channels.  The height will be significantly reduced from the DUNE \dword{apa} height of \SI{6}{m} to about \SI{1.25}{m}, and the wire lengths will be reduced by the same factor.  The \dword{pdsp} CR boards will be used, and  the \dword{apa} will accommodate a single half-length \dword{pd}.  The TPC will have the \dword{apa} in the center and a cathode on either end, creating two drift volumes with drift distance \SI{0.3}{m} each.  The TPC will be installed in the cryostat with the wire planes parallel to the floor to optimize the orientation of the cosmic ray tracks.  It will be instrumented with a full readout chain of \dword{pdsp} electronics, specifically  the cables, \fdth flange, \dword{wib}, \dword{ptc}, and warm interface crate.  Initially, \dword{pdsp} \dwords{femb} will be used for commissioning, and later \dwords{femb} with prototype \dwords{asic} will be swapped in.

The TPC electronics and photodetector will be read out through a slice of the \dword{pdsp} \dword{daq}.  This system will provide a low-noise environment that will allow one to make detailed comparisons of the performance of the new \dwords{asic}. It will also enable the study of interactions between the TPC readout and other systems, including the \dword{pd} readout and the \dword{hv} distribution, to exclude the possibility that one system generates noise on another one when both are being operated.
 
The \dword{apa} will be housed in a new \lar cryostat at \fnal located in the Proton Assembly Building that complies with the DUNE grounding and shielding requirements, and connected to an existing recirculation system for argon purification.  The cryostat will be a vertical cylinder, with an inner depth of \SI{185}{cm} and an inner diameter of \SI{150}{cm}.  With the cryogenic connections on the upper portion of the cylinder, the flat top plate will have penetrations dedicated to readout and cryogenic instrumentation.  The target date for the fabrication of new \dword{fe} motherboards is fall of 2018, and these will house the latest version of the \dword{fe} \dword{asic}, the first prototype of \dword{coldata}, and various \dword{adc} prototypes (including the SLAC CRYO \dword{asic}).  The new \dword{apa} and the new cryostat will be completed on the same timescale so that tests in both the \dword{pdsp} cold box at CERN and the test TPC at \fnal can be completed and analyzed prior to the submission of the \dword{tdr}.

\subsubsection{Additional Test Facilities}
\label{sec:fdsp-tpc-elec-qa-facilities-other}
For \dword{ce} development, testing prototypes at room temperature is the first step, as many problems can be identified quickly and without the expense of cryogens.  A quick access test stand with the \dword{femb} connected to an \dword{apa} inside a shielded environment that is in the same location as the \dword{femb} and \dword{asic} development is invaluable for rapid progress.  Two such facilities are available to DUNE: the shielded room at \fnal and the \num{40}\,\% \dword{apa} test stand at BNL.  In addition, a test dewar design developed by Michigan State University, referred to as the Cryogenic Test System (CTS), allows for additional testing of the \dwords{femb} and \dwords{asic} in LN$_2$.

The shielded room at \fnal (see Figure~\ref{fig:shieldedroom}) is \SI{2.5}{m} tall, and \SI{2}{m} on each side, with 
a double layer of copper mesh in the walls, floor and ceiling, plus a solid metal plate in the floor all electrically connected to create a Faraday cage.  A flexible AC distribution and isolated grounding configuration offers the ability to easily ground the shielded room and refer the associated electronics to either a building ground or a detector ground.  In addition
to evaluating different grounding schemes for the \dwords{apa}, capacitive coupling issues can be studied by varying the
distance between the floor and a copper plate positioned underneath.   
This room has uniquely easy access to the setup through a shielded door,
and a person can remain inside safely with the door closed and probe the electronics directly while operating
in a shielded environment.  Currently mounted inside are two \dwords{apa} from the \dword{35t} 
with adapter boards to connect \dword{pdsp} electronics.  This installation satisfies the \dword{pdsp} grounding and shielding guidelines.  A rough demonstration of the shielding adequacy for our purposes  is the measured noise level of 800~\dword{enc} for \dword{pdsp} prototype \dwords{femb}.  The same noise level was measured at room temperature in the \num{40}\,\% \dword{apa} at BNL for the same \dwords{femb}.

\begin{dunefigure}
[Picture of the shielded room at \fnal.]
{fig:shieldedroom}
{Picture of the shielded room at \fnal.}
\includegraphics[angle=270,width=0.4\linewidth]{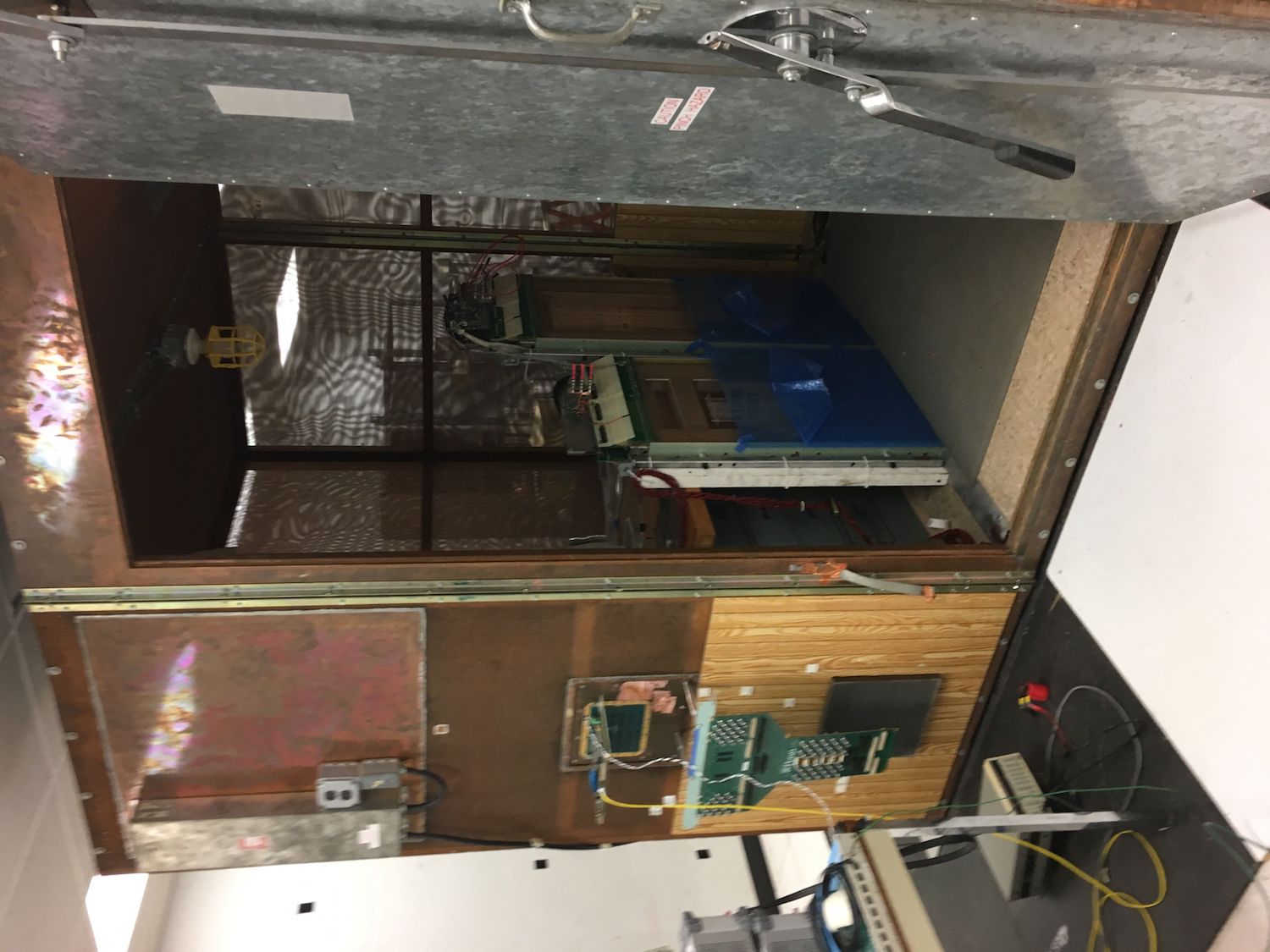}
\end{dunefigure}

The \num{40}\,\% \dword{apa} at BNL is a \SI{2.8}{m}~$\times$~\SI{1.0}{m} three-plane \dword{apa} with two layers of \num{576} wrapped ($U$ and $V$) wires and one layer of \num{448} straight ($X$) wires. It is read out by eight \dword{pdsp} \dwords{femb} with the full \SI{7}{m} \dword{pdsp} length data and \dword{lv} power cables, four on the top and four on the bottom. The readout uses the full \dword{ce} system for \dword{pdsp}, with a prototype \dword{ce} flange and \dword{wiec}, two \dwords{wib} and one \dword{ptc}, as shown in Figure~\ref{fig:tpcelec_40apa}. Detailed integration tests of the \dword{ce} readout performance while following the DUNE grounding and shielding guidelines have been done at the \num{40}\,\% \dword{apa}. Additional input capacitance (equivalent to longer wire length) have been added to a subset of channels to project the \dword{enc} performance from the 40\% \dword{apa} teststand to the \dword{pdsp} and SBND detectors. The results from the \num{40}\,\% \dword{apa} indicate that, if the new \dword{adc} performs as expected, the full \dword{ce} system as installed on the test  stand at BNL will have a noise level in \lar around \num{500}\,e$^-$ and \num{600}\,e$^-$ for the collection and induction plane channels, respectively, in line with the CERN cold box tests described in Section~\ref{sec:fdsp-tpc-elec-qa-facilities-coldbox}.

\begin{dunefigure}
[One side of the \num{40}\,\% \dword{apa} with four \dwords{femb} and the full \dword{ce} \fdth and flange.]
{fig:tpcelec_40apa}
{Left: one side of the \num{40}\,\% \dword{apa} with four \dwords{femb}.  Right: the full \dword{ce} \fdth and flange.}
\includegraphics[width=0.72\linewidth]{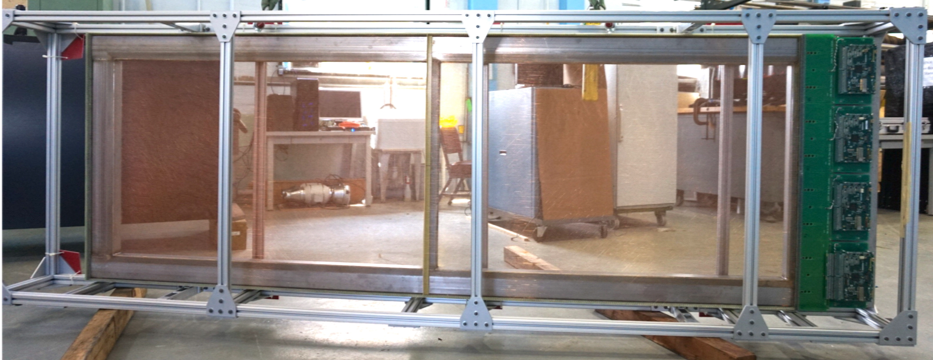}
\hspace{3mm}
\includegraphics[width=0.2\linewidth]{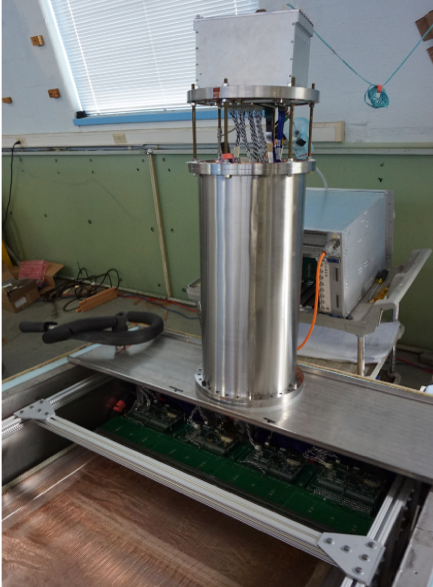}
\end{dunefigure}


To facilitate testing of individual components and printed circuit boards, members of the Michigan State group have developed the CTS (see Figure~\ref{fig:CTS}).  This system allows a device under test to be cooled down in nitrogen gas, immersed in LN$_2$ for testing, and then warmed back to room temperature in a nitrogen gas.  This process avoids the condensation of water from air that can otherwise interfere with the tests or damage the test equipment.  A total of nine CTSs will be built and used at consortium member institutions.

\begin{dunefigure}
[The Cryogenic Test System (CTS)]
{fig:CTS}
{Cryogenic Test System: an insulated box is mounted on top of a commercial LN$_2$ dewar.  Simple controls allow the box to be purged with nitrogen gas and LN$_2$ to be moved from the dewar to the box and back to the dewar.}
\includegraphics[width=0.4\linewidth]{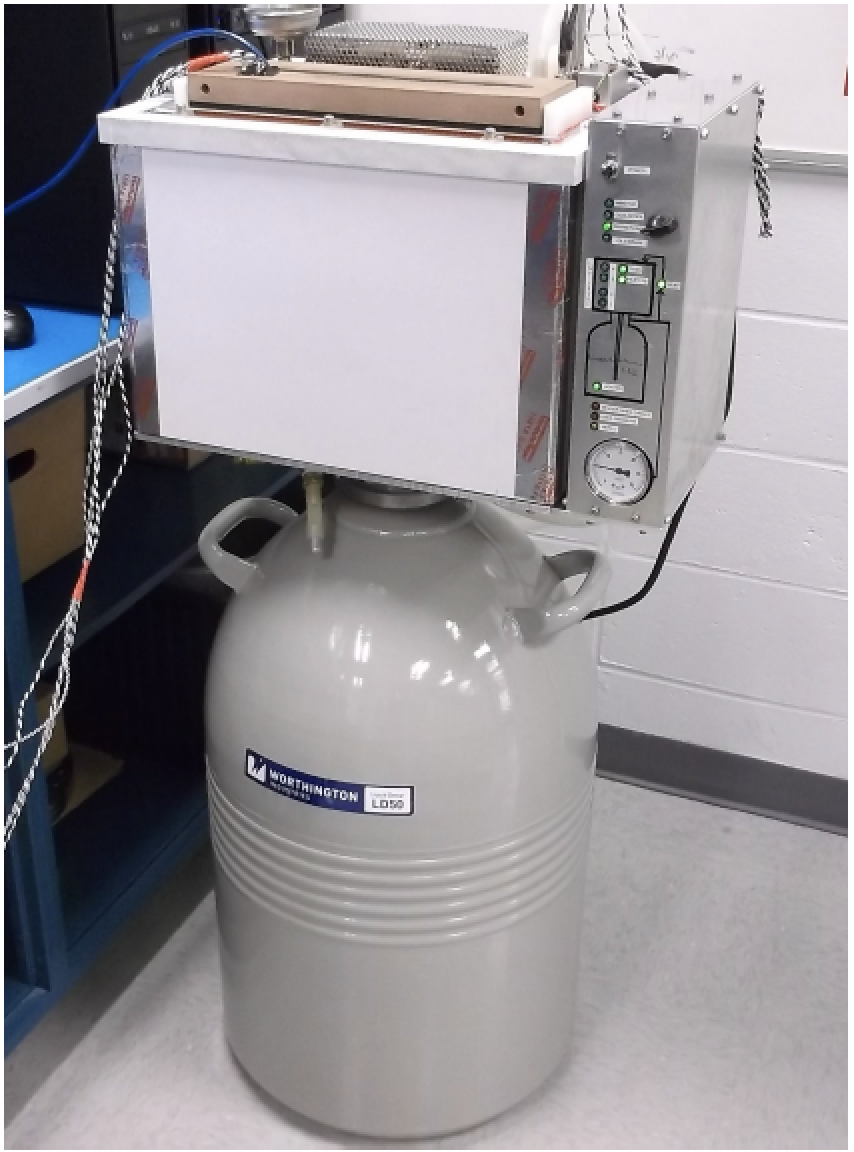}
\end{dunefigure}

\section{Quality Control}
\label{sec:fdsp-tpc-elec-qc}
Once there is a final design that clearly satisfies the requirements and constraints, given all of the testing at the device and system level described above, it will be necessary to put in place procedures and controls to ensure that the production \dword{ce} parts will continue to fully satisfy the requirements and constraints. This set of \dword{qc} procedures, based on the experience gained with \dword{pdsp}, can only be sketched at this point without yet knowing, for instance, yield statistics on the final \dwords{asic}. Nevertheless, it is possible to put forward a general plan based upon long-established good practice.

All of the custom \dwords{asic} will be packaged by a commercial vendor. 
All custom printed circuit boards will be produced by qualified vendors to at least IPC class 2\footnote{ "Acceptability of Printed Boards", IPC(R), IPCA-600F, Association Connecting Electronics Industries\texttrademark{}, \url{http://www.ipc.org}.} standards; the same standard will apply to all commercial assembly of those boards. 
All commercial parts will be procured from known, reliable vendors and manufacturers and qualified for use at LN temperatures. Depending upon the actual yield of the custom packaged chips it may or may not be necessary to individually test those chips prior to assembly onto printed circuits, and the use of an automated cryogenic test station is being considered. In any event, all custom printed circuits delivered by the assembler will be run through a full functional test sequence at a DUNE institution and then subjected to a powered burn-in period of one or more weeks at elevated temperature followed by a second functional test sequence. Depending upon experience with pre-production assemblies, this burn-in may also include a temperature cycling step. All cables will be fully tested for continuity and lack of shorts either by the assembler or at a DUNE institution prior to installation. Cable testing may also include impedance and electrical length verification. All discrete \dword{ce} parts will be serial-numbered and a production database will record the specifics for each part and each test sequence.

However, as much of the \dword{spmod} 
TPC electronics will operate in a cryogenic environment, which differs from typical commercial experience, it will be necessary to add additional cryogenic testing steps to the above list for all of the \dwords{asic} and circuit boards that will be subject to those cryogenic conditions. Based upon current experience, the expectation is that repeating the functional test sequence with the board under test immersed in LN$_2$ should be sufficient. It may be useful to subject a small sample of such boards to the stress of multiple cycles between room and LN$_2$ temperatures; this will allow one to understand how much margin is available against failure due to stress caused by thermal cycling for each production batch.

Handling of \dword{ce} components at DUNE institutions must follow standard IPC class 2 good practice for cleanliness, electrostatic discharge protection, and environmental conditions. Boards, cables, and other \dword{ce} components must be stored in qualified storage facilities and all shipments of boards and other components will be in qualified shipping containers and shipped via qualified shippers.

\section{Installation, Integration, and Commissioning}
\label{sec:fdsp-tpc-elec-install}

\subsection{Installation and Integration with APAs}
\label{sec:fdsp-tpc-elec-install-apa}
The installation and commissioning of the detector components built or
purchased by the \dword{ce} consortium takes place both prior to
and after the insertion of the \dwords{apa} in the
detector cryostat, with testing performed after each step, to avoid the
need for rework that could cause significant delays. The installation and
initial commissioning of the \dword{ce} electronics is likely to be on the critical
path for the completion of the \dword{spmod} and the amount of
time available for testing and possibly repairing or replacing components
after their installation is going to be very limited. Cold tests of the complete
\dwords{apa} for the entire detector require the
availability of at least two independent cold boxes at the integration facility.
To reduce the number of failures in these tests, all \dword{ce} components will
be qualified for operation in \lar prior to their installation.

After the completion of the cryostat, the spool pieces that house all of
the cables for the \dword{ce}, \dword{pds}, and \dword{hv} consortia (with the exception of the
high-voltage \fdth for the cathode planes) are installed and leak tested.
This includes the installation of the crossing tube cable support and of
the flanges that provide the cold-to-warm interface for all of the cables.
In parallel, the racks that house the \dword{ce} and \dword{pds} electronics components on
the top of the cryostat and the corresponding \dword{cisc} and detector safety
system monitoring, controlling, and interlock hardware can be installed.
Cable trays between different spool pieces belonging to a set of six \dwords{apa}
and two \dwords{cpa} can also be put in place. Readout fiber bundles between the
spool pieces and the central utility cavern used for the readout of the wire
information from the \dwords{apa} can also be put in place. In the meantime, inside
the cryostat, all of the cable trays used to support the \dword{ce} and \dword{pds} cables, and
to accommodate the slack of the cables, can be put in place.

In parallel, the \num{20} 
\dwords{femb} required to read out the
wires from one \dword{apa} are installed with their shielding box onto the \dword{apa} and
connected to the CR boards. This work will be performed at the integration
facility (or facilities), because there will not be enough space to perform
this type of work in parallel on multiple \dwords{apa} in the detector cavern. Once
the \dwords{femb} are installed, a temporary set of cables will be
used to connect each \dword{femb} to a temporary power, control, and readout system
to ensure that all the wires can be properly read out. All \dwords{apa}
will then be inserted into a cold box similar
to the one used at CERN for \dword{pdsp} and tested at a temperature of
$\sim$\SI{150}{K}. Once these tests
are completed, the temporary cables are disconnected and the \dword{apa} is prepared
for shipment to \surf. It is not considered feasible to transport the
top \dword{apa} from the integration facility to the detector cryostat with the
\dword{ce} cables already installed. For the bottom \dword{apa} the final cables can be
installed only after the two \dwords{apa} are joined together in the so-called \textit{toaster} area
just outside of the detector cryostat.

Further work on the \dword{ce} components installed on the \dwords{apa} is performed after
the \dwords{apa} are transported to the clean area outside the cryostat at
\surf. All of the cables that provide power and control and are used
to read out the \num{20} \dwords{femb} associated with the top and bottom \dwords{apa} need to be
installed; the cables that connect to the SHV boards that are used to
distribute the bias voltage to the \dword{apa} wires, electron diverters, and \dword{fc}
termination electrodes need to be installed as well. The cables for the bottom
\dword{apa} need to be routed through the frames of the \dwords{apa}, an operation that
can be performed only after the two \dwords{apa} are mechanically coupled inside
the toaster area. Quick tests are performed after the installation of the
final cables to ensure that the detector has not been damaged in the transport
and that all cable connections have been performed correctly. Only then 
can the \dwords{apa} be moved to their final positions inside
the detector cryostat. At that point, the \dword{ce} and \dword{pds} cables can be routed
through the spool piece and connected to the respective flanges and strain
reliefs. The flanges are then moved to the final position on the spool pieces
and leak tests and electrical connectivity tests can be performed. Then the
bottom plate of the crossing tubes, along with its additional strain relief for
the \dword{ce} and \dword{pds} cables, can be put in place; the final slack of the cables
are also arranged in the cable trays attached to the supports inside the cryostat.
After the cabling work is completed, the \dwords{wiec} for a pair of \dwords{apa} can be
installed on the spool piece and more testing of the entire power, control,
and readout chain can be performed: first with local control, and later
after connection of the readout and timing and control fibers to the \dwords{wib}
using the final \dword{daq} system. The installation of the next row of \dwords{apa} will
begin only when all of these tests have been completed, with a requirement
that all \dwords{femb} are properly read out from the \dword{daq}, allowing at most for
a \num{0.1}\,\% fraction of non-working channels.

A total of seven months is available in the schedule for the installation
of the \num{25} rows of \dwords{apa} and \dwords{cpa}. The current plans foresee that six \dwords{apa}
and two \dwords{cpa} are installed and
tested in one week before moving on to the next set. As soon as another
set of \dwords{apa} and \dwords{cpa} is in place, any replacement of components or rework
of connections inside the cryostat becomes very difficult or impossible.
This requires that all readout tests be performed very quickly after
each step in the installation. The schedule foresees four work days each
week for the installation work (in two eight-hour shifts). The testing
activities may require that work is performed in parallel, i.e., that
the \dword{ce} components of a pair of \dwords{apa} are tested while another pair is
being installed and/or connected to the powering, control, and readout
system. It is also likely that the testing work may require a more
extended working schedule (six or seven working days per week). More
complex tests, involving the reading out of multiple rows of \dwords{apa}, will continue
throughout the entire period (eight months) during which all the detector
components are installed inside the cryostat. Final tests should be
performed reading out the entire detector through the \dword{daq} system
prior to the closure of the \dword{tco} and before
starting to fill the cryostat with argon and cooling down. There will be a hiatus
in the commissioning activities during the filling of the cryostat, during
which the conditions of the detector will continue to be monitored. The
commissioning at \lar temperature most likely will be possible only
after the cryostat is completely full of \lar.

\subsection{Commissioning and Calibration}
\label{sec:fdsp-tpc-elec-install-calib}
Directly following the installation of the instrumented \dwords{apa} into the cryostat, commissioning of the
TPC electronics will commence, and will be carried out both before and after the \lar fill;
the former establishes whether or not the installation procedure led to impairment of the
electronics, and the latter checks that the cryogenic electronics do not experience failures
in the cold \lar.

As described in Section~\ref{sec:fdsp-tpc-elec-install-apa}, the electronics checkout prior to
\lar fill will begin as soon as the first \dword{apa} is installed; 
that is,
installation and testing will proceed in parallel.  This has the advantage of informing the 
 installation of subsequent \dwords{apa}, and thus minimizing the total number of electronics channels lost
during the installation process.  Items to be checked during the commissioning process include
the noise level on every channel, dead or noisy (high \rms) channels,
cross-talk across neighboring wires or neighboring channels in the electronics,
pick-up noise (including spatial dependence in the detector) if present, and
noise coherent across channels sharing common electronics (e.g.,~the same \dword{femb}).
These measurements will be performed first during the tests in the cold box.  Repeating these
measurements prior to filling the cryostat with \lar will allow for final repairs or replacements
for all the \dword{ce} components prior to the closure of the \dword{tco}. This will also give an opportunity
to intervene on noise or electronics issues that would become evident when reading out
the entire detector at once, before proceeding proceeding with the \lar fill.

Once every \dword{apa} is installed and tested as described above, the cool-down process and \lar fill
will be carried out.  Monitoring of noise levels and dead/noisy channel count will be
done continuously during this process.  After the \lar fill, another electronics checkout
similar to the one described above will be carried out.  Any electronics issues identified
during this procedure will be fixed before continuing, if possible.  The wire bias \dword{hv} and
cathode \dword{hv} will then be brought up, with noise studies repeated after each subsystem is
turned on.  With the wire bias \dword{hv} and cathode \dword{hv} up, one can then utilize ionization signals
(e.g.,~from $\mathrm{{}^{39}Ar}$ beta decays) to distinguish between different possible issues
that might impair the readout of a given channel, such as a short between the wire planes
(which would alter the wire field response on nearby wires) and a problem with the electronics.

Both during the commissioning phase of the experiment and during normal operations, it may
be desired to perform an in situ calibration of one or more parts of the TPC electronics chain.
These calibrations will utilize a combination of noise data and data collected while a
calibration pulser (internal DAC on the FE \dword{asic}) is periodically injecting charge into the TPC
electronics channels.  Calibrations of interest include determining the gain and shaping time of
every electronics channel in the TPC (as the FE \dwords{asic} may experience changes in these quantities
in the cold) and characterizing the linearity of the \dword{adc} \dwords{asic} in the cold.  With the updated \dword{adc} \dword{asic}
design, it is not expected that nonlinearity of the \dwords{adc} will be a significant issue (see
Section~\ref{sec:fdsp-tpc-elec-design-femb-adc} for a discussion of the on-chip calibration
that is used), but it is important to verify this with data collected in the experiment.
Specific in situ calibration algorithms are being studied at \dword{pdsp}, and will be further
developed there and in the other electronics test facilities described in
Section~\ref{sec:fdsp-tpc-elec-qa-facilities}.  Experience from \microboone and other running
\lartpc experiments will be very useful in informing TPC electronics calibration procedures for
the DUNE \dwords{spmod}.

\section{Safety}
\label{sec:fdsp-tpc-elec-safety}
The TPC electronics will be built and handled in such a way as to ensure the safety of both personnel and equipment.  The team will work closely with the project Technical Coordination organization to make sure that all applicable safety procedures are followed and documented.

The instrumentation of the TPC electronics will include multiple printed circuit boards and cabling.  The cabling includes the high-voltage wire bias distribution, low-voltage power and signals.  Each of these elements will require attention to relevant safety standards and solutions will be subject to the review of the project Technical Coordination organization.

All printed circuit boards will be designed such that the connectors and copper-carrying traces are rated to sustain the maximum current load.  In the case of the low-voltage warm electronics, all boards will be fused following prescribed safety standards.  The cold electronic low-voltage boards inside the cryostat will not be fused because these boards will be inaccessible during operations and fire is not a danger once the cryostat is filled with \lar.  Special precautions must be taken during installation and commissioning of the \dword{ce} prior to the cryostat being filled with \lar.  The TPC electronics group will work with the project Technical Coordination to implement this.

All cabling and connectors will be selected such that they meet or exceed the possible ampacity and voltage ratings of the connected power supplies.  In the case of \dword{hv} wire bias distribution, all accessible warm connectors will be SHV type connectors, which limit the possibility of a touch potential that could shock a person.  In the cold, many of the \dword{hv} connections will be open soldered connections. Care will be taken to ensure personnel safety should these connections need to be energized while the \dword{apa} is exposed.  However, it is not anticipated that \dword{apa} wire bias will be powered by more than \SI{50}{V} unless the \dword{apa} is enclosed within a Faraday shield.

Finally, the safety of the equipment must be taken into account during production, initial checkout, installation, and cabling.  Proper electrostatic discharge (ESD) procedures will be followed at all stages, including use of ESD safe bags for storage and ESD wrist straps used by personnel when handling the cards.  The TPC electronics will also make use of shorting connectors on all cables which are attached to the printed circuit cards, but not attached at the opposite end.  During installation, multiple long cables will be attached to front-end boards, but will not be attached to the connectors on the flanges for some period of time.  Detailed procedures for the use of cable-shorting connectors will be written and used.

Finally, the handling of the \dword{apa} and attached front-end electronics must follow ESD safe handling procedures whenever the \dword{apa} is moved from one ground reference to another, or after it has been left in a ``floating'' state for any period of time.  Whenever the \dword{apa} is moved and could encounter a step potential, a connection must be made through a slow discharge path that equalizes the \dword{apa} frame potential to the new environment.  Again, ESD safe-handling rules will be documented and followed.

\section{Organization and Management}
\label{sec:fdsp-tpc-elec-org}

\subsection{Single-Phase TPC Electronics Consortium Organization}
\label{sec:fdsp-tpc-elec-org-consort}
For the moment the \dword{ce} consortium does not have a formal substructure
with coordinators appointed to oversee specific areas. This is in part due
to the current focus on \dword{asic} development; informal subgroups exist that
are following the design of the various \dwords{asic}. 
A BNL collaborator is currently leading the
design of the new version of \dword{larasic}, and in parallel following the studies
of commercial \dwords{adc} for SBND, which is using the same \dword{asic}.  
Collaborators from LBNL, BNL, and \fnal are working on the design of a new \dword{adc} \dword{asic} with 65 nm technology. 
Collaborators from \fnal and SLAC, respectively, 
are overseeing the development of the new \dword{coldata} \dword{asic} and 
the adaptation of the nEXO CRYO \dword{asic} 
for use in DUNE. 
A new working group is tasked with studying reliability 
issues in the \dword{ce} components and preparing recommendations for the choice
of \dwords{asic}, the design of printed circuit boards, and testing. This working group
will consider past experience from cryogenic detectors operated for a long
time (ATLAS \lar calorimeters, NA48 liquid krypton calorimeter, HELIOS),
from space-based experiments (FERMI/GLAST), and the lessons learned from
\dword{pdsp} construction and commissioning. Input from other fields will
also be sought. Later this working group will develop the \dword{qc} program for the \dword{ce} detector
components, starting from the \dword{pdsp} experience. It is planned to
reassess the structure of the group in a few months, with a likely split
between components inside and outside the cryostat, a new group responsible
for testing, and various contact people for calibration, physics, software and
computing, and integration and installation.
 
The main decision that the consortium has to face in the next \numrange{12}{18} months
is the choice of \dwords{asic} to be used in the DUNE \dwords{femb}. A first decision will
be taken early in Summer 2018, when it will be determined whether or not system
tests, beyond those planned by the SBND collaboration, should be performed for 
commercial \dword{adc} chips and for the ATLAS \dword{adc}. In February 2019, following tests
performed with a \dword{pdsp} \dword{apa} in the cold box at CERN and with a small TPC in
\lar at \fnal, a list of options will be prepared for presentation in the \dword{tdr}.
Only \dwords{asic} that satisfy the DUNE performance and reliability requirements
will be included in this list of options. A final choice for the \dwords{asic} to
be used in DUNE should be taken in summer 2019, prior to the DOE CD-2/CD-3b
review. Physics performance, reliability, and power constraint considerations
will be taken into account when making this choice, which will go through
the Executve Board approval procedure.

\subsection{Planning Assumptions}
\label{sec:fdsp-tpc-elec-org-assmp}
Plans for the \dword{ce} consortium are based on the overall schedule for DUNE
that assumes that the first \dwords{apa} will be fully populated with electronics
and tested in spring 2022, with the installation of
the \dwords{apa} inside the cryostat beginning in May 2023. Plans are being made
for replacing three of the six \dwords{apa} of \dword{pdsp} with the final DUNE 
\dwords{apa} including final \dwords{asic} and \dwords{femb}, and for a second period of
data-taking in 2021-2022. The integration of the
\dwords{apa} for the first cryostat with the electronics should be finished by
October 2023, and their installation in the cryostat by January 2024.
This requires integrating two \dwords{apa} per week over a span of 21 months,
which allows for a ramp-up period at the beginning and a contingency of
two to three months at the end. This defines the time window for the completion of
the R\&D program on the \dwords{asic}. A set of \dwords{asic} (or a single \dword{asic}) meeting
all the DUNE requirements has to be fully qualified by fall 2020, such
that pre-production \dwords{asic} and the corresponding \dwords{femb} can be assembled and
tested in spring 2021, launching the full production in summer 2021.

Meeting this timeline requires that the development of the \dwords{asic}, and in particular
of the newly designed ones (the SLAC CRYO \dword{asic}, the joint LBNL-BNL-\fnal cold \dword{adc},
and \dword{coldata}) are prototyped by the end of summer 2018, with testing
completed by the end of 2018. This would allow for a second round of prototyping,
if necessary, in the first half of 2019. It also leaves room for a possible
third design iteration and qualification of \dwords{asic} and \dwords{femb} between the
end of 2019 and fall 2020. The \dwords{femb} used for \dword{pdsp} will
likely have to be redesigned to house a new \dword{adc} and to replace the \dword{fpga} used
in \dword{pdsp} with the \dword{coldata} \dword{asic}. Multiple variants of this board will be
necessary, depending on the success of the various \dword{adc} R\&D projects being currently
pursued. These design changes will be made in the second half of 2018.
Additional \dword{femb} prototypes will be designed and fabricated depending on
the outcome of the initial testing of the SLAC CRYO \dword{asic}. A second iteration
of \dword{femb} prototype(s) will be necessary in 2019, when the final \dwords{asic} (that
may have a different channel count from the first round of prototypes) will become
available. 

It is assumed that apart from the \dwords{asic}, where rapid development is still
required, and the \dwords{femb}, which have to be redesigned to accommodate the
new \dwords{asic}, most of the detector components to be delivered by the \dword{ce} consortium
will require only minor changes relative to the \dword{pdsp} components. For
this reason the modifications of these other detector components will
be delayed until 2019 or 2020, which will also help with the funding profile.
Exceptions will be made for further development in test stands, for cabling
studies, and for conceptual studies of automated testing assemblies, rack space
assignment, and the interface to the \dword{daq} system.

\subsection{WBS and Responsibilities}
\label{sec:fdsp-tpc-elec-org-wbs}
A preliminary \dword{wbs} has been prepared for the activities
of the \dword{ce} consortium. The \dword{wbs} is split in a time-ordered fashion
between activities related initially to design, R\&D, and engineering, then
to production setup, and finally to the production, integration, and installation
phases for the \dword{spmod} to be installed in the first DUNE cryostat.
The latter three sets of activities could be repeated for the construction
of additional \dwords{spmod}.  
Physics and simulation
activities are to proceed in parallel to the detector design and construction activities.
Within each phase (starting with the design and ending with the installation),
the \dword{wbs} foresees work packages that cover system engineering, installation of the detector
components inside the cryostat (including all \dwords{asic}, \dwords{femb},
cables, and corresponding support structures and cryostat penetrations),
and the detector components installed on top of the cryostat (including the warm
interface electronic crates with their boards, the \dword{lv} and bias-voltage
power supplies with their crates, and all of the associated cables and infrastructure). This
matches the current plan for the future group structure of the \dword{ce} consortium. 
In addition, a separate work package covers the development and support of the testing facilities and the related software in order to provide these activities with the effective supervision that is required in order to meet the reliability requirements of the DUNE experiment.

All of the institutions currently interested and committed to the construction of
the detector components that are a responsibility of the \dword{ce}
consortium are from the USA and are supported by a single
funding agency, the Department of Energy. For this reason, the exact role of
the individual institutions in the activities of the consortium has not been
defined, except for the currently ongoing \dword{asic} development. The role of
each institution will be defined prior to the submission of the \dword{tdr}.

\subsection{Timeline and Key Milestones}
\label{sec:fdsp-tpc-elec-org-timeline}
A preliminary list of milestones indicating the current planning for the completion
of the design, R\&D, and engineering phase, and then later for the production setup
and the production, integration, and installation activities is shown in Table~\ref{tab:ce-milestones}.
This list of milestones corresponds to a scenario in which only two iterations in
the design of the \dwords{asic} and \dwords{femb} are required, and results in a float of nine months
for the integration of the \dword{ce} with the \dwords{apa}. In the case that a third
iteration is required, the availability of the \dwords{asic} and \dwords{femb} could introduce a 
delay in the overall schedule of the \dword{spmod}. 
A detailed schedule of all of the activities will be prepared in the coming months,
with the goal of having a better estimate of the critical path for the project
under different assumptions on the number of design iterations for the \dwords{asic} and \dwords{femb}.

\begin{dunetable}
[\dword{ce} consortium milestones]
{ll}
{tab:ce-milestones}
{Milestones of the Cold Electronics consortium.}
\textbf{Date} & \textbf{Milestone} \\ \toprowrule
Jun 2018 & Submission of first version of all custom \dwords{asic} \\ \colhline
Oct 2018 & Bench test of first version of all custom \dwords{asic} \\ \colhline
Feb 2019 & System tests of first version of all \dwords{asic} and \dwords{femb} \\ \colhline
Feb 2019 & Conceptual design of fibers and cabling plant \\ \colhline
Feb 2019 & Demonstrate cable routing through \dword{apa} frames \\ \colhline
May 2019 & Submission of second version of all custom \dwords{asic} \\ \colhline
Dec 2019 & System tests of second version of all \dwords{asic} and \dwords{femb} \\ \colhline
Mar 2020 & Revise design of \dwords{wib} and crates 
\\ \colhline
Jun 2020 & Revise design of cryostat penetrations \\ \colhline
Sep 2020 & Revise design of detector components outside the cryostat \\ \colhline
Nov 2020 & Launch pre-production of \dwords{asic} and \dwords{femb} \\ \colhline
Jun 2021 & Integrate \dword{ce} with pre-production \dwords{apa} \\ \colhline
Jul 2021 & Availability of all test stands for \dwords{asic} and \dwords{femb} \\ \colhline
Jul 2021 & Availability of vertical slice test with final production components \\ \colhline
Oct 2021 & Launch pre-production of all detector components \\ \colhline
Mar 2022 & Begin integration of \dword{ce} on production \dwords{apa} \\ \colhline
Nov 2022 & Integration of \dword{ce} on \dwords{apa} \num{50}\,\% complete \\ \colhline
Nov 2022 & Launch production of cryostat penetrations \\ \colhline
Nov 2022 & Launch production of warm interface electronic crates and boards \\ \colhline
Jun 2023 & Start detector installation in the cryostat at \surf \\ \colhline
Oct 2023 & Complete integration of \dword{ce} on \dwords{apa} \\ \colhline
Jan 2024 & Complete \dword{apa} installation at \surf \\ \colhline
Feb 2024 & Complete initial tests of detector prior to \dword{tco} closing \\
\end{dunetable}

\cleardoublepage

\chapter{High Voltage System}
\label{ch:fdsp-hv}

\section{High Voltage System Overview}
\label{sec:fdsp-hv-ov}

\subsection{Introduction}
\label{sec:fdsp-hv-intro}

A \dword{lartpc} requires an equipotential cathode plane at \dword{hv} and a precisely regulated interior \efield{} to drive 
electrons from particle interactions to sensor planes.  In the case of the DUNE \dlong{sp} technology, 
this requires vertical cathode planes, called \dwords{cpa}, held at \dword{hv}; vertical anode planes, called \dwords{apa}, described in Chapter~\ref{sec:fdsp-apa-intro}; and formed sets of conductors at graded voltages surrounding the
 the central drift volume, collectively called the \dlong{fc}. The \dword{fc} consists of portions on the top and bottom  
of the drift volume (called the \dword{topfc} and \dword{botfc}, respectively), and along the sides, called \dwords{ewfc}.

\fixme{\dword{cpa} by itself is the only form of \dword{cpa} not defined. What is the cathode plane ``assembly'' as opposed to the \dword{cpa} plane, panel, array, unit...? RKP: Believe this is clear as written.}

The SP \dword{tpc} construction is shown in Figure~\ref{fig:dune_sp_fd}.
The  drift fields transport the ionization electrons 
towards the \dwords{apa} at sides and center.
One should note that the \dword{hv} systems for the \single and \dual TPC concepts share components with similar designs. These include the \dwords{fc} profiles and supporting FRP beams as well as the voltage divider boards. More details can be found in \voltitledpfd Chapter 4.

\begin{dunefigure}[One unit of the \dword{spmod}]
{fig:dune_sp_fd}
{
A schematic of a \dword{spmod} showing the three \dword{apa} arrays (at the far left and right and in the center, spanning the entire \dword{detmodule} length) and the two \dword{cpa} arrays, occupying the intermediate second and fourth positions. The top and bottom \dword{fc} modules are shown in blue; the \dwords{ewfc} are not shown. 
On the right, the front top and bottom \dword{fc} modules are shown folded up against the \dword{cpa} panels to which they connect, as they are positioned for shipping and insertion into the cryostat.  The \dwords{cpa}, \dwords{apa} and \dwords{fc} together define the four drift volumes of the \dword{spmod}.}
\includegraphics[width=0.8\textwidth]{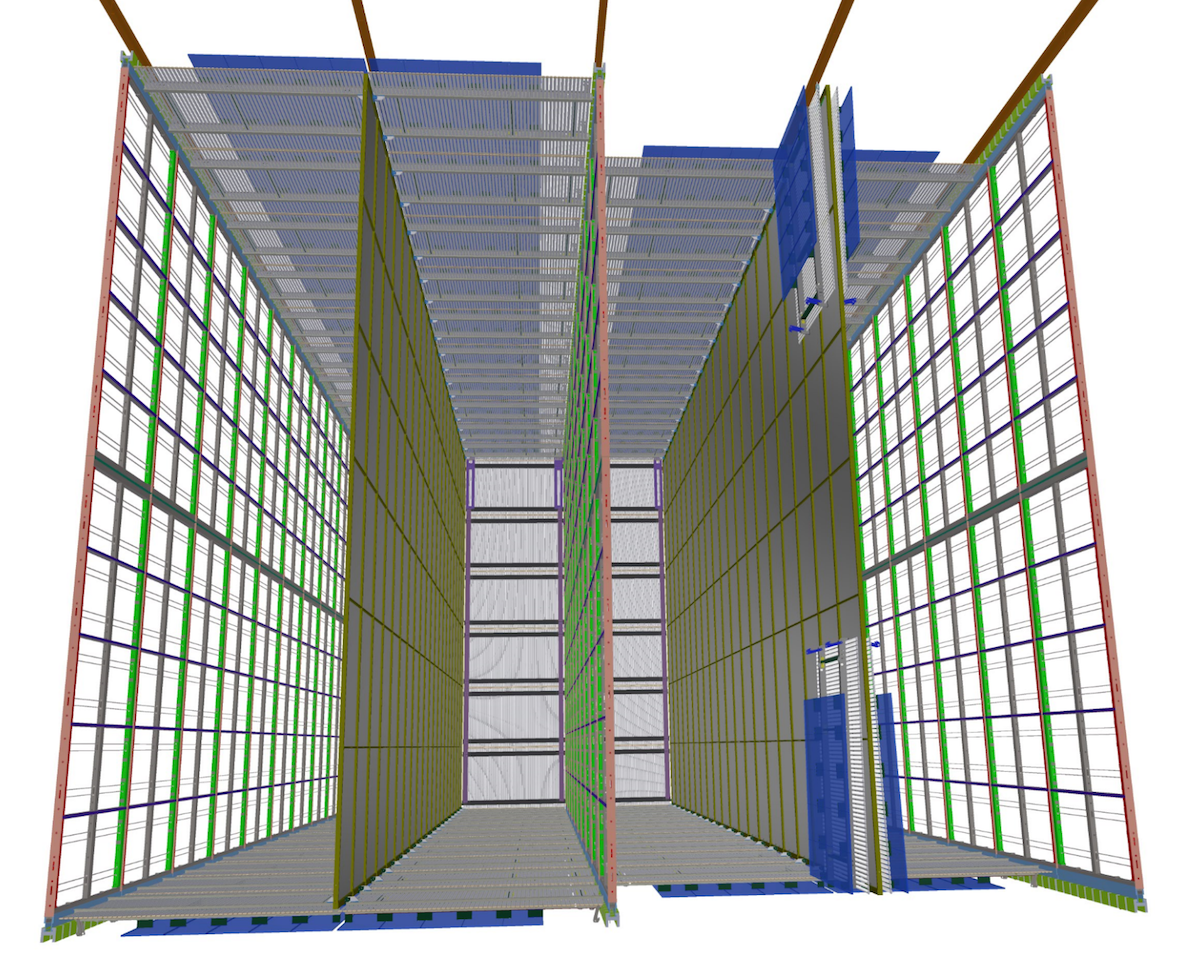}
\end{dunefigure}

The \dword{hv} consortium provides systems that operate at the full range of voltages, 
maximum to ground, inside the \dword{tpc} volume. As a result, its systems constitute a large fraction of the total internal structures of the \dword{tpc} itself, the principal exception being the \dwords{apa} and the \dword{pds}. In addition, the \dword{hv} system largely bounds the useful fiducial volume of the experiment, and thus plays a key role in determining the event rate for all DUNE physics processes. Mechanical and structural concerns are integrated with electrical design to meet the requirements. 


Two possible anode-cathode plane configurations exist for the fixed DUNE maximum electron drift length: cathode planes facing cryostat wall (C-A-C-A-C) or anode planes facing the cryostatic wall (A-C-A-C-A).  The latter configuration keeps most of the cathode plane surfaces far away from the grounded cryostat walls, reducing electrostatic breakdown risks and decreasing the total energy stored in the electric field to \SI{800}{J}.

In this configuration, energy is mostly stored in the high \efield{} region between the field cage and the nearby grounded conductors.  In an unexpected \dword{hv} breakdown, the entire \SI{400}{J} associated with one cathode could be released into a small volume of material, possibly causing physical damage.
It is difficult to predict the distribution of energy along a discharge path. A conservative approach treats this energy as a risk to the TPC and the cryostat membrane.  
Mitigating this risk entails slowing down the energy release as much as possible to minimize the potential damage by subdividing the \dword{fc} into electrically isolated modules, and constructing the cathode with highly resistive material.

Previous large \lartpc{}s (ICARUS, \microboone) have used continuous stainless steel tubes as electrodes. 
Electrically, linking such electrodes to span more than \SI{100}{\m} in total length increases the stored energy each electrode has, and increases the risk of damaging the field cage components in a HV discharge. 

Having the \dword{fc} divided into mechanically and electrically independent modules eases the construction and assembly of the \dword{fc}, and also greatly restricts the extent of drift field distortion caused by a resistor failure on the divider chain of a \dword{fc} module.

If the cathode is made of metal, a \dword{hv} discharge can cause the electrical potential of the entire cathode surface to swing from its nominal bias (e.g., \SI{-180}{kV}) to \SI{0}{V} in nanoseconds. This would induce a large current into the analog \dword{fe} amplifiers connected to the sensing wires on the \dwords{apa} (mostly to the first induction wire plane channels). Internal study (docdb 1320) has shown that this surge of current would overwhelm the internal ESD protection in the \dword{fe} ASICs.  To reduce this induced current, we chose to construct the cathode out of material with high resistivity.  Figure~\ref{fig:cpa-frame-discharge} shows the release of stored energy in time and the voltage distribution of a section of the cathode at one moment in time.  To minimize the induced current to the amplifiers, the surface resistivity should be raised until the ionization current from the TPC starts to cause significant voltage drop along the cathode.  In the DUNE FD the current is dominated by $^{39}$Ar decay, and we can tolerate surface resistivity well above \SI{1}{\giga\ohm/sq}.

\begin{dunefigure}[Simulated \dword{cpa} discharge event]
{fig:cpa-frame-discharge}
{Bottom: Simulated \dword{cpa} Discharge event on a highly resistive cathode surface (1\,G$\Omega$/$\Box$), showing the voltage distribution on a section of the cathode (2.3\,m $\times$ 12\.m) 0.2\,ms after the discharge. Top: Time dependence of removal of stored energy.}
\centering
\includegraphics[width=0.6\textwidth,trim=2mm 2mm 2mm 2mm,clip]{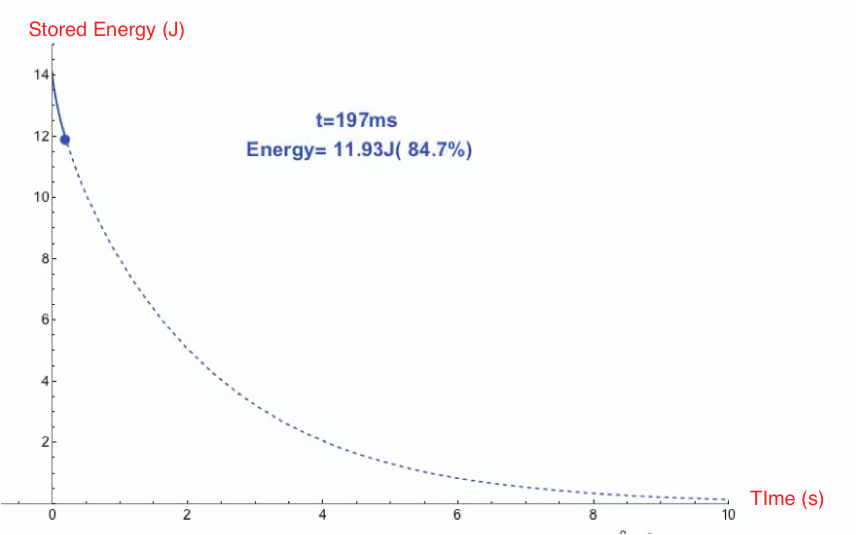} \\ \vspace{30pt}    
\includegraphics[width=0.5\textwidth,trim=2mm 2mm 2mm 2mm,clip]{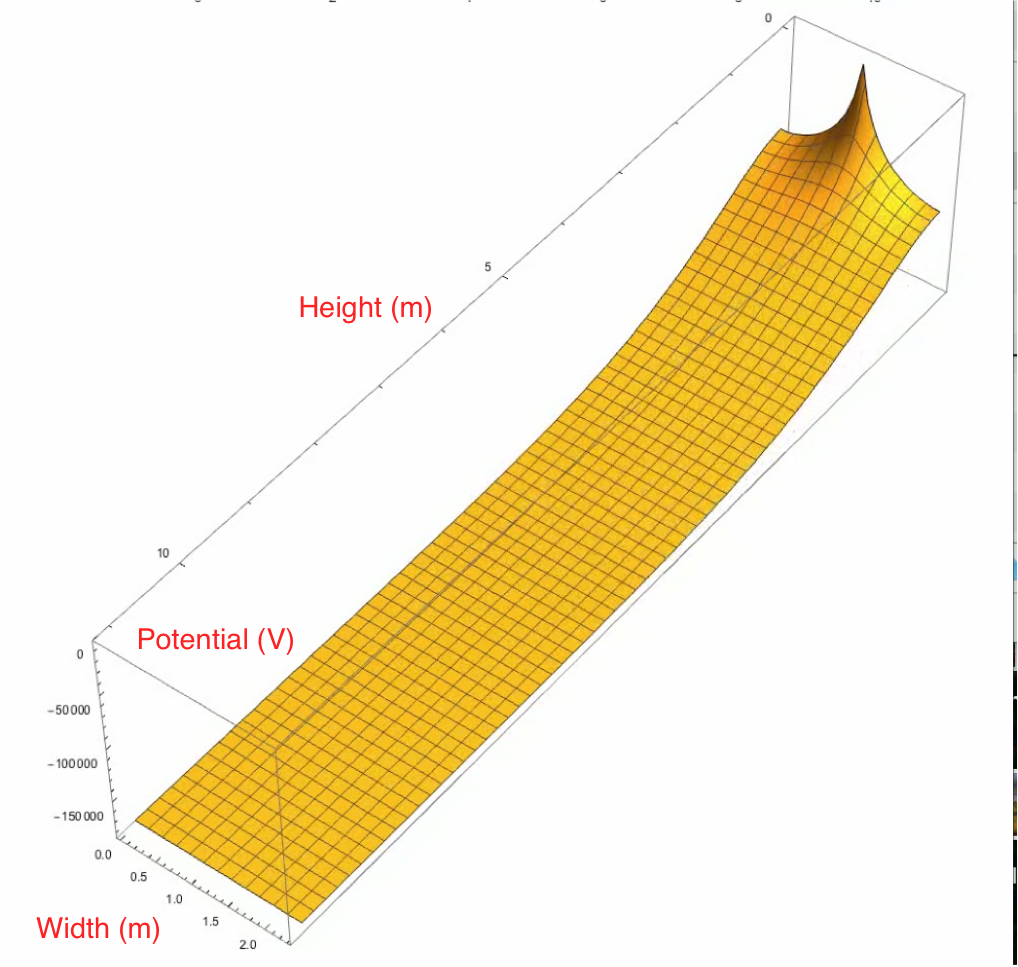}
\end{dunefigure}

The SP \dword{hv} system may have modifications if problems are identified in the present design in \dword{pdsp}. Issues identified in earlier testing form the basis of an ongoing R\&D program.

\subsection{Design Requirements}
\label{sec:fdsp-hv-des-consid}
 
The 
\dword{hv} system is designed to meet the physics requirements of the DUNE experiment. These are both physical requirements (such as \efield{}s that allow robust event reconstruction) and operational (a system that avoids over-complication so as to maximize the time that the detector can collect neutrino events). An important collection of the requirements affecting \dword{hv} is shown in 
Table~\ref{tab:hvphysicsreqs}.

\begin{dunetable}
[HV system requirements]
{p{0.05\textwidth}p{0.2\textwidth}p{0.35\textwidth}p{0.15\textwidth}p{0.15\textwidth}}
{tab:hvphysicsreqs}
{\dword{hv} system requirements}   
No. & Requirement & Physics Requirement Driver & Requirement & Goal \\ \toprowrule
1 & Establish uniform minimum \efield{} in \dword{tpc} drift volume. & Limit recombination, diffusion, and space charge impacts $e$, $\mu$, $p$ particle ID.  Establish constant drift velocity and adequate \dword{s/n} on induction planes for tracking. & >\SI{250}{V/cm} & \spmaxfield \\ \colhline
 2 & Do not exceed maximum \efield{} in \dword{lar} volume. 
 & Avoid damage to detector to enable data collection over long periods. & \SI{30}{kV/cm} & \dword{alara} \\  \colhline
3 & Minimize power supply ripple & Keep readout electronics free from external noise, which confuses event reconstruction.  & 0.9~mV & 0.9~mV\\ \colhline
4 &  Maximize power supply stability. & Maintain ability to reconstruct data taken over long period.  Maintain high operational uptime to maximize experimental statistics. \\ \colhline
5 & Provide adequate resistivity to create acceptable decay constant for discharge of the cathode surface and \dword{fc}.  & Avoid discharge damage to \dword{detmodule} or electronics to enable data collection over long periods. 
& \SI{1}{\mega\ohm}/sq & \SI{1}{\giga\ohm}/sq \\ \colhline
6 & Provide redundancy in all \dword{hv} connections. & Avoid single point failures in \dword{detmodule} that interrupt data collection. & Two-fold & Four-fold \\ 
\end{dunetable}

\subsection{Scope}
\label{sec:fdsp-hv-scope}

The scope of the \dword{hv} system includes the selection and procurement of materials for, and the fabrication, testing, delivery and installation of systems to generate, distribute, and regulate the voltages that
create a stable and precise \efield{} within a DUNE \dword{spmod}. 

The \dword{hv} system consists of components both exterior and interior to the cryostat. The voltage generated at the \dword{hv} power supplies passes through the cables, filters, and the \dword{hv} feedthrough into the cryostat. From the voltage delivery into the cryostat, it is further distributed by components that form part of the \dword{tpc} structure. 
These components are:
\begin{itemize}
\item \dword{hv} power supply;
\item \Dword{topfc}, \dword{botfc}, and \dwords{gp};
\item \Dword{ewfc}.
\end{itemize}

\fixme{Here I propose a figure that illustrates how these cpa and fc pieces relate to each other, could be based off of figure 11. Anne}

The \dword{tpc} has two \dwords{cpa} \textit{arrays}, that span the length and height of the \dword{spmod}, as shown in Figure~\ref{fig:dune_sp_fd}. Given the modular design, each array is assembled a set of 25 adjacent \dword{cpa} \textit{planes}, which in turn are constructed of smaller pieces.  Each plane is a set of two adjacent \textit{panels} (full height, half length, as measured along the \dword{detmodule} length). Each panel consists of three stacked \textit{units}, approximately \SI{4}{\m} high by \SI{1.2}{\meter} long. The units each consist of two half-height vertically stacked resistive panels enclosed within a FR4\footnote{NEMA grade designation for flame-retardant glass-reinforced epoxy laminate material, multiple vendors, National Electrical Manufacturers Association\texttrademark{},  \url{https://www.nema.org/pages/default.aspx}.} frame.

An  installation rail supports the panels from above through a single mechanical link.

The sides of the drift volumes on both sides of the \dword{cpa} plane are covered by the \dword{fc} and \dword{ewfc}  modules to define a uniform drift field of \spmaxfield{}, with a increasing potential over \SI{3.5}{m} from the \dword{hv} \dword{cpa} (\SI{-180}{kV}) to ground potential at the \dword{apa} sensor planes. The cathode bias is provided by an external \dword{hv} power supply through an \dword{hv} \fdth connecting to the \dword{cpa} plane inside the cryostat.
The \dword{fc} modules come in two distinct types: the top and bottom (\dword{fc}), which run the full length of the \dword{detmodule}, and the \dwords{ewfc}, 
which complete the detector at either end. The modules of both systems are constructed from an array of extruded aluminum open profiles supported by FRP\footnote{Fiber-reinforced plastic, a composite material made of a polymer matrix reinforced with fibers, many vendors.} (fiber-reinforced plastic) structural beams. A resistive divider chain connects adjacent metal profiles to provide a linear voltage gradient between the cathode and anode planes.  The \dword{topfc} and \dword{botfc} modules are nominally  \SI{2.3}{\meter} wide by  \SI{3.5}{\meter} long. At the ends, the \dword{ewfc} modules are \SI{3.5}{\meter} wide by \SI{1.5}{\meter} in height.

Structurally, the frames of the cathode and field cages are made from materials with similar thermal expansion coefficients, minimizing issues of differential thermal expansion. The field cage frames support aluminum profiles but these are restrained at only one location and are allowed to float within the frame.
The \dwords{ewfc} modules, each \SI{1.5}{\m} high by \SI{3.5}{\m} wide (along the drift volume dimension), stack eight units high to cover the \tpcheight{} height of the \dword{tpc}.  
Extensive tests have been performed of mechanical and electrical properties of materials used in the \dword{hv} system.  These are fully documented elsewhere\fixme{Add to bib: \textit{CPA Electrical Connections Cold Test}, DUNE DocDB 2338; \textit{CPA and FC Design}, DUNE DocDB 1504; \textit{Technical Review: Mechanical Specifications for \dword{pdsp} \dword{fc} test in \dword{35t} at \fnal}, DUNE DocDB 1601.
}

The \dwords{cpa} and \dwords{apa} support the \dword{topfc} and \dword{botfc} modules, whereas
installation rails above the \dwords{apa} and \dwords{cpa} support the \dword{ewfc} modules. 
A \dlong{gp} consisting of tiled, perforated stainless steel sheets 
runs along the outside surface of each of the 
\dword{topfc} and \dword{botfc}, with a \SI{20}{\centi\meter} clearance. 

Tables~\ref{tab:cpaparts} and~\ref{tab:fcparts} contain summaries of terminology and parts.

\begin{dunetable}
[\dword{hv} \dword{cpa} components]
{p{0.4\textwidth}p{0.12\textwidth}
p{0.12\textwidth}p{0.32\textwidth}}
{tab:cpaparts}
{\dword{hv} Cathode Plane Components} 
Component and Quantity &  Length (z) & Height (y) & Per \dword{spmod} \\ \toprowrule
\dword{cpa} array (2 per \dword{spmod}) & \SI{58}{\meter} & \SI{12}{\meter} & 2  \\ \colhline
\dword{cpa} plane (25 per \dword{cpa} array)  & \SI{2.3}{\meter}  &\SI{12}{\meter} & 50  \\ \colhline
\dword{cpa} panel (2 per \dword{cpa} plane)  & \SI{1.2}{\meter}   & \SI{12}{\meter} & 100  \\ \colhline
\dword{cpa} unit (3 per \dword{cpa} panel)  & \SI{1.2}{\meter}  & \SI{4}{\meter} & 300 \\ \colhline
\dword{cpa} RP (2 per \dword{cpa} unit)  & \SI{1.2}{\meter}  & \SI{2}{\meter} & 600 \\
\end{dunetable}

\begin{dunetable}
[\dword{hv} \dword{fc} components]
{p{0.37\textwidth}p{0.07\textwidth}p{0.07\textwidth}p{0.07\textwidth}p{0.07\textwidth}
p{0.1\textwidth}p{0.15\textwidth}}
{tab:fcparts}{\dword{hv} Field Cage Components}
Component & Count & Length (z) & Width (x) & Height (y) & Submodules & Grand Total \\ \toprowrule
\dword{fc} (Top/Bottom Field Cage) & 200 & 2.3 m & 3.5 m & - & 57 & 200 \\ \colhline
\dword{fc}-Profiles (per \dword{fc}) & 57 & 2.3 m & - & - & - & 11400 \\ \colhline
Ground Plane Modules (per \dword{fc}) & 5 & 2.3 m & 0.7 m & - & - & 1000 \\ \colhline
EW-Plane (Endwall Field Cage) & 2 & - & 14.4 m & 12 m & 4 & 2 \\ \colhline
EW (per EW-Plane) & 4 & - & 3.5 m & 12 m & 8 & 8 \\ \colhline
EW-Modules (per EW) & 8 & - & 3. m & 1.5 m & 57 & 64 \\ \colhline
EW-Profiles (per EW-Module) & 57 & - & - & 1.5 m & - & 3648 \\
\end{dunetable}
\section{HV System Design}
\label{sec:fdsp-hv-design}

\subsection {High Voltage Power Supply and Feedthrough}
The \dword{hv} delivery system consists of
\begin{itemize}
\item two power supplies,
\item \dword{hv} cables,
\item filter resistors, and
\item \dword{hv} feedthroughs into the cryostat.
\end{itemize}

For \dword{hv} delivery, two power supplies are used to generate the voltage, one for each \dword{cpa} array. 
This separated setup more easily accommodates different running conditions and helps isolate any instabilities. 
The cryostat design has two feedthrough ports for each \dword{cpa} array, one at each end of the cryostat. The spare, downstream port provides redundancy against any failure of the primary \dword{hv} delivery system. 

Each \dword{cpa} connects to two drift volumes in parallel, presenting a net resistance of \SI{1.14}{\giga\ohm} to each supply. At the nominal \SI{180}{kV} cathode voltage, each power supply must provide \SI{0.16}{mA}.

The planned power supply model for the \dword{spmod} is similar to the power supply\footnote{Heinzinger, PNC HP200000 \dword{hv} power supply, Heinzinger\texttrademark{} Power Supplies, \url{http://www.heinzinger.com/}.} used on \dword{pdsp}, with a maximum output voltage of \SI{200}{kV} and a maximum current draw of \SI{0.5}{mA}.  An 
option is an existing \SI{300}{kV}, \SI{0.5}{mA} model from the same vendor.
The \dword{hv} cables are commercially available models compatible with the selected power supplies. 


Filter resistors are placed in between the power supply and the feedthrough.  Along with the cables, these resistors reduce the discharge impact by partitioning the stored energy in the system.  The resistor-cable assembly also serves as a low-pass filter reducing the \SI{30}{kHz} voltage ripple on the output of the power supply.  With filtering, such supplies have been used successfully in other \lartpc experiments, such as \microboone and ICARUS.

Figure~\ref{fig:ps_filter_ft_schematic} provides a sample schematic of the \dword{hv} supply circuit.

\begin{dunefigure}[A schematic showing the \dword{hv} delivery system to the cryostat.]  
{fig:ps_filter_ft_schematic}
{Right:  A schematic showing the \dword{hv} delivery system to the cryostat (Credit:  SEL). 
One of the two filters sits near the power supply; the other sits near the feedthrough. Left:   
A Heinzinger power supply (Credit: H.~Wang).}
\begin{minipage}{6in}
  \centering
  $\vcenter{\hbox{\includegraphics[width=0.2\textwidth]{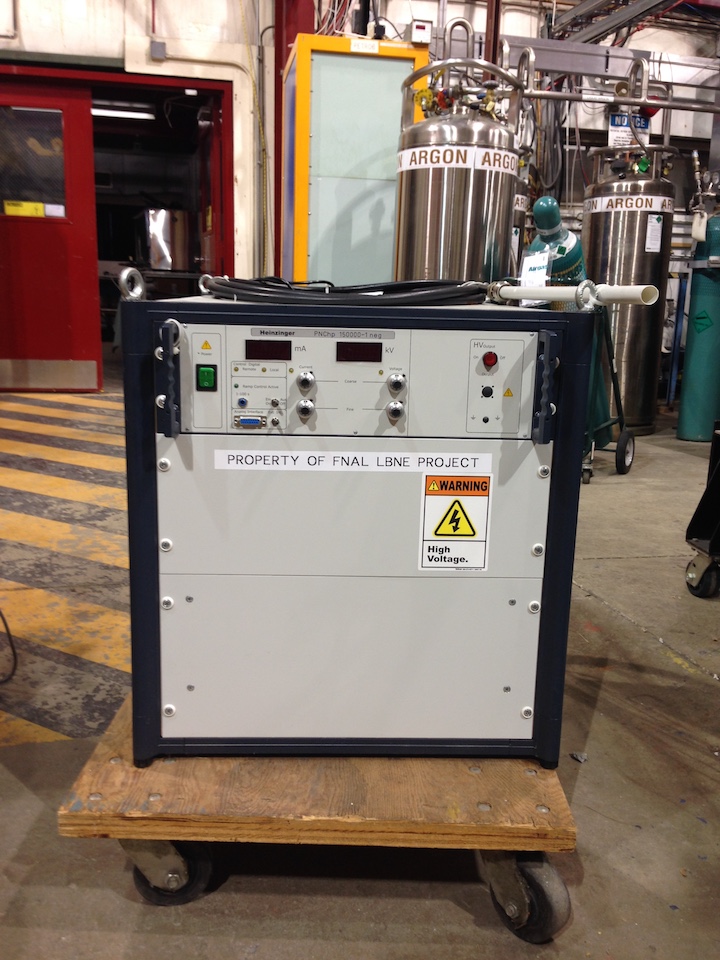}}}$
  \hspace*{.2in}
  $\vcenter{\hbox{\includegraphics[width=0.7\textwidth]{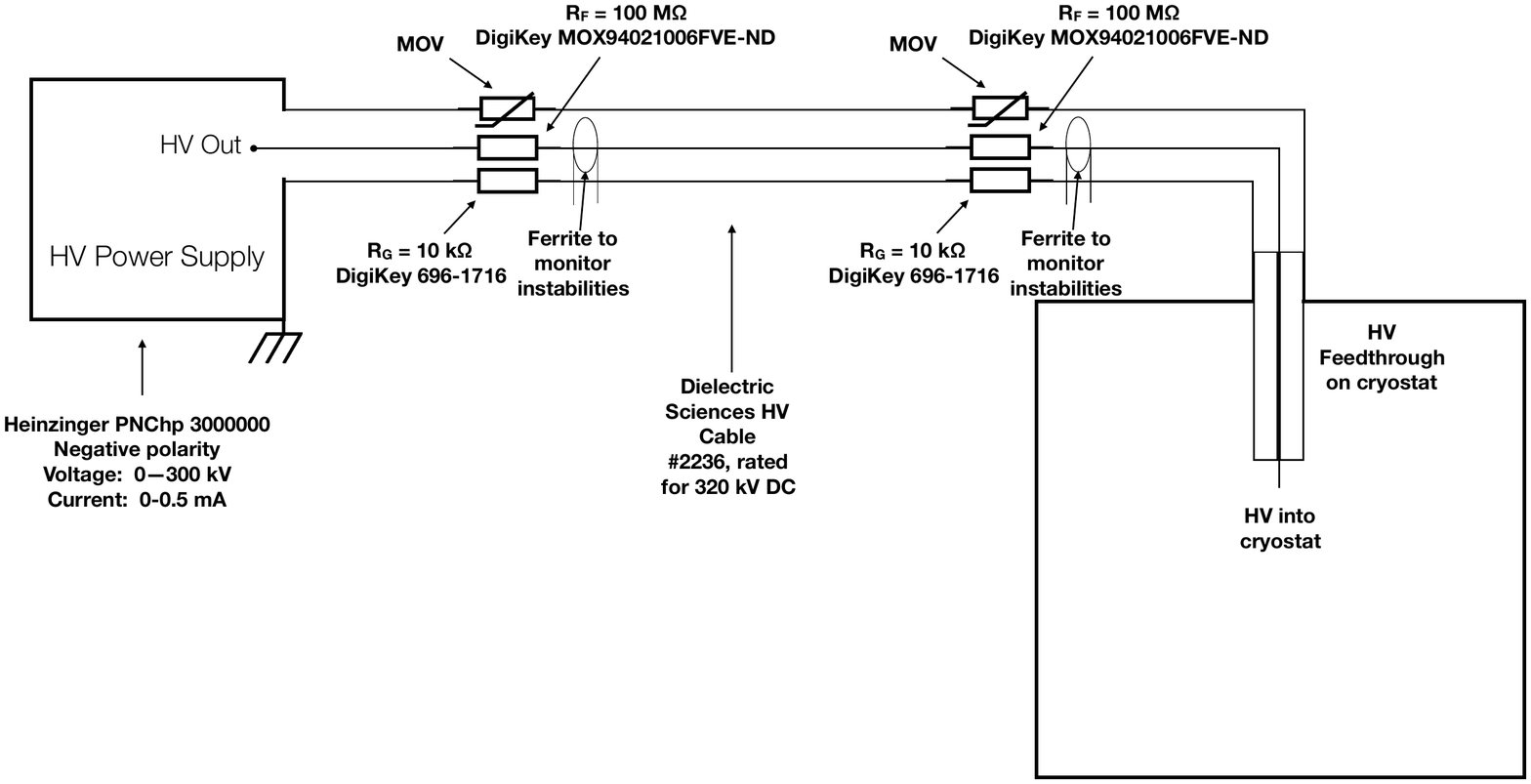}}}$
\end{minipage}
\end{dunefigure}
The requirement 
on low electronics noise sets the upper limit of residual voltage ripple on the cathode to be \SI{0.9}{mV}.  Typically, commercial supplies 
specify  that the ripple variation is limited to 
\SI{.0001}{\%} around an absolute precision in nominal voltage of plus or minus \SI{50}{mV}.
Assuming cable lengths of \SI{30}{m} and \SI{3}{m} between the filters themselves, and between the filter and \fdth, respectively, resistances as low as a few \si{\mega\ohm} yield the required noise reduction according to calculations and experience. 

The current plan for the filters is a cylindrical design.  
Here each end of an \dword{hv} resistor is electrically connected to a cable receptacle. 
The resistor 
must withstand a large over-power condition.  Radially out from the resistor is an insulator,  
for which other designs have used transformer oil or ultra-high molecular weight polyethelene (UHMWPE).  The outer case of the filter is a grounded stainless-steel shell. The current filter design is shown in Figure~\ref{fig:filterAndFeedthrough}.

The \dword{hv} feedthrough 
is based on the successful ICARUS design, \fixme{ref} which has been adapted for \dword{pdsp}.  The voltage is transmitted by a stainless steel center conductor.  On the warm side of the cryostat, this conductor mates with a cable end.  Inside the cryostat, the end of the center conductor has a spring-loaded tip that 
contacts a receptacle cup mounted on the cathode, delivering \dword{hv} to the field cage.  The center conductor of the \fdth is surrounded by UHMWPE. A drawing is shown in Figure~\ref{fig:filterAndFeedthrough}.

\begin{dunefigure}[Drawings of the \dword{hv} filters and feedthroughs]{fig:filterAndFeedthrough}
{Left:  Drawing of a \dword{hv} filter (Credit:  A.~Renshaw). Right:  \dword{hv} feedthrough drawing (Credit:  (F.~Sergiampietri).}
\includegraphics[width=0.3\textwidth]{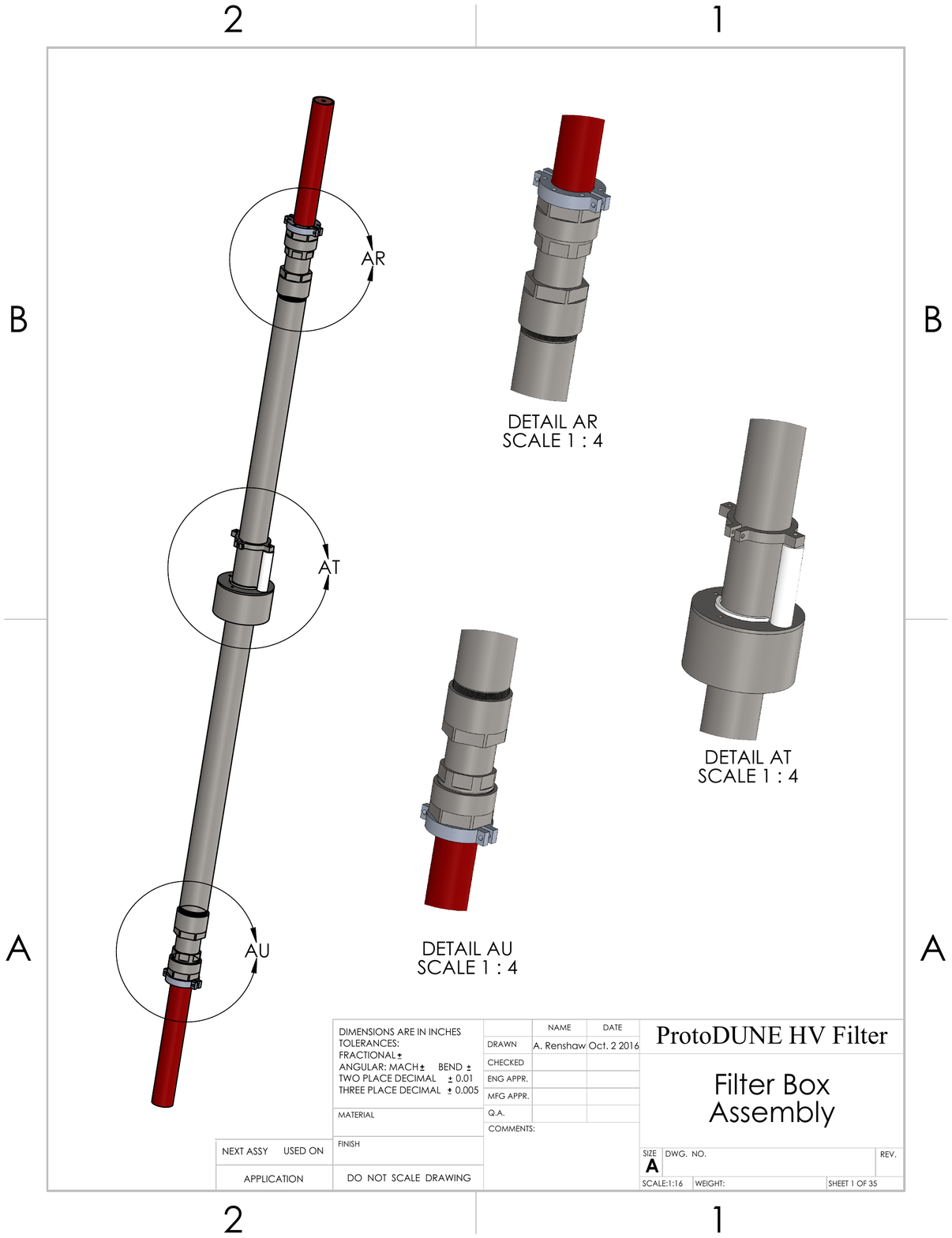}
\includegraphics[width=0.6\textwidth]{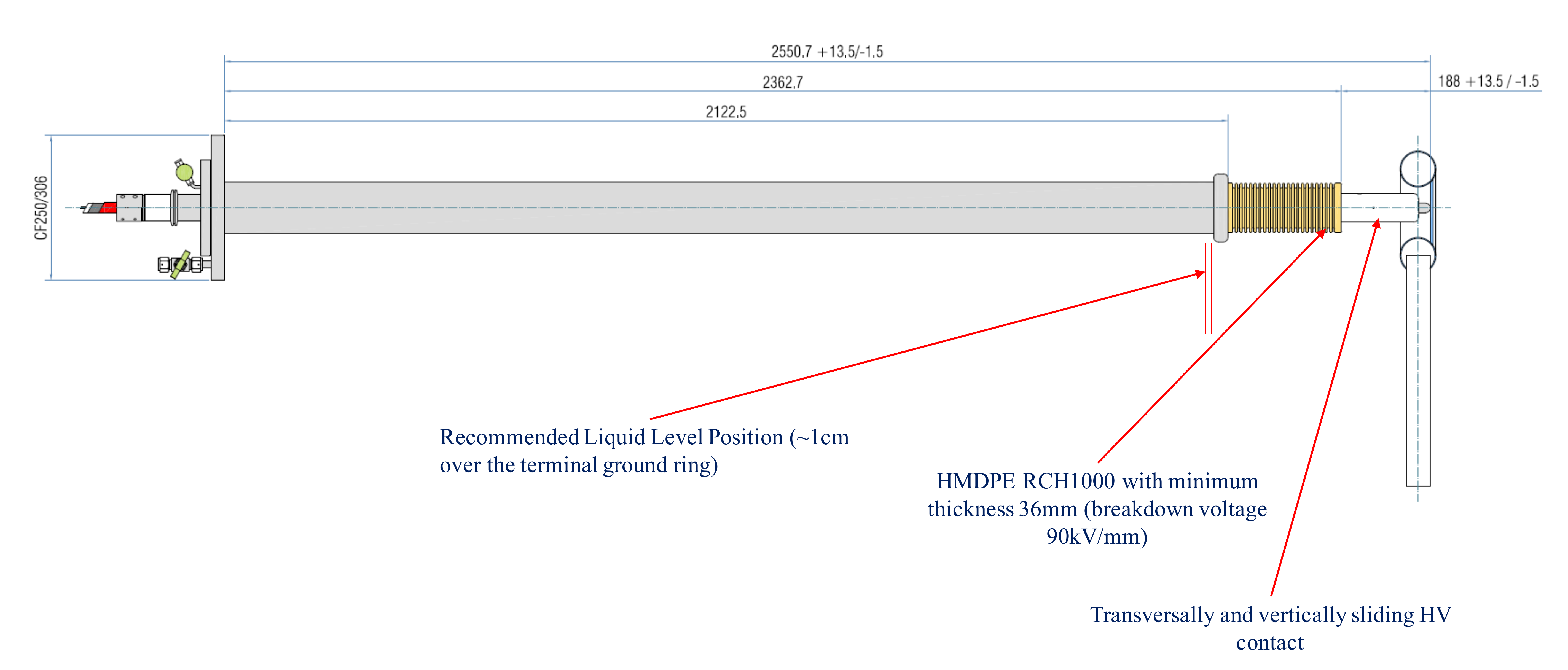}
\end{dunefigure}

The upper bound of operating voltage on a feedthrough is, to first order, set by the maximum \efield{} on the feedthrough.  This \efield{} is reduced by increasing the insulator radius.  For the target voltage, the feedthrough uses a UHMWPE cylinder of approximately \SI{15}{cm} diameter.  In the gas space and into at least \SI{15}{\centi\meter} of the liquid, the insulator is surrounded by a tight-fitting stainless steel ground tube.  The ground tube has a \SI{25}{\centi\meter}  Conflat (an industry standard) flange  welded on for attachment to the cryostat.


Outside of the cryostat, the \dword{hv} power supply and cable-mounted toroids will monitor the \dword{hv}.    The power supplies 
typically have sensitivities down to tens of \si{\nano\ampere} in current read-back capability 
\fixme{`in read-back `mode'? or `with' read-back capability? RKP: Believe language clear as written.} and are able to sample the current and voltage every \SI{300}{\ms}.  The cable-mounted toroids are sensitive to fast changes in current; 
the polarity of a toroid's signal 
indicates the location of the current-drawing feature as either upstream or downstream of it.  Experience from the \dword{35t} installation suggested sensitivities to changing currents with a timescale between \SIrange{0.1}{10}{\micro\s}, providing information on the timescale of any current changes.

Inside the cryostat, pick-off points near the anode will monitor the current 
in each resistor chain.  Additionally, the voltage of the \dwords{gp} above and below each drift region can be equipped to diagnose problems via a high-value resistor connecting the \dword{gp} to the cryostat.  In the \dword{35t}, such instrumentation provided useful information on \dword{hv} stability and where any stray charge was flowing.

Both commercial and custom \dword{hv} components must be rated for sufficient voltage and satisfy tests to meet the requirements summarized in Table~\ref{tab:hvphysicsreqs}.  Further details on these tests are in Section~\ref{sec:fdsp-hv-qc}.

The resistances in the filters, in combination with the capacitances between the \dword{hv} system and the cathode,
 determine the attenuation of the tens of \si{\kilo\hertz} ripple from the power supply.  The filters 
are designed such that the ripple is reduced to an acceptable level when installed in the complete system, thus satisfying requirement (3) that the power supply ripple is minimized.
\fixme{`minimize ripple'' is not a validatable requirement. What is the acceptable level? Or what dictates it? Anne. RKP: Addressed in collaboration comments.}

\subsection{Cathode Plane Assembly (CPA)}

The \dword{cpa} provides a constant potential surface at \SI{-180}{\kV} for the \dword{spmod}.  It receives its \dword{hv} from the feedthrough that makes contact with the \dword{hv} bus mounted on the \dword{cpa} frame through an attached donut assembly attached to the frame, as shown in Figure~\ref{fig:donut_cpa}. 
The \dword{cpa} also provides \dword{hv} to the first profile on the top and bottom \dword{fc} elements and to the \dwords{ewfc} as well. 
Details on the electrical connections are found in Section ~\ref{sec:fdsp-hv-design-interconnect}.

\begin{dunefigure}[\dword{hv} input donut connection to \dword{cpa}]{fig:donut_cpa}{\dword{hv} input donut connection to \dword{cpa}.}
\includegraphics[width=0.6\textwidth]{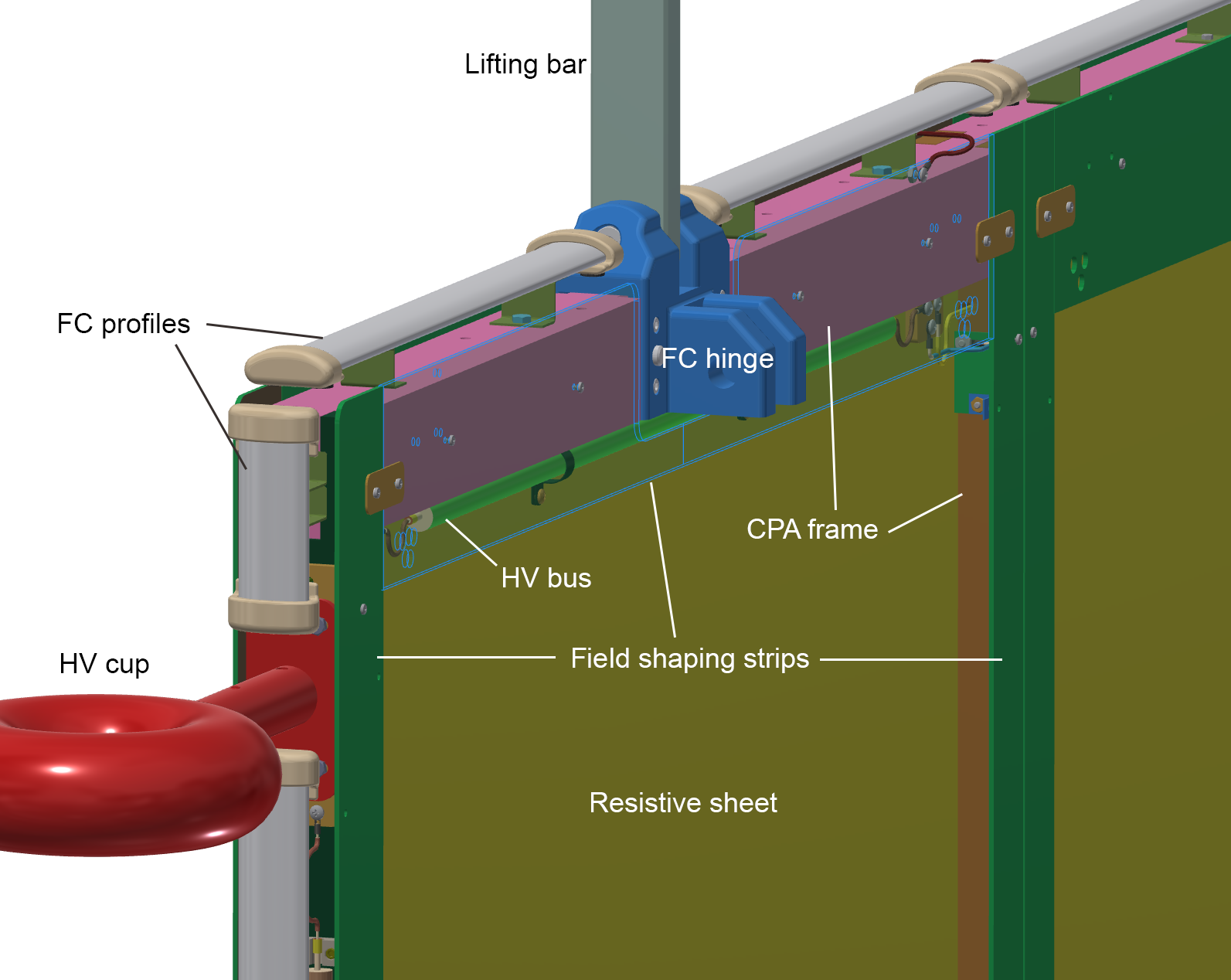} 
\end{dunefigure}

Ideally, the cathode would be constructed from a large thin resistive sheet.  However, inside the cryostat, there is a moderate convective flow of the \lar that can produce a small pressure difference across the cathode surface.  To maintain the position and flatness of the cathode, the cathode surface must be reinforced for stiffness.  This is accomplished using \SI{6}{cm} thick FR-4 frames at \SI{1.2}{m} intervals. Since FR4 is a good insulator at cryogenic temperature with a different dielectric constant that \lar, the presence of the frame causes a local \efield distortion that can become pronounced if the frame surface charges up as a reult from ionization in the TPC.  To minimize this distortion, a resistive field shaping strip is placed on the cathode frame and biased at a different potential.  Figure~\ref{fig:fss_concept} illustrates the drift field uniformity improvement with the field shaping strips.

\begin{dunefigure}[\dword{fss} concept]{fig:fss_concept}{A comparison of three cathode cross sections to illustrate the benefit of the \dword{fss}. Both equipotential lines (horizontal) and \efield{} lines (vertical) are shown.  The amplitude of the \efield{} is shown as color contours. Each color contour is a 10\% step of the nominal drift field.  The gray rectangles represent the frame and the resistive sheet in each case. Left: a conductive/resistive frame similar to that of ICARUS or SBND; Middle: an insulating frame with the insulating surfaces charged to an equilibrium state; Right: an insulating frame covered with a field shaping strip (purple) and biased at the optimum potential. }
\includegraphics[width=0.9\textwidth]{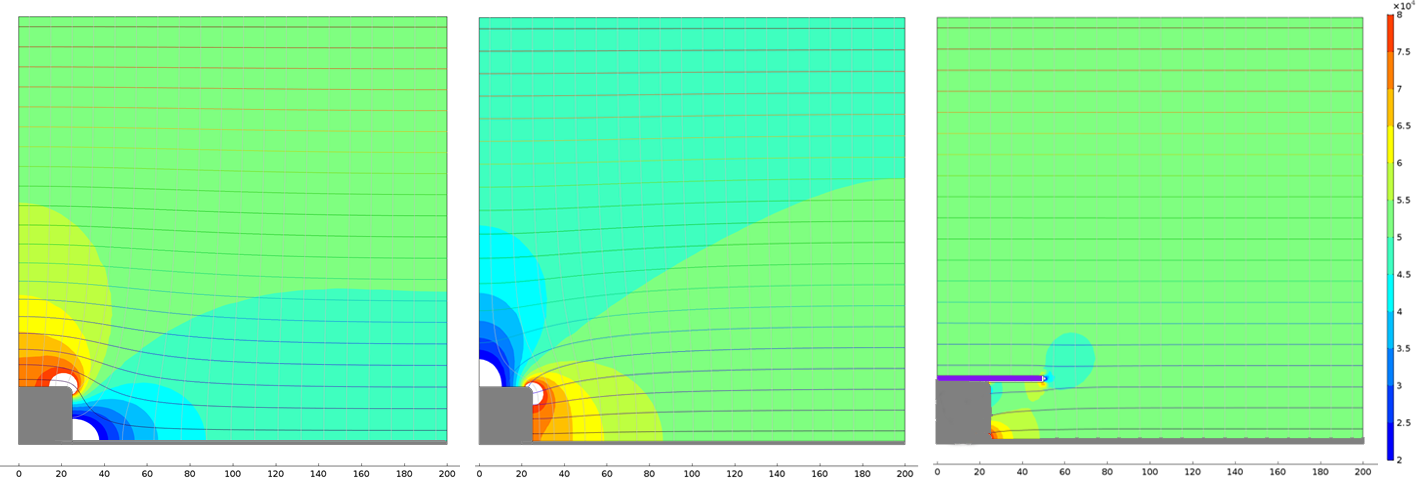} 
\end{dunefigure}

The \dwords{cpa}' constant potential surfaces are resistive panels (\dword{cpa} RPs) composed of a thin layer of carbon-impregnated Kapton\footnote{DuPont\texttrademark{}, Kapton\textsuperscript{\textregistered} polymide film,  E. I. du Pont de Nemours and Company,  \url{http://www.dupont.com/}.} 
 laminated to both sides of a \SI{3}{\milli\meter} thick FR4 sheet of \SI{1.2}{\meter}  $\times$ \SI{2}{\meter} size.  The surface resistivity of the \dword{cpa} RPs is required to be greater than 1 M\si{\ohm}/square in order to provide for slow reduction of accumulated charge in the event of a discharge.  A goal of 1 G\si{\ohm}/square for DUNE \dword{cpa} RPs extends the protection from discharges in the condition of anticipated higher stored energy at DUNE, compared to prototypes. Other \dword{hv} components of the \dword{cpa} include \dword{fss} mounted to the \dword{cpa} frames, edge aluminum profiles to act as the first elements of the field cage, and cable segments forming the \dword{hv} bus. Careful inspection of these items during the assembly process ensures that no sharp points or edges are present. The surface resistivity of the \dword{cpa} RPs and the \dword{fss} are checked multiple times during assembly -- first when the resistive panels and strips are received and after assembly into \dword{cpa} units on the table.  Coated parts that do not meet the minimum surface resistivity requirement are replaced.  This ensures that requirement (5) on Table~\ref{tab:hvphysicsreqs} is satisfied.  Figure~\ref{fig:cpa_panel-complete} shows a completed \dword{pdsp} \dword{cpa} panel on the production table ready for lifting into vertical position for mounting on its trolley.

\begin{dunefigure}[Completed \dword{pdsp} \dword{cpa} panel on production table]{fig:cpa_panel-complete}{Completed \dword{pdsp} \dword{cpa} panel on production table.}
\includegraphics[width=0.5\textwidth]{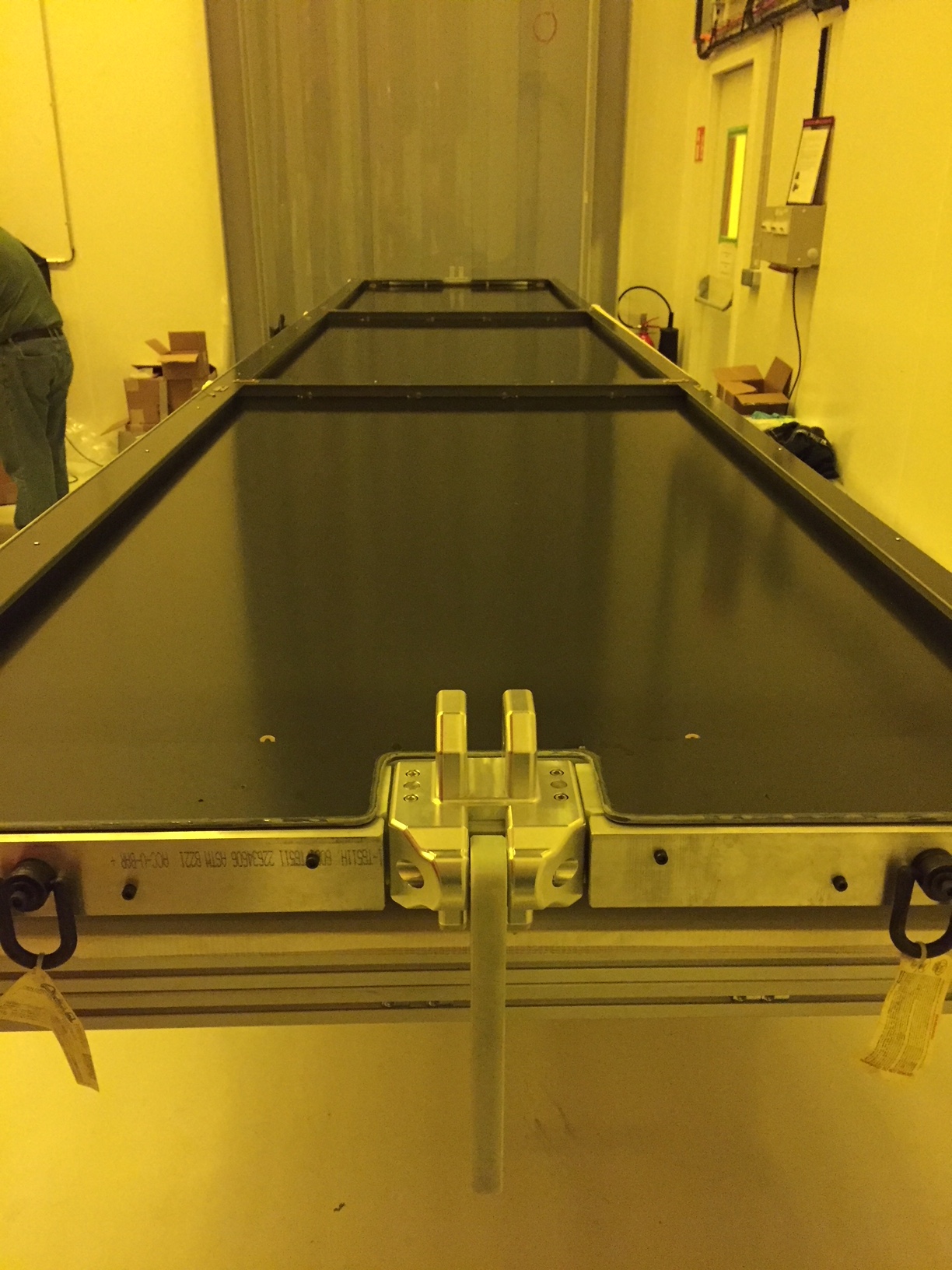}
\end{dunefigure}

All electrical connections on the \dword{cpa} and between the \dword{cpa} and other \dword{hv} system components (top, bottom, and \dwords{ewfc}) are redundant by at least a factor of two.  Connections between RPs in a \dword{cpa} unit are four-fold redundant.  The \dword{hv} connection from the \dword{hv} power supply is a closed loop around the \dword{cpa} which can sustain at least one broken connection without loss of the cathode plane \dword{hv}.  This ensures compliance with 
requirement (6) of Table~\ref{tab:hvphysicsreqs}.

The \dword{cpa} frames are required to support, in addition to the \dword{hv} components, the \dword{topfc} and \dword{botfc} units attached to both sides of the \dword{cpa} panel. The arrangement and deployment of these components will be the same as in \dword{pdsp}.  
\subsection{Field Cages}
\subsubsection{General Considerations}

A uniform \efield{} is required to drift ionization electrons towards the \dwords{apa}. The \dwords{fc} consisting of field shaping electrodes form a band surrounding the top, bottom, and ends of the active drift volume. The electrodes are biased at different potentials to establish a uniform field inside the \dword{lar} volume.
The \dword{spmod} will use extruded aluminum profiles as a cost-effective way to establish the equipotential surfaces. 

For safe and stable operation of the \lar cryogenic system, the cryostat must have a small fraction of its volume filled with gaseous argon, commonly referred to as the ullage. Since we want to make good use of the \lar in the cryostat, the top boundary of the TPC, the upper \dword{fc}, is not very far from the ullage. There are many grounded metallic components in the ullage with sharp features.  The \efield near these conductors could easily exceed the breakdown strength of gaseous argon. To prevent such breakdowns in the argon gas, a \dword{gp} is added above the upper \dword{fc} electrodes at a safe distance and below the liquid surface to shield the high \efield from entering the gas ullage.  The need for such shielding diminishes toward the \dword{apa} end of the \dword{fc} due to the lower voltages on the \dword{fc} profiles in that region. Therefore the \dword{gp} on the top only covers about \SI{70}{\%} of the \dword{cpa} side of the \dword{fc}, leaving extra room for cable routing near the \dwords{apa}.
On the bottom of the cryostat, a similar set of \dwords{gp} is planned to prevent breakdown, in the liquid, to cryogenic pipings and other sensors with sharp features.  No \dwords{gp} are planned beyond the two \dwords{ewfc} since there is sufficient clearance in those regions.  

The shape of the electrodes is critical as it determines the strength of the \efield{} between a given profile and its neighboring profiles, as well as
other surrounding parts, including the \dword{apa}, which is electrically at ground. 
Electric fields need to be well below \SI{30}{\kilo\volt/\centi\meter} 
to satisfy design requirement (2) and enable safe \dword{tpc} operation~\cite{Blatter:2014wua}. 

The commercially available profiles used for \dword{pdsp}, and forming the \dword{spmod} design, are estimated to lead to \efield{}s of up to \SI{12}{\kilo\volt\per\centi\meter} under the 
configuration and operating voltage assumptions.
Figure~\ref{fig:profile-e-field} illustrates results from an \efield{} calculation.
\begin{dunefigure}
[\efield map and equipotential contours of profiles at \SI{-180}{\kV}]{fig:profile-e-field}
{\efield map (color) and equipotential contours of an array of roll formed profiles biased up to \SI{-180}{\kV} and a ground clearance of \SI{20}{\cm}(Credit: BNL CAD model).} \includegraphics[width=0.8\textwidth]{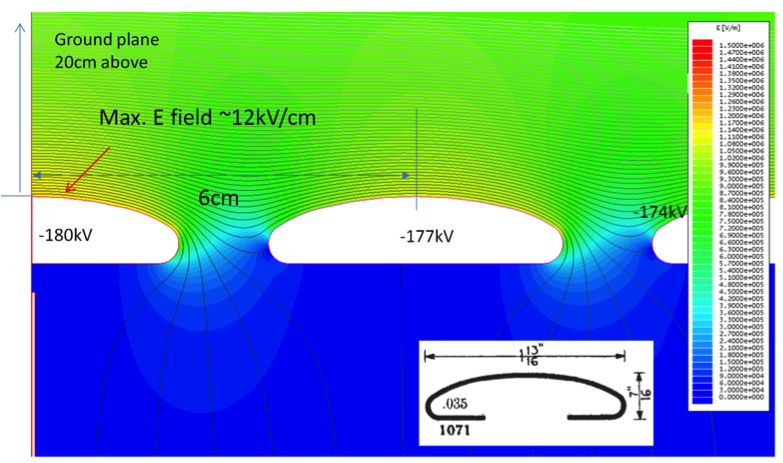}
\end{dunefigure}

The profiles ends are equipped with UHMW polyethylene caps to reduce the risk of arc formation.  These caps are designed to have sufficient wall thickness (\SI{6}{\milli\m}) to withstand the full voltage across their walls.

The aluminum profiles are attached to fiber-reinforced plastic (FRP) pultruded structural elements, including I-beams and box beams.  
Pultruded FRP material is non-conductive and strong enough to withstand the \dword{fc} loads  in the temperature range of \SI{-150}{C} and \SI{23}{C}, as certified by vendors. Testing of the FRP joints were conducted at liquid nitrogen temperatures (see DocDb 1504).  The strength of the material increased over room temperature tests which provides confidence in the material behavior at \lar  temperature. Tests of FRP joints at LN temperature\fixme{ add citations in bib: CPA and FC Design, DUNE DocDB 1504} showed that the strength of the material increases at cryogenic temperature relative to room temperature, providing confidence in FRP material behavior at LAr temperature.
The FRP material meets class A standards for fire and smoke development established by the International Building Code characterized by ASTM E84\footnote{\textit{Standard Test Method for Surface Burning Characteristics of Building Materials}, ASTM International, \url{https://compass.astm.org/EDIT/html_annot.cgi?E84+18}.}

As discussed in Section~\ref{sec:fdsp-hv-intro}, 
the field cage modules are of two types: the top and bottom \dword{fc} and the \dword{ewfc}, both of which are described below. 
A resistive divider chain interconnects all the aluminum profiles to provide a linear voltage gradient between the cathode and anode planes.  The top and bottom modules are nominally \SI{2.3}{\m} wide by \SI{3.5}{\m} long. A \dword{gp}, in the form of tiled, perforated stainless steel sheet panels, is mounted on the outside surface of the T/B field cage module with a \SI{20}{\cm} clearance. The top and bottom \dword{fc} modules are supported by the \dwords{cpa} and \dwords{apa}. The \dword{ewfc} modules are \SI{1.5}{\m} tall by \SI{3.5}{\m} long. They are stacked eight units high (\SI{12}{\m}), and are supported by the installation rails above the \dwords{apa} and \dwords{cpa}.

The \dwords{fc} are designed to produce a uniform field with understood characteristics.
The current (i.e., \dword{pdsp}) \dword{fc} design has a large gap between the \dword{ewfc} module and its neighboring top and bottom modules to allow the latter to swing pass the \dword{ewfc} during the \dword{fc} deployment. This gap causes the largest known distortion in the drift field in the \dword{tpc}. Figure~\ref{fig:fc-distortion} shows the extent of the distortion in this limiting scenario. In \dword{pdsp}, the gap produces two regions (of total \lar mass \SI{20}{kg}) in the TPC near both bottom corners that suffer \SI{5}{\%} \efield distortions.

\begin{dunefigure}[Electric field at edge of \dword{fc}]
{fig:fc-distortion}
{\efield at a corner between the bottom and endwall \dword{fc} modules, showing effects of a \SI{7}{cm} gap. Left: the extent of \num{5}\% \efield{} non-uniformity boundary (black surface, contains less than \SI{10}{kg} of \lar) and \num{10}\% non-uniformity boundary (white surface, contains $\sim$ \SI{6}{kg} of \lar) inside the TPC's active volume. The inset is a view from the CAD model.  Right: electron drift lines originating from the cathode surface.}
\includegraphics[width=0.9\textwidth]{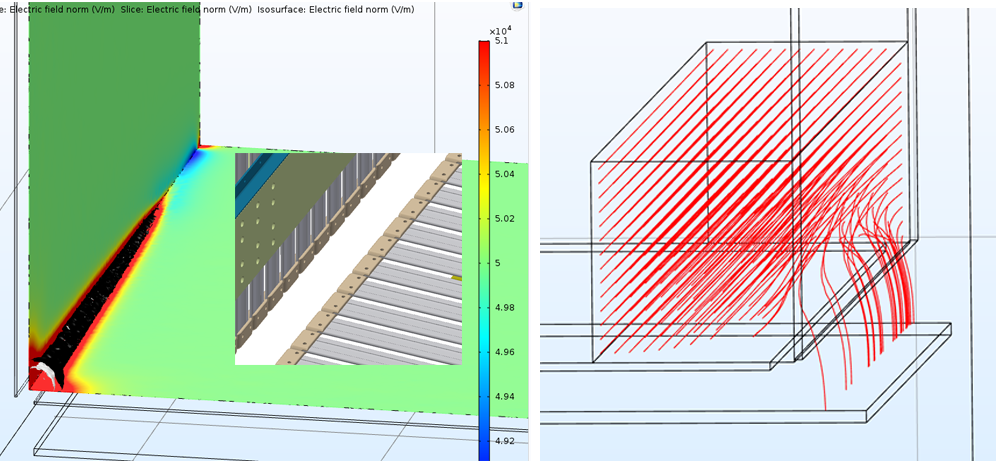}
\end{dunefigure}

The \dwords{fc} 
are designed to 
meet the system requirements specified in Table~\ref{tab:hvphysicsreqs}. All components other than the aluminum profiles, \dwords{gp}, and electronic divider boards are made of insulating FRP and FR4 materials, and the end of each profile is covered with a UHMWPE end cap, to allow the system to reach the design \dword{tpc} \efield{} (requirement 1). The profiles have been carefully modeled to study the resulting \efield{}, and small-scale laboratory tests have been conducted to ensure that the maximum \efield{} does not approach \SI{30}{\kV\per\cm} (requirement 2). These design features are expected to avoid sparking, and thus to draw very small stable currents, which should produce a consistent load on the power supply (requirements 3, 4, and 5). Finally, all voltage divider boards provide redundant paths for establishing the profile-to-profile potential differences, and two redundant boards provide the connection from the \dwords{fc} modules to the \dword{cpa} (requirement 6).

\subsubsection{Top and bottom field cages}

The \dword{topfc} and \dword{botfc} modules are \spfcmodlen{} long, which is set by the length of the two \SI{15.2}{\cm} (\SI{6}\,in) FRP I-beams that form the primary support structure of the modules. The I-beams are connected to each other by three  \SI{7.6}{\cm} (\SI{3}\,in) FRP cross beams. The connections between the longitudinal and cross I-beams are made with L-shaped FRP braces that are attached to the I-beams with FRP spacer tubes, and secured with FRP threaded rods, FRP hex-head nuts, and custom-machined FR4 washer plates.

The modules are \SI{2.3}{\m} wide, which corresponds to the length of the aluminum profiles, including the UHMW polyethylene end caps. Profiles are secured to the FRP frame using custom-machined double-holed stainless steel slip nuts that are slid into and electrically in direct contact with the Al profiles such that they straddle the webbing of the \SI{15}{\cm} I-beams, and are held in place with screws that penetrate the I-beam flanges. The profile offset with respect to the FRP frame is different for modules closest to the \dwords{ewfc}, 
and modules in the center of the active volume.

Each \dword{topfc} and \dword{botfc} module holds five ground planes, which are connected to the outside (i.e., the non-drift side) of the module. The \dwords{gp} are positioned $\sim$\SI{20.5}{\cm} above the profiles, and are pushed to the \dword{cpa} side of the module, leaving the last 14 profiles (\SI{88}{\cm}) on the \dword{apa} side of the module exposed. Between the \dwords{gp} and the \SI{15}{\cm} I-beams are standoffs made of short sections of \SI{10.2}{\cm} (4\,in)  FRP I-beams, which are connected with FRP threaded rods and slip nuts. The electrical connection between the ground planes is made with copper strips.

The connections between the top and bottom modules and the \dwords{cpa} are made with aluminum hinges, \SI{2.54}{\cm} (1\,in) in thickness, that allow the modules to be folded in to the \dword{cpa} during installation. The hinges are electrically connected to the second profile from the \dword{cpa}. The connections to the \dwords{apa} are made with stainless steel latches that are engaged once the top and bottom modules are unfolded and fully extended toward the \dword{apa}.

The voltage drop between adjacent profiles is established by voltage divider boards that are screwed into the drift volume side of the profiles. A custom-machined nut plate is used that can be inserted into the open slot of each profile and twisted \SI{90}{\degree} 
to lock into position. Two additional boards to connect the modules to the \dwords{cpa} were screwed into the last profile on the \dword{cpa}-side of the module. This system is also described in Section~\ref{sec:fdsp-hv-design-interconnect}. A fully assembled module is shown in Figure~\ref{fig:tbfc1-2}.

\begin{dunefigure}[Top and bottom field cage modules]{fig:tbfc1-2}{The fully assembled modules with ground planes are shown (left), as well as a close up of a \dword{cpa} end as viewed from the bottom (drift) side of the module.}
\includegraphics[width=0.65\textwidth]{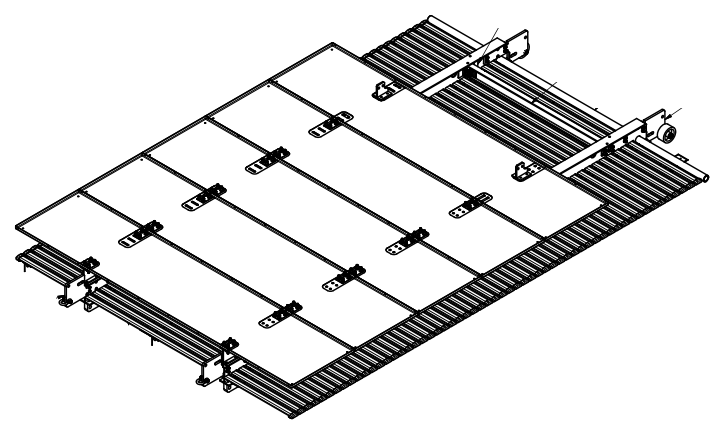}
\includegraphics[width=0.33\textwidth]{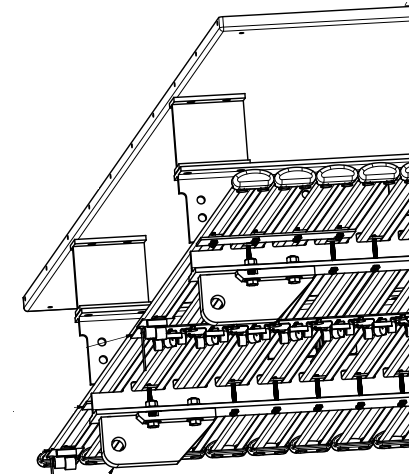}
\end{dunefigure}

Between any two adjacent nodes of the resistor divider chain are two \SI{5}{\giga\ohm} resistors, and three serially connected metal oxide varistors (MOVs) in parallel.  The nominal voltage drop is \SI{3}{kV} between each node.
An open resistor on the divider chain would approximately double the voltage across the remaining resistor to \SI{6}{kV}.  This will force the varistors in parallel to that resistor into conduction mode, resulting in a voltage drop of roughly \SI{5}{kV} (\SI{1.7}{kV} $\times$ \num{3}), while the rest of the divider chain remains linear, with a slightly lower voltage gradient.
Because the damage to the divider would be local to one module, its impact to the \dword{tpc} drift field is limited to region near this module.  This is part of the intention of the modular design.
An example of a simulated \efield{} distortion which would be caused by a failed resistor is shown in Figure~\ref{fig:fc-broken-resistor}. 

\begin{dunefigure}[\efield distortion from broken voltage divider path]{fig:fc-broken-resistor}{Simulated \efield{} distortion from one broken resistor in the middle of the voltage divider chain on one bottom field cage module, emphasizing the need for redundancy. Left: Extent of \efield{} non-uniformity in the active volume of the TPC. the green planes mark the boundaries of the active volume inside the field cage. The partial contour surfaces represent the volume boundaries where \efield{} exceeds 5\% (dark red, contains less than 100\,kg of LAr) and 10\% (dark blue, contains less than 20\,kg of LAr) of the nominal drift field. Units are \si{\volt\per\m} in the legend. Right: electron drift lines connecting the \dword{cpa} to \dword{apa} in a bottom/end wall field cage corner.  The maximum distortion to the field line is about 5\,cm for electrons starting at mid drift at the bottom edge of the active volume.}
\includegraphics[width=0.9\textwidth]{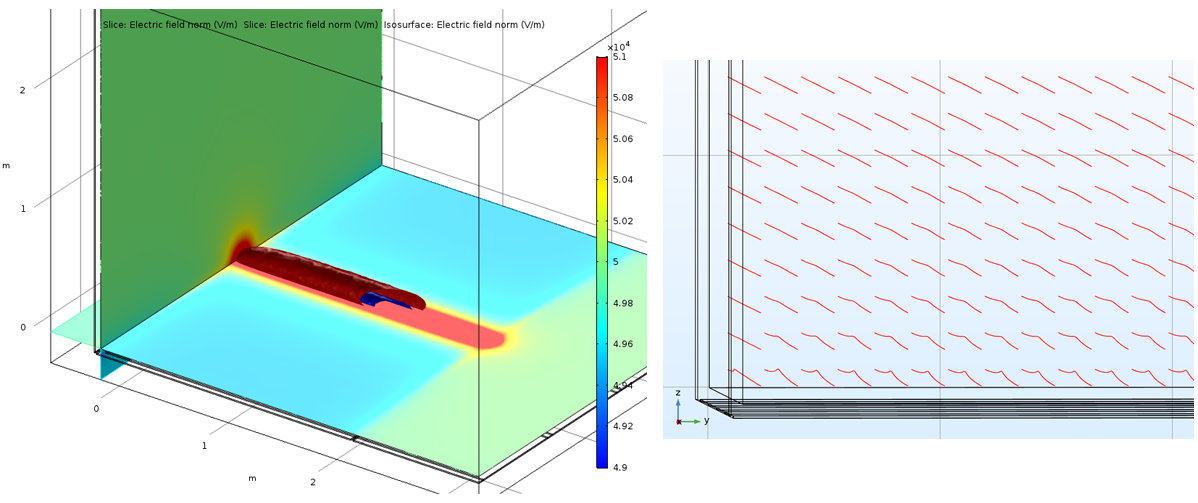}
\end{dunefigure}
The effect of the non-uniformity in resistor values can also be scaled from this study.  A 2\% change in a resistor value (1\% change from the 2R in parallel) would give about 1.5\% of the distortion from a broken resistor, i.e. less than 1\,mm of transverse distortion in track position, with no noticeable drift field amplitude change inside the active volume.

\subsubsection{Endwall field cages (EWFC)}


Each of the four drift volumes has two \dwords{ewfc}, one on each end. Each \dword{ewfc} is in turn composed of eight \dword{ewfc} modules.
There are two different types of \dword{ewfc} modules, each of which comes in a \textit{regular} and in a \textit{mirrored} configuration to account for 
mounting constraints and to match the detector geometry. Figure~\ref{fig:fc_endwall_panels} illustrates the layout for the topmost 
and the other panels, respectively.

\begin{dunefigure}[Endwall \dword{fc} panels]{fig:fc_endwall_panels}{Left: Uppermost panel of the \dword{ewfc}. Right: Non-uppermost \dword{ewfc} panel.}
\includegraphics[width=0.48\textwidth]
{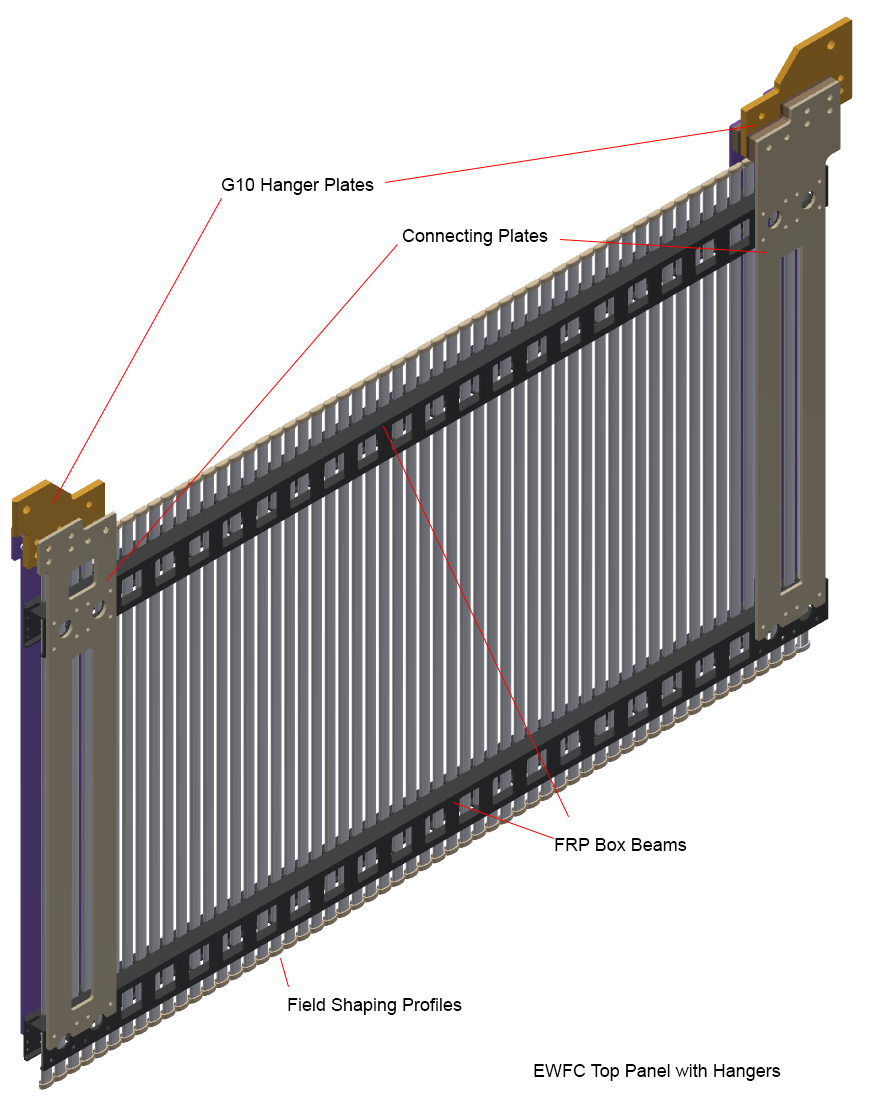}
\includegraphics[width=0.48\textwidth]
{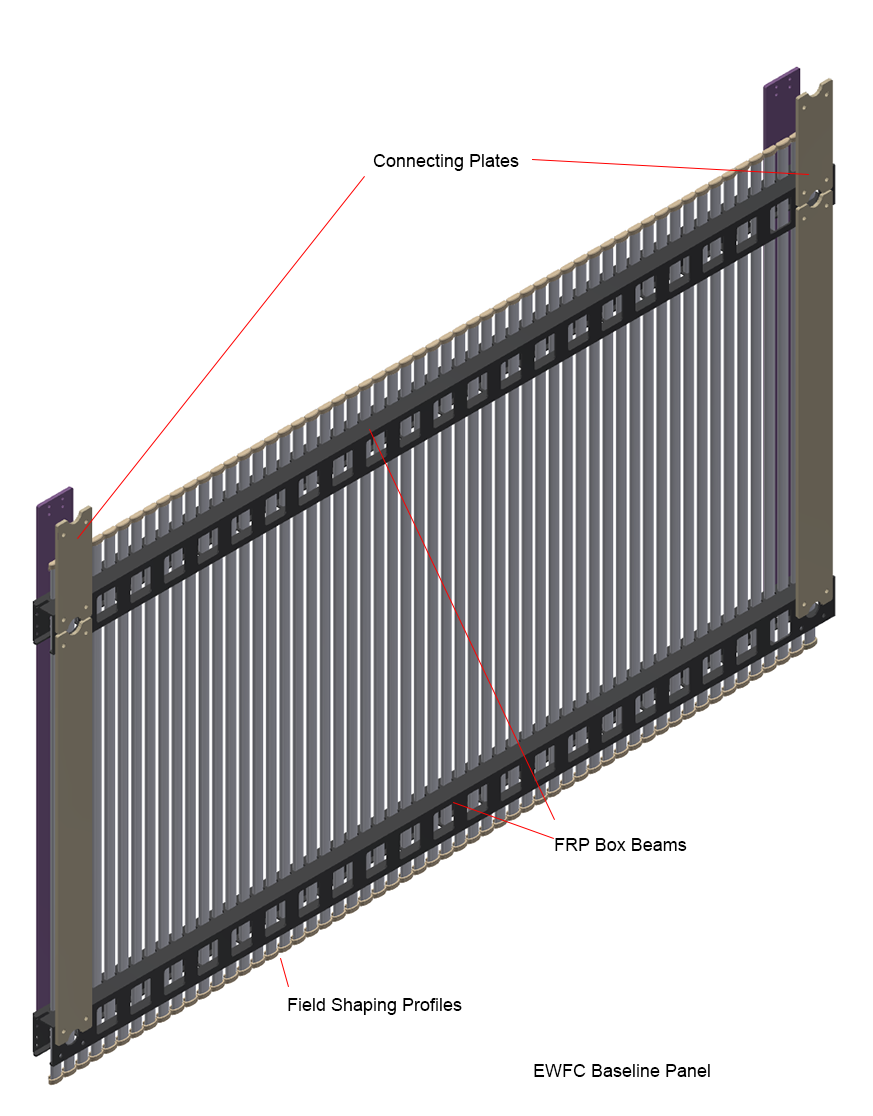}
\end{dunefigure}

Each \dword{ewfc} module is constructed of two FRP box beams which are \SI{3.5}{\m} long. The box beam design also incorporates cutouts on the outside face to minimize charge build up. Box beams are connected using \SI{1.27}{\cm} (\num{0.5}\,in) thick FRP plates. The plates are connected to the box beams using a shear pin and bolt arrangement. The inside plates facing the active volume are connected using special stainless steel slip nuts and stainless steel bolts. The field-shaping profiles are connected to the top box beam using stainless steel slip nuts, an FRP angle, and two screws each. The profiles are connected to the bottom box beam with a slip nut that is held in place by friction.


\subsection{Electrical Interconnections} 
\label{sec:fdsp-hv-design-interconnect}

\begin{dunefigure}[\dword{hv} interconnection topology]{fig:fdsp-hv-design-interconnect-concept}
  {High-level topology of the \dword{hv} interconnections}
  \includegraphics[width=0.7\textwidth]{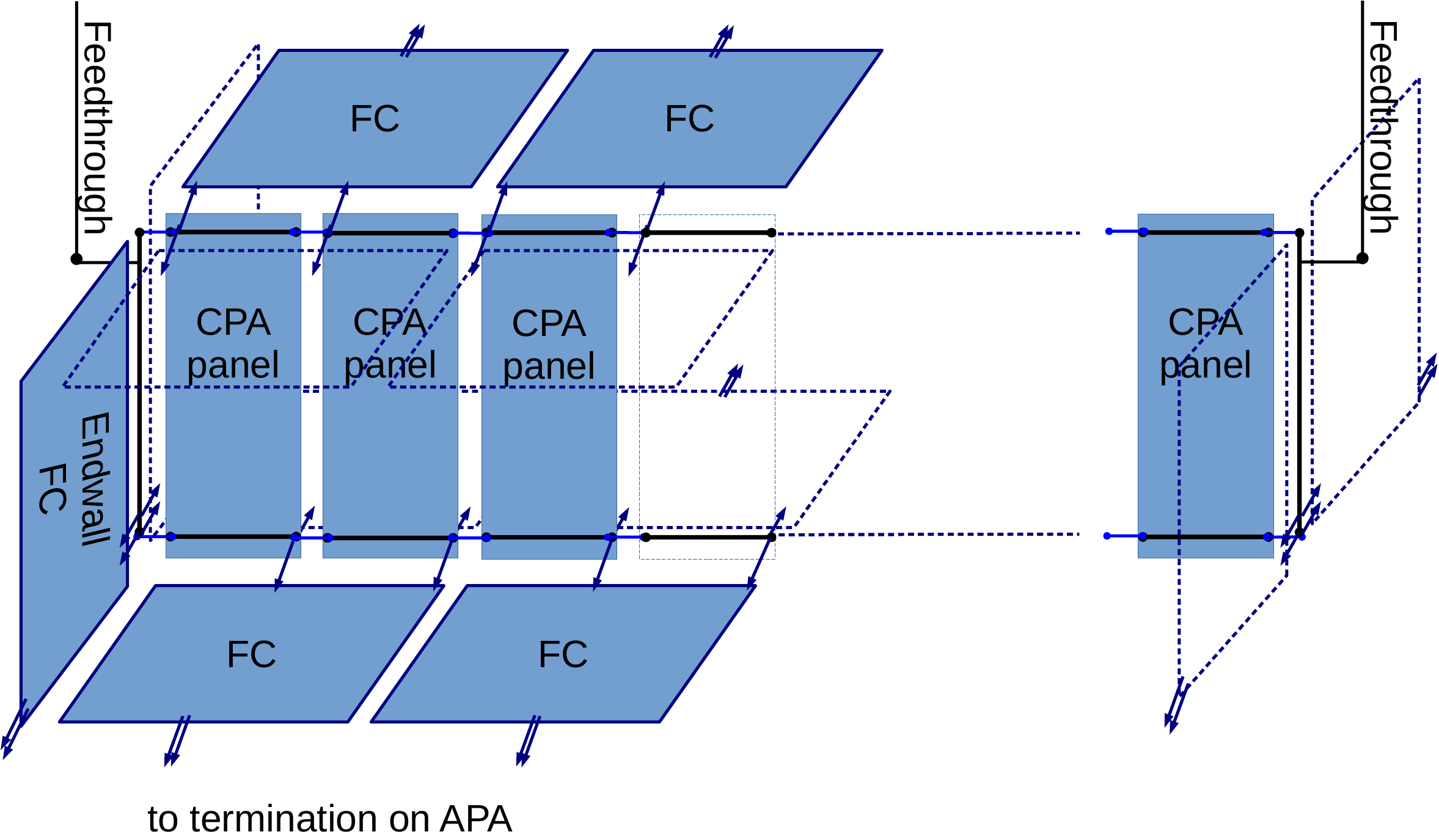}
\end{dunefigure}

Electrical interconnections are needed among the \dword{hv} delivery system, \dword{cpa} planes, \dword{fc} modules, and termination
boards on the \dword{apa} modules, as well as between resistive dividers and
the field-forming elements on the \dwords{cpa} and \dwords{fc}.  Redundancy is
needed to avoid single points of failure. 
Some connections must be
insulated in order to avoid creating a discharge path that might
circumvent the discharge mitigation provided by the resistive \dword{cpa}
surface and \dword{fc} partitioning.  Certain connections must be
flexible in order to allow for \dword{fc} deployment, thermal
contraction, and motion between separately supported \dwords{cpa}.  Figure~\ref{fig:fdsp-hv-design-interconnect-concept} shows a high-level
overview of the interconnections between the \dword{hv}, \dword{cpa}, and \dword{fc} modules.

High voltage feedthroughs connect to cups mounted on the \dword{cpa} frame
that attach to a \dword{hv} bus running through the \dwords{cpa}.  \Dword{hv} bus connections
between \dword{cpa} panels are made by flexible wires through holes in the
\dword{cpa} frame. The \dword{hv} bus is a loop in order to mitigate risk of a single
point failure; feedthroughs at each end of each \dword{cpa} plane mitigate
risk of a double-break failure.  Voltage dividers on each \dword{cpa} panel
bias the field shaping strips and the resistive dividers on the top
and bottom \dwords{fc}.  \dword{cpa}-to-\dword{fc} connections are made using
flexible wire to accommodate \dword{fc} deployment.  To further
increase redundancy, two \dword{cpa} panels connect to each top or bottom
field cage, and two connections are also made to each \dword{ewfc}. Resistor divider boards attach directly to the interior side of
the \dword{fc} profiles with screws.   A redundant pair of flexible wires
connects a circuit board on the last profile of each \dword{fc} to a
bias-and-monitoring board mounted on the corresponding \dword{apa}.

Short sections of flexible wire at the ends of each \dword{hv} bus segment
attach to screws in brass tabs on the \dword{cpa} resistive panels (\dword{cpa} RPs).
Vertical \dword{hv} bus segments on the outer ends of each \dword{cpa} plane connect
the top and bottom \dword{hv} buses to complete the loop.  Solid wire is used
to connect resistive panels within a \dword{cpa} panel.

Each \dword{fc} module is as electrically independent as possible in order to
mitigate discharge.  However, only the bottom module of each endwall
can make connections to the \dword{hv} bus and \dword{apa}, so each endwall module
is connected to its upper neighbor at its first and last profiles
using metal strips.

All flexible wires have ring or spade terminals and are secured by
screws in brass tabs.  Spring washers are used with every electrical
screw connection in order to maintain good electrical contact with
motion and changes of temperature.

Table \ref{tab:sp-hv-interconnects} summarizes the interconnections
required.

\begin{dunetable}
[\dword{hv} system interconnections]
{p{0.35\linewidth}p{0.62\linewidth}}
{tab:sp-hv-interconnects}
{\dword{hv} System Interconnections}   
 Connection & Method \\ \toprowrule
 \dword{hv} cup to \dword{hv} bus & wire to screw in \dword{hv} cup mount on \dword{cpa} frame \\ \colhline
 \dword{hv} bus between \dword{cpa} panels & wire between screws in brass tabs \\ \colhline
 \dword{hv} bus to \dword{fss} & wire to circuit board mounted on \dword{fss} \\ \colhline
 \dword{fss} to \dword{topfc} and \dword{botfc} & wire to circuit board on first \dword{fc} profile, two per \dword{fc} module \\ \colhline
 \dword{hv} bus to endwall \dword{fc} & wire to circuit board mounted on first \dword{fc} profile, two per endwall \\ \colhline
 \dword{fc} divider circuit boards & directly attached to profiles using screws \\ \colhline
 \dword{fc} to bias and monitoring termination & redundant wires from board mounted on last \dword{fc} profile \\ \colhline
 \dword{hv} bus to \dword{cpa} panels & brass tab on \dword{cpa} resistive panel \\ \colhline
 \dword{cpa} RP interconnections & solid wire between screws in brass tabs \\ \colhline
 Endwall \dword{fc} module interconnections & metal strips, first and last profiles only
 \\ \colhline
\end{dunetable}

The redundancy in electrical connections described above meets requirement (6).
The \dword{hv} bus and interconnections are all made in low field regions in order to meet requirement (2).
The \dword{hv} bus cable is rated at the full cathode \dword{hv} such that even in case of a rapid discharge of the \dword{hv} system no current can flow to the cathode or \dword{fc} except at the intended contact points, preserving the ability of the resistive cathode and \dwords{fc} to meet requirement (5).


\section{Production and Assembly}
\label{sec:fdsp-hv-prod-assy}

\subsection{Power Supplies and Feedthroughs}
\label{sec:fdsp-hv-supplies-feedthroughs}

Power supplies will be commercially procured, for example through Heinzinger. The \dword{hv} cable is commercially available.

The power supply is tested extensively along with the controls and monitoring software.  Features to be included in the software are:
\begin{itemize}
\item The ability to ramp, or change the voltage.  The rate and an ability to pause the ramp shall be included.  In previous installations, the ramp rate was typically between 60 to 120 V/s.
\item An input for a user-defined current limit.  This parameter is the current value at which the supply reduces the voltage output to stay below the current limit.  The current-limiting is done in hardware.
\item An input for a trip threshold.  At this current reading, the program would reduce the voltage output through software.  In previous experiments, the trip function in software would set the output to \SI{0}{kV}.
\end{itemize}
\noindent Additionally, the software should record the current and voltage read-back values with a user-defined frequency, as well as any irregular current or voltage events.

The \dword{hv} feedthroughs, filters, and splitter are custom devices.  One feedthrough option is to use the \dword{pdsp} design and similar procurement.  \Dword{hv} splitters are already on hand if they are desired.

\subsection{Cathode Plane Assemblies}
\label{sec:fdsp-hv-prod-cpa}

The component parts of the \dword{cpa} 
are produced by commercial vendors for the following items:
\begin{itemize}
\item manufactured FR4 RP frames packed into three \dword{cpa} unit kits making up a \dword{cpa} panel,
\item carbon-impregnated Kapton coated resistive panels (RPs) and \dword{fss},
\item \dword{hv} cable segments and wire jumpers making up the \dword{cpa} \dword{hv} bus and RP interconnects,
\item machined brass tabs for connecting RPs, \dword{hv} Bus, and \dword{fss}, and
\item top, bottom, and exterior edge profiles and associated connection hardware.
\end{itemize}
The above items are packaged into \dword{cpa} panel kits by the vendors and are sent to the assembly factories, the locations of which will be determined later.  The basic construction unit for an assembly factory is a pair of \dword{cpa} panels so that shipment to \surf from an assembly factory consists of two \dword{cpa} panels that are paired on site to form a \dword{cpa} plane.

The most basic element of the \dword{cpa} is an RP mounted in a machined slot in the top, bottom and sides of FR4 frames.  There are three different types of these \dword{cpa} RP elements -- an upper, which has as its top frame the \dword{cpa} mounting bracket and \dword{topfc} hinge, a middle, and a lower, which has as its bottom frame a \dword{botfc} hinge.  Two such \dword{cpa} RP elements are bolted together and pinned to form a shipment \dword{cpa} unit of size \SI{1.2}{\m} $\times$ \SI{4}{\m}.  These \dword{cpa} units are assembled horizontally on a smooth, flat table to meet the dimensional requirements of \dword{cpa} construction.  In addition to the frames and RPs, 
\dword{fss} strips are mounted on the exposed sides of the FR4 frames, aluminum profiles are attached to the top and bottom of the upper and lower elements, and cables are attached to the RPs to form segments of the \dword{hv} bus.  The shipment \dword{cpa} unit comes in three varieties in order to make a full \SI{12}{\m} tall \dword{cpa} Panel.  These are : (1) an upper \dword{cpa} RP element attached to a middle element, (2) two middle elements connected, and (3) a middle element attached to a lower element.  The \dword{cpa} unit order in the shipping crate from top to bottom is middle-and-lower, middle-and-middle, and upper-and-middle.  For the \SI{10}{\kt} \dword{spmod}, there are 100 upper elements, 100 lower elements and 400 middle elements that make up the 100 \dword{cpa} panels of the \dword{tpc}.
A comparison of a \SI{6}{\m} \dword{pdsp} \dword{cpa} panel and a \SI{12}{\m} \dword{pdsp} panel is shown at Ash River Laboratory in Minnesota, USA,  in Figure~\ref{fig:12m-cpa}.

\begin{dunefigure}[\dwords{cpa} at Ash River]{fig:12m-cpa}{A \SI{12}{\m} DUNE-SP \dword{cpa} mockup panel and a smaller \SI{6}{\m} \dword{pdsp} panel mockup at Ash River.}  
\includegraphics[width=0.35\textwidth]{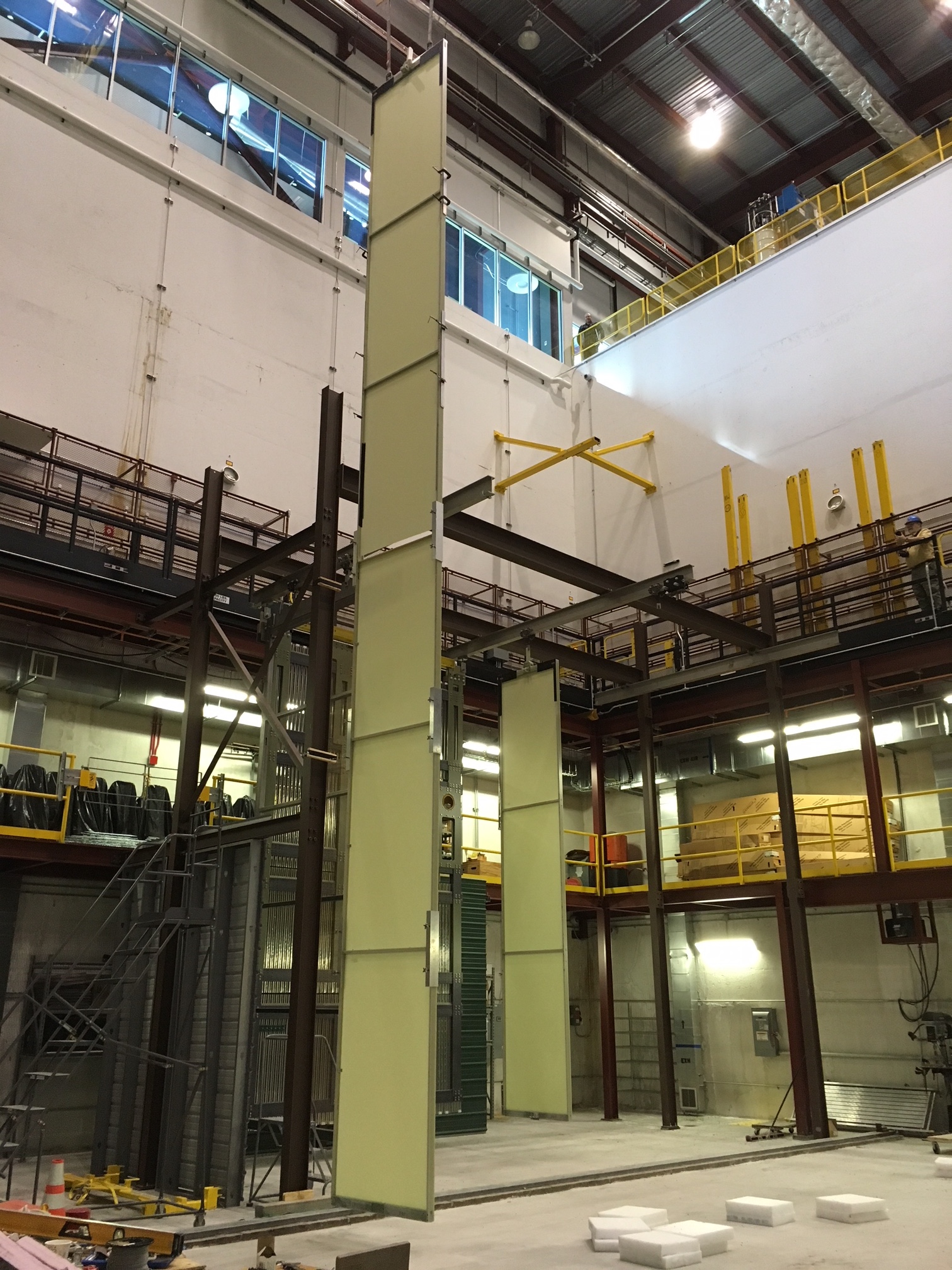}
\end{dunefigure}

\subsection{Field Cages}
\label{sec:fdsp-hv-prod-fc}


\subsubsection{Top and Bottom Field Cages}

The FRP and FR4 components of the  \dwords{topfc} and \dwords{botfc} will be commercially produced by firms that specialize in the machining of fiberglass components for electrical applications, as was successfully done for \dword{pdsp}. All parts are machined in the absence of water and cleaned with a lacquer thinner. Machined edges, other than small circular holes, are coated with translucent epoxy. The stainless steel and aluminum components will be produced in local university and commercial machine shops. Voltage divider boards and \dword{fc} and \dword{cpa} connection boards will likely be fabricated by university groups.

The FRP frame assembly primarily consists of fastening together FRP I-beams with FRP threaded rods and hex nuts, which are secured with a limited and specified torque, to avoid damage to the threads. A detailed view of one of these connections is shown in Figure~\ref{fig:tbfc3}.

\begin{dunefigure}[Top and bottom \dword{fc} module frame assembly]{fig:tbfc3}{The above figure shows the procedure for connecting the cross beams to the main I-beams for the \dword{topfc}. Left: The components of each connection, which (from top to bottom) are the threaded rods, the spacer tubes, washer plates, the hexagonal nuts, and an L-shaped FRP brace. An intermediate stage (middle) and final stage (right) of the assembly are also shown."}
\includegraphics[width=0.70\textwidth]{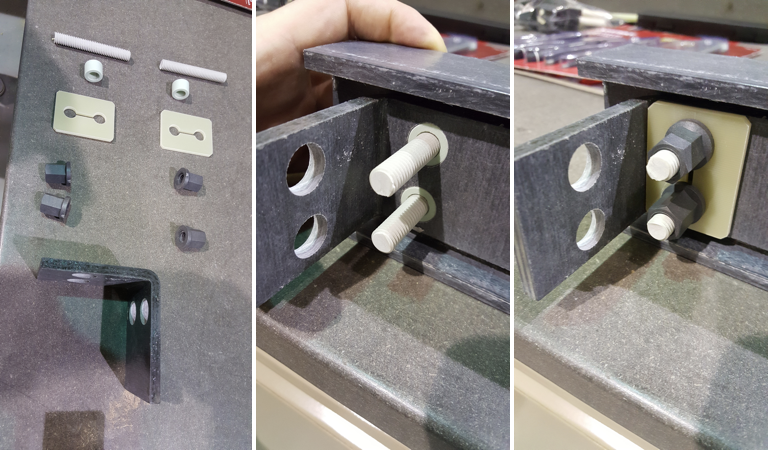}
\end{dunefigure}

Prior to sliding each profile into the FRP frame, the holes should be covered with Kapton tape to avoid damage to the profile coating. An end cap is attached to each profile using plastic rivets, and then the profiles are aligned against an aligment fixture running the length of the \dword{fc}. After securing each profile to the frame, the tension in the mounting screws is adjusted to remove any angular deflection in the extended portion of the profile.

The ground planes are attached to the 10 cm stand-off I-beam sections with threaded rods and a machined plate. The copper strips are connected to adjacent modules at the same locations. Care must be taken to avoid bending the corners of the \dwords{gp} toward the profiles, particularly on the \dword{cpa} side of of the module.

\subsubsection{Endwall Field Cages}

All FRP plates are commercially cut to shape by water jet. The cut outs in the FRP box beams are also cut by water jet. Holes that accommodate G10 bushings are reamed in a machine shop. FRP frames are pre-assembled to ensure proper alignment of all FRP parts and matching of holes. The profiles are not inserted at this stage. The FRP modules are hung off of each other by means of interconnecting FRP plates to ensure accurate alignment.

Next, parts are labeled and the frames are taken apart. All components are cleaned by pressure washing or ultrasonic bath. All cut FRP surfaces are then coated with polyurethane, which contains the same main ingredient as the FRP resin, allowing it to bond well to the FRP fibers. Final panels are constructed from cleaned and inspected parts. In order to ease assembly, which requires access to both sides of a module,
a dedicated assembly table has been manufactured that allows convenient module rotation. 

Figure~\ref{fig:endwall_assy_rot_table} shows a partially assembled \dword{ewfc} FRP frame on the assembly table.
\begin{dunefigure}[Endwall assembly table]{fig:endwall_assy_rot_table}{Assembly table with partially assembled \dword{ewfc} module (Credit: LSU)}
 \includegraphics[width=0.8\textwidth]{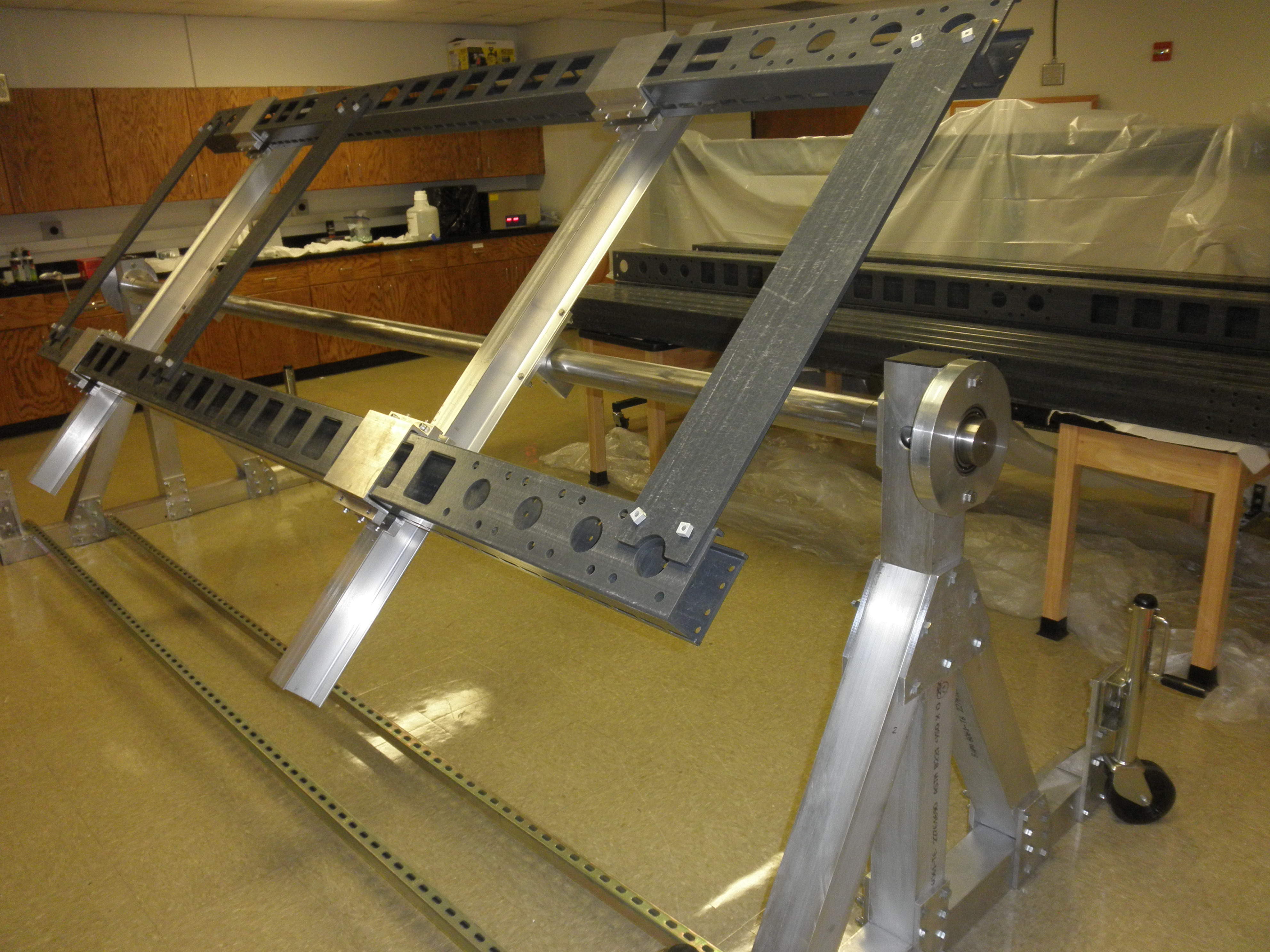}
 \end{dunefigure}

The FRP box beams are sandwiched between \SI{1.27}{\cm} (\num{0.5}\,in) thick FRP panels which are held on one side by means of G10 bushings and rods with square nuts
as shown in Figure \ref{fig:endwall_assy_detail}. One the other side M10 stainless steel bolts engage with large slip nuts that are inserted into the Al profiles. The profiles 
are pulled towards a \SI{2.5}{\cm} thick FRP plate (red in Figure~\ref{fig:endwall_assy_detail}) on the inside of the box beam.

\begin{dunefigure}[Endwall assembly detail]{fig:endwall_assy_detail}{
Top and center \dword{ewfc} module frames hanging. (Credit: LSU)}
\includegraphics[width=0.5\textwidth]{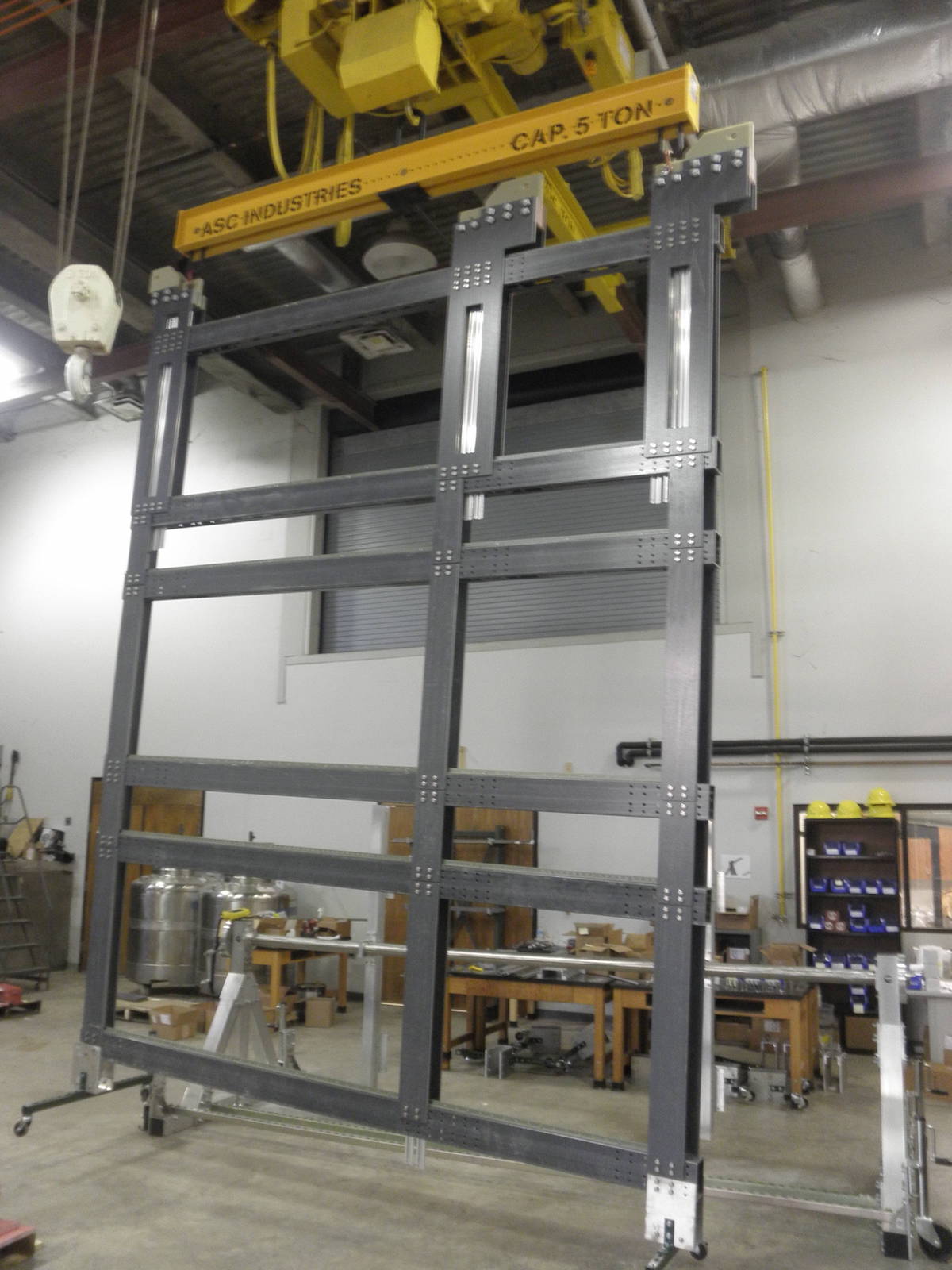}
\end{dunefigure}

Aluminum profiles are inserted into the cutouts of the box beams and attached with screws and stainless steel slip nuts to brackets that are mounted on the 
FRP box beams. After this, the resistive divider boards are mounted to the profiles using brass screws that engage with stainless steel slip nuts inside 
the profiles.

\subsection{Electrical Interconnections}
\label{sec:fdsp-hv-prod-interconnect}

All electrical fasteners and wires used on the \dword{cpa} and \dword{fc} are produced
to specification by commercial vendors and packaged with the \dword{cpa} or \dword{fc} modules.  
As discussed above (\ref{sec:fdsp-hv-prod-cpa}, \ref{sec:fdsp-hv-prod-fc}), 
this includes the \dword{hv} cable segments, as well as wire jumpers, machined brass
tabs, etc.

Circuit boards for 
\dword{hv} interconnections are produced and tested at the  university shops according to the same design used for \dword{pdsp}.  The \dword{fc} voltage dividers were produced for \dword{pdsp} at Louisiana State University, and the boards for \dword{cpa} frame bias and \dword{cpa}-\dword{fc} connections were produced at Kansas State University.
Both institutions have created custom test apparatus for verifying proper operation of the boards at full voltage and over-voltage conditions.  Production and testing could be scaled up by the required order of magnitude at these institutions, or shared with other institutions, whichever best meets the needs of the project. Each board is free of solder flux and flux-remover. 
 
\section{Installation, Integration and Commissioning}
\label{sec:fdsp-hv-install}

\subsection{Transport and Handling}
\label{sec:fdsp-hv-install-transport}

The power supply, cables, filters, and feedthroughs are sent to the site in standard shipping crates.  Handlers wear gloves when handling insulators that are between \dword{hv} and ground.  Surfaces can be cleaned with alcohol and allowed to dry.

\dword{cpa} panels are shipped in crates to the \surf site.  The \SI{12}{\m} \dword{cpa} panels are disassembled into their three \dword{cpa} units, loaded into the crates with all hardware needed to complete the \dword{cpa} panel assembly at the \surf site.  Each shipment should consist of two crates which contain the two \dword{cpa} panels that will be paired to form a \dword{cpa} plane. There will be very little room for storage at the \surf site, so it is important to ship \dword{cpa} panels in this way so that final assembly, integration, and installation can proceed as soon as components are received.

\Dword{topfc} and \dword{botfc} modules will either be fully assembled at university production sites and shipped to SURF ready for installation into the mine, or the components will fabricated and \dword{qc} inspected at university sites before being shipped to \surf for final assembly, as was done for the \dword{pdsp} modules at CERN. Crate design for \dword{topfc} and \dword{botfc} modules will depend strongly on whether the \dwords{gp} remain attached to the modules, as in the \dword{pdsp} design, or if the \dwords{gp} are, instead, connected directly to the cryostat. In the former case, if fully assembled \dwords{topfc} and \dwords{botfc} modules with attached \dwords{gp} are transported to the underground area, the complexity of the crates is significantly enhanced, as it is difficult to fully support a module on all sides while allowing for tipping of the crate. Hence, it may be necessary for \dwords{gp} to be installed underground in either design scenario. Thus far, only single fully assembled \dwords{topfc} and \dwords{botfc} modules have been crated for shipment, but more complex designs that will allow for multiple modules (without installed ground planes) will be developed.

The \dwords{ewfc} sections each consisting of eight modules are shipped in two separate shipping crates each containing four modules in an upright and vertical orientation.
As was done for ProtoDUNE each \dwords{ewfc} module will be individually wrapped in plastic 
and can be extracted from the crate by means of an overhead crane and spreader bar.
Extracted modules will be rotated into horizontal position using the ledge on a dedicated assembly table. The module side showing steel bolts should be on top for final QC.

The \dword{hv} bus segments, \dword{fss} bias boards, and  interconnection wires will be integrated with the \dword{cpa} units before shipment, while \dword{cpa} panel interconnections and \dword{fss} connection tabs will be packaged and shipped with the \dword{cpa} panels for integration on site.
\dword{fc} divider boards will be attached to \dword{fc} modules during \dword{fc} assembly as described in \ref{sec:fdsp-hv-prod-fc}.
The \dword{cpa}-to-\dword{fc} and \dword{fc}-to-\dword{apa} boards will be shipped separately and attached just before each \dword{fc} is hoisted onto its \dword{cpa} panel in order to avoid risk of damage to exposed boards on the ends of the \dwords{fc} during shipment and handling.

\subsection{Installation and Integration}
\label{sec:fdsp-hv-integration}
Upon arriving at the SURF site, The \dword{cpa} shipping crate is unpacked with the middle/lower \dword{cpa} unit placed in a vertical stand on the floor of the so-called toaster region, an area in the DUNE detector cavern between the cryostat endwall and the cavern wall set aside for installation functions.  Next, the middle-middle \dword{cpa} unit is removed and vertically attached to the middle-lower Unit.  Finally, the upper-middle \dword{cpa} unit is removed and attached.  This assembly makes up a \dword{cpa} panel.  The \dword{cpa} panel is lifted and vertically attached to its trolley with an FR4 hangar.  Two \dword{cpa} panels are paired to form a \dword{cpa}  plane which then forms the unit for attachment of the \dwords{fc}.  Figure~\ref{fig:cpas-in-cryostat} shows on the left a two-Panel \dword{cpa} plane mounted on its trolleys in the clean room at \dword{pdsp}, waiting for installation of the \dword{topfc} and \dword{botfc} units.

It should also be noted that the \dword{cpa} panels on each end of each \dword{cpa} array are special -- they have aluminum profiles along the exterior side to form the first element of the field cage.  They also contain parts of the \dword{hv} bus mounted on the \dword{cpa} frames.  The \dword{hv} bus forms a continuous loop from the \dword{hv} Input through the top of the 25 \dword{cpa} planes, down the far external side, back along the bottom and up the  external side back to the \dword{hv} Input.

\begin{dunefigure}[\Dword{pdsp} \dword{cpa} plane before and after \dword{fc} attachment]{fig:cpas-in-cryostat}{Left: Completed \dword{pdsp} \dword{cpa} plane ready for \dword{fc} attachment. Right: Two completed \dword{cpa}-\dword{fc} assemblies in the \dword{pdsp} cryostat. The top and bottom \dwords{fc} with their \dwords{gp} attached are visible to the right of the cathode plane in their folded-up pre-installation position.}
\includegraphics[width=0.45\textwidth]{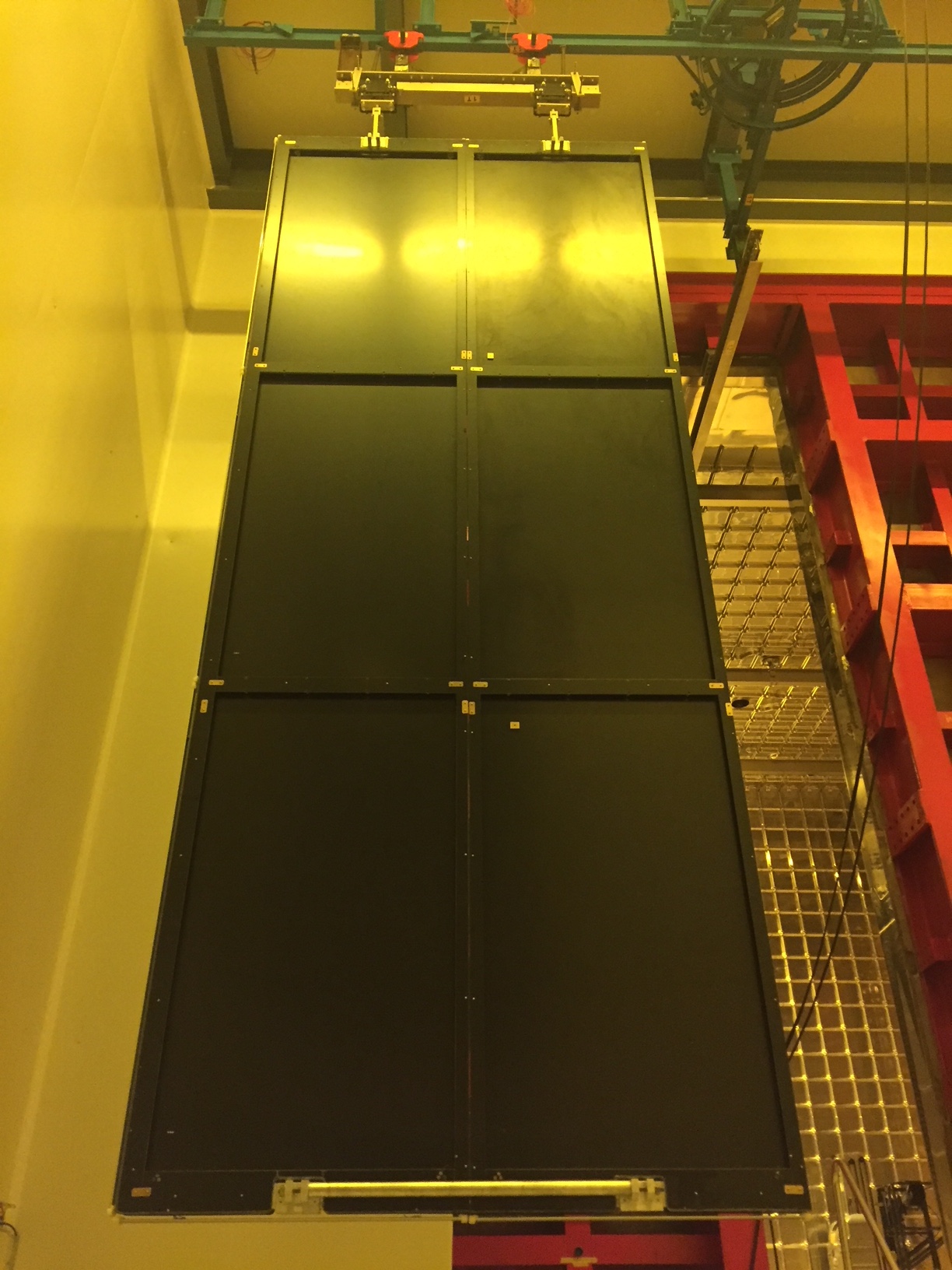}
\includegraphics[width=0.45\textwidth]{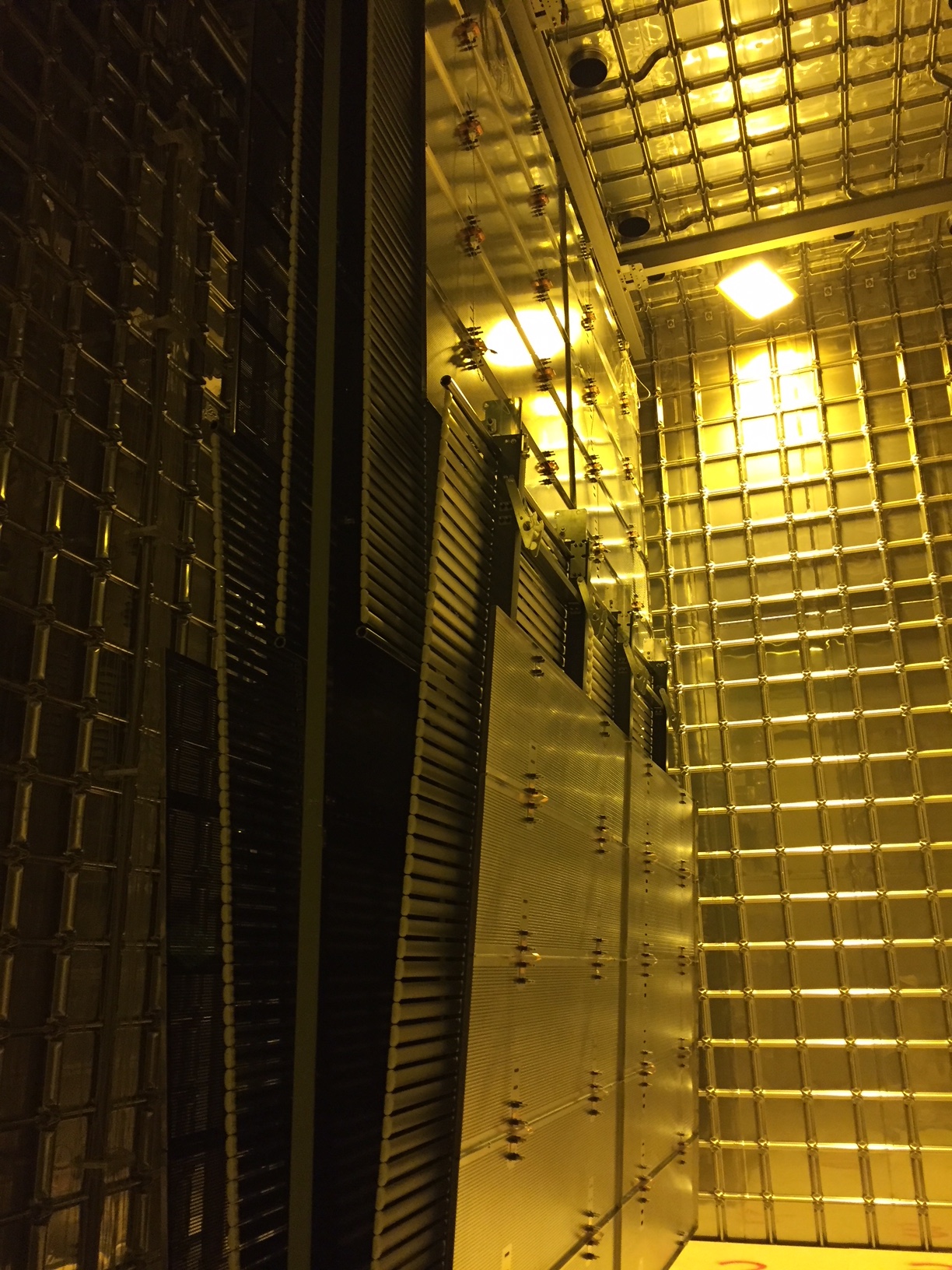}
\end{dunefigure}

There are a total of 50 \dword{cpa} panel pairs (\dword{cpa} planes) arranged in the 
\dword{spmod} in two rows of 25 each.  After the panels are paired, \dword{fc} units are attached folded against the \dword{cpa} and the full \dword{cpa}-\dword{fc} assembly is placed in the \dword{spmod} cryostat through the access door.  Figure~\ref{fig:cpas-in-cryostat} shows on the right two completed \dword{cpa}-\dword{fc} assemblies on their beam in the \dword{pdsp} cryostat. Upon deployment of the folded top and bottom \dword{fc} units, and after the final \dword{ewfc} installations, the \dword{tpc} \dword{fc} is complete.

To assemble the \dword{ewfc}, the modules are rotated vertically in the assembly area, then placed in a stand designed to hold the \SI{120}{\kg} pieces vertically. The \dword{topfc} is lifted and moved above the next module in the vertical stand where the two are bolted together and any electrical connections are made. The two modules are then lifted up and aligned above the following module.  The modules are again bolted, electrically connected and raised and aligned above the next piece.  This process is repeated until all eight modules are hanging together.  Once the bolting and electrical connections have been completed on the last piece, the continuity of all the grounding connections of the \dword{ewfc} are checked.
 
The completed \dword{ewfc} section is then moved over to the installation beam where its spreader bar is attached to the installation system.  It is then rolled into the cryostat and located on the appropriate beam for installation into the \dword{tpc}.

\begin{dunefigure}[\Dword{pdsp} \dword{ewfc} installation]{fig:EndwallInstall}{Completed endwall in process of installation into \dword{pdsp} cryostat.}
\includegraphics[width=0.45\textwidth]{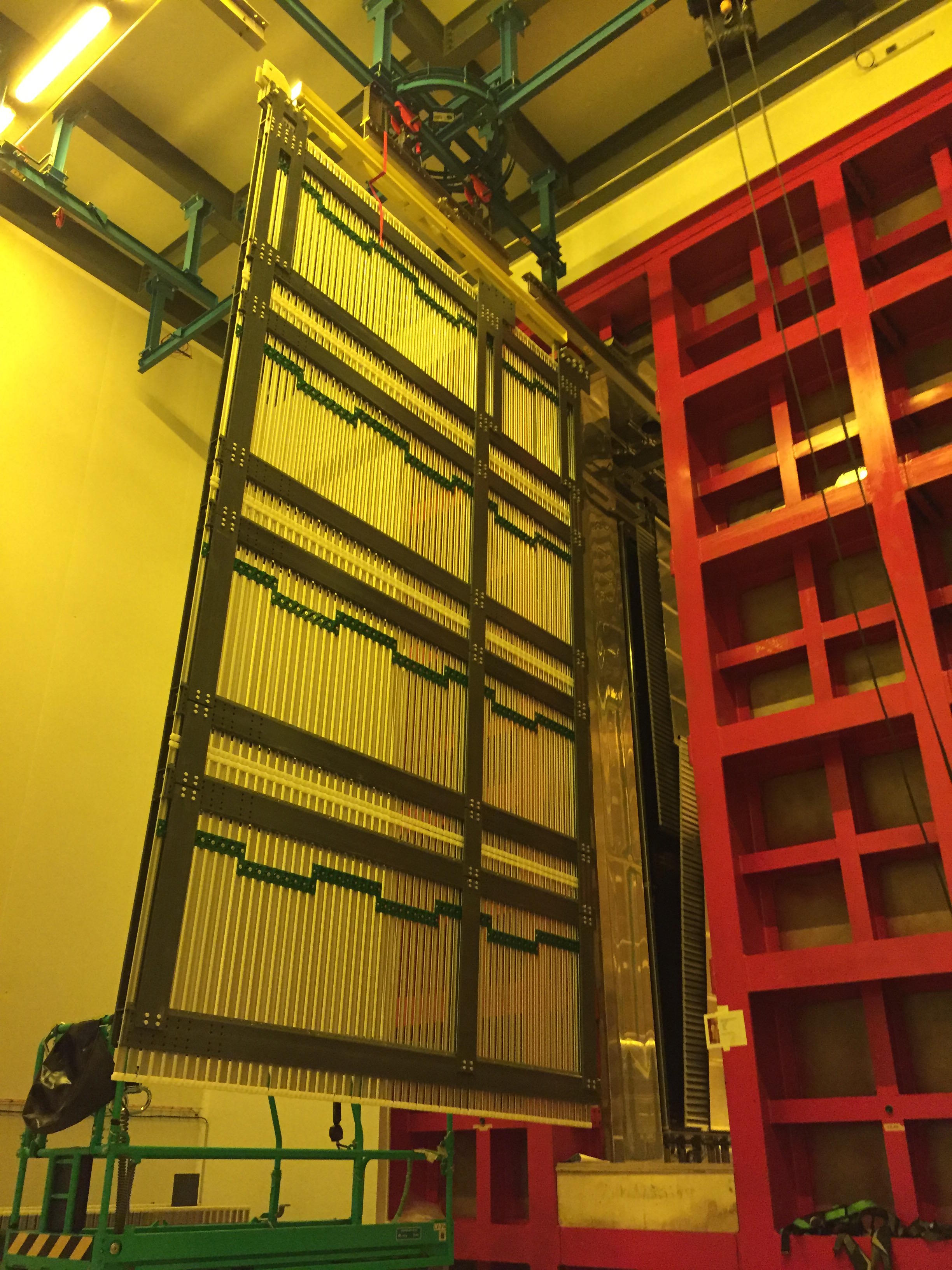}
\end{dunefigure}

\subsection{Interfaces}
\label{sec:fdsp-hv-interface}

Among \dword{tpc} components there are internal interfaces between the \dword{cpa} and \dword{fc}; the \dword{cpa} and \dword{hv}, and  the \dword{fc} and \dword{hv}. In addition there are significant interfaces with systems from other consortia.

The interfaces with the \dword{hv} system include the electrical connections with the \dword{hv} cup, \dword{hv} bus, and \dword{fc}, interconnecting the panels within the \dword{cpa}, and propagating the \dword{hv} bus between adjoining \dwords{cpa}.  A selection of the mechanical interfaces are as follows:
\begin{itemize}
\item Metal contact plates on the cathode surface at each corner, with two captive screws in each, used for making electrical contact with the \dword{hv} bus and the resistor board for the frame field strips;
\item Through holes for the \SI{3}{\milli\m} diameter cable in the sides of each frame adjacent to the contact plates to allow interconnection of the \dword{hv} bus between \dwords{cpa};
\item A means of securing the \dword{hv} bus cable in place;
\item Through-holes near the centers of the top and bottom frames for connecting the \dword{topfc} and \dword{botfc} elements to the cathode, with corresponding metal contact plate on the cathode surface;
\item Between the \dword{cpa} and the \dword{fc} the interface at the hinge joint ensures that the \dword{cpa} and \dword{fc} are properly located relative to each other.
\end{itemize}
Since these interfaces occur within the design group that is constructing the \dword{tpc} there is close coordination and communication between the \dword{cpa}, \dword{fc} and \dword{hv} groups to ensure that all requirements are met and that the components all fit together.

The interfaces between the \dword{cpa} system and the \dword{hv} system, and the interfaces among the \dword{cpa}-\dword{fc}-\dword{hv} systems are shown in integration drawings and documented in the project documentation system. \fixme{reference properly. RKP: Requires reference policy, under discussion.}


The various components in the \dword{hv} system also have interfaces with components from other systems.  These interfaces are formally defined by interface documents. 
 
Key interfaces between the \dword{hv} and other systems are summarized in Table~{\ref{tab:hvinterfaces}}.



\begin{dunetable}
[\dword{hv} system interfaces]
{p{0.2\linewidth}p{0.75\linewidth}}
{tab:hvinterfaces}
{Key \dword{hv} System Interfaces to Other Systems} 
Interface to & Description \\ \toprowrule
DSS  &  Support, positioning, and alignment of all \dword{cpa}, \dword{fc} modules inside the cryostat both warm and cold\\ \colhline
\dword{apa} & \dword{fc} support (top, bottom, and end wall) on \dword{apa} frames; Mounting of field cage termination filter boards and \dword{fc} failsafe terminations; Mounting of the electron diverter boards.
\\ \colhline
\dword{tpc} Elec. & \dword{fc} termination wire connectors on CE feedthrough flange, \dword{fc} termination wires routed with CE cables
 \\ \colhline
PDS & Mounting of PD calibration flash diffusers and routing of their fibers to \dwords{cpa}; Possible \dword{tpc} coated reflector foil on \dword{cpa}s.
 \\ \colhline
Facility & Locations and specifications of the \dword{hv} Feedthrough ports; gas and \dword{lar} flow velocities and patterns. \\ \colhline
Calibration & \dword{fc} openings for the calibration laser heads \\ \colhline
Cryogenic Instrumentation \& Slow Control. & \dword{hv} vs. \dword{lar} level interlock, sensor locations in high field regions, cold/warm camera coverage, \dword{hv} signal monitoring, etc. 
 \\ \colhline
Integration Facility & Storage buffer, inspections/tests, repackage for underground delivery
 \\ \colhline
Physics / Software & Requirements: range of operating drift field, uniformity of the drift field; Supply detector geometry and \efield{} map.
 \\ 
\end{dunetable}

\section{Quality Control (QC)}
\label{sec:fdsp-hv-qc}

Power supplies used in a \dword{spmod} will be tested before installation.  Output voltages and currents must be checked on a known load. 

The feedthrough and filters should be tested at the same time, preferably with the planned power supply.  The feedthrough must be tested to hold the required voltage in \dword{tpc}-quality \dword{lar} ($\tau\geq$\SI{1.6}{ms}) for several days.  The ground tube submersion and \efield{} environment of the test setup should be comparable to the real field cage setup or more challenging (e.g., the test liquid level can be lower than 
that in the \dword{spmod} but not higher).  Additionally, the feedthrough must be leak-tight to satisfy cryogenics requirements.

\Dword{cpa} \dword{qc} consists of forms that are filled out as part of the various procedures from initial assembly at factories to the final testing of connections in the DUNE cryostat.  A group of \dword{qc} forms filled out during initial panel unit assembly travels in the shipping crate with the panel.  At SURF, \dword{cpa} panel and \dword{cpa} plane assemblies have their own set of \dword{qc} forms for \dword{cpa} assembly and for \dword{cpa}-\dword{fc} integration.  Finally, before top and bottom \dword{fc} deployment, a final check of all electrical connections on the \dword{cpa} and between the \dword{cpa} and the top, bottom, and endwall \dwords{fc} is completed.  

Before assembly and shipment of \dword{topfc} and \dword{botfc}, each FRP component is subjected to inspection.
Each of the beams is checked for flatness, torsion, and structural and surface defects, such as cracks or fractures, layer separation, thermal decomposition, and evidence of water-induced bloating. Any scratching or grooving of the surface layer can be remediated by coating with commercially available epoxy.


For the \dwords{ewfc}, each box beam and all FRP stock will be inspected prior to assembly for dimensions, deformations, surface scratches and delamination defects. If the inspection is not passed, the parts will be rejected.
Preliminary assembly of \dwords{ewfc} modules will take place in a high-bay area under full crane access. After assembly, individual \dwords{ewfc} modules will be hung off each other in pairs of two to test the interconnections of the modules. Hanging will be performed in a top-down sequence consistent with space constraints (maximum three at once). This process will be repeated for all \dwords{ewfc}. All parts will be labeled to uniquely identify their position. After disassembly, all parts are cleaned before being moved into the clean room, where they will be inspected again. Final assembly will occur in a dedicated clean laboratory space.

Each panel will undergo mechanical and quality control tests  before shipping. 
\fixme{lsu - do we want to specify?} Electrical \dword{qc} tests will be performed after assembly of all aluminum profiles and resistor divider chains and again prior to installation of the \dwords{ewfc} into the cryostat. These tests will be performed with a pico-ammeter to measure the nominal resistance between profiles while grounding neighboring profiles not involved in any particular measurement.


Once at SURF, all \dwords{ewfc} panels shall be inspected visually after unpacking and prior to installation. Electrical tests will be repeated at various stages of the staging and installation process.
The tests will be again be performed with a pico-ammeter to measure the nominal resistance between profiles while grounding neighboring profiles.

During electrical interconnections local assembly, quality control inspection checklists for \dword{cpa}, \dword{fc}, and
interconnections will be adapted from \dword{pdsp} checklists for \dword{cpa}
panels, \dword{hv} bus segments, interconnection wires, and resistor boards.
These checklists include physical inspection for defects and
resistance tests.  For maximum traceability, checklists may be both
filed in an electronic logbook and sent in hard copy along with the
components tested.

For the electrical interconnections \dword{qc} at SURF, expected resistances have been modeled at all stages of
\dword{cpa}+\dword{fc} integration.  The height of the \dwords{cpa} make it best to catch a
faulty connection as early as possible.  This will be achieved by
quickly checking resistances for expected values between \dword{cpa} panels and between \dword{cpa} panels and field strips as \dword{cpa} unit subassemblies are added.  Profile-to-profile field cage resistances
will be checked upon reception and rechecked immediately before
attaching to \dwords{cpa} underground.  Once all four \dwords{fc} are attached to a
\dword{cpa}, resistances between selected \dword{fc} profiles and the \dword{hv} bus will be
measured, checked against expected values, and recorded. Continuity of
the \dword{hv} bus between \dwords{cpa} will be checked at top and bottom as each \dword{cpa}
is connected to its neighbor inside the cryostat.

\section{Safety}
\label{sec:fdsp-hv-safety}

Safety is central to the design of the \dword{hv} system. In all phases including fabrication, installation, and operations, safety will be the highest priority. There will be documented assembly, testing, transport, and installation procedures. Particular attention was paid to these topics in the design of  
\dword{pdsp}  with explicit concern to a design that is identical to the \dword{spmod} design, the most critical of which are also noted in the preliminary \dword{hv} risk assessment, which is under development. 

The structural and electrical designs for the \dword{sp} \dword{hv} are based on designs that were vetted and validated in the \dword{pdsp} construction, which is currently in its final phase of deployment at CERN. Previously, Fermilab \dword{hv} tests implemented a full-voltage and full-scale \dword{hv} feedthrough, power supply, filtering, and monitoring system, along with the \dword{hv} connection cup and arm, after completing full safety reviews. These devices worked as designed and are essentially reproduced in both \dword{pdsp} and the \dword{spmod}. 

When operating the \dword{fc} at its full operating voltage there is a substantial amount of stored energy. The design of the \dword{cpa} is centered around storing charge  at the highest voltage on a resistive surfaces to limit the power dissipated during a power supply trip or other failure which unexpectedly drops the \dword{hv}. This design has been successfully tested at full voltage over \num{2}\,m$^2$ surfaces at full voltage and will soon be tested at larger scale in \dword{pdsp}.  

Integral to the \dword{sp} \dword{fc} design, both in \dword{pdsp} and the \dword{spmod}, is the concept of pre-assembled modular panels of field-shaping conductors with individual voltage divider boards. The structural design and installation procedures used in \dword{pdsp} were selected to be compatible with use at the Far Detector site and were vetted by project engineers, engineering design review teams, and CERN's safety engineers. Some revisions to these designs are expected based on lessons learned in installation and operations; these revisions will be reviewed both within the Project and by Fermilab \dword{esh} personnel. The overall design is on solid footing. 

Assembly of the \dword{fc} panels and resistor-divider boards will involve collaboration technical, scientific, and student labor and  does not present unusual industrial hazards. The \dword{hv} consortium will work closely with each assembly site to ensure that procedures meet both Fermilab and institutional requirements for safe procedures, personal protective equipment, environmental protection, trained materials handling, and training. The vast majority of production part fabrication will be carried out commercially and shipping will be contracted through approved commercial shipping companies. Prior to approving a site as a production venue, each site will be visited and reviewed by an external safety panel to ensure best practices are in place and maintained. 

%



\section{Organization and Management}
\label{sec:fdsp-hv-org}

\subsection{HV System Consortium Organization}
\label{sec:fdsp-hv-org-consortium}

The consortium consolidates all the institutions that are participating in the design, construction and assembly of the \dword{hv} systems for both \dword{pdsp}  and \dword{pddp}. It is currently composed of US institutions and CERN, presently the only non-USA participant. As in the case of \dword{protodune}, CERN is heavily committed to a significant role in terms of funding, personnel, 
 and the provision of infrastructure for R\&D and detector optimization. Moreover, CERN will be responsible for a significant fraction of subsystem deliverables; as such it is the intention of CERN to attract additional European institutions into the consortium. The consortium organization structure includes a leader (currently from CERN), a technical lead (currently from BNL), a \dword{tdr} editor (currently from \fnal), and a HVS design and integration lead (currently from ANL). 



In the \dword{hv} consortium organization, each institution is naturally assuming the same responsibilities as for the developments of \dword{pdsp}. The consortium is organized into working groups addressing the design and  R\&D phases and the hardware production and installation.

\begin{itemize}
\item WG1. Design optimization for \dword{spmod} and \dword{dpmod}; assembly, system integration, detector simulation, physics requirements for monitoring and calibrations. 
\item WG2. R\&D activities, R\&D facilities. 
\item WG3. \dword{sp}-\dword{cpa}: Procurement, in situ \dword{qc}, resistive panels, frame strips, electrical connections of planes; \dword{qc}, assembly, shipment to assembly site; \dword{qc}. 
\item WG4. \dword{dp} cathode.
\item WG5. \dword{fc} modules. 
\item WG6. \dword{hv} supply and filtering, \dword{hv} power supply and cable procurement, R\&D tests, filtering and receptacle design and tests. 
\end{itemize}

\noindent Merging of \dword{sp} and \dword{dp} groups is envisaged for the working groups where synergies are being identified: \dword{hv} feedthroughs, voltage dividers, aluminum profiles, FRP beams, and assembly infrastructure.

\subsection{Planning Assumptions}
\label{sec:fdsp-hv-org-assmp}
The present baseline design for all elements of the \dword{spmod} (\dword{cpa}, top/bottom \dword{topfc}, \dword{botfc}, \dword{ewfc} and \dword{hv} distribution) follows the \dword{pdsp} design as it has been produced and is being assembled.  It is also assumed that no major issues in the \dword{hv} operation of \dword{pdsp} will be encountered and therefore that the basic \dword{hv} concepts are sound.

However some design modifications/simplifications are envisaged to be implemented to take into account the different height of the \dword{cpa}  and the \dword{ewfc} modules and to adapt the installation procedure to the underground environment.

Additional design modifications could be expected if the \dword{pdsp} test run (as well as tests at Fermilab using the \SI{35}{\tonne} cryostat) identifies weaknesses in the present baseline option.


\dword{pdsp} is the testbed to understand and optimize detector element assembly, installation sequence, integration as well as requirements in manpower, space and tooling, and schedule. 



%
%
%
\subsection{High-level Cost and Schedule}
\label{sec:fdsp-hv-org-cs}




\begin{dunetable}
[\dword{hv} system R\&D program and milestones to lead to CD-2 approval]
{p{0.07\linewidth}p{0.55\linewidth}p{0.10\linewidth}p{0.10\linewidth}p{0.10\linewidth}}
{tab:HVschedule}
{DRAFT- \dword{hv} system R\&D program and Milestones to lead to CD-2 approval.}   
WBS&Task Name&Start&Finish \\ \toprowrule
1.5& CD-2 DOE Review& 10/4/19& 10/4/19 \\ \colhline
7& \dword{hv} system& & \\ \colhline
7.1& Finalize \single \dword{fc} design& 06/27/18& 09/30/19 \\ \colhline
7.2& Finalize \single cathode design& 06/27/18& 09/30/19 \\
7.3& Run \dword{sp} \dword{hv} design integration test& 01/01/18& 12/31/19 \\ \colhline
7.4& \dword{hv} \dshort{tdr} - submit for internal review& 03/29/19& 03/29/19 \\ \colhline
7.5& \dword{cpa} procurement& 09/21/21& 12/06/22\\ \colhline
7.6& \dlong{gp} procurement& 08/08/22& 12/06/22\\ \colhline
7.7& Assemble and test voltage dividers& 08/08/22& 12/06/22\\ \colhline
7.8& \dword{fc}  procurement&  03/11/22& 12/06/22\\ \colhline
7.9& Production readiness reviews& 01/02/23& 01/07/23\\ \colhline
7.10& Cryostat  ready for TPC installation& 05/01/23& 05/01/23 \\ \colhline
7.11& \dword{cpa} assembly& 01/31/23& 07/25/23\\ \colhline
7.12& Top-bottom \dword{fc} assembly&01/31/23& 07/25/23\\ \colhline
7.13& \Dword{ewfc} assembly & 01/05/23& 04/23/23\\
\end{dunetable}

\cleardoublepage

\chapter{Photon Detection System}
\label{ch:fdsp-pd}
%

\section{Photon Detection System (PDS) Overview}
\label{sec:fdsp-pd-ov}


\subsection{Introduction}
\label{sec:fdsp-pd-intro}

The \dword{pds} is an essential subsystem of a DUNE \dword{spmod}. The detection of the prompt scintillation light signal, emitted in coincidence with an ionizing event inside the active volume, allows the determination of the time of occurrence of an event of interest with much higher precision than charge collected from ionization in the TPC. This capability is most critical for the primary DUNE science objectives that are uncorrelated with the timing signal from the neutrino source at \fnal, such as proton decay and neutrinos from a \dword{snb}, and for the ancillary science program including measurements of neutrino oscillation phenomena using atmospheric neutrinos. A number of scientific and technical issues impact the \single and \dual \dword{pds} in a similar way, and the consortia for these two systems cooperate closely. See \voltitledpfd{}, Chapter~5.

Timing information from the \dword{pd} and TPC systems allows determination of the drift time of the ionizing particles. Knowledge of the drift time provides localization of the event inside the active volume and provides the ability to correct the measured charge for effects that depend on the drift path length, purity of \lar, or for specific locations in the detector if there are non-uniformities.  This correction is important for the reconstruction of the energy deposited by the ionizing event. In addition to allowing optimum track reconstruction, scintillation light measured by the system may also be used as a trigger and for improved calorimetric measurement in combination with charge measurement.

Table~\ref{tab:pds-sys-req} summarizes the high-level system performance requirements for the \dword{pds} necessary to achieve the DUNE science objectives. The first row provides a requirement to ensure high efficiency and good energy resolution for proton decay and atmospheric neutrinos.  The second row targets core collapse \dword{snb} neutrinos, but specifies only a timing measurement for event localization (in conjunction with the TPC drift time measurement). 
However, a consensus has emerged in the collaboration that the current requirements do not fully exploit the potential  for \lar scintillation light to contribute to the energy reconstruction of events, in particular for lower energy events such as from \dwords{snb} 
(\num{10}-\SI{100}{MeV}). In response, the third row is a proposed requirement to measure the energy in scintillation light from \dword{snb} events near the peak of the spectrum ($\sim$\SI{10}{MeV}) with a precision similar to that of the ionization measurement. The combined measurement of ionization charge and scintillation light has been shown to improve the determination of the energy deposition of an event. 
Table~\ref{tab:pds-det-req} shows the corresponding photon detection light yield, timing and spatial separation requirements. To achieve a \num{10}\%  calorimetric measurement with light requires approximately ten times higher light yield for the \dword{pds} than the original requirement. 



To achieve these requirements, there is an ongoing intense R\&D program to investigate methods that maximize the photon detection efficiency of the \dword{pds} within the constraints of the \single TPC design .  All three of the photon collector options described in this chapter could meet the original performance requirements, albeit with different event efficiency, but one, the ARAPUCA, has highest potential to perform the low energy calorimetric measurements at the desired precision. It also has the highest efficiency for \dword{snb} events and provides higher spatial granularity for background rejection.




\begin{dunetable}
[PDS performance requirements to achieve the primary science objectives.]
{p{0.45\textwidth}p{0.45\textwidth}}
{tab:pds-sys-req}
{\dword{pds} performance requirements to achieve the primary science objectives (under review).} 
Requirement  	& Rationale \\ \toprowrule
The  \dword{fd} \dword{pds} shall detect sufficient light from events depositing visible energy >\SI{200}{MeV} to efficiently measure the time and total intensity. 
			& This is the region for nucleon decay and atmospheric neutrinos. The time measurement is needed for event localization for optimal energy resolution and background rejection.			\\ \colhline
The  \dword{fd} \dword{pds} shall detect sufficient light from events depositing visible energy <\SI{200}{MeV} to provide a time measurement.  The efficiency of this measurement shall be adequate for \dword{snb} events. 
			& Enables low energy measurement of event localization for \dword{snb} events. The efficiency may vary significantly for visible energy in the range \SI{5}{MeV} to \SI{100}{MeV}. 		\\ \colhline
(Proposed) The  \dword{fd} \dword{pds} shall detect sufficient light from events depositing visible energy of  \SI{10}{MeV} to provide an energy measurement with a resolution of 10\%. 
			& Enables energy measurement for \dword{snb} events with a precision similar to that from the TPC ionization measurement. \\ \colhline
The \dword{fd} \dword{pds} readout electronics shall record time and signal amplitude from the photosensors with sufficient precision and range to achieve the key physics parameters. 
			& The resolution and dynamic range needs to be adjusted so that a few-\phel signal can be detected with low noise.  The dynamic range needs to be sufficiently high to measure light from a muon traversing a TPC module.  \\ 
\end{dunetable}

\begin{dunetable}
[Preliminary \dword{pds} performance requirements.]
{p{0.45\textwidth}p{0.45\textwidth}}
{tab:pds-det-req}
{\dword{pds} performance requirements (under review). } 
Parameter  					& Value \\ \toprowrule
(Current) Minimum detector response per MeV energy deposition (Light Yield).
  							& \SI{1}{pe/MeV} for events at the center of the TPC and no less than \SI{0.5}{pe/MeV}  at all points in the fiducial volume. \\ \colhline
(Proposed) Minimum detector response per MeV energy deposition (Light Yield).
  							& \SI{10}{pe/MeV} for events at the center of the TPC and no less than \SI{5}{pe/MeV}  at all points in the fiducial volume. \\ \colhline
Minimum requirements on energy deposition, spatial separation, and temporal separation from other events, for which the system must associate a unique event time (\textit{flash matching}). 
							& \SI{10}{MeV}, \SI{1}{m}, \SI{1}{ms}  respectively. \\
\end{dunetable}

\subsection{Design Considerations}
\label{sec:fdsp-pd-des-consid}

Scintillation Light: \lar is known to be an abundant scintillator and emits about \SI{40}{photons/keV} when excited  by minimum ionizing particles\cite{Doke:1990rza},
in the absence of external \efield{}s. In the presence of \efield{}s the yield is reduced due to recombination; for the nominal DUNE \dword{spmod} field of \SI{500}{V/cm} the yield is approximately \SI{24}{photons/keV.}~\cite{PhysRevB.20.3486}. 

As depicted in Figure~\ref{fig:scintAr}, the passage of ionizing radiation in \lar produces excitations and ionization of the argon atoms that ultimately results in the formation of the excited dimer Ar$^*_2$.  
Photon emission proceeds through the de-excitation of the lowest lying singlet and triplet excited states, $^{1}\Sigma$ and 
$^{3}\Sigma$ to the dissociative ground state. The de-excitation from the $^{1}\Sigma$ state is very fast and has a characteristic time of the order of 
$\tau_{fast}$ $\simeq$ \SI{6}{ns}. The de-excitation from the $^{3}\Sigma$, state is much slower with a characteristic time of $\tau_{slow}$ $\simeq$ \SI{1.3}{$\mu$sec}, since it is forbidden by the selection rules. 
In both decays, photons are emitted in a \SI{10}{nm} band centered around \SI{127}{nm}, which is in the Vacuum Ultra-Violet (\dword{vuv}) region of the electromagnetic spectrum~\cite{Heindl:2010zz}.
The relative intensity of the  fast and slow components is related to the ionization density of \lar and depends on the ionizing particle: \num{0.3} for electrons, \num{1.3} for alpha particles and \num{3} for neutrons~\cite{PhysRevB.27.5279}.
This phenomenon is the basis for the  particle discrimination capabilities of \lar exploited by many experiments that have the capacity to separate the two components, for DUNE the greatest significance relates to a pending decision on treatment of light signals.

\begin{dunefigure}[Schematic of scintillation light production in argon.]{fig:scintAr}
{Schematic of scintillation light production in argon.}
\includegraphics[width=0.8\columnwidth]{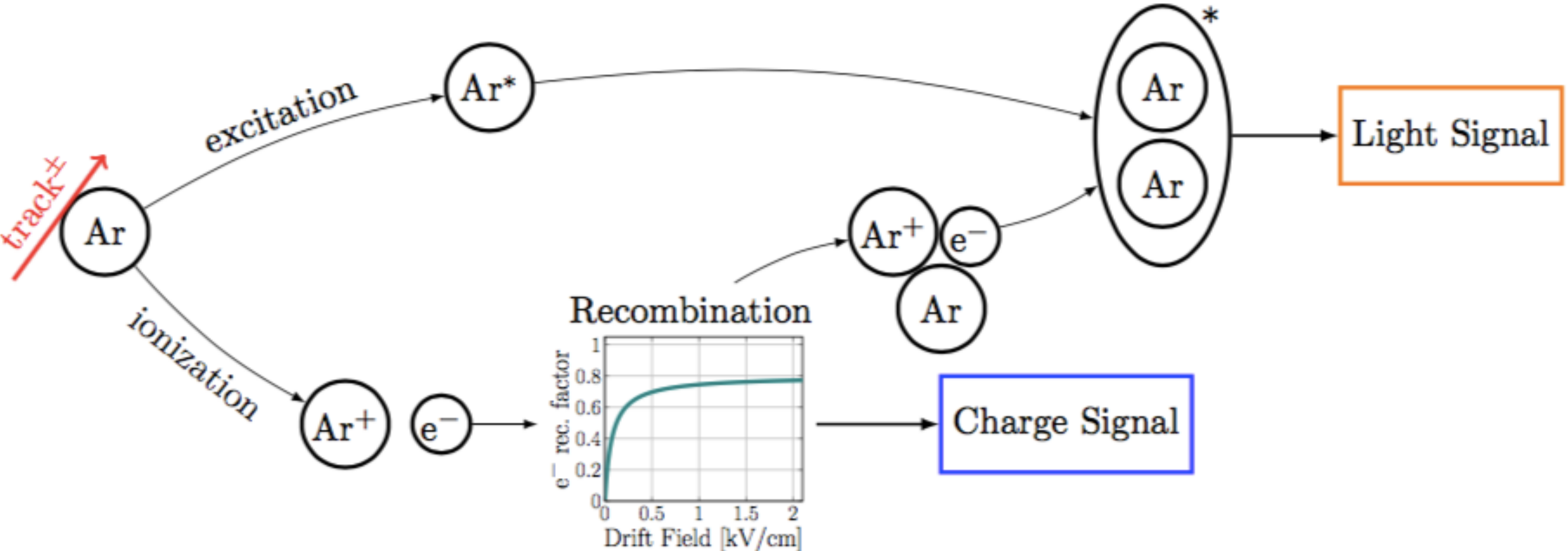}
\end{dunefigure}

In massive \lartpc{}s, a cost-effective approach is to use photon collector systems that collect light from large areas and attempt channel it in an efficient way towards much smaller photosensors that produce an electrical signal.
This paradigm for the detection of \lar scintillation light depends on the use of chemical wavelength shifters since most currently available commercial (cryogenic) large area photosensors are not directly sensitive to \dword{vuv} radiation, primarily due to the lack of transparency of fused silica and glass optical windows. 

\emph{\Ar39:}  The long-lived cosmogenic radioisotope \Ar39 has a natural abundance with an activity of approximately \SI{1}{Bq/kg} and
undergoes beta decay with a mean beta energy of \SI{220}{keV} with an endpoint of \SI{565}{keV}. In the \SI{10}{kt} \dword{fd} modules this leads to a rate of more than \SI{10}{MHz} of very short ($\sim$\SI{1}{mm}) tracks uniformly distributed throughout the detector module, each of which produces several thousand \dword{vuv} scintillation photons. This continuous background impacts the \dword{daq}, trigger and spatial granularity required of the \dword{pds}.

Wavelength Shifter: The most widely used wavelength shifter used in combination with \lar is \dword{tpb}\footnote{1,1,4,4-Tetraphenyl-1,3-butadiene, supplier: Sigma-Aldrich\textregistered{}, \url{https://www.sigmaaldrich.com/}.}, which absorbs \dword{vuv} photons and re-emits them with a spectrum centered around \SI{420}{nm}, close to where most commercial photosensors have their maximum quantum efficiency for photoconversion. 
Though \dword{tpb} has been utilized quite extensively with great success, there are recent publications that warrant caution. For example, until recently the conversion efficiency of \dword{tpb} in coating was taken to be high, approaching or even exceeding \num{100}\% (possible by multi-photon emission), however a recent arXiv paper~\cite{Benson:2017vbw} refutes this previous frequently referenced result. Using much of the same equipment but replacing a damaged reference photodiode, the authors (including an author of the previous paper) report a measurement for the quantum efficiency of \num{40}\% for incident \SI{127}{nm} light. 
Another recent paper\cite{Asaadi:2018ixs} reports that some methods used to coat surfaces with \dword{tpb} suffered loss of the \dword{tpb} coating in \lar, whereas there is no measurable effect if the fluor is dissolved in a polymer matrix. These developments will be followed carefully and highlight the importance of the ongoing R\&D and prototype program.


Physical Constraints: The physical dimension of the \dword{pd} system is constrained by the need to fit within the innermost wire planes of the \dwords{apa} and to be installed through slots in the \dword{apa} mechanical frame after it is wound (see Section~\ref{sec:fdsp-apa-design}). 
Individual \dword{pd} modules will be restricted to be within an envelope in the form of a long, thin box. At the time of preparation of this proposal the dimensions were \SI{14.6}{cm} $\times$ \SI{9.6}{cm} $\times$ \SI{212.7}{cm}, but it is anticipated that the size of the slot in the \dword{apa} will be increased by about \num{25}\% (the long dimension of the modules will remain the same). There will be ten \dword{pd} modules per \dword{apa}, for a total of \num{1500} modules.


\subsubsection{Photon Collectors} 
\label{sssec:photoncollectors}

The core modular elements of the \dword{pds} are the large area photon collectors that convert incident \SI{127}{nm} scintillation photons into photons in the visible range (>\SI{400}{nm}), which in turn are converted to an electrical signal by compact (\dwords{sipm}). 
As detailed in Section~\ref{sec:fdsp-pd-design}, since the size and cost of currently available \dwords{sipm} are not well matched to meeting the performance requirements in the large-volume  \dword{spmod}, the photon collector design aims to maximize the active \dword{vuv}-sensitive area of the \dword{pd} module while minimizing the necessary photocathode (\dword{sipm}) coverage. 
In the following we will distinguish between the terms photon \textit{collection} efficiency and photon \textit{detection} efficiency (PDE). Collection efficiency is the number of visible photons delivered to the \dwords{sipm} divided by the number of \dword{vuv} photons incident on the \dword{pd} module active area; this parameter is used to report results of calculations or simulations of predicted device performance independent of the \dword{sipm} used.  Detection efficiency is the number of detected \phel{}s from the \dword{sipm}(s) divided by the number of \dword{vuv} photons incident on the \dword{pd} module active area; this is generally the result of a direct measurement unless the detailed performance of the \dword{sipm} is known and divided out. The effective area of a \dword{pd} module is another useful figure-of-merit that is defined to be the photon detection efficiency multiplied by the photon collecting area of a \dword{pd} module. 

Three different designs of \dword{pd} photon collector modules have been developed and are being considered by the \single \dword{pd} consortium. The baseline design, ARAPUCA\footnote{\textit{Arapuca} is the name of a simple trap for catching birds originally used by the Guarani people of Brazil.}, is a relatively new concept that is scalable and has the potential for the best performance of the three designs by a significant factor. It is functionally a light trap that captures wavelength-shifted photons inside boxes with highly reflective internal surfaces where they are eventually detected by \dwords{sipm}.  There are also two alternative designs based on the use of wavelength-shifters and light guides coupled to \dwords{sipm}. Both have undergone more development than ARAPUCA, but their performance meets the basic physics requirements with only a limited safety margin and are not easily scalable within the geometric constraints of the \dword{spmod}.
Figure~\ref{fig:3dtpc-pd} shows a \threed model of the \single TPC with a zoom in to the anode plane where the three candidates photon collector technologies are visible for illustration -- in the final \dword{detmodule} there will be a single type.

\begin{dunefigure}[\threed model of \dwords{pd} in the \dword{apa}.]{fig:3dtpc-pd}
{\threed model of \dwords{pd} in the \dword{apa}. The model on the left shows the full width of the TPC with the configuration APA-CPA-APA-CPA-APA. The figure on the right shows a zoom in to the top far side of the TPC where three candidates photon collector technologies are visible for illustration -- in the final detector there will be a single type.}
\includegraphics[height=6cm]{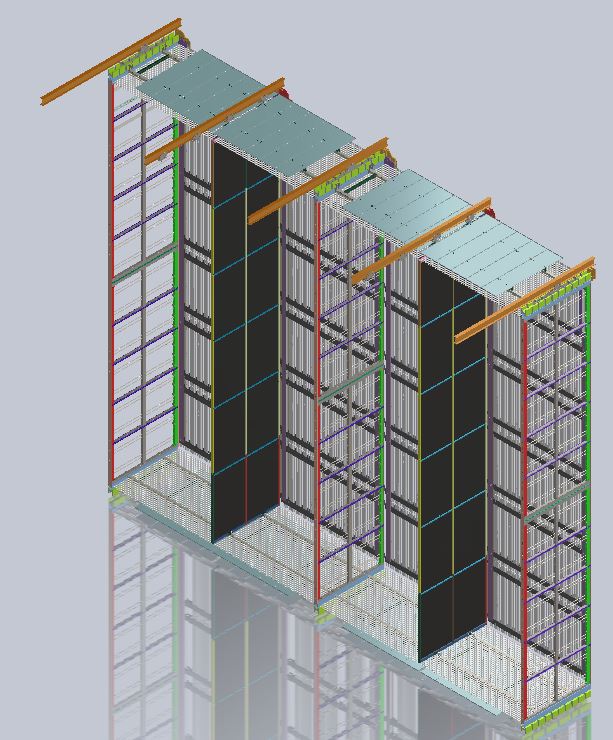}
\includegraphics[height=6cm]{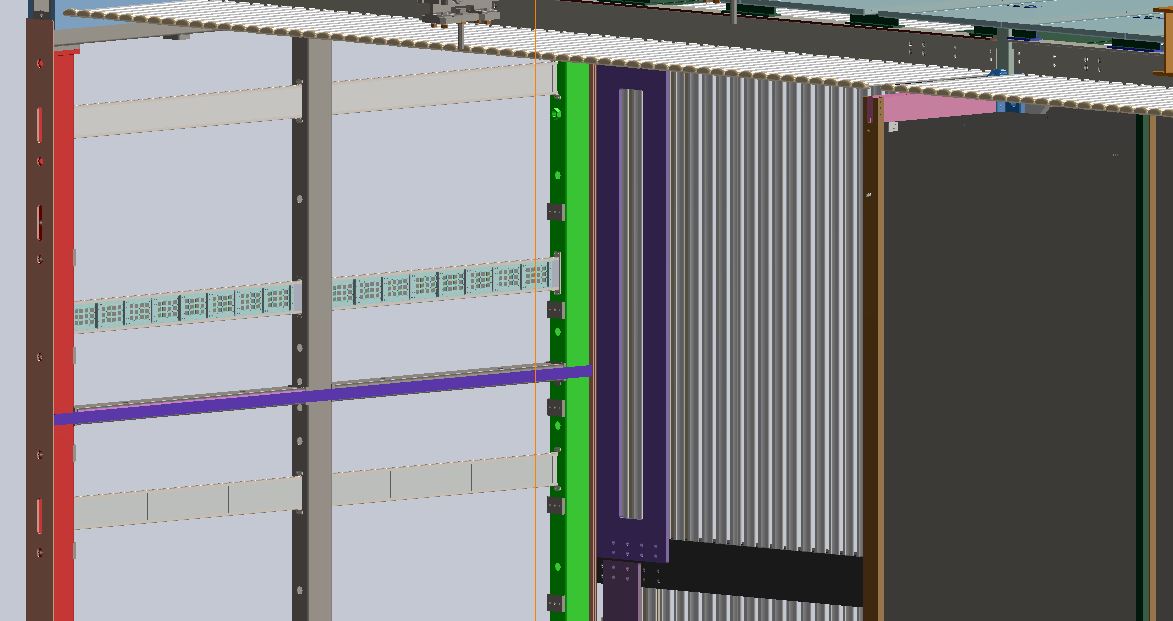}
\end{dunefigure}

\textit{\bf ARAPUCA Option:} The first large-scale implementation of an ARAPUCA module, in \dword{pdsp}, is composed of an array of sixteen ARAPUCA cells each one acting as an individual detector element. This configuration allows for finer spatial segmentation along the detector bar than is the case for the light guide designs. 
The \dword{pdsp} ARAPUCA design collects light from one side of the box through an optical window formed  
by a dichroic filter deposited with a layer of pTP\footnote{p-TerPhenyl,  supplier: Sigma-Aldrich\textregistered.}
wavelength shifter on the external surface. This shifts the incident \dword{vuv} light to a near-visible frequency that is able to pass through the filter plate to the interior of the box.  

In the \dword{pdsp} version of the device, the inner surface of the box opposite the window houses an array of \dwords{sipm} that covers a small fraction of the area of the window (2.8-5.6\%), surrounded by a foil of a highly reflective material coated with a second wavelength shifter, \dword{tpb}. The \dword{tpb}
converts the light passing through the filter to a wavelength that is reflected by the filter. It has been shown in simulation and in prototypes that a large fraction of these trapped photons, reflecting from the filter and the lined walls of the box, will eventually fall on a \dword{sipm} and be detected.
The X-ARAPUCA described in Section~\ref{sssec:x-arapuca}, is a promising variant of the concept that uses a wavelength shifter-doped plate between two dichroic filter windows with \dwords{sipm} on the narrow sides of the cell; in addition to viewing scintillation light from both sides as needed for the central \dword{apa}, it is expected to provide a higher light collection efficiency.


The ARAPUCA concept is relatively recent -- it was first proposed in 2015 and accepted for installation in \dword{pdsp} in mid-2016. A series of tests in \lar have been performed with an evolving prototype design that resulted in detection efficiency measurements ranging from  \num{0.4}\% to \num{1.8}\%, demonstrating the potential for substantially higher performance than the light guide designs. \dword{mc} simulations show that detection efficiencies at the level of several per cent could be reasonably reached with improvements to the basic design. 
While the results of the experimental tests are encouraging, a deeper understanding of the optical phenomena involving emission and scattering on wavelength-shifter coated surfaces is needed to optimize the design.

\textit{\bf Light Guide Options:} The fundamental idea of this approach is to convert \dword{vuv} scintillation light to visible wavelengths on (or near) the surface of an optical light guide, which then guides some fraction of the converted light by total internal reflection to \dwords{sipm} mounted on one or both ends of the guide.

Several approaches were investigated and narrowed down to the two most promising ones based a set of comparative measurements taken simultaneously in \lar~\cite{Whittington:2015rkr}. These two have been improved over several years and have reached a reasonable level of maturity and reliability.
The \textit{ dip-coated} light guides are pre-treated commercially-cut acrylic bars that are dip-coated with a solution of \dword{tpb}, acrylic, and toluene. When the toluene evaporates it leaves a thin film of \dword{tpb} embedded in the acrylic matrix on the surface of the bar.  In the \textit{double-shift} light guides, the conversion and guiding processes of the photons are decoupled. The first conversion is in a radiator plate, which is an ultraviolet transmitting UVT-acrylic plate coated with pure \dword{tpb} through a spraying process. It is positioned 
just above a commercial WLS-doped bar that absorbs the blue light produced by \dword{tpb} and re-emits it in the green; a fraction of this green light propagates down the bar. 

Both bar designs have demonstrated attenuation lengths for the  trapped light along the long dimension of the bar comparable to the length of the bars themselves, which ensures a reasonable uniformity along the beam direction. Preliminary measurements of both designs with readout at just one end indicate an photon detection efficiency range of  \num{0.1}\% to \num{0.25}\% averaged along the length of the bar. 
Up to a factor of four higher detection efficiency might be achieved with straightforward enhancements, such as:  \dword{sipm} at both ends of the bars; higher PDE \dword{sipm}; and coating the long edges of the bars with reflective foils. 

\subsubsection{Wavelength Shifter-Coated Cathode Plane} 
Since the \dword{pd} modules are installed only on the anode plane, light collection is not uniform over the entire active volume of the TPC. A possible solution to improve this is to install a reflective foil coated with wavelength shifter on the cathode.
This would increase the light yield of the detector and could enable calorimetric measurements based on light emitted by the ionizing particles. It may also be possible to remove the \Ar39 background through \dword{pd}-supplied timing cuts, a background that may otherwise cause a huge counting rate for events near the anode plane. This option would require good visible light sensitivity for the photon collectors, which is not the case for current light collector options. The capability could be incorporated in a variety of ways but with an impact on the direct light measurement. This option has yet to be formally adopted by the \dword{pd} consortium but is under study through \dword{mc} simulations and the mechanical feasibility is being discussed with the \dword{hv} consortium.

\subsubsection{Silicon Photosensors} 
In each photon collector concept, the final stage of converting a visible wavelength photon into an electrical signal is performed by a \dword{sipm}. The device must operate reliably for many years at \lar temperatures.
Experience with a promising early candidate that failed in later batches, due to an unadvertised change in the fabrication process, emphasizes the importance of a multi-source approach with active engagement of potential vendors to develop a device expressly for cryogenic operation. Currently, there are ongoing investigations of MPPCs (multi-pixel photon counters) produced by Hamamatsu\footnote{Hamamatsu\texttrademark{} Photonics K.K., \url{http://www.hamamatsu.com/}.} (Japan) including a model specifically designed for cryogenic operation, and a device developed for operation in \lar by FBK\footnote{Fondazione Bruno Kessler\texttrademark{}, \url{https://www.fbk.eu}.} (Italy) in collaboration with the DarkSide experiment.

\subsubsection{Readout Electronics} 
For prototype development and for \dword{pdsp}, a waveform digitizer has been developed that enables a thorough investigation of the photosensor signals, particularly as we investigate the impact of electrically ganging multiple \dwords{sipm}. The design of the readout electronics for the final system will be strongly influenced by the outcomes of \dword{mc} simulations that are in progress. Of particular interest is the extent to which pulse  shape capabilities are important to maximizing sensitivity to low energy neutrino interactions from \dwords{snb}. 
Initial \dword{mc} simulations suggest that it may not be necessary to fully digitize the \dword{sipm} waveforms in order to achieve the \dword{pd} performance requirements.  Charge integration electronic readout systems, which offer the promise of significantly lower cost and smaller cabling harnesses, are under investigation and are expected to be the baseline solution.
A lower-cost waveform digitization based on lower sampling rate commercial electronics will be investigated as a potential backup option in case our evolving understanding of the requirements necessitates collecting waveform data from the \dwords{sipm}.

The size of currently available \dwords{sipm} is far smaller than the spatial granularity required for the experiment so the output of individual devices will be electrically summed (ganged) to reduce the electronics channel count. This will be achieved either by simply connecting together the output of multiple devices, passive ganging, or using active components, active ganging, if the signal is too degraded for the passive approach. Both approaches are under investigation. 

\subsubsection{R\&D Priorities} 
Since the light guide designs are comparatively well understood, the need for an improved understanding of the potential ARAPUCA performance drives the strategy for the R\&D program that will be carried out before the \dword{tdr} (mid-2019). 
An intense effort is underway to demonstrate that an implementation of the ARAPUCA concept will increase the light yield of the \dword{spmod} by a factor of five to ten with respect to the light guides; resources (personnel and funding)  are being sought by the consortium to achieve this.  Since the ARAPUCA approach demands a larger number of \dwords{sipm} than the light guides, a related high priority is demonstration of active ganging of a sufficient number of devices with adequate \dword{s/n} properties.

It is anticipated that by the time of the \dword{tdr}, the consortium will present ARAPUCA as a baseline design for the photon collector with one alternative design for risk mitigation.  


\subsection{Development and Evaluation Plans}

The performance of the different photon collection options will be evaluated at several facilities available to the consortium. 
Relative and absolute measurements will be performed at both room and cryogenic temperatures.

The most comprehensive set of data will come from the fully instrumented modules in the \dword{pdsp} experiment currently 
under construction at CERN, which will start operations in the last third of \num{2018}.
All  three photon collector designs are present in \dword{pdsp}: \num{29} 
double-shift guides, \num{29} dip-coated guides, and two ARAPUCA arrays. 
The TPC will provide precise reconstruction in \threed of the track of any ionizing event inside the active volume and matching  
the track with the associated light signal will enable an accurate comparison of the relative detection efficiencies of the different \dword{pd} modules. 
In principle, absolute calculations are possible using \dword{mc} simulations, but currently some of the optical parameters that 
regulate \dword{vuv} light propagation in \lar are poorly known,
which will limit the precision of this approach. 
A plan will be developed to address this limitation.

\dword{pdsp} will also provide a long-term test of full-scale \dword{pd} modules for the first time so it may be possible to quantify any deterioration in their performance such as the loss of \dword{tpb} from the coating noted previously. 
More broadly, aging effects in various detectors technologies, such as scintillator and \dwords{pmt}, are well documented, and knowledge of such effects is required at the design stage so that the photon detection performance will meet minimum requirements for the whole life of the experiment.

An R\&D program will be executed in parallel with the \dword{pdsp} operation since additional comparative measurements will be needed, particularly for the newer ARAPUCA concept, prior to establishing the baseline design for the \dword{tdr}.
Several facilities are accessible to the consortium that will allow testing of smaller scale prototypes of the modules (or sections of them). 
These include: the cryogenic facilities 
at Fermilab,  Colorado State University and Universidade Estadual de Campinas (UNICAMP); and facilities for precision optical measurements and cryogenic testing of photosensors at Fermilab, Indiana University, Northern Illinois University, University of Iowa, Syracuse University, UNICAMP, and Institute of Physics in Prague. 


A critical issue for large experiments are the interfaces between the subsystems. \dword{pd} modules and interfaces with the \dword{apa} system and cold electronics will be conducted using cryogenic gaseous nitrogen in cold box studies at CERN, using a test stand developed for testing of \dword{pdsp} components prior to installation into the detector.  A full-scale \dword{pdsp} \dword{apa} has been fabricated, 
and will be instrumented with \dword{ce} and \dwords{pd}, allowing the interfaces to be carefully studied.

In addition, a small-scale TPC is planned for cold electronics testing at \dword{fnal}, and will be instrumented with as many as three 1/2-length \dword{pd} modules to provide triggering information for the TPC and to continue interface studies with the \dword{apa} and cold electronics.  It is envisioned that up to three test cycles will be performed prior to the \dword{tdr}, allowing testing and continued development of the ARAPUCA concept.


%

\section{Photon Detector Efficiency Simulation}
\label{sec:fdsp-pd-simphys}

The potential physics performance of \dword{pd} designs will be evaluated using a full simulation, reconstruction, and analysis chain developed for the \larsoft framework. The goal is to evaluate the performance in physics deliverables for each of the photon collector designs under consideration. The metrics evaluated will include efficiency for determining the time of the event ($t_0$), timing resolution, and calorimetric energy resolution for three physics samples: \dword{snb} neutrinos, nucleon decay events, 
and beam neutrinos. However, the development of analysis tools to take advantage of this full simulation chain is fairly recent, so this proposal will only include one test case: $t_0$-finding efficiency for \dword{snb} neutrinos versus the effective area of the photon collectors (see Section~\ref{sssec:photoncollectors}).

The first step in the simulation specific to the \dword{pds} is the simulation of the production of light and its transport within the volume to the \dwords{pd}. Argon is a strong scintillator, producing \SI{24000}{$\gamma$s/MeV} at our nominal drift field. Even accounting for the efficiency of the \dwords{pd}, it is prohibitive to simulate every optical photon with \dword{geant4} in every event. So, prior to the full event simulation, the detector volume is voxelized and many photons are produced in each voxel. The fraction of photons from each voxel reaching each photosensor is called the visibility, and these visibilities are recorded in a 4-dimensional library (akin to the photon maps used in the \dword{dpmod} simulation described in \voltitledpfd~Chapter 6.
This library includes Rayleigh scattering length ($\lambda=$ \SI{55}{cm}\cite{Grace:2015yta}), absorption length ($\lambda=$ \SI{20}{m}), and the measured collection efficiency versus position of the double-shift light-guide bars. When a particle is simulated, at each step it produces charge and light. The light produced is distributed onto the various \dwords{pd} using the photon library as a look-up table and the early (\SI{6}{ns}) plus late (\SI{1.6}{$\mu$s}) scintillation time constants are applied. Transport time of the light through the \lar is not currently simulated, but is under development.

The second step is the simulation of the electronics response. For now, the \dword{sipm} signal processor (\dword{ssp}) readout electronics used for \dword{pd} development and in \dword{pdsp} is assumed (see Section~\ref{sec:fdsp-pd-pde}). 
Waveforms are produced on each channel by adding an \dword{sipm} single-\phel response shape for each true photon. In addition, other characteristics of the \dword{sipm} are included such as dark noise, crosstalk and afterpulsing, based on data from device measurements. 
Dark noise, at a rate of \SI{10}{Hz} for each of the three \dword{sipm}s on each channel is include by the addition of extra single-\phel waveforms. Crosstalk (where a second cell avalanches when a neighbor is struck by a photon generated internal to the silicon) is introduced by adding a second \phel \num{16.5}\% of the time when an initial \phel is added to the waveform. Additional uncorrelated random noise is added to the waveform with an RMS of 
\SI{0.1}{\phel}. The response of the SSP self-triggering algorithm, based on a leading-edge discriminator, is then simulated to determine if and when a \SI{7.8}{$\mu$s} waveform will be read out, or in the case of the simulation it will be stored and passed on for later processing.

The third step is reconstruction, which proceeds in three stages. The first is a ``hit finding'' algorithm that searches for peaks on individual waveforms channel-by-channel, identifying the time based on the time of the first peak and the total amount of light collected based on the integral until the hit goes back below threshold. The second step is a ``flash finding'' algorithm that searches for coincident hits across multiple channels. All the coincident light is collected into a single object that has an associated time (the earliest hit), an amount of light (summed from all the hits), and a position on the plane of the \dwords{apa} ($y$-$z$) that is a weighted average of the positions of the photon collectors with hits in the flash. 
The final step is to ``match'' the flash to the original event by taking the largest flash within the allowed drift time that is within \SI{240}{cm} in the $y$-$z$ plane. Since the TPC reconstruction is still in active development, especially for low-energy events, we match to the true event 
vertex of the event in the analyses presented here. This is a reasonable approximation since the position resolution of the TPC will be significantly better than that of the \dword{pds}. 

\begin{dunefigure}[Preliminary estimates of the efficiency for finding $t_0$ for SNB events.]{fig:pds-snefficiency}
{Preliminary estimates of the efficiency for finding $t_0$ for core collapse \dword{snb} events vs. the effective area (top), distance from the anode plane (bottom-left), and neutrino energy (bottom-right).}
  \includegraphics[width=0.4\columnwidth]{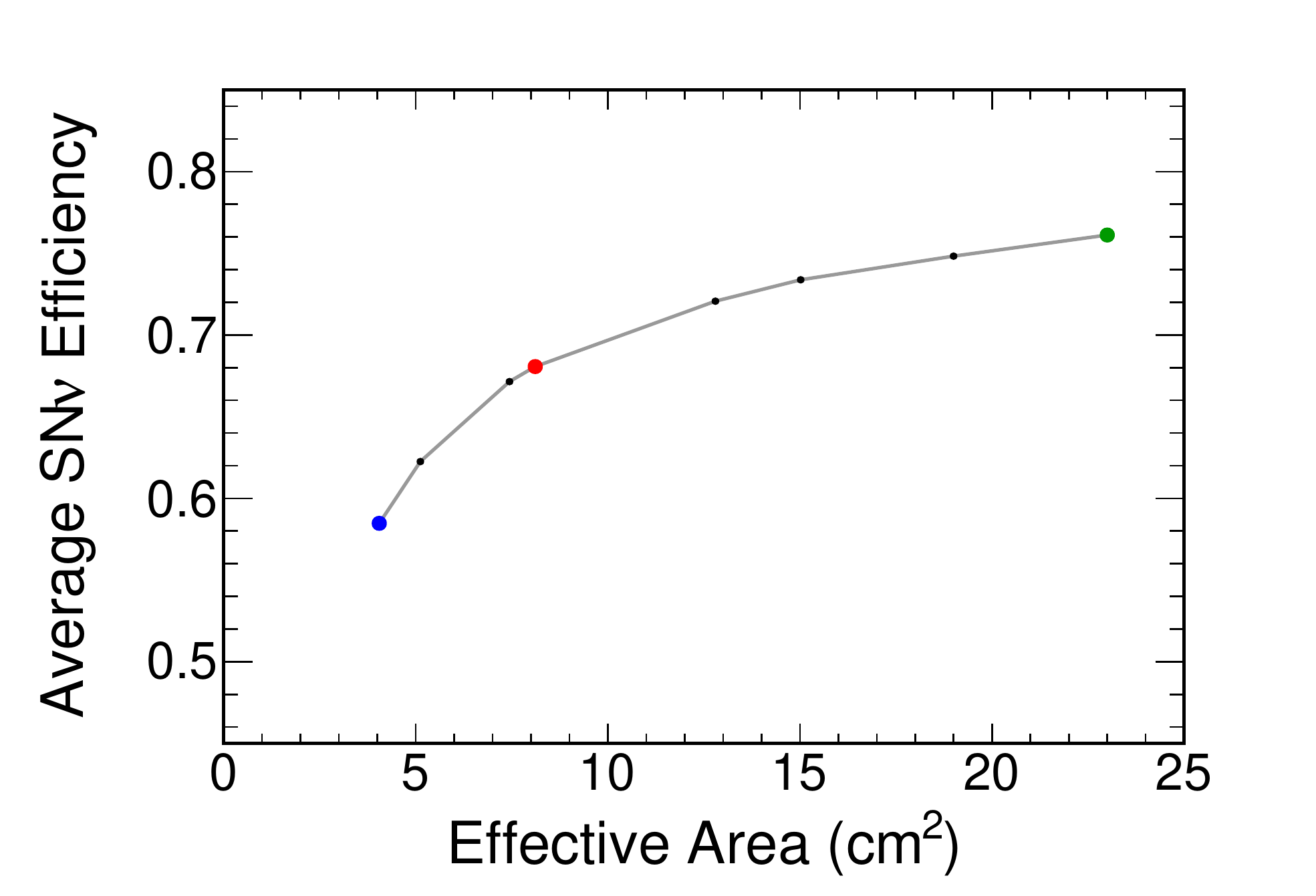}\\
  \includegraphics[width=0.4\columnwidth]{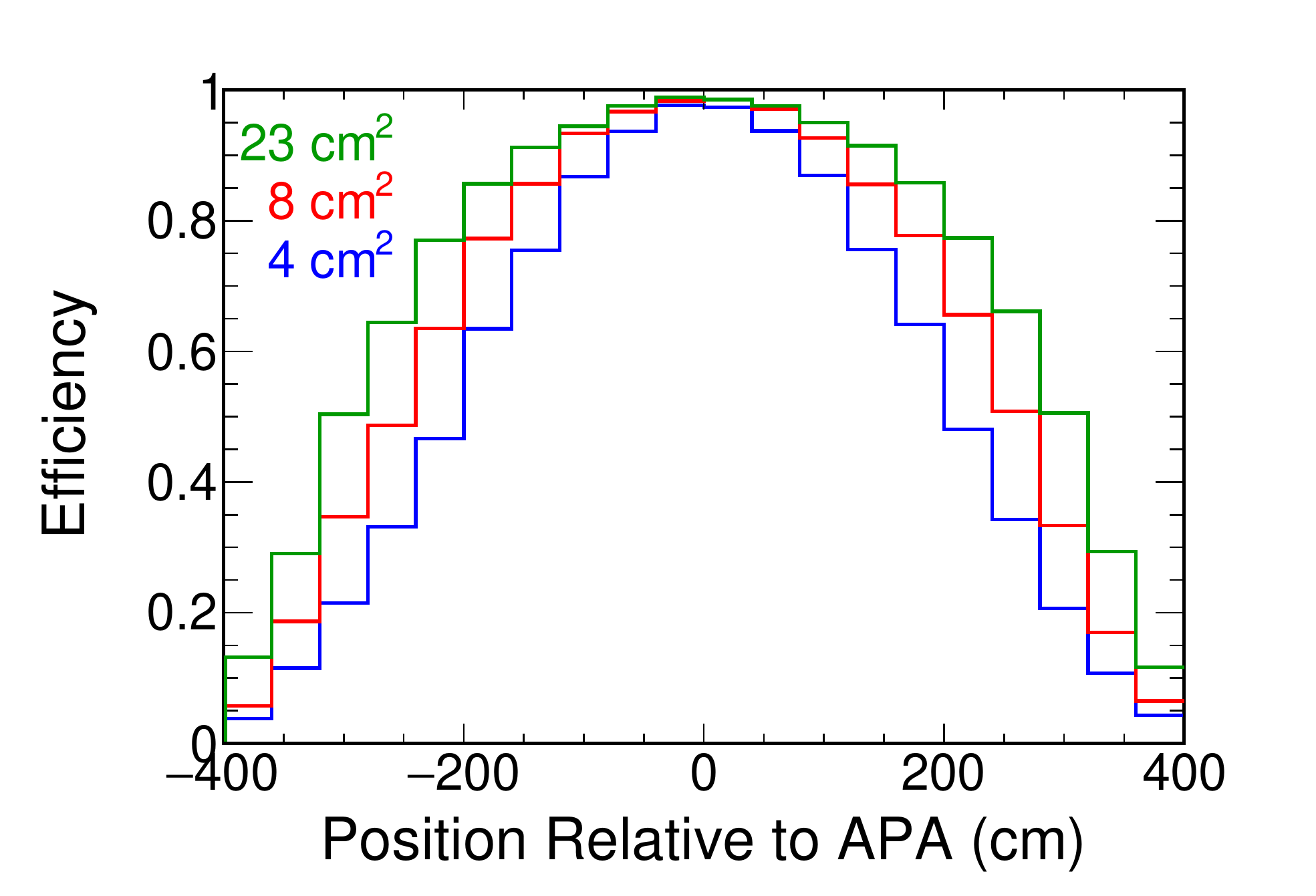}
  \includegraphics[width=0.4\columnwidth]{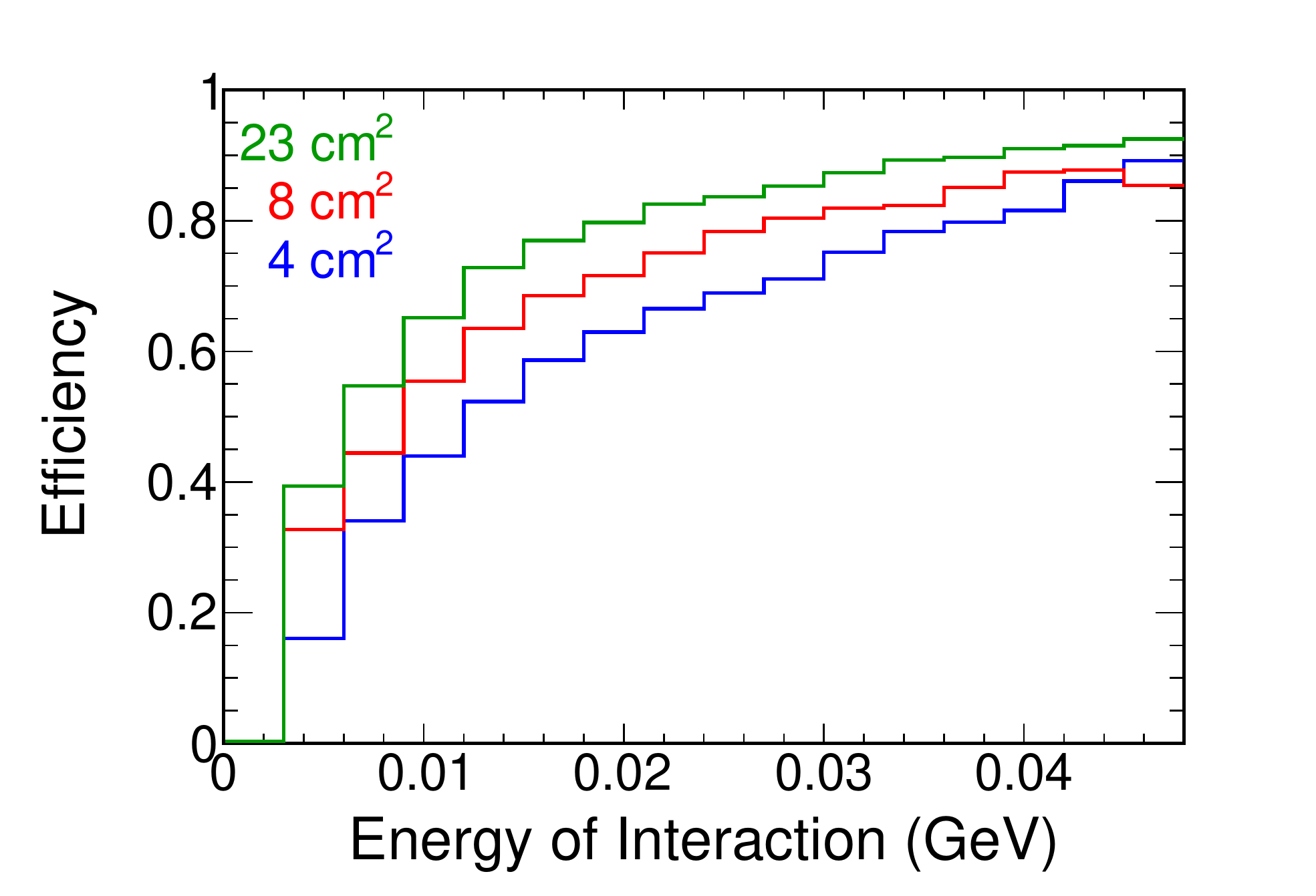}
\end{dunefigure}


Figure~\ref{fig:pds-snefficiency}~(top) shows the efficiency for determiningt $t_0$ for events in a typical \dword{snb} spectrum using the tools above. The changes in effective areas that would correspond to alternative photon collection designs are achieved by simply scaling up the total efficiency of the simulated double-shift light guide design shown here. The differences in attenuation behaviors within the bars are a second-order effect relative to the total amount of light collected. The efficiency for finding $t_0$ for these events increases, but less than linearly as the performance of the light collectors is improved. Figures~\ref{fig:pds-snefficiency}~(bottom-left) and \ref{fig:pds-snefficiency}~(bottom-right) show how the efficiency varies as a function of neutrino energy and distance from the anode plane for three chosen points.  These algorithms are still in development so there is potential for improvement in the performance as development continues.

\begin{dunefigure}[Resolution on $t_0$ for \dword{snb} events.]{fig:pds-snt0}
{Resolution on $t_0$ for \dword{snb} events. These are based on simulation with effective light collector area of \SI{4}{cm^2}, which corresponds to the photon detection efficiency of 0.23\% measured for a double-shift light guide module.}
  \includegraphics[width=0.4\columnwidth]{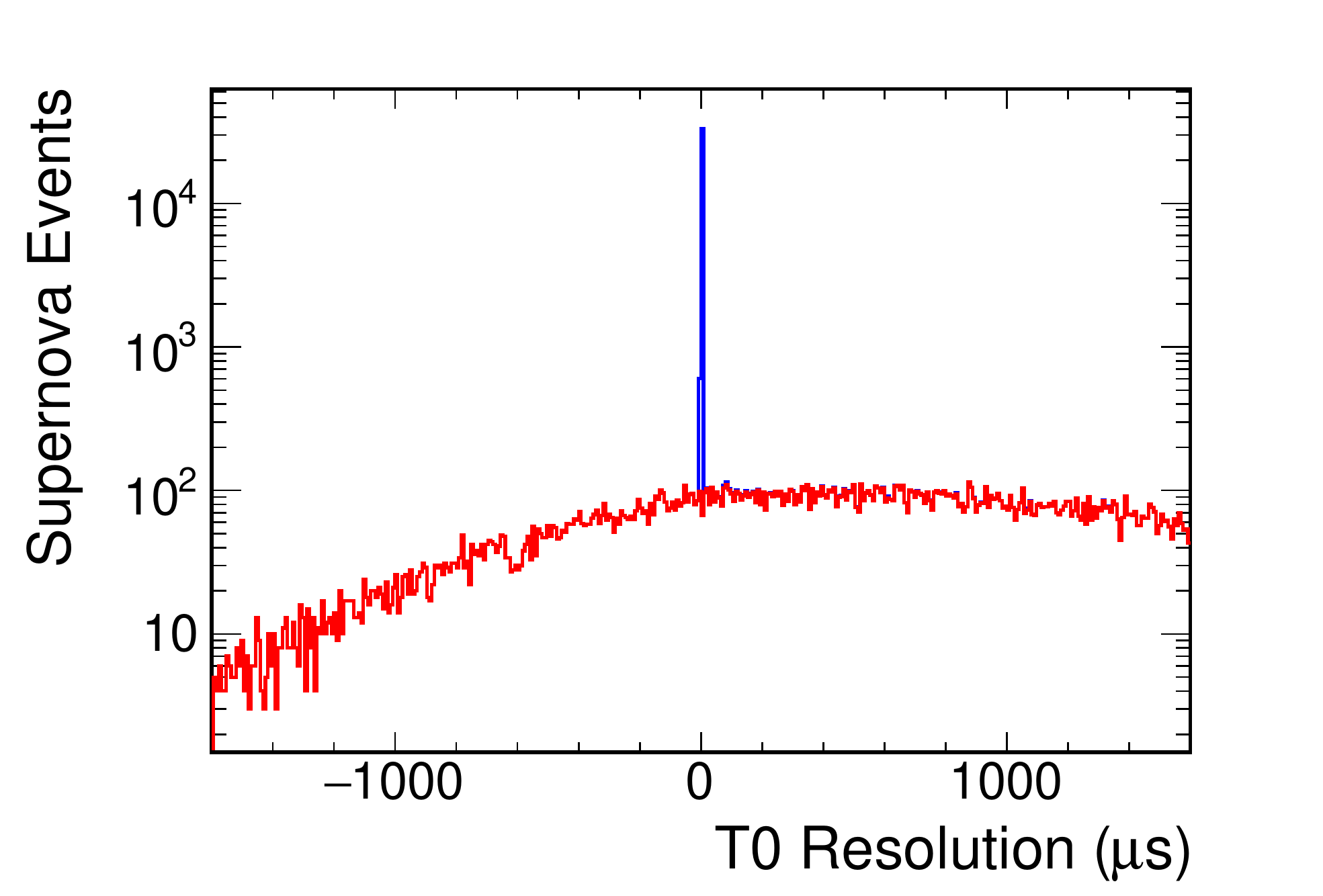}\\
  \includegraphics[width=0.4\columnwidth]{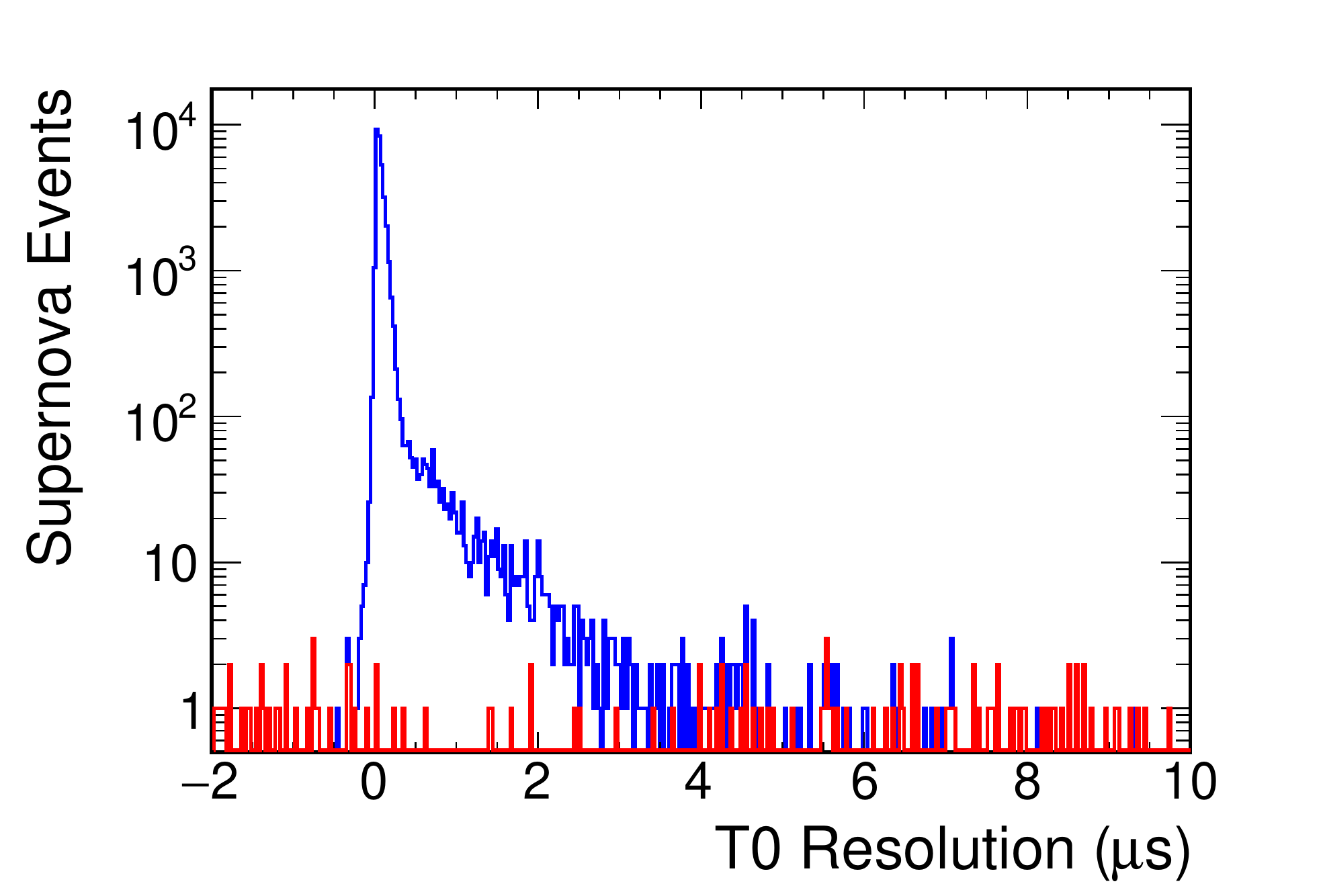}
\end{dunefigure}

If the correct flash is identified for a \dword{snb} event, the resolution on $t_0$ is excellent, as shown in Figure~\ref{fig:pds-snt0} -- for 95\% of the events, the time is identified from the prompt light and the timing resolution is better than \SI{100}{ns}, well within a single TPC time tick. In the remaining 5\% of cases where a correct flash is matched, the $t_0$ is biased towards later times (with respect to the true event time) by a few $\mu$s, driven by the late light time constant. If the wrong flash is identified the $t_0$ found is essentially randomly distributed in the drift time.

This preliminary study shows that each of the \dword{pd} options, with effective area per module currently estimated to be in the range  \numrange{4}{22}\si{cm$^2$}, will be able to determine the event $t_0$ with reasonable efficiency and it illustrates the benefit of higher photon detection efficiency. 

\section{Photon Detection System Design}
\label{sec:fdsp-pd-design}


The principal task of the \dword{spmod}\dword{pds} is to measure the \dword{vuv} scintillation light produced by ionizing tracks in the TPC within the geometrical constraints of the \dword{apa} structure. A commercially available compact solution for photon measurement is the \dword{sipm}, however, the response of the devices, which typically peaks in the visible range (>\SI{400}{nm}) is not well-matched to incident \SI{127}{nm} scintillation photons, so a wavelength shifter or some sort must be employed. 
In addition, even though production cost and key performance parameters have improved significantly in recent years, the cost of the readout electronics (channel count) and the \dword{sipm}s needed to meet the physics requirements of the \dword{pds} would be prohibitive. 

The photon collector must optimize the costs of various components of the system while meeting the performance requirements.  In practice, this consists of collecting \dword{vuv} photons over an area of hundreds of square-meters (viewing the entire \SI{10}{kt} \lar fiducial mass), converting the photons to longer wavelengths and guiding them onto \dwords{sipm} that are typically O(cm$^2$). 
For reference, an array of \num{48} \dwords{sipm} demonstrated a detection efficiency of \SI{13}{\%}, corresponding to an effective area of \SI{2.2}{cm$^2$}. This array, tested in the \fnal \dword{tallbo} \lar facility, consisted of twelve \num{4}$\times$\num{4} units of \SI{3}{mm}$\times$\SI{3}{mm} sensL C-series coated with \SI{100}{$\mu$g/cm$^2$} of \dword{tpb}. 


A challenge for the \dword{pds} is that a full set of requirements is not yet fully defined for one of the priority physics topics, \dword{snb} neutrinos. So the designs strive to demonstrate that at minimum the requirements for the accelerator neutrino program, atmospheric neutrinos and nucleon decay will be met, while maintaining the flexibility to adjust to the greater demands for the SBN physics.    
 

At the time of the \dword{tp} there are three photon collector options under consideration; Figure~\ref{fig:3dtpc-pd} shows how they are incorporated into the TPC anode plane assembly by an identical modular mounting scheme. In the following we summarize the design and development status for each photon collector option\footnote{For the \dword{tdr} there will be a baseline design and at most one alternative.}.


\subsection{Photon Collector: ARAPUCA}
\label{ssec:fdsp-pd-pc-arapuca}

The ARAPUCA is a device based on a new approach to \lar scintillation photon detection where the effective photon detection area is increased by trapping photons inside a box with highly reflective internal surfaces until reflections guide them to a much smaller \dword{sipm}~\cite{arapuca_jinst}. 

Photon trapping is achieved through a novel use of wavelength-shifting and the technology of the dichroic shortpass optical filters. These commercially available filters are created by using multilayer thin films that in combination have the property of being highly transparent to photons with a wavelength below a tunable cut-off while being almost perfectly reflective to photons with wavelength above the cut-off.  Such a filter coated with either one or two different wavelength-shifters, depending on the detailed implementation,  forms the entrance window to a flat box with internal surfaces covered by highly reflective acrylic foils
except for a small fraction of the surface that is occupied by active photosensors (\dwords{sipm}).

To act as a photon trap, the wavelength-shifter deposited on the outer face of the dichroic filter must have its emission wavelength \textit{less} than the cut-off wavelength of the filter, below which transmission is typically greater than 95\%. These photons pass through the filter where they encounter a second wavelength-shifter, either on the inner surface of the filter or coated on the reflecting inner surfaces of the box,
with an emission spectrum greater than the cut-off wavelength. For these photons the reflectivity of the filter is typically greater than 98\%, so they will reflect off the filter surface (and the inner walls) and so be trapped inside the box with a high probability to be incident on an \dword{sipm} before being lost to absorption. The concept is illustrated in Figure \ref{fig:arapuca}, in this example the filter cut-off is \SI{400}{nm}.


The net effect of the ARAPUCA is to amplify the active area of the \dword{sipm} used to readout the trapped photons. It is easy to show that, for small values of \dword{sipm} coverage of the internal surface, the amplification factor is equal to $A=1/(2(1-R))$,
where R is the average value of the reflectivity of the internal surfaces; for an average reflectivity of 0.95 the amplification factor is equal to ten.

\begin{dunefigure}[Schematic representation of the ARAPUCA operating principle.]{fig:arapuca}
{Schematic representation of the ARAPUCA operating principle.}
  \includegraphics[height=7cm]{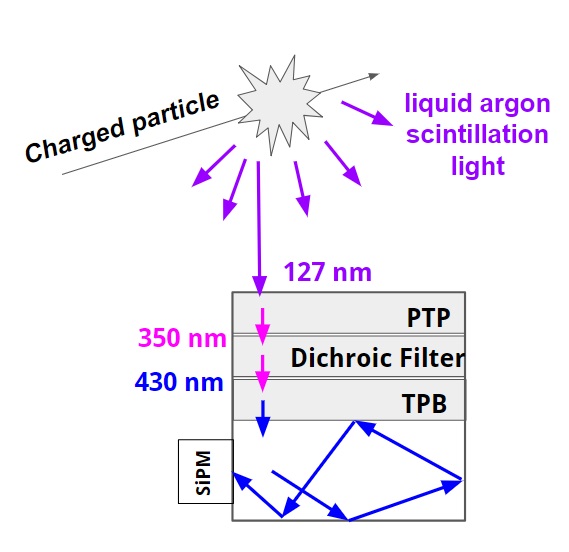}   
\end{dunefigure}

\subsubsection{Prototype Measurements}
\label{subsec:testlnls}

ARAPUCA prototypes with different configurations have been tested in \lar at multiple facilities. In each case, the first wavelength shift of \SI{127}{nm} scintillation photons down to \SI{350}{nm} that can pass through the filter substrate was performed by p-TerPhenyl (pTP) evaporated onto the outside of a dichroic filter window. 

The first prototype was made of PTFE with internal dimensions of \SI{3.5}{cm}$\times$\SI{2.5}{cm}$\times$\SI{0.6}{cm} with a window formed from a dichroic filter with dimensions of \SI{3.5}{cm}$\times$\SI{2.5}{cm} and wavelength cut-off at \SI{400}{nm}. 
TetraPhenyl-Butadiene (\dword{tpb}) was evaporated onto the internal side of the filter where it absorbs the shifted \SI{350}{nm} photons and reemits around \SI{430}{nm}. Trapped light is detected by a single \SI{0.6}{cm}$\times$\SI{0.6}{cm} sensL \dword{sipm} mod C60035\footnote{http://sensl.com/products/c-series/}.
The device was installed inside a vacuum tight stainless-steel cylinder closed by two CF100 flanges. The cylinder was deployed inside a \lar open bath, vacuum pumped down to a pressure around  10$^{-6}$~\si{mbar} and then filled with one liter of ultra-pure \lar\footnote{Argon 6.0, less than \SI{1}{ppm} total residual contamination.}.

Scintillation light emission was produced by an alpha source\footnote{A $^{238}$U-Al alloy in the form of a metallic foil, with alpha particle emission of 4.267 MeV.} installed in front of the ARAPUCA immersed in \lar. Signals were read out through an Aquiris\footnote{Aquiris High-Speed Digitizer products; http://www.acqiris.com/.} PCI board and stored on a computer.
Figure \ref{LNLS_test} shows photographs of the ARAPUCA and cryogenic system 
at the Brazilian Synchrotron Light Laboratory (LNLS). 

\begin{dunefigure}[ARAPUCA test at the Brazilian Synchrotron Light Laboratory.]{LNLS_test}
{ARAPUCA test at the Brazilian Synchrotron Light Laboratory} 
	\includegraphics[height=6cm]{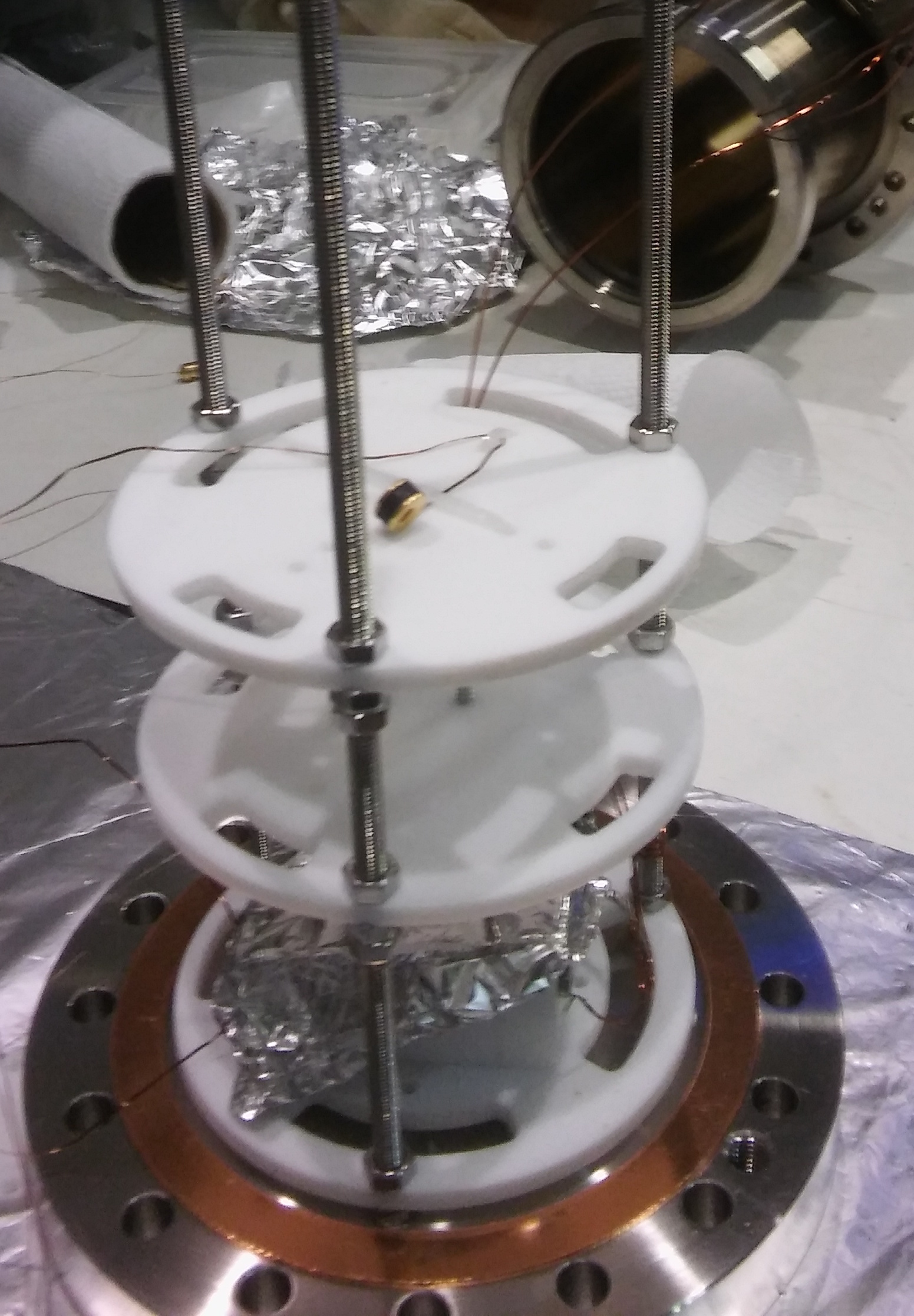} \quad
	\includegraphics[height=6cm]{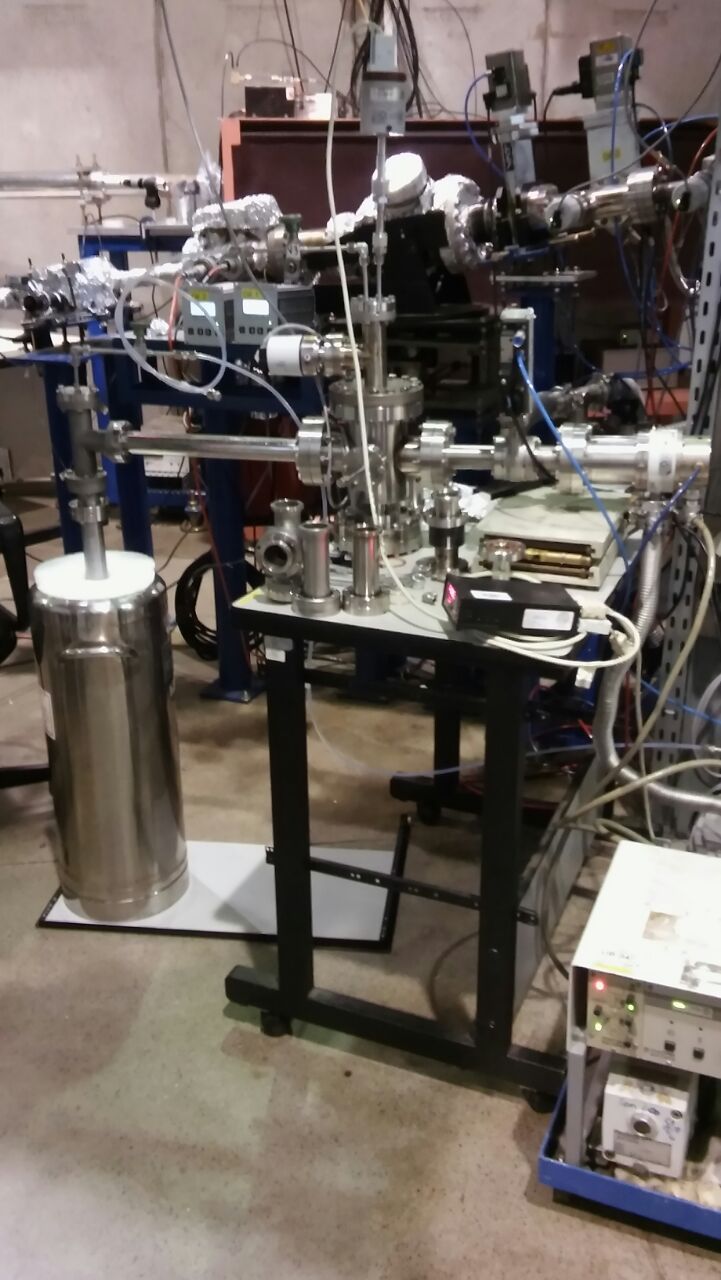}\quad
	\includegraphics[height=6cm]{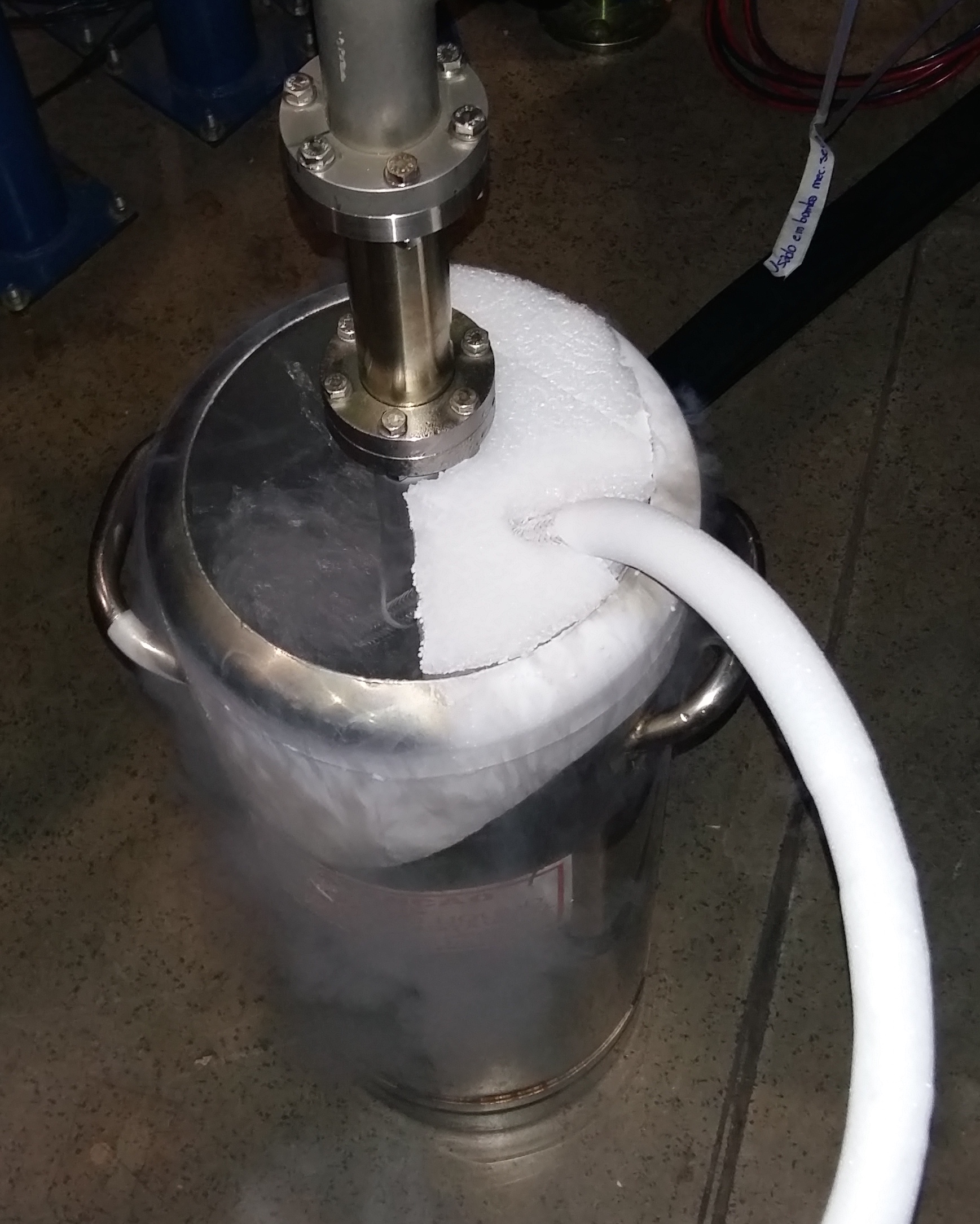}
\end{dunefigure}

The detection efficiency of the ARAPUCA was calculated by  determining the number of photoelectrons detected corresponding to the end point of the $\alpha$ spectrum and comparing it with the expected number of photons impinging on the acceptance window for that  particular energy value ($\sim$4.3 MeV).  This depends only on known properties of \lar and on the solid 
angle subtended by the ARAPUCA window. A detection efficiency at the level of 1.8\%  was measured, consistent with \dword{mc} expectations for the this configuration\cite{Marinho:2018doi}.


The next several prototypes were tested under cryogenic conditions at Fermilab.  
The first, performed in mid-2016 at the Proton Assembly Building (PAB) at the ScENE cryogenic test facility, had dimensions of  \SI{5.0}{cm}$\times$\SI{5.0}{cm}$\times$\SI{1.0}{cm} with a dichroic window of \SI{5.0}{cm}$\times$\SI{5.0}{cm} with a cut-off of \SI{400}{nm} which was deposited with pTP and \dword{tpb}. However, in this case, two of sensL \dwords{sipm} mod C60035 were installed inside the box.  The ARAPUCA was again deployed inside a vacuum-tight cryostat filled with ultrapure \lar. An $^{241}$Am alpha source was positioned in front of the window of the device \SI{5}{cm} from its center. The efficiency of the ARAPUCA was estimated taking into account that the alpha particles from this source have a  monochromatic energy of about \SI{5.4}{MeV}. 
The estimated efficiency in this case was approximately 1\%, a factor two below the expected value; this is attributed to the sub-optimal quality and uniformity of the pTP and \dword{tpb} films, and to the lack of reflectivity of the inner PTFE surfaces in this early prototype.

The next set of tests was performed at the beginning of 2017 at the PAB, but using the \dword{tallbo} facility, which is large enough to allow testing of several devices at a time. Eight different ARAPUCA cells with filters from different manufacturers, different reflectors, and different dimensions were tested.  Scintillation light was again produced by alpha particles emitted by an $^{241}$Am  source mounted on a holder that could be moved with an external manipulator in order to place it in 
front of each prototype. The detection efficiencies of these ARAPUCAs ranged from 0.4\% to 1.0\%.

The most recent measurements were performed in the \dword{tallbo} facility at the end of 2017 with an array of eight ARAPUCAs together with two reference bars (double-shift light guide design. The data analysis for the ARAPUCA array is currently underway. 


\subsubsection{ARAPUCA in ProtoDUNE-SP}

Two arrays of ARAPUCA modules will be operated inside \dword{pdsp} to test the devices in a large-scale experimental environment and allow direct comparison of their performance with the light guide designs. 
 
Each \dword{pdsp} ARAPUCA module array is composed of sixteen cells where each cell is an ARAPUCA box with dimensions of \SI{8}{cm}$\times$\SI{10}{cm}; half of the cells have twelve \dwords{sipm} installed on the bottom side of the cell and  half have six \dwords{sipm}. The \dwords{sipm} have active dimensions \SI{0.6}{cm}$\times$\SI{0.6}{cm} and account for 5.6\% (\num{12} \dwords{sipm}) or \num{2.8}\% (\num{6} \dwords{sipm}) of the area of the window (\SI{7.8}{cm}$\times$\SI{9.8}{cm}).
The \dwords{sipm}  are passively ganged together, so that only one readout channel is needed for each ARAPUCA grouping of \num{12} \dwords{sipm} (the boxes with six \dwords{sipm} are ganged together to form \num{12}-\dword{sipm} units) so a total of \num{12} channels is required per array. Studies are underway to investigate active ganging that would permit combining signals from multiple boxes, as required to reduce the number of electronics channels and cables under the working assumption that the \single \dword{pds} is restricted to four readout channels per \dword{pd} module. 
The total width of a module is \SI{9.6}{cm}, while the active width of an ARAPUCA is \SI{7.8}{cm}, the length is the same as the light guide modules (\SI{210}{cm}). 
The first ARAPUCA array installed in \dword{pdsp} is shown in Figure~\ref{fig:arapuca_array}. If the ARAPUCA cells achieved the same detection efficiency as earlier prototypes (1.8\%), the effective area of an ARAPUCA module will be approximately \SI{23}{cm$^2$}.


\begin{dunefigure}[Full-scale ARAPUCA for \dword{pdsp} during assembly.]{fig:arpk}
{Full-scale ARAPUCA for \dword{pdsp} during assembly. \dwords{sipm} are visible in the sixteen cells before the installation of reflecting foils, coated filter windows, and readout cabling. } 
\includegraphics[height=9cm]{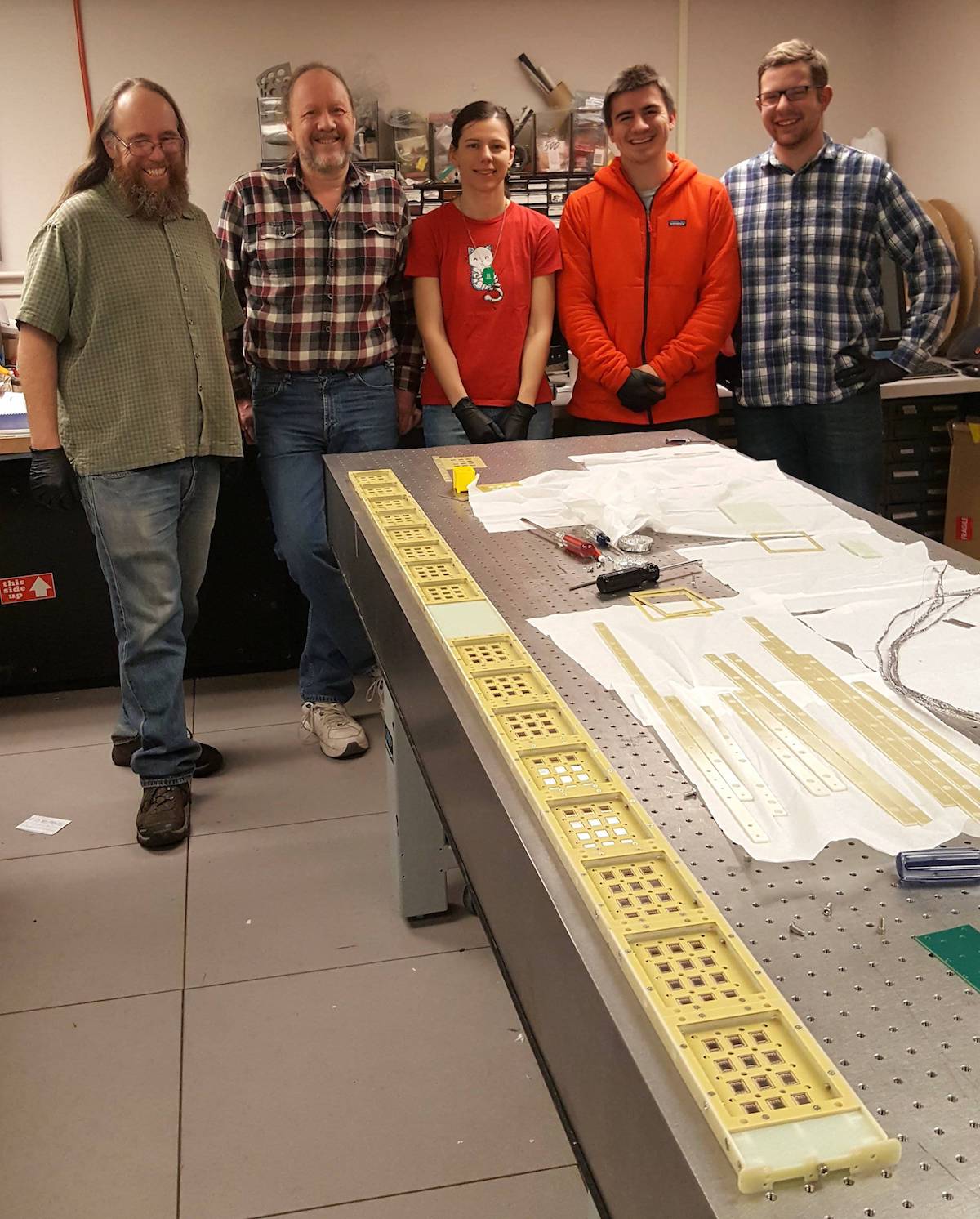}
\end{dunefigure}

\begin{dunefigure}[ARAPUCA array in \dword{pdsp}.]{fig:arapuca_array}
{ARAPUCA array in \dword{pdsp}.} 	
\includegraphics[height=8cm]{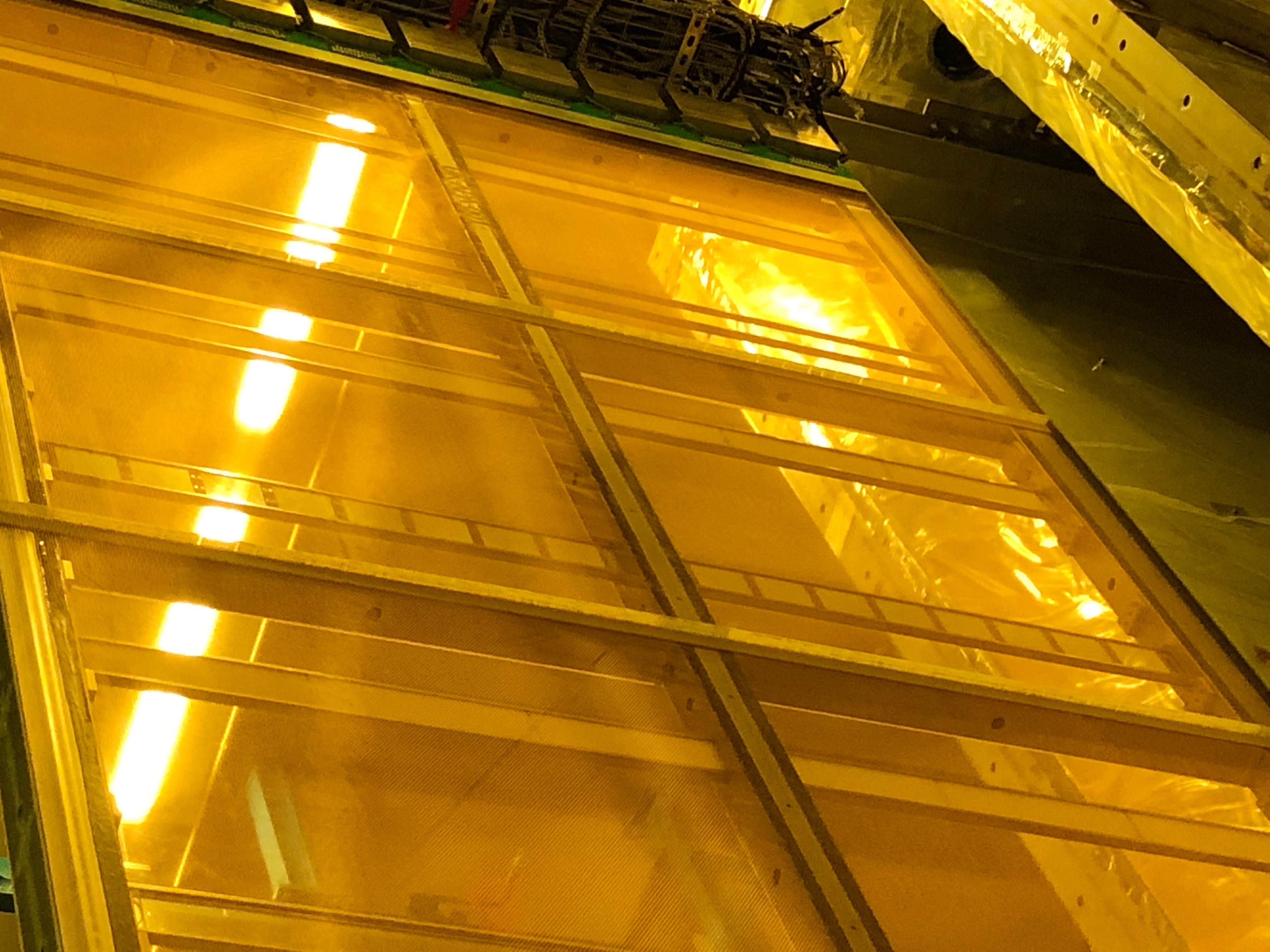} 
\end{dunefigure}


\subsubsection{X-ARAPUCA} 
\label{sssec:x-arapuca}
X-ARAPUCA represents  an alternative line of development with the aim of further improving the collection efficiency, while retaining the same working principle, mechanical form factor and active  photo-sensitive coverage. X-ARAPUCA is effectively a hybrid solution between the ARAPUCA and the wavelength-shifting light guide concepts, where photons trapped in the ARAPUCA box are shifted and transported to the readout via total internal reflection in a short light guide placed inside the box.
This solution minimizes the number of reflections on the internal surfaces of the box and thus the probability of photon loss. Simulations suggest that this modification will lead to a significant increase of the collection efficiency, to around 60\%, so the photon detection efficiency including the \dword{sipm} response could approach 20\%. 

 \begin{dunefigure}[X-ARAPUCA design: assembled cell (left),  exploded view (right).]{fig:pds-x-arapuca-cell}
{X-ARAPUCA design: assembled cell (left),  exploded view (right). The size and aspect ratio of the cells can be adjusted to match the spatial granularity required for a \dword{pd} module.}
  \includegraphics[height=.25\textheight]{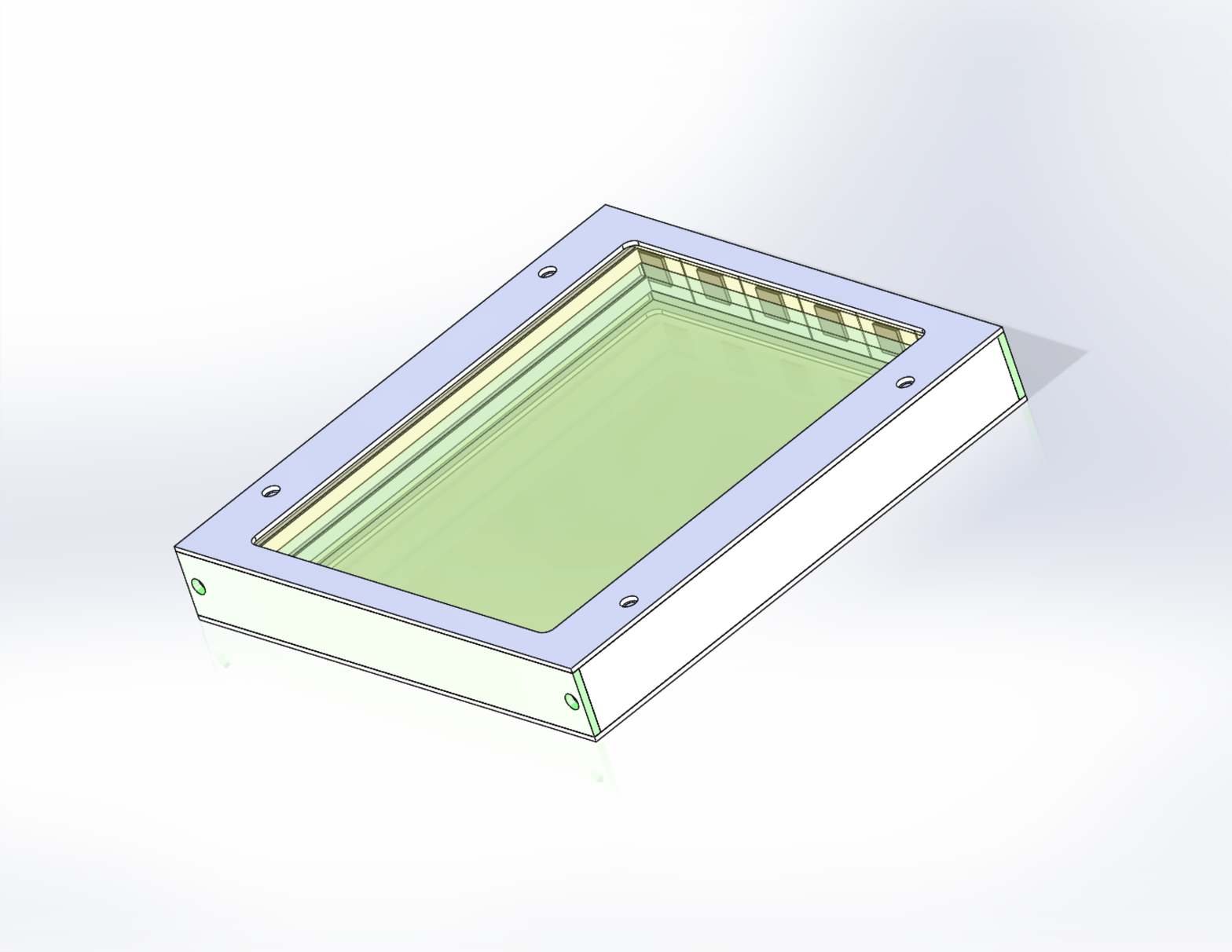}
  \includegraphics[height=.25\textheight]{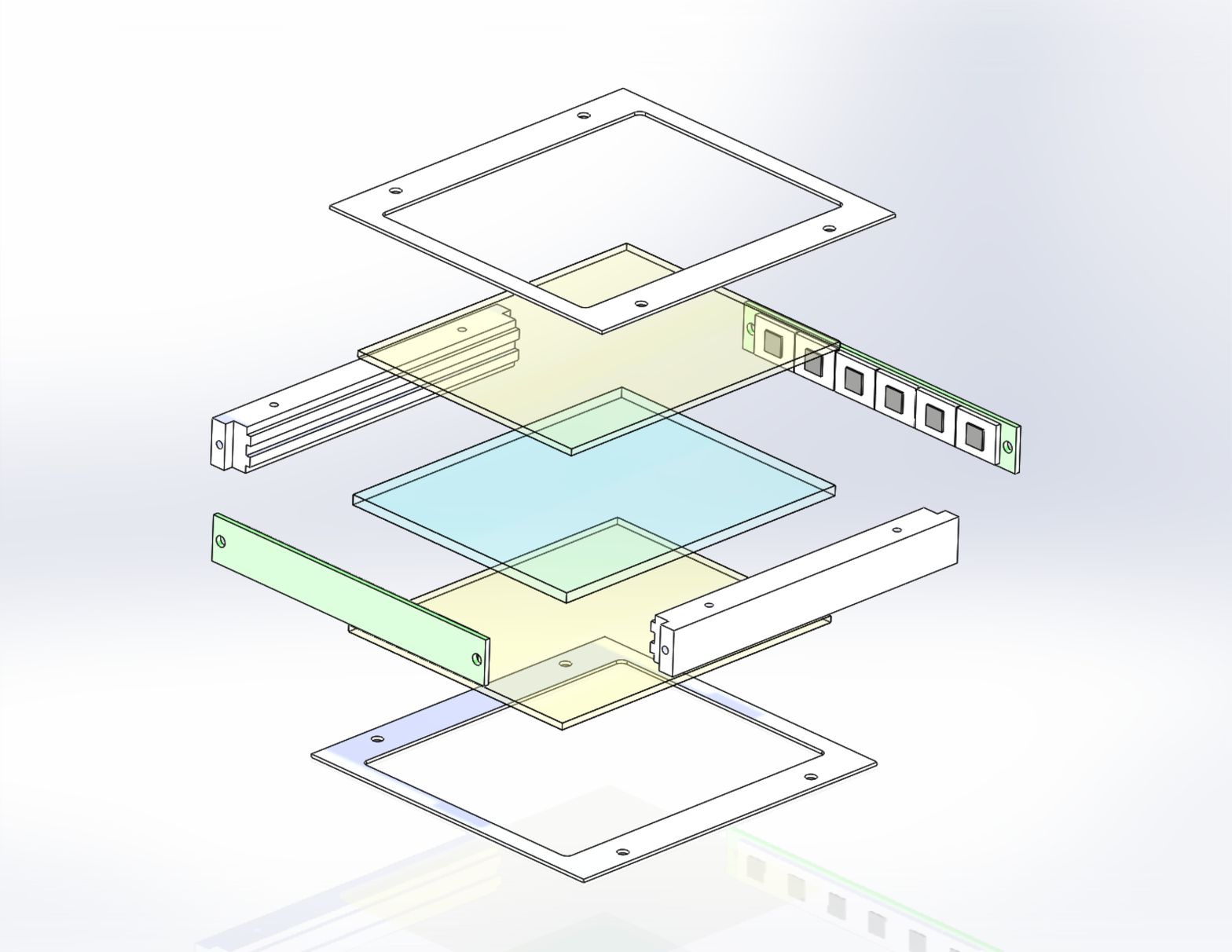}
\end{dunefigure}

In the X-ARAPUCA design, Figure \ref{fig:pds-x-arapuca-cell}, the inner shifter coating/lining over the reflective walls of the box is replaced by a thin wavelength-shifting light guide slab inside the box, of the same dimensions of the acceptance filter window and parallel to it. The \dword{sipm} arrays are installed vertically on the sides of the box, parallel to the light guide thin ends. 
 In this way a fraction of the photons will be converted inside the slab and guided to the readout, other photons,  e.g,. those at small angle of incidence below the critical angle of the light guide slab, after conversion at the slab surface will remain trapped in the box and eventually collected as for the standard ARAPUCA.
 
 A full-sized X-ARAPUCA prototype is under development. The light guide is a \SI{3}{mm} thick \dword{tpb}-doped acrylic plate. Two readout boards, each with several passively ganged \dwords{sipm} in a strip configuration, are mounted along the thin edges of the box and their ganged signals are combined into a single channel readout. 
 The aspect ratio of the cells can be adjusted to match the required spatial granularity for the \dword{pd} module.
 
 %
%
 
\subsubsection{ARAPUCA Configuration in DUNE \SI{10}{kt} }
\label{sssec:arapuca-dune}

The modular arrangement of the \dword{spmod} TPC calls for a configuration across the width of the cryostat starting with an \dword{apa} plane against one cryostat wall, and following with \dwords{apa} and \dwords{cpa} arranges as follows:  APA-CPA-APA-CPA-APA. This means that the central \dword{apa} will collect charge and see scintillation light from \lar volumes on both sides, whereas those by the wall collect from only one side.  
While the ARAPUCA modules deployed in \dword{pdsp} collect light from only one direction, several ARAPUCA configurations under development are capable of collecting light from both sides (including the X-ARAPUCA concept).
The optimal configuration of ARAPUCA modules has not yet been determined, but the basic design allows for both single-sided and double-sided cells with no impact on the \dword{apa} design.


\subsection{Photon Collector: Dip-Coated Light Guides}
\label{ssec:fdsp-pd-pc-bar1}


The dip-coated light guide design is mechanically the simplest of the three options. Figure~\ref{fig:pds-dippedbarpic} illustrates the process by which \lar scintillation photons are converted and detected in this approach.  \dword{vuv} scintillation photons incident on the bar are absorbed and wavelength-shifted to blue ($\sim$\SI{430}{nm}) by the \dword{tpb}-based coating on the surface of the bar.  A portion of this light is captured in the bar and guided to one end through total internal reflection where it is detected by an array of \dwords{sipm}, whose PDE is well-matched to the blue light.  Dimensions of the bars in \dword{pdsp} are: \SI{209.3}{cm}$\times$\SI{8.47}{cm}$\times$\SI{0.60}{cm}.
Since the bar is coated on all sides, in the \dword{spmod} it can be employed both in the wall \dwords{apa} as well as in the center \dword{apa} array where scintillation light approaches from two drift volumes.

\begin{dunefigure}[Schematic of scintillation light detection with dip-coated light guide bars.]{fig:pds-dippedbarpic}
{Schematic of scintillation light detection with dip-coated light guide bars.}
  \includegraphics[width=0.8\columnwidth]{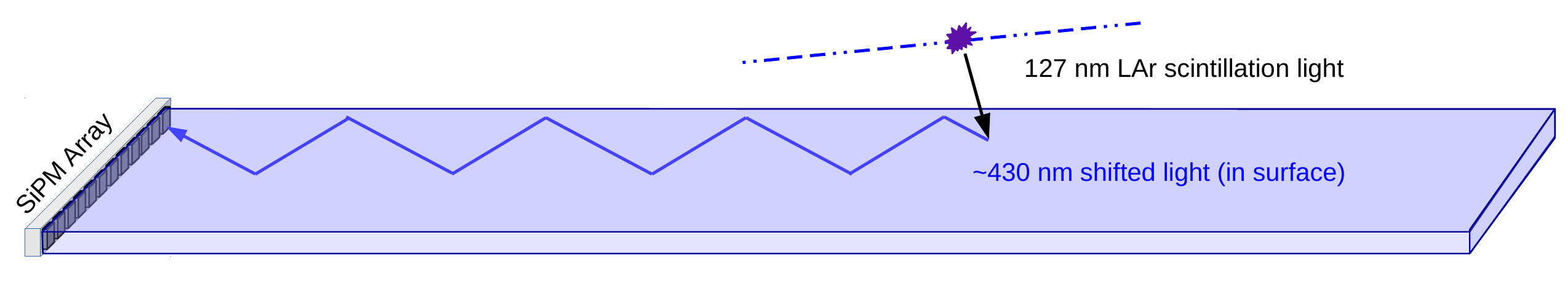}
\end{dunefigure}


The dipping process and coating formula have undergone a series of development iterations~\cite{Moss:2014ota},
with the bars undergoing extensive testing at both room and cryogenic temperatures.
As a part of the production process, the attenuation of each dip-coated light guide bar is measured at room temperature in a dark box with a UV LED; 80\% of the bars measured have an attenuation length in air of \SI{6}{m} or greater.   
Attenuation measurements on full-length bars have not yet been performed in \lar, but a model presented in ~\cite{Moss:2014ota} predicts effective attenuation lengths greater than \SI{2}{m}~\cite{Moss:2016yhb}.  The general features of this model were validated by measurements using $^{210}$Po alpha sources in the \dword{tallbo} cryogenic test stand at \dword{fnal}.

A further validation of the bar performance came from a set of measurements taken in the \dword{tallbo} cryostat containing the four initial candidate photon collector technologies using alpha sources and tracked cosmic ray muons, allowing side-by-side comparisons~\cite{Whittington:2015rkr}.  As a result, the two approaches that showed the highest promise, dip-coated and double-shift light guides (described in the next section), were continued for further development since these had similar photon detection efficiency, $\sim$0.1\%, that was significantly higher than the other two. 

 A simple improvement to the bar performance is to read out both ends of the bar rather than a single end as is the case for bars deployed in \dword{pdsp}.  
In addition, test bars have been produced with a higher \dword{tpb}-to-acrylic ratio, which may have a higher conversion efficiency without introducing a reduction in attenuation length. These improvements could increase the photon detection efficiency of the dip-coated light guide by more than a factor of two.

\subsection{Photon Collector: Double-Shift Light Guides}
\label{ssec:fdsp-pd-pc-bar2}

In the early implementations of the dip-coated light guide development there was a strong dependence of the light yield along the length of the bar, presumed to be due to the impact of the coating on the total internal reflection efficiency.  In an effort to mitigate this effect, the double-shift light guide design decouples the process of converting \dword{vuv} photons to optical
wavelengths from the transportation of photons along the bars. This is achieved by positioning an array of acrylic plates coated with \dword{tpb} in front of a high-quality commercial polystyrene light guide doped with a second wavelength-shifting compound.


Figure~\ref{fig:pds-doubleshiftlg-cartoon} illustrates the double-shift light guide concept. \dword{vuv} scintillation photons incident on the acrylic plates are converted to blue wavelengths ($\sim$\SI{430}{nm}) and a fraction of these blue photons penetrate the light guide and are converted to green ($\sim$\SI{490}{nm}). The re-emission of these green photons, taken to be a Lambertian distribution (isotropic luminance), leads to some becoming trapped by total internal reflection within the light guide and  transported to the end of the light guide where they are detected by an array of \dwords{sipm}.

\begin{dunefigure}[Schematic of double-shift light guide concept]{fig:pds-doubleshiftlg-cartoon}
{Schematic of the double-shift light guide concept.}
  \includegraphics[width=0.8\columnwidth]{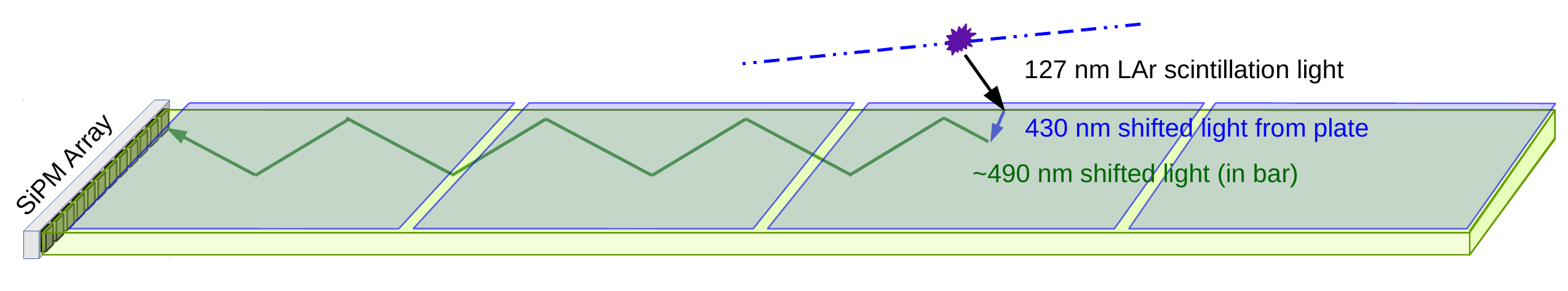}
\end{dunefigure}

%


The radiator plates are formed by spray-coating \dword{tpb} on the outer surface of acrylic plates, as described in Section~\ref{ssec:fdsp-pd-pc-prod-bar2}.
A full-scale double-shift light collector module consists of six radiator plates mounted on each face of the wavelength-shifting light guide (12 plates total). The \SI{210}{cm}$\times$\SI{8.6}{cm} light guide is fabricated by Eljen Technologies\footnote{http://www.eljentechnology.com} and consists of a polystyrene bar doped with the EJ-280 wavelength shifter. 
EJ--280 features an absorption spectrum that is well matched to the \dword{tpb} emission spectrum so wavelength-shifted photons emitted from the plates are absorbed with good efficiency. 

For most of the testing and the \dword{pdsp} modules,  the \dword{sipm} array used \SI{0.6}{cm}$\times$\SI{0.6}{cm} sensL C-series MicroFC-60035-SMT \dwords{sipm}. These were originally selected since they were a good match for the \dword{tpb} emission spectrum on the dip-coated bars. However, they have a photon detection efficiency between 20\%--35\% across the emission spectrum of the EJ-280 wavelength shifter, compared to up to 40\% at the peak, so selecting a different device with a better-matched photon detection efficiency would improve the  performance of the double-shift design.


The double-shift light guide design has undergone a series of development iterations to improve its performance, carried out at Indiana University (IU) and at Fermilab's cryogenic and vacuum test facility in the PAB. Comparative testing of light guide designs at PAB in mid-2015 demonstrated the viability of the double-shift light guide concept~\cite{bib:JINST-11-C05019}. 
An improved design similar to that deployed at \dword{pdsp} was studied at the Blanche test stand at Fermilab in September of 2016 with a complementary component-wise analysis program at IU afterward, as detailed in \cite{bib:DoubleShiftLG-NIM-171113}. The attenuation characteristics of this light guide were measured at IU, while the  detection efficiency for incident \lar scintillation photons was measured with a vacuum-ultraviolet (\dword{vuv}) monochromator at IU and using scintillation light from cosmic rays at the Blanche test stand.


Analysis of the double-shift light guide's attenuation properties determined an attenuation profile in \lar characterized by a double-exponential function of the form $f(z) = A \exp(-z/\lambda_{A}) + B \exp(-z/\lambda_B)$ with $z$ the distance from the instrumented end and parameters $A = $0.29, $\lambda_A = $4.3~cm, $B = $0.71, and $\lambda_B = $225~cm~\cite{bib:DoubleShiftLG-NIM-171113}. The effective attenuation length of \SI{2.25}{m} is comparable to the width of an \dword{apa} when the double-shift light guide is deployed in \lar.


Using both direct measurement with the monochromator and scintillation light, the photon detection efficiency of this detector was determined to be 0.48\% at the end close to the \dword{sipm} readout. The total effective area for detecting \dword{vuv} scintillation photons in this module can be determined by integrating the product of this efficiency and the attenuation function over the area of the detector, 
yielding an effective area $A_{eff}=4.1$.
This corresponds to an effective area for detecting \dword{vuv} scintillation photons of 4.1 cm$^{2}$ per module per drift volume, which corresponds to overall 0.23\% photon detection efficiency for events occurring on one side of the \dword{apa}.
Since the radiator plates are deployed on both faces of the light guide, modules in the center \dword{apa} array are sensitive to scintillation light from the two drift volumes on either side. 




There are several ways that the current design could be improved. The double-shift light guide deployed in the \dword{pdsp} \dwords{apa} is constrained to read out at a single end. Proposed changes to the \dword{apa} size and cabling routing scheme for the \dword{spmod} would allow for a second array of \dwords{sipm} at the opposite end of the light guide, which would almost double the performance of the photon detection system.
A \dword{sipm} with a wavelength-dependent PDE that is better matched to the EJ-280 emission spectrum would also improve the efficiency. Simulations of the transport of light within the light guide suggest that applying a highly reflective coating to the long, narrow inactive sides of the light guide would improve the attenuation function and increase the effective area of the light guide module. These effects combined lead to a potential increase of the effective area up to four times the current prototypes, approaching 1\% detection efficiency.




\subsection{Additional Techniques to Enhance Light Yield}
\label{sec:fdsp-pd-enh}


Though we anticipate that the designs described in the previous sections will meet the \dword{pd} performance requirements we do not yet have final designs and so we have also considered options for enhancing the light yield if that becomes necessary. Some of the initial ideas, such as deploying a large array of Winston-cone style reflectors focusing light onto \dwords{sipm} throughout the entire area enclosed by the \dword{apa} frame, would require a significant change in the \dword{apa} production and assembly planning and so will become increasing untenable.  However, one option being investigated in parallel with the photon collector modules design is to convert the scintillation light falling on the cathode plane into the visible wavelengths, which in turn illuminates  photon detectors embedded in the \dword{apa}, as is currently envisioned.

A motivation for this approach is that, due to geometric effects, the baseline PDS design will result in some non-uniformity of light collection along the drift direction. Light emitted from interactions close to the \dwords{apa} has an order of magnitude larger chance of being detected compared to interactions close to the \dword{cpa}. This effect can be mitigated by installing wavelength-shifter (\dword{tpb}) coated dielectric reflector foils on the cathode planes.
Light impinging on these foils is wavelength-shifted into visible wavelengths and reflected from the underlying foils. This light can subsequently be detected by photon detectors placed in the \dwords{apa} provided they are sensitive to visible light (which is not the case for the current photon collector modules). Fig. \ref{fig:ly_with_foils}, shows that if the \dword{apa} photon collectors are capable of recording both direct scintillation light and the visible light from the \dword{cpa}, there is an enhancement of the total light collection close to the cathode (black points), which will increase the detection efficiency in that region. 
Another benefit is the increase in uniformity - this can enable calorimetric reconstruction with scintillation light, which would enhance the charge-based energy reconstruction as well as increase the efficiency of triggering on low energy signals. Introducing the foils on the cathode may also enable drift position resolution  using only scintillation light. This requires the photon detectors to be able to differentiate direct \dword{vuv} light from re-emitted visible light (e.g. two different \dword{pd} detector types) and good enough timing of arrival of first light.

Coated reflector foils are manufactured through low-temperature evaporation of \dword{tpb} on dielectric reflectors e.g. 3M DM2000 or Vikuiti\texttrademark\  ESR. Foils prepared in this manner have been successfully used in dark matter detectors such as WArP\cite{Acciarri:2008kv}. Recently they have been shown to work in LArTPCs at neutrino energies, namely  in the LArIAT test-beam detector \cite{Garcia-Gamez:2017cmu}. In LArIAT they have been installed on the field-cage walls and, during the last run, on the cathode.  

The necessity to record both \dword{vuv} and visible photons in the photon collectors would require a change in the current design but is conceptually possible. For example, if the cathode plane were coated with tTP,  some of the ARAPUCA modules could be constructed without the pTP coating on the outer surface of the filter and benefit from the same photon trapping effect but these cells would no longer be sensitive to direct scintillator light.   
Understanding the impact of these competing effects on the physics is under study by the simulation group and the feasibility of coating the cathode with a dielectric medium is being investigated with the DUNE HV consortium.

\begin{dunefigure}[Predicted light yield with WLS-coated reflector foils on the \dword{cpa}.]{fig:ly_with_foils}
{Predicted light yield in with WLS-coated reflector foils on the \dword{cpa}. Blue points represent direct \dword{vuv} light impinging on the \dwords{pd} assuming a 0.42\% photon detection efficiency and 70\% wire mesh transmission; red stars - represent scintillation light that has been wavelength-shifted and reflected on the \dword{cpa} assuming the same photon detection efficiency. Black points show the sum of these two contributions (which would require twice the number of \dword{pd} modules in the current \dword{apa} configuration).}
\includegraphics[width=0.6\columnwidth]{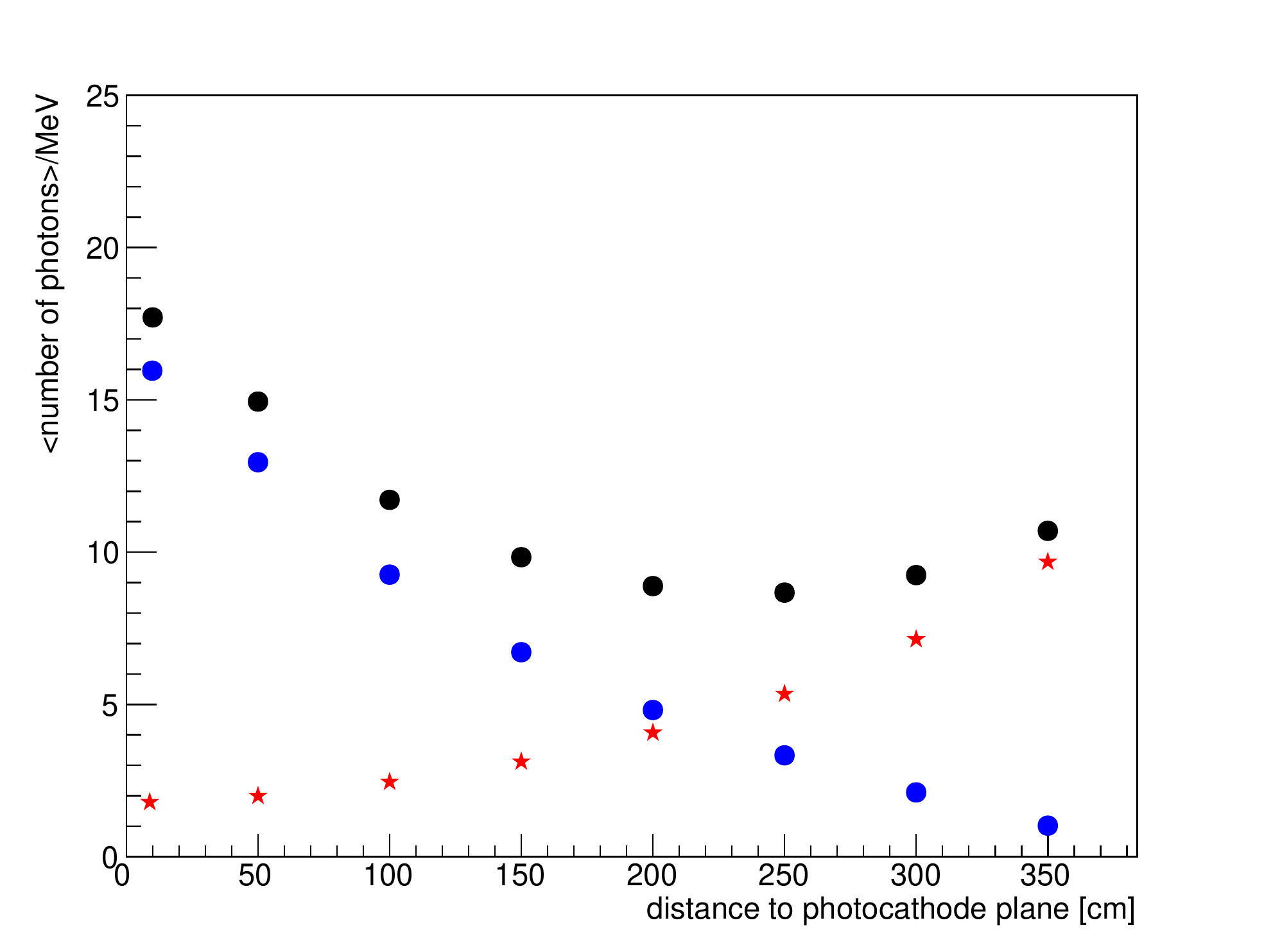}
\end{dunefigure}

\subsection{Silicon Photosensors}
\label{sec:fdsp-pd-ps}

The \dword{spmod} \dword{pds} uses a multi-step approach to scintillation light detection with final stage of conversion into electrical charge performed by silicon photomultipliers (\dword{sipm}). Robust photon detection efficiency, low operating voltages, small size and ruggedness make their use attractive in the \single design where the photon detectors must be accommodated inside the \dword{apa} frames. 
As implemented in \dword{pdsp}, there are twelve \num{6}$\times$\SI{6}{mm$^2$} \dwords{sipm} per bar and \numrange{6}{12} per ARAPUCA box.
With this configuration, a \nominalmodsize \dword{spmod} with \num{150} \dwords{apa}, each with \num{10} \dword{pd} modules, would contain \num{18000}-\num{36000} (single or double-ended readout) \dwords{sipm} for the light guide designs and 10-20 times more for the higher granularity ARAPUCA design. This corresponds to approximately \num{1}-\SI{13}{m$^2$} of active \dword{sipm} surface area.

The following summarizes the most salient guiding principles and requirements for this \dword{sipm}-based photodetection system.

\begin{itemize}

\item  The full suite of \dword{sipm} requirements (number of devices, spectral sensitivity, dynamic range, triggering, 
zero-suppression threshold etc.) is determined by the physics goals and the photon collection implementation.  As discussed in Section~\ref{sec:fdsp-pd-intro},  the requirements for \dword{snb} neutrinos are not yet fully established however, 
R\&D carried out to date indicates that devices from several vendors have the 
performance characteristics close to that needed for the \dword{pds} (see Table~\ref{tab:photosensors}). 
Nearly one thousand of several types of these devices are used in the \dword{pdsp} \dword{pd}\footnote{\dword{pdsp} \dword{pds} uses 516 sensL MicroFC-60035C-SMT, 288 Hamamatsu MPPC 13360-6050CQ-SMD with cryogenic packaging, 180 Hamamatsu MPPC 13360-6050VE.}, which will provide an excellent test bed for evaluating and monitoring \dword{sipm}
performance in a realistic environment over a period of months.

\item A key requirement is to ensure the mechanical and electrical integrity of these devices in a cryogenic environment. However, currently, catalogue devices for most vendors are certified for operation only down to \num{-40}$^\circ$C. It is essential to be in close communication with  vendors in the design, fabrication and \dword{sipm} packaging certification stages to ensure that the device will be robust and reliable for long-term operation in a cryogenic environment. 
Two sources have expressed interest to engage with the consortium in this fashion with the goal of having the vendor warranty the product for our application: Hamamatsu Photonics K.K., a large well-known commercial vendor in Japan and Fondazione Bruno Kessler (FBK) in Italy, an experienced developer of solid state photosensors that typically licenses its technology and which is partnering with the DarkSide collaboration to develop a devices with very similar requirements as DUNE. 
Contact with other vendors and experiments using this technology in a similar environment is being pursued. 

\item Comparative performance evaluation of promising \dword{sipm} candidates from
multiple vendors will need to be carried out in parallel over the next year. This evaluation will need to
address inherent device characteristics (gain, dark rate, x-talk, after-pulsing etc), which are common to all three photon collector options, along with ganging performance, form factor, spectral response, and mechanical mounting options that may have different optimization for the two light guide design and ARAPUCA.
Experience acquired from \dword{pdsp} construction and operation will inform QA/QC plans for the full detector, which will need to be delineated in detail.

\item The optimal \dword{sipm} may depend on the photon collector option selected.  All 
options currently being considered involve shifting the \SI{127}{nm} \lar scintillation light to 
longer wavelengths, but each may present a different  spectral distribution to the \dword{sipm}. 
In this case, final selection of the \dword{sipm} might be delayed to allow an optimal match to the photon collector. 
However, we would not expect this fine-tuning to be more than a 15-20\% effect, so it is not a driving factor.

\item For the light guide photon collector designs, the \dword{sipm} packaging should allow for tileable arrays to be constructed to facilitate high packing efficiency across the end of the bars and efficient space utilization inside the \dword{apa} frame. 

\item Current candidate \dwords{sipm} have an area of less than \SI{1}{cm$^2$}, a much finer granularity than needed. In addition, the cold feedthrough size and space in the \dwords{apa} for cable runs limits  the number of \dword{pd} signal and power cables. These constraints, and other considerations, imply that the signal output of \dwords{sipm} must be electrically ganged. The degree of ganging depends on the photon collectors technology and currently ranges from six \dwords{sipm} for the light guides to \num{48} or more for the ARAPUCA modules. Whether simple passive-ganging (wiring the outputs together) will suffice or if active-ganging (with active components) is under investigation as a joint responsibility of the photon sensor and electronics working groups (see Section~\ref{sec:fdsp-pd-elec-intro} for more details).

\item The terminal capacitance of the sensors strongly affects the signal-to-noise when devices are ganged in parallel and so is a factor in \dword{sipm} selection, and may ultimately determine the maximum number of individual sensors that can be ganged this way. 

\end{itemize}

\begin{dunetable}[Candidate photosensors characteristics.]
{p{0.18\textwidth}p{0.18\textwidth}p{0.18\textwidth}p{0.18\textwidth}p{0.18\textwidth}}
{tab:photosensors}
{Candidate Photosensors Characteristics.}
	                         &Hamamatsu                           & sensL                                 & KETEK                       & Advansid                    \\ \toprowrule
Series part \#            & S13360                                 & DS-MicroC                         & PM33                          & NUV-\dwords{sipm}                \\ \colhline
Vbr range                 & 48 V to 58 V                           & 24.2 V to 24.7 V                & 27.5 V                         & 24 V to 28 V               \\ \colhline
Vop range                 & Vbr + 3 V                               & Vbr +1 V to +3 V               & Vbr+2V to +5 V           & Vbr +2 V to +6 V         \\ \colhline
Temp. dependence   & 54 mV/K                                & 21.5 mV/K                         & 22 mV/K                      & 26 mV/K                      \\ \colhline
Gain                           & $1.7 \times 10^6$                  & $3 \times 10^6$                & $1.74 \times 10^6$      & $3.6 \times 10^6$       \\ \colhline
Pixel size                   & 50 $\mu$m                            & 10 $\mu$m to 50 $\mu$m & 15 $\mu$m to 25 $\mu$m     & 40 $\mu$m       \\ \colhline
Sizes                          & 2x2 mm                                 & 1x1                                    & 3x3                               & 4x4                             \\ \colhline
                                  & 3x3 mm                                 & 3x3                                    &                                      & 3x3                             \\ \colhline
                                  & 6x6 mm                                 & 6x6                                    &                                      &                                    \\ \colhline
Wavelength                & 320 to 900 nm                       & 300 to 950 nm                  & 300 to 950 nm              & 350 to 900 nm            \\ \colhline
PDE peak wavelength  & 450 nm                               & 420 nm                             & 430 nm                         & 420 nm                       \\ \colhline
PDE @ peak              & 40\%                                     & 24\% to 41\%                    & 41\% at Vov=5 V          & 43\%                           \\ \colhline
DCR @0.5PE             & 2 to 6 MHz                            & 0.3 kHz to 1.2 MHz          & 100 kHz  at Vovr=5 V   & 100 kHz/mm$^2$       \\ \colhline
Crosstalk                    &< 3\%					& 7\%                                  & 15\%                             &  < 4\% (correlated noise) \\ \colhline
Afterpulsing                &                                               & 0.20\%                             & \textless 1\%                 & \textless4\%               \\ \colhline
Terminal capacitance & 1300 pF                                 & 3400 pF                           & 750 pF                          &800 pF                         \\ \colhline
Lab experience          & Good experiences from Mu2e and ARAPUCA & Crack at LN2 temps. after specifications change&         &     \\         
\end{dunetable}

\subsection{Electronics}
\label{sec:fdsp-pd-pde}

\subsubsection{Introduction}
\label{sec:fdsp-pd-elec-intro}

The \dword{pd} design requires the readout system to collect and process electrical signals from photosensors reading out the light collector bars, 
to provide interface with trigger and timing systems to support data reduction and classification, and to enable data transfer 
to offline storage for physics analysis.

The readout system must enable the measurement of the t$_0$ of non-beam events with deposited energy above \SI{10}{MeV}. 
This capability will also enhance beam physics, by recording interaction time of events within 
beam spill to help separate against potential cosmic background interactions. Two main methods of data collection are currently considered:  self-triggered integrated charge readout and wave form digitization.  Charge integration appears to be a likely candidate at this point in our development, as it offers the potential for a simpler, commercially available charge integration circuit and perhaps a smaller, less-expensive cable plant to read it out.  Physics simulation studies are currently underway to determine if pulse-shape discrimination will be required, which would provide the capability to record both prompt and delayed components of scintillation light (characteristic times of \SI{6}{ns} and \SI{1.3}{$\mu$s}), the latter consisting mostly of single photoelectrons and thus place stringent requirements on signal-to-noise performance. The photon detector collects a limited amount of light, so it could be beneficial to collect the light from both excited states. 
Since this requirement has not yet been established the option is kept open in the electronics design.

All photon collector options require some level of electrical ganging of the \dwords{sipm}, either passive direct connection of the \dword{sipm} outputs or active (cold signal summing and possibly amplification).  To that end we desire a system where the ganging is maximized to minimize the electronics channel count while maintaining adequate redundancy and granularity, as well as readout system performance.  This represents a significant interface between the electronics, photosensor and light collector designs, and will be a main focus of our development and optimization work up to the \dword{tdr}.

Technical factors that affect performance of the ganging system are the characteristic capacitance of the \dword{sipm} 
and the number of \dwords{sipm} connected together, which together dictate the signal to noise ratio and affect the system 
performance and design considerations. Selection of the ganging option will include passive or active solutions, where the active 
circuitry may require cold components such as an amplifier in the \lar volume. Design options with active cold components will need 
to address issues of power dissipation and potential risks of single-point failures of multi-channel devices inside the cryostat.
In the case of passive ganging, analog signals are transmitted outside of the cryostat for processing and digitalization. 
Successful demonstrations of passive ganging at \lar temperatures have been made for groups of four and twelve 6x6 mm Micro-FC-60035C-SMT C series, and groups of 2, 4, 8, and 12  Hamamatsu MPPCs (S13360-6050PE) at  \num{25}$^\circ$C, - \num{70}$^\circ$C and \SI{77}{K}. Active ganging has been demonstrated for an array of 12 sensL \num{4}$\times$\num{4} arrays of \SI{3}{mm}$\times$\SI{3}{mm} sensL C-series \dwords{sipm} (48 in all) and  72 \dwords{sipm} mounted in a hybrid combination of passive and active ganging using \SI{6}{mm}$\times$\SI{6}{mm} MPPCs with a low noise operational amplifier--this design combines 12 active branches into the op-amp, where each branch has six MPPCs in a parallel passive-ganging configuration.

Typically, arrival time and total charge are the key parameters to be obtained from a detector. Extraction of these parameters 
is possible using analog or digital systems. Charge preamplifiers will be connected to the output of the detector to integrate 
current producing a charge proportional output. In the case of digital systems an amplifier is needed to adjust the detector output signal 
level to the input of an \dword{adc}.  In both systems, performance parameters related  to sampling rate, number of bits, 
power requirements, signal to noise ratio, and interface requirements should be evaluated to arrive to selected solution.  
Pulse shapes can be fully analyzed to improve detection of a new physics but it will have an important impact on the digitalization frequency.



\subsubsection{ProtoDUNE-SP Electronics}


A dedicated photon-detector readout system, presented schematically  in Figure~\ref{fig:fig-pds-readout}(left), was developed for \dword{pdsp}, which will be operational in the second half of \num{2018}.  Twenty-four custom \dword{sipm} Signal Processor (\dword{ssp}) units were produced to read out the 58 light guide and 2 ARAPUCAs photon collectors.  
An \dword{ssp} contains of twelve readout channels packaged in  a self-contained 1U module; four \dwords{ssp} are shown in Figure~\ref{fig:fig-pds-readout}(right).

A passive ganging scheme with three \dwords{sipm} ganged together was chosen for the light guides (4 \dword{ssp} channels for each bar) and groups of twelve \dwords{sipm} are passively ganged for the two ARAPUCA modules (12 \dword{ssp} channels per module).  The unamplified analog signals from the \dwords{sipm} are transmitted to outside the cryostat for processing and digitization over an approximately \SI{25}{m} cable to the \dword{ssp} outside the cryostat. 
 Each channel receives the \dword{sipm} signal into a termination resistor that matches the characteristic impedance of the signal cable followed by a fully-differential voltage amplifier and a \num{14}-bit, \num{150}-MSPS \dword{adc} that digitizes the \dword{sipm} signal waveforms. 
  

 \begin{dunefigure}[\dword{pdsp} \dword{pd} module readout.]
 {fig:fig-pds-readout}
 {Block diagram of the \dword{pdsp} \dword{pd}  readout module (left). \Dword{pd}  readout system operational at \dword{pdsp} (right). }
\includegraphics[angle=0,width=8.4cm,height=6cm]{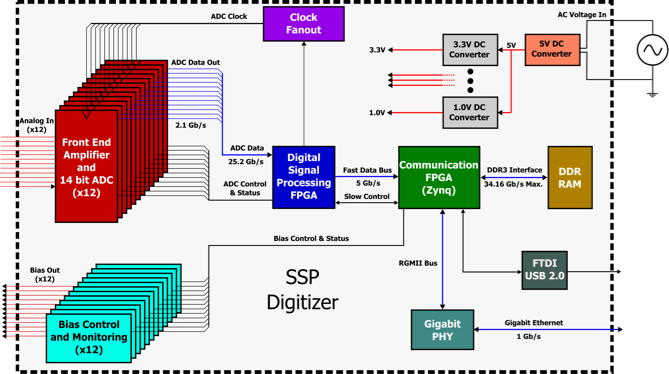}\includegraphics[angle=0,width=8.4cm,height=6cm]{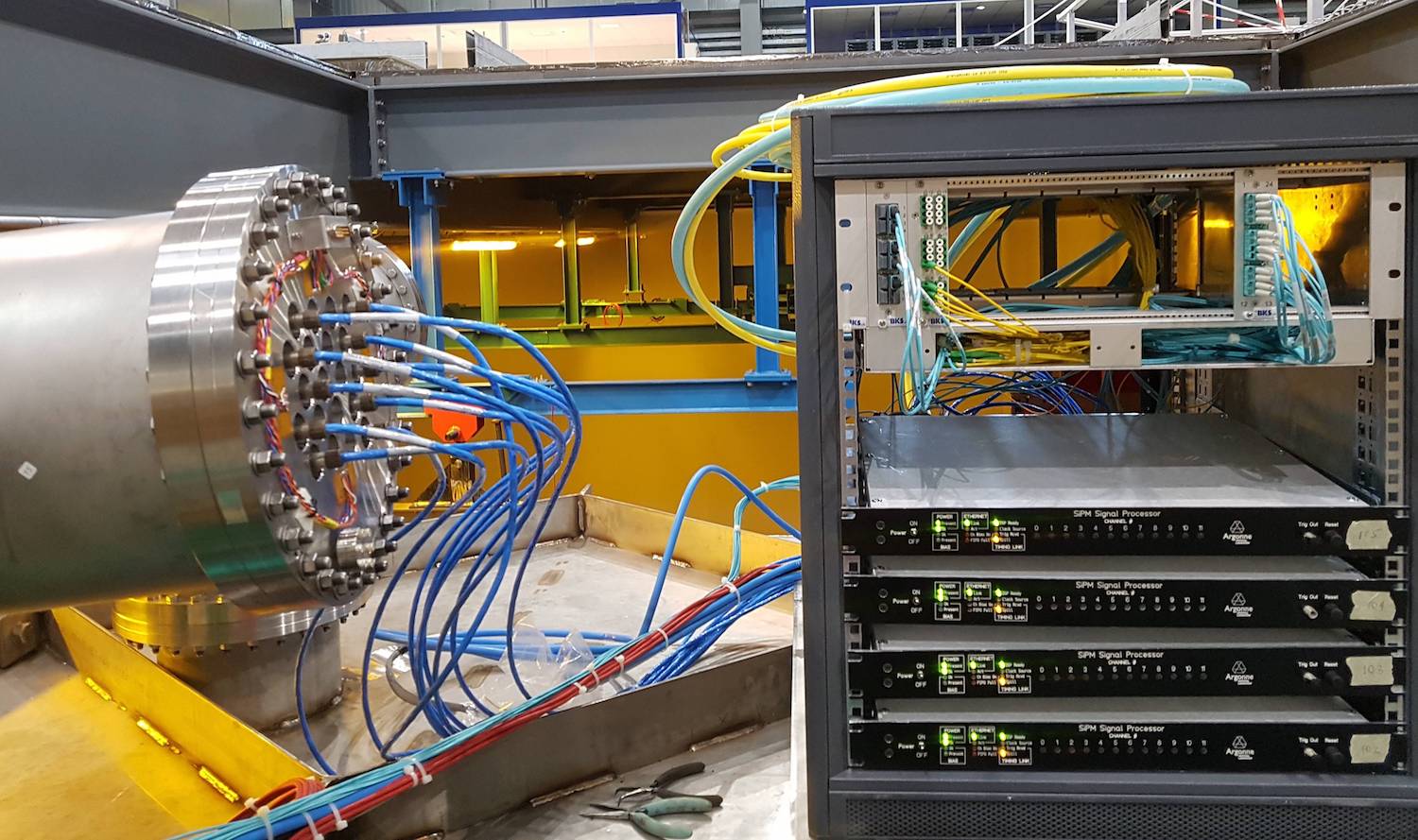}
\end{dunefigure}

In the standard mode of operation, the module performs waveform capture, using either an external or internal trigger. In the latter case the 
module self-triggers to capture only waveforms with an amplitude greater than a specified threshold. In \dword{pdsp} the photon readout 
is configured to read waveforms when triggered by a beam event, or to provide header information when self-triggered by cosmic muons.
The header portion summarizes pulse amplitude, integrated charge, and time-stamp information of events. The \dword{ssp} for \dword{pdsp} uses \si{Gb} Ethernet 
communication implemented over an optical interface. The \SI{1}{Gb/s} Ethernet supports full TCP/IP protocol.  

The module includes a separate \num{12}-bit high-voltage DAC for each channel to provide bias to each \dword{sipm}. Currently there are two DAC options: one with a voltage range of \SIrange{0}{30}{V}, used with the sensL \dwords{sipm} (\num{17} of the \num{24} \dword{ssp} units); and the other with a range \SIrange{0}{60}{V} for use with the Hamamatsu MPPCs (seven of the 24 \dword{ssp} units). The \dword{ssp} provides a trigger output signal from internal discriminators in firmware based on programmable coincidence logic, with a standard ST fiber interface to the central trigger board (CTB).
Input signals are provided to CTB from the beam instrumentation, the \dwords{ssp}, and the beam TOF system. The CTB receives timing information from 
the \dword{pdsp} timing system and the CTB trigger inputs are distributed to the experiment via the timing system.
To that end, the \dword{ssp} implements the timing receiver/transmitter endpoint hardware to receive trigger inputs and clock signals from the timing system.

\subsubsection{Electronics Next Steps}

Although the requirements for the electronics system are not all fully established, it not expected that the system requires novel high-risk techniques and can be developed and fabricated well within the schedule for the \dword{pds}.
In the latter half of CY18, \dword{pdsp} test beam and cosmic-ray muon data analysis will provide evaluation of the readout system implemented in \dword{pdsp} and the \dword{pd} Photon Sensor and Simulation groups will provide essential guidance on optimization of performance and cost.

As identified in Section~\ref{sec:fdsp-pd-ps}, the most important near term R\&D program will be to optimize the ganging scheme including choice of \dword{sipm} and cable types. 
The first objective is to demonstrate that an ensemble of \numrange{48}{72} Hamamatsu \SI{6}{mm}$\times$\SI{6}{mm} MPPCs can be summed into a single channel by a combination of passive and active ganging. This board will also measure the photoelectron collection efficiency when the \dwords{sipm} are coated with \dword{tpb} as a reference for ARAPUCA measurements with a similar ganging level (the summing  board is the same size as the \dword{pdsp} ARAPUCA backplane to facilitate the comparisons).
Charge processing requires a charge preamplifier ideally located within the cold environment, so the design must take into consideration the failure risks and the power dissipated into the environment.

The timing resolution, minimum threshold and dynamic range requirements for the system are dictated by the physics requirements. These are well known for the higher energy physics (>\SI{200}{MeV}) but, as noted elsewhere in this document, are still evolving for lower energy. Currently, 
a timing resolution of 1$\mu$s is called for and the sampling rate and number of sample bits is estimated based on this. For this task 
some digital process such as a sample interpolation may be proposed, enhancing the recorded raw sample time precision.
The light sensitivity and the dynamic range requirement will determine the number of bits and the sample rate required by either waveform or charge collection methods. In both cases, the signal to noise ratio and the power consumption must be estimated.  
With this data from \dword{pdsp} and the ganging studies, the choice between waveform readout and integrated charge readout will be made taking into account DAQ  readout and trigger requirements.


\section{Production and Assembly}
\label{sec:fdsp-pd-prod-assy}


\subsection{Photon Collector Production}
\label{sec:fdsp-pd-prod-pc}

\subsubsection{ARAPUCA}
\label{ssec:fdsp-pd-pc-prod-arapuca}
Although the individual cell dimensions may differ, the basic design of the ARAPUCA-based \dword{pd} modules for the first  \dword{spmod} is similar to that of the two prototypes produced for \dword{pdsp}. Here we describe the production and assembly envisaged based on that experience.

Each \dword{pdsp} ARAPUCA module is shaped as a bar with external dimensions of 
\SI{207.3}{cm}$ \times$ \SI{9.6}{cm} $\times$ \SI{1.46}{cm}, which allows for it to be inserted between the wire planes through 10 slots the \dword{apa}. The module currently contains sixteen basic ARAPUCA cells, each one with an optical window with an area of \SI{7.8}{cm} $\times$ \SI{9.8}{cm} and internal dimensions of approximately \SI{8}{cm} $\times$ \SI{10}{cm} $\times$ \SI{0.6}{cm}. The \dwords{sipm} are mounted on the backplane of the cell, which in \dword{pdsp}, allows for  two different configurations of either 12 or 6 passively-ganged \dwords{sipm}.

\begin{dunefigure}[\dword{pdsp} ARAPUCA modules during assembly.]{fig:arap-prod01}
{\dword{pdsp} ARAPUCA modules during assembly prior to installation of the dichroic filters; \dwords{sipm} and TPB coated reflector are visible.}
  \includegraphics[height=8cm]{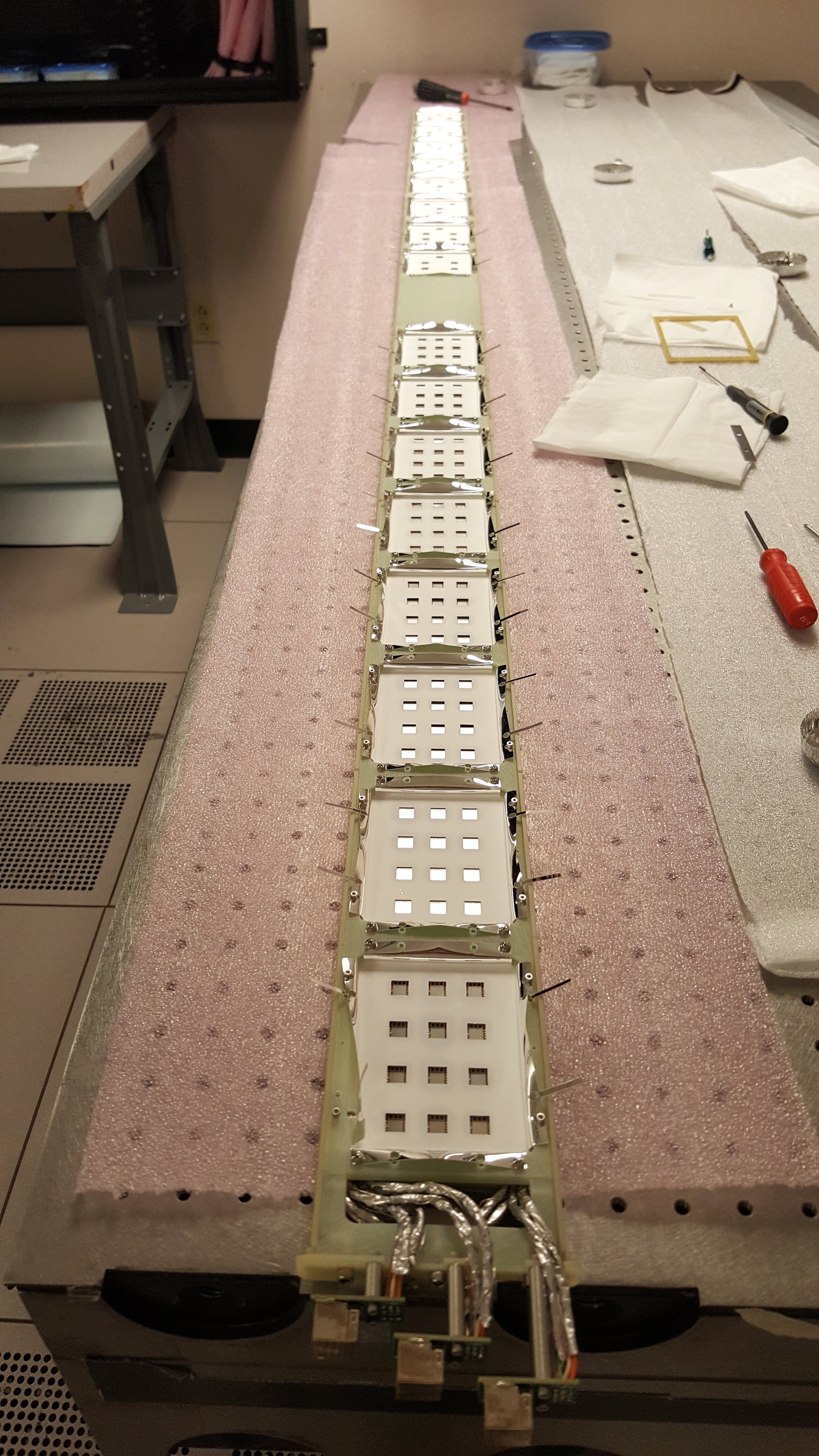}
\end{dunefigure}

The internal surface of the box is lined with a dielectric mirror foil\footnote{3M Vikuiti\texttrademark\  ESR - http://multimedia.3m.com/mws/media/193294O/vikuiti-tm-esr-application-guidelines.pdf} laser cut with openings at the locations of the \dwords{sipm}; these are visible in Figure~\ref{fig:arap-prod01}, which shows an ARAPUCA during assembly prior to installation of the optical windows. The backplane \dword{sipm} boards for the \dword{pdsp} modules were designed at CSU and produced by an external USA vendor\footnote{Advanced Circuits Inc.; www.4pcb.com.}; the \dwords{sipm} were soldered on the boards using a reflow oven at CSU. Before mounting into the ARAPUCA module they were tested at room and LN2 temperatures. It is anticipated that the production of the boards for  \dwords{spmod} will be done outside the USA. 

The optical window of each box is a dichroic filter with cut-off at \SI{400}{nm}. While the filters used for the \dword{pdsp} prototypes have been acquired from Omega Optical Inc.\footnote{http://www.omegafilters.com/}, other vendors are being considered for the DUNE production\footnote{ASHAI -Japan, Andover-USA, Edmunds Optics-USA}.
Prior to coating, the filters are cleaned according to the procedures given by the manufacturer using isopropyl alcohol. Since the most likely vector for scratching/damaging the coating is dragging contaminated wipes across the surface, new clean lint free wipes are used for each  cleaning pass on the surface. Clean filters are then baked at 100$^\circ$C for \SI{12}{hours}. The Vikuiti\texttrademark\  foils do not need to be cleaned and baked since they have a protective film that is removed just before the evaporation.
   
The filters are coated on the external side facing the \lar active volume with pTP,
while the internal dielectric mirror side is coated with TPB.
The coatings for the \dword{pdsp} modules have been made at the Thin Film facility at Fermilab using a vacuum evaporator. Each coated filter was dipped in LN2 to check the stability of the evaporated coating at cryogenic temperature. For \dword{pdsp} \dword{pd} production the evaporation process will be performed at UNICAMP in Brazil, where a large vacuum evaporator with an internal diameter of one meter is now available. The  conversion efficiency of the film deposited on the filters or on the Vikuiti\texttrademark\   foils will be measured with a dedicated set-up that will use the \SI{127}{nm} light produced by a VUV monochromator.

\subsubsection{Dip-Coated Light Guides}
\label{ssec:fdsp-pd-pc-prod-bar1}

To produce the full-size \dword{pdsp}  dip-coated light guide bars, the production methods initially developed at MIT for \SI{50.8}{cm} (\SI{20}{in}) dip-coated light guide bars were scaled up for a facility at \dword{fnal}. For \dword{pdsp} production four steps will remain essentially the same:

\begin{enumerate}
\item Cut and polished UVT acrylic bars are annealed at $180^{\circ}$F in a temperature-controlled oven to prevent subsequent crazing.
\item The TPB-based coating mixture is prepared in a fume hood and poured into an upright vessel located inside a larger enclosed volume.
\item A mechanized system dips the annealed bars into the coating solution where they soak and are then hung to dry in a low humidity environment established through a dry nitrogen purge of the enclosed volume.
\item The coated acrylic bars are placed in a dark box and their attenuation length in air is measured with a UV LED that is scanned along the length of the bar.
\end{enumerate}

The coating solution consists of four components in the following ratios: 100 mL 99.9\% pure toluene; 25 mL 200 proof ethanol; 0.2 g UVT acrylic pellets; 0.2 g scintillation grade TPB. The TPB and UVT acrylic pellets are first dissolved in a flask filled with toluene and mixed overnight with a teflon-coated magnetic stir bar.  Then the ethanol is mixed into the coating solution before it is poured into the dipping vessel.  This will produce an optically transparent, TPB-embedded coating, which adheres well to the surface of the bar and has a smooth surface.

A picture of the oven used for annealing, the fume hood used for mixing the coating, the vessel used for dipping, and the dark box used for measuring attenuation lengths for the production of dip-coated light guides for \dword{pdsp} is shown in Figure~\ref{fig:dipprod}.  These same production methods would also be used for the  \dwords{spmod}, 
but scaled up to dip and scan multiple bars at the same time.  Additionally, multiple production sites would be built to produce dip-coat light guide bars for the  \dwords{spmod}.

\begin{dunefigure}[\dword{pdsp} dip-coated light guide bars production.]{fig:dipprod}
{\dword{pdsp} dip-coated light guide bars production: annealing oven (left); dark box (right)}
  \includegraphics[height=7cm]{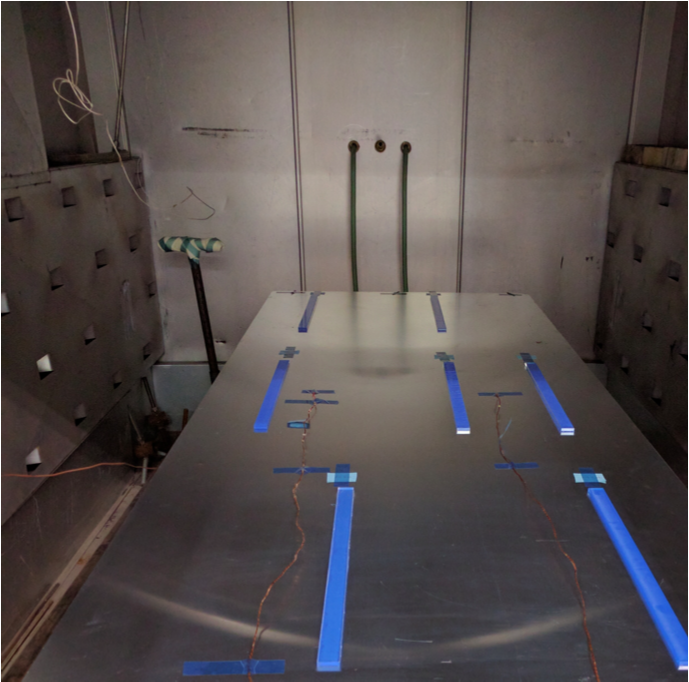}
  \includegraphics[height=7cm]{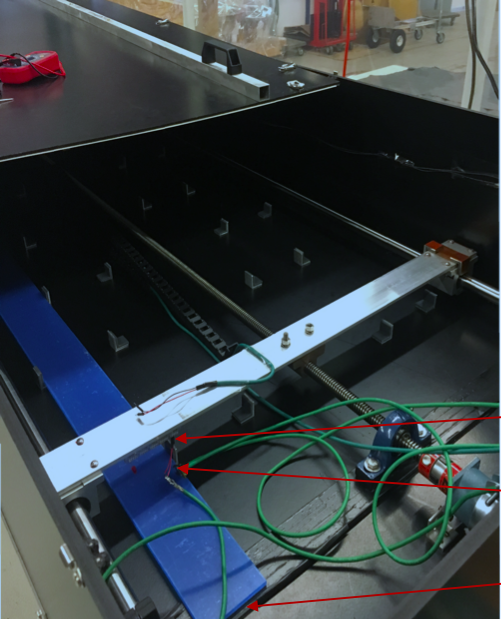}
\end{dunefigure}


\subsubsection{Double-Shift Light Guides}
\label{ssec:fdsp-pd-pc-prod-bar2}

The production and assembly of the double-shift light guide modules has two main components; the wavelength-shifting plates and the EJ-280 light guides. Many of the production, quality assurance, and assembly procedures developed for the double-shift light guide design deployed at \dword{pdsp} would remain the same for the  \dwords{spmod}.
													
\paragraph*{WLS Plates}

Sheets of \SI{0.166}{cm} (\SI{0.0625}{in}) thick UVT acrylic purchased from McMaster-Carr\footnote{\url{https://www.mcmaster.com}.} are laser-cut into \SI{77}{cm}$\times$\SI{9}{cm} templates with two \SI{34.2}{cm}$\times$\SI{8.6}{cm} plates per template (Figure~\ref{fig:pds-doubledhiftlg-plate}~(top)). Each template also includes three small \SI{3.81}{cm}$\times$\SI{2.54}{cm} pop-out tabs on either side of and between the two plates. After the acrylic templates are coated with TPB these tabs are separated from the plate and tested in a VUV monochromator to determine the quality of the coating on the two associated plates.

\begin{dunefigure}[Dip-coated acrylic plates.]
{fig:pds-doubledhiftlg-plate}
{A laser-cut acrylic template holding two plates and three test tabs (top). TPB-coated acrylic plates after spraying during fabrication of parts for \dword{pdsp} (bottom).}
    \includegraphics[width=0.7\columnwidth]{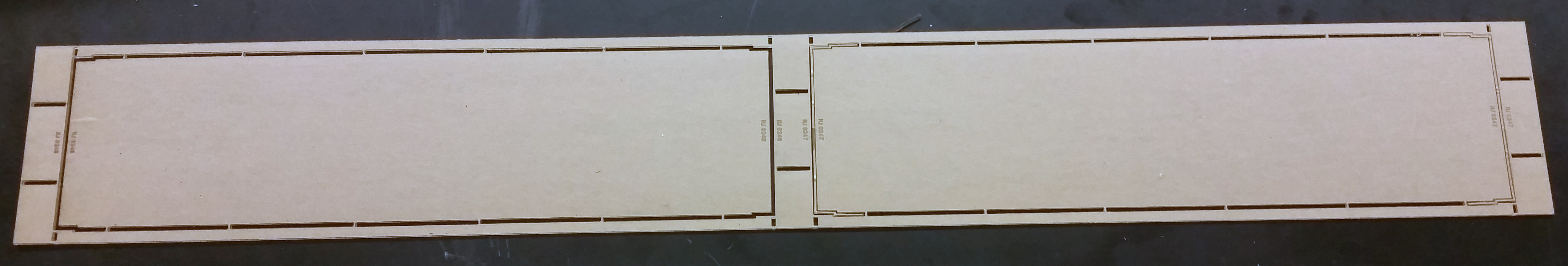}\\
    \vspace{0.3cm}
    \includegraphics[width=0.7\columnwidth]{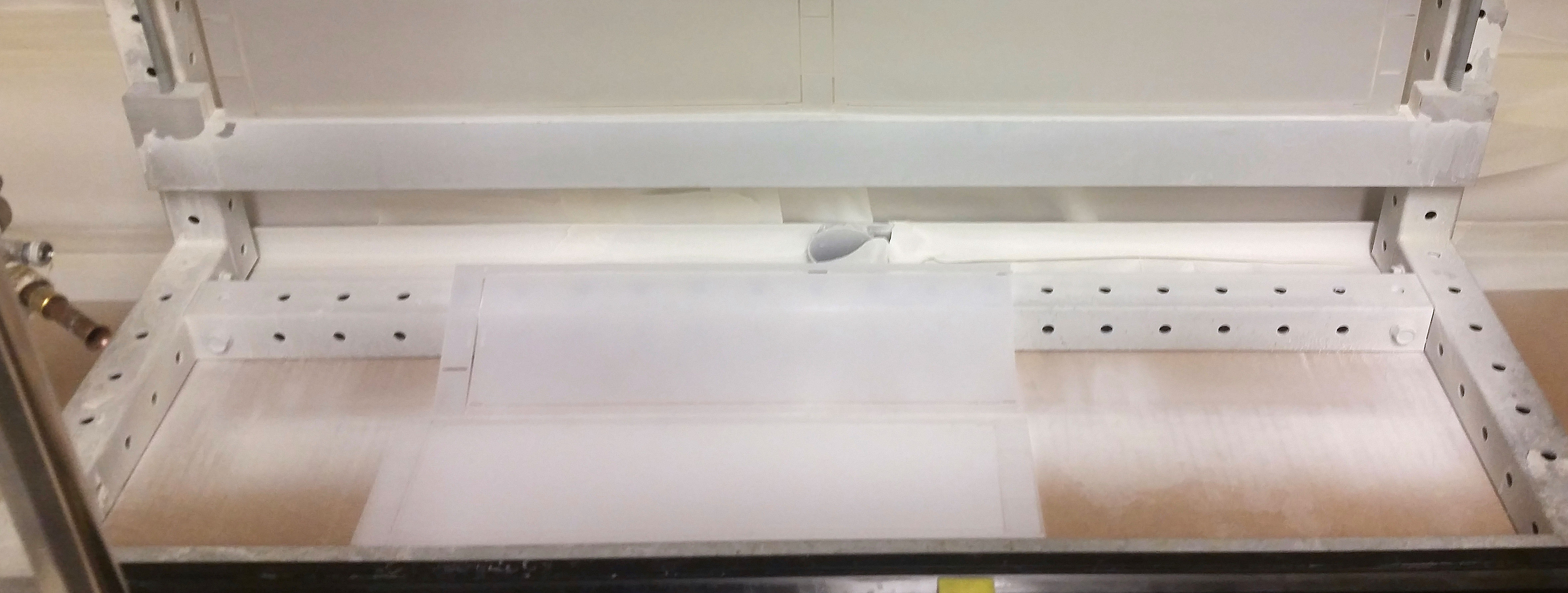}
\end{dunefigure}

Scintillation grade ($\ge 99$\%) TPB is dissolved in dichloromethane (DCM) at a ratio of \SI{5}{g} TPB per \SI{1000}{g} DCM. The solution is applied to the templates using a high-volume low-pressure (HVLP) sprayer system under a fume hood. The relatively small number of plates manufactured for \dword{pdsp} were sprayed by a technician to approximate an established standard coating thickness measured to have an acceptably high VUV photon conversion efficiency. Figure~\ref{fig:pds-doubledhiftlg-plate}(bottom) shows the HVLP spray-coating mount with a coated acrylic template. Two plates and three test tabs can be seen in the HVLP mounting frame. A second sprayed template has been broken at one of the midpoint cuts and positioned in the photo. For \dword{pdsp} production, the spray-coating process will be automated or commercialized to accommodate the large-scale production necessary for the  \dwords{spmod}.

After spraying, the acrylic templates are baked in a vacuum oven at 80$^{\circ}$C overnight, just below the glass transition point of acrylic. The softened acrylic partially absorbs the TPB into the surface, better affixing the wavelength-shifting coating. Uneven heating during the baking process described above can deform the coated plates, but this is minimized by careful oven fixturing to ensure even heating. After baking, the dimensions of the samples are measured and only plates within the production tolerance are accepted for further testing.
To ensure adequate and uniform performance of the coated plates, the conversion efficiency is tested using a VUV monochromator. During \dword{pdsp} production, plates were fabricated and tested 
using a McPherson\footnote{\url{http://mcphersoninc.com}.} VUV monochromator with deuterium lamp source to study performance at \SI{127}{nm}. The full plates were too large for the sample chamber
VUV monochromator system, so the testing tabs on either side of each plate were used to constrain the plate's performance. 
Only plates that exhibit a relative efficiency above an acceptance threshold are shipped to the assembly facility for deployment along a light guide. For \dword{pdsp}, a threshold was chosen to accept plates that were comparable or superior to those studied at the Blanche test stand~\cite{bib:DoubleShiftLG-NIM-171113} described previously.

\paragraph*{WLS-Doped Light Guides}

The EJ-280 light guides are fabricated and cut to length by Eljen Technologies. Upon receipt, each light guide is unpacked, visually inspected for defects, checked for dimensional tolerance, and scanned using a \SI{430}{nm} LED to determine its attenuation length in air (Figure~\ref{fig:pds-doubleshiftlg-ej280}). 
Since the index of refraction for $\sim$\SI{500}{nm} light in \lar is larger than in air and the critical angle for trapping by total internal reflection is correspondingly lower, so attenuation scans of sample light guides were made in both air and \lar (using a movable Am-241 $\alpha$ source) to quantify the correlation between measurements in air and in \lar.  Attenuation lengths longer than $\sim$5~meters measured in the darkbox correspond to attenuation lengths in \lar longer than $\sim$2~meters. An acceptance threshold of 5~meters measured at both ends of an EJ-280 light guide in the darkbox ensures adequate attenuation performance for the modules deployed in the \dwords{spmod}.

\begin{dunefigure}[EJ-280 light guide in darkbox for attenuation scan QA.]{fig:pds-doubleshiftlg-ej280}
{EJ-280 light guide in a dark box for attenuation scan QA at Indiana University (prepared for \dword{pdsp}).}
  \includegraphics[width=0.6\columnwidth]{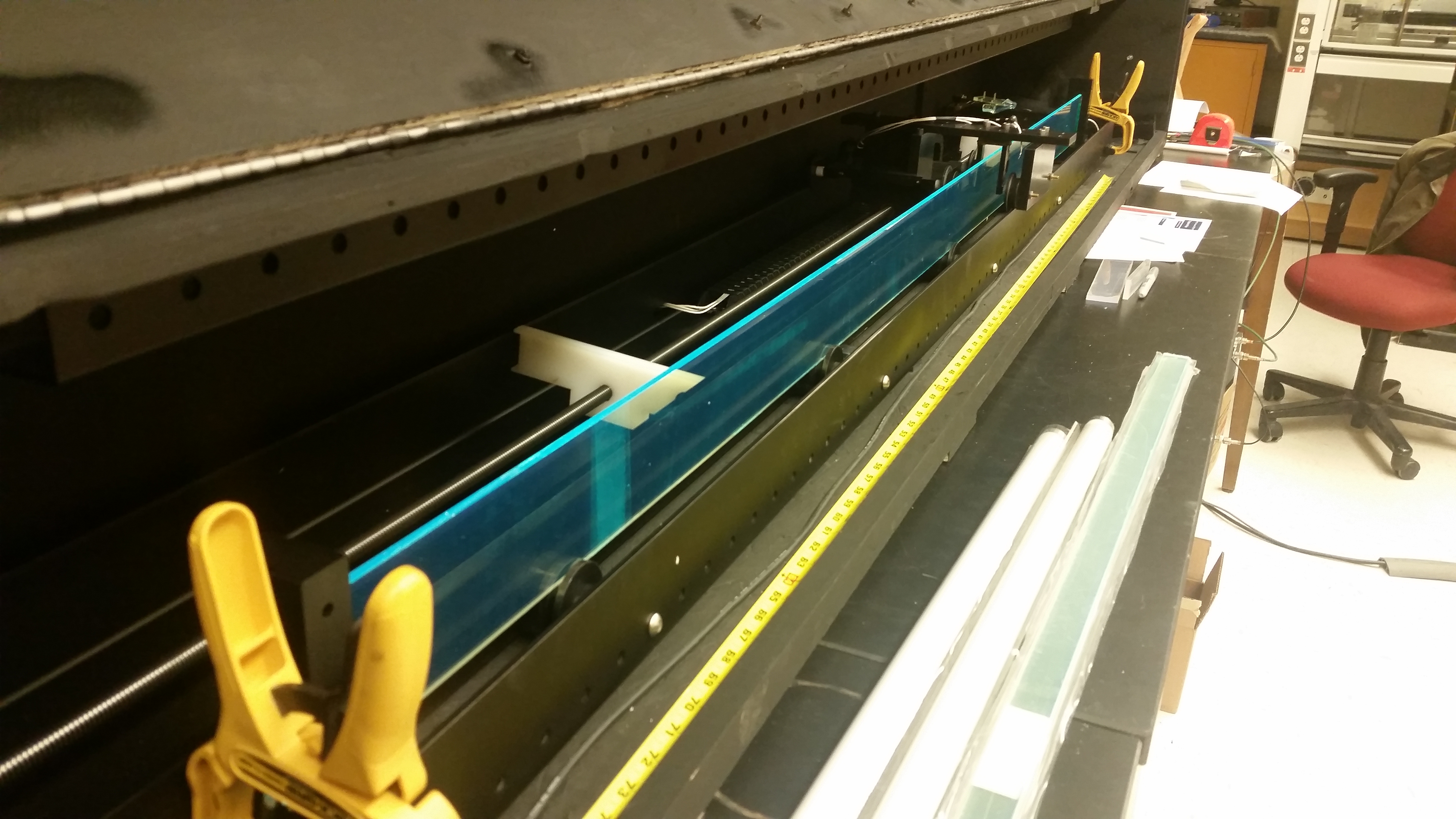}
\end{dunefigure}

Visual inspection of light guides received for \dword{pdsp} found multiple instances of fogging or mottling on the surface and within the bulk of some light guides. However, these features did not appear to impact the attenuation properties or uniformity during darkbox scans. Acceptance of light guides for shipment to the assembly facility was based on the metrology and attenuation results.





\subsection{Photon Detector Module Assembly}

Final assembly planning for \dword{pd} modules is guided by the assembly of \num{60} \dword{pdsp} \dword{pd} modules (representing multiple units of all three varieties) at the Colorado State University assembly facility.  \dword{pdsp} assembly will occur at one or more assembly facilities, to be determined prior to submission of the \dword{tdr}.  Several features of this are common to all three types of module, and these aspects will be covered in this section.

\subsection{Incoming Materials Control}

All materials for \dword{pd} module assembly will be delivered with a previously generated QC traveller (in the case of materials custom fabricated for DUNE) or will have an incoming materials traveller generated immediately upon receipt of the component (for commercial components).  These travelers will be scanned upon receipt at the assembly facility, and the data stored in the DUNE QC database.  Materials will either arrive with a pre-existing DUNE inventory control batch/lot number, or will have one assigned prior to entering the assembly area.  Bar code labels attached to storage containers for all components in the assembly area will facilitate traceability throughout the assembly process.

Immediately upon receipt all materials will undergo an incoming materials inspection, including confirmation of key dimensional tolerances as specified on the incoming materials documentation for that component.  The results of these inspections will be included on the traveller for that batch/lot and entered into the database.

In the case of discrepancy, the deviation from nominal will be recorded in an exception section of the traveller, as well as the resolution of the discrepancy.

\subsection{Assembly Area Requirements}

Assembly will occur in a class \num{100000} or better clean assembly area.  Photosensitive components (TPB coated surfaces) are sensitive to near-UV light exposure, and will be protected by blue-filtered light in the assembly area (>\SI{400}{nm} or better filters\footnote{For example, GAMTUBE T1510 from GAM Products, Inc., \url{http://www.gamonline.com/catalog/gamtube/index.php}.}); it has been determined that this level of filtering is sufficient to protect coated surfaces during  exposures of up to several days. For exposures of weeks or months, such as in the ProtoDUNE cryostat assembly area, a higher cut-off yellow filter is used\footnote{F007-010 Amber with Adhesive - http://www.epakelectronics.com/uv\_filter\_materials\_flexible.htm.}. 
The requirement for light exposure is to be revisited prior to \dword{pdsp} production.


Exposure of photosensitive components will be strictly controlled.  Work flow will be restricted to ensure no component exceeds a total exposure of \SI{8}{hours} to filtered assembly area lighting (including testing time).

\subsection{Component Cleaning}

All components will be cleaned  as appropriate, following manufacturer's specifications and DUNE materials test stand recommendations.  Cleaning procedures will be written for all incoming materials, and completion of these procedures noted in the appropriate travelers.

\subsection{Assembly Procedures}

Following the example of the \dword{pdsp} experience, detailed, step-by-step written procedure documents will be followed for each module, and a QC traveller for each module is completed and recorded in the database.  \dword{pdsp} experience suggests that a two-person assembly team is necessary and sufficient for all three currently-considered versions of the light collector modules.  Our current assembly plan envisions two 2-person teams operating at the same time, with a fifth person acting as shift leader.  The shift leader is not directly involved in assembly, but rather acts as a QC officer responsible primarily for ensuring distributing materials to the assembly teams (documenting the batch/lot numbers for each detector on the relevant module travelers) and ensuring that documented assembly procedures are followed.

Assembly fixtures mounted to \SI{2.4}{m} long flat optical tables will be used to support and align \dword{pd} components during assembly.  All workers handling \dword{pd} components will wear gloves, hair nets, shoe covers, and clean room disposable lab jackets at all times.





\subsection{Post-Assembly Quality Control}

Post-assembly QC planning is currently based on \dword{pdsp} experience, modified as appropriate for larger-scale production.  Each module will go through a series of go-no gauges designed to control tolerances of critical interface points.  Following this, each module will be inserted into a test \dword{apa} support model, representing the tightest slot allowed by \dword{apa} mechanical tolerances. Next, each module will be scanned at a fixed set of positions (to be determined prior to the TDR) with \SI{275}{nm} UV LEDs.  The detector response at each position will be read out using \dword{pd} readout electronics, and the data compared to pre-established criteria.  Figure~\ref{fig:pds-pd-scanner} is a photograph of the scanner used for \dword{pdsp} modules. These performance data will serve as a baseline for the module, and will be compared against those taken in an identical scanner shortly before installation into the \dword{apa}, as in the \dword{pdsp} experience.  All data collected will be recorded to the module traveler and to the DUNE QC database.
As a final QC check, post-assembly immersion into a LN2 cryostat followed by a repeat scan of each \dword{pd} module (as in \dword{pdsp}) is being considered.

\begin{dunefigure}[\dword{pd} module scanner.]{fig:pds-pd-scanner}
{\dword{pd} module scanner.}
  \includegraphics[width=0.6\columnwidth]{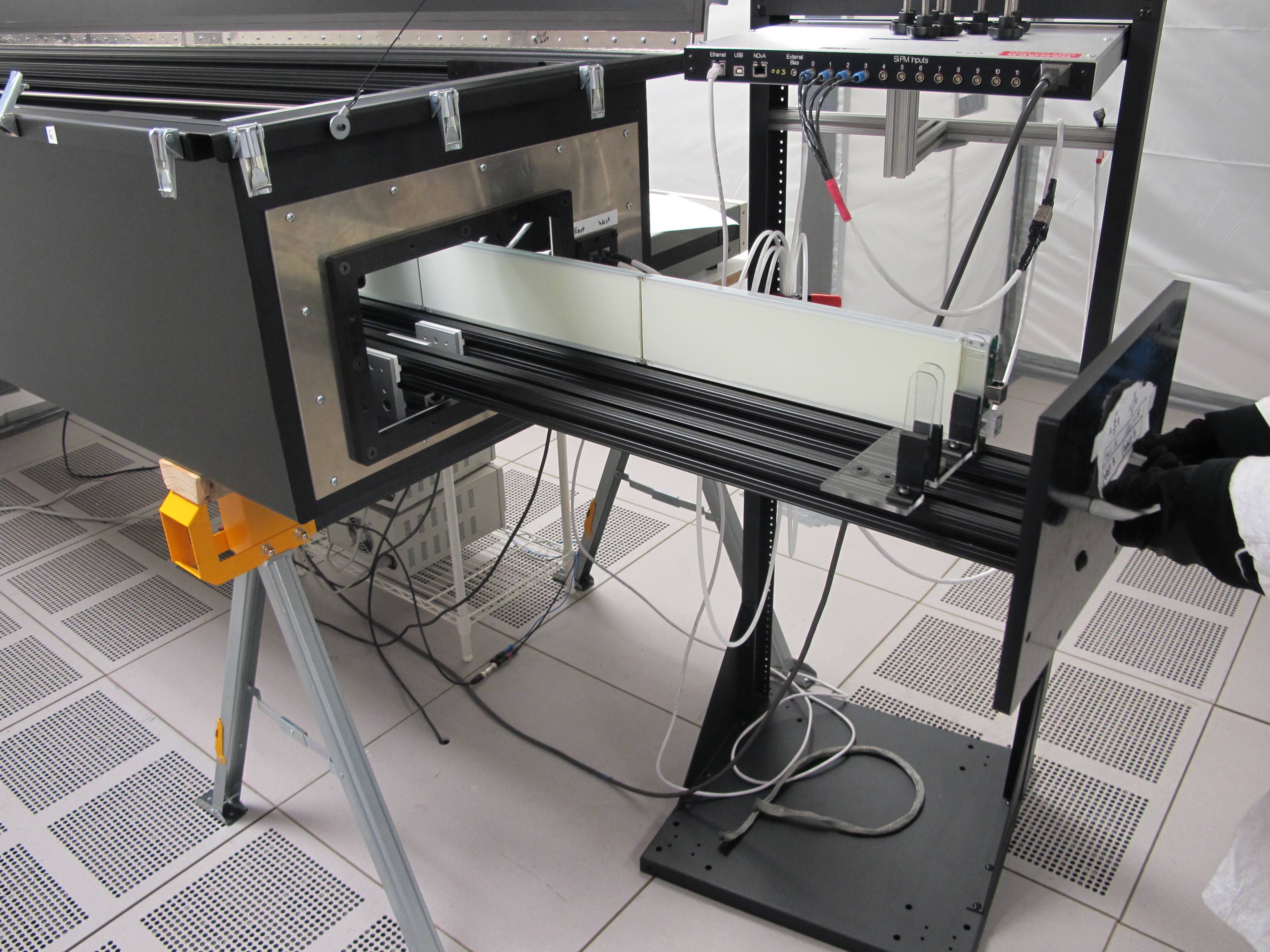}
\end{dunefigure}


\subsection{APA Frame Mounting Structure and Module Securing}	
\label{sec:fdsp-pd-assy-frames}

\Dword{pd} modules are inserted into the \dword{apa} frames through ten slots 
(five on each side of the \dword{apa} frame) and are supported inside the frame by 
stainless steel guide channels.  The slot dimensions for the \dword{pdsp} \dword{apa} frames 
were \SI{108.0}{mm}$\times$\SI{19.2}{mm} wide (see Figure~\ref{fig:pds-pd-mounting}(top)). These dimensions are expected to increase by about \SI{20}{\%} in the next round of \dword{apa} design revisions, allowing for larger \dword{pd} modules.   
The guide channels are pre-positioned into the \dword{apa} frame prior to applying the wire shielding mesh to the \dword{apa} frames, and are
not accessible following wire wrapping. Following insertion, the \dword{pd} modules are fixed in place in the \dword{apa} frame using
 two stainless steel captive screws, as shown in Figure~\ref{fig:pds-pd-mounting}(bottom).

\begin{dunefigure}[\dword{pd} mounting rails in \dword{apa} frame.]{fig:pds-pd-mounting}
{\dword{pd} mounting in \dword{apa} frame: Rails (top) and securing to the frame with captive screws  (bottom).}
	\includegraphics[height=7cm]{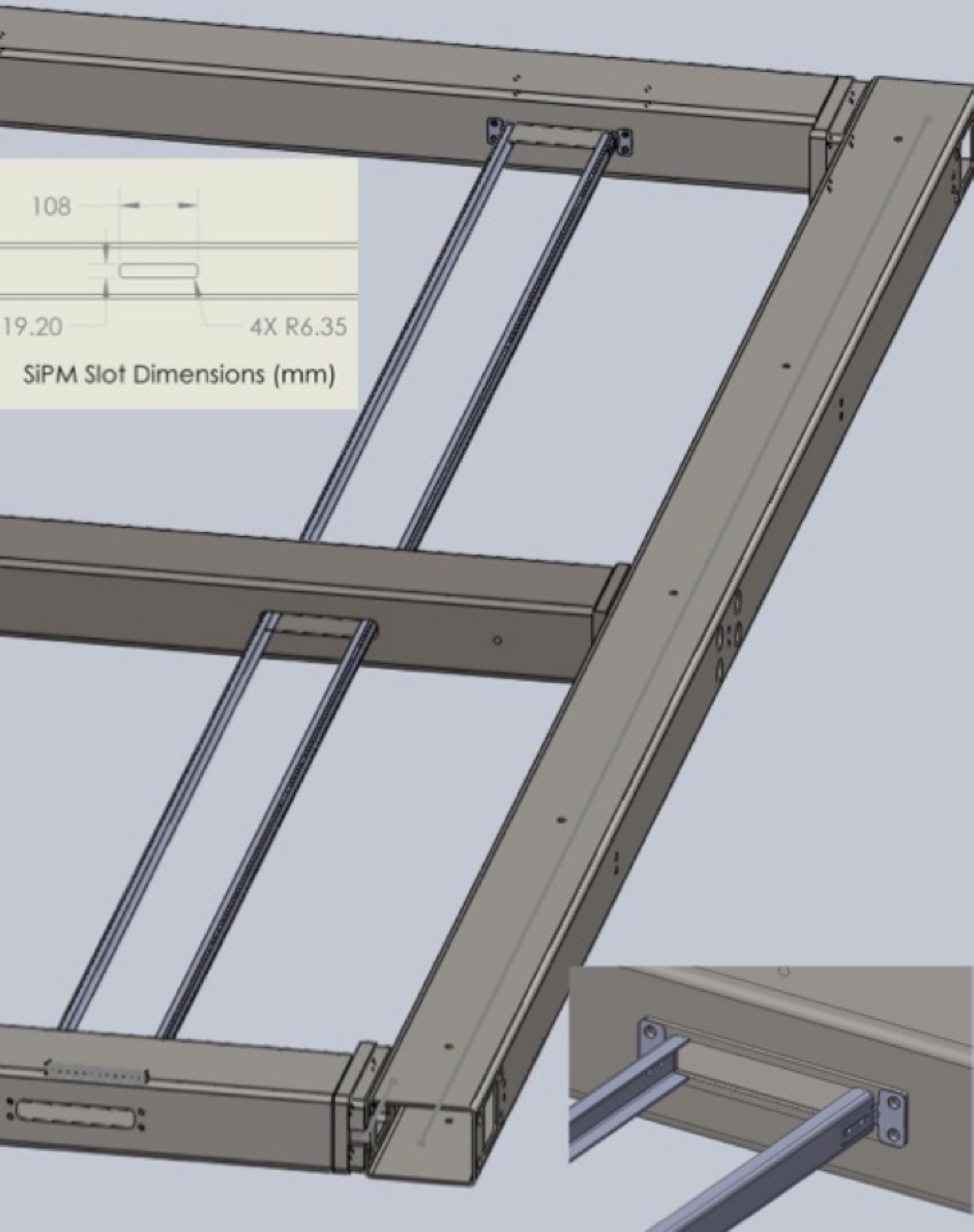}
	\includegraphics[height=5cm]{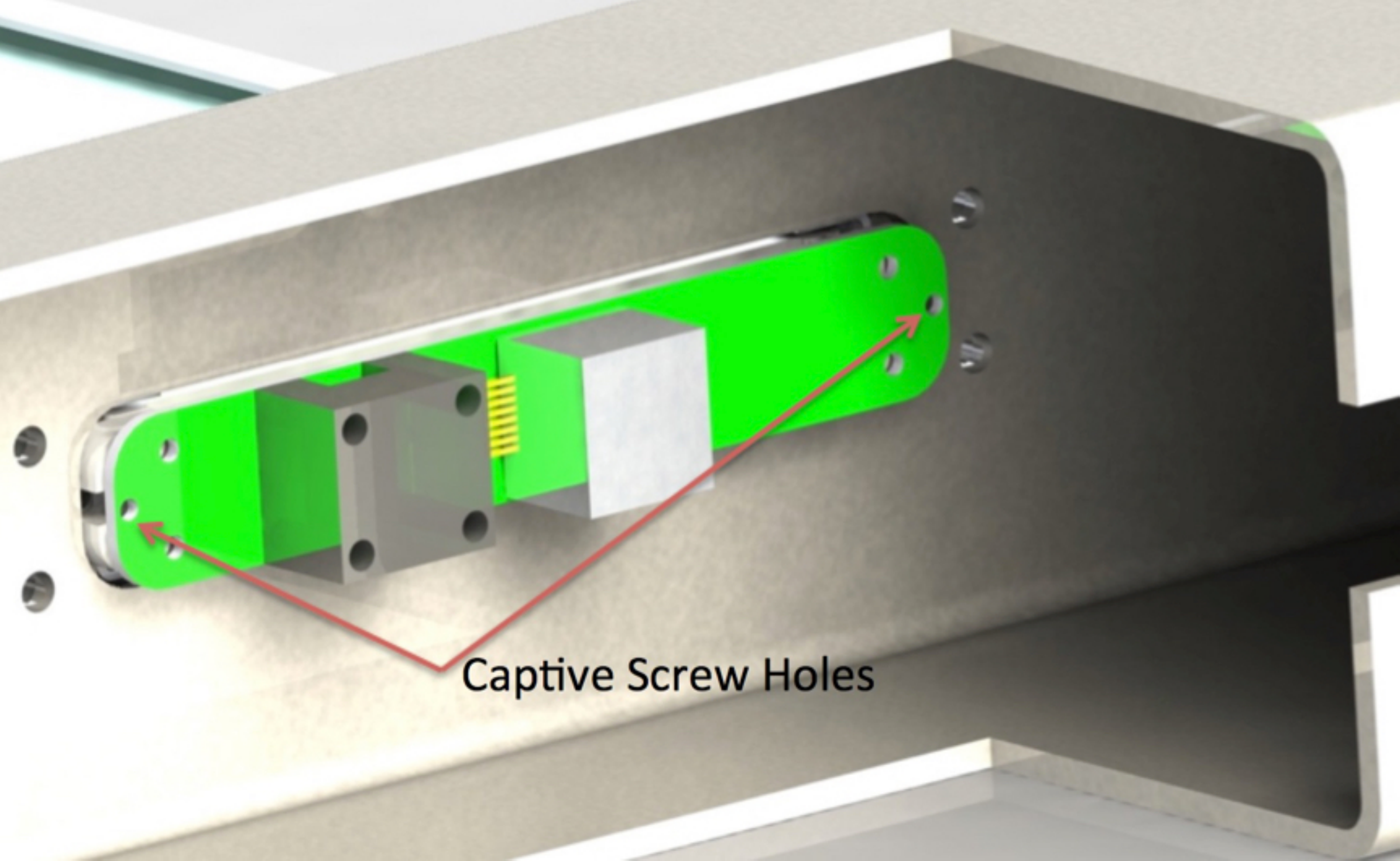}
\end{dunefigure}


\subsubsection{Cryogenic thermal contraction}

Bar-style \dword{pd} modules are structurally composed of primarily polycarbonate, polystyrene and 
acrylic, which have significantly different shrinkage factors compared to the 
stainless steel \dword{apa} and \dword{pd} support frames (see Table~\ref{tbl:fdsfpdshrink}).

\begin{dunetable}[Shrinkage of \dword{pd} materials.]
{lc}
{tbl:fdsfpdshrink}
{Shrinkage of \dword{pd} module materials for a $206^{\circ}$C temperature drop}
Material 			 & Shrinkage Factor (m/m)\\ \toprowrule
Stainless Steel (304) & $2.7\times10^{-3}$\\ \colhline
FR-4 G-10 (In-plane) & $2.1\times10^{-3}$\\ \colhline
Polystyrene (Average) & $1.5\times10^{-2}$\\ \colhline
Acrylic and Polycarbonate (Average) & $1.4\times10^{-2}$\\ 
\end{dunetable}

These differences in thermal expansion (or contraction, in this case) are an important factor during design of the \dword{pd} module supports.  Mitigation of these contractions is detailed in Table~\ref{tbl:fdsfpdshrinkeffects}.

Thermal expansion coefficients (CTEs) for the fused-silica filter plates ($1.1\times10^{-4}$~\si{m/m} for a 206$^\circ$C temperature drop) informed the materials selection for the ProtoDUNE ARAPUCA modules.  The frame components for the ARAPUCA were fabricated from FR-4 G-10, resulting in a shrinkage of the stainless steel frame structure relative to the frame of approximately \SI{1.2}{mm} along the long ($\sim$\SI{2}{m}) axis of the bar.  The shrinkage of the frame relative to the filter plate is <\SI{0.2}{mm}.  Both these relative shrinkage factors are accounted for in the dimensions and tolerances of the design.

\begin{dunetable}[Relative shrinkage of \dword{pd} components and \dword{apa} frame]
{p{0.2\textwidth}p{0.2\textwidth}p{0.5\textwidth}}
{tbl:fdsfpdshrinkeffects}
{Relative Shrinkage of \dword{pd} components and \dword{apa} frame, and mitigations.}
\textbf{Interface} & \textbf{Relative shrinkage} & \textbf{Mitigation} \\ \toprowrule
\dword{pd} Length to \dword{apa} width & \dword{pd} shrinks \SI{25.7}{mm} Relative to \dword{apa} frame & \dword{pd} affixed only at one end of \dword{apa} frame, free to contract at other end \\ \colhline
Width of \dword{pd} in \dword{apa} Slot & \dword{pd} shrinks \SI{1.2}{mm}  relative to slot width & \dword{pd} not constrained in C-channels. C channels and tolerances designed to contain module across thermal contraction range \\ \colhline
Width of \dword{sipm} mount board ({\it Hover board}) to stainless steel frame & Stainless frame shrinks \SI{0.06}{mm}  more than PCB & Diameter of shoulder screws and FR-4 board clearance holes selected to allow for motion \\ \colhline
Width of \dword{sipm} mount board relative to polycarbonate mount block & Polycarbonate block shrinks \SI{1}{mm} more than PCB & Allowed for in clearance holes in \dword{sipm} mount board \\ 
\end{dunetable}

\subsubsection{\dword{pd} Mount frame deformation under static \dword{pd} load}

\Dword{fea} modeling of the \dword{pd} support structure was conducted to study static deflection 
prior to building prototypes.  Modeling was conducted in both the vertical
 orientation (\dword{apa} upright, as installed in cryostat) and also horizontal orientation.  
Basic assumptions used were fully-supported fixed end conditions for the rails, 
with uniform loading of 3X \dword{pd} mass (\SI{5}{kg}) along rails.  
Figure~\ref{fig:pds-rail} illustrates the rail deflection for the \dword{apa} in the horizontal (left) and vertical (right) orientations.
Prototype testing confirmed these calculations.

\begin{dunefigure}[\dword{pd} mechanical support analysis.]{fig:pds-rail}
{\dword{pd} mechanical support analysis: Rail deflection for the \dword{apa} in the horizontal (left) and vertical (right) orientations.}
	\includegraphics[height=4.5cm]{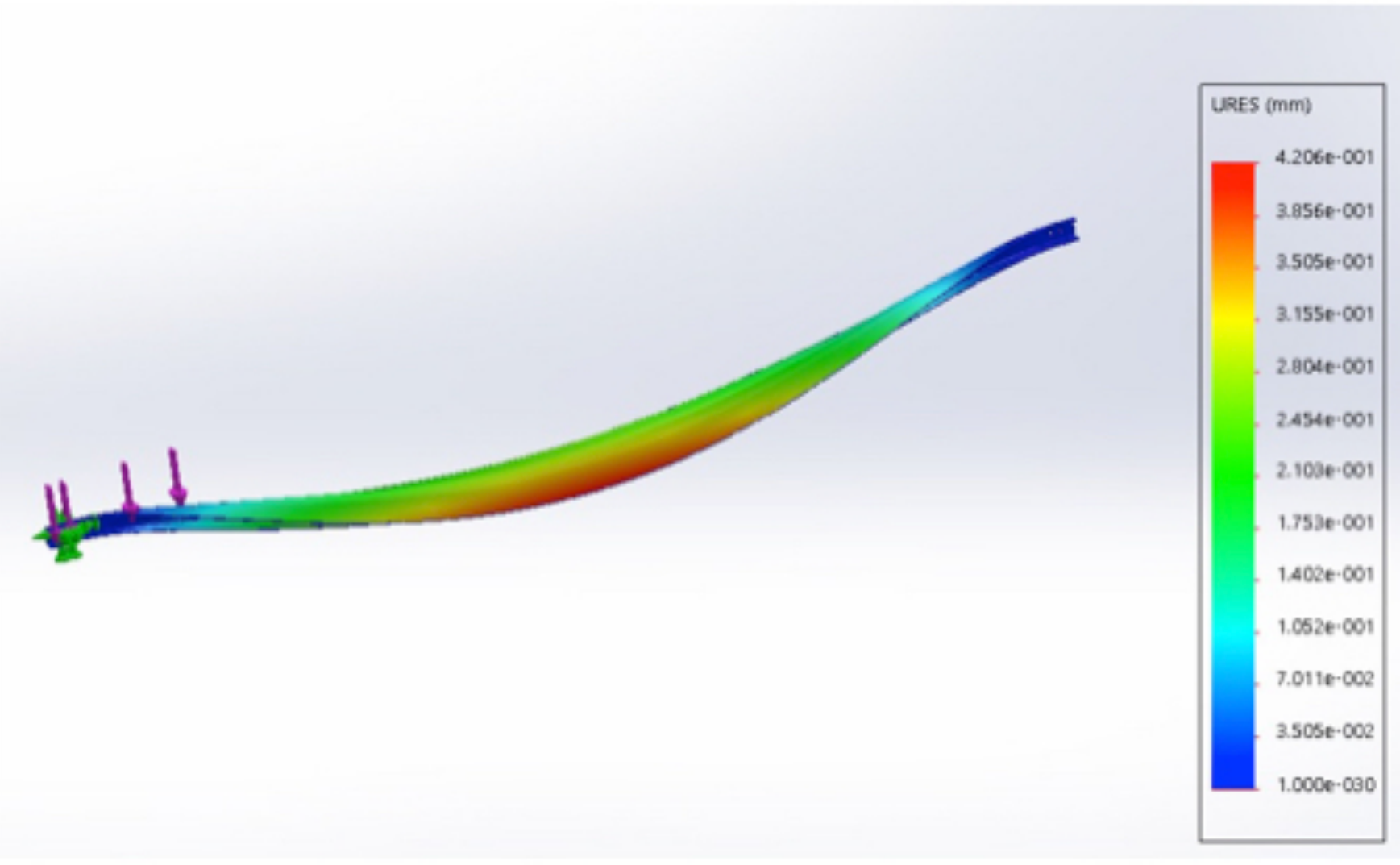} 
	\includegraphics[height=4.5cm]{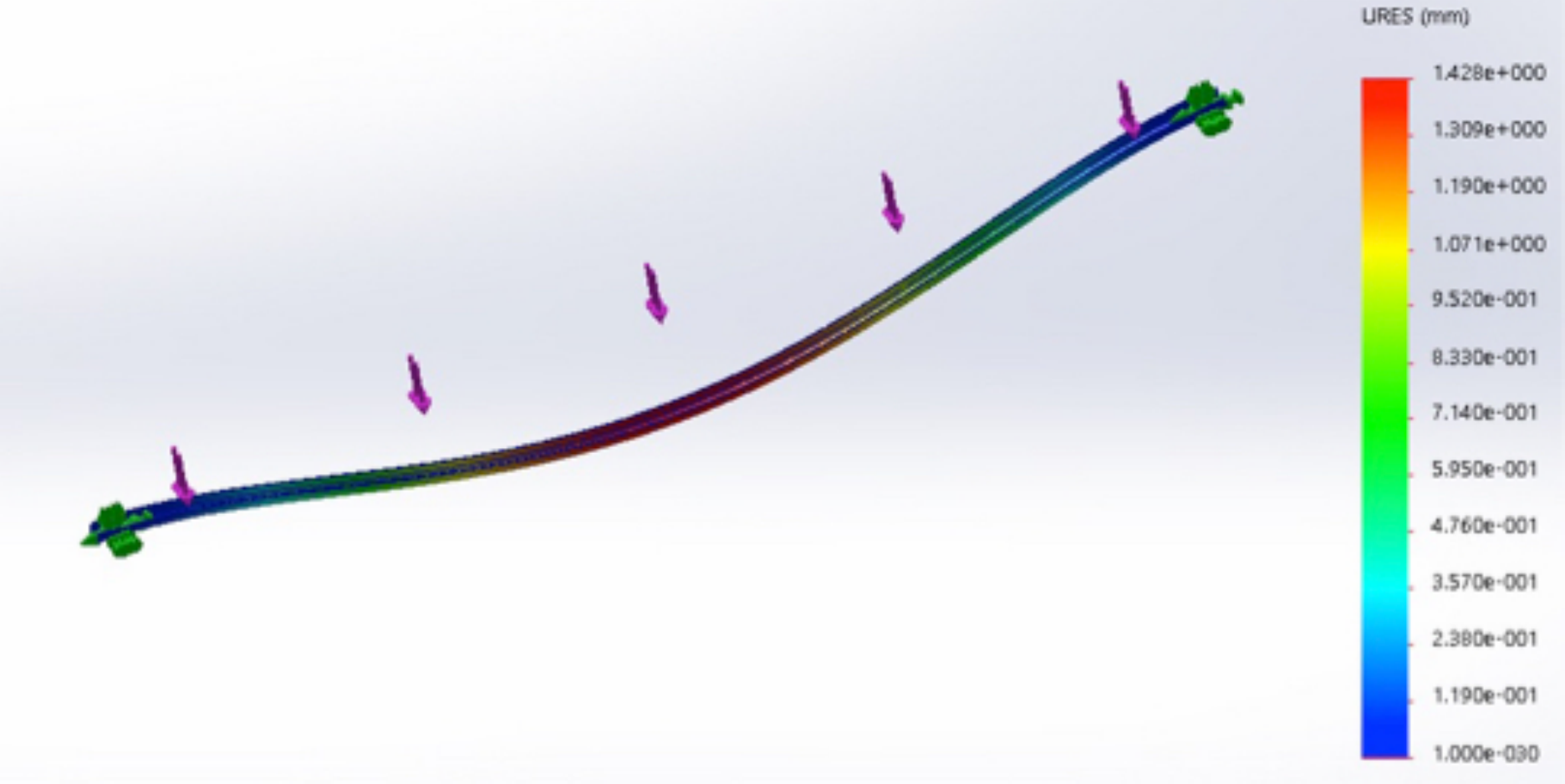}\\
\end{dunefigure}

\subsection{Photosensor Modules}
\label{sec:fdsp-pd-assy-psm}

Depending on the photon collector  technology selected, the \dword{sipm} analog signal will be ganged in groups of 6-48 in close proximity to the sensors inside the
\lar volume; both {\it passive} and {\it active} ganging schemes are under consideration.  Passive ganging (sensors in parallel) implemented with traces on the \dword{sipm} mounting board (module) and has been implemented for \dword{pdsp}.  The \dwords{sipm} are mounted using a pick-and-place machine and standard surface mount device soldering procedures. 
 The ganged analog signals are then brought out via long cables (approximately \SI{25}{m}) for digitization outside the cryostat.
 \dword{pdsp} will provide essential operational experience with a passive ganging board and signal transport provided by Teflon ethernet CAT6 cables.
It is already apparent that R\&D is needed to optimize the connectors used to couple the cable to the board;  it is a priority to understand the 
mechanical stresses involved in the \dword{sipm}-PCB-Connector system (with different CTEs) as it is cooled (or cycled) to cryogenic temperatures.
 
A basic level of active ganging locates summing circuitry on the board carrying the photosensors or on a separate PCB also mounted on the \dword{pd} module.
A more complex scheme is being considered that would include cold amplifiers and ADCs. This solution would provide more flexibility in the level of photosensor ganging and also obviate the need for carrying analog signals of long cables. Production of the board would follow standards practices but the complexity introduces concerns with reliability and long-term stability issues related to cold electronics.  Basic active ganging prototypes are under study with high priority but the design is not yet at a mature stage. 



\subsection{Electronics}
\label{sec:fdsp-pd-assy-pde}

Extensive experience of manufacturing processes was gained during the development of the SSPs under current use on \dword{pdsp}, a general description of the readout system of \dword{pdsp} can be seen in the section \ref{sec:fdsp-pd-pde}. Compatibility between elements designed by different institutions is guaranteed when standard procedures are followed so the circuit design must be done in accordance with mutually agreed-upon specification documents.  A sufficient  number of units needs to be produced to allow local testing and for testing in the central facility -- for example, in \dword{pdsp} \num{5} 12-channel  SSPs were produced and delivered to CERN for integration testing prior. Twenty-four were fabricated for \dword{pdsp} operation. 

The readout electronics of the photon detection system will be designed and produced with similar tools and protocols used for  \dword{pdsp}. For example: printed circuit board (PCB) layout is performed in accordance with IPC\footnote{IPC\texttrademark{}, Association Connecting Electronics Industries, \url{http://www.ipc.org/}.} specifications. Bare PCB manufacturing requirements are embedded within the Gerber file fabrication documents (e.g., layers, spacing, impedance, finish, testing, etc.). Components are assembled on to circuit boards using either trained \dword{pd} consortium technical staff or by external assembly vendors, based upon volume, in accordance with per-design assembly specification documents. Testing occurs at labs and universities within the collaboration in accordance with a per-design test procedure that typically includes a mix of manual, semi-automatic and automated testing in an engineering test bench followed by overall characterization in a system- or subsystem-test stand.
Other considerations and practices relevant to readout electronics production and assembly are itemized here:

\begin{itemize}

\item Components: Schematic capture is done using appropriate tools (such as OrCAD 16.6.\footnote{OrCAD\texttrademark{} schematic design tool for PCB design http://www.orcad.com} or similar toolset) available within design facility. Design is hierarchical with common front-end page referenced multiple times to ensure that all input channels are identical. The schematic contains complete bill-of-materials (BOM) including all mechanical parts. Subversion repository is typically used for version control and backup. Multiple internal design reviews held before schematic is released to layout. The bill of materials is stored directly within schematic, extracted to spreadsheet when ordering parts. Every part is specified by both manufacturer and distributor information. Distributor information may be overridden by a technician at order time due to price or availability. Standard search engines such as Octopart\footnote{Octopart https://octopart.com/}, ECIA\footnote{ ECIA https://www.eciaauthorized.com} and PartMiner\footnote{PartMiner https://www.part-miner.com/} are used to check price or availability across all standard distributors. A parts availability check review is performed prior to handoff from schematic to layout; as required obsolete or long lead time parts were removed from design and replaced. BOM information will include dielectric, tolerance, temperature coefficient, voltage rating and size (footprint) to ensure all parts are fully described.

\item Boards: There are standard tools (such as the Allegro\footnote{Cadence Allegro\textregistered PCB design solution https://www.cadence.com} toolset) available for the printed-circuit-board (PCB) layout. Conventional PCBs are realized as multi-layer, controlled impedance board with many sets of delay-matched nets. In usual practice the complete impedance and delay characteristics are calculated within layout tool and crosschecked by PCB vendor prior to manufacture. In usual practice, a competitive bid between multiple previously qualified vendors is used, with a full electrical and impedance testing required. Multiple internal design reviews are held prior to release of the design.

\item Cable plant: Cabling will be designed taking into consideration the \dword{apa} space and in close collaboration with the TPC electronics group to avoid cross-talk effects. A final decision on cable procurement will be taken based on the possibility of cable manufacturing in an institution belonging to the photon detection consortium and the cost of a commercial solution.  

\item Manufacturer list: In addition to the general laboratory procedures for quality assurance, the general practice will be to use only printed circuit board manufacturers and external assembly vendors whose workmanship and facilities have been personally inspected by experienced production team members. All external assemblers are required to quote in accordance with an assembly specifications document describing the IPC class and specific solder chemistry requirements of the design. The bill of materials document will show selected and alternate suppliers for every component of the front-end boards.

\item \Dword{fe} electronics firmware: This will be specified and updated iteratively in collaboration with other systems. The electronics working group will be responsible for responding to requests for additional firmware development, including for example, modifications to timing interface, modifications to trigger interface, and implemented sensitivity to in-spill vs. not-in-spill conditions. Documents describing firmware architecture for each major change will be written and distributed to \dword{pd} and DAQ working groups before implementation. \Dword{fe}  electronics user's manual containing all details of new firmware will be distributed with production units when manufactured.

\item Mechanical assembly: With the mechanical assembly of electronics readout boards it is common practice to use AutoCAD\footnote{AutoDESK AutoCAD\textregistered computer aided design software application https://www.autodesk.com/} with Allegro (as PCB layout tool). All relevant dimensions of the PC board including connector and indicator placement is extracted from Allegro as base DXF file from which overall exploded mechanical diagram of chassis and other mechanical parts is made. Mechanical items such as shield plates will be provided as well. It is assumed that the front-end chassis will made by external vendors (one for chassis, one for front/back panels) from AutoCAD drawings provided by the consortium.

\end{itemize}

\section{System Interfaces}
\label{sec:fdsp-pd-intfc}



This section describes the interface between the 
\dword{spmod} \dword{pds} and several other consortia, task forces (TF) and subsystems listed below:

\begin{itemize}
\item \dword{apa},
\item{\fdth{}s},
\item TPC \dword{ce}, 
\item{\dword{cpa} / \dword{hv} System -- if the coated-reflector foils option is implemented},
\item \dword{daq},
\item Calibration and monitoring.
\end{itemize}
%
The contents of the section are focused on what is needed to complete the design, fabrication, installation of the related subsystems, and are organized by the elements of the scope of each subsystem at the interface between them.

\subsection{Anode Plane Assembly}
\label{sec:fdsp-pd-intfc-apa}

The \dword{pd} is integrated in the \dword{apa} frame to form a single unit for the detection of both ionization charge and scintillation light. 
The hardware interface between \dword{apa} and \dword{pds} is both mechanical and electrical: 
\begin{itemize}
\item Mechanical: (1) supports for the \dword{pds} detectors; (2) access slots for installation of the detectors;
 (3) access slots for the cabling of the \dword{pd} detectors; d) routing of the \dword{pds} cables inside the side beams of the \dword{apa} frame.
\item Electrical: grounding scheme and electrical insulation, to be defined together with the \dword{ce} consortium, given the \dword{ce} strict requirements on noise.
\end{itemize}

\subsection{Feedthroughs}
\label{sec:fdsp-pd-intfc-feed}

Several \dword{pd} \dword{sipm} signals are summed together into a single readout channel. A long multi-conductor cable with four twisted pairs read out the \dword{pd} module. 
Analog signals from the \dwords{sipm} are transmitted directly by cables to the appropriate flanges to outside the cryostat. 
All cold cables originating from the inside the cryostat connect to the outside warm electronics through PCB board feedthroughs installed in the signal flanges that are distributed along the cryostat roof.

All technical specifications for the feedthroughs should be provided by the photon detector group.

\subsection{TPC Cold Electronics}
\label{sec:fdsp-pd-intfc-ce}

The hardware interfaces between the \dword{ce} and \dword{pd} occur in the \fdth{}s and the racks mounted on the top of the cryostat which house low and high voltage power supplies for \dword{pd}, low and bias voltage power supplies for \dword{ce}, as well as equipment for the \dword{cisc}, and possibly \dword{daq} consortia. There should be no electrical contact between the \dword{pd}S and \dword{ce} components except for sharing a common reference voltage point (ground) at the \fdth{}s. An additional indirect hardware interface takes place inside the cryostat where the \dword{ce} and \dword{pd} components are both installed on the \dword{apa} (responsibility of the single phase far detector \dword{apa} consortium, \dword{apa} in the following), with cables for \dword{ce} and \dword{pd} that may be physically located in the same space in the \dword{apa} frame, and where the cables and fibers for \dword{ce} and \dword{pd} may share the same trays on the top of the cryostat. These trays are the responsibility of the facility and the installation of cables and fibers will follow procedures to be agreed upon in consultation with the underground installation team, \dword{uit} in the following.

In the current design \dword{ce} and \dword{pd} use separate flanges for the cold-to-warm transition and each consortium is responsible for the design, procurement, testing, and installation, of their flange on the \fdth{}, together with LBNF, who is responsible for the design of the cryostat. 
The installation of the racks on top of the cryostat is a responsibility of the facility, but the exact arrangement of the various crates inside the racks will be reached after common agreement between the \dword{ce}, \dword{pd}, \dword{cisc}, and possibly \dword{daq} consortia. The \dword{pd} and \dword{ce} consortia will retain all responsibilities for the selection, procurement, testing, and installation of their respective racks, unless for space and cost considerations an agreement is reached where common crates are used to house low voltage or high/bias voltage modules for both \dword{pd}S and \dword{ce}. Even if both \dword{ce} and \dword{pd} plan to use floating power supplies, the consequences of such a choice on possible cross-talk between the systems needs to be studied. 

Various test stands and integration facilities will be developed. In all cases the \dword{ce} and \dword{pd} consortia will be responsible for the procurement, installation, and initial commissioning of their respective hardware in these common test stands. The main purpose of these test stands is study the possibility that one system may induce noise on the other, and the measures to be taken to minimize this cross-talk. For these purposes, it is desirable to repeat noise measurements whenever new, modified detector components are available for one or the other consortium. This requires that the \dword{ce} and \dword{pd} consortia agree on a common set of tests to be performed and that the \dword{ce} consortium can operate the \dword{pd}S detectors within a pre-determined range of operating parameters, and vice versa, without the need of providing personnel from the \dword{pd}S consortium when the \dword{ce} consortium is performing tests or vice versa. Procedures should be set in place to decide the time allocation to tests of the components of one or the other consortium.

\subsection{Cathode Plane Assembly and High Voltage System}
\label{sec:fdsp-pd-intfc-le}

The \dword{pd} and the \dword{hv} systems interact in the case that the former includes wavelength-shifting reflector foils mounted on the cathode plane array (\dword{cpa}). An additional interface is addressed in Section~\ref{sec:fdsp-pd-intfc-calib}.

The purpose of installing the wavelength-shifting (\dword{wls}) foils is to allow enhanced detection of light from events near to the cathode plane of the detector. The \dword{wls} foils consist of a wavelength shifting material (such as TPB) coated on a reflective backing material. The foils would be mounted on the surface of the \dword{cpa} in order to enhance light collection from events occurring nearer to the \dword{cpa}, and thus greatly enhancing the spatial uniformity of the light collection system as detected at the \dword{apa} mounted light sensors. The foils may be laminated on top of the resistive Kapton surface of the \dword{cpa} frames, with the option of using metal fasteners or tacks that would also serve to define the field lines. 

Production of the FR4+resistive Kapton \dword{cpa} frames are the responsibility of the \dword{hv} consortium. Production and \dword{tpb} coating of the \dword{wls} foils will be the responsibility of the Photon Detection consortium. The fixing procedure for applying the \dword{wls} foils onto the \dword{cpa} frames and any required hardware will be the responsibility of the Photon Detection consortium, with the understanding that all designs and procedures will be pre-approved by the \dword{hv} consortium. 

This new detector component is not being tested in \dword{pdsp}, however its integration in the present \dword{spmod} \dword{hv} system could imply performance and stability degradation (due for example to ion accumulation at the \dword{cpa} surface); the assembly procedure of the \dword{cpa}/FC module could become more complex due to the presence of delicate \dword{wls} foils. Intense R\&D will be required before deciding on its implementation.

An integration test stand will likely be employed to verify the proper operation of the \dword{cpa} panels with the addition of \dword{wls} foils under high voltage conditions. Light performance (wavelength conversion and reflectivity efficiency) will also be verified. The \dword{hv} consortium will be responsible for \dword{hv} aspects of the test stand and the \dword{pd} consortium will be responsible for the light performance aspects.

\subsection{Data Acquisition}
\label{sec:fdsp-pd-intfc-daq}

The main system interfaces include:


\textbf{Data Physical Links: }Data are passed from the \dword{pd} to the \dword{daq} on optical links conforming to an IEEE Ethernet standard. The links run from the \dword{pd} readout system on the cryostat to the \dword{daq} system in the Central Utilities Cavern (\dword{cuc}).

\textbf{Data Format:} Data are encoded using a data format based on UDP/IP. The data format is derived from the one used by the Dual Phase TPC readout. Details will be finalized by the time of the \dword{daq} \dword{tdr}.

\textbf{Data Timing:} The data must contain enough information to identify the time  at which it was taken.

\textbf{Trigger Information:} The \dword{pd} may provide summary information useful for data selection. If present, this will be passed to the \dword{daq} on the same physical links as the remaining data.

\textbf{Timing and Synchronization: }Clock and synchronization messages will be propagated from the \dword{daq} to the \dword{pd} using a backwards compatible development of the \dword{pdsp} Timing System protocol (DUNE docdb-1651). There will be at least one timing fiber available for each data links coming from the \dword{pd}S. Power-on initialization and Start of Run setup:  The \dword{pd}S may require initialization and setup on power-on and start of run. Power on initialization should not require communication with the \dword{daq}. Start run/stop run and synchronization signals such as accelerator spill information will be passed by the timing system interface.

\textbf{Interaction with other groups: }Related interface documents describe the interface between the \dword{ce} and LBNF, \dword{daq} and LBNF, \dword{daq} and Photon and both \dword{daq} and \dword{ce} with Technical Coordination. The cryostat penetrations including through-pipes, flanges, warm interface crates and feedthroughs and associated power and cooling are described in the LBNF/\dword{pd}S interface document.  The rack, computers, space in the \dword{cuc} and associated power and cooling are described in the LBNF/\dword{daq} interface document. Any cables associated with photon system data or communications are described in the \dword{daq}/Photon interface document. Any cable trays or conduits to hold the \dword{daq}/\dword{ce} cables are described in the LBNF/Technical Coordination interface documents and currently assumed to be the responsibility of Technical Coordination.

\textbf{Integration:} Various integration facilities are likely to be employed, including vertical slice tests stands, \dword{pd}S test stands, \dword{daq} test stands and system integration/assembly sites. The \dword{daq} consortia will provide hardware and software for a vertical slice test. The \dword{pd} consortium will provide \dword{pd} emulators and \dword{pd} readout hardware for \dword{daq} test stands. (The \dword{pd} emulator and \dword{pd} readout hardware may be the same physical object with different configuration). Responsibility for supply and installation of \dword{daq}/\dword{pd} cables in these tests will be defined by the time of the \dword{daq} \dword{tdr}.

\subsection{Calibration and Monitoring}
\label{sec:fdsp-pd-intfc-calib}

This subsection defines the internal calibration system for the \single \dword{pds}. It may be interfaced to a calibration consortium later.

It is proposed that the \single \dword{pds} gain and timing calibration system, a pulsed UV-light system, also be used for \dword{pd} monitoring purposes during both commissioning and experimental operation. The hardware consists of warm and cold components.  

By placing light sources and diffusers on the cathode planes designed to illuminate the anode planes, the \dwords{pd} embedded in the \dwords{apa} can be illuminated. Cold components (diffusers and fibers) interface with \dword{hv} and are described in a separate interface document. \fixme{ref?} Warm components include controlled pulsed-UV source and warm optics. These warm components interface calibration and monitoring  with the \dword{cisc} and \dword{daq} subsystems, and are described in corresponding documents. \fixme{ref?} Optical \fdth is a cryostat interface. 

Hardware components will be designed and fabricated by SP-\dword{pd}S. Other aspects of hardware interfaces are described in the following. The CTF and \dword{pd}S groups might share rack spaces that needs to be coordinated between both groups. There will not be dedicated ports for all calibration devices. Therefore, multi-purpose ports are planned to be shared between various groups. CTF and SP-\dword{pd} will define ports for deployment. It is possible that SP-\dword{pd} might use Detector Support Structure (DSS) ports or TPC signal ports for routing fibers. The CTF in coordination with other groups will provide a scheme for interlock mechanism of operating various calibration devices (e.g. laser, radioactive sources) that will not be damaging to the \dword{pd}. 



\section{Installation, Integration and Commissioning}
\label{sec:fdsp-pd-install}


\subsection{Transport and Handling}
\label{sec:fdsp-pd-install-transport}

Following assembly and testing of the \dword{pd} modules they will be packaged and shipped to the \dword{fd} site for checkout and any final testing prior to installation into the cryostat. Handling and shipping procedures will depend on the environmental requirements determined for the photon detectors, and will be specified prior to the \dword{tdr}.

A testing plan will be developed to determine environmental requirements for photon detector handling and shipping. The environmental conditions apply for both surface and underground transport, storage and handling. Requirements for light (UV filtered areas), temperature, and humidity exposure will also be developed.

Handling procedures that ensure environmental requirements are met will be developed. This will include handling at all stages of component and system production and assembly, testing, shipping, and storage. It is likely that \dword{pd} modules and components will be stored for periods of time during production and prior to installation into the \dword{fd} cryostats. Appropriate storage facilities need to be constructed at locations where storage will take place. Shipping and storage containers need to be designed and produced. Given the large number of photon detector modules to be installed in the \dword{fd}, it will be cost effective to take advantage of reusable shipping containers.

Documentation and tracking of all components and \dword{pd} modules will be required during the full production and installation schedule. Well defined procedures will be in place to ensure that all components/modules are tested and examined prior to, and after, shipping. Information coming from such testing and examinations will be stored in a hardware database.

An Integration and Test Facility (ITF) will be constructed at a location to be decided by the collaboration/project for the integration of the \dword{pd}s into \dwords{apa}. Transportation to and from ITF should be carefully planned. The \dword{pd}S units will be shipped from the production area in quantities compatible with the \dword{apa} transport rates.
    
Operations: The \dword{pd}S deliveries will be stored in temperature and humidity controlled storage area. Their mechanical status will be inspected.

Transportation to SURF: The delivery to SURF will be such that the storage time before integration will be at most two weeks.

\subsection{Integration with APA and Installation}
\label{sec:fdsp-pd-install-pd-apa}





\dword{pd} modules integration into the \dword{apa} frame will happen at the Integration facility. Experts from both groups will work with the installation team. 
An electrical test with \dword{apa}/\dword{pd}S/\dword{ce} will be performed at the integration facility in a cold box, after the integration of \dword{pd}S and \dword{ce} on the \dword{apa} frame has been completed.

The \dword{apa} consortium will be responsible for the transportation of the integrated \dword{apa} frames from the integration facility to the LBNF/SURF facility. 
The \dword{uit} team, under supervision of the \dword{apa} group, will be responsible to move the equipment into the clean room. 
Work on the 2-\dword{apa} connection and inspection underground, prior to installation in the cryostat, is performed by the \dword{apa} group.
Work on cabling during this assembly process is performed by \dword{pd}S and \dword{ce} groups under supervision of the \dword{apa} group.
Once the \dwords{apa} will be moved inside the cryostat, the \dword{pd}S and \dword{ce} consortia will be responsible for the routing of the cables in the trays hanging from the top of the cryostat.

\subsection{Installation into the Cryostat and Cabling}
\label{sec:fdsp-pd-install-pd-cryo}




The \dword{pd} modules are installed into the \dwords{apa}. There are ten \dword{pd}'s per \dword{apa}, inserted into alternating sides of the \dword{apa} frame, five from each direction. Once a \dword{pd} is inserted, it is attached mechanically to the \dword{apa} frame  and cabled up with a single power/readout cable. Following \dword{pd} installation cold electronics (\dword{ce}) units are installed at the top of the \dword{apa} frame.

After the \dword{apa} has been integrated with the \dword{pd}S and \dword{ce}, it will be moved via the rails in the clean room to the integrated cold test stand for testing and be moved into the cryostat. The two anode planes of the TPC will be assembled inside the cryostat, each of the fully tested \dwords{apa} mechanically linked together. Signal cables from the TPC readout and the \dword{pd} modules are routed up to the feedthrough flanges on the cryostat top side. The cables from each of the \dword{ce} and \dword{pd}'s on the \dword{apa} are then routed and connected to the final flanges on the cryostat.

\subsection{Calibration and Monitoring}
\label{sec:fdsp-pd-install-calib}

Commissioning of the \dword{spmod} \dword{pds} will rely heavily on the readout electronics, \dword{daq}, and calibration and monitoring system.  Deployment and testing of the readout electronics separately from the in situ installation of photon detector modules in the \dword{apa} is important to establish their proper functioning before connection to the photon detectors or their flanges.  Careful checking at each step of the integration process will help to find unexpected problems early enough to be corrected before individual units are mounted into the larger systems (first in the \dword{apa} then after installation in the cryostat). 

Once the electronics are read out out via the \dword{daq} system, it will be appropriate to add the \dword{pd} modules and continue commissioning of the installed system.  In order to be properly tested the \dword{pd} modules will have to be in the dark.  Making sure it is possible to make this check frequently enough to catch problems early is critical. This will have to be balanced with the needs of installation, as work progresses.  

Once the basic operation of the readout system is established, the calibration and monitoring system will be of great use during the commissioning.  While the background signals from the warm photon detectors may make calibration difficult, the monitoring system will be able to flash UV light to excite the \dword{pd} modules.  These light signals can be used to determine that cabling is connected, and connected properly by looking at light from different UV emitters.  Once the detector is beginning to cool down, the operation of the calibration and monitoring system will become even more important as the monitoring of the individual channels should be a good indication of their proper operation, and again, the proper cabling and interface.

%





\section{Quality Assurance and Quality Control}
\label{sec:fdsp-pd-qaqc}



\subsection{Design Quality Assurance}
\label{sec:fdsp-pd-designqa}

\dword{pd} design \dword{qa} focuses on ensuring that the detector modules meet the following goals:
\begin{itemize}
\item Physics goals as specified in the DUNE requirements document,
\item Interfaces with other detector subsystems as specified by the subsystem interface documents,
\item Materials selection and testing to ensure non-contamination of the \lar volume.
\end{itemize}

The \dword{pds} consortium will perform the design and fabrication of the components in accordance with the applicable requirements of the LBNF/DUNE Quality Assurance Plan. If the institute (working under the supervision of the consortium) performing the work has a documented \dword{qa} program the work may be performed in accordance with their own program.

Upon completion of the \dword{pd}S design and \dword{qa}/\dword{qc} plan there will be a preliminary design review process, with the reviewers charged to ensure that the design demonstrates compliance with the goals above.

\subsection{Production and Assembly Quality Assurance}
\label{sec:fdsp-pd-prodqa}

The photon detector system will undergo a \dword{qa} review for all components prior to completion of the design and development phase of the project.  The \dword{pdsp} test will represent the most significant test of near-final \dword{pd} components in a near-DUNE configuration, but additional tests will also be performed.  The \dword{qa} plan will include, but not be limited to, the following areas:

\begin{itemize}
\item Materials certification (in the \dword{fnal} materials test stand and other facilities) to ensure materials compliance with cleanliness requirements
\item Cryogenic testing of all materials to be immersed in \lar, to ensure satisfactory performance through repeated and long-term exposure to \lar.  Special attention will be paid to cryogenic behavior of plastic materials (such as radiators and light guides), \dwords{sipm}, cables and connectors.  Testing will be conducted both on small-scale test assemblies (such as the small test cryostat at CSU) and full-scale prototypes (such as the full-scale CDDF cryostat at CSU). 
\item Mechanical interface testing, beginning with simple mechanical go-nogo gauge tests, followed by installation into the \dword{pdsp} system, and finally full-scale interface testing of the \dword{pds} into the final pre-production TPC system models
\item Full-system readout tests of the \dword{pd} readout electronics, including trigger generation and timing, including tests for electrical interference between the TPC and \dword{pd} signals.
\end{itemize}

Prior to the release of the \dword{tdr} the \dword{pds} will undergo a final design review, where these and other \dword{qa} tests will be reviewed and the system declared ready to move to the pre-production phase.

\subsection{Production and Assembly Quality Control}
\label{sec:fdsp-pd-prodqc}

Prior to the start of fabrication, a manufacturing and \dword{qc} plan will be developed detailing the key manufacturing, inspection and test steps.  The fabrication, inspection and testing of the components will be performed in accordance with documented procedures. This work will be documented on travelers and applicable test or inspection reports. Records of the fabrication, inspection and testing will be maintained. When a component has been identified as being in noncompliance to the design, the nonconforming condition shall be documented, evaluated and dispositioned as use-as-is (does not meet design but can meet functionality as is), rework (bring into compliance with design), repair (will be brought into meeting functionality but will not meet design) and scrap. For products with a disposition of accept as is or repair, the nonconformance documentation shall be submitted to the design authority for approval.   

All \dword{qc} data  (from assembly and pre- and post-installation into the \dword{apa}) will be directly stored to the DUNE database for ready access of all \dword{qc} data.  Monthly summaries of key performance metrics (TBD) will be generated and inspected to check for quality trends.

Based on the \dword{pdsp} model, we expect to conduct the following production testing:
\begin{itemize}
\item Dimensional checks of critical components and completed assemblies to insure satisfactory system interfaces.
\item Post-assembly cryogenic checkouts of \dword{sipm} mounting PCBs (prior to assembly into \dword{pd} modules).
\item Cryogenic testing of completed modules (in CSU CDDF or similar facility) to provide a final pre-shipping module test.
\item Warm scan of complete module using motor-driven \dword{led} scanner (Or UV \dword{led}  array).
\item Complete visual inspection of module against a standard set of inspection points, with photographic records kept for each module.
\item End-to-end cable continuity and short circuit tests of assembled cables.
\item \Dword{fe} electronics functionality check.
\end{itemize}

\subsection{Installation Quality Control}
\label{sec:fdsp-pd-prodqc}

\dword{pd}S pre-installation testing will follow the model established for \dword{pdsp}.  Prior to installation in the \dword{apa}, the \dword{pd} modules will undergo a warm scan in a scanner identical to the one at the \dword{pd} module assembly facility and the results compared.  In addition, the module will undergo a complete visual inspection for defects and a set of photographs of all optical surfaces taken and entered into the \dword{qc} record database.  Following installation into the \dword{apa} and cabling an immediate check for electrical continuity to the \dwords{sipm} will be conducted.

It is expected that following the mounting of the TPC cold electronics and the photon detectors the entire \dword{apa} will undergo a cold system test in a gaseous argon cold box, similar to that performed during \dword{pdsp}.  During this test, the \dword{pd}S system will undergo a final integrated system check prior to installation, checking dark and \dword{led}-stimulated \dword{sipm} performance for all channels, checking for electrical interference with the cold electronics, and confirming compliance with the detector grounding scheme.

\section{Safety}
\label{sec:fdsp-pd-safety}

Safety management practices will be critical for all phases of the photon system assembly and testing.  Planning for safety in all phases of the project, including fabrication, testing and installation will be part of the design process.  The initial safety planning for all phases will be reviewed and approved by safety experts as part of the initial design review.  All component cleaning, assembly, testing  and installation procedure documentation will include a section on safety concerns relevant to that procedure, and will be reviewed during the appropriate pre-production reviews.

Areas of particular importance to the \dword{pd}S include:
\begin{itemize}
\item Hazardous chemicals (particularly WLS chemicals such as TPB used in radiator bar dipping and spraying) and cleaning compounds:  All chemicals used will be documented at the consortium management level, with MSDS and approved handling and disposal plans in place.

\item Liquid and gaseous cryogens used in module testing:  Full hazard analysis plans will be in place at the consortium management level for all module or module component testing involving cryogenic hazards, and these safety plans will be reviewed in the appropriate pre-production and production reviews

\item High voltage safety:  Some of the candidate \dwords{sipm} require bias voltages above \SI{50}{VDC}, which may be a regulated voltage as determined by specific labs and institutions.  Fabrication and testing plans will demonstrate compliance with local HV safety requirements at the particular institution or lab where the testing or operation is performed, and this compliance will be reviewed as part of the standard review process.

\item UV and VUV light exposure:  Some \dword{qa} and \dword{qc} procedures used for module testing and qualification may require use of UV and/or VUV light sources, which can be hazardous  to unprotected operators.  Full safety plans must be in place and reviewed by consortium management prior to beginning such testing.

\item Working at heights, underground:  Some aspects of \dword{pd}S module fabrication, testing and installation may require working at heights, or deep underground.  Safety considerations will be taken into consideration during design and planning for these operations, all procedures will be reviewed prior to implementation, and all applicable safety requirements at the relevant institutions will be observed at all times.

\end{itemize}

\section{Organization}
\label{sec:fdsp-pd-org}

\subsection{Consortium Organization}
\label{sec:fdsp-pd-org-consortium}

The \single \dword{pds} consortium follows the typical organizational structure of DUNE consortia:
\begin{itemize}
\item A consortium lead 
provides overall leadership for the effort, and attends meetings of the DUNE Executive and Technical Boards.
\item A technical lead 
provides technical support to the consortium lead, attends the Technical Board and other project meetings, oversees the project schedule and \dword{wbs}, and oversees the operation of the project working groups.  In the case of the \dword{pds}, the technical lead is supported by a deputy technical lead. 
\end{itemize}

\begin{dunefigure}[PDS consortium organization chart]{fig:pds-org.pdf}
{\Dword{pds} consortium organization chart.}
 \includegraphics[width=0.8\columnwidth]{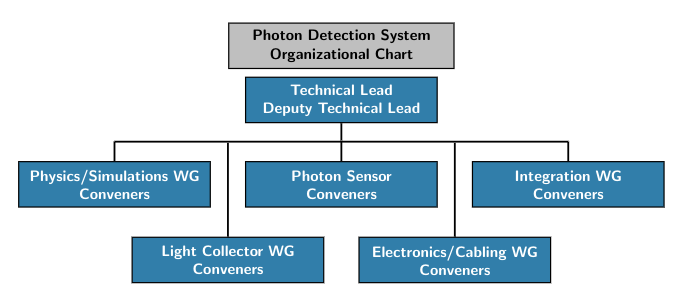} 
\end{dunefigure}

Below the leadership, the consortium is divided up into five working groups, each led by two or three working group conveners as shown in Figure~\ref{fig:pds-org.pdf}. 
Each working group is charged with one primary area of responsibility within the consortium, and the conveners report directly to the technical lead regarding those responsibilities.  As the consortium advances to a more detailed \dword{wbs} and project schedule, it is envisioned that each working group will be responsible for one section of those documents.

The working group conveners are appointed by the \dword{pds} project lead and technical lead, and the structure may evolve as the consortium matures and additional needs are identified. 

\subsection{Planning Assumptions}
\label{sec:fdsp-pd-org-assmp}

Plans for the \dword{pds} consortium are based on the overall schedule for the DUNE \dword{fd}. In particular, the \dword{apa} schedule defines the time window for the completion of the final development program for the light collectors: A final down-select to a baseline light collector option, photosensors, and \dword{fe} electronics must be made by late February 2019.  Due to the early stage of development for the ARAPUCA light collector system, we may maintain an alternate light collector option up to the pre-production review in September of 2020, but all other systems must be defined prior to the \dword{tdr}.

For planning purposes, we assume that the \dword{pds} modules will undergo final assembly and testing at one or more \dword{pds} assembly facilities, with an initial  assembly rate of approximately twenty modules per week, accelerating to forty modules per week in the second half of module fabrication.

We further assume that the modules will be shipped from the fabrication facilities to a detector integration facility, \fixme{to the \dword{itf}?} 
at a site to be determined later, to be integrated along with the \dword{ce} into the \dword{apa} frames and cold tested in a cryogenic test facility.  We plan for an initial rate of two \dwords{apa} per week, with the possibility of accelerating to four \dwords{apa} per week as production lessons are learned.  \Dword{pds} personnel will be present at the integration facility to oversee the installation and testing.

Meeting this timeline requires that the development of the ARAPUCA system be aggressively pursued throughout 2018, with a goal of testing near-final prototypes in the late fall of 2018 and allowing technology comparisons between the ARAPUCA and the light guide technologies in winter of 2019.

Additional development efforts prior to the \dword{tdr} will focus on:

\begin{itemize}
\item Identifying and selecting reliable cryogenic photosensor (\dword{sipm}) candidates,
\item Reducing cost and optimizing performance of \dword{fe} electronics,
\item Solidifying \dword{pds} performance requirements from additional physics simulation efforts.
\end{itemize}

We assume that apart from these items, where rapid development is still required, most of the detector components to be delivered by the \dword{pd} consortium will require only minor changes relative to the \dword{pdsp} components. For this reason, modifications of these other detector components will be delayed until 2019, which will also help with the funding profile. Exceptions will be made for further development in test stands with regard to cabling studies, and for the interface engineering required to ensure satisfactory integration of the \dword{pd} with the \dword{apa} and \dword{ce}  systems.


\subsection{High-Level Schedule}
\label{sec:fdsp-pd-org-cs}



\begin{dunetable}[Pre-\dword{tdr} key milestones.]
{ll}
{fig:pds-pretdrkeymilestones}
{Pre-\dword{tdr} key milestones.}
Milestone													&	Date	       \\ \toprowrule
Preliminary \dword{pd} technology selection criteria determined				&	03/21/18	\\ \colhline
Results from final prototype light collector studies available			&	02/21/19	\\ \colhline
Final \dword{pd} technology selection criteria available						&	02/21/19	\\ \colhline
Down-select to primary (and potential alternate) light collector technology	&	02/22/19	\\ \colhline
Submit initial \dword{tdr} draft for internal review							&	03/29/19	\\ 
\end{dunetable}

High-level post-\dword{tdr} milestones are listed in Table~\ref{fig:pds-posttdrkeymilestones}.


\begin{dunetable}[Post-\dword{tdr} key milestones.]
{ll}
{fig:pds-posttdrkeymilestones}
{Post-\dword{tdr} key milestones.}
Milestone											&	Date	       \\ \toprowrule
\dword{pd} pre-production review(s) complete					&	03/2020 	\\ \colhline
Initial \dword{pd} module fabrication begins						&	09/2020	\\ \colhline
Final \dword{pd} production review based on initial production \dword{qa}		&	02/2021	\\ \colhline
First  \dword{pd} modules delivered for installation				&	05/2021	\\ \colhline
Installation into \dwords{apa} begins							&	06/2021     \\ \colhline
\dword{pd} fabrication complete (first \dword{spmod})			&	07/2023	\\ 
\end{dunetable}

\cleardoublepage

\chapter{Data Acquisition System}
\label{ch:fdsp-daq}

\section{Data Acquisition (DAQ) System Overview}
\label{sec:fd-daq-ov}

\metainfo{DP/SP shared.  Georgia Karagiorgi and Dave Newbold. 2 Pages - largely
  generic but some highlighting of SP-specifics. 
  Focus on describing to HEP but non-DAQ expert. 
  Include how design is resilient in the face of potential
  uncertainties such as excess noise or the need to reduce drift HV
  (just two examples, maybe there are more).}

\subsection{Introduction}
\label{sec:fd-daq-intro}

The DUNE \dword{fd} \dword{daq} system must enable the readout,
triggering, processing and distribution to permanent storage of data
from all \dwords{detmodule}, which includes both their electrical
\dword{tpc} and optical \dword{pds} signals.  
The final output data must retain, with very high efficiency and low
bias, a record of all activity in the detector that pertains to the
recognized physics goals of the DUNE experiment. 
The practical constraints of managing this output requires that the
\dword{daq} achieve these goals while reducing the input data volume by almost four
orders of magnitude.

The current generation of \dword{lartpc} \dwords{daq}, such as used in
\dword{protodune} and \microboone, produce data spanning a fixed window of
time that is chosen based on the acceptance of an external trigger. 
The DUNE \dword{daq} faces several major challenges beyond those of the
current generation. 
Foremost, it must accept data from about two orders of magnitude more
channels and from that data it must form its own triggers.
This self-triggering functionality requires immediate processing of
the full-stream data from a large portion of all TPC channels with a
throughput of approximately one terabyte per second per
\dword{detmodule}. 
From this data stream, triggers must be raised based on two very
different patterns of activity. 
The first is activity 
localized in a small region of one
\dword{detmodule}, such as due to beam neutrino interactions or the
passage of relatively rare cosmic-ray muons. 
This activity tends to correspond to a relatively large deposition of
energy, around \SI{100}{\MeV} or more. 
The second pattern that must lead to a trigger is lower energy activity
dispersed in both time and spatial extent of the \dword{detmodule}, such as due to a 
\dword{snb}.

The 
\dword{daq} must also contend with a higher order of
complexity compared to the current generation. 
The \dword{fd} is not monolithic but ultimately will consist
of four \dwords{detmodule} each of \nominalmodsize fiducial mass. 
Each module will 
implement somewhat different 
technologies and the
inevitable asymmetries in the details of how data are read out from
each must be absorbed by the unified \dword{daq} at its front end. 
Further, each \dword{detmodule} is not monolithic but has at least one
layer of divisions, here generically named \dwords{detunit}. 
For example, the \dword{sp} \dword{detmodule} has \dwords{apa} each
providing data from a number of \dwords{wib} and the \dword{dp} \dword{detmodule} has
\dword{cro} and \dword{lro} units associated with specific electronics
crates.
In each \dword{detmodule}, there are on the order of \num{100} \dwords{detunit}
(\num{150} for \dword{sp} and \num{245} for \dword{dp}) and each unit has a
channel count that is of the same order as that of an entire \lartpc
detector of the current generation.
The DUNE \dword{daq}, composed of a cohesive collection of \dword{daq} instances
called
\dwords{daqpart}, must run on a subset of all possible
\dwords{detunit} for each given \dword{detmodule}. 
Each instance effectively runs independently of all the others, however
some instances indirectly communicate through the exchange of
high-level trigger information. 
This allows, for example, each \dword{detmodule} to take data in
isolation. It also allows for all \dwords{detmodule} to contribute to forming and
accepting global \dword{snb} triggers, and to simultaneously run small portions -- consisting of a few \dwords{detunit} -- separately in
order to debug problems, run calibrations or 
 perform other activities while not interfering with nominal data taking in order to maintain high uptime.

Substantial computing hardware is required to provide the processing
capability needed to identify such activity while keeping up with the
rate of data.
The nature of various technical, financial and physical constraints
leads to the need for much of the computing hardware 
required for this processing
to reside underground, near the \dwords{detmodule}. 
In such an environment, power, cooling, space, and access is far more
costly than in typical data centers. 

Past \lartpc and \dword{lbl} neutrino detectors have successfully
demonstrated external triggering using information related to their beam. 
The DUNE \dword{fd} \dword{daq} will accept external information on recent
times of Main Injector beam spills from \fnal. 
This will assure triggering with high efficiency to capture activity
pertaining to interactions from the produced neutrinos. 

However, even if the DUNE experiment were interested only in 
neutrinos from 
beam spills, an external beam
trigger alone would not be sufficient. 
Absent any other information, such a trigger must inevitably call for
the readout of all possible data from the \dword{fd} 
over at least one \lartpc drift time.
This would lead to an annual data volume approaching an exabyte
($10^{18}$ bytes), the vast majority of which would consist of just noise. 
This entire data volume would have to be saved to permanent storage
and then processed offline in order to get to the signals.

DUNE's physics goals of course extend beyond beam-related interactions, including
cosmic-ray muons, which provide an important
source of detector calibration, and atmospheric neutrino interactions,
which give a secondary source from which to measure neutrino
properties. 
Taken together, 
recording their activity will
dominate the data rate.
The \dword{daq} must also record data with sensitivity to rare interactions
(both known and hypothetical) such as nucleon decay, other baryon
number violating processes (such as neutron-antineutron oscillation),
and interactions from the products of \dwords{snb} as well as possibly
being able to observe isolated low-energy interactions from solar
neutrinos and diffuse supernova neutrinos. 

Some of these events, while rare in themselves, produce patterns of
activity that can be mimicked by other higher-rate backgrounds, particularly
in the case of \dwords{snb}. 
While the exact processes involved in \dwords{snb} are not fully understood,
it is expected that a prolonged period of activity of many tens of
seconds will occur over which their neutrino interactions may be
observed. 
Individually, these interactions will be of low energy (relative to
that of beam neutrino interactions, for example), and will be spread
over time and over the bulk of the \dwords{detmodule}. 
Because of their signature and their importance, special attention is
required to first ascertain that a \dword{snb} may be occurring and to save as
much data as possible over its duration.

Thus the \dword{daq} must greatly reduce the full-stream of its input data
while using the data itself to do so. 
It must do this efficiently both in terms of recording essentially all
activity important to the physics goals of DUNE and in terms of a rate of data output 
that is manageable.  
To perform these primary duties the \dword{daq} 
provides run
control, configuration management, monitoring of both its processes
and the general health of the data, and a user interface for these activities.

\subsection{Design Considerations}
\label{sec:fd-daq-des-consid}

The different \dwords{detmodule} vary in terms of their
readout technology and schemes, timing systems, channel counts and data
throughput and format.
These aspects determine the nature of the digital data input
to the \dword{daq}. 
The design of the \dword{daq} strives to contain the unique layers that adapt
to the variation in the \dwords{detmodule} toward its front end in
order to allow as many of its back end components to remain as identical across
the \dwords{detmodule} as possible. 
In particular, the \dword{daq} must present a unified interface to the
ultimate consumer of its data, DUNE offline computing.
It must also accept and process the data from a variety of other
sources including the accelerator, various calibration systems
(including laser, \dword{ce}, \dwords{pd}, and potentially
others) as well as trigger sources external to DUNE.
The modular nature of the DUNE \dword{fd} implies that the \dword{daq} instances running
on each module must also exchange trigger information. 
In particular, exchanging module-local \dword{snb} trigger information
will allow higher efficiency for this important physics.
The \dword{daq} must be optimized for the above while also retaining
the flexibility to scale to handle risks such as excess noise,
changes in \dword{hv}, cut network connectivity and other issues that could arise. 

\begin{dunefigure}[DAQ overview]{fig:daq-overview}
  {The high-level, \textit{nominal} design for the DUNE \dword{fd} \dword{daq} in
    terms of data (solid) and trigger (dashed) flows between one
    \dword{daqfrag} \dword{fe} and the trigger processing and event
    building back end for one \dword{daqpart}. 
    Line thickness indicates relative bandwidth requirements.
    Blue indicates where the full data flow for the \dword{daqfrag} is
    concentrated to one endpoint.
    Green indicates final output of normally triggered (non-\dword{snb}) data.
    Red indicates special handling of potential \dword{snb}. 
    Each detector module has specialized implementation of some of these
    high level components, particularly toward the upstream \dword{fe}
    as described in the text. 
    The grayed boxes are not in the \dword{daq} scope.
  }
  \includegraphics[width=\textwidth{}, trim={1cm 0 1cm 0},clip]{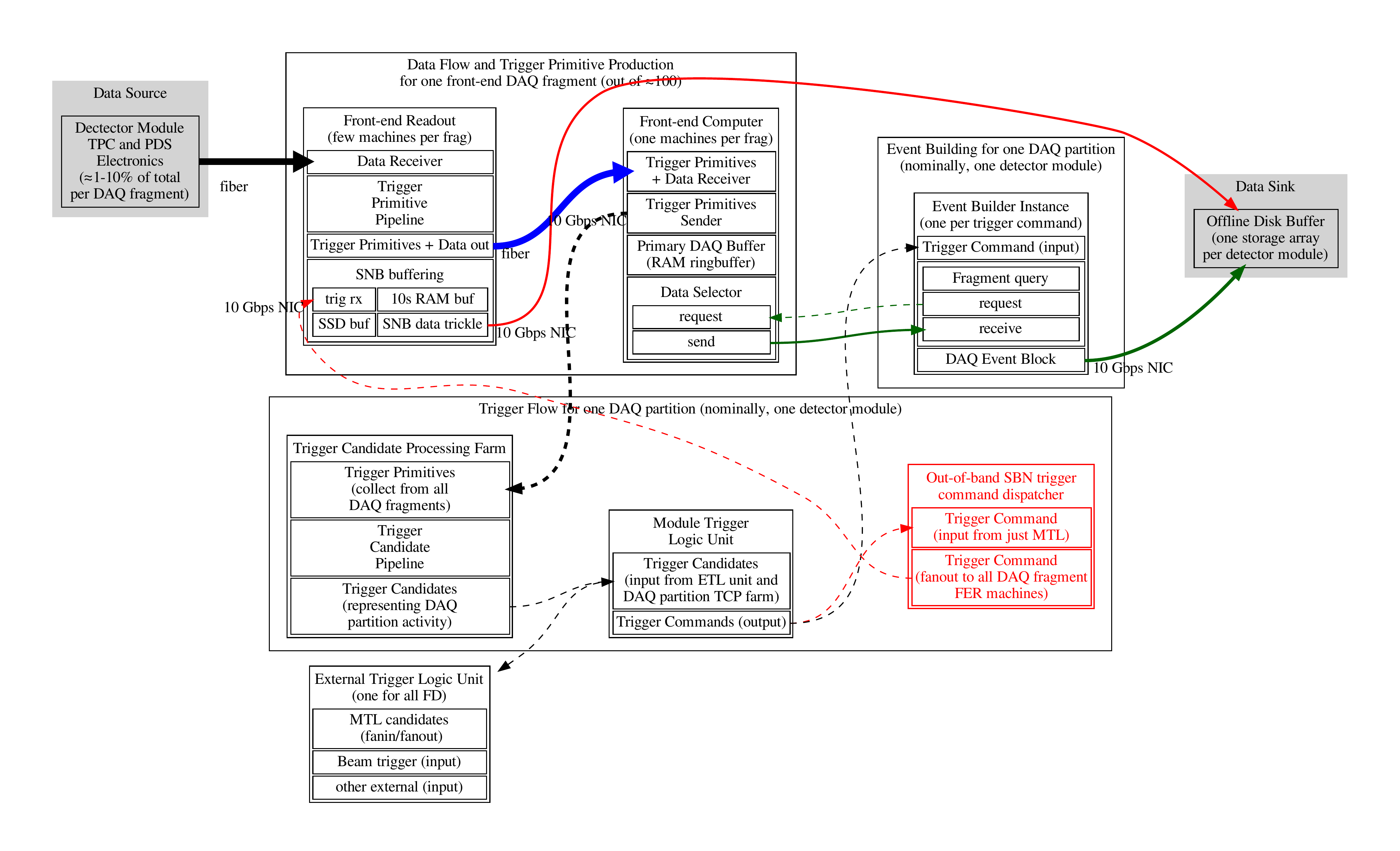}%
\end{dunefigure}

\begin{dunefigure}[\dword{daq} overview]{fig:daq-overview-alt}
  {The high-level, alternate design for the DUNE \dword{fd}\dword{fd} \dword{daq} in
    terms of data (solid) and trigger (dashed) flows between one
    \dword{daqfrag} \dword{fe} and the trigger processing and event
    building back end for one \dword{daqpart}. 
    Line thickness indicates relative bandwidth requirements.
    Blue indicates where the full data flow for the \dword{daqfrag} is
    concentrated to one endpoint.
    Green indicates final output.
    Note, except for a longer readout, \dword{snb} is handled
    symmetric to normal data.
    Each detector module has specialized implementation of some of
    these high level components, particularly toward the upstream
    front-end as described in the text. 
    The grayed boxes are not in the \dword{daq} scope.
  }
  \includegraphics[width=\textwidth{}, trim={1cm 0 1cm 0},clip]{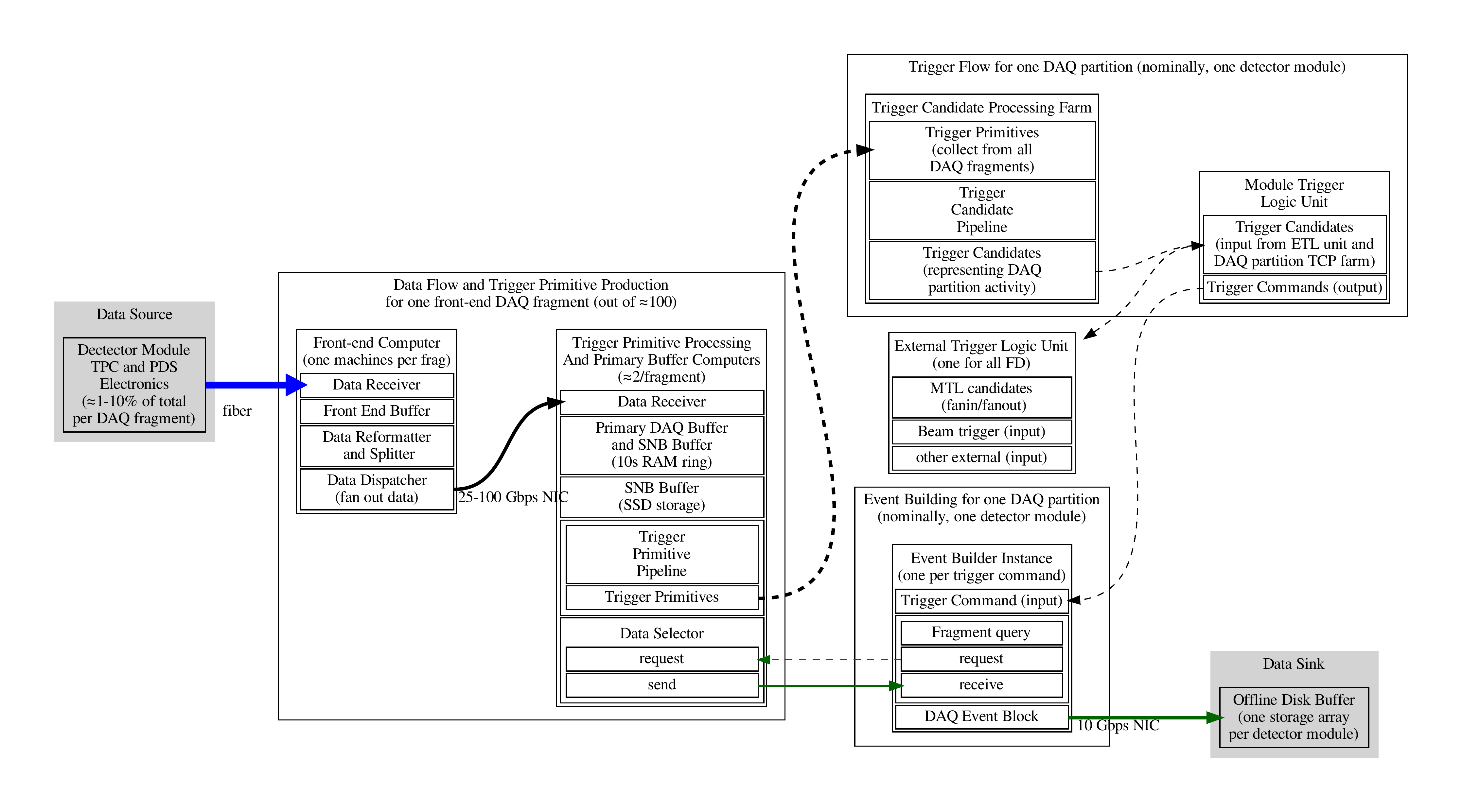}%
\end{dunefigure}

Currently, two major variations for the DUNE \dword{daq} are under consideration. 
The eventual goal is to reduce this to a single high-level design
which will service both \single and \dual \dwords{detmodule} and be reasonably
expected to support the third and forth modules to come.
The first design, designated in this proposal as \textit{nominal}, is
illustrated in a high-level way in terms of its data and trigger flow
in Figure~\ref{fig:daq-overview}. 
The second design, designated as the alternate, is similarly
illustrated in Figure~\ref{fig:daq-overview-alt}. 
The two variants differ largely at their \dwords{fe} in terms of the
order in which they buffer the data received from the \dword{detmodule} 
electronics and use it to form \dwords{trigprimitive}. 
They also differ in how they treat triggering and data flow due to a
potential \dword{snb}. 
As their \dwords{fe} are also sensitive to differences between the
\dword{detmodule} electronics, this further variation for each general
design is described below in Sections~\ref{sec:fd-daq-fero}
and~\ref{sec:fd-daq-fetp} for the \dword{detmodule}  specific to this volume.

At this general high level, the two designs are outlined. 
For both, the diagrams are 
centered on one \dword{daqfrag}
\dword{fe}, which is a portion of the entire \dword{daqpart} servicing a
\dword{detmodule} that has one \dword{fec} accepting about
\numrange{10}{20}~\si{\Gbps} of data (uncompressed rate) from some integral
number of \dwords{detunit}. 
Each of the participating \dwords{daqfrag} do the following: 
\begin{itemize}
\item Accept TPC and \dword{pds} data from the \dwords{detunit} associated with the \dword{daqfrag}.
\item Produce and emit a stream of per-channel \dwords{trigprimitive}.
\item Buffer the full data stream long enough for the \dword{trigdecision} to complete (at least \snbpretime as driven by \dword{snb} requirements).
\item Accept data selection requests and return corresponding \dwords{datafrag}.
\end{itemize}

All participating \dwords{daqfrag} in the particular \dword{daqpart}
(i.e., the \dword{daq} instance) servicing a portion of one \dword{detmodule}
communicate with one trigger processing and event building system.
The trigger processing system must:
\begin{itemize}
\item Receive the stream of per-channel \dwords{trigprimitive} from all \dwords{daqfrag}.
\item Correlate the primitives in time and spatially (across channels), and otherwise use them to form higher-level \dwords{trigcandidate}.
\item Exchange \dwords{trigcandidate} with the \dword{etl}.
\item From them form \dwords{trigcommand}, each of which describes a
  portion of the data in time and a channel to be read out, such that
  no two trigger commands overlap.
\item Dispatch these commands as required (in general to the \dword{eb}).
\end{itemize}
The event building system is responsible for performing the following actions:
\begin{itemize}
\item Accept a trigger command and allocate one \dword{eb} instance to dispatch it.
\item 
Interpret and execute the command by making
  data selection requests to referenced \dwords{daqfrag}.
\item 
Accept the returned \dword{datafrag} from each
  \dword{daqfrag} and combine them into a \dword{rawevent}.
\item Write the result to the \dword{diskbuffer}, which is the boundary
  shared with DUNE offline computing.
\end{itemize}

The nominal and the alternate \dword{daq} designs differ largely in where
the \dword{trigprimitive} and \dword{snb} buffering exist. 
The \textit{nominal} design places these functions in machines
comprising a \dword{daqfer}, which is upstream of the \dword{fec}. 
This then requires the \dword{snb} data and trigger handling to be
different than that for normal (non-\dword{snb}) data. 
When a \dword{snb} trigger command is raised it is forwarded to the
\dword{daqoob} which sends it down to the \dwords{daqfer}. 
After the \dword{snb} data is dumped to \dwords{ssd} it is
``trickled'' out via a path separate from the normal data to the
\dword{diskbuffer}. 
The \textit{alternate} design, on the other hand, places these
functions downstream of the \dword{fec} in trigger processing and data
buffering nodes.
The RAM of these nodes is used to provide the primary \dword{daq} buffer for
normal triggering as well as the deeper buffers needed for \dword{snb}. 
This design handles the \dword{snb} data somewhat
symmetrically with normal data. 
When an \dword{eb} makes a request for \dword{snb} data, it differs
only in its duration, spanning tens of seconds of instead just a few
milliseconds. 
The \dword{fe} buffering nodes, instead of directly attempting to
return the full \dword{sbn} data immediately, streams it to local
\dword{ssd} storage. 
From that storage, the data is sent to the \dword{eb} as low
priority (i.e., also trickled out).
Since the \dword{mtl} ensures no overlapping commands, the buffer
nodes may service subsequent requests from post-dump data that is 
in the RAM buffer.
Since each trigger command is handled by an individual \dword{eb}
instance, the trickle proceeds asynchronously with respect to any subsequent
trigger command handled by another \dword{eb} instances.

Further description of 
these designs
is
given in Section~\ref{sec:fd-daq-design}.

The most critical requirements for the DUNE \dword{fd} \dword{daq} 
are summarized in Table~\ref{tab:daqrequirements}.

\begin{dunetable}[Important requirements on the \dword{daq} system design]
{p{0.2\textwidth}p{0.6\textwidth}}
{tab:daqrequirements}
{Important requirements on the \dword{daq} system design}   
Requirement  & Description \\ \toprowrule
Scalability & The DUNE \dword{fd} \dword{daq} shall be capable of receiving and
buffering the full raw data from all four \dwords{detmodule} \\ \colhline 
Zero deadtime & The DUNE \dword{fd} \dword{daq} shall operate without deadtime under
\textit{normal} operating conditions \\ \colhline
Triggering & The DUNE \dword{fd} \dword{daq} shall provide full-detector triggering
functionality as well as self-triggering
functionality; the data selection shall maintain high efficiency to
physics events while operating within a total bandwidth of \offsitepbpy
for all operating \dwords{detmodule} \\ \colhline
Synchronization & The DUNE \dword{fd} \dword{daq} shall provide synchronization of
different \dwords{detmodule} to within \SI{1}{$\mu$s}, and of different subsystems
within a module to within \SI{10}{ns}\\ 
\end{dunetable}

The input bandwidth and processing needs of the \dword{daq} are expected to be
dominated by the rate of data produced by the TPC system of each
\dword{detmodule}.
These rates vary between the modules and their estimations are summarized in
Table~\ref{tab:daq-input-bandwidth}.
\begin{dunetable} [Pre-trigger data rates from the  \dword{fd} TPCs and
  into \dword{daq} front end.]
  {lll} {tab:daq-input-bandwidth} {The parameters governing the
    pre-trigger data rate from units of each \dword{detmodule} TPC
    \dwords{ce} and the aggregate throughput into the \dwords{fec} of
    the \dword{daq} \dwords{daqfrag}. 
    Compression is an estimate and will be reduced if excess noise is
    introduced.  
  }
  Parameter & \dlong{sp} & \dlong{dp} \\
  \colhline
  TPC unit & \dword{apa} & \dword{cro} crate \\ \colhline
  Unit multiplicity & \num{150} & \num{240} \\ \colhline
  Channels per unit & \num{2560} (\num{800} collection) & \num{640} (all collection) \\ 
  ADC sampling & \SI{2}{\MHz} & \SI{2.5}{\MHz} \\
  ADC resolution & \num{12}\,bit & \num{12}\,bit \\ \specialrule{1.5pt}{1pt}{1pt}
  Aggregate from \dword{ce} & \SI{1440}{\GB/\s} & \SI{576}{\GB/\s} \\
  Aggregate with compression & \SI{288}{\GB/\s} (5$\times$) & \SI{58}{\GB/\s} (10$\times$)  \\
  \colhline
\end{dunetable}

The ultimate limit on the output data rate of the DUNE \dword{fd} \dword{daq} is
expected to be provided by the available bandwidth to the tape,
disk and processing capacity of \fnal. 
An ample guideline has been established that places this limit at
about \offsitepbpy or \offsitegbps.
Extrapolating to four \dwords{detmodule}, this requires a \dword{daq} data reduction factor of almost four orders of magnitude. 
This is achieved through a simple self-triggered readout strategy.

An overestimate of the annual triggered but uncompressed data volume
for one \nominalmodsize  \dword{detmodule} is summarized in
Table~\ref{tab:daq-data-rates}. 
It assumes a very generous and simple trigger scheme whereby the data
from the entire \dword{detmodule} is saved for a period longer than
two drift times around the trigger time.
This essentially removes any selection bias at the cost of
recording a substantial amount of data that will simply contain noise.
Detailed trigger efficiency studies still remain to be performed. 
Initial understanding indicates that trigger efficiency should be near
\SI{100}{\%} for localized energy depositions of at least \SI{10}{\MeV}. 
Sub-\si{\MeV} signals can be ascertained from noise in existing \lartpc{}s
so the effective trigger threshold may be even lower with high
efficiency. 
Of course, data rates rise quickly when the threshold drops into the
range of an \si{\MeV}. 
Additional simulation and use of early data will be used to better
optimize this threshold.

\begin{dunetable} [Uncompressed data rates for one \dword{spmod}.]
  {p{0.30\textwidth}p{0.13\textwidth}p{0.4\textwidth}}
  {tab:daq-data-rates} {Anticipated annual, uncompressed data rates
    for a single \dword{spmod}. The rates for normal (non-\dword{snb} triggers)
    assume a readout window of \SI{5.4}{\ms}. 
    For planning purposes these rates are assumed to apply to a  
    \dword{dpmod} as well, which has a longer readout time but fewer channels. 
    In reality, application of lossless compression is expected
    to provide as much as a $5\times$ reduction in data volume for the \dword{spmod}
    and as much as $10\times$ for the \dword{dpmod}.}   
  Event Type  & Data Volume \si{\PB/year} & Assumptions \\ \toprowrule
  Beam interactions & \num{0.03} & \num{800} beam and \num{800} dirt muons; \SI{10}{\MeV} threshold in coincidence with beam time; include cosmics\\ \colhline
  Cosmics and atmospherics & \num{10} &  \SI{10}{\MeV} threshold, anti-coincident with beam time \\ \colhline
	 Front-end calibration & \num{0.2} & Four calibration runs per year, \num{100} measurements per point \\ \colhline
 Radioactive source calibration & \num{0.1} & Source rate $\le$\SI{10}{Hz}; single fragment readout; lossless readout \\ \colhline
 Laser calibration & \num{0.2} & \num{1e6} total laser pulses, lossy readout \\ \colhline
 Supernova candidates & \num{0.5} & \num{30} seconds full readout, average once per month \\ \colhline
 Random triggers & \num{0.06} & \num{45} per day\\ \colhline
 Trigger primitives & $\le$\num{6} &  All three wire planes; \num{12} bits per primitive word; \num{4} primitive quantities; $^{39}$Ar-dominated\\ \colhline
\end{dunetable}

The data volume estimates also assume that any excess noise beyond
what is expected due to intrinsic electronics noise will not lead to
an increase in trigger rates. 
If, for example, excess noise occurs such that it frequently mimics
more than about \SI{10}{\MeV} of localized ionization, this would
lead to an increase in various types of triggers and subsequently more
data.
However, at the same time, these estimates do not take into account
that some amount of lossless compression of the TPC data will be
achieved. 
In the absence of excess noise it is expected that a compression
factor of at least $5\times$ can be achieved with the \single data and up
to $10\times$ may be achieved with the \dual data, although the actual  
factor achieved will ultimately depend on
the level of excess noise experienced in each \dword{detmodule}. 
Studies using data from the DUNE \dword{35t} and early \microboone
running have shown that a compression factor of at least $4\times$ can
be expected even in the case of rather high levels of excess noise.

One category that will be particularly sensitive to excess noise is the trigger primitives. 
As discussed further in Section~\ref{sec:fd-daq-fetp}, their primary
intended use is as transient objects produced and consumed locally
and directly by the \dword{daq} in the \dword{trigdecision} process. 
However, as their production is expected to be dominated by $^{39}$Ar
decays (absent excess noise) they may carry information that proves
very useful for calibration purposes. 
Future studies with simulation and with early data will determine 
 the most feasible methods to exploit this data. 
These may include committing all or a portion to permanent storage or
potentially developing processes that can summarize their data while
still retaining information salient to calibration.

Finally, it is important to note that early data will be used to
evaluate other selection criteria. 
It is expected that efficient and bias-free selections can be
developed and validated that save a subset of the entire
\dword{detmodule} for any given trigger type. 
For example, a cosmic-muon trigger command for a \dword{spmod} will indicate which \dwords{apa}
contributed to its formation (i.e., which ones had local ionization activity). 
This command can then direct reading out these \dwords{apa}, possibly also
including their neighbors, while discarding the data from all other
\dwords{apa}. 
This may reduce the estimated \SI{10}{\PB/year} for cosmics and
atmospherics by an order of magnitude. 
A similar advanced scheme can be applied to the 
\dword{dpmod} by retaining data for the given readout window from
only the subset of \dword{cro} crates (and again, potentially their
nearest neighbors) that contributed to the formation of the given
trigger.

\subsection{Scope}
\label{sec:fd-daq-scope}


The nominal scope of the \dword{daq} system is illustrated in
Figure~\ref{fig:daq-overview} by the white boxes. 
It includes the continued procurement of materials for, and the
fabrication, testing, delivery and installation of the following
systems:

\begin{itemize}
\item \dword{fe} readout (nominal design) or trigger farm (alternate
  design) hardware and firmware or software development for
  \dword{trigprimitive} generation.
\item \dword{fe} computing for hosting of \dword{daqdr}, \dword{daqbuf} and \dword{daqds}.
\item Back-end computing for hosting \dword{mtl}, \dword{eb} and the \dword{daqoob} processes.
\item External trigger logic and its host computing.
\item Algorithms to generate trigger commands that perform data selection.
\item Timing distribution system.
\item \dword{daq} data handling software including that for receiving and building 
  events.
\item The \dword{om} of \dword{daq} performance and data content.
\item Run control software, configuration database, and user interface
\item Rack infrastructure in the \dword{cuc} for readout
  electronics, \dword{fe} computing, timing distribution, and data
  selection.
\item Rack infrastructure on surface at \surf for back-end computing.
\end{itemize}

\section{DAQ Design}
\label{sec:fd-daq-design}

\metainfo{16 Pages.  This section is mostly DP/SP shared but with some subsections broken out.  This file is \texttt{far-detector-single-phase/chapter-fdsp-daq/design.tex}}

\subsection{Overview}
\label{sec:fd-daq-overview}

The design for the \dword{daq} has been driven by finding a cost-effective solution that satisfies the requirements. Several design
choices have 
been made and two major variations remain to
be studied. 
From a hardware perspective, the \dword{daq} design follows a standard HEP
experiment design, with customized hardware at the upstream, feeding
and funnelling (merging) and moving the data into computers. 
Once the data and triggering information are in computers, a
considerable degree of flexibility is available;  the processing
proceeds with a pipelined sequence of software operations, involving
both parallel processing on multi-core computers and switched
networks. The flexibility allows the procurement of computers and
networking to be done late in the delivery cycle of the DUNE
\dwords{detmodule}, to benefit from increased capability of commercial devices
and falling prices.

Since DUNE will operate over a number of decades, the \dword{daq} has been
designed with upgradability in mind. 
With the fall in cost of serial links, a guiding principle is to
include enough output bandwidth to allow all the data to be passed
downstream of the custom hardware.
This allows the possibility for a future very-fast farm of computing
elements to accommodate new ideas in how to collect the DUNE data. 
The high output bandwidth also gives a risk mitigation path in case
the noise levels in a part of the detector are higher than specified
and higher than tolerable by the baseline trigger decision mechanism;
it will allow additional data processing infrastructure to be added
(at additional cost).

Digital data will be collected from the TPC and \dword{pd}
readout electronics of the \single and \dual
\dwords{detmodule}. 
These categories of data sources are viewed as essentially four types
of \dwords{submodule} within the \dword{daq} and follow the same overall
data collection scheme as shown for the nominal design in
Figure~\ref{fig:daq-overview} and for the alternate design in
Figure~\ref{fig:daq-overview-alt}. 
The readout is arranged to allow making a \dword{trigdecision} 
in a hierarchical manner. 
Initial inputs are formed at the channel level, then combined at the
\dword{detunit} level and again 
combined at the
\dword{detmodule} level.
In addition, the \dword{trigdecision} process combines 
information at this level that may come from the other \dwords{detmodule} as well as
information from sources external to the \dword{daq}. 
This hierarchical structure in forming and consuming triggers 
allows safeguards to be developed so that any problems in one cavern or
in one \dword{detunit} of one \dword{detmodule} need not overwhelm the
entire \dword{daq}.
It also allows a \dword{snb} to be recorded in all
operational parts of the detector while others may be down for
calibration or maintenance.

Generally speaking, the \dword{daq} consists of data flow and trigger flow.
The trigger flow involved in self-triggering originates from
processing a portion of the data flow. 
The trigger flow is then consumed back by the \dword{daq} in order to govern
what portion of the data flow is finally written out to permanent
storage. 
The nominal and alternate designs differ in where in the data flow
the trigger flow originates. 

In both designs, a single \dword{daqfrag} associates an integral
number of \dwords{detunit} with one \dfirst{fec}.
This fragment forms one conceptual unit of the \dword{fe} \dword{daq}.
The processing on a \dword{fec} is kept minimal such that each has a
throughput limited by I/O bandwidth. 
The recently released PCIe v4 doubles the bandwidth from the prior
version and thus we assume that $\approx$\SI{20}{\GB/\s} throughput (out of
a theoretical \SI{32}{\GB/\s} max) can be achieved based on tests
using PCIe v3.
In principle then, this allows one \dword{fec} to accept the data
from: two (if uncompressed) or ten (if $5\times$ compressed) of the
\num{150} \single \dwords{apa}, ten of the \num{240} \dual \dword{cro} crates
given their nominal $10\times$ compression or the uncompressed data
from all five \dword{dp} \dword{lro} crates.

In the nominal design, the data enters the \dword{daq} via the fragment's
\dfirst{daqfer} component.
In the \dword{sp} the \dword{daqfer} consists of eight \dwords{rce}
and in the \dword{dp} it consists of a number of \dword{bow}
computers, (see Section~\ref{sec:fd-daq-fero} in each respective \dword{detmodule} volume).
The \dword{daqfer} is responsible for accepting that data and from it
producing channel level \dwords{trigprimitive}.
It is also responsible for forwarding compressed data and the
primitives to the \dfirst{daqdr} in the corresponding \dword{fec}.
The \dword{daqfer} is also responsible for supplying transient memory
(RAM) and non-volatile buffer in the form of \dword{ssd} sufficient
for \dword{snb} triggering and readout.
The \dword{daqdr} accepts the full data stream and transfers it to the
\dlong{daqbuf} of its \dword{daqfrag}. 
There it is held awaiting a query from the \dfirst{eb}. 
When the \dword{eb} receives a \dword{trigcommand} it uses the
included information to query all appropriate \dwords{daqds} and from
their returned \dwords{datafrag} an \dword{rawevent} is built and
written to file on the \dword{diskbuffer}. 
From there the data becomes responsibility of the offline group to
transfer to \fnal for permanent storage and further processing.

In the alternate design, the data is accepted directly by the
\dfirst{daqdr} in a \dword{fec} from the detector electronics
for the particular \dword{detmodule}.
The data then flows into the \dword{daqbuf} and the portion required
for forming trigger primitives is dispatched to the trigger computers
of the fragment for the production of \dwords{trigprimitive}.
Current \dword{ssd} technology may allow \dword{ssd} to be directly mounted to the
\dword{fec} to provide for the \dword{snb} dump buffer. 
Another solution, which puts less pressure on write throughput, is to
distribute the \dword{ssd} for the \dword{snb} dumps to the trigger computers. 
In order to supply enough CPU for trigger primitive pipelines it is
expected that at least two hosts per \dword{fec} will be needed.
While their CPUs are busy finding trigger primitives, their I/O
bandwidth will be relatively unused and thus they provide synergistic,
cost-effective hosting for the \dword{ssd}s.

Regardless of where the \dwords{trigprimitive} are produced in either
the nominal or alternate design, they are further processed at the
\dword{daqfrag} level to produce \dwords{trigcandidate}. 
At this level, they represent possible activity localized in time and
by channel to a portion of the overall \dword{detmodule}.
The \dwords{trigcandidate} emitted by all \dwords{daqfrag} are sent to
the \dfirst{mtl} associated with the \dword{daqpart}.
There, they are time ordered and otherwise processed to form
\dwords{trigcommand}.
At this level they represent activity localized across the
\dword{detmodule} and over some period of time.

The \dword{daqpart} (or \dword{daq} instance) just introduced is the cohesive
collection of \dword{daq} parts. 
One \dword{daqpart} operates essentially independently from any other,
and there is typically one per \dword{detmodule}. 
In some cases multiple \dwords{daqpart} may operate simultaneously in
a \dword{detmodule}, such as when some fraction of \dwords{detunit}
are undergoing isolated testing or calibration.

Each \dword{trigcommand} is consumed by a single \dword{eb} instance
in order to query back to the \dwords{daqfrag} of its \dword{daqpart}
as described above.
In addition, the \dword{mtl} of one module is exchanging messages in
the form of \dwords{trigcandidate} with the others. 
For example, one module may raise a local \dword{snb}
\dword{trigcandidate} and forward it to all other modules.
Each module is also emitting candidates to sinks and accepting them
from sources of external trigger information.

The exact implementation of some of these high-level functions,
particularly those near the \dword{fe}, depends on the particular
\dword{detmodule}. 
The required specialization and in general, more implementation-level
details are described in the following sections.
Subsequent description proceeds toward the \dword{daq} back end including
processes handling dataflow, triggering, event building and data
selection.

\subsection{Front-end Readout and Buffering}
\label{sec:fd-daq-fero}

\metainfo{Giles Barr \& Giovanna Miotto \& Brett Viren, this is SP-specific.}

\begin{dunefigure}[Nominal \single \dword{fe} \dword{daq} fragment]{fig:daq-readout-buffering-baseline}
  {Illustration of data (solid arrows) and trigger (dashed) flow for
    one \single \dword{daq} fragment (two APAs) in the nominal design. 
    Black arrows indicate normal data and trigger flow and red indicate special flow
    for handling of a potential \dword{snb}.  } 
  \includegraphics[width=0.95\textwidth]{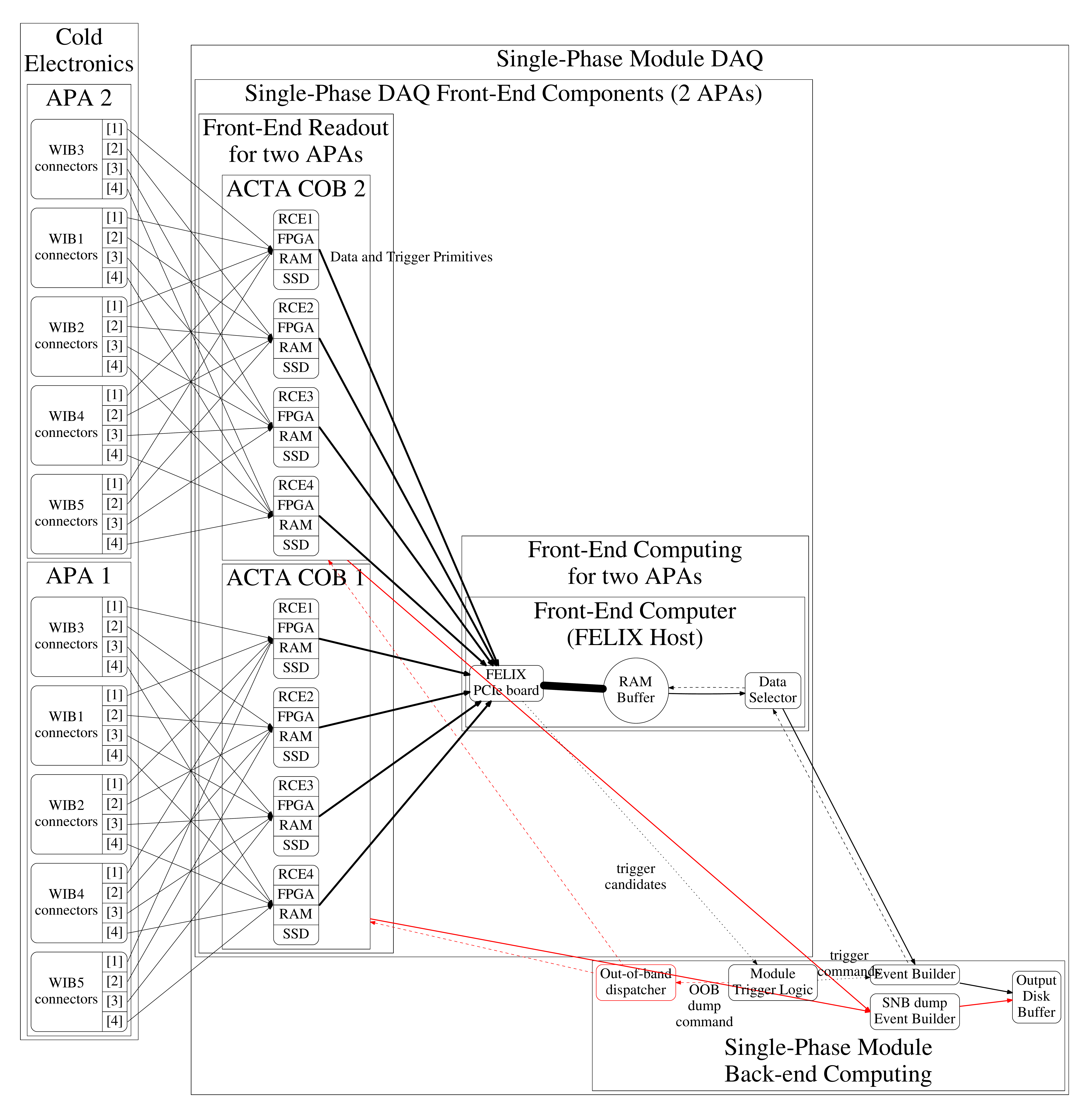}%
\end{dunefigure}

Figure~\ref{fig:daq-readout-buffering-baseline} illustrates the
\dword{sp}-specific \dword{daqfrag} specializing from the generic,
nominal design illustrated in Figure~\ref{fig:daq-overview}.  
Starting from the left, it shows the fiber optic connectivity pattern
between the four connectors of five \dword{sp} \dwords{wib} associated
with each \dword{apa} and the elements of one \dword{sp} \dfirst{daqfer}. 
In total, the \dword{daqfer} is associated with two \dwords{apa}, each of which
is serviced by one \dshort{atca} \dfirst{cob} hosting four compute
units called \dfirst{rce}. 
Each \dword{rce} provides \dword{fpga}, RAM and \dword{ssd} resources.
Its primary functions include:
\begin{itemize}
\item Receive data from \dwords{wib},
\item Produce trigger primitives from collection channels (see Section~\ref{sec:fd-daq-fetp}),
\item Compress the data,
\item Forward data and trigger primitives to the \dword{fec},
\item Buffer data in RAM, 
\item Stream data from RAM to \dwords{ssd} on receipt of a \textit{dump}
  trigger command (such as is raised by an \dword{snb} candidate),
\end{itemize}

The data and trigger primitives from the eight \dwords{rce} are
aggregated to a single \dword{felix} PCIe board residing in the
\dword{fec}. 
This is done via \num{16} \SI{10}{\Gbps} optical fibers. 
With no data compression performed in the \dwords{rce} the bandwidth
of these fibers will be close to saturated. 
If excess noise is not greater than experienced by \microboone{}, then a
lossless compression factor of at least five may be expected.
If achieved, the total throughput into \dword{felix} from the two \dwords{apa}
is expected to be about \SI{4}{\GB/s}

The firmware running on the \dword{felix} \dword{fpga} transfers the data
and trigger streams to the host system RAM. 
This type of transfer has been demonstrated by ATLAS with a PCIe v3
\dword{felix} board at a throughput up to \SI{10}{\GB/\s}. 
The next generation of \dword{felix} based on PCIe v4 
is expected to
obtain about a factor of two improvement.
The \dwords{trigprimitive} are combined across channels to form 
\dwords{trigcandidate} that will be sent to the \dword{mtl}.
Meanwhile, the data stream streams into a \dword{daqbuf}. 
This buffer will be sized sufficient to retain the full data stream
for the period of time needed for a \dword{trigdecision} to be made. 
As described above, that decision culminates in a \dword{trigcommand}
that is sent to an \dword{eb}. 
Based on its information, the \dword{eb} makes a request to the
\dword{daqds} representing the \dword{daqfrag}, and 
the \dword{daqds}  replies with a 
\dword{datafrag} build from the data available in the \dword{daqbuf}.

An \dword{snb} \dword{trigcommand} is formed via the usual trigger
hierarchy, as described in 
Section~\ref{sec:fd-daq-fetp},
and is consumed by the \dlong{daqoob} component. 
This component simply dispatches the command back down to the \num{600}
\dwords{rce} in order to relieve the duty from the \dword{mtl}, 
thus avoiding a source of trigger latency.
This means an \dword{snb} trigger command is serviced differently than
are all the other types of trigger commands. 
The RAM on board the \dwords{rce} is used to store the full data
stream long enough for an \dword{snb} \dword{trigcommand} to be formed
and distributed. 
It has been estimated that the rise time to detect an \dword{snb} in
the \dword{sp} \dword{detmodule} is about \SI{1}{\s}, so the RAM
must be sized to buffer at least this much data.
\Dword{snb} models differ on the total duration over which significant
neutrino interactions may be expected, as well as to their possible
time profiles. 
Some allow for \dword{snb} neutrino interactions to occur for some
time but at a rate not sufficient to rise above a trigger threshold
determined by \dword{snb} backgrounds. 
It is assumed that an \dword{snb} dump should start about \snbpretime
before the \dword{snb} trigger and should span a total \snbtime.
During the dump, all data is sent to both the \dword{ssd} storage
distributed among the \dwords{rce}. 
Data also continues to flow to the associated \dword{daqfer} as during
nominal running, which ensures  that no dead time is suffered for all other
non-\dword{snb} triggers. 
The multiplicities of \dwords{rce} and \dwords{ssd} are such that
uncompressed throughput of the \dword{snb} dump will just saturate
current technology. 
Given a lossless compression factor of five the throughput to each \dword{ssd} 
is expected to be \SI{500}{\MB/\s}.

Given the infrequency of detectable \dwords{snb} the average
\dword{snb} trigger rate is effectively governed only by a chosen
threshold and by the rate of background from radiological decays,
neutrons from cosmic-ray muons and fluctuations of noise, especially
any coherent excess noise. 
The threshold must be tuned to maintain high efficiency for a broad
class of \dword{snb} models while also not flooding the \dword{daq} and
potentially offline computing. 
This needs further study, but for the purposes of illustration a
nominal false positive rate of one \dword{snb} dump per month is
assumed. 
Uncompressed, this results in \SI{540}{\TB/\year}. 
Each dump will take up \SI{75}{\GB} on an \dword{ssd}, each of which is
expected to provide \SI{500}{\GB} of storage; this is enough for
at least six dumps.
Again, lossless compression is expected to achieve a factor of five.

The \dword{snb} dumps are expected to remain on the \dword{ssd}
storage for some time in order to perform checks on the data to either
rule out and delete the data or accept the candidate and migrate its
data to permanent storage. 
This data migration is done out-of-band of the connection to the
\dword{felix} board using a local network connection to the
\dshort{atca} crate. 
With the assumed average of one dump per month, if all were saved
uncompressed it would require an average bandwidth aggregated over the
entire \dword{spmod} of just over \SI{100}{\Mbps}.

In the alternate \dword{sp} \dword{daq} design (not diagrammed), which
corresponds to the generic Figure~\ref{fig:daq-overview-alt}, the ten
\dwords{wib} shown in Figure~\ref{fig:daq-readout-buffering-baseline}
are directly connected to one \dword{felix} board via \SI{10}{\Gbps} fiber
optic.  The readout and buffering then follow the generic design.

\subsection{Front-end Trigger Primitive Generation}
\label{sec:fd-daq-fetp}

\metainfo{Josh Klein \& J.J. Russel \& Brett Viren.  This section is SP-specific.}

\Dwords{trigprimitive} are generated inside the \dword{fe} readout
hardware associated with each \dword{apa} from TPC data on a per-channel basis.
They are sent along with the waveform data to the \dword{fe} \dword{daq}
computing.
In the alternate design, the data is directly sent from the \dwords{apa} to
the \dwords{fec} and the data are sent to the trigger processing
computers.  
In both designs, the primitives from one \dword{daqfrag} are further
processed to produce \dwords{trigcandidate}. 
As such they represent a localization of activity on the corresponding
\dwords{apa}  for a given period of time. 
These candidates are emitted to the \dword{mtl}, which may consider
candidates from other \dwords{detmodule} or external sources before
generating trigger commands.
Section~\ref{sec:fd-daq-sel} describes the selections involved in this
triggering.

Only the \num{480} collection channels associated with each \dword{apa} face are
used for forming \dwords{trigprimitive}. 
Reasons for this limitation include the fact that collection
channels:

\begin{itemize}
\item have higher signal to noise ratios compared to induction channels;
\item are fully and independently sensitive to activity on their \dword{apa} face;
\item have unipolar signals that give direct approximate measures
  of ionization charge without the costly computation that would be needed to 
  deconvolve the field response functions required for the the induction channels;
\item can be easily divided into smaller, independent groups in order
  to better exploit parallel processing.
\end{itemize}

Figure~\ref{fig:daq-readout-buffering-baseline} illustrates the connectivity between the
four connectors on each of the five \dwords{wib} and the \dword{fe} readout hardware.
The data is received via eighty \SI{1}{Gbps} fiber optical links by four \dwords{rce}
in the \dword{atca} \dword{cob} system. 

Due to the pattern of connectivity between \dwords{wib} and \dwords{rce}, each \dword{rce} receives the data from the collection channels that cover one contiguous half of one \dword{apa} face. 
Each \dword{rce} has two primary functions. 
The first is transmission of all data as described in
Section~\ref{sec:fd-daq-hlt}. 
The second is to produce \dwords{trigprimitive} from its portion of the
collection channel data.
The algorithms to produce the \dwords{trigprimitive} still require
development but can be broadly described, as follows.

\begin{enumerate}
\item On a per channel basis, calculate a rolling baseline and spread
  level that characterizes recent noise behavior such that the result
  is effectively free of influence from actual ionization signals.
\item Locate contiguous runs of \dword{adc} samples that are above a threshold
  defined in terms of the baseline and noise spread.
\item Emit their time bounds and total charge as a \dword{trigprimitive}.
\end{enumerate}

Each \dword{trigprimitive} represents ionization activity localized
(relatively) along the drift direction by the times at which the
signal crosses threshold and by two planes parallel to the collection
wire and located midway between the wire and each of its neighboring
wires.
Depending on the threshold set, these \dwords{trigprimitive} may be
numerous due to $^{39}$Ar decays and noise fluctuations.
If their rate cannot be sustained, the threshold may be raised or
further processing may be done, still at the \dword{apa} level, that 
considers more global information.
This may be performed either in the \dwords{rce} or later in the
\dword{fe} computing hosts. 
In either case, the results are in the form of \dwords{trigcandidate},
which are sent to the \dword{mtl}.

Sources of \dword{rf} emission inside the cryostat are minimized by
design. 
Any residual \dword{rf} is expected to be picked up coherently across
some group of channels. 
Depending on its intensity, additional processing of the collection
waveforms must be employed to mitigate this coherent noise and this
must occur before the data is sent for \dword{trigprimitive}
production. 
If the required mitigation algorithms outgrow the nominally specified
\dword{rce} \dword{fpga} it is possible to double the number of
\dwords{cob} per \dword{apa}, which would require a redistribution of fibers. 
Alternatively, or in addition, the higher number of
\dwords{trigprimitive} produced as a result of excess noise can be
passed along for further processing in the \dword{fe} computing. 
This would require reprocessing the raw waveform data.

\subsection{Dataflow, Trigger and Event Builder}
\label{sec:fd-daq-hlt}

\metainfo{Giles Barr \& Josh Klein \& Giovanna Miotto \& Kurt Biery \& Brett Viren.  This is a DP/SP shared section.  It's file is \texttt{far-detector-generic/chapter-generic-daq/design-flow.tex}}

In the general data and trigger flow diagrams for the nominal
(Figure~\ref{fig:daq-overview}) and alternate
(Figure~\ref{fig:daq-overview-alt}) designs, the dataflow, trigger and
event builder functions take as input data from the \dword{detmodule}
electronics and culminate in files deposited to the \dword{diskbuffer} for
transfer to permanent storage by offline computing processes.  
The continuous, uncompressed data rate of the input from one
\dword{detmodule} is on the order of \SI{1}{\TB/\s}. 
The final output data rate, for all \dwords{detmodule} operating at
any given time is approximately limited to \offsitegbyteps. 

To accept this high data-inflow rate and to apply the substantial
processing needed to achieve the required reduction factor, which is on the 
order of \num{1000}, the \dword{daq} follows a distributed design.
The units of distribution for the front end of the \dword{daq} must match up
with natural units of the \dword{detmodule} providing the data. 
This unit is called the \dword{daqfrag} and each accepts input at a
rate of about \numrange{10}{20}\,\si{\GB/\s}. 
The exact choice maps to some integral number of physical
\dword{detmodule} units (e.g., \dword{sp} \dwords{apa} or \dword{dp}
\dwords{cro} and \dwords{lro}).

As described in the previous sections, the nominal and alternate
designs differ essentially in the order and manner in which the
\dword{snb} buffering occurs and the \dwords{trigprimitive} are
formed. 
The overall data flow, higher level triggering and building of
\textit{event} data blocks for final writing are conceptually very similar.
This processing begins with the data being received by the
\dword{felix} PCIe board hosted in the \dword{fec}. 
The \dword{felix} board performs a DMA transfer of the data into the
\dword{daqbuf} for the \dword{daqfrag}, which resides in the
\dword{fec} host system RAM.  
This buffer is sized to hold ten seconds of data assuming the maximum 
uncompressed input rate associated with the fragment.
While data is being written to the buffer, a delayed portion is
also being read in order to dispatch it for various purposes.
Any and all requests to further dispatch a subset of this data from
the \dword{daqbuf} must arrive within this buffer time.
In the nominal design, the only dispatching will be from a request
made by an \dword{eb} (described more below) upon receipt of a
\dword{trigcommand}. 
In the alternate design, a suitable fraction of the data is also
dispatched via high bandwidth (at least \numrange{25}{50}\,\si{\Gbps} simplex, less
if data is compressed at this stage) network connections to a trigger
farm so that \dwords{trigprimitive} may be formed. 
Whether the primitives are formed in this manner or extracted from the
stream sent by the \dword{daqfer} (as in the nominal design) these
trigger primitives from one \dword{daqfrag} are collectively sent for
further processing in order to be combined across channels and  to then 
produce \dwords{trigcandidate}. 
These are finally combined for one \dword{detmodule} in the
\dword{mtl}. 
It is in the \dword{mtl} where \dwords{trigcandidate} from additional
sources are also considered, as described in
section~\ref{sec:fd-daq-sel}.

In both the nominal and alternate designs the dispatch of data
initiated by normal (non-\dword{snb}) \dwords{trigcommand} is
identical. 
This dispatch, commonly termed \textit{event building} involves collection
of data spanning an identical and continuous period of time from
multiple \dwords{daqbuf} across the \dword{daq}.
As introduced above, each \dword{trigcommand} is consumed by an
\dword{eb} process. 
It  uses fragment address information in the \dword{trigcommand} to
query the \dword{daqds} process representing each referenced
\dword{daqfrag} and accepts the returned a \dword{datafrag}.
In the exceptional case that the delay of this request is so large
that the \dword{daqbuf} no longer contains the data, then an error
return is supplied and recorded by the \dword{eb} in place of the
lost data. 
Such failures lead to indicators displayed by the detector operation
monitoring system.
The \dword{eb} finally assembles all responses into a
\dword{rawevent} and writes it to file on the \dword{diskbuffer} where
it becomes the responsibility of DUNE offline computing.

The \dword{daqds} and \dword{eb} services are implemented using
 the general-purpose \fnal data acquisition framework \dword{artdaq} for
distributed data combination and processing. 
It is designed to exploit the parallelism that is possible with
modern multi-core and networked computers, and has been used in \dword{protodune} and other experiments.
The \dword{artdaq} framework is the principal architecture that will be used for the DUNE \dword{daq} back-end computing.
The authors of \dword{artdaq} have accommodated DUNE-specific 
requests for feature additions. Also, a number of libraries have been developed based on
existing parts of \dword{artdaq} used to handle incoming data from data
sources. 
It is likely that future DUNE extensions will be made by one of these two
routes.

Unlike the dispatch of data initiated by a normal \dwords{trigcommand},
a command formed to indicate the possibility of a \dword{snb} is
handled differently between the nominal and alternate designs. 
Such a command is interpreted to save all data from all channels
for a rather extended time of \snbtime starting from \snbpretime
before the time associated with the \dword{trigcommand}. 
As no data selection is being performed, given the required bandwidth, special buffering to nonvolatile storage, in the form of \dword{ssd}, is required.  
Today's technology supplies individual \dword{ssd} in the M.2 expansion card form factor,
which supports individual write speeds up to \SI{2.5}{\GB/\s}. 
The two designs differ as to the location of and data source for these
buffers.

In the nominal design, these \dwords{ssd} reside in the \dword{daqfer}
as described in Section~\ref{sec:fd-daq-fero}. 
In that location, due to larger granularity of computing units, the
data rate into any one \dword{ssd} is within the quoted write
bandwidth. 
However, and as shown in Figure~\ref{fig:daq-overview}, the data and
trigger flow for \dword{snb} in the nominal case takes a special
path. 
Instead of an \dword{eb} consuming the \dword{trigcommand} as
described above, it is sent to the \dfirst{daqoob}, which dispatches
it to each \dword{daqfer} unit hosting an \dword{ssd}. 
This component is used 
to immediately free up the \dword{mtl}
to continue to process normal triggers.
When the command is received, each host must begin to stream data from
its local RAM, supplying at least \snbpretime of buffer to the
\dword{ssd}, and continue until the full \snbtime has elapsed. 
While it is performing this dump it must continue to form
\dwords{trigprimitive} and pass them and the full data stream to the
connected \dword{fec}.

In the alternate design the same \dword{daqbuf} provides the
\snbpretime of pre-trigger \dword{snb} buffering. 
As in the nominal case, it must rely on fast, local \dword{ssd}
storage to sink the dump. 
Current \dword{ssd} technology allows four M.2 \dword{ssd} devices to
be hosted on a PCIe board. 
Initial benchmarks of this technology show that such a combination can
achieve \SI{7.5}{\GB/\s} write bandwidth, which is short of linear
scaling. 
To support the maximum of \SI{20}{\GB/\s}, three such boards would be
required.
The alternate design presents a synergy between the need to dump
high-rate data and the need to provide CPU to form the
\dwords{trigprimitive}. 
With current commodity computing hardware it is expected that each
\dword{fec} will need to be augmented with about two computers in the trigger
farm. 
These trigger processors will need to accept the entire \dual and
three-eighths of the \single data stream from their \dword{daqfrag}. 
If they instead accept the entire stream, they can also provide
RAM buffering and split up the data rate, which must be sunk to
\dword{ssd} buffers.

In both designs, the data dumped to \dword{ssd} may contain precious
information about a potential \dword{snb}. 
It must be extracted from the buffer, processed and either discarded
or saved to permanent storage. 
The requirements on these processes are not easy to determine.
The average period between actual \dwords{snb} to which DUNE is
sensitive is measured in decades. 
However, to maintain high efficiency for capturing such important
physics, the thresholds will be placed as low as feasible, limited
only by the ability to acquire, validate and (if validated) write out the
data to permanent storage. 
Notwithstanding, the (largely false positive) \dword{snb} trigger rate is
expected to be minuscule relative to normal triggers.
Understanding the exact rate requires more study, including using
early data, but for planning purposes it is simply assumed that one
whole-detector data dump will occur per month on average.
Using the \dword{spmod} as an example, and choosing the
nominal time span for the dump to be \snbtime{}, about \spsnbsize of
uncompressed data would result.
In the nominal \dword{sp} \dword{daq} design, this dump would be spread over
\num{600} \dword{ssd} units leading to \SI{75}{\GB} per \dword{ssd} per dump.
Thus, typical \dwords{ssd} offer storage to allow any given dump to be
held for at least one half year before it must be purged to assure
storage is available for subsequent dumps.
If every dump were to be sent to permanent storage, it would represent
a sustained \SI{0.14}{\Gbps} (per \dword{detmodule}), which is a small
perturbation on the bandwidth supplied throughout the \dword{daq} network. 
Saved to permanent storage this rate integrates to \SI{0.5}{\PB/\year}, 
which while substantial, is a minor fraction of the total data budget.
The size of each dump is still larger than is convenient to place into
a single file, so the \dword{snb} event-building will likely differ from
that for normal triggers in that the entire dump is not held in a
single \dword{rawevent}. 
Finally, it is important to qualify that these rates assume
uncompressed data. 
At the cost of additional processing elements, lossless compression
can be expected to reduce this data rate by \numrange{5}{10}\,$\times$ or
alternatively allow lower thresholds that lead to the same factor of 
more
dumps.
Additional study is required to optimize the costs against the expected
increase in sensitivity.


\subsection{Data Selection Algorithms}
\label{sec:fd-daq-sel}

\metainfo{Josh Klein \& Brett Viren.  This is a shared DP/SP section.  It's file is \texttt{far-detector-generic/chapter-generic-daq/design-sel.tex}}

Data selection follows a hierarchical design. 
It begins with forming \dword{detunit}-level \dwords{trigcandidate}
inside the \dword{daqfrag} \dword{fe} computing using channel-level
\dwords{trigprimitive}. 
These are then used to form \dword{detmodule} \dwords{trigcommand}
in the \dword{mtl}.
When executed, they lead to readout of a small subset of the total
data. 
In addition, \dwords{trigcandidate} are provided to the
\dword{mtl} from external sources such as the \dword{etl} in order to
indicate external events such as beam spills, or \dword{snb} candidates
detected by the other \dwords{detmodule}. 
In addition to supplying triggers to \dword{snews}, triggers from
\dword{snews} or other cosmological detector sources such as LIGO and
VIRGO can be accepted in order to possibly record low-energy or
dispersed activity that would not pass the self-triggering. 
The latency of arrival for these sources must be less than the nominal
\snbpretime buffers used to capture low-level early \dword{snb}
activity.
A \dword{hlt} may also be active within the \dword{mtl}. 
The hierarchical approach is natural from a design standpoint and 
it allows for vertical slice testing and running multiple
\dwords{daqpart} simultaneously during commissioning of the system or
when debugging of individual \dwords{detunit} is required.

As discussed in Sections~\ref{sec:fd-daq-fero}
and~\ref{sec:fd-daq-fetp}, \dwords{trigprimitive} are generated in
either in \dwords{daqfer} (in the nominal design) or in trigger
processing computers (in the alternate design). 
In both designs, and for both \dword{sp} and \dword{dp}
\dwords{detmodule}, only data from TPC collection channels (three-eighths of \single and all of \dual channels) feed
the self-triggering, as their waveforms directly supply a measure of
ionization activity without computationally costly signal processing.
The \dwords{trigprimitive} contains summary information for each 
channel, such as the time of any threshold-crossing pulse, its
integral charge, and time over threshold. 
A channel with an associated \dword{trigprimitive} is said to be
\textit{hit} for the time spanned by the primitive. 
Trigger primitives from one \dword{detunit} are then further processed
to produce a \dword{trigcandidate}. 
The candidate represents a cluster of hits across time and
channel, localized to the \dword{detunit}.
The candidates from all \dwords{daqfrag} are passed to the
\dword{mtl}.

The \dword{mtl} arbitrates between various trigger types, determines
trigger priority and ultimately the time range and detector coverage
for a \dword{trigcommand}, which it emits back to the \dwords{fec}.
The \dword{mtl} assures that no \dwords{trigcommand} are issued that 
overlap in time or in detector channel space.
It also may employ a \dword{hlt} to reduce or aggregate triggers into
fewer \dwords{trigcommand} so as to optimize the subsequent readout. 
For example, aggregating many small readouts into fewer but larger
ones may allow for more efficient processing.   This can be particularly
important during periods of high-rate activity due to e.g., various
backgrounds or instrumental effects.

When activity leads to the formation of a \dword{trigcommand} this
command is sent down to the \dwords{fec} instructing which slice of
time of its buffered data should be saved. 
The \dword{trigcommand} information is saved along with this data. 
At the start of DUNE data taking, it is anticipated that for any given
single-interaction trigger (a cosmic-ray track, for example), waveforms
from all channels in the \dword{detmodule} will be recorded over a one
\dword{readout window} (nominally, \spreadout for \dword{sp} and
\dpreadout for \dword{dp}, chosen to be two drift times  plus an
extra \SI{20}{\%}). 

Such an approach is clearly very generous in terms of the amount of
data saved, but it ensures that associated low-energy physics (such as
captures of neutrons produced by neutrino interactions or cosmic rays)
are recorded without any need to fine-tune \dword{detunit}-level
triggering, and does not depend on the noise environment across
\dwords{detunit}. 
In addition, the wide \dword{readout window} ensures that the data of
all associated activity is recorded.
As generous as it is, it is estimated that this \dword{readout window}
will not produce an unmanageable volume of data.
As shown in Table~\ref{tab:daq-data-rates}, the uncompressed selected
data from the \dword{spmod} will fill about half of the
nominal annual data budget. 
The longer \dual drift and its fewer channels will give approximately the
same data rate. 
However, once a modest amount of lossless compression is applied, the
nominal data budget can be met. 
Early running will allow experience to be gained and more advanced
data selection algorithms to be validated allowing the \dword{daq} to discard
the many \dwords{datafrag} in each trigger consistent
with just electronics noise. 
This has the potential for a reduction of at least another factor of ten.

Other trigger streams -- calibrations, random triggers, and prescales
of various trigger thresholds -- are also generated at the
\dword{detmodule} level, and filtering and compression can be applied
based upon the trigger stream. 
For example, a large fraction of random triggers may have \dword{zs}
applied to their waveforms, reducing the data volume substantially, as
the dominant data source for these will be $^{39}$Ar events.
Additional signal-processing can also be done on particular trigger
streams if needed and if the processing is available, such as fast
analyses of calibration data.

At the \dword{detmodule} level, a decision can also be made on whether
a series of interactions is consistent with an \dword{snb}. 
If the number of \dword{detunit}-level, low-energy
\dwords{trigcandidate} exceeds a threshold for the number of such
events in a given time, a trigger command is sent from the \dword{mtl}
back to the \dwords{daqfer}, which store up to \SI{10}{\s} of full
waveform data. 
That data is then streamed to non-volatile storage to allow for
subsequent analysis by the \dword{snb} working group, perhaps as an
automated process. 
If not rejected, it is sent out of the \dword{daq} to permanent offline
storage.

In addition, the \dword{mtl} passes \dwords{trigcandidate} up to a
detector-wide \dword{etl}, which among other functions, can decide
whether, integrated across all modules, enough \dwords{detunit} have
detected interactions to qualify as an \dword{snb}, even if within a
particular module the threshold is not exceeded. 
\Dwords{trigcandidate} from the \dword{etl} are passed to the
\dword{mtl} for dispatch to the \dwords{fec} (or \dwords{daqfer} in the
case of \dword{snb} dump commands in the nominal design). 
That is, to the \dword{mtl}, an \dword{externtrigger} looks like just
one more \textit{external} trigger input.

\Dword{detunit} level \dwords{trigcandidate} are generated within
the context of one \dword{daqfrag}, specifically in each \dword{fec}. 
The trigger decision is based on the number of nearby channels
hit in a given fragment within a time window (roughly \SI{100}{\micro\s}),
the total charge collected in these adjacent channels, and possibly the
union of time-over-threshold for the \dwords{trigprimitive} in the
collection plane.
Studies show that even for low-energy events (roughly
\SIrange{10}{20}{\MeV}) the reduction in radiological backgrounds is
extremely high with such criteria.
The highest-rate background, $^{39}$Ar, which has an overall rate of
\SI{10}{MBq} within a \nominalmodsize volume of argon, has an endpoint 
of \SI{500}{keV} and requires significant pileup in both space and time to get near
a \SI{10}{\MeV} threshold.
One important background source is $^{42}$Ar, which has a \SI{3.5}{MeV}
endpoint and an overall rate of \SI{1}{kBq}. 
$^{222}$Rn decays via a 
 \SI{5.5}{MeV} kinetic energy $\alpha$  and is
also an important source of background.
The radon decays to $^{218}$Po, which within a few minutes leads to a
\SI{6}{MeV} kinetic energy $\alpha$, and ultimately to a $^{214}$Bi daughter (many
minutes later), which has a $\beta$ decay with its endpoint near \SI{3.5}{MeV}  kinetic energy. 
The $\alpha$ ranges are short, resulting in charge being collected on one or two anode wires at most,
but the charge deposit can be large, and therefore the charge
threshold must be well above the $\alpha$ deposits plus any
pileup from $^{39}$Ar and noise.

At the level of one \dword{detunit}, two kinds of local
\dwords{trigcandidate} can be generated.
One is a high-energy trigger that indicates local ionization
activity corresponding to more than than \SI{10}{\MeV}. 
The per-channel thresholds on total charge and time-over-threshold
will be optimized to achieve at least \SI{50}{\%} efficiency at this energy
threshold, with efficiency increasing to \SI{100}{\%} via a turn-on curve
that ensures at least \SI{90}{\%} efficiency at \SI{20}{\MeV}. 
The second type of trigger candidate 
generated is for
low-energy events between \SI{5}{\MeV} and \SI{10}{\MeV}. 
In isolation, these candidates do not lead to formation of a
\dword{trigcommand}. 
Rather, at the \dword{detmodule} level they are combined, time
ordered and their aggregate rate compared against a threshold based on
fluctuations due to noise and backgrounds in order to form an
\dword{snb} \dword{trigcommand}.

The \dword{mtl} takes as input \dwords{trigcandidate} (both low-energy
and high-energy) from the participating \dwords{daqfrag}, as well as
\dword{externtrigger} sources, such as the \dword{etl}, which includes
global, detector-wide triggers, external trigger sources such as
\dword{snews}, and information about the time of a \fnal beam
spill. 
The \dword{mtl}  also generates \dwords{trigcommand} for internal
consumption, such as random triggers and calibration triggers (for
example, telling a laser system to fire at a prescribed time). 
The \dword{mtl} can also generate \dwords{trigcommand} from a
prescaled fraction of trigger types that otherwise do not generate
such commands on their own. 
For example, a prescaled fraction of single, low-energy trigger
commands could be allowed to generate a trigger command, even though
those candidates normally only result in a trigger command when
aggregated (i.e., as they would be for an \dword{snb}).

The \dword{mtl} is also responsible for checking candidate triggers
against the current \dword{rc} trigger mask: in some runs, for
example, we may decide that only random triggers are accepted, or that
certain \dwords{trigcandidate} streams should not be considered
because their \dwords{daqfrag} have been producing unreasonably large
rates in the recent past (such as may be due to noise spikes, flaky
hardware or buggy software).
In addition, the \dword{mtl} counts low-energy trigger candidates,
and based upon their number and distribution over a long time interval
(e.g., \SI{10}{\s}), decides to generate an \dword{snb} trigger command.
The trigger logic will be optimized to record the data due to at least
\SI{90}{\%} of all Milky Way supernovae, and studies of simple low-energy
trigger criteria show that a much higher efficiency can likely be
achieved.

The \dword{hlt} can also be applied at this level, particularly if
there are unexpectedly higher rates from instrumental or low-energy
backgrounds that require some level of reconstruction or pattern
recognition. 
An \dword{hlt} might also allow for efficiently triggering on
lower-energy single interactions, or allow for better sensitivity for
supernovae originating outside the Milky Way galaxy, by employing a weighting scheme to
individual \dwords{trigcandidate} -- higher-energy
\dwords{trigcandidate} receiving higher weights. 
Thus, for example, two \dwords{trigcandidate} consistent with
\SI{10}{\MeV} interactions in \SI{10}{\s} might be enough to create a
\dword{snb} candidate trigger, while a hundred \SI{5}{\MeV} trigger
candiates in \SI{10}{\s} might not.
Lastly, the \dword{hlt} can allow for dynamic thresholding; for
example, if a trigger appears to be due to a cosmic-ray muon, the
threshold for single interactions can be lowered (and possibly
prescaled) for a short time after that to identify spallation
products. 
In addition, the \dword{hlt} could allow for a dynamic threshold after
a \dword{snb}, to extend sensitivity beyond the \SI{10}{\s}
\dword{snb} \dword{readout window}, while not increasing the data
volume associated with \dword{snb} candidates linearly. 

All low-energy \dwords{trigcandidate} are also passed upwards to the
\dword{etl} so that they may be integrated across all \SI{10}{\kton}
\dwords{detmodule} in order to determine that a \dword{snb} may be
occurring. 
This approach increases the sensitivity to trigger on \dwords{snb} by
a factor of four (for \SI{40}{\kton}), thus extending the burst
sensitivity to a distance twice as far as for a single \nominalmodsize
\dword{detmodule}.

The \dword{mtl} is also responsible for including in the
\dword{trigcommand} a global timestamp built from its input
\dwords{trigcandidate}, and information on what type of trigger was
created. 
Information on \dwords{trigcandidate} is also kept, whether or not
they contribute to the formation of a \dword{trigcommand}. 
As described above, the \dword{readout window} for nominal
\dwords{trigcommand} (those other than for \dword{snb} candidates)
is somewhat more than two times the maximum drift time. 
Further, a nominal readout spans all channels in a \dword{detmodule}. 
The \dword{mtl} is also responsible for sending the trigger commands
that tell the \dwords{daqfer} to stream all data from the past
\snbpretime and for a total of \snbtime in hopes to catch
\dwords{snb}.
This command may be produced based on \dwords{trigcandidate} from
inside the \dword{mtl} itself or it may be produced based on an external
\dword{snb} \dword{trigcandidate} passed to the \dword{mtl} by the
\dword{etl}.

\subsection{Timing and Synchronization}
\label{sec:fd-daq-timing}

\metainfo{David Cussans \& Kostas Manolopoulos.  This is an SP-specific section.  It's file is \texttt{far-detector-single-phase/chapter-fdsp-daq/design-timing.tex}}


All components of the \dword{spmod} are
synchronized to a common clock.  In order make full use of the
information from the \dword{pds}, the common clock must be
aligned within a single \dword{detunit} with an accuracy of $O\,1$\si{ns}.
In order to form a common trigger for \dword{snb} between
\dwords{detmodule}, the timing between them must be aligned with an
accuracy of $O\,1$\si{ms}.  However, a tighter constraint is the need to
calibrate the common clock to universal time (derived from GPS) in
order to adjust the data selection algorithm inside an accelerator
spill, which requires an absolute accuracy of $O\,1$\si{\micro\s}.

\single and \dual \dwords{detmodule} use different timing systems,
driven by the different technical requirements and development history
of the two technologies. 
A \dword{spmod} has many more
timing end points than a \dword{dpmod} and many of the end points
are simpler than the end points in the \dual{}, for example a \dword{wib}
versus \dword{utca} crate. Both systems have been sucessfully prototyped.

The DUNE \dword{spmod} uses a development of the \dword{pdsp} 
timing system. Synchronization messages are transmitted over a serial
data stream with the clock embedded in the data. The format is
described in DUNE DocDB-1651~\cite{docdb-1651}. Figure~\ref{fig:daq-readout-timing}
shows the overall arrangement of components within the \single
Timing System (SPTS). 
\fixme{need a gloss term?}
A stable master clock, disciplined with a \SI{10}{\MHz}
reference is used in the SPTS. A \dword{pps} is
also received by the system and is time-stamped onto a counter clocked
by the SPTS master clock, however the periodic synchronization
messages distributed to the \dword{sp} \dword{detmodule} are an exact number
of clock cycles apart even if there is jitter in the \dword{pps}.

The GPS signal is encoded onto optical fiber and transmitted to the
\dword{cuc}, where it is converted back to an RF signal on coaxial cable and
used as the input to a GPS displined oscillator. The oscillator module
also houses a IEEE 1588 (PTP) grandmaster and an NTP server. The PTP
grandmaster provides a timing signal for the \dword{dp} \dword{wr}
timing network. The NTP server provides an absolute time for the
\dword{pps}. The SPTS relates its time counter onto GPS time by
timestamping the \dword{pps} onto the SPTS time counter and reading
the time in software from the NTP server.

The latency from the GPS antenna on the surface to the GPS receiver in
the \dword{cuc} will be measured by optical time domain reflectometry at
installation. Given the modest absolute time accuracy required
(sufficient to select data within an accelerator spill) dynamic
monitoring of this delay is not required.

The \dword{wr} synchronization signals from the \dword{dp} \dword{detmodule} are
time-stamped onto the SPTS clock domain and the SPTS synchronization
signals are time stamped onto the \dword{dp} clock domain. This allows
the timing in the \dword{sp} and \dword{dp} \dwords{detmodule} to be
aligned. A similar scheme is used to relate the \dword{pdsp}
 timing domain to the beam instrumentation
\dword{wr} time domain.

In order to provide redundancy, and also the ability to easily detect
issues with the timing path, two independent GPS systems are used. One
with an antenna at the head of the Yates Shaft, the other with an
antenna at the head of the Ross Shaft. The two independent timing
paths are brought together in the same rack in the \dword{cuc}. Using 1:2
fiber splitters one SPTS unit can be left as a hot spare while the
other is active. This also allows testing of new firmware and software
during comissioning without the risk of losing the SPTS if a bug is
introduced.

\begin{dunefigure}[Arrangement of components in DUNE timing system]{fig:daq-readout-timing}
  {Illustration of the components in the DUNE timing system.}
\includegraphics[width=0.8\textwidth]{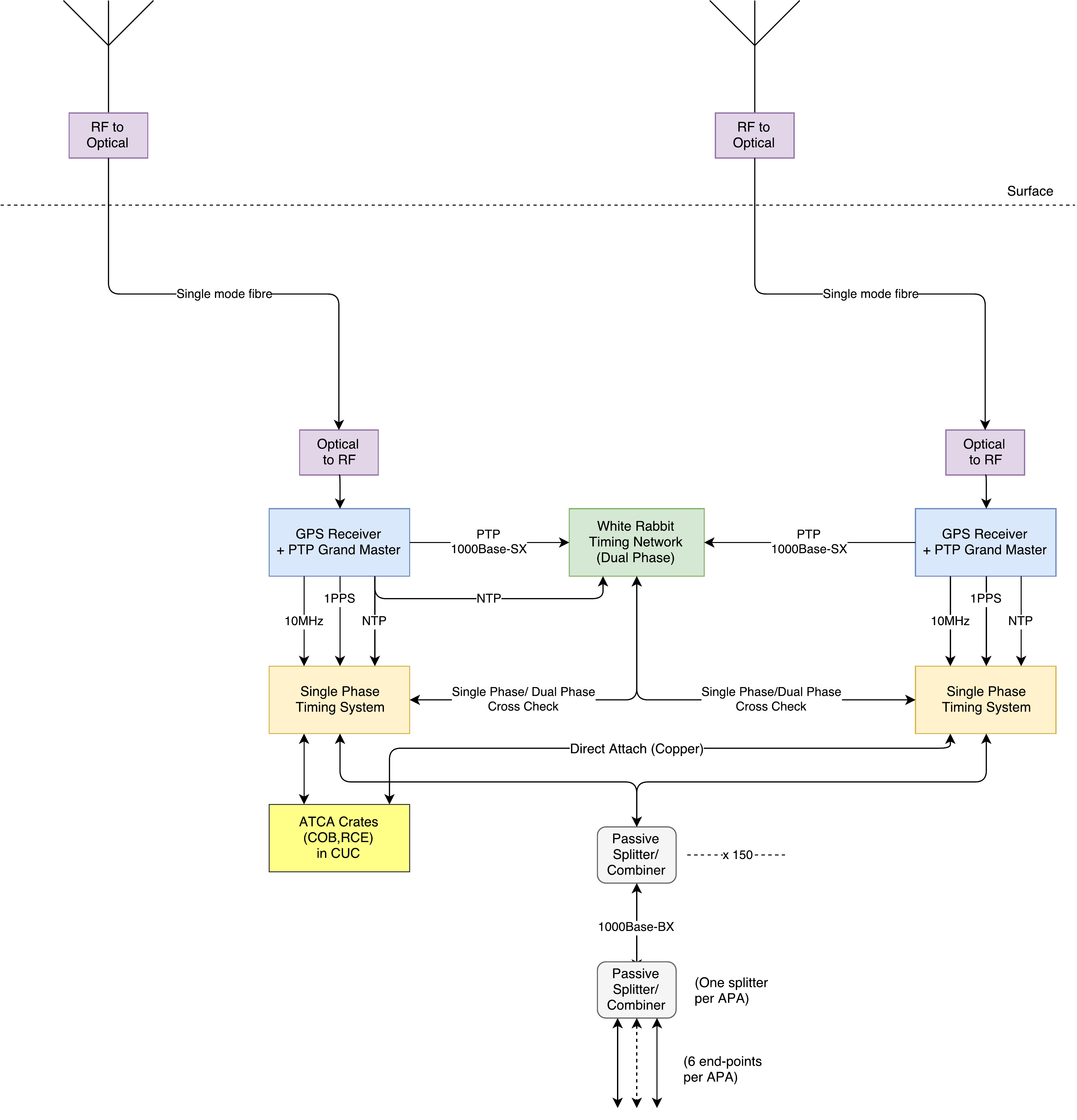}
\end{dunefigure}

All the custom electronic components for the SPTS are contained in two
\dword{utca} shelves. At any one time one is active and the other is a
hot spare. The \SI{10}{\MHz} reference clock and the \dword{pps} are received
by a single width \dword{amc} at the center of the \dword{utca} shelf. This
master timing \dword{amc} produces the SPTS signals and encodes them onto a
serial data stream. This serial datastream is distributed over a
standard star-point backplane to the fanout \dwords{amc}, which each drive the
signal onto up to \num{13} SFP cages. The SFP cages are either occupied by
1000Base-BX SFPs, each of which connects to a fiber running to an \dword{apa},
or to a Direct Attach cable which connects to systems elsewhere in the
\dword{cuc},  i.e., the \dword{rce} crates and the data selection system. This
arrangement is shown in Figure~\ref{fig:daq-readout-sp-timing}

\begin{dunefigure}[Arrangement of components in \dlong{sp} timing system]{fig:daq-readout-sp-timing}
  {Illustration of the components in the \single timing system.}
  \includegraphics[width=0.8\textwidth]{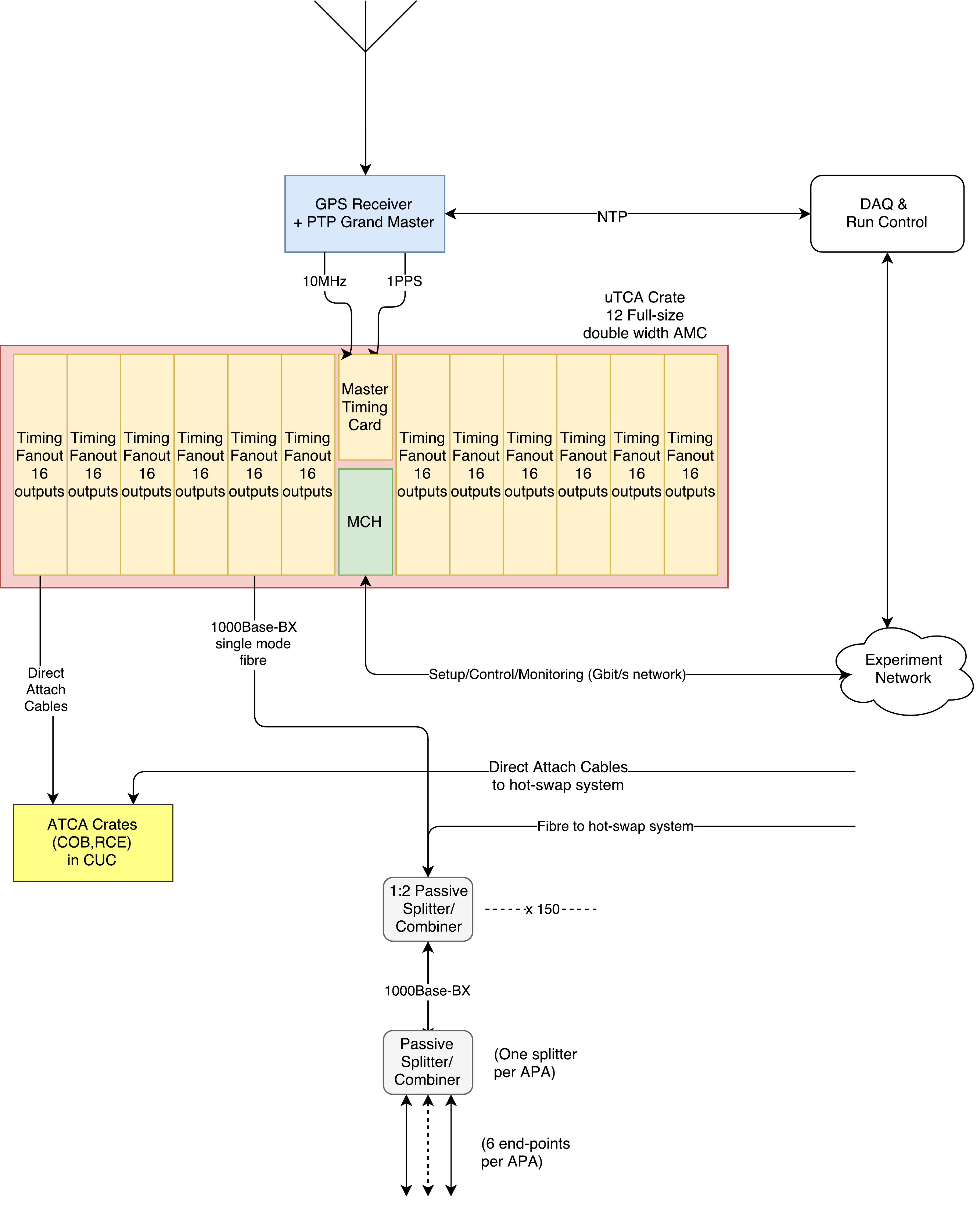}
\end{dunefigure}

\subsubsection{Beam timing}
\label{sec:fd-daq-design-beamtiming}

The neutrino beam is produced at the Fermilab accelerator complex in
spills of \SI{10}{\us} duration. 
A \dword{sls} at the Far Detector site will locate the time periods in
the data when beam could be present, based on network packets received
from Fermilab containing predictions of the GPS-time of spills soon to
occur or absolute time stamps of recent spills. 
Experience from MINOS and \nova shows that this can provide beam
triggering with high reliability with some small fraction of late or
dropped packets.
To improve reliability further, the system outlined here contains an extra layer
of redundancy in the prediction process. 
Several stages of prediction based on recent spill behavior will be applied, aiming
for an accuracy of better than 10\% of a readout
time (sub-\si{\ms}) in time for the data to be selected from
the \dword{daq} buffers. 
Ultimately, an offline database will match the actual time of the
spill with the data, thus removing any reliance on real-time network
transfer for this crucial stage of the oscillation measurements. The
network transfer of spill-timing information is simply to ensure that a
correctly located and sufficiently wide window of data is considered
as beam data. This system is not required, and is not designed to
provide signals accurate enough to measure neutrino time-of-flight.

The precision to which the spill time can be predicted at \fnal
improves as the acceleration process of the protons producing the
spill in question advances.  The spills currently occur at intervals
of \SI{1.3}{\s}; the system will be designed to work with any interval, and
to be adaptable in case the sequence described here changes.  For
redundancy, three packets will be sent to the far detector for each
spill.  The first is approximately \SI{1.6}{\s} before the spill-time, which
is at the point where a \SI{15}{\Hz} booster cycle is selected; from this
point on, there will be a fixed number of booster cycles until the
neutrinos and the time is subject to a few ms of jitter.  The second
is about \SI{0.7}{\s} before the spill, at the point where the main injector
acceleration is no longer coupled to the booster timing; this is
governed by a crystal oscillator and so has a few \si{\us} of jitter.
The third will be at the so called `\texttt{\$74}' signal generated before the beamline kicker magnet fires
to direct the protons at the LBNF target; this doesn't improve the
timing at the Far Detector much, but serves as a cross check for
missing packets.  This system is enhanced compared to that of
MINOS-\nova, which only use a third of the above timing signals.  The
reason for the larger uncertainty in the time interval from \SI{1.6}{\s} to
\SI{0.7}{\s} is that the booster cycle time is synchronized to the
electricity supply company's \SI{60}{\Hz}, which has a variation of about
1\%.

Arrival-time monitoring information from a year of MINOS data-taking
was analyzed, and it was found that 97\% of packets arrived within
\SI{100}{\ms} of being sent and 99.88\% within \SI{300}{\ms}.

The \dword{sls} will therefore have estimators of the GPS-times of
future spills, and recent spills with associated data contained in the
\dwords{daqbuf}. These estimators will improve in precision as
more packets arrive.  The \dword{daq} will use data in a wider window than
usual, if, at the time the trigger decision has to be made, the
precision is lower 
due to missing or late packets.  From the
MINOS monitoting analysis, this 
expected to be very rare.

\subsection{Computing and Network Infrastructure}
\label{sec:fd-daq-infra}

\metainfo{Kurt Biery \& Babak Abi.  This is a DP/SP shared section.  It's file is \texttt{far-detector-generic/chapter-generic-daq/design-compnet.tex}}

The computing and network infrastructure that will be used in each
of the four \dwords{detmodule} is similar, if not identical.
It supports the buffering, data selection, event
building, and data flow functionality described
above, and it includes computing elements that consist of servers that:

\begin{itemize}
\item buffer the data until a \dword{trigdecision}
  is received;
\item host the software processes that
  build the data fragments from the relevant
  parts of the detector into complete events;\fixme{whole detector or module?}
\item host the processes that make the
  \dword{trigdecision};
\item host the data logging processes and
  the disk buffer where the data is written;
\item host the real-time \dlong{dqm} processing;
\item host the control and monitoring processes.
\end{itemize}

The network infrastructure that connects these computers has the following components:

\begin{itemize}
\item subnets for transferring triggered data from the buffer
  nodes to the event builder nodes; these need to
  connect underground and above-ground computers;
\item a control and monitoring subnet that connects all
  computers in the \dword{daq} system and all \dword{fe}
  electronics that support Ethernet communication; this
  sub-network must connect to underground and
  above-ground computers;
\item a subnet for transferring complete events from the
  event builder servers to the storage servers; this subnet
  is completely above-ground.
\end{itemize}

\subsection{Run Control and Monitoring}
\label{sec:fd-daq-tcm}

\metainfo{Giovanna Miotto \& Jingbo Wang.  THis is a DP/SP shared section.  The file is \texttt{far-detector-generic/chapter-generic-daq/design-ctrmon.tex}}


The online software constitutes the backbone of the DUNE \dword{daq}
system and provides control, configuration and monitoring of the data taking in
a uniform way.
It can be subdivided logically into four subsystems: the run control,
the management of the \dword{daq} and \dword{detmodule} electronics configuration, the monitoring, and the non-physics data archival.
Each of these subsystems has a distinct 
function, but their
implementation will share underlying technologies and tools.

In contrast to experiments in which data taking sessions, i.e., runs, are
naturally subdivided into time slots by external conditions (e.g., a
collider fill, a beam extraction period), the DUNE experiment aims to
take data continuously.
Therefore, a classic run control with a coherent state machine and a
predefined and concurrently configured number of active detector and
\dword{daq} elements does not seem adequate. 

The DUNE online software is thus structured according to the
architecture principle of loose coupling: each component has as
little knowledge as possible of other components.
While the granularity of the back-end \dword{daq} components may match the 
individual software processes, for the front-end \dword{daq} a minimum
granularity must be defined, balancing fault tolerance and
recovery capability against the requirement of data consistency.
The smallest independent component is called a \dword{daqfrag}, which
is made up of the \dwords{detunit} associated with a single
\dlong{fec}.
In the nominal design, this corresponds to two \dword{sp} \dwords{apa}
and about ten \dword{dp} \dword{cro} crates.

The concept of a \textit{run} represents a period of time in which the same
\dword{fe} elements are active or the same data selection criteria
are in effect (possibly with maximum lengths for offline processing
reasons). 
More than just orchestrating data taking, the run control
provides the mechanisms allowing \dword{daq} applications to publish
their availability, subscribe to information, and exchange messages. 
In addition, the online software provides a configuration service
for \dword{daq} elements to store their settings and a conditions
archive, keeping track of varying detector electronics settings and
status.

Another important aspect of the online software is the monitoring
service.
Monitoring can be subdivided into two main domains: the monitoring of
the data taking operations (rates, number of \dwords{datafrag}
in flight, error flags, application logs, network bandwidth, 
computing and network infrastructure) and the monitoring of the
physics data.
Both are essential to the success of the experiment and must be 
designed and integrated into the \dword{daq} system from the start.
In particular, for such a large and distributed system, the information sharing and archival system is very important, as are 
scalable and easily accessible data visualization tools, which will evolve during the lifetime of the experiment.
The online software provides the glue that holds the
\dword{daq} applications together and enables 
data taking.
Its architecture guides the
approach to \dword{daq} application design and also shapes the view
that the operators will have of the experiment.

\section{Interfaces} 
\label{sec:fd-daq-intfc}

\metainfo{5 Pages.  This is a shared SP/DP section.  If there are assymetries simply describe both detector modules.}

\fixme{Include an image of each interface in appropriate section.  Can maybe refer to Figure~\ref{fig:daq-overview} but it currently lacks some of the interfaces.}

\subsection{TPC Electronics}
\label{sec:fd-daq-intfc-elec}

Details about the interfaces between the \dword{daq} and the TPC electronics
are documented for the \dword{sp} \dwords{detmodule}
in~\cite{docdb-6742}.

In the case of the \dword{spmod}, data from the \dword{ce} 
\dwords{femb} are 8b10 encoded and sent to the
\dwords{wib} on copper cables at a bit rate of \SI{1.28}{\Gbps}. There
are two options being considered for the \dwords{wib}. In one, the
data are simply converted to optical signals and transmitted to the
\dword{daq} in the \dword{cuc} on \SI{1.28}{\Gbps} optical links with a total of \num{80}
fibres per \dword{apa}. In the second option, the \dwords{wib} aggregate the
data onto links running at $\approx$\SI{10}{\Gbps} before transmission
to the \dword{daq}, with a total of ten fibres per \dword{apa}. In both cases the data
are received on rear transition modules connected to the \dword{cob}
\dword{atca} boards (see Section~\ref{sec:fd-daq-fero}).

\subsection{PD Electronics}
\label{sec:fd-daq-intfc-photon}

Details about the interfaces between the \dword{daq} and \dword{sp}
\dlong{pds} are documented in~\cite{docdb-6727}.


For the \dword{sp} \dword{pds} the \dword{s/n} ratio of the
\dword{sipm} signals is high enough to allow \dlong{zs} to be safely
applied to the data. This reduces the data flow so that a bandwidth of
\SI{8}{\Gbps} per \dword{apa} is sufficent to transfer it to the \dword{daq},
with an order of magnitude safety factor. The link from the \dword{sp}
cryostat to the \dword{cuc} will be implemented as either eight 1000Base-SX
links or a single 10GBase-LR link per \dword{apa}.
The data on the links will be encoded using UDP/IP.

\subsection{Offline Computing} 
\label{sec:fd-daq-intfc-fnal-cmptg}

The interface between the \dword{daq} and offline computing is
described in~\cite{docdb-7123}.
The \dword{daq} team is responsible for reducing the data volume
to the level that is agreed upon by all interested parties, and the
raw data files are transfered from \surf to \fnal using a
dedicated network connection.
A disk buffer is provided by the \dword{daq} on or near the \surf
site to hold 
several days worth of data 
so that the
operation of the experiment is not 
affected if there happens to
be a network disruption between \surf and \fnal.

During stable running, the data volume produced by the
\dword{daq} systems of all four \dwords{detmodule} will be no larger
than \offsitepbpy.
The maximum data rate is expected to be independent of the number of
\dwords{detmodule} that are operational.
During the construction of the second, third, and fourth
\dwords{detmodule}, the extra rate per \dword{detmodule} will be used
to gather data to aid in the refinement of the data selection
algorithms.
During commissioning, the data rate is expected to be higher than
nominal running and it is anticipated that  
a data volume corresponding to (order) one year will be necessary to commission a \dword{detmodule}.

The disk buffer at \surf is planned to be \SI{300}{\TB} in size.
The data link from \surf to \fnal will support \surffnalbw
(\offsitepbpy corresponds to about \offsitegbps).
The offline computing team is responsible for developing the
software to manage the transfer of files from \surf to \fnal.
The \dword{daq} team is responsible for producing a reference
implementation of the software that is used to access and decode the
raw electronics data.
The offline group is also responsible for providing the framework
for real-time \dword{dqm}. 
This monitoring is distinct from the \dfirst{om}.
Developing the payload jobs that run various algorithms to
summarize the data is the joint responsibility of the \dword{daq}, offline,
reconstruction and other groups.
The \dword{dqm} system includes a visualization system that can be
accessed from the Internet and shows specifically where operation shifts are
performed.

\subsection{Slow Control}
\label{sec:fd-daq-intfc-sc}
\label{sec:fd-daq-intfc-sc}

The \dword{cisc} systems monitor detector hardware and conditions not
directly involved in taking the data described above.
That data is stored both locally (in \dword{cisc} database servers in the
\dword{cuc}) and offline (the databases will be replicated back to \fnal)
in a relational database indexed by timestamp.
This allows bi-directional communications between the \dword{daq} and \dword{cisc} by
reading or inserting data into the database as needed for non
time-critical information.  

For prompt, time sensitive status information such as \textit{run is in
progress} or \textit{camera is on}, a low-latency software status register
is available on the local network to both systems.

There is no hardware interface. However, several racks of \dword{cisc} servers are in the counting room of the \dword{cuc}, and rack monitors in \dword{daq} racks are read out into the \dword{cisc} data stream.

Note that life and hardware safety-critical items will be hardware
interlocked 
according to \fnal standards, and fall outside the scope of this interface.

\subsection{External Systems} 
\label{sec:fd-daq-intfc-ext}

\fixme{Need to receive information on beam spills (Giles) , SNEWS (Alec).}


The \dword{daq} is required to save data based on external triggers, e.g., when a pulse of beam neutrinos  arrives at the \dword{fd}; or upon notice of an interesting astrophysical event by \dword{snews}~\cite{snews} or LIGO. This could involve going back 
to save data that has already been buffered (see Section~\ref{sec:fd-daq-fero}), or changing the trigger or zero suppression criteria for data taken during the interesting time period.

\subsubsection{Beam Trigger} 

The method for predicting and receiving the time of the beam spill is described in
Section~\ref{sec:fd-daq-design-beamtiming}.
Once that time is known to the \dword{daq}, a high-level trigger can be issued
to ensure that the necessary full data can be saved from the buffer
and saved as an event.

\subsubsection{Astrophysical Triggers} 

\dfirst{snews} is a coincidence
network of neutrino experiments that are individually sensitive to
an \dword{snb} 
observed from a core-collapse
supernova somewhere in our galaxy.
While DUNE must be sensitive to such a burst on its own, and 
is expected to be able to contribute to the coincidence network (Section~\ref{sec:fd-daq-sel}) via a TCP/IP socket, the capability to save data based on other observations provides an additional opportunity to ensure capture of this rare and valuable data. 
A \dword{snews} alert is formed when two or more neutrino experiments
report a potential \dword{snb} signal within \SI{10}{\s}.
A script running on the \dword{snews} server at BNL, provided by a given experiment that wishes to receive an alert,  sends out a message with the earliest time in the coincidence.
The latency from the neutrino burst is set by the response time of the
second fastest detector to report to \dword{snews}. This could be as
short as seconds, but could be tens of seconds.
At latencies larger than \SI{10}{\s}, full data might not be
available, but selected data is expected to be manageable. 


Other astrophysical triggers are available to which DUNE alone is unlikely to have sensitivity, except in rare cases, or if the triggers are taken as an ensemble. 
 Among these are gravitational wave triggers 
 (the details are being worked
out during the current LIGO shutdown), and high-energy photon
transients, most notably gamma ray bursts.
In fact, the use of network sockets on the timescale of seconds
enabled cooperation between LIGO, VIRGO, the Gamma Ray Coordinates
Network (GCN)~\footnote{Described in detail at
  \url{https://gcn.gsfc.nasa.gov/gcn_describe.html}}, and a number of
automated telescopes to make the discovery that \textit{short/hard} gamma ray bursts are caused by colliding neutron stars~\cite{kilonova}.

\section{Production and Assembly}
\label{sec:fd-daq-prod-assy}

\subsection{DAQ Components}

The \dword{fd} \dword{daq} system comprises the classes of components listed below. In each case, we outline the production, procurement, \dword{qa}, and \dword{qc} strategies.

\subsubsection{Custom Electronic Modules}

Custom electronic modules, specified and designed by the \dword{daq}
consortium, are used for two functional components in the \dword{daq}
\dword{fe}. 
The first is to interface the \dword{detmodule} electronics to the \dword{daq} \dword{fec} systems, which are likely to be based on the \dword{felix}
PCIe board.
The other is for real-time data processing (particularly for the
\dword{spmod}), which will likely be based on the
\dword{cob} \dshort{atca} blade.
\Dword{pdsp} currently implements both designs, and new designs optimized according to
DUNE requirements will be developed.
It is possible that we will make use of commercially-designed hardware
in one or other of these roles. \dword{daq} consortium institutes have
significant experience in the design and production of high-performance digital electronics for previous experiments.
Our strategy is therefore to carry out design in-house, manufacturing
and \dword{qa} steps in industry, and testing and \dword{qc} procedures at a number of
specialized centers within the DUNE collaboration. 
Where technically and economically feasible, modules will be split
into subassemblies (e.g., carrier board plus processing
daughter cards), allowing production tasks to be spread over more
consortium institutes.

DUNE electronic hardware will be of relatively high performance by commercial standards, and will contain high-value subassemblies such as large \dwords{fpga}. Achieving a high yield will require significant effort in design verification, prototyping and pre-production tests, as well as in tendering and vendor selection. The production schedule is largely driven by these stages and the need for thorough testing and integration with firmware and software before installation, rather than by the time for series hardware manufacture. This is somewhat different from the majority of other DUNE \dword{fd} components.

\subsubsection{Commercial Computing}

The majority of procured items will be standard commercial computing equipment, in the form of compute and storage servers. Here, the emphasis is on correct definition of the detailed specification, and the tracking of technology development, in order to obtain the best value 
during the tendering process. Computing hardware will be procured in several batches, as the need for \dword{daq} throughput increases during the construction period. 

\subsubsection{Networking and data links}

The data movement system is a combination of custom optical links (for data transmission from the cryostats to the \dword{cuc}) and commercial networking equipment. The latter items will be procured in the same way as other computing components. The favored approach to procurement of custom optics is purchase of pre-manufactured assemblies ready for installation, rather than 
on-site fiber preparation and termination. Since transmission distances and latencies in the underground area are not critical, the fiber run lengths do not need to be of more than a few variants. It is assumed that fibers will not be easily accessible for servicing or replacement during the lifetime of the experiment, meaning that procurement and installation of spare \textit{dark} fibers (including, if necessary, riser fibers up the \surf hoist shafts) is necessary.

\subsubsection{Infrastructure}

All \dword{daq} components will be designed for installation in \SI{48.3}{cm} (standard \SI{19}{in}) rack infrastructure, either in the \dword{cuc} or above ground. Standard commercial server racks with local air-water heat exchangers are likely to be used. These items will be specified and procured within the consortium, but will be pre-installed (along with the necessary electrical, cooling and safety infrastructure) under the control of \dword{tc} before \dword{daq} beneficial occupancy.

\subsubsection{Software and firmware}

The majority of the \dword{daq} construction effort will be invested in the production of custom software and firmware. Based on previous experiments, these projects are likely to use tens to hundreds of staff-years of effort, and will be significant projects even by commercial standards, mainly due to the specialized skills required for real-time software and firmware. A major project management effort is required to guide the specification, design, implementation and testing of the necessary components, especially as developers will be distributed around the world. Use of common components and frameworks across all areas of the \dword{daq} is mandatory. Effective \dword{daq} software and firmware development has been a demonstrated weakness of several previous experiments, and substantial work is required 
 in the next two years to put in place the necessary project management regime.

\subsection{Quality Assurance and Quality Control}

High availability is a basic requirement for the \dword{daq}, and this rests upon three key principles:

\begin{itemize}
	\item A rigorous \dword{qa} and \dword{qc} regime for components (including software and firmware);
	\item Redundancy in system design, to avoid single points of failure;
	\item Ease of component replacement or upgrade with minimal downtime.
\end{itemize}

The lifetime of most electronic assemblies or commercial computing components will not match the \dunelifetime lifespan of the DUNE experiment. It is to be expected that essentially all components will therefore be replaced during this time. Careful system design will allow this to take place without changes to interfaces. However, it is intended that the system  run for at least the first three to four years without substantial replacements, and \dword{qa} and \dword{qc}, as well as spares production, will be steered by this goal. Of particular importance is adequate burn-in of all components before installation underground, and careful record-keeping of both module and subcomponent provenance, in order to identify systematic lifetime issues during running.

\subsection{Integration testing}

Since the \dword{daq} will use subcomponents produced by many different teams,
integration testing is a key tool in ensuring compatibility and
conformance to specification. This is particularly important in the
prototyping phase before the design of final hardware. Once
pre-production hardware is in hand, an extended integration phase will
be necessary in order to perform final debugging and performance tuning
of firmware and software. In order to facilitate ongoing optimization in
parallel with operations, and compatibility testing of new hardware or
software, we envisage the construction of one or more permanent
integration test stands at \dword{daq} institutions. These will be in locations
convenient to the majority of consortium
members, i.e., at major labs in Europe and the USA. A temporary \dword{daq}
integration and testing facility near \surf will also be required as
part of the installation procedure.

\section{Installation, Integration and Commissioning}
\label{sec:fdsp-daq-install}

\subsection{Installation}
\label{sec:fdsp-daq-install-transport}

The majority of \dword{daq} components will be installed in a dedicated and partitioned area of the \dword{cuc} as shown in Figure~\ref{fig:daq-install-controlroom}, starting as soon as the consortium has beneficial occupancy of this space. 
The \dword{cf} is responsible for running fiber from the \dword{spmod}'s \dwords{wib} to the \dword{daq}, and from the \dword{daq} to the surface. This is currently projected to take place eighteen months before \dwords{apa} are installed in the \dword{spmod}, allowing time for final component acceptance testing in the underground environment, and to prepare the \dword{daq} for detector testing and commissioning. Some \dword{daq} components (event builder, storage cluster and WAN routers, plus any post-event-builder processing) will be installed above ground.

A total of \SI{500}{kVA} of power and cooling will be available to run the computers in the counting room. 
Twelve \SI{48.3}{cm} (standard \SI{19}{in}) server racks (of up to 58U height) per module have initially been allocated for each detector module, with two more each for facilities and \dword{cisc}. An optimized layout, including the necessary space for hardware installation and maintenance, plus on-site spares, will be developed once the \dword{daq} design is finalized. The racks will be water cooled with local air-to-water heat exchangers. To allow expanded headroom for initial testing, development, and commissioning throughput, the full complement of rack infrastructure and network equipment for four \dwords{detmodule} will be installed from the start.  

\begin{dunefigure}[CUC control room layout]{fig:daq-install-controlroom}
  {Floor plan for the \dword{daq} and control room space in the \dword{cuc}.  The \dword{daq}
    Room has space for at least \num{52} racks of servers and routers.
    Fiber from the \dwords{wib} in the detector caverns enter in the upper
    right of this room, terminate in a breakout panel, and are
    distributed to the \dwords{rce} in these racks, then to \dword{felix} servers (also
    in this room) as outlined in
    Figure~\ref{fig:daq-readout-buffering-baseline}.  Fibers to the
    surface enter this room from the lower left.}
\includegraphics[width=0.8\textwidth]{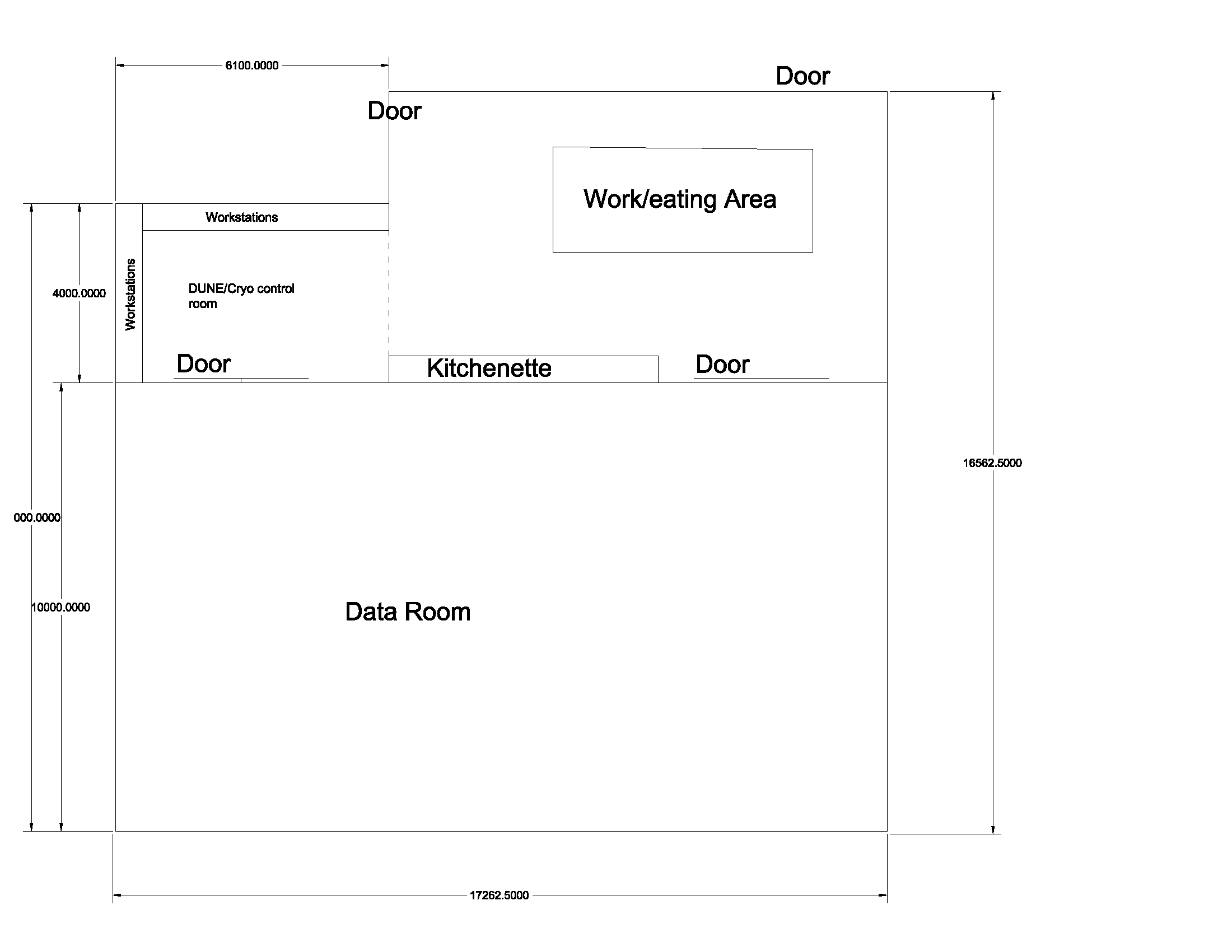}
\end{dunefigure}

The counting room is similar to a server room at a university or national lab in terms of the need for cleanliness, ventilation, fire protection, drop flooring, and access control. Networking infrastructure and fiber breakout will take up some of the rack space, but 
very little of the power budget. Power to individual machines and crates must to be controlled remotely via power distribution units, since it is desirable to minimize \dword{daq} workers' presence underground if there is work that can be done from the surface or remotely.  Some uninterruptible power supply (UPS) capacity is needed to allow for an orderly shutdown of computers, but only networking equipment requires longer-duration backup power, this is to enable remote recovery from short-term power failures.  

\subsection{Integration with Detector Electronics}
\label{sec:fdsp-daq-install-transport}

Basic technical integration with detector electronics will take place before installation, during a number of integration exercises in the preceding years. We anticipate that the consortium will supply and support small-scale instances of the \dword{daq} system for testing of readout hardware at the production sites, based on prototype or pre-production hardware. Full-scale \dword{daq} testing will have been completed with artificial data sources during internal integration. The work to be done during installation is therefore essentially channel-by-channel verification of the final system as it is installed, on a schedule allowing for any rectifying work to be carried out on the detector immediately (i.e., the \dword{daq} must gather and present data in effectively real time). This implies the presence of a minimal but sufficient functional \dword{daq} system before detector installation commences, along with the timing and fast control system, and the capability to permanently record data for offline analysis. However, it does not require triggering, substantial event building or data transfer capacity. The \dword{daq} installation schedule is essentially driven by this requirement.

In addition, the data pipeline from event builder, via the storage buffer and WAN, to the offline computing facilities, must be developed and tested. We anticipate this work largely happening at \fnal in parallel with detector installation, and the full-scale instances of these components being installed at \surf in preparation for start of data-taking.

\subsection{Commissioning}
\label{sec:fdsp-daq-commissioning}

System commissioning for the \dword{daq} comprises the following steps:

\begin{itemize}
	\item Integration with detector subsystems of successive \dwords{detmodule};
	\item Final integration and functional testing of all \dword{daq} components;
	\item Establishment of the necessary tools and procedures to achieve high-efficiency operation;
	\item Selection, optimization and testing of trigger criteria;
	\item Ongoing and continuous self-test of the system to identify actual or imminent failures, and to assess performance.
\end{itemize}

Each of these steps will have been carried out at the integration test stands before being used on the final system. The final steps are to some extent continuous activities over the experiment lifetime, but which require knowledge of realistic detector working conditions before final validation of the system can take place. We anticipate that these steps will be carried out during the cryostat filling period, and form the major focus of the \dword{daq} consortium effort during this time.

\section{Safety}
\label{sec:fd-daq-safety}

Two overall safety plans will be followed by the \dword{fd} \dword{daq}. General work underground will comply with all safety procedures in place for working in the detector caverns and \dword{cuc} underground at \surf. \Dword{daq}-specific procedures for working with racks full of electronics or computers, as defined at \fnal, will be followed, especially with respect to electrical safety and the fire suppression system chosen for the counting room. For example, a glass wall between the server room space and the other areas in Figure~\ref{fig:daq-install-controlroom} will be necessary to prevent workers in the server room from being unseen if they are in distress, and an adequate hearing protection regime must be put in place.

There are no other special safety items for the \dword{daq} system not already covered by the more general safety plans referenced above. The long-term emphasis is on remote operations capability from around the world, limiting the need for physical presence at \surf, and with underground access required only for urgent interventions or hardware replacement.

\section{Organization and Management}
\label{sec:fd-daq-org}

At the time of writing, the \dword{daq} consortium comprises \num{30} institutions, including universities and national labs, from five countries. Since its conception, the \dword{daq} consortium has met on roughly a weekly basis, and has so far held two international workshops dedicated to advancing the  \dword{fd} \dword{daq} design. The current \dword{daq} consortium leader is 
from U. Bristol, UK.

Several key technical and architectural decisions have been made in the last months, that have formed an agreed basis for the \dword{daq} design and implementation presented in this document.

\subsection{DAQ Consortium Organization}
\label{sec:fd-daq-org-consortium}

The DUNE \dword{daq} consortium is currently organized in the form of five active
Working Groups (WG) and WG leaders:
\begin{itemize}
\item Architecture, current WG leaders are from: U. Oxford and CERN;
\item Hardware, current WG leaders are from: U. Bristol and SLAC;
\item Data selection, current WG leader is from: U. Penn.;
\item Back-end, current WG leader is from: \fnal;
\item Integration and Infrastructure, current WG leader is from: U. Minnesota Duluth.
\end{itemize}

During the ongoing early stages of the design, the architecture and hardware WGs have been holding additional meetings focused on aspects of the design related to architecture solutions and costing. In parallel, the \dword{daq} Simulation Task Force effort, which was in place at the time of the consortium inception, has been adopted under the data selection WG, and simulation studies have continued to inform design considerations. This working structure is expected to remain in place through at least the completion of the \dword{tp}. During the construction phase of the project we anticipate a new organization, built around major subsystem construction and commissioning responsibilities, and drawing also upon expertise build up during the \dword{protodune} projects.

\subsection{Planning Assumptions}
\label{sec:fd-daq-org-assmp}

The \dword{daq} planning is based the assumption of a \dword{spmod} first, followed by a \dword{dpmod}. The schedule is sensitive to this assumption, as the \dword{daq} requirements for the two module types are quite different. Five partially overlapping phases of activity are planned (see Figure~\ref{fig:daq-schedule}):

\begin{itemize}
	\item A further period of R\&D activity, beginning at the time of writing, and culminating in a documented system design in the \dword{tdr} around July 2019;
	\item Production and testing of a full prototype \dword{daq} slice of realistic design, culminating in an engineering design review;
	\item Preparation and fit out of the \dword{cuc} counting room with a minimal \dword{daq} slice, in support of the first module installation;
	\item Production and delivery of final hardware, computing, software and firmware for the first module;
	\item Production and delivery of final hardware, computing, software and firmware for the second module.
\end{itemize}

This schedule assumes beneficial occupancy of the \dword{cuc} ounting room by end of the first quarter of 2022, and the availability of facilities to support an extended large-scale integration test in 2020 (e.g., CERN or \fnal). We assume the availability of resources for installation and commissioning of final \dword{daq} hardware (e.g., surface control room and server room facilities) from around the first quarter of 2023, and the \dword{itf} from the second quarter of 2022. The majority of capital resources for \dword{daq} construction will be required from the second quarter of 2022, with a first 
portion of funds for the minimal \dword{daq} slice from the first quarter of 2021.



\subsection{High-level Cost and Schedule}
\label{sec:fd-daq-org-cs}

The high-level \dword{daq} schedule, which is based upon the current DUNE \dword{fd} top-level schedule, is shown in Figure~\ref{fig:daq-schedule}.

\begin{dunefigure}[\dword{daq} high-level schedule]{fig:daq-schedule}
  {\dword{daq} high-level schedule}
\includegraphics[width=1.0\textwidth]{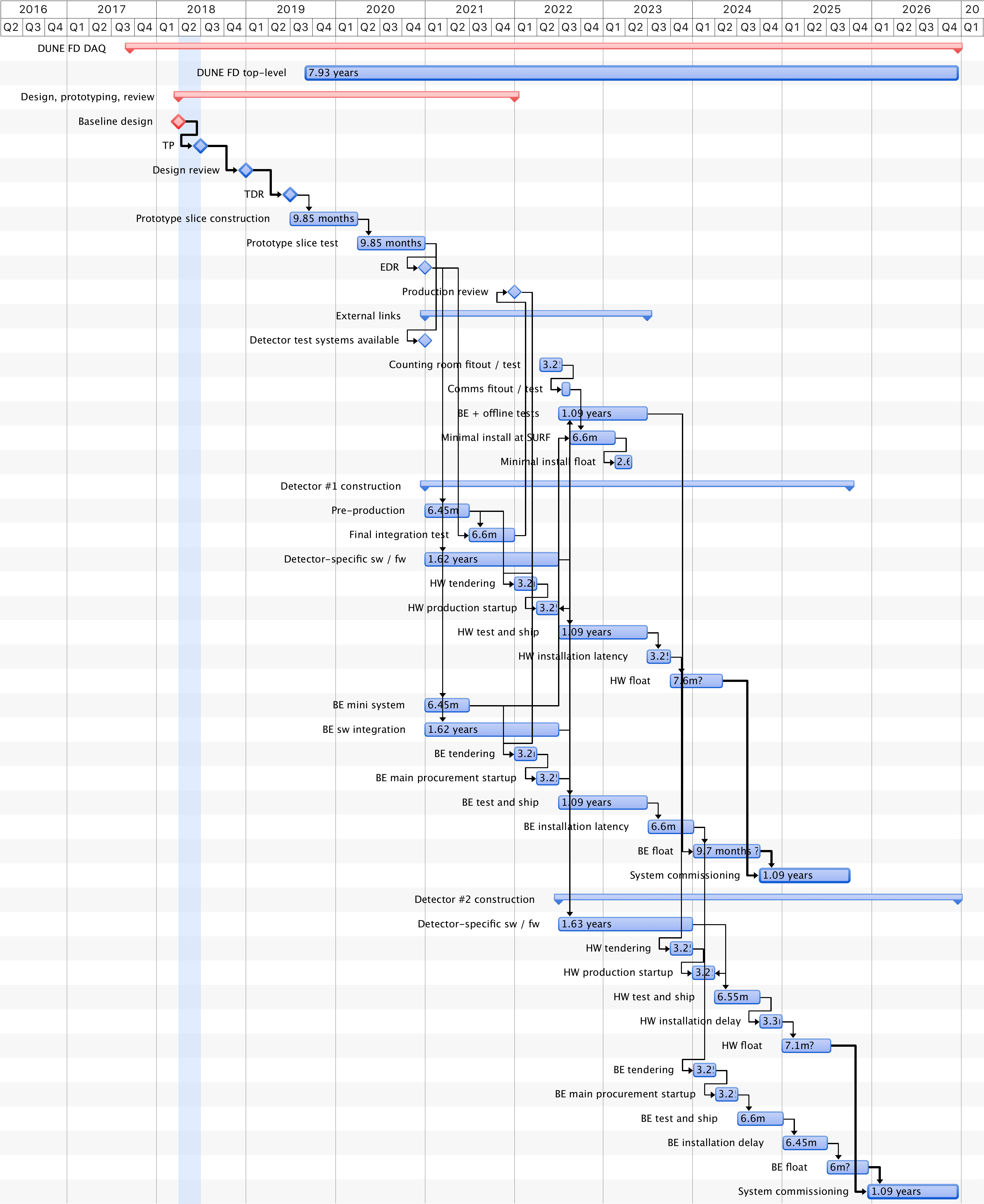}
\end{dunefigure}



\cleardoublepage

\chapter{Slow Controls and Cryogenics Instrumentation}
\label{ch:fdsp-slow-cryo}

\section{Slow Controls and Cryogenics Instrumentation Overview}
\label{sec:fdsp-slow-cryo-ov}





\subsection{Introduction}
\label{sec:fdsp-slow-cryo-intro}


The \dword{cisc} system provides comprehensive monitoring for all \dword{detmodule} components as well as for the \lar quality and behavior, both being crucial
to guarantee high-quality data. Beyond passive monitoring, \dword{cisc} also provides a control system for some of the detector components. 
The structure of the \dword{cisc} consortium is quite complex. A subsystem chart
for the \dword{cisc} system is shown in Figure~\ref{fig:sp-slow-cryo-subsys}. 

\begin{dunefigure}[CISC subsystems]{fig:sp-slow-cryo-subsys}
{\dword{cisc} subsystem chart}
\includegraphics[width=0.5\textwidth,trim=20mm 30mm 30mm 70mm,clip]{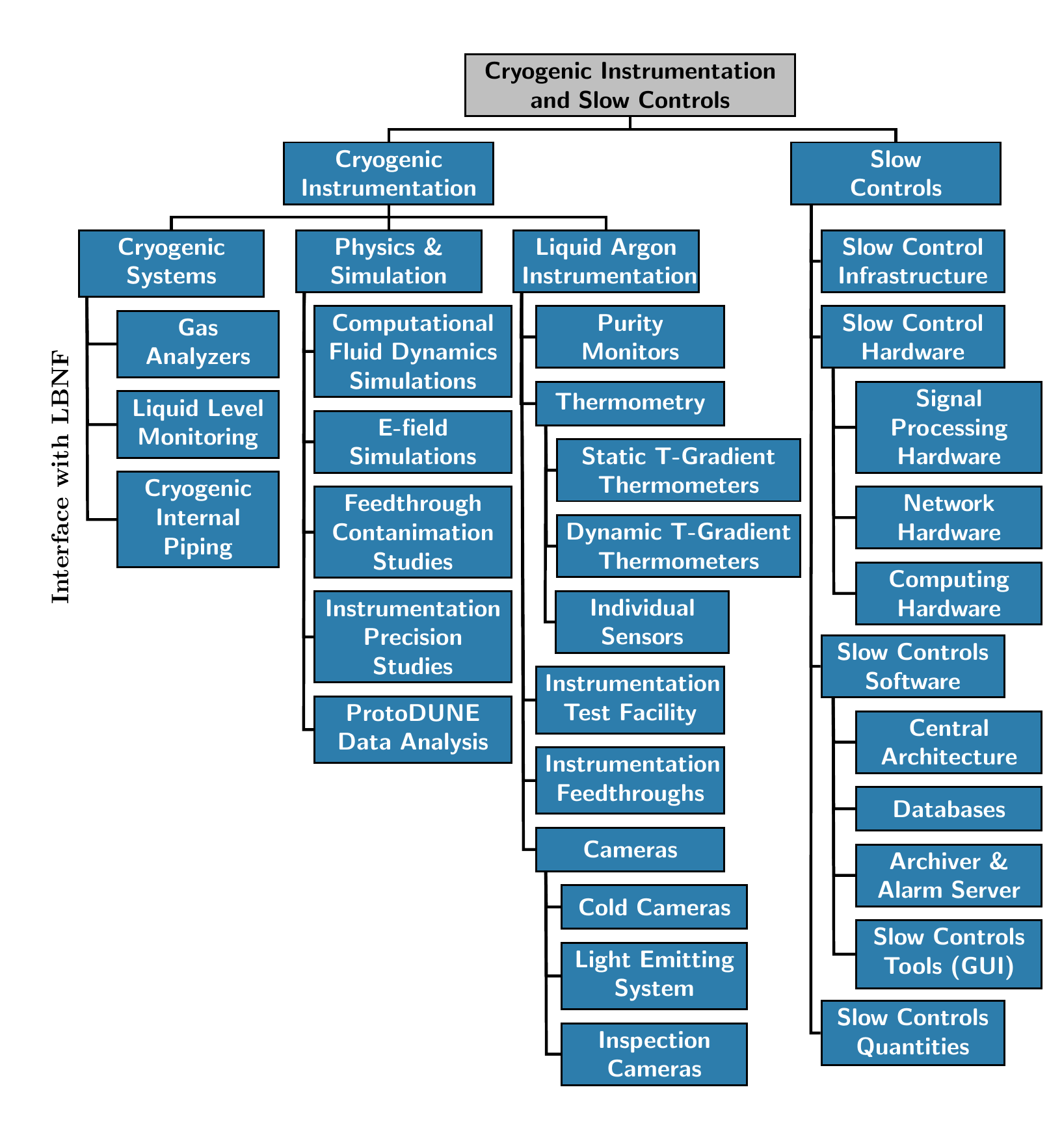}  
\end{dunefigure}

Two main branches can be distinguished: cryogenics instrumentation and slow controls. The former includes a set of devices 
to monitor the quality and behavior of the \lar volume in the cryostat interior, ensuring the correct functioning of
the full cryogenics system and the suitability of the \lar for good quality physics data. Those devices are 
purity monitors, temperature monitors, gas analyzers, \lar level monitors, cameras with their associated
light emitting system.

Cryogenics instrumentation also requires significant physics and
simulation work such as \efield simulations and cryogenics modeling
studies using \dfirst{cfd}. \efield simulations
are required to identify desirable locations for instrumentation
devices in the cryostat so that they are not in regions of high \efield and
that their presence does not induce large field distortions. \dshort{cfd}
simulations are needed to understand the expected temperature,
impurity and velocity flow distributions and guide the placement and
distribution of instrumentation devices inside the cryostat.

From the organizational point of view
cryogenics instrumentation has been divided into three main parts: (1) cryogenics systems, which includes all components directly related to the external cryogenic system as
liquid level monitoring, gas analysers and internal cryogenics piping, all having substantial interfaces with LBNF; (2) \lar  instrumentation, which includes all
other instrumentation devices, and (3) physics and simulation.

The second branch of \dword{cisc} is the slow controls (SC) system, in charge of monitoring and controlling most detector elements, as power supplies, electronics, racks, instrumentation devices,
calibration devices, etc. It includes four main components: hardware, infrastructure,
software, and firmware. The slow controls hardware and infrastructure consists of
networking hardware, signal processing hardware, computing hardware, and relevant
rack infrastructure. The slow controls software and firmware is needed for
signal processing, alarms, archiving, and control room displays.

Two other systems have been included by the DUNE management as part of the \dword{cisc} consortium,
a test facility for the instrumentation devices and the cryogenics piping inside the cryostat.
Those are included inside the cryogenics instrumentation branch.


\subsection{Design Considerations}
\label{sec:fdsp-slow-cryo-des-consid}


For all \lar instrumentation devices, \dword{pdsp} designs are
considered as the baseline, and requirements for most design
parameters are extrapolated from \dword{pdsp}. Hence a critical step for
the \dword{cisc} consortium is to analyze data from \dword{pdsp} when available
to validate the instrumentation designs and understand their
performance. For example, a crucial design parameter, which should be evaluated in \dword{pdsp},
is the maximum noise level induced by instrumentation devices on the readout electronics that can be tolerated to avoid confusing event reconstruction. 

Some of the common design considerations for
instrumentation devices include stability, reliability and longevity
such that the devices can survive for a period of at least 
\dunelifetime{}.  Since it is uncommon for any device
to have such a long lifetime, provisions are made in the overall
design to allow replacement of devices where possible.

As for any other element inside the cryostat, 
the \efield on the instrumentation devices is 
required to be less than \SI{30}{kV\per\cm},
so that the risk of dielectric breakdown in \lar is minimized.
This requirement imposes stringent constraints on the location and mechanical 
design of some devices. Electrostatic simulations 
will be performed to compute the expected field on the boundaries of 
instrumentation devices and to design the appropriate \efield shielding
in the case the \efield approaches the limit. 

Another common consideration for all instrumentation devices is their support structure
design, which is expected to be substantially different from the one used in \dword{pdsp}.

For slow controls, the system 
is designed to be robust enough to support a large number of monitored variables and a broad range of
monitoring and archiving rates and has 
to interface
with a large number of systems to establish two-way communication for
control and monitoring. Table \ref{tab:sp-cisc-requirements} shows
some of the important \dword{cisc} system design requirements.

\begin{dunetable}
[Important design requirements on the \single CISC system design]
{p{0.22\textwidth}p{0.17\textwidth}p{0.32\textwidth}p{0.19\textwidth}}
{tab:sp-cisc-requirements}
{Important design requirements on the single phase \dword{cisc} system design}
Design Parameter & Requirement & Motivation & Comment
\\ \toprowrule
Electron lifetime measurement precision
 & $<\SI{1.4}{\%}$ at \SI{3}{ms}
 & Per DUNE-FD Task Force\,\cite{fdtf-final-report}, needed to keep the bias on the charge readout in the \dshort{tpc} to below \SI{0.5}{\%} at \SI{3}{ms}
 & Purity monitors do not directly sample \dshort{tpc}: see sec.\ \ref{sec:fdgen-slow-cryo-purity-mon}
\\  \colhline
Thermometer precision
 & $<\SI{5}{mK}$
& Driven by \dshort{cfd} simulation validation; based on \dword{pdsp}-SP design
& Expected \dword{pdsp} performance \SI{2}{mK}
\\ \colhline
Thermometer density
 & \(>2/\si{m}\) (vert.), \(\sim\)~0.2/\si{m} (horiz.)
 & Driven by \dshort{cfd} simulation
 & Achieved by design
\\ \colhline
Liquid level meters precision (\single)
 & \SI{0.1}{\%} over \SI{14}{m}
& Standard sensitivity; two level meters for redundancy
& \dword{pdsp} design
\\  \colhline
 Cameras
 & \multicolumn{3}{p{0.64\textwidth}}{--- multiple requirements imposed by interfaces: see Table \ref{tab:fdgen-cameras-req} ---}
 \\ \colhline
Cryogenic Instrumentation Test Facility cryostat volumes
 & \num{0.5} to \SI{3}{m^3}
& Based on filling costs and turn around times
& Under design
\\  \colhline
 Max.\ \efield on instrumentation devices
 & \(<\SI{30}{kV/cm}\)
 & The mechanical design of the system should be such that \efield is below this value, 
 to minimize the risk of dielectric breakdown in \lar
 & \dword{pdsp} designs based on electrostatic simulations
\\ \colhline
 Noise introduced into readout electronics
 & Below significant levels
 & Keep readout electronics free from external noise, which confuses event reconstruction
 & To be evaluated at \dword{pdsp}
\\ \colhline
Total no.\ of variables
 & \num{50}k to \num{100}k
& Expected number based on scaling past experiments; requires robust base software model that can handle large no. of variables.
& Achievable in existing control systems; DUNE choice in progress.
\\  \colhline
Max.\ archiving rate per channel
 & \SI{1}{Hz} (burst), \SI{1}{\per\minute} (avg.)
& Based on expected rapidity of interesting changes; impacts the base software choice; depends on data storage capabilities
& Achievable in existing control system software; DUNE choice in progress.
\\
%
%
%
\end{dunetable}


\subsection{Scope}
\label{sec:fdsp-slow-cryo-scope}


As described above, and shown schematically in Figure~\ref{fig:sp-slow-cryo-subsys},
the scope of the \dword{cisc} system spans a broad range of activities.  In the
case of cryogenics systems (gas analyzers, liquid level monitors and
cryogenic internal piping), LBNF provides the needed expertise and
is responsible for the design, installation, and commissioning activities
while the \dword{cisc} consortium provides the resources and supplements the labor as
needed. In the case of \lar Instrumentation devices (purity monitors,
thermometers, cameras and light-emitting system; and their associated \fdth{}s) and instrumentation
test facility, \dword{cisc} is responsible from design to commissioning in
the \dwords{fd}.

From the slow controls side, \dword{cisc} provides control and monitoring of
all detector elements that provide data on the health of the
\dword{detmodule} or conditions important to the experiment.
The scope of systems that slow controls includes is listed below:

\begin{itemize}
\item {\bf Slow Controls Base Software and Databases}: provides the central tools needed to develop control and monitoring for various detector systems and interfaces.
  \begin{itemize}
  \item Base input/output software,
  \item Alarms; archiving; display panels; operator interface tools,
  \item Slow controls system documentation and operations guidelines.
  \end{itemize}
\item {\bf Slow Controls for External Systems}: export data from systems external to the detector and provide status to operators and archiving.
  \begin{itemize}
  \item Beam status; cryogenics status; \dword{daq} status; facilities systems status,
  \item For the systems above, import other interesting monitoring data as needed (e.g., pumps data from cryogenics system, heaters data from facility systems, etc.),
  \item Building controls; detector hall monitoring; ground impedance monitoring,
  \item Interlock status bit monitoring (but not the actual interlock mechanism).
  \end{itemize}
\item {\bf Slow Controls for Detector Hardware Systems}: develop software interfaces for detector hardware devices
  \begin{itemize}
  \item Monitoring and control of all power supplies,
  \item Full rack monitoring (rack fans, thermometers and rack protection system),
  \item Instrumentation and calibration device monitoring (and control to the extent needed),
  \item Power distribution units monitoring; computer hardware monitoring,
  \item High voltage system monitoring through cold cameras,
  \item Detector components inspection through warm cameras.
  \end{itemize}
\end{itemize}

In terms of slow controls hardware, \dword{cisc} will develop, install and
commission any hardware related to rack monitoring and control. While
most power supplies might only need a cable from the device to an
Ethernet switch, some power supplies might need special cables (e.g., 
GPIB or RS232) for communication. The \dword{cisc} consortium is responsible for
providing such control cables.

In addition to the listed activities, \dword{cisc} also has activities that span
outside the scope of the consortium and require interfacing with other
groups. This is discussed in Sec. \ref{sec:fdgen-slow-cryo-intfc}.



\section{Cryogenics Instrumentation}
\label{sec:fdsp-cryo-instr} 
\label{sec:fddp-cryo-instr} 
\label{sec:fdgen-cryo-instr} 

Instrumentation inside the cryostat must ensure that the condition of the \dword{lar} is adequate for operation of the \dshort{tpc}.
This instrumentation includes devices to monitor the impurity level of the argon, e.g., the purity monitors, which provide high-precision electron lifetime measurements,
and gas analyzers to ensure that the levels of atmospheric contamination drop below certain limits during the cryostat purging, cooling and filling.
The cryogenics system operation is monitored by temperature sensors deployed in vertical arrays and at the top and bottom of the detector, providing a 
detailed \threed temperature map that can help to predict the \dword{lar} purity across the entire cryostat. The cryogenics instrumentation also includes \lar level monitors and
a system of internal cameras to help in locating sparks in the cryostat and for overall monitoring of the cryostat interior. 
As mentioned in the introduction, cryogenics instrumentation requires simulation work to identify the proper location for these devices inside the cryostat and
for the coherent analysis of the instrumentation data. 

Figure~\ref{fig:sp-slow-cryo-ports} shows the current map of cryostat ports for the \dword{spmod}, highlighting the ones assigned to instrumentation devices,
as well as the preliminary location for some of these devices. Vertical temperature profilers are located behind the \dwords{apa} ($T_S$) and behind the east end wall ($T_D$).
They are complemented by a coarser \twod grid of sensors at the top and bottom of the cryostat (not shown in the figure). Purity monitors and level meters are planned
in each detector side, behind the two front end walls. Inspection cameras will use some of the multipurpose instrumentation ports, but their exact locations are yet to be decided.

\begin{dunefigure}[Cryostat ports]{fig:sp-slow-cryo-ports}
{Cryostat ports and preliminary location of some instrumentation devices. }
\includegraphics[width=0.95\textwidth]{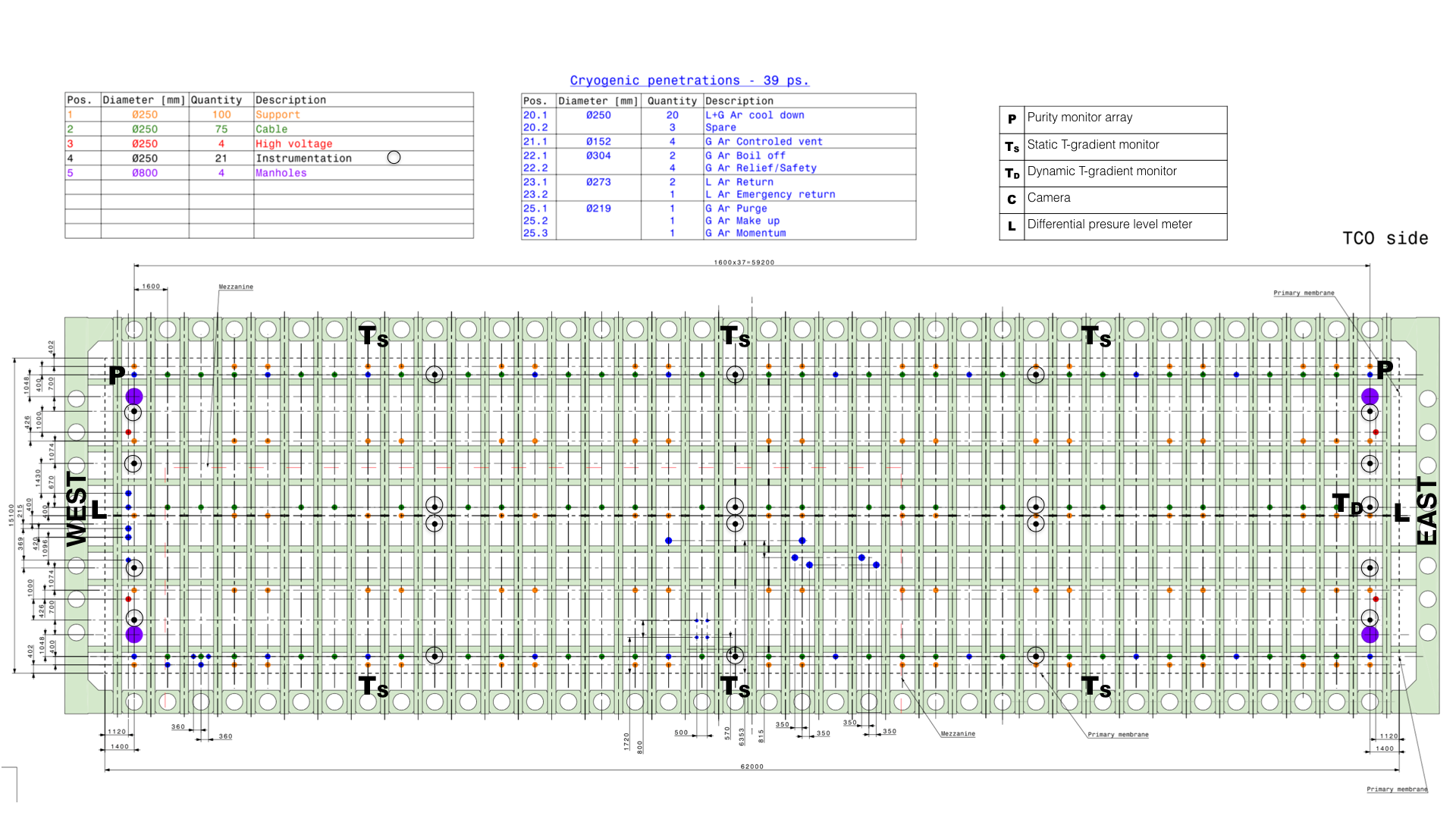}
\end{dunefigure}

\subsection{Fluid Dynamics Simulations}
\label{sec:fdgen-slow-cryo-cfd}

Proper placement of purity monitors, thermometers, and liquid level monitors within the \dword{detmodule} requires knowledge of how \dword{lar} behaves within the cryostat in terms of its fluid dynamics, heat and mass transfer, and distribution of impurity concentrations. 
Fluid motion within the cryostat is driven primarily by small changes in density from thermal gradients, although pump flow rates and inlet and outlet locations also contribute. 
Heat sources include exterior heat from the surroundings, interior heat from the electronics, and heat flow through the pump inlet.

The fluid flow behavior can be determined through simulation of \dword{lar} flow within the detector using ANSYS CFX\footnote{ANSYS\texttrademark{}, \url{https://www.ansys.com/products/fluids/ansys-cfx}.}, a commercially available \dfirst{cfd} code. Such a model must include proper definition of the fluid characteristics, solid bodies and fluid-solid interfaces, and a means for measuring contamination, while still maintaining reasonable computation times. 
Although simulation of the \dword{detmodule} presents challenges, there exist acceptable simplifications for accurately representing the fluid, the interfacing solid bodies, and variations of contaminant concentrations. Because of the magnitude of thermal variation within the cryostat, modeling of the \dword{lar} is simplified through use of constant thermophysical properties, calculation of buoyant force through use of the Boussinesq Model (using constant a density for the fluid with application of a temperature dependent buoyant force), and a standard shear stress transport turbulence model. Solid bodies that contact the \dword{lar} include the cryostat wall, the cathode planes, the anode planes, the ground plane, and the \dword{fc}. As in previous \dshort{cfd} models of the DUNE \dword{35t} and \dword{protodune} by South Dakota State University (SDSU)\cite{docdb-5915}, the \dword{fc} planes, anode planes, and \dword{gp} can be represented by porous bodies. Since impurity concentration and electron lifetime do not impact the fluid flow, these variables can be simulated as  passive scalars, as is commonly done for smoke releases~\cite{cfd-1} in air or dyes released in liquids.

Significant discrepancies between real data and simulations can have potential impacts on detector performance, as simulation results contribute to decisions about where to locate sensors and monitors, as well as definitions of various calibration quantities. However, methods of mitigating such risks include well established convergence criteria, sensitivity studies, and comparison to results of previous \dshort{cfd} simulation work by SDSU and \fnal. Additionally, the simulation will be improved with input from temperature measurements and validation tests. 

\begin{dunefigure}[\dshort{cfd} example]{fig:cfd-example}
  {Distribution of temperature on a plane intersecting an inlet (right) and halfway between an inlet and an outlet (left), as predicted by SDSU \dshort{cfd} simulations \cite{docdb-5915}. (See Figure\ \ref{fig:cfd-example-geometry} for geometry.)}
  \includegraphics[height=0.4\textwidth]{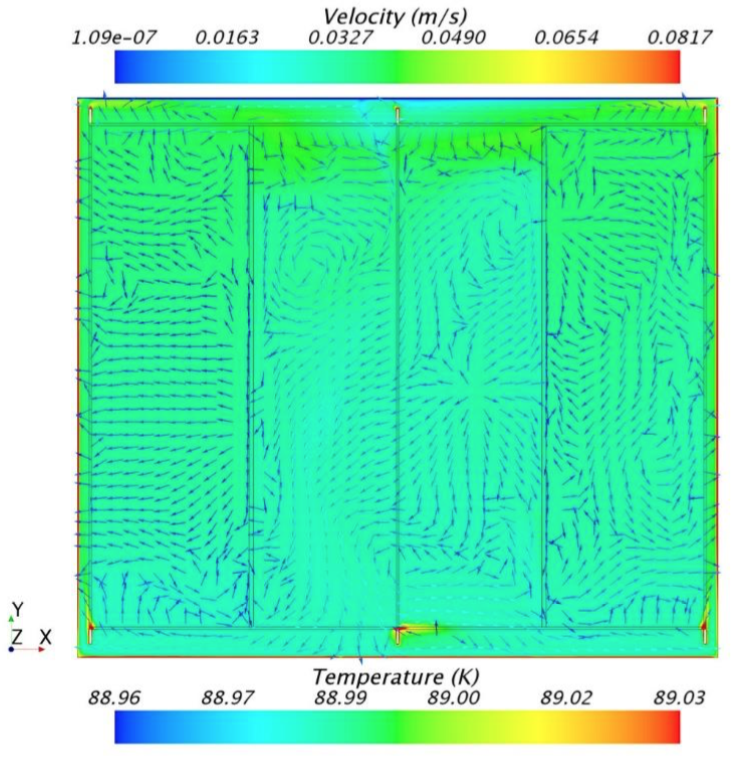}
  \includegraphics[height=0.4\textwidth]{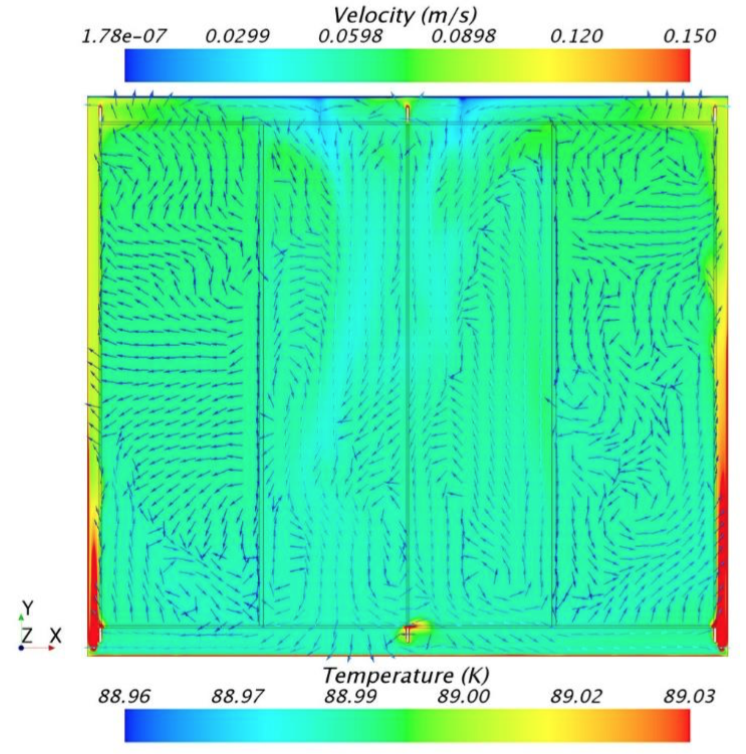}
\end{dunefigure}

Figure~\ref{fig:cfd-example} shows an example of the temperature
distribution on a plane intersecting a \dword{lar} inlet and at a
plane halfway between an inlet and an outlet; the geometry used for
this simulation is shown in Figure~\ref{fig:cfd-example-geometry}.
Note the plume of higher temperature \dword{lar} between the walls and
the outer \dword{apa} on the inlet plane.  The current locations of instrumentation in
the cryostat as shown in Figure~\ref{fig:sp-slow-cryo-ports} were
determined using the temperature and impurity distributions from these
previous simulations.

\begin{dunefigure}[\dshort{cfd} example geometry]{fig:cfd-example-geometry}
  {Layout of the \dshort{tpc} within the cryostat (top) and positions of
    \dword{lar} inlets and outlets (bottom) as modeled in the SDSU
    \dshort{cfd} simulations \cite{docdb-5915}.
    The Y axis is vertical and the X axis is parallel to the \dword{tpc}
    drift direction.
    Inlets are shown in green and outlets are shown in red.}
  \includegraphics[width=0.8\textwidth]{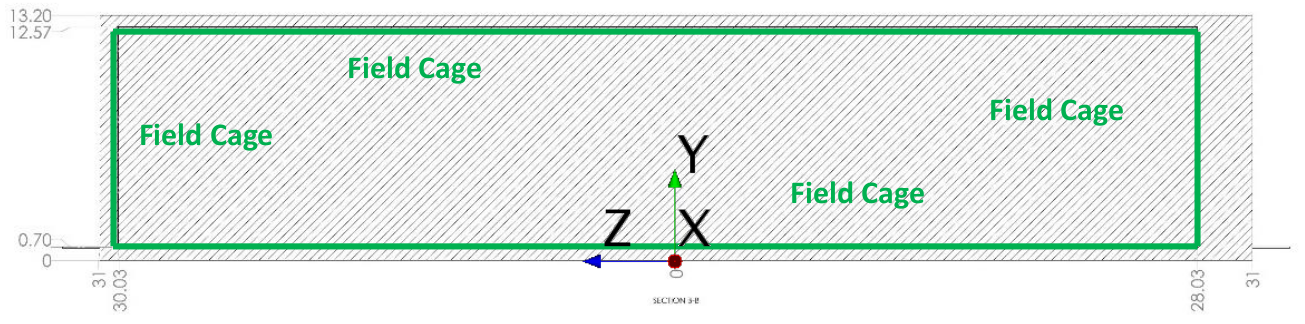}
  \includegraphics[width=0.8\textwidth]{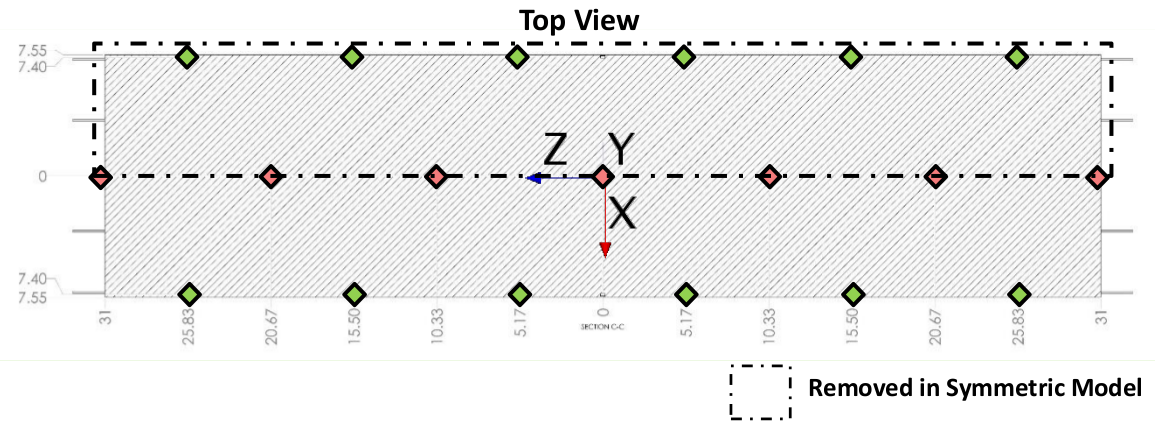}
\end{dunefigure}

The initial strategy for the future \dword{cfd} simulation effort is to understand the performance of \dword{protodune} cryogenics system and model the \dwords{fd} to derive requirements for instrumentation devices.
The following is a prioritized set of studies planned to help drive the requirements for other systems:
\begin{enumerate}
\item Review the DUNE \dword{fd} cryogenics system design and verify the current implementation in simulation; this is important to ensure that the model represents what will be built.
\item Model the \dword{pddp}  liquid and gas regions with the same precision as the \dword{fd}. Presently only the liquid model exists. The liquid model is needed to interpret the thermometer data, and the gas model is needed to understand how to place thermometers in the ullage and verify the design of the gaseous argon purge system.
\item Perform a \dword{cfd} study to determine the feasibility of a wier for \dual; this helps to determine if it can be used to clean the \lar surface before the extraction grid is submerged in the \dword{dpmod}.
\item Verify the \dword{spmod} \single \dword{cfd} model in simulation performed by LBNF; this defines the requirements for instrumentation devices (e.g., thermometry).
\item Model the \dword{pddp} liquid and gas regions with the same precision as the \dword{fd}. \fixme{same as a previous bullet}
\end{enumerate}


\subsection{Purity Monitors} 
\label{sec:fdgen-slow-cryo-purity-mon}
\label{sec:fdsp-slow-cryo-purity-mon} 
\label{sec:fddp-slow-cryo-purity-mon} 
A fundamental requirement of a \dword{lar} \dshort{tpc} is that ionization electrons drift over long distances in \dword{lar}. Part of the charge is inevitably lost due to the presence of electronegative impurities in the liquid. To keep such loss to a minimum, purifying the \dword{lar} during operation is essential, as is the monitoring of impurities.

Residual gas analyzers are an obvious choice when analyzing argon gas and can be exploited for the monitoring of the gas in the ullage of the tank. Unfortunately, commercially available and suitable mass spectrometers have a detection limit of \num{\sim10}\dword{ppb}, whereas DUNE requires a sensitivity down to the \dword{ppt} level. Instead, specially constructed purity monitors measure \lar purity in all the phases of operations, and enable the position-dependent purity measurements necessary to achieve DUNE's physics goal. 

Purity monitors also serve to mitigate \lar contamination risk.  
The large scale of the \dwords{detmodule} increases the risk of failing to notice a sudden unexpected infusion of contaminated \lar being injected back into the cryostat.   
If this condition were to persist, it could cause irreversible contamination to the \dword{lar} and terminate useful data taking.  Strategically placed purity monitors mitigate this risk. 

Purity monitors are placed inside the cryostat, but outside of the detector \dshort{tpc}, as well as outside the cryostat within the recirculation system before and after filtration. 
Continuous monitoring of  the \dword{lar} supply lines to the \dword{detmodule} provides a strong line of defense against contaminated \lar. Gas analyzers (described in Section~\ref{sec:fdgen-slow-cryo-gas-anlyz}) provide a first line of defense against contaminated gas.  Purity monitors inside the \dword{detmodule} provide a strong defense against all sources of contamination in the \lar volume and contamination from recirculated \lar. 
Furthermore, multiple purity monitors measuring lifetime with high precision at carefully chosen points can provide key inputs to \dshort{cfd} models of the detector, such as vertical gradients in impurity concentrations.

Purity monitors have been deployed in the ICARUS and \microboone detectors and in the \dword{35t} detector at Fermilab. In particular during the first run of the \dword{35t}, two out of four purity monitors stopped working during the cooldown, and a third was intermittent. It was later found out that this was due to poor electrical contacts of the resistor chain on the purity monitor. A new design was then implemented and successfully tested in the second run. 
The \dword{pdsp} and \dword{pddp} employ purity monitors based on the same design principles. \dword{pdsp} utilizes a string of purity monitors similar to that of the \dword{35t}, enabling measurement of the electron drift lifetime as a function of height.  A similar system design is exploited in the DUNE \dword{fd}, with modifications made to accommodate the instrumentation port placement relative to the purity monitor system and the requirements and constraints coming from the different geometric relations between the \dshort{tpc} and cryostat. 

\subsubsection{Physics and Simulation}

A purity monitor is a small ionization chamber that can be used to independently  infer the effective free electron lifetime in the \lartpc.  The operational principle of the purity monitor consists of generating a known electron current via illumination of a cathode with UV light, followed by collecting at an anode the current that survives after drifting a known distance.  The  attenuation of the current can be related to the electron lifetime.
The electron loss can be parameterized as
\(N(t) = N(0)e^{-t/\tau},\)
where $N(0)$ is the number of electrons generated by ionization, $N(t)$ is the number of electrons after drift time $t$, and $\tau$ is the electron lifetime.

For the \dword{spmod}, the 
\fixme{max?} drift distance is \spmaxdrift and the \efield is \SI{500}{\volt\per\centi\meter}. Given the drift velocity at this field of approximately \SI{1.5}{\milli\meter\per\micro\second}, the time to go from cathode to anode is around \SI{\sim2.4}{\milli\second} \cite{Walkowiak:2000wf}.
The \dword{lar} \dshort{tpc} signal attenuation, \([N(0)-N(t)]/N(0)\), is to be kept less than \SI{20}{\percent} over the entire drift distance \cite{fdtf-final-report}. The corresponding electron lifetime is $2.4/[-\ln(0.8)] \simeq \SI{11}{ms}$.

For the \dword{dpmod}, the maximum drift distance is \dpmaxdrift{}, therefore the requirement on the electron lifetime is much higher.

The 
\dword{35t} at Fermilab was instrumented with four purity monitors. The data taken with them during the first part of the second phase is shown in Figure~\ref{fig-35t-prm} and clearly shows the ability to measure the electron lifetime between \SI{100}{\micro\second} and \SI{3.5}{\milli\second}.  

\begin{dunefigure}[Electron lifetimes measured in the purity monitors in the \dword{35t}]{fig-35t-prm}
  {The measured electron lifetimes in the four purity monitors as a function of time at Fermilab \dword{35t}.}
  \includegraphics[width=0.6\textwidth]{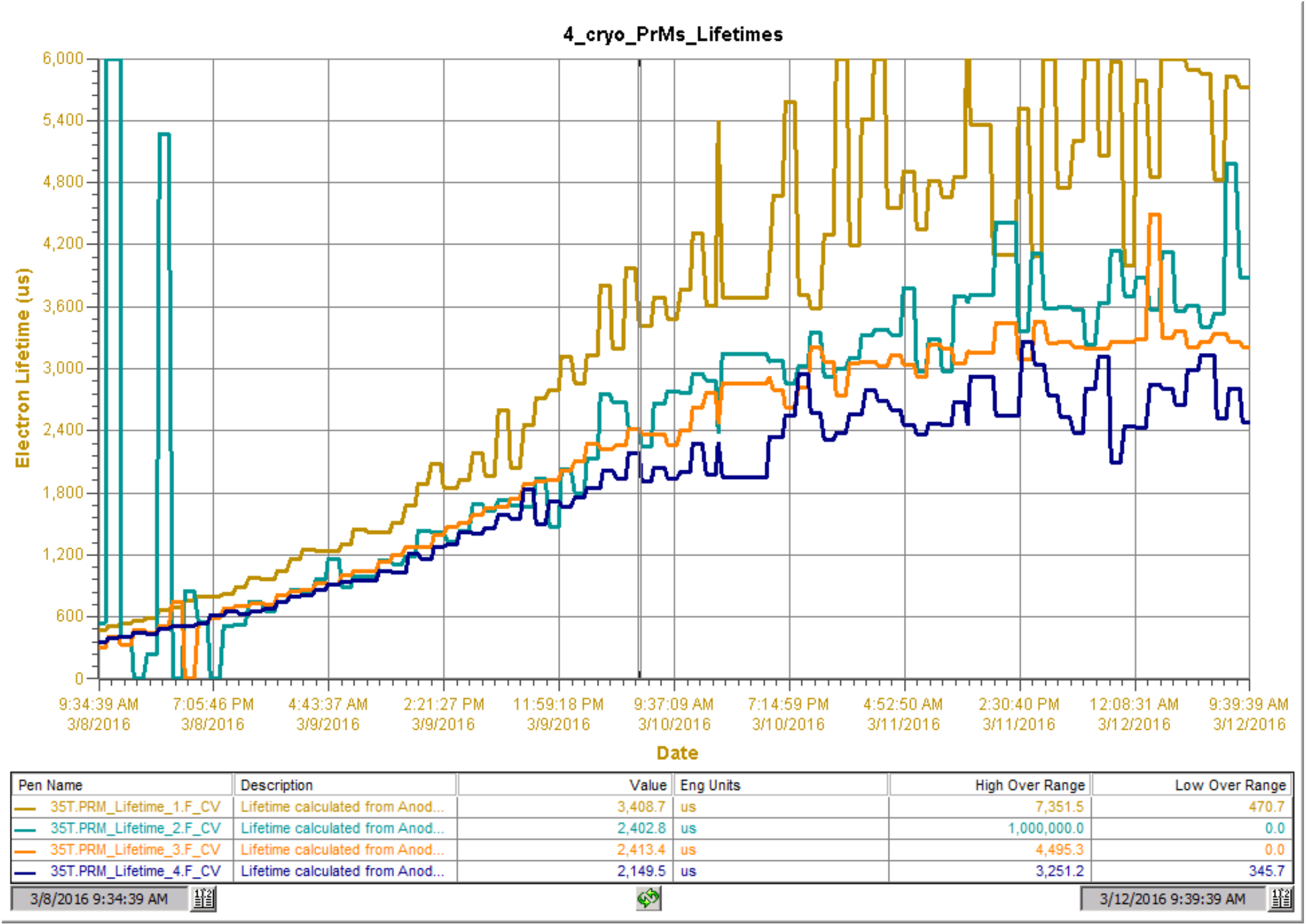}%
\end{dunefigure}

\subsubsection{Purity Monitor Design}

The basic design of a purity monitor is based on those used by the ICARUS experiment (Figure~\ref{fig:prm})\cite{Adamowski:2014daa}. It is a double-gridded ion chamber immersed in the \lar volume.   The purity monitor consists of four parallel, circular electrodes: a disk holding a photocathode, two grid rings (anode and cathode), and an anode disk. The cathode grid is held at ground potential. The cathode, anode grid, and anode are electrically accessible via modified vacuum grade high-voltage \fdth{}s and separate bias voltages held at each one.  
The anode grid and the field shaping rings are connected to the cathode grid by an internal chain of \SI{50}{\mega\ohm} resistors to ensure the uniformity of the \efield{}s in the drift regions. A stainless mesh cylinder is used as a Faraday cage to isolate the purity monitor from external electrostatic backgrounds. 

The purity monitor measures the electron drift lifetime between its anode and cathode. The electrons are generated by the purity monitor's UV-illuminated gold photocathode via the photoelectric effect. As the electron lifetime in \lar is inversely proportional to the electronegative impurity concentration, the fraction of electrons generated at the cathode that arrive at the anode ($Q_A/Q_C$) after the electron drift time $t$ gives a measure of the electron lifetime $\tau$:
\( Q_A/Q_C \sim e^{-t/\tau}.\)
%

It is clear from this formula that the purity monitor reaches its sensitivity limit once the electron lifetime becomes much larger than the drift time $t$. For $\tau >> t$ the anode to cathode charge ratio becomes $\sim\,1$. But, as the drift time is inversely proportional to the \efield, by lowering the drift field one can in principle measure any lifetime no matter the length of the purity monitor (the lower the field, the lower the drift velocity, i.e., the longer the drift time). 
In practice, at very low fields it is hard to drift the electrons all the way up to the anode. Currently, specific sensitivity limits for purity monitors with a drift distance of the order of $\sim$\SI{20}{\centi\meter} are still to be determined in a series of tests. If the required sensitivity is not achieved by these ``short'' purity monitors, longer ones may be developed.

\begin{dunefigure}[Purity monitor diagram]{fig:prm}
  {Schematic diagram of the basic purity monitor design~\cite{Adamowski:2014daa}.}
  \includegraphics[width=0.5\textwidth]{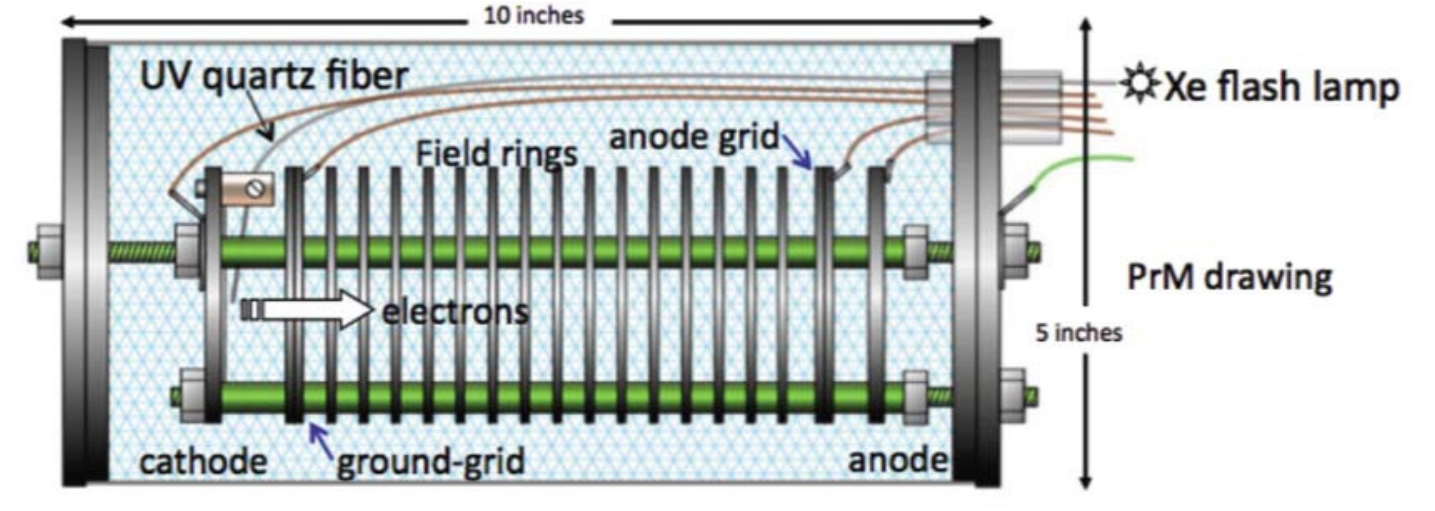}
\end{dunefigure}

The photocathode that produces the \phel{}s is an aluminum plate coated with \SI{50}{\angstrom} of titanium and \SI{1000}{\angstrom} of gold and attached to the cathode disk. A xenon flash lamp is used as the light source in the baseline design, although this could potentially be replaced by a more reliable and possibly submersible light source in the future, perhaps LED driven. The UV output of the lamp is quite good around $\lambda=$ \SI{225}{\nano\meter}, which is close to the work function of gold (\SIrange{4.9}{5.1}{\eV}). Several UV quartz fibers are used to carry the xenon UV light into the cryostat to illuminate the gold photocathode.   Another quartz fiber is used to deliver the light into a properly biased photodiode outside of the cryostat to provide the trigger signal for when the lamp flashes. 

\subsubsection{Electronics, DAQ and Slow Controls Interfacing}
The purity monitor electronics and \dword{daq} system consist of \dword{fe} electronics, waveform digitizers, and a \dword{daq} PC.  The block diagram of the system is shown in Figure~\ref{fig:cryo-purity-mon-diag}.

The baseline design of the \dword{fe} electronics is the one used for the purity monitors at the \dword{35t}, LAPD, and \microboone. The cathode and anode signals are fed into two charge amplifiers contained within the purity monitor electronics module.
The amplified outputs of the anode and cathode are recorded with a waveform digitizer that interfaces with a \dword{daq} PC.

\begin{dunefigure}[Purity monitor block diagram]{fig:cryo-purity-mon-diag}
  {Block diagram of the purity monitor system.}
  \includegraphics[width=0.7\textwidth]{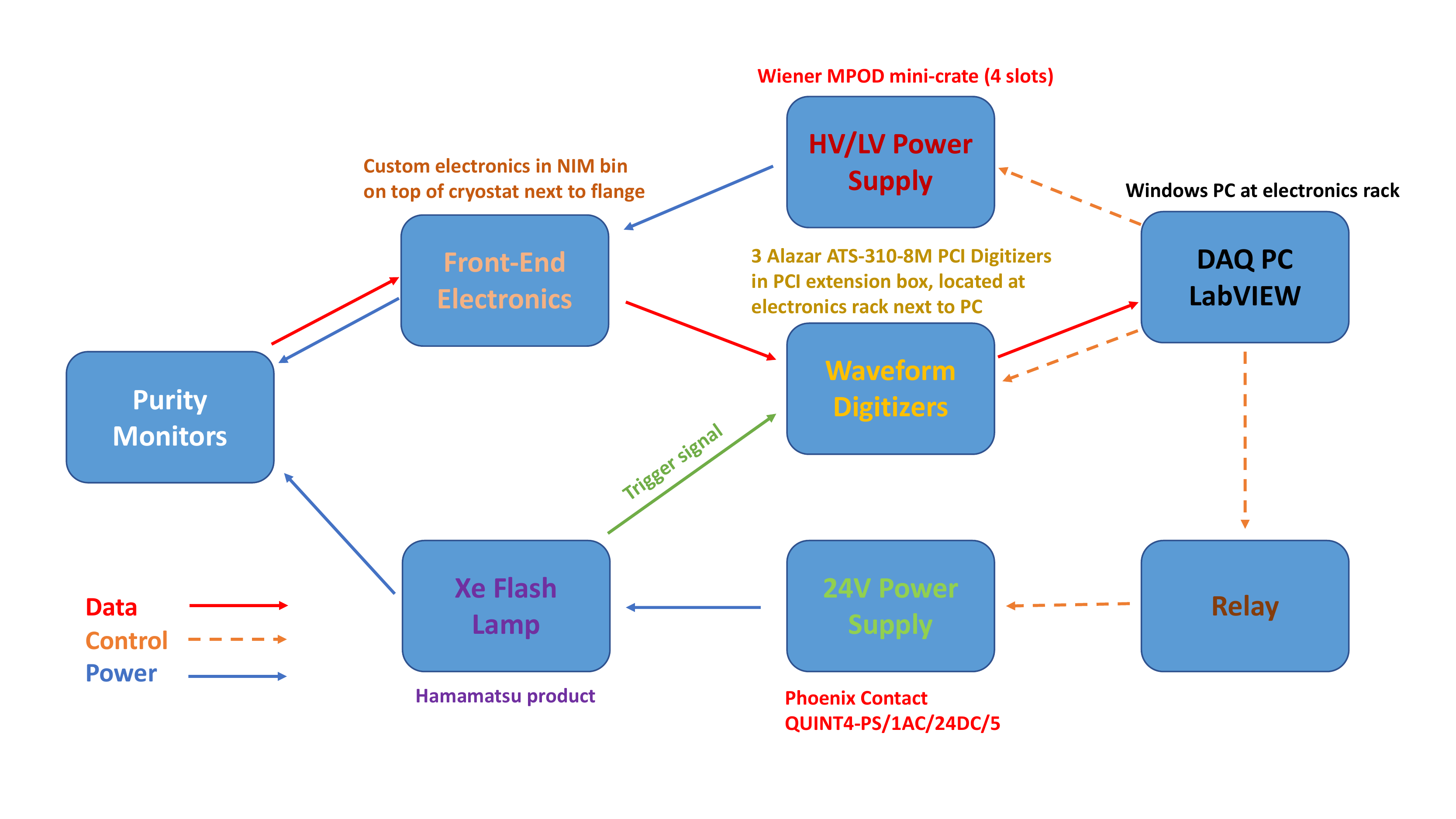}%
\end{dunefigure}

A custom LabVIEW application running on the \dword{daq} PC is developed and 
executes two functions: it controls the waveform digitizers and the power supplies, and it monitors the signals and key parameters. The application configures the digitizers to set the sampling rate, the number of waveforms to be stored in the memory, pre-trigger data, and a trigger mode. A signal from a photodiode triggered by the xenon flash lamp is directly fed into the digitizer as an external trigger to initiate data acquisition. The LabVIEW application automatically turns on the xenon flash lamp by powering a relay at the start of data taking and then turns it off when finished.
The application continuously displays the waveforms and important parameters, such as measured electron lifetime, peak voltages, and drift time of electrons in the purity monitors, and shows these parameters over time.

The xenon flash lamp and the \dword{fe} electronics are installed close to the purity monitor flange, to reduce light loss through the optical fiber and prevent signal loss. Other pieces of equipment are mounted in a rack separate from the cryostat. They distribute power to the xenon flash lamp and the \dword{fe} electronics, as well as collect data from the electronics. The slow control system communicates with the purity monitor \dword{daq} software and has control of the \dword{hv} and \dword{lv} power supplies of the purity monitor system. As the optical fiber has to be very close to the photocathode (less than \SI{0.5}{\milli\meter}) for efficient \phel extraction, no interference with the \dword{pds} is expected. Nevertheless light interference will be evaluated more precisely at \dword{protodune}.

Conversely the electronics of purity monitors may induce noise in the \dshort{tpc} electronics, largely coming from the current surge in the discharging process of the main capacitor of the purity monitor xenon light source when producing a flash.  This source of noise can be controlled by placing the xenon flash lamp inside its own Faraday cage, allowing for proper grounding and shielding; the extent of mitigation will be evaluated at \dword{protodune}.
If an unavoidable interference problem is found to exist, then software can be implemented to allow the \dword{daq} to know if and when the purity monitors are running and to veto purity monitor measurements in the event of a \dword{snb} alert or trigger.

\subsubsection{Production and Assembly}
\label{sec:PrMon-Production-Assembly}
Production of the individual purity monitors and their assembly into the string that gets placed into the \dword{detmodule} cryostat follows the same methodology that is being developed for \dword{protodune}.  Each of the individual monitors is fabricated, assembled and then tested in a smaller test stand.  After confirming that each of the individual purity monitors operates at the required performance, they are assembled together via the support tubes used to mount the system to the inside of the cryostat such that three purity monitors are grouped together to form one string, as shown in Figure~\ref{fig:PrMon-SystemString}.
Each monitor is assembled as the string is built from the top down, and in the end 
three individual purity monitors 
hang from a single string.  The assembly of the string concludes once the purity monitors are each in place, but with the Faraday cages removed and the \dword{hv} cables and optical fibers yet to be run.  This full string assembly is then 
shipped to the \dword{fd} site for installation into the cryostat.

\begin{dunefigure}[Purity monitor string]{fig:PrMon-SystemString}
  {Design of the purity monitor string that will contain three purity monitors.}
  \includegraphics[width=\textwidth]{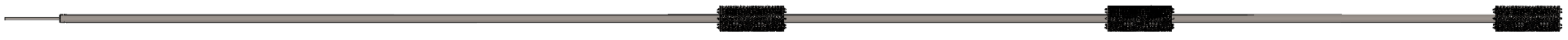}
\end{dunefigure}



\subsection{Thermometers}
\label{sec:fdgen-slow-cryo-therm}

A detailed \threed temperature map is important to monitor the correct functioning of the cryogenics system and the \lar uniformity.
Given the complexity and size of purity monitors, those can only be installed on the cryostat sides to provide a local measurement of
the \lar purity. While a direct measurement of the \lar purity across the entire cryostat is not viable, a sufficiently detailed \threed temperature map
can be used to predict the \lar purity using \dword{cfd} simulations. Measuring the vertical temperature profile is especially important since this is closely related to
the \lar recirculation and uniformity. 

High-precision temperature sensors are distributed near the TPC walls in two ways:
(1) in high-density (\(>2\) sensors/\si{m}) vertical arrays (the T-gradient monitors), and (2) in coarser ($\sim$ 1 sensor/\SI{5}{m}) \twod arrays 
at the top and bottom of the detector, which are the most sensitive regions (the individual sensors).   

Since temperature variations inside the cryostat are expected to be very small ($\SI{0.02}{K}$, see Figure~\ref{fig:cfd-example}), to properly measure the \threed temperature map 
sensors must be cross-calibrated to better than $\SI{0.005}{K}$. Most sensors are calibrated in the laboratory, prior to installation,
as described in Section~\ref{sp-cisc-thermom-static-t}.  This is in fact the only viable method for sensors in areas where the available space is restricted: on the long sides of the detector
(behind the \dwords{apa} for SP, and behind the lateral 
\dword{ewfc} for \dual) and top/bottom of the detector.
Given the precision required and the unknown longevity of the sensors (which could require a new calibration after some time), a complementary method
is used for T-gradient monitors behind the front endwalls, at least for the \dword{spmod}.
In those areas there is sufficient space for a movable system that can be used to cross-calibrate in situ
the temperature sensors. 

In the baseline design for all three systems mentioned above, three elements are common: sensors, cables and readout system.
Platinum sensors with \SI{100}{\ohm} resistance, PT100 series, produced by Lakeshore\footnote{Lakeshore\texttrademark{}, \url{https://www.lakeshore.com/Pages/Home.aspx}.},
are adequate for the temperature range of interest, \SIrange{83}{92}{K}, since in this range those sensors have $\sim\SI{5}{mK}$ reproducibility 
 and absolute temperature accuracy of \SI{100}{mK}.
In addition, using four-wire readout greatly reduces the issues related to the lead resistance, any parasitic resistances,
connections through the flange, and general electromagnetic noise pick-up. The Lakeshore PT102 sensors
have been previously used in the \dword{35t} and \dword{pdsp} detector,
giving excellent results. For the inner readout cables a custom cable made by Axon\footnote{Axon\texttrademark{}, \url{http://www.axon-cable.com/en/00_home/00_start/00/index.aspx}} is the baseline. It consists of four AWG28 teflon-jacketed copper wires forming two twisted pairs, with a metallic external shield
and an outer teflon jacket.
The readout system is described below in  Section~\ref{sec:fdgen-slow-cryo-therm-readout}. 



Another set of lower-precision sensors is used to monitor the filling of the cryostat in its initial stage. Those sensors are epoxied into the cryostat bottom membrane with
a density to be determined, not to exceed one sensor every \SI{5}{m}. 
Finally, the inner walls and roof of the cryostat are instrumented with the same type of sensors in order to monitor their temperature during cooldown and filling.
The baseline distribution has three vertical arrays of sensors epoxied to the membrane: one behind each of the two 
front \dwords{ewfc} and the third one in the middle of the cryostat
(behind the \dwords{apa} for \single and behind the lateral 
\dwords{ewfc} for \dual). 


\subsubsection{Static T-gradient Monitors}
\label{sp-cisc-thermom-static-t}

Several vertical arrays of high-precision temperature sensors cross-calibrated in the laboratory are installed near the lateral walls
(behind the \dwords{apa} for \single and behind the lateral 
\dwords{ewfc} for \dual). 
For the \dword{spmod}, since the electric potential is zero behind the \dwords{apa}, no \efield shielding is required, simplifying enormously the mechanical design.
This does not apply for the \dword{dpmod}, for which the proper shielding must be provided.

Sensors are cross-calibrated in the lab using a well controlled environment and a high-precision readout system, described below in Section~\ref{sec:fdgen-slow-cryo-therm-readout}. 
Although the calibration procedure will certainly improve, the one currently used for \dword{pdsp} is described here.
Four sensors are placed as close as possible (such that identical temperature can be assumed for all of them) inside a small cylindrical aluminum capsule,
which protects the sensors from thermal shocks and helps in minimizing convection.
One of the sensors acts as reference while the other three are 
calibrated. Five independent calibrations
are performed for each set of three sensors, such that the reproducibility of each sensor can be computed. For each calibration 
the capsule is introduced in a \threed printed polylactic acid (PLA) box of size \(9.5\times9.5\times\SI{19}{cm^3}\), with two concentric independent volumes of \lar
and surrounded by a polystyrene box with \SI{15}{cm} thick walls. A small quantity of \lar is used to slowly
cool down the capsule to $\sim\SI{90}{K}$, avoiding thermal shocks that could damage the sensors.
Then the capsule is immersed in  \lar such that it penetrates
inside, fully covering the sensors. Once the temperature stabilizes to the 1-\SI{2}{mK} level (after 5-15 minutes) measurements are taken. Then the capsule is taken out from \lar
and exposed to room temperature until it reaches \SI{200}{K}. As mentioned above, this procedure is repeated five times, before going to the next set of three sensors.  
As shown in Figure~\ref{fig:Trepro} a reproducibility (\rms of the mean offset in the flat region) of $\sim \SI{2}{mK}$ has been achieved in the \dword{pdsp} design.  

The baseline design for the mechanics of the \single system consists of two stainless strings anchored at top and bottom corners of the cryostat
using the available M10 bolts (see Figure~\ref{fig:sensor-support}). One of the strings is used to route the cables while the other,
separated by \SI{340}{mm}, serves as support for temperature sensors.
Given the height of the cryostat, the need of intermediate anchoring points is under discussion. For the \dword{dpmod} no baseline design exists yet,
since additional complications due to the required \efield shielding must be taken into account. Figure~\ref{fig:sensor-support} shows the baseline design of the
PCB support for temperature sensors, with an IDC-4 male connector. It has a size of $52\times \SI{15}{mm^2}$. Each four-wire cable from the sensor to the flange has an IDC-4 female connector
on the sensor side; on the other side, it is directly soldered into the inner pins of male SUBD-25 connectors on the flanges. The CF63 side ports on the \dword{dss}/cryogenic ports are 
used to extract the cables. 

\begin{dunefigure}[Cryostat bolts and temperature sensor support]{fig:sensor-support}
  {Left: bolts at the bottom corner of the cryostat. Right: Lakeshore PT102 sensor mounted on a PCB with an IDC-4 connector.}
  \includegraphics[height=0.2\textwidth]{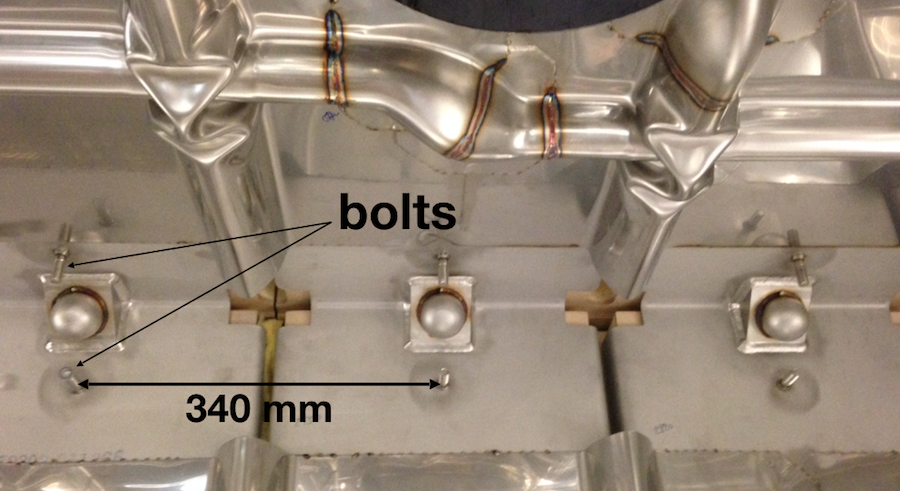}%
    \hspace{1cm}%
  \includegraphics[height=0.2\textwidth]{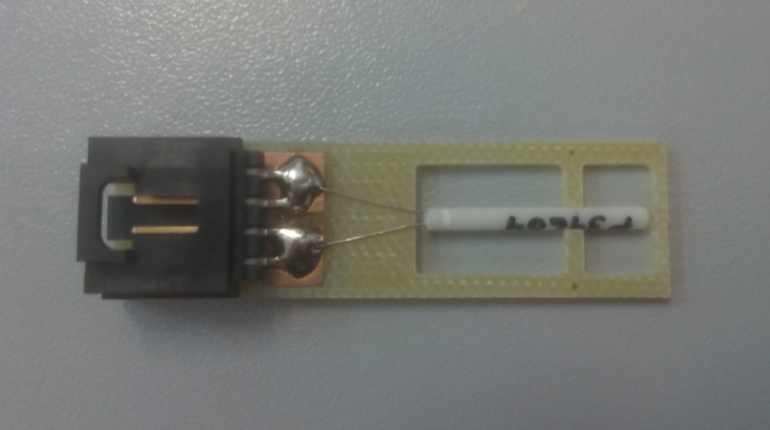}%
\end{dunefigure}

\begin{dunefigure}[Temperature sensor resolution and reproducibility]{fig:Trepro}
  {Temperature offset between two sensors as a function of time for five independent inmersions in \lar. The reproducibility of those sensors,
    defined as the RMS of the mean offset in the flat region, is $\sim \SI{2}{mK}$,
    The resolution for individual measurements, defined as the RMS of one of the offsets in the flat region, is better than \SI{1}{mK}.}
  \includegraphics[width=0.5\textwidth]{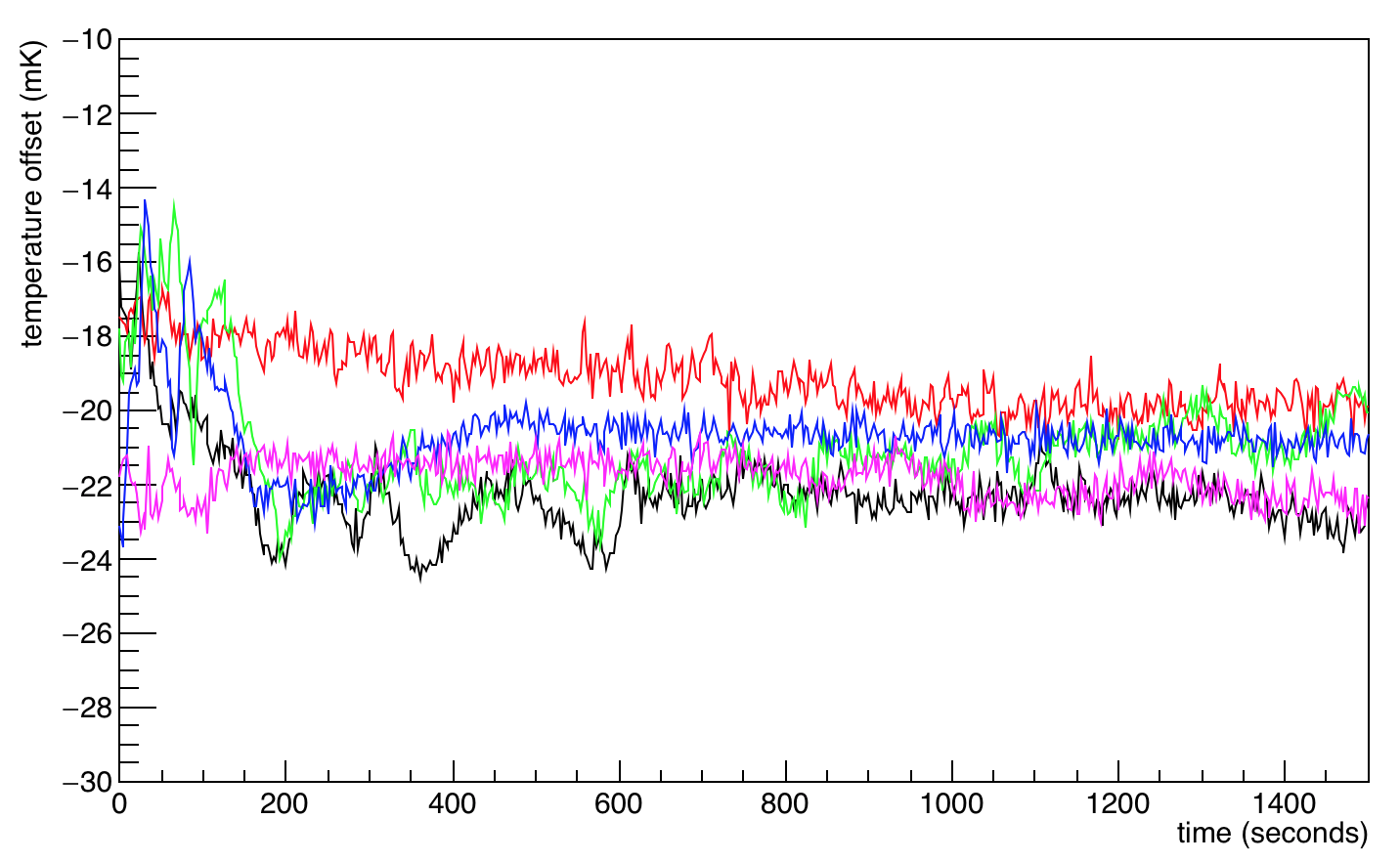}%
\end{dunefigure}

\subsubsection{Dynamic T-gradient Monitors}

The dynamic temperature monitor is a vertical array of high precision temperature sensors with the goal of measuring vertical temperature gradient with precision of few \si{mK}. The design of the system is driven by two factors:
\begin{itemize}
\item
A few-\si{mK} uncertainty in the measured vertical temperature profile over the entire detector height is required in order to monitor \lar purity and provide useful feedback of efficiency of cryogenic recirculation and purification.
\item
Simulations of the cryogenic recirculation predict very slow change in temperature at meter scale except at the bottom and top of the cryostat. Thus, sensors are placed every \SI{50}{cm} along the \dword{detmodule} height with increased frequency in the first \SI{50}{cm}, closest to the bottom of the cryostat and the last \SI{50}{cm}, closest to the top of the cryostat, where spacing between sensors is reduced to \SI{10}{cm}.
 \end{itemize}


 
 In order to address concerns related to possible differences in the sensor readings prior to 
 and after installation in a \dword{detmodule}, a dynamic temperature monitor allows cross-calibration of sensors in situ. 
 \fixme{difference in voltage, or differences in the sensor reading that may happen? I think the latter... (anne)}
 Namely, this T-gradient monitor  can move vertically while installed in the \dword{detmodule}, which allows for precise cross-calibration between the sensors in situ at predefined locations, as well as in between them. \fixme{not clear}
 The procedure for cross-calibrations is the following: the temperature reading is taken at the lowest position with all sensors. The stepper motor then moves the carrier rod up \SI{50}{cm},  
 putting all sensors in the previous location of their neighbor that was \SI{50}{cm} above them. 
 Then the second reading is taken. In this manner, except for the lowest position we have temperature measurement at each location with two adjacent sensors, and by linking the temperature offsets between the two readings at each location, temperature readings from all sensors are cross-calibrated in situ, cancelling all offsets due to electromagnetic noise or any parasitic resistances that may have prevailed despite the four point connection to the sensors that should cancel most of the offsets. These measurements are taken with very stable current source, which ensures high precision of repeated temperature measurements over time. The motion of the dynamic T-monitor is stepper motor operated, 
delivering measurements with high spatial resolution. 

\subsubsection{Dynamic T-gradient Monitor Design}

A dynamic T-gradient monitor consists of three distinct parts: a carrier rod on which sensors are mounted, an enclosure above the cryostat housing the space that allows vertical motion of the carrier rod 1.5\,m above its lowest location, and the motion mechanism. The motion mechanism consists of a stepper motor connected to a gear and pinion through a ferrofluidic dynamic seal. The sensors have two pins that are soldered to a printed circuit board (PCB). Two wires are soldered to the common soldering pad for each pin, individually.   There is a cutout in the PCB around the sensor that allows free flow of argon for more accurate temperature reading.  Stepper motors typically have very fine steps allowing high-precision positioning of the sensors.  Figure~\ref{fig:fd-slow-cryo-dt-monitor-overview} shows the overall design of the dynamic T-gradient monitor with the sensor carrier rod, enclosure above the cryostat and the stepper motor mounted on the side of the enclosure. The enclosure consists of two parts connected by a six-way cross flange. One side of this cross flange is used to for the signal wires, another side is used as a viewing window, while the two other ports are spares. Figure~\ref{fig:fd-slow-cryo-sensor-mount} (left) shows the mounting of the PCB board on the carrier rod and mounting on the sensor on the PCB along with the four point connection to the signal readout wires. Finally, Figure~\ref{fig:fd-slow-cryo-sensor-mount} (right) shows the stepper motor mounted on the side of the rod enclosure. The motor is kept outside, at room temperature, and its power and control cables are also kept outside.
 \fixme{above pgraph needs some work}

\begin{dunefigure}[Dynamic T-gradient monitor overview]{fig:fd-slow-cryo-dt-monitor-overview}
  {An overview of the dynamic T-gradient monitor.}
 \includegraphics[width=0.11\textwidth,angle=90]{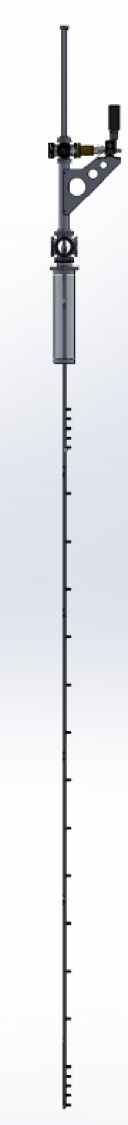}
\end{dunefigure}
\begin{dunefigure}[Sensor-cable assembly for dynamic T-gradient monitor]{fig:fd-slow-cryo-sensor-mount}
  {Left: Sensor mounted on a PCB board and PCB board mounted on the rod. Right:
    The driving mechanism of the dynamic T-gradient monitor. It consists of a stepper motor driving the pinion and gear linear motion mechanism. }
  \includegraphics[width=0.40\textwidth]{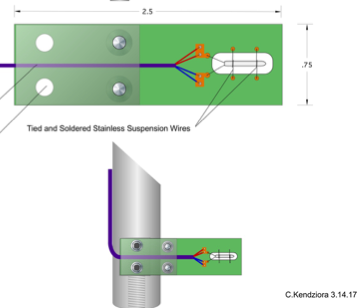}
  \hspace{3cm}%
  \includegraphics[width=0.12\textwidth]{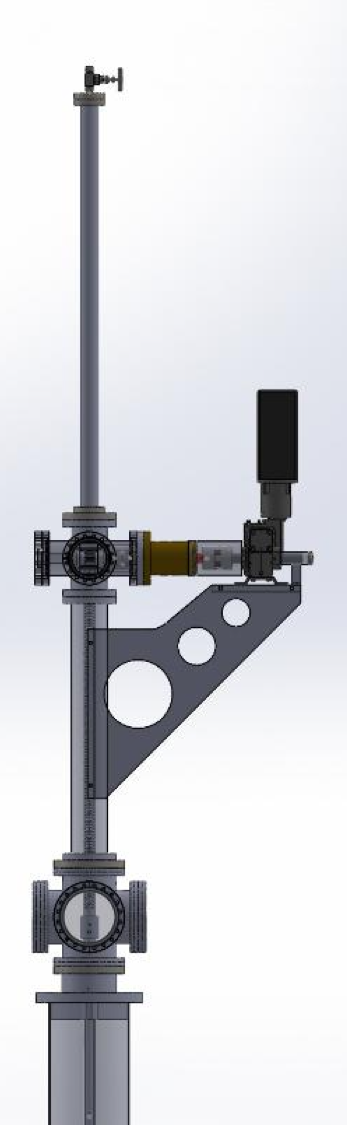}
\end{dunefigure}

\subsubsection{Individual Temperature Sensors}

T-Gradient monitors provide a vertical temperature profiling outside the TPCs. They are complemented by a coarser \twod array at the top and bottom of the
detector. Sensors, cables and readout system are the same as for the T-gradient monitors. 

In principle, a similar distribution of sensors is used at top and bottom.
Following the \dword{pdsp} design, bottom sensors use the cryogenic pipes as a support structure, while top sensors are anchored to the \dwords{gp}.
Teflon pieces (see Figure~\ref{fig:cable-support}) are used to route cables from the sensors to the CF63 side ports on \dword{dss}-cryogenics ports, which are used to extract the cables.
The PCB sensor's support, cables and connection to the flanges are the same as for the static T-gradient monitors. 

\begin{dunefigure}[Cable supports for individual temperature sensors]{fig:cable-support}
  {Left: support for two cables on ground planes. Right: Supports for three cables  mounted on cryogenics pipes using split clamps}
  \includegraphics[width=0.3\textwidth]{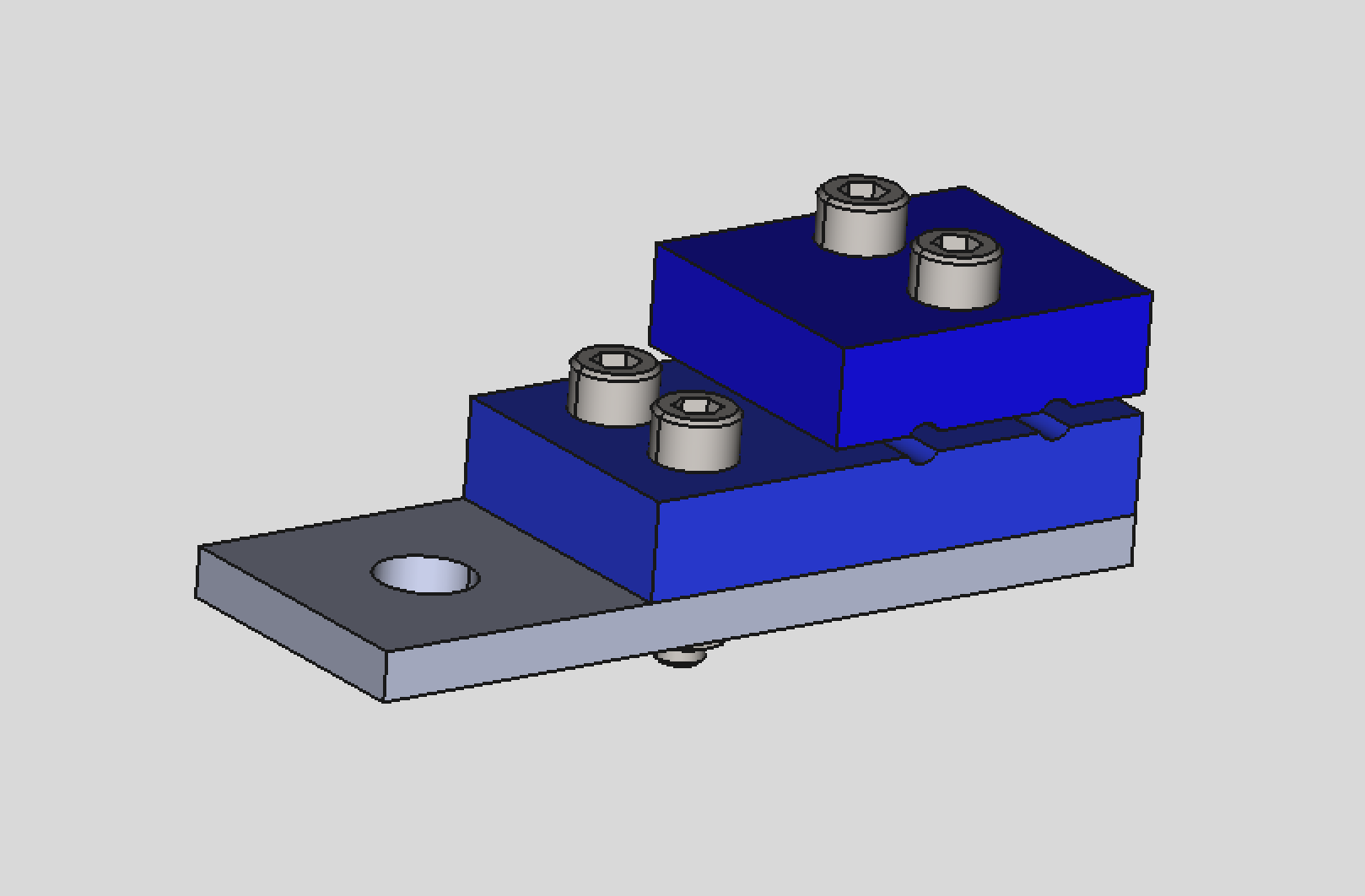}
  \includegraphics[width=0.315\textwidth]{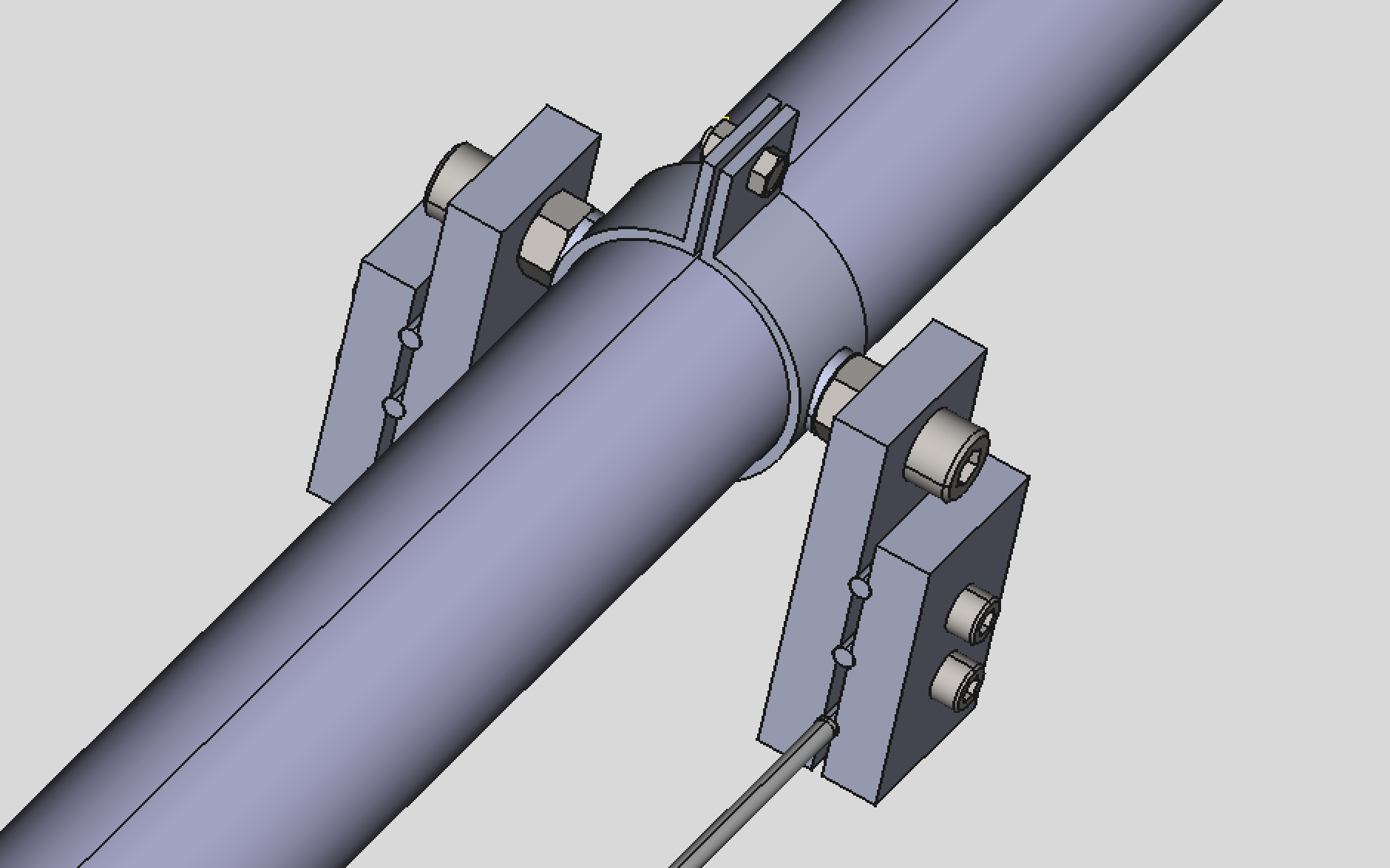}
\end{dunefigure}

\subsubsection{Readout System for Thermometers}
\label{sec:fdgen-slow-cryo-therm-readout}

A high precision and very stable system is required to achieved the design precision of $<\SI{5}{mK}$.
The proposed readout system is the one used in \dword{pdsp}, which is based on a variant of an existing mass PT100 temperature readout system developed at
CERN for one of the LHC experiments. The system consists of three parts:
\begin{itemize}
\item An accurate current source for the excitation of the temperature sensors, implemented by a compact electronic circuit using a high-precision voltage reference from Texas Instruments~\footnote{Texas Instruments\texttrademark{}, \url{http://www.ti.com/}.};
\item A multiplexing circuit based on an Analog Devices ADG707\footnote{Analog Devices\texttrademark{},\url{http://www.analog.com/media/en/technical-documentation/data-sheets/ADG706_707.pdf}.} multiplexer electronic device;
\item A high-resolution and accuracy voltage signal readout module based on National Instruments~\footnote{National Instruments\texttrademark{}, \url{http://www.ni.com/en-us.html/}.} NI9238, which has \SI{24}{bit} resolution over a \SI{1}{V} range.
  This module is inserted in a National Instruments compact RIO device that distributes the temperature values to the main slow control software
  through the standard OPC UA protocol. The Ethernet \dword{daq} also includes the multiplexing logic.
\end{itemize}

The current mode of operation averages over \num{2000} samples taken every second for each sensor. 
As inferred from Figure~\ref{fig:Trepro} the system has a resolution better than
\SI{1}{mK}, the \rms of one of the offsets in the stable region.


\subsection{Liquid Level Monitoring}
\label{sec:fdgen-slow-cryo-liq-lev}

The goals for the level monitoring system are basic level sensing when filling, and precise level sensing during static operations. 

For filling the \dword{detmodule} the differential pressure between the top of
the detector and known points below it can be converted to depth using
the known density of \lar.  The temperatures of \dwords{rtd} at known
heights may also be used to determine when the cold liquid reaches 
each \dword{rtd}.

During operation, the purpose of liquid level monitoring is twofold:
the cryogenics system uses it to tune the \lar flow, and 
the \dword{spmod} uses it to guarantee that the top \dwords{gp} are always
submerged (otherwise the risk of dielectric breakdown is high).
Two differential pressure level meters are installed as part of
the cryogenics system, one on each side of the \dword{detmodule}.  They 
have a precision of \SI{0.1}{\%}, which corresponds to \SI{14}{mm} at the
nominal \lar surface.  This precision is sufficient for the  \dword{spmod}, since the plan is to keep the \lar surface at least \SI{20}{cm} above the \dwords{gp} (this is the value used for the \dword{hv}
interlock in \dword{pdsp}); thus, no additional level meters are
required for the \single. 
However, in the \dual \lar system the surface level should be controlled at the millimeter level, which can be accomplished with capacitive monitors. Using the same capacitive monitor system in each \dword{detmodule} reduces design differences and provides a redundant system for the \single.  Either system could be used for the \dword{hv} interlock.

Table \ref{tab:fdgen-liq-lev-req} summarizes the
requirements for the liquid level monitor system.

\begin{dunetable}
[Liquid level monitor requirements]
{p{0.45\linewidth}p{0.50\linewidth}}
{tab:fdgen-liq-lev-req}
{Liquid level monitor requirements}   
Requirement & Physics Requirement Driver \\ \toprowrule
 Measurement accuracy (filling) \(\sim \SI{14}{mm}\) & Understand status of detector during filling \\ \colhline
 Measurement accuracy (operation, \dual) \(\sim \SI{1}{mm}\) & Maintain correct depth of gas phase. (Exceeds \single requirements) \\ \colhline
 Provide interlock with \dword{hv} & Prevent damage to \dword{detmodule} from \dword{hv} discharge in gas \\
\end{dunetable}

Cryogenic pressure sensors will be purchased from commercial sources.
Installation methods and positions will be determined as part of the
cryogenics internal piping plan.  Sufficient redundancy will be designed in
to ensure that no single point of failure compromises the level measurement.

Multiple capacitive level sensors are deployed along the top of
the fluid to be used during stable operation and checked against each
other.

During operations of the \dword{wa105}, the cryogenic programmable logic controller (PLC) continuously checked the measurements from one level meter on the charge readout plane (\dword{crp}) in order to regulate the flow from the liquid recirculation to maintain a constant liquid level inside the cryostat. Continuous measurements from the level meters around the drift cage and the \dword{crp} demonstrated the stability of the liquid level within the \SI{100}{\micro\meter} intrinsic precision of the instruments. The observation of the level was complemented by live feeds from the custom built cryogenic cameras, thereby providing qualitative feedback on the position and flatness of the surface.

In addition to the installed level meters, the liquid height in the extraction region of the \dword{crp} could be inferred by measuring the capacitance between the grid and the bottom electrode of each \dword{lem}. Averaging over all \num{12} \dwords{lem} the measured values of this capacitance typically ranged from \SI{150}{pF} with the liquid below the grid to around  \SI{350}{pF}  when the \dwords{lem} are submerged.This method offers the potential advantage of monitoring the liquid level in the \dword{crp} extraction region with a \num{50}$\times$\SI{50}{\cm^2} granularity and could be used for the \dword{crp} level adjustment in a DUNE \dword{dpmod} where, due to the space constraints, placement of the level meters along the \dword{crp} perimeter is not possible.


\subsection{Gas Analyzers}
\label{sec:fdgen-slow-cryo-gas-anlyz}

 Gas analyzers are commercially produced modules that measure trace quantities of specific gases contained within a stream of carrier gas. The carrier gas for DUNE is argon, and the trace gases of interest are oxygen ($\text{O}_2$), water ($\text{H}_2\text{O}$), and nitrogen ($\text{N}_2$). Oxygen and water impact the electron lifetime in \dword{lar}, while $\text{N}_2$ impacts the efficiency of scintillation light production. In the \dword{lar} environment, these trace gases represent contaminants that need to be kept at levels below \SI{0.1}{ppb}.
The argon is sampled from either the argon vapor in the ullage or from the \dword{lar} by the use of small diameter tubing run from the sampling point to the gas analyzer. Typically the tubing runs from the sampling points are connected to a switchyard valve that is used to route the sample points to the desired gas analyzers. Figure~\ref{fig:GA-switchyard} is a photo of such a switchyard.

\begin{dunefigure}[Gas Analyzer switchyard]{fig:GA-switchyard}
  {A Gas Analyzer switchyard that routes sample points to the different gas analyzers.}
  \includegraphics[width=0.35\textwidth]{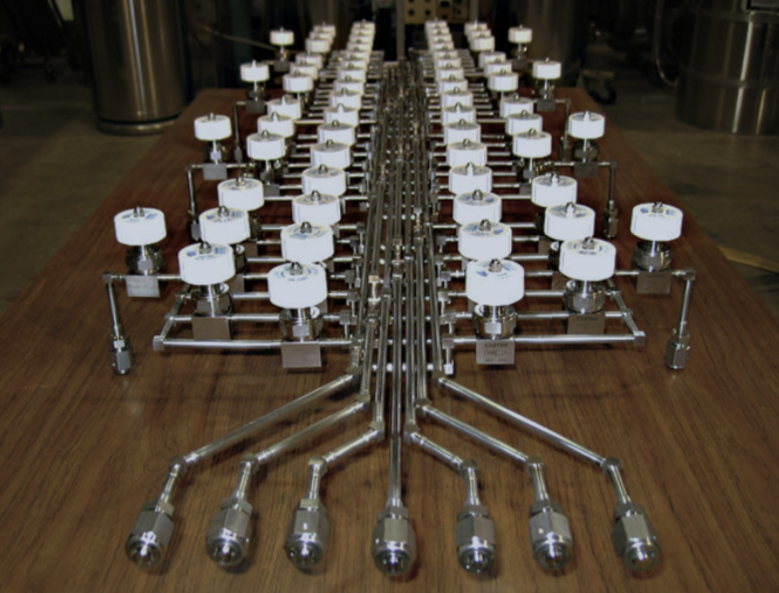}%
\end{dunefigure}

Gas analyzers can be used to:

\begin{enumerate}
\item Eliminate the air atmosphere from the cryostat after detector installation to levels low enough to begin cooldown is an argon piston purge followed by a recirculation of the remaining argon gas through the filtration system. This process is described more fully in Section~\ref{sec:fdgen-slow-cryo-install}. Figure~\ref{fig:GA-purge} shows the evolution of the $\text{N}_2$, $\text{O}_2$, and $\text{H}_2\text{O}$ levels from gas analyzer data taken during the purge and recirculation stages of the DUNE \dword{35t} 
phase 1 run.

\item Track trace $\text{O}_2$ and $\text{H}_2\text{O}$ contaminants from the $\>$tens of ppb to the hundreds of ppt. This is useful when other means of monitoring the impurity level (e.g., purity monitors, or \dshort{tpc} tracks) are not yet sensitive. Figure~\ref{fig:GA-O2} shows an example plot of the $\text{O}_2$ level at the beginning of \dword{lar} purification from one of the later \num{35}\si{t} 
\dword{hv} runs.

\item Monitor the tanker \dword{lar} deliveries purity during the cryostat-filling period. This allows tracking the impurity load on the filtration system and rejecting any deliveries that are out of specifications. Likely specifications for the delivered \dword{lar} are in the \SI{10}{ppm} range per contaminant.

\end{enumerate}

\begin{dunefigure}[Gas analyzer purge]{fig:GA-purge}
  {Plot of the $\text{O}_2$, $\text{H}_2\text{O}$, and $\text{N}_2$ levels during the piston purge and gas recirculation stages of the \num{35}\si{t} 
  phase 1 run.}
  \includegraphics[width=0.65\textwidth]{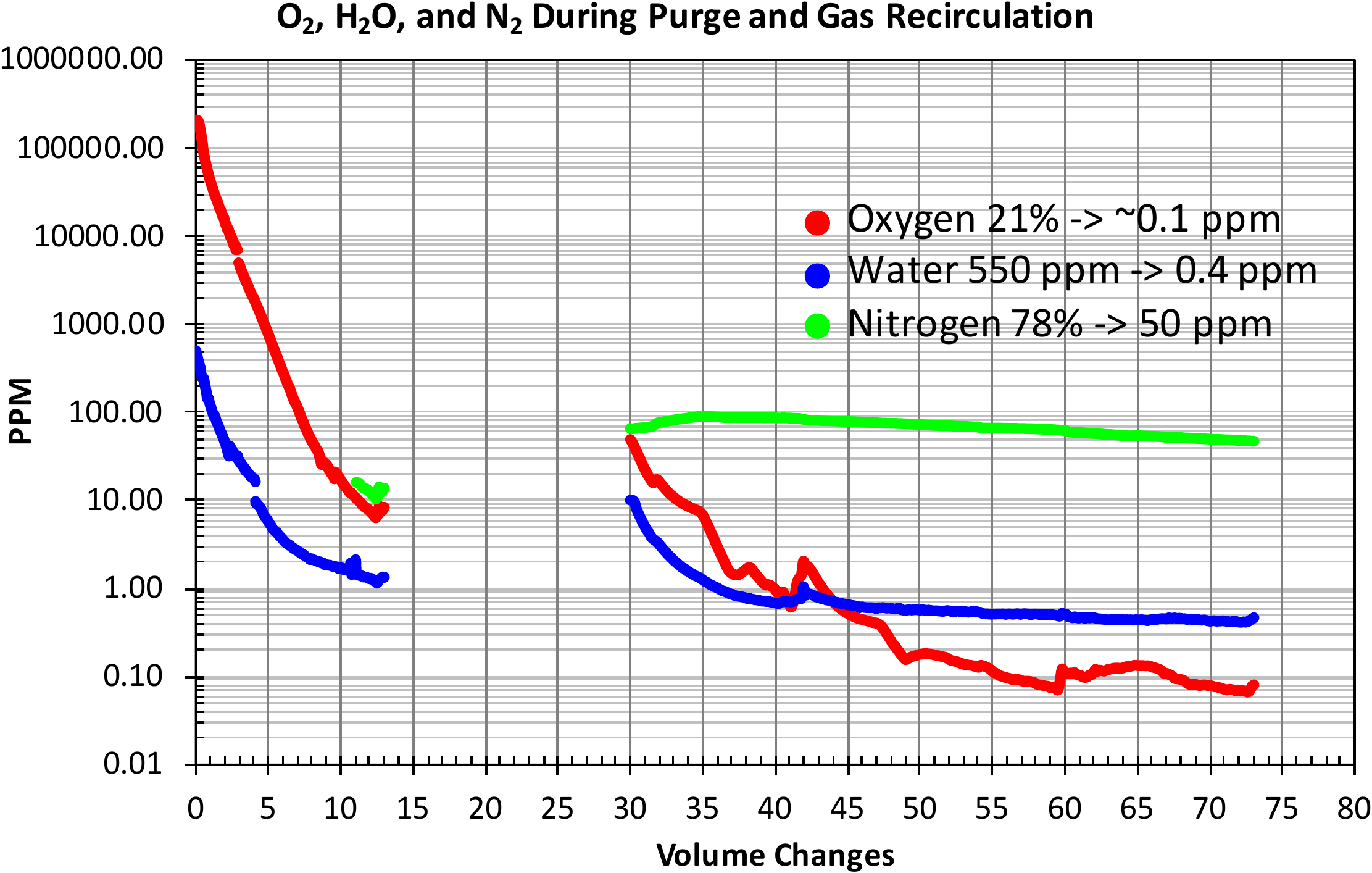}%
\end{dunefigure}

\begin{dunefigure}[Gas analyzer $O_2$ level after \dword{lar} filling]{fig:GA-O2}
  {$\text{O}_2$ as measured by a precision $\text{O}_2$ analyzer just after the \dword{35t} 
  was filled with \dword{lar}, continuing with the \dword{lar} pump start and beginning of \dword{lar} recirculation through the filtration system. As the gas analyzer loses sensitivity, the purity monitor is able to pick up the impurity measurement. Note that the purity monitor is sensitive to both $\text{O}_2$ and $\text{H}_2\text{O}$ impurities giving rise to its higher level of impurity.}
  \includegraphics[width=0.7\textwidth]{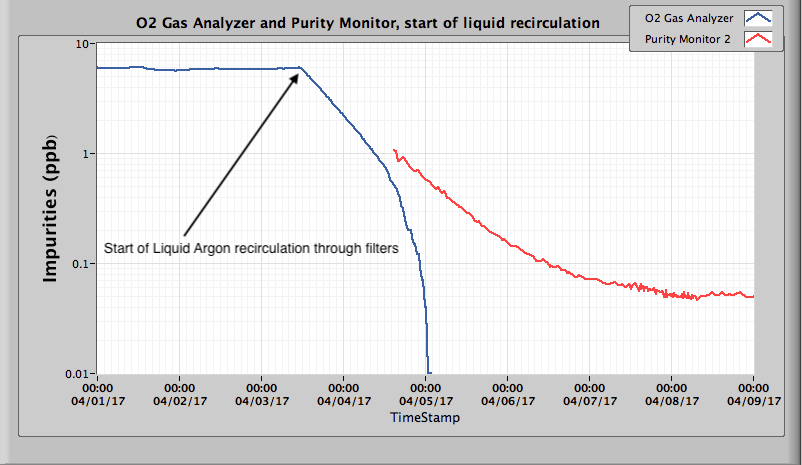}%
\end{dunefigure}

As any one gas analyzer covers only one contaminant species and \numrange{3}{4} orders of magnitude of range, multiple units are needed both for the three contaminant gases and to cover the ranges that are seen between the cryostat closure to the beginning of \dshort{tpc} operations:
\SI{20}{\percent} to $\lesssim 100$~ppt for $\text{O}_2$,
\SI{80}{\percent} to $\lesssim 1$~ppm for $\text{N}_2$, and
$\sim \SI{1}{\percent}$ to $\lesssim 1$~ppb for $\text{H}_2\text{O}$.
Since the total cost of these analyzers exceeds $\SI{100}[\textdollar]{k}$, it is useful to be able to  sample more than a single location or cryostat with the same gas analyzers. At the same time, the tubing run lengths from the sample point should be as short as possible in order to keep the response of the gas analyzer timely. This puts some constraints on the sharing of devices since, for example, the argon deliveries are at the surface, perhaps necessitating a separate surface gas analyzer.


\subsection{Cameras}
\label{sec:fdgen-slow-cryo-cameras}

Cameras provide direct visual information about the state of the
\dword{detmodule} during critical operations and when damage or unusual
conditions are suspected.  Cameras in the \dword{wa105} allowed spray from cool-down
nozzles to be seen and the level and state of the \lar to be
observed as it covered the \dword{crp} \cite{Murphy:20170516}.  A camera was
used in the Liquid Argon Purity Demonstrator
cryostat\cite{Adamowski:2014daa} to study \dword{hv} discharges in
\lar, and in EXO-100 during operation of a TPC
\cite{Delaquis:2013hva}.  Warm cameras viewing \lar from a distance
have been used to observe \dword{hv} discharges in \lar in
fine detail \cite{Auger:2015xlo}.  Cameras are commonly used during
calibration source deployment in many experiments (e.g., the
\kamland ultra-clean system \cite{Banks:2014hra}).

In DUNE, cameras are used to verify the stability, straightness,
and alignment of the hanging TPC structures during cool-down and
filling; to ensure that there is no bubbling near the \dwords{gp}
(\single) or \dwords{crp} (\dual); to inspect the
state of movable parts in the \dword{detmodule} (calibration devices, dynamic
thermometers) as needed; and to closely inspect parts of the TPC as
necessary following any seismic activity or other unanticipated
occurrence.  These functions are performed using a set of fixed
\textit{cold} cameras permanently mounted at fixed points in the cryostat
for use during filling and commissioning, and a movable, replaceable
\textit{warm} inspection camera that can be deployed through any free
instrumentation flange at any time throughout the life of the
experiment.  Table \ref{tab:fdgen-cameras-req} summarizes the
requirements for the camera system.

\begin{dunetable}
[Camera system requirements]
{p{0.45\linewidth}p{0.50\linewidth}}
{tab:fdgen-cameras-req}
{Camera system requirements}   
 Requirement & Physics Requirement Driver \\ \toprowrule
 \multicolumn{2}{l}{\bf General} \\ \specialrule{1.5pt}{1pt}{1pt}
 No component may contaminate the \lar{}. & High \lar purity is required for TPC operation. \\ \colhline
 No component may produce bubbles in the liquid argon if the \dword{hv} is on. & Bubbles increase risk of \dword{hv} discharge. \\ \colhline
 No point in the camera system shall have a field greater than \SI{15}{kV/cm} when the drift field is at nominal voltage. & Fields must be well below \SI{30}{kV/cm} to avoid risk of \dword{hv} discharge.\\ \colhline
The camera system shall not produce measurable noise in any detector system. & Low noise is required for TPC operation. \\ \colhline
 Cameras provide the viewing functionality as agreed upon with the other subsystems for viewing, as documented in the ICDs with the individual systems. \\ \toprowrule
\multicolumn{2}{l}{\bf Cold cameras}\\ \specialrule{1.5pt}{1pt}{1pt}
Minimize heat dissipation when camera not in operation. & Do not generate bubbles when \dword{hv} is on. \\ \colhline
Longevity must exceed \num{18} months. & Cameras must function throughout cryostat filling and detector commissioning. \\ \colhline
Frame rate \(\geq\SI{10}{\per s}\). & Observe bubbling, waves, detritus, etc. \\ \toprowrule
\multicolumn{2}{l}{\bf Inspection cameras}\\ \specialrule{1.5pt}{1pt}{1pt}
Keep heat transfer to \lar low when in operation. & Do not generate bubbles, some use cases may require operation when \dword{hv} is on. \\ \colhline
Deploy without exposing \lar to air. & Keep \lar free of N2 and other electronegative contaminants. \\ \colhline
Camera enclosure must be replaceable. & Replace broken camera, or upgrade, throughout life of experiment. \\ \colhline
{\bf Light emitting system} \\ \colhline
No emission of wavelengths shorter than \(\SI{400}{nm}\) & Avoid damaging \dword{tpb} waveshifter. \\ \colhline
Longevity must exceed \num{18} months. & Lighting for fixed cameras must function throughout cryostat filling and detector commissioning. \\ \colhline
\end{dunetable}

The following sections describe the design considerations for the cold
and warm cameras and the associated lighting system.  The same basic
design may be used for both the single and dual phase detectors.

\subsubsection{Cryogenic Cameras (Cold)}

The fixed cameras 
monitor the following items during filling:
\begin{itemize}
\item Positions of corners of \dword{apa} or \dword{crp}, \dword{cpa} or cathode, \dwords{fc}, \dwords{gp} (\SI{1}{mm} resolution);
\item Relative straightness and alignment of \dword{apa}/\dword{crp}, \dword{cpa}/cathode, and \dword{fc} (\(<\sim\SI{1}{mm}\));
\item Relative position of profiles and endcaps (\SI{0.5}{mm} resolution);
\item State of \lar surface: e.g., the presence of bubbling or debris.
\end{itemize}

There are published articles and unpublished presentations describing
completely or partially successful operation of low-cost,
off-the-shelf \dword{cmos} cameras in custom enclosures immersed in cryogens.
(e.g., EXO-100: \cite{Delaquis:2013hva}; DUNE \dword{35t} test
\cite{McConkey:2016spe}; \dword{wa105}: \cite{Murphy:20170516}.)  Generally
it is reported that such cameras show poor performance and ultimately
fail to function below some temperature of order \SIrange{150}{200}{K}, but some report that their cameras recover fully after
being stored (not operated) at temperatures as low as \SI{77}{K} and
then brought up to minimum operating temperature.

However, as with photon sensors, experience has also shown that it is
non-trivial to ensure reliable and reproducible mechanical and
electrical integrity of such cameras in the cryogenic environment 
(e.g., \cite{McConkey:2016spe} and
\cite{Valencia-Rodriquez:20180130}).  Off-the-shelf cameras and camera
components are generally only specified by the vendors and original
manufacturers for operation down to \SI{-40}{\celsius} or \SI{-50}{\celsius}.
In addition, many low-cost cameras use digital interfaces not intended
for long distance deployment, such as USB (\(2\sim\SI{5}{m}\)) or CSI (circuit
board scale), leading to signal degradation and noise problems.

The design for the DUNE fixed cameras uses an enclosure based on
the successful EXO-100 design\cite{Delaquis:2013hva}, which was also
used successfully in
LAPD (see Figure~\ref{fig:gen-fdgen-cameras-enclosure}). The enclosure is
 connected to a stainless steel gas line to allow it to be
flushed with argon gas at low enough pressure to prevent
liquification, using the same design as the gas line for the beam plug
tested in the \dword{35t} \dword{hv} test and in \dword{protodune}.  A thermocouple in the
enclosure allows temperature monitoring, and a heating element
provides temperature control.  The camera transmits its video
signal using either a composite video signal over shielded coax or
Ethernet over optical fiber.  Most importantly, the DUNE \dword{cisc}
consortium must work with vendors to design camera circuit boards that
are robust and reliable in the cryogenic environment.

\begin{dunefigure}[A camera enclosure]{fig:gen-fdgen-cameras-enclosure}
  {CAD exploded view of vacuum-tight camera enclosure suited for cryogenic applications from \cite{Delaquis:2013hva}.
    (1) quartz window, (2 and 7) copper gasket, (3 and 6) flanges, (4) indium wires, (5) body piece, (8) signal \fdth.
  }
  \includegraphics[width=0.6\textwidth]{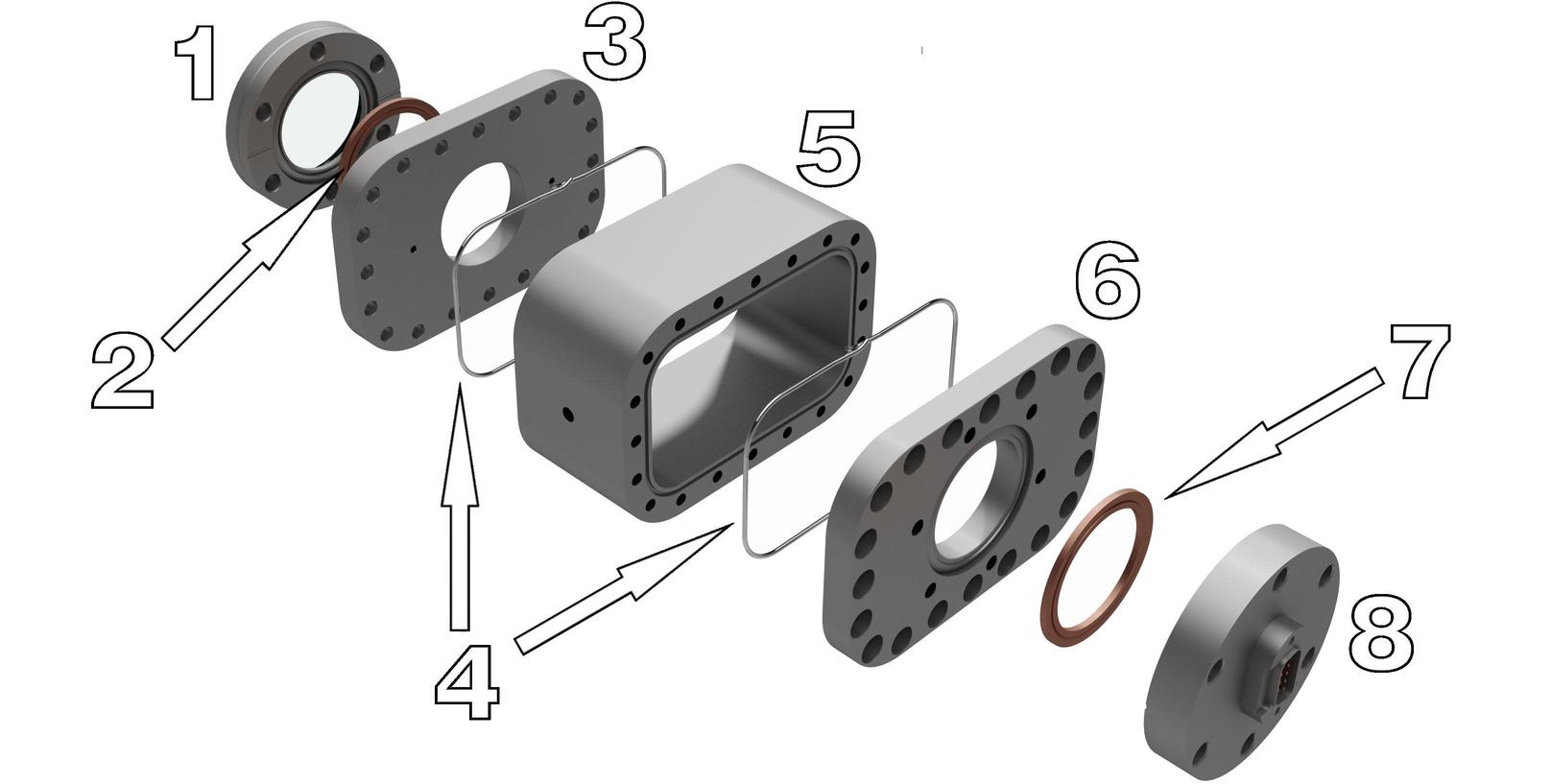}%
\end{dunefigure}

\subsubsection{Inspection Cameras (Warm)}

The inspection cameras are intended to be as versatile as possible.
However, the following locations have been identified as likely
to be of interest:
\begin{itemize}
\item Status of \dword{hv} \fdth and cup;
\item Status of \dword{fc} profiles, endcaps (\SI{0.5}{mm} resolution);
\item $y$-axis deployment of calibration sources;
\item Status of thermometers, especially dynamic thermometers;
\item \dword{hv} discharge, corona, or streamers on \dword{hv} \fdth, cup, or \dword{fc};
\item Relative straightness and alignment of \dword{apa}/\dword{crp}, \dword{cpa}/cathode, and \dword{fc} (\SI{1}{mm} resolution);
\item Gaps between \dword{cpa} frames (\SI{1}{mm} resolution);
\item Relative position of profiles and endcaps (\SI{0.5}{mm} resolution);
\item Sense wires at top of outer wire planes in \single \dword{apa} (\SI{0.5}{mm} resolution).
\end{itemize}

Unlike the fixed cameras, the inspection cameras need operate only as
long as inspection lasts, as the camera can be replaced in case of failure.  It
is also more practical to keep the cameras continuously \textit{warm}
(above \SI{-150}{\celsius}) during deployment; this offers 
more options for commercial cameras, e.g., 
the same model camera used successfully to observe discharges
in \lar from \SI{120}{cm} away \cite{Auger:2015xlo}.

\begin{dunefigure}[Inspection camera design]{fig:gen-fdgen-cameras-movable}
  {An overview of the inspection camera design.}
  \includegraphics[width=0.3\textwidth]{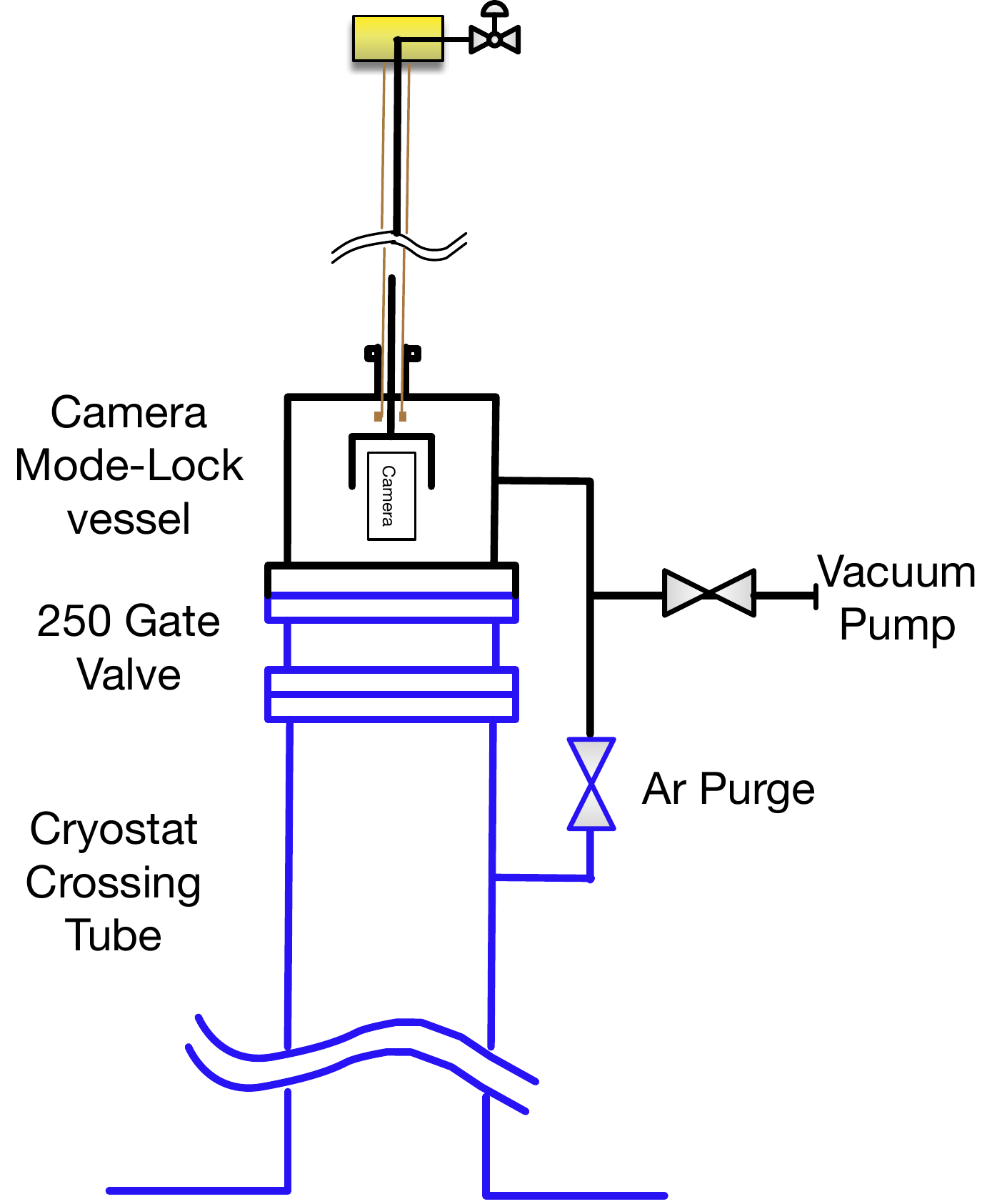}%
\end{dunefigure}

The design for the inspection camera system employs the same basic
enclosure design as for cold cameras, but mounted on an insertable
fork using a design similar to the dynamic temperature probes. See
Figure~\ref{fig:gen-fdgen-cameras-movable} and
Figure~\ref{fig:fd-slow-cryo-sensor-mount}.  The entire system is sealed to
avoid contamination with air. In order to avoid contamination, the
camera can only be deployed through a \fdth equipped with a gate
valve and a purging system, similar to that used for the vertical axis
calibration system at \kamland~\cite{Banks:2014hra}. The entire system
is  purged with pure argon gas before the gate valve is opened.

Motors above the flange allow rotation and vertical movement of the fork. 
 A chain drive system, with motor
mounted on the end of the fork, allows tilting of the camera assembly, 
creating a point-tilt mount with vertical motion capability.
Taking into account the room above the cryostat flanges and the
thickness of the cryostat insulation, a vertical range of motion of
\SI{1}{m} inside the cryostat is achievable.
The motors for rotation and vertical motion are located outside the sealed
volume, coupled mechanically using ferrofluidic seals, thus reducing
contamination risks and allowing for manual rotation of the vertical
drive in the event of a motor failure.  A significant protyping and
testing effort is needed to finalize and validate this design.

\subsubsection{Light-emitting system}
The light-emitting system is based on \dwords{led} with the capability of illuminating the interior with selected
wavelengths (IR and visible) that are suitable for detection by the
cameras.  Performance criteria for the light-emission system are based
on the efficiency of detection with the cameras, in conjunction with
adding minimal heat to the cryostat. The use of very high-efficiency
\dwords{led}  
helps reduce heat generation; as an
example, one \SI{750}{nm} \dword{led} has a specification of
\SI{32}{\%} conversion of electrical input power to light.

While data on the performance of \dwords{led} at cryogenic temperatures is sparse,
some studies related to NASA projects~\cite{Carron:2017zzz} 
indicate that \dword{led} efficiency increases with reduced temperature,
and that the emitted wavelengths may change, particularly for blue \dwords{led}.
The wavelength changes cited would have no impact on illumination, however, since
in order  to avoid degradation of wavelength-shifting materials in the \dword{pds},
such short wavelength \dwords{led} would not be used.

\begin{dunefigure}[\dword{led} chain for illumination]{fig:gen-cisc-LED}
  {Suggested \dword{led} chain for lighting inside the cryostat, with
    dual-wavelength and failure-tolerant operation.}
\includegraphics[width=0.7\textwidth]{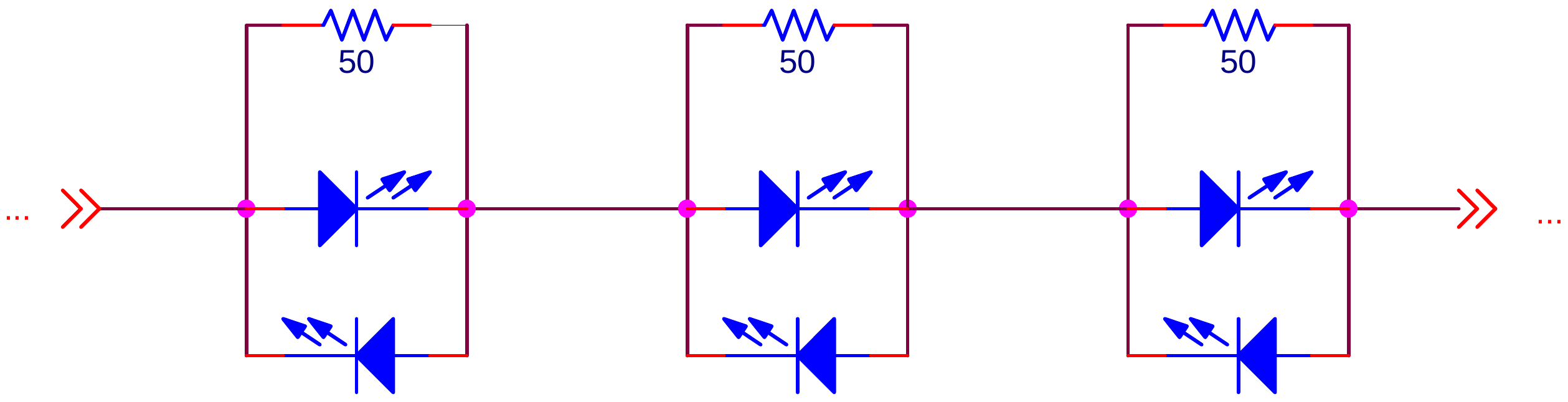}
\end{dunefigure}

\fixme{shoulds and woulds...}
A \textit{chain} of \dwords{led} should be connected in series and driven with a
constant-current circuit. It would be advantageous to pair each
\dword{led} in parallel with an opposite polarity \dword{led} and a resistor
(see Figure~\ref{fig:gen-cisc-LED}).
This allows two different wavelengths of illumination with a single installed
chain (by changing the direction of the drive current) and 
continued use of an \dword{led} chain even if individual \dwords{led} fail.

The \dwords{led} should be placed as a \textit{ring light} around the outside of each
camera lens, pointing in the same direction as the lens, to 
illuminate the part of the \dword{detmodule} within the field of
view of the camera. Commercially available \dwords{led} can be obtained with
a range of angular spreads, and can be matched to the needs of the
cameras without additional optics.


\subsection{Cryogenics Test Facility}
\label{sec:fdgen-slow-cryo-test-facil}
The cryogenics test facility is intended to provide the access to a small ($<$ \num{1} ton) to intermediate ($\sim$ \num{4} tons) volumes of purified TPC-grade \lar{}. Hardware that needs liquid of purity this high include any device intending to drift electrons for millisecond time periods. Not all devices require purity this high, but some may need a relatively large volume to provide the needed prototyping environment. Of importance is a relatively fast turn-around time of approximately a week for short prototyping runs.

Figure~\ref{fig:CryTest-Blanche} shows the Blanche test stand cryostat at \fnal.

\begin{dunefigure}[CryTest Blanche Test]{fig:CryTest-Blanche} 
  {Blanche Cryostat at \fnal. This cryostat holds $\sim 0.75$ tons of \lar{}.}
  \includegraphics[width=0.35\textwidth]{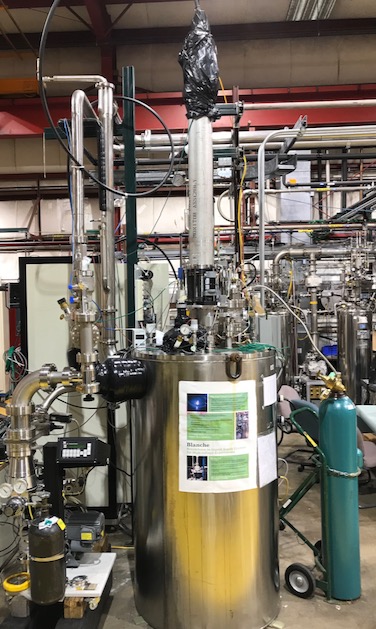}%
\end{dunefigure}


\subsection{Cryogenic Internal Piping}
\label{sec:fdgen-slow-cryo-int-piping}
\label{sec:fdsp-slow-cryo-int-piping}
\label{sec:fddp-slow-cryo-int-piping}


The cryogenic internal piping comprises several manifolds to
distribute the liquid and gaseous argon inside the cryostat during all
phases (e.g., gaseous purge, liquid distribution, cool down) and
various pipe stands to return argon to the outside (e.g.,
boil-off gaseous argon).  Vacuum-insulated pipe stands are needed to
transition from inside to outside in a way that does not affect the
purity and does not introduce a significant heat load.

LBNF has the expertise for engineering design and installation of the
detector internal piping, while the \dword{cisc} consortium has the expertise
on the physics requirements, the relevant risk registries, and the
interfaces with other detector systems. Ultimate responsibility for
costing the internal cryogenic piping system also lies with the \dword{cisc}
consortium. It is important for these two groups to interact closely
to ensure that the system enables achievement of the physics, 
avoids interference with other detector systems, and mitigates
risks.

DUNE has formed a cryogenics systems working group with conveners from
both the \dword{cisc} consortium and LBNF. This group has both LBNF and
\dword{cisc} members and provides an official forum where we interface and
establish the final design.

The initial design for the cryogenic internal piping calls for some
\SI{750}{m} of pipe per cryostat for purging and filling, laid out as
shown in Figure~\ref{fig:fd-slow-cryo-int-piping}-Left, and 20 flange-pipes assemblies, as the one shown
on the right pannel of Figure~\ref{fig:fd-slow-cryo-int-piping}, with a CF DN250 flange penetrated by two $\sim$ \SI{2.2}{m} long pipes.

\begin{dunefigure}[Cryogenic internal piping]{fig:fd-slow-cryo-int-piping}
  {Left: Cryogenic internal piping for purging (red) and filling (blue). Right: Cool-down pipes, \lar in blue (vacuum jacketed) and gaseous argon in red. }
  \includegraphics[height=0.3\textwidth]{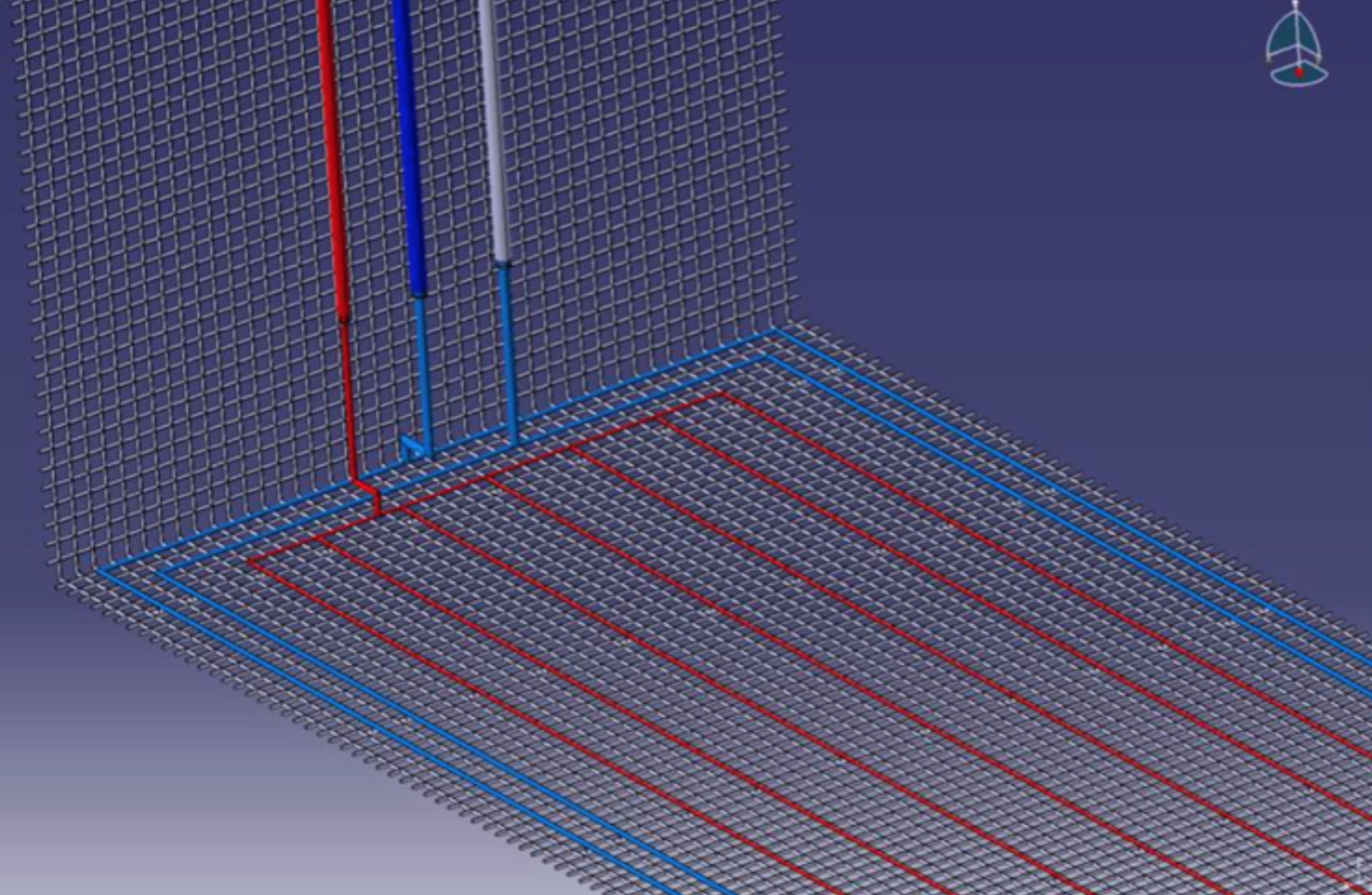}
  \includegraphics[height=0.3\textwidth]{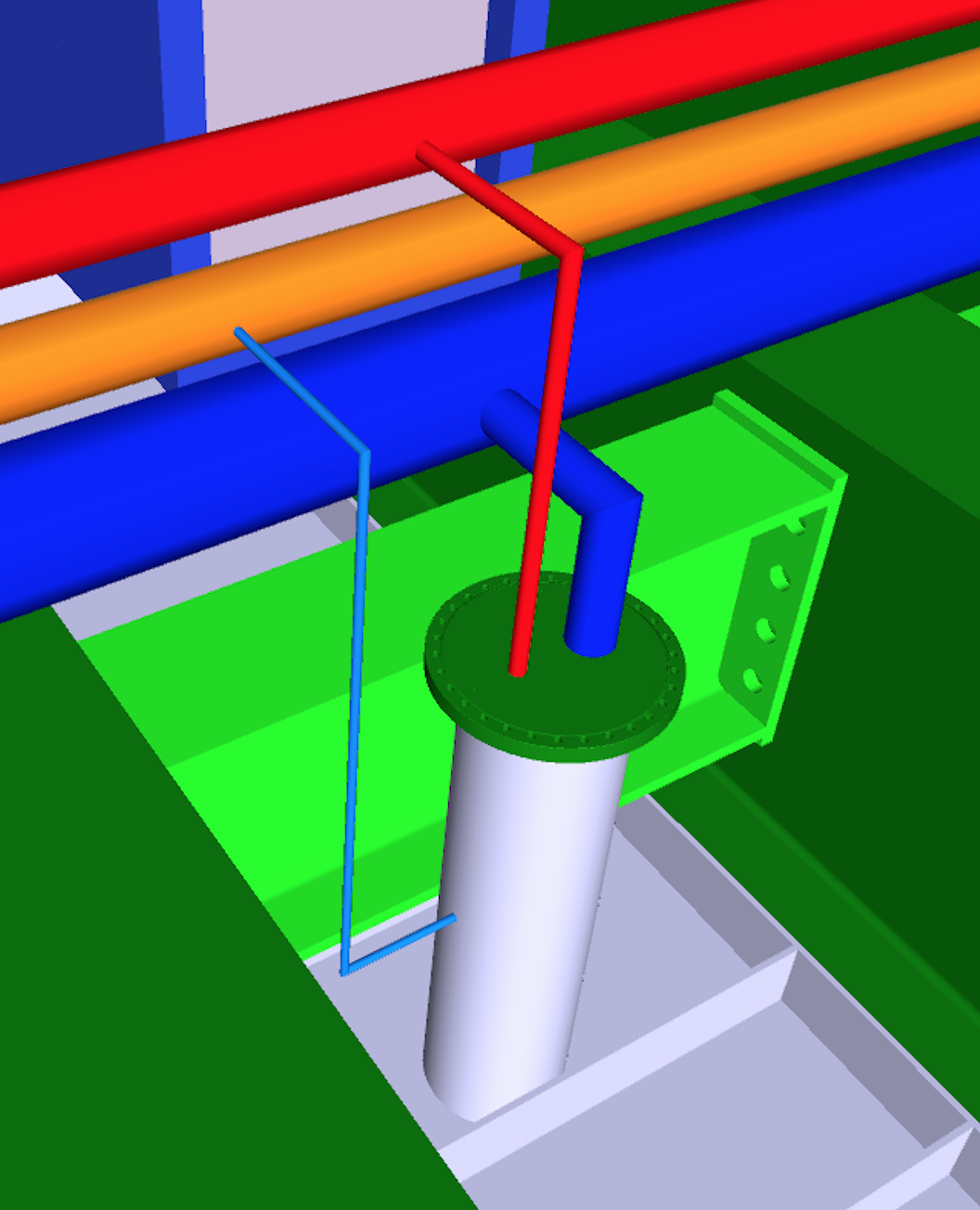}
\end{dunefigure}

\section{Slow Controls}
\label{sec:fdgen-slow-cryo-ctrl}


The slow controls system collects, archives, and displays data from
a broad variety of sources, and provides real time alarms and
warnings for detector operators. Data is acquired via network
interfaces.  Figure \ref{fig:gen-slow-controls-diagram} shows the
connections between major parts of the slow controls system and other
systems.  

\begin{dunefigure}[Slow Controls connections and data]{fig:gen-slow-controls-diagram}
{Typical Slow Controls system connections and data flow}
\includegraphics[width=0.7\textwidth]{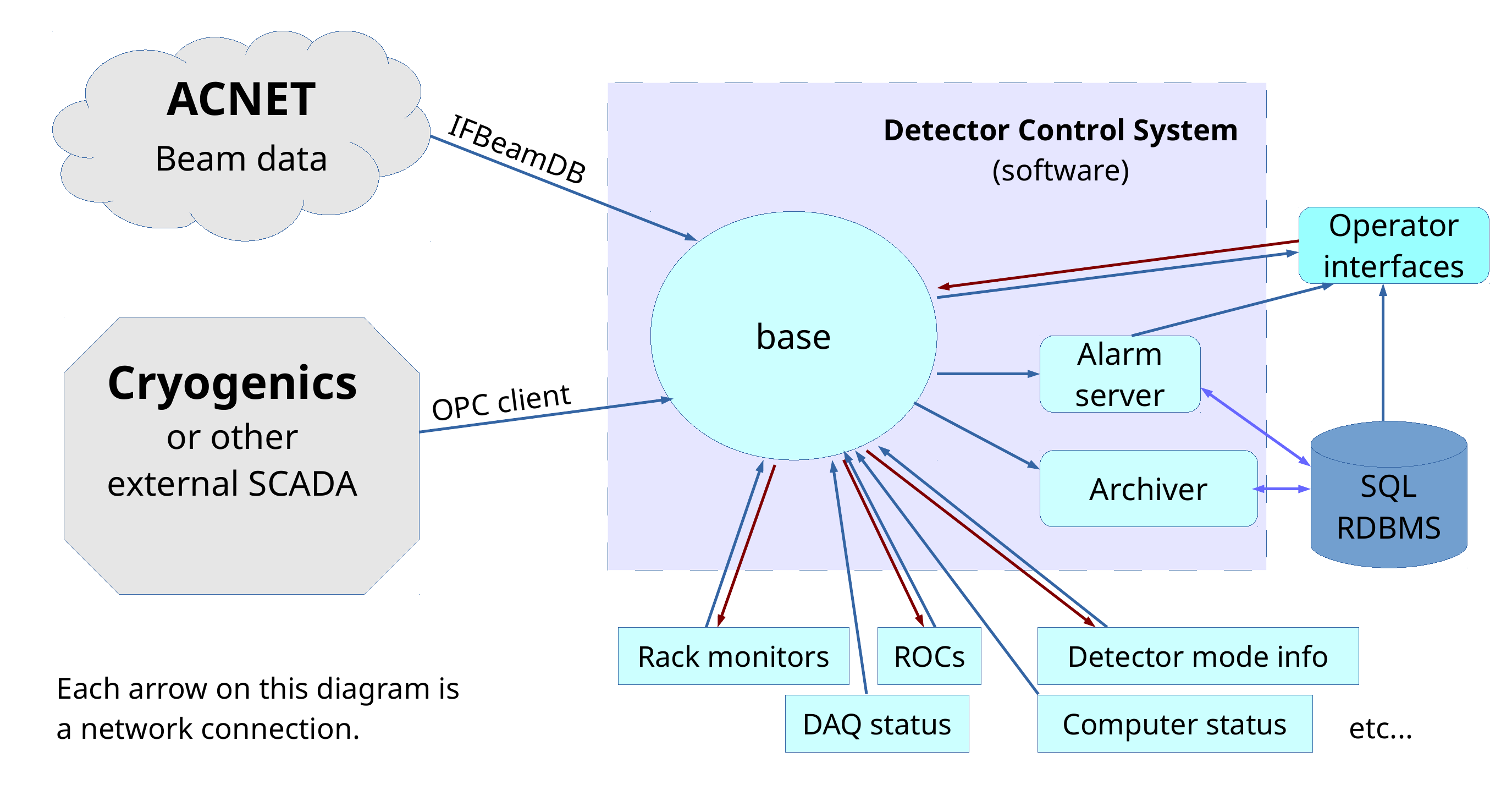}
\end{dunefigure}

\subsection{Slow Controls Hardware}
\label{sec:fdgen-slow-cryo-hdwr}

The slow controls will always require a small amount of dedicated network and
computing hardware as described below.  It also relies on common
infrastructure, as described in
Section~\ref{sec:fdgen-slow-cryo-slow-infra}.

\subsubsection{Slow Controls Network Hardware}
\label{sec:fdgen-slow-cryo-slow-network}
The slow controls data originates from the cryogenic instrumentation discussed in
Section~\ref{sec:fdgen-cryo-instr} and from other systems,
and is collected by software running on servers
(Section~\ref{sec:fdgen-slow-cryo-slow-compute})
housed in the underground data room in the \dlong{cuc},
where data is archived in a central \dword{cisc} database.
The instrumentation data is transported over
conventional network hardware from any sensors located in the cryogenic
plant.  However, the readouts that are located in the racks on top the
cryostats must take care withgrounding and noise.  Therefore, each
rack on the cryostat has a small network switch that sends
any network traffic from that rack to the \dshort{cuc} via a fiber transponder.
This is the only network hardware specific to slow controls;
network infrastructure requirements are described in
Section~\ref{sec:fdgen-slow-cryo-slow-infra}.

\subsubsection{Slow Controls Computing Hardware}
\label{sec:fdgen-slow-cryo-slow-compute}

Two servers (a primary server and a replicated backup) suitable for the needed relational database discussed
in Section~\ref{sec:fdgen-slow-cryo-sw} are located in the \dshort{cuc} data
room, with an additional
two servers to perform \dword{fe} monitoring interface services: for
example, assembling dynamic \dword{cisc} monitoring web pages from the adjacent
databases.  Any special purpose software, such as iFix or EPICS, would
also run here. It is expected that one or two more servers will accommodate
these programs.
Replicating this setup on a per-module basis would allow for easier
commissioning and independent operation, accommodate different module
design (and the resulting differences in database tables), and ensure
sufficient capacity.  Including four sets of networking hardware, this
would fit tightly into one rack or very comfortably into two.


\subsection{Slow Controls Infrastructure}
\label{sec:fdgen-slow-cryo-slow-infra}

The total number of slow controls quantities and the update rate are low enough
that the data rate will be in the tens of kilobytes per second range
(Section~\ref{sec:fdgen-slow-cryo-quant}), placing minimal requirements
on the local network infrastructure.
Network traffic out of \surf to \fnal will be primarily database calls
to the central \dword{cisc} database: either from monitoring applications, or from
database replication to the offline version of the \dword{cisc} database.  This
traffic is of a low enough bandwidth that the proposed general purpose
links both out of the mine and back to \fnal can accommodate it.

Up to two racks of space and appropriate power and cooling are
available in the \dshort{cuc}'s \dword{daq} server room for \dword{cisc} usage.
Somewhat less space than that is currently envisioned, as described in
\ref{sec:fdgen-slow-cryo-slow-compute}.

\subsection{Slow Controls Software}
\label{sec:fdgen-slow-cryo-sw}


The slow controls software includes the following components in order 
to provide complete monitoring and control of detector subsystems:
\begin{itemize}
 \item{Control systems base} that performs input and output operations
  and defines processing logic, scan conditions, alarm conditions,
  archiving rate, etc.;
 \item{Alarm server} that monitors all channels and sends alarm
  messages to operators; 
 \item{Data archiver} that performs automatic sampling and storage of
  values for history tracking;
 \item{Integrated operator interface} that provides display panels for
  controls and monitoring.
\end{itemize}

An additional requirement for the software is the ability to indirectly
interface with external systems (e.g., cryogenics control
system) and databases (e.g., beam database) to export data into
slow controls process variables (or channels) for archiving and status
displays. This allows integrating displays and warnings into one
system for the experiment operators, and 
provides integrated
archiving for sampled data in the archived database. In this case, one
can imagine an input output controller (IOC) running on a central \dword{daq}
server provides soft channels for these data.
Figure~\ref{fig:gen-slow-controls-diagram} shows a typical workflow of a
slow controls system.

In terms of key features of the software, a highly evolved software is
needed that is designed for managing real-time data exchange, scalable
to large number of channels and high bandwidth if needed. The software
should be well documented, supported, and known to be reliable. The base
software should also allow easy access of any channel by name. The
archiver software should allow data storage in an SQL database with
adjustable rates and thresholds such that one can easily retrieve data
for any channel by using channel name and time range. Among the key
features, the alarm server software should remember state, support
arbitrary number of clients, and provide logic for delayed alarms and
acknowledging alarms. As part of the software, a standard naming
convention for channels is followed to aid dealing with large
number of channels and subsystems.

\subsection{Slow Controls Quantities}
\label{sec:fdgen-slow-cryo-quant}


The final set of quantities to monitor will ultimately be determined
by the needs of the subsystems being monitored, as documented in
appropriate  interface control documents (ICDs), and continually revised based on operational
experience.  The total number of quantities to monitor has been very
roughly estimated by taking the total number of quantities monitored
in \microboone and scaling by the detector length and the number of
planes, giving a number in the range of \numrange{50}{100} thousand.
Quantities are expected to update on average no faster than once per minute.
The subsystems
to be monitored include the 
cryogenic instrumentation
described in this chapter, the other detector systems, and relevant
infrastructure and external devices. Table \ref{tab:gen-slow-quant}
lists the kind of quantities expected from each system.

\begin{dunetable}
[Slow controls quantities]
{p{0.3\textwidth}p{0.6\textwidth}}
{tab:gen-slow-quant}
{Slow controls quantities}
System & Quantities \\ \toprowrule
\multicolumn{2}{l}{\bf Detector Cryogenic Instrumentation } \\ \specialrule{1.5pt}{1pt}{1pt}
Purity monitors & Anode and cathode charge, bias voltage and current, flash lamp status, calculated electron lifetime \\ \colhline
Thermometers & Temperature, position of dynamic thermometers \\ \colhline
Liquid level & Liquid level \\ \colhline
Gas analyzers & Purity level readings \\ \colhline
Cameras & Camera voltage and current draw, temperature, heater current and voltage, lighting current and voltage \\ \colhline
Cryogenic internal piping & \fdth gas purge flow and temperature \\ \toprowrule
\multicolumn{2}{l}{\bf Other Detector Systems } \\ \specialrule{1.5pt}{1pt}{1pt}
\dword{hv} systems & Drift \dword{hv} voltage, current; end-of-field cage current, bias; ground plane currents \\ \colhline
TPC electronics & Voltage and current to electronics \\ \colhline
\dword{pd} & Bias, current, electronics \\ \colhline
\dword{daq} & Warm electronics currents and voltages; run status; \dword{daq} buffer sizes, trigger rates, data rates, GPS status, etc.; computer and disk health status; other health metrics as defined by \dword{daq} group \\ \colhline
\dword{crp} / \dword{apa} & Bias voltages and currents \\ \toprowrule
\multicolumn{2}{l}{\bf Infrastructure and external systems } \\ \specialrule{1.5pt}{1pt}{1pt}
Cryogenics (external) & Status of pumps, flow rates, inlet and return temperature and pressure (via OPC or similar SCADA interface) \\ \colhline
Beam status & Protons on target, rate, target steering, beam pulse timing (via IFBeamDB) \\ \colhline
Near detector & Near detector run status (through common slow controls database) \\ \colhline
Racks power and status & PDU current and voltage, air temperature, fan status if applicable, interlock status (fire, moisture, etc.) \\
\end{dunetable}

\subsection{Local Integration}
\label{sec:fdgen-slow-cryo-slow-loc-integ}


The local integration for the slow controls consists entirely of software
and network interfaces with systems outside of the scope of the \dword{detmodule}.
This includes the following:
\begin{itemize}
\item readings from the LBNF-managed external cryogenics systems, for status of pumps, flow rates, inlet and return temperature and pressure, which are implemented via OPC or a similar SCADA interfaces;
\item beam status, such as protons on target, rate, target steering, beam pulse timing, which are retrieved via IFBeamDB;
\item near detector status, which can be retrieved from a common slow controls database.
\end{itemize}

This integration occurs after both the slow controls and non-detector
systems are in place.  The LBNF-\dword{cisc} interface is managed by the
cryogenics systems working group described in Section~\ref{sec:fdgen-slow-cryo-int-piping}.  IFBeamDB is already well established.
An internal near-detector--\dword{fd} working group may be established
to coordinate detector status exchange between near and far sites interfacing.

%
%
%

%
%


\section{Interfaces}
\label{sec:fdgen-slow-cryo-intfc}


The \dword{cisc} consortium interfaces with all other consortia, task forces (calibration), working groups (physics, software/computing) and technical coordination.
This section provides a brief summary; further details can be found in references~\cite{bib:docdb6745}-\cite{bib:docdb7018}.

There are obvious interfaces with detector consortia since \dword{cisc} provides full rack monitoring (rack fans, thermometers and rack protection system),
interlock status bit monitoring (not the actual interlock mechanism) and monitoring and control for all \pwrsupps. The  \dword{cisc} consortium must maintain close contacts with all other consortia to ensure that specific hardware choices have acceptable slow controls (SC)  solutions.  
Also, installation of instrumentation devices interferes with other devices and must be coordinated with the respective consortia.  
On the software side \dword{cisc} must define, in coordination with other consortia, the quantities to be monitored and controlled by slow controls and the corresponding alarms,
archiving and GUIs. 



A major interface is the one with the cryogenics system. As mentioned in Section~\ref{sec:fdgen-slow-cryo-purity-mon} purity monitors and gas analyzers are essential
to mitigate the liquid argon contamination risk. The appropriate interlock mechanism to prevent the cryonenics system from irreversible contamination
must be designed and implemented. 

Another important interface is the one with the \dword{hv} system \cite{bib:docdb6787} since several aspects related with safety must be taken into account. 
For all instrumentation devices inside the cryostat, electric field simulations are needed to guaranty proper shielding is in place.
Although this is a \dword{cisc} responsibility, input from \dword{hv} is crucial.
During the deployment of inspection cameras, generation of bubbles must be avoided when \dword{hv} is on, as it can lead to discharges.

There are also interfaces with the \dword{pds}~\cite{bib:docdb6730}. Purity monitors and the light-emitting system for cameras both emit light that might damage \dwords{pd}.
Although this should be understood and quantified, \dword{cisc} and the \single \dword{pds} may have to define the necessary hardware interlocks
that avoid turning on any other light source accidentally when \dwords{pd} are on.

The \dword{daq}-\dword{cisc} interface~\cite{bib:docdb6790} is described in Section~\ref{sec:fd-daq-intfc-sc}. 
\dword{cisc} data is stored both locally (in \dword{cisc} database servers in the
\dword{cuc}) and offline (the databases are replicated back to \fnal)
in a relational database indexed by timestamp.
This allows bidirectional communications between \dword{daq} and \dword{cisc} by
reading or inserting data into the database as needed for non-time-critical information.



\dword{cisc} also interfaces with the beam and cryogenics group since at least the status of these systems must be monitored.


Assuming that the scope of software \& computing \dword{swc} group includes scientific computing support to project activities, there are substancial interfaces with that group~\cite{bib:docdb7126}. 
The hardware interfaces resposibility of the \dword{swc} include networking installation and maintenance,
maintenance of SC servers  and any additional computing hardware needed by instrumentation devices.
\dword{cisc} provides the needed monitoring for power distribution units (PDUs). Regarding software interfaces the \dword{swc} group  provides:
(1) SC database maintenance, (2) API for accessing the SC database offline,
(3) UPS packages, local installation and maintenace of software needed by \dword{cisc}, and (4) \dword{swc} creating and maintaining computer accounts on production clusters. 
Additionally  \dword{cisc} provides the required monitoring and control of \dword{swc} quantities including alarms, archiving, and GUIs, where applicable.

The \dword{cisc} consortium may have additional hardware interfaces with the not-as-yet formed calibration consortium \cite{bib:docdb7072}. Indeed, since 
the shared ports are multi-purpose to enable deploying various devices,
both \dword{cisc} and calibration must interact in terms of flange design and sharing space around the ports. Also, \dword{cisc} might use calibration ports to extract cables from \dword{cisc} devices. 
At the software level, \dword{cisc} is responsible for calibration device monitoring (and control to the extent needed) and 
monitors the interlock bit status for laser and radioactive sources.

\dword{cisc} indirectly interfaces to physics through the shared devices. One specific need for physics is to extract 
instrumentation or slow controls data to correlate high-level quantities to low-level or calibration data.
This requires tools to extract data from the slow controls database (see \dword{cisc}-\dword{swc} interface document \cite{bib:docdb7126}).
A brief list of what \dword{cisc} data is needed by physics is given in the \dword{cisc}-Physics interface document \cite{bib:docdb7099}. 

Interfaces between \dword{cisc} and technical coordination are detailed in the corresponding interface documents for the facility \cite{bib:docdb6991}, installation \cite{bib:docdb7018}
and integration facility~\cite{bib:docdb7045}.

\fixme{Interfaces with technical coordination has to be expanded}








\section{Installation, Integration and Commissioning}
\label{sec:fdsp-slow-cryo-install}    
\label{sec:fddp-slow-cryo-install} 
\label{sec:fdgen-slow-cryo-install} 

\subsection{Cryogenics Internal Piping}
\label{sec:fdgen-slow-cryo-install-pipes}

The installation of internal cryogenic pipes occurs soon after the cryostat is completed or towards the end of the cryostat completion, depending on how the cryostat work proceeds. A concrete installation plan will be developed by the company designing the internal cryogenics. It depends on how they address the thermal contraction of the long horizontal and vertical runs. We are investigating several options, which each have different installation sequences. 
All involve delivery and welding together of prefabricated spool pieces inside the cryostat, and vacuum insulation of the vertical lines. The horizontal lines are bare pipes. 

The cool-down assemblies are installed in dedicated cool-down \fdth{}s at the top, arranged in
two rows of ten each in the long direction of the cryostat. Each one features a \lar line connected to a gaseous argon line via a mixing nozzle and a gaseous argon line with spraying nozzles. The mixing nozzles generate droplets of liquid that are circulated uniformly inside the cryostat by the spraying nozzles. They are prefabricated at the vendor's site and delivered as full pieces, 
then mounted over the \fdth{}s.

The current \threed model of the internal cryogenics is developed and archived at CERN as part of the full cryostat model. CERN is currently responsible for the integration of the detector cavern: cryostat, detector, and proximity cryogenics in the detector cavern, including cryogenics on the mezzanine and main \lar circulation pumps.

The prefabricated spool pieces and the cool down nozzles undergo testing at the vendor  before delivery. The installed pieces are helium leak-checked before commissioning, but no other integrated testing or commissioning is possible after the installation, because the pipes are open to the cryostat volume. The internal cryogenics are commissioned once the cryostat is closed.

\subsection{Purity Monitors}
\label{sec:fdgen-slow-cryo-instal-pm}

The purity monitor system is built in a modular way, such that it can be assembled outside of \dword{detmodule} cryostat.  The assembly of the purity monitors themselves occurs outside of the cryostat and includes everything described in the previous section.  The installation of the purity monitor system can then be carried out with the least number of steps inside the cryostat.  The assembly itself is transported into the cryostat with the three individual purity monitors mounted to the support tubes but before installation of \dword{hv} cables and optical fibers. The support tube at the top and bottom of the assembly is then mounted to the brackets inside the cryostat that could be attached to the cables trays or the detector support structure. \fixme{brackets attached to trays or DSS depending on SP vs DP?} In parallel to this work, the \dword{fe} electronics and light source can be installed on the top of the cryostat, along with the installation of the electronics and power supplies into the electronics rack.  

Integration begins by running the \dword{hv} cables and optical fibers to the purity monitors, coming from the top of the cryostat.  The \dword{hv} cables are attached to the \dword{hv} \fdth{}s with enough length to reach each of the respective purity monitors.  The cables are run through the port reserved for the purity monitor system, along cable trays inside the cryostat until they reach the purity monitor system, and are terminated through the support tube down to each of the purity monitors.  Each purity monitor has three \dword{hv} cables that connect it to the \fdth, and then along to the \dword{fe} electronics.  The optical fibers are run through the special optical fiber \fdth, into the cryostat, and guided to the purity monitor system either using the cables trays or guide tubes.  Whichever solution is adopted for running the optical fibers from the \fdth to the purity monitor system, it must protect the fibers from accidental breakage during the remainder of the detector and instrumentation installation process.  The optical fibers are then run inside of the purity monitor support tube and to the respective purity monitors,  terminating at the photocathode of each. 

Integration  continues with the connection of the \dword{hv} cables between the \fdth and the system \dword{fe} electronics, and then the optical fibers to the light source.  The cables connecting the \dword{fe} electronics and the light source to the electronics rack are also run and connected at this point.  This allows for the system to turn on and the software to begin testing the various components and connections.  Once it is confirmed that all connections are successfully made, the integration to the slow controls system is made, first by establishing communications between the two systems and then transferring data between them to ensure successful exchange of important system parameters and measurements.  

The purity monitor system is formally commissioned 
once the cryostat is purged and a gaseous argon atmosphere is present.  At this point the \dword{hv} for the purity monitors is ramped up without the risk of discharge through the air, and the light source is turned on.  Although the drift electron lifetime in the gaseous argon is very large and therefore not measurable with the purity monitors themselves, comparing the signal strength at the cathode and anode gives a good indication of how well the light source is generating drift electrons from the photocathode and whether they drift successfully to the anode. 

\subsection{Thermometers}
\label{sec:fdgen-slow-cryo-instal-th}

Individual temperature sensors on pipes and cryostat membrane are installed prior to any detector component, right after the installation of the pipes.
First, all cable supports are anchored to pipes. Then each cable is routed individually starting from the sensor end (with IDC-4 female connector but no sensor)
towards the corresponding cryostat port. Once a port's cables are routed, 
they are cut to the same length such that they can be properly soldered
to the pins of the SUBD-25 connectors on the flange. 
To avoid damage, the sensors are installed at a later stage, just before unfolding the bottom \dwords{gp}.

Static T-gradient monitors are installed before the outer \dwords{apa}, 
after the installation of the pipes
and before the installation of individual sensors. This proceeds in several steps: (1) installation of the two stainless steel strings to the bottom and top corners of the cryostat,
(2) tension and verticality checks, (3) installation of cable supports in one of the strings, (4) installation of sensor supports in the other string, (5) cable routing starting from
the sensor end towards the corresponding cryostat port, (6) cutting all cables at the same point in that port, and (7) soldering cable wires to the pins of the SUBD-25 connectors on the flange. Then, at a later stage, just before moving corresponding \dword{apa} into its final position, (8) the sensors are plugged into IDC-4 connectors. 

For the \single{}, individual sensors on the top \dword{gp} must be integrated with the \dwords{gp}. For each \dshort{cpa} (with its corresponding four \dshort{gp} modules)
going inside the cryostat, cable and sensor supports are anchored to the \dshort{gp} threaded rods as soon as possible.
Once the \dshort{cpa} is moved into its final position and its top \dword{fc} is ready to be unfolded, sensors on those \dwords{gp} are installed. Once unfolded, cables 
exceeding the \dshort{gp} limits can be routed to the corresponding cryostat port either using neighboring \dwords{gp} or \dshort{dss} I-beams. 
\fixme{prev pgraph needs work}

Dynamic T-gradient monitors are installed after the completion of the detector.
The monitor comes in several segments with sensors and cabling already
in place. Additional slack is provided at segment joints to ease the
installation process. Segments are fed into the flange one at a 
time. The segments being fed into the \dword{detmodule} are held at the top
with a pin that prevents the segment from sliding in all the way. Then the next
segment is connected. The pin is removed, and the
segment is pushed down until the next segment top is held with the
pin at the flange. Then this next segment is installed. The
process  continues until the entire monitor is in its place
inside the cryostat. Use of a crane is foreseen to facilitate the process.
Extra cable slack at the top is provided again in order to ease  the
connection to the D-Sub standard connector flange and to allow  vertical movement of the
entire system. Then, a four-way cross flange with electric \fdth{}s on
one side and a window on the other side. The wires are connected to
the D-sub connector on the electric flange \fdth on the side. On the
top of the cross, a moving mechanism is then installed with a crane.
The pinion is connected to the top segment. The moving mechanism will
come reassembled with motor on the side in place and pinion and gear
motion mechanism in place as well. The moving mechanism enclosure  is then connected to top part of the cross and this completes the
installation process of the dynamic T-gradient monitor.
\fixme{prev pgraph needs work}

Commissioning of all thermometers proceeds in several steps. Since in the first stage only cables are installed,
the readout performance and the noise level inside the cryostat are
tested with precision resistors. Once sensors are installed the entire chain is checked again at room temperature.
The final commissioning phase occurs during and after cryostat filling.

\subsection{Gas Analyzers}
\label{sec:fdgen-slow-cryo-install-ga}
 
Prior to the piston purge and gas recirculation phases of the cryostat commissioning, the gas analyzers are installed near the 
tubing switchyard. This minimizes tubing runs and is  convenient for switching the sampling points and gas analyzers. Since each is a standalone module, a single rack with shelves, should be adequate to house the modules.

Concerning the integration, the gas analyzers typically have an analog output (\numrange{4}{20} \si{mA} or \numrange{0}{10}\si{V}) that maps to the input range of the analyzers. They also usually have a number of relays that indicate the scale they are currently running. These outputs can be connected to the slow controls for readout. However, using a digital readout is preferred since this directly gives the analyzer reading at any scale. Currently there are a number of digital output connections, ranging from RS-232, RS-485, USB, and Ethernet. At the time of purchase, one can choose the preferred option, since the protocol is likely to evolve. The readout usually responds to a simple set of text commands. Due to the natural time scales of the gas analyzers, and lags in the gas delivery times (depending on the length of the tubing runs), sampling on timescales of a minute most likely is adequate. \fixme{``sampling on timescales of a minute...'' otherwise `minute' may strike reader as pronounced ``minoot''}

Before the beginning of the gas phase of the cryostat commissioning, the analyzers must be brought online and calibrated. Calibration varies for the different modules, but often requires using argon gas with both zero contaminants (usually removed with a local inline filter) for the zero of the analyzer, and argon with a known level of the contaminant to check the scale. Since the start of the gas phase  begins with normal air, the more sensitive analyzers are valved off at the switchyard to prevent overloading their inputs and potentially saturating their detectors. As the argon purge and gas recirculation progress, the various analyzers are valved back in when the contaminant levels reach the upper limits of the analyzer ranges. 

\subsection{Liquid Level Monitoring}
\label{sec:fdgen-slow-cryo-install-llm}

%
%

Multiple differential pressure level monitors are installed in the
cryostat, connected both to the side
penetration of the cryostat at the bottom and to dedicated
instrumentation ports at the top.

The capacitance level sensors are installed at the top of the
cryostat in coordination with the \dword{tpc} installation.  Their
placement relative to the upper ground plane (single phase) or
\dshort{crp} (dual phase) is important as these sensors will be used for a
hardware interlock on the \dword{hv}, and, in the case of the \dword{dpmod}, to measure the \lar level at the millimeter level as required
for \dual operation.
Post installation in situ testing of the capacitive level sensors can be
accomplished with a small dewar of liquid.

\subsection{Cameras and Light-Emitting System}
\label{sec:fdgen-slow-cryo-install-c}

Fixed camera installation is in principle simple, but involves a
considerable number of interfaces. Each camera enclosure has 
threaded holes to allow bolting it to a bracket. A mechanical
interface is required with the cryostat wall, cryogenic internal
piping, or \dword{dss}. Each enclosure is attached
to a gas line for maintaining appropriate underpressure in the fill
gas; this is an interface with cryogenic internal piping. Each camera has a
cable for the video signal (coax or optical), and a multiconductor
cable for power and control, to be run through cable trays to flanges
on assigned instrumentation \fdth{}s.

The inspection camera is designed to be inserted and removed on any
instrumentation \fdth equipped with a gate valve at any time
during operation.  Installation of the gate valves and purge system
for instrumentation \fdth{}s falls under cryogenic internal
piping.

Installation of fixed lighting sources separate from the cameras would
require similar interfaces as fixed cameras.  However, the current
design has lights integrated with the cameras, which do not require separate
installation.

\subsection{Slow Controls Hardware}
\label{sec:fdgen-slow-cryo-install-sc-hard}

Slow controls hardware installation includes multiple
servers, network cables, any specialized cables needed
for device communication, and possibly some custom-built rack
monitoring hardware. The installation sequence is interfaced and
planned with the facilities group and other consortia. The network
cables and rack monitoring hardware are common across many racks
and are installed first as part of the basic rack installation, 
led by the facilities group. The installation of
specialized cables needed for slow controls and servers is done
after the common rack hardware is installed, and will be coordinated
with other consortia and the \dword{daq} group respectively.

\subsection{Transport, Handling and Storage}
\label{sec:fdgen-slow-cryo-install-transport}

Most instrumentation devices are shipped to \surf in pieces and mounted in situ. 
Instrumentation devices are in general small except the support structures for purity monitors and T-gradient monitors,
which will cover the entire height of the cryostat. Since the load on those structures is relatively small
 (\(<\SI{100}{kg}\)) they can be fabricated in parts of less than \SI{3}{m},
which can be easily transported to \surf. These parts are also easy to transport down the shaft and through the tunnels.
All instrumentation devices except the dynamic T-gradient monitors, which are introduced into the cryostat through a dedicated cryostat port, 
\fixme{above something?}
can be
moved into the cryostat without the crane.

Cryogenic internal piping needs special treatment given the number of pipes and their lengths.
Purging and filling pipes will be most likely pre-assembled by the manufacturer as much as possible, using the largest  
size that can be shipped and transported down the shaft. Assuming \SI{6}{m} long sections,
pipes could be grouped in bunches of \numrange{10}{15} pipes and stored in five pallets or boxes of about \SI{6.2}{m} $\times$ \SI{0.8}{m} $\times$ \SI{0.5}{m}. 
These would be delivered to the site, stored, transported down to the detector cavern,
 and stored again before they are used.
Depending on when they are installed, they could be stored inside the cryostat itself or in one of the drifts. 
Cool-down pipes are easier to handle. They could be transported in \num{20} boxes of \SI{2.2}{m} $\times$ \SI{0.6}{m} $\times$ \SI{0.6}{m}, although
there is room for saving some space using a different packaging scheme. 
Once in the cavern they could be stored on top of the cryostat.







\section{Quality Control}
\label{sec:fdgen-slow-cryo-qc}
A series of tests should be done by the manufacturer and the institute in charge of the device assembly. The purpose of  \dword{qc} is to ensure that the equipment is capable of performing its intended function. The \dword{qc} includes post-fabrication tests and also tests to run after shipping and installation. In case of a complex system, the whole system performance will be tested before shipping. 
Additional \dword{qc} procedures can be performed at the \dword{itf} and underground after installation if possible. The planned tests for each subsystem are described below.  
\fixme{the organization of prev pgraph is confusing}

\subsection{Purity Monitors}
\label{sec:fdgen-slow-cryo-qc-pm}

The purity monitor system undergoes a series of tests to ensure the performance of the system.  This  starts with testing the individual purity monitors in vacuum after each one is fabricated and assembled.  This test looks at the amplitude of the signal generated by the drift electrons at the cathode and the anode.  This ensures that the photocathode is able to provide a sufficient number of photoelectrons for the measurement 
with the required precision, and that the field gradient resistors are all working properly to maintain the drift field and hence transport the drift electrons to the anode.  A follow-up test in \lar is then performed for each individual purity monitor, ensuring that the performance expected in \lar is met.  

The next step 
is to assemble the entire system and make checks of the connections along the way.  Ensuring that the connections are all proper during this time reduces the risk of having issues once the system is finally assembled and ready for the final test.  
The assembled system is placed into the shipping tube, which serves as a vacuum chamber, and tested. 
If an adequately sized \lar test facility is available, a full system test can be performed at \lar temperature prior to installation. 
\fixme{and what if no \lar facility exists there? Orig sentence: If there is a \lar test facility with the height or length required for the full purity monitor system and it is available for use, then a final full system test would be made there ensuring that the full system operates in \lar and achieves the required performance.}


\subsection{Thermometers}
\label{sec:fdgen-slow-cryo-qc-th}

\subsubsection{Static T-Gradient Thermometers}
\label{sec:fdgen-slow-cryo-qc-thst}

Three type of tests are carried out at the production site prior to installation. First, the mechanical rigidity of the system is tested such that swinging is minimized (< \SI{5}{cm})
to reduce the risk of touching the \dwords{apa}. This is done with a \SI{15}{m} stainless steel string, strung horizontal anchored to two points; its tension is controlled and measured. 
Second, 
all sensors are calibrated in the lab, as explained in Section~\ref{sec:fdgen-slow-cryo-therm}.
The main concern is the reproducibility of the results since sensors could potentially change their resistance (and hence their temperature scale)
when undergoing successive immersions in \lar{}. In this case the \dword{qc} is given by the calibration procedure itself since five independent measurements
are planned for each set of sensors. Sensors with reproducibility (based on the \rms of those five measurements) beyond the requirements (\SI{2}{mK} for \dword{pdsp}) are discarded.  
The calibration serves as \dword{qc} for the readout system (similar to the final one) and of the PCB-sensor-connector assembly. Finally, the cable-connector assemblies are tested: sensors must measure the expected values with no additional noise introduced by the cable or connector. 

If the available \lar test facility has sufficient height or length to test a good portion of the system, an integrated system test is conducted there ensuring that the system
operates in \lar and achieves the required performance. Ideally, the laboratory sensor calibration will be compared with the in situ calibration
of the dynamic T-gradient monitors by operating both dynamic and static T-gradient monitors simultaneously.   

The last phase of \dword{qc} takes place after installation. 
The verticality of each array is checked and the tensions in the horizontal strings are adjusted as necessary.
Before soldering the wires to the flange, the entire readout chain is tested with temporary SUBD-25 connectors. 
This allows testing the sensor-connector assembly, the cable-connector assembly and the noise level inside the cryostat.
If any of the sensors gives a problem, it is replaced. If the problem persists, the cable is checked and replaced if needed.

\subsubsection{Dynamic T-Gradient Thermometers}
\label{sec:fdgen-slow-cryo-qc-thdy}

The dynamic T-gradient monitor consists of an array of high-precision temperature sensors mounted on a vertical rod. The rod can move vertically in order to perform cross-calibration of the temperature sensors in situ. Several tests are foreseen to ensure that the dynamic T-gradient monitor delivers vertical temperature gradient measurements with precision at the level of a few \si{mK}.

\begin{itemize}
\item
Before installation, temperature sensors are tested in LN to verify correct operation and to set the baseline calibration for each sensor with respect to the absolutely caibrated reference sensor. 
\item
Warm and cold temperature readings are taken with each sensor after mounting on the PCB board and soldering 
the readout cables.
\item
The sensor readout is taken for all sensors after the cold cables are connected to electric \fdth{}s on the flange and the warm cables outside of the cryostat are connected to the temperature readout system.
\item 
The stepper motor is tested before and after connecting to the gear and pinion system.
\item
The fully assembled rod is connected to the pinion and gear, and moved with the stepper motor on a high platform many times to verify repeatability, possible offsets and uncertainty in the positioning. Finally, by repeating the test a large number of times, the sturdiness of the system will be verified.
\item
The full system is tested after installation in the cryostat: both motion and sensor operation are tested by checking 
sensor readout and vertical motion of the system.
\end{itemize} 

\subsubsection{Individual Sensors}
\label{sec:fdgen-slow-cryo-qc-is}

The method to address the quality of individual precision sensors is the same as for the static T-gradient monitors.
The \dword{qc} of the sensors is part of the laboratory calibration. After mounting six sensors with their corresponding cables, a
temporary SUBD-25 connector will be added and the six sensors tested at room temperature. All sensors should work and give values within specifications.  
If any of the sensors gives problems, it is replaced.  If the problem persists the cable is checked and replaced if needed.

\subsection{Gas Analyzers}
\label{sec:fdgen-slow-cryo-qc-ga}

The gas analyzers will be guaranteed by the manufacturer. However, once received, the gas analyzer modules are checked for both \textit{zero} and the \textit{span} values using a gas-mixing instrument. This is done using two gas cylinders with both a zero level of the gas analyzer contaminant species and a cylinder with a known percentage of the contaminant gas. This should verify the proper operation of the gas analyzers. When eventually installed at \surf, this process is repeated before the commissioning of the cryostat. It is also important to repeat the calibrations at the manufacturer-recommended periods over the gas analyzer lifetime.

\subsection{Liquid Level Monitoring}
\label{sec:fdgen-slow-cryo-qc-llm}

The differential pressure level meters will require \dword{qc} by the manufacturer.
While the capacitive 
sensors can be tested with a modest sample of \lar in the lab,
the differential pressure level meters require testing over a greater range.  While they do not
require testing over the whole range,  lab tests  in \lar 
done over a meter or two can ensure operation
at cryogenic temperatures.  Depth tests can be accomplished using a
pressurization chamber with water.


\subsection{Cameras}
\label{sec:fdgen-slow-cryo-qc-c}

Before transport to \surf, each cryogenic camera unit (comprising the enclosure, camera, and internal thermal control and monitoring) is checked for correct operation of all operating features, for recovery from \SI{87}{K} non-operating mode, for no leakage, and for physical defects. Lighting systems are similarly checked for operation. Operations tests will include verification of correct current draw, image quality, and temperature readback and control. The movable inspection camera apparatus is inspected for physical defects, and checked for proper mechanical operation before shipping. A checklist is completed for each unit, filed electronically in the DUNE logbook, and a hard copy sent with each unit. 

Before installation, each fixed cryogenic camera unit is inspected for physical damage or defects and checked in the cryogenics test facility  for correct operation of all operating features, for recovery from \SI{87}{K} non-operating mode, and for no contamination of the \lar{}. Lighting systems are similarly checked for operation. Operations tests include correct current draw, image quality, and temperature readback and control. After installation and connection of wiring, fixed cameras and lighting are again  checked for operation. The movable inspection camera apparatus is inspected for physical defects and, after integration with a camera unit, tested in facility for proper mechanical and electronic operation and cleanliness, before installation or storage. A checklist is completed for each \dword{qc} check and filed electronically in the DUNE logbook. 

\subsection{Light-emitting System}
\label{sec:fdgen-slow-cryo-qc-les}

The complete system is checked before installation to ensure the functionality of the light emission. 
\fixme{to ensure it meets requirements?}
Initial testing of the light-emitting system (see Figure~\ref{fig:gen-cisc-LED}) is done by first
measuring the current when a low voltage (\SI{1}{V}) is applied, to check
that the resistive \dword{led} failover path is correct. Next, measurement
of the forward voltage is done with the nominal forward current applied, to
check that it is within \SI{10}{\%} of the nominal forward voltage drop of
the \dwords{led}, that all of the \dwords{led} are illuminated, and that each of the
\dwords{led} is visible over the nominal angular range. If the \dwords{led} are
infrared, a video camera with IR filter removed is used for the
visual check. This procedure is then duplicated with the current
reversed for the \dwords{led} oriented in the opposite direction.  

These tests are duplicated during
installation to make sure that the system has
not been damaged in transportation or installation. However, once
the \dwords{led} are in the cryostat a visual check could be difficult or impossible.

\subsection{Slow Controls Hardware}
\label{sec:fdgen-slow-cryo-qc-sc-hard}

Networking and computing systems will be purchased commercially, requiring \dword{qa}. However, the new servers are tested after delivery to confirm no damage during shipping. The new system is allowed to \textit{burn in} overnight or for a few days, 
running a diagnostics suite on a loop. This should turn up anything that escaped the manufacturer's \dword{qa} process.

The system can be shipped directly to the underground, 
\fixme{maybe ``once the system arrives on-site, it can be transported down to the 4850L''}
where an on-site
expert peforms the initial booting of systems and basic
configuration. Then the specific configuration information is pulled over
the network, after which others may log in remotely to do the final
setup, minimizing the number of people underground.


\section{Safety}
\label{sec:fdgen-slow-cryo-safety}


Several aspects related to safety must be taken into account for the different phases of the \dword{cisc} project, including R\&D, laboratory calibration and testing, mounting tests and installation. 
The initial safety planning for all phases is reviewed and approved by safety experts as part of the initial design review, and always prior to implementation. 
All component cleaning, assembly, testing  and installation procedure documentation includes a section on safety concerns
relevant to that procedure, and is reviewed during the appropriate pre-production reviews.

Areas of particular importance to \dword{cisc} include:
\begin{itemize}
\item Hazardous chemicals (e.g., epoxy compounds used to attach sensors to cryostat inner membrane) and cleaning compounds:
  All chemicals used are documented at the consortium management level, with an MSDS (Material safety data sheet) and approved handling and disposal plans in place.

\item Liquid and gaseous cryogens used in calibration and testing: LN and \lar are used for calibration and testing of most of the instrumentation devices.
  Full hazard analysis plans will be in place
  \fixme{are being developed?}  at the consortium management level for all module or
  module component testing involving cryogenic hazards, and these safety plans will be reviewed in the appropriate pre-production and production reviews

\item \dword{hv} safety:  Purity monitors operate at $\sim$\SI{2000}{V}. Fabrication and testing plans will demonstrate compliance with local
  \dword{hv} safety requirements at the particular institution or lab where the testing or operation is performed, and this compliance will be reviewed as part of the standard review process.


\item Working at heights: Some aspects of the fabrication, testing and installation of \dword{cisc} devices require working at heights. This is the 
  case of T-gradient monitors and purity monitors, which are quite long. 
  Temperature sensors installed near the top cryostat membrane and cable routing for all instrumentation devices
  require working at heights as well. The appropriate safety procedures including lift and harness training will be designed and reviewed. 
  
\item Falling objects: all work at height comes with associated risks of falling objects. The corresponding safety procedures, including the proper helmet ussage 
  and the observation of  well delimited safety areas, will be included in the safety plan. 
\end{itemize}


\section{Organization and Management}
\label{sec:fdgen-slow-cryo-org}

\subsection{Slow Controls and Cryogenics Instrumentation Consortium Organization}
\label{sec:fdgen-slow-cryo-org-consortium}


The organization of the \dword{cisc} consortium is shown in
Figure~\ref{fig:gen-slow-cryo-org}. The \dword{cisc} consortium board is
currently formed from institutional representatives from \num{17} institutes. 
The consortium leader
acts as the spokesperson for the consortium and is responsible for the
overall scientific program and management of the group. The technical
leader of the consortium is responsible for the project management for
the group.  Currently five working groups are envisioned in the
consortium (leaders to be appointed):



\begin{description}
 \item[Cryogenics Systems] gas analyzers, liquid level
  monitors and cryogenic internal piping; CFD simulations.
 \item[\lar Instrumentation] purity monitors, thermometers,
   cameras and lightemitting system, and instrumentation test facility;
   feedthroughs; \efield simulations;
   instrumentation precision studies;
   \dword{protodune} data analysis coordination efforts.
 \item [Slow Controls Base Software and Databases]  Base software, alarms and archiving databases, and monitoring tools;
   variable naming convention and slow controls quantities.
 \item [Slow Controls Detector System Interfaces] Signal processing software and hardware interfaces (e.g., power supplies);
   firmware; rack hardware and infrastructure.   
 \item [Slow Controls External Interfaces] Interfaces with external detector systems (e.g., cryogenics system, beam, facilities, \dword{daq}).
\end{description}

\begin{dunefigure}[\dword{cisc} consortium organization]{fig:gen-slow-cryo-org}
{\dword{cisc} consortium organizational chart}
\includegraphics[width=0.6\textwidth,trim=20mm 80mm 30mm 70mm,clip]{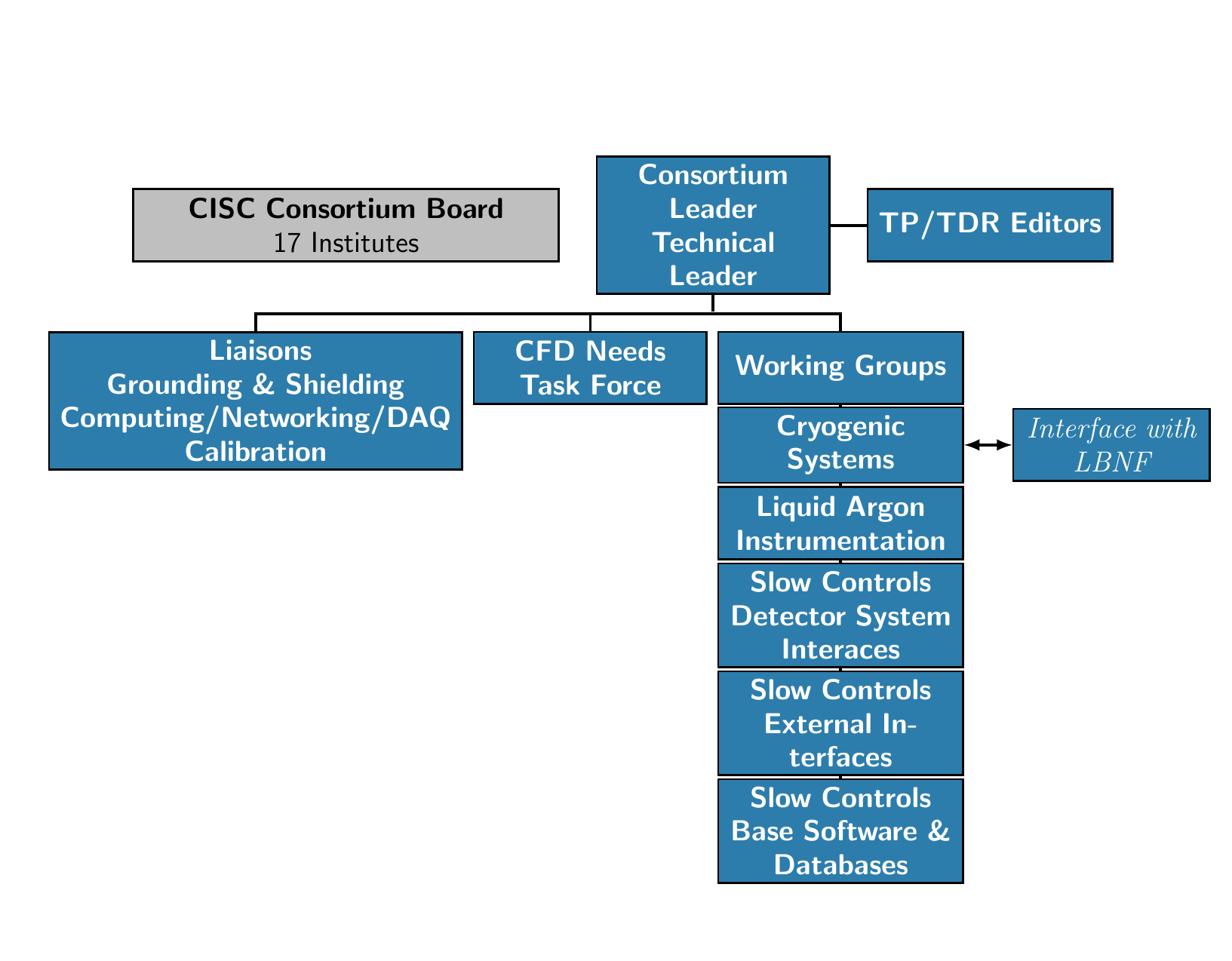}  
\end{dunefigure}

Additionally, since the \dword{cisc} consortium broadly interfaces with other
groups, liaisons have been identified for various roles as listed in
Figure~\ref{fig:gen-slow-cryo-org}. A short-term focus group was
recently formed to understand the needs for cryogenics modeling for the
consortium.  Currently members from new
institutes are added to the consortium based on consensus from the
consortium board members.


\subsection{Planning Assumptions}
\label{sec:fdgen-slow-cryo-org-assmp}

The slow controls and cryogenic instrumentation is a joint effort for \single and \dual{}.
A single slow controls system will be implemented to serve both \single and \dual{}.

Design and installation of cryogenics systems (gas analyzers, liquid level monitoring, internal piping) is coordinated with LBNF, with the consortium providing resources, and effort and expertise provided by LBNF.
\dword{protodune} designs for \lar instrumentation (purity monitors, thermometers, cameras, test facility) provide the basis for DUNE designs. Design validation, testing, calibration, and performance will be evaluated through \dword{protodune} data.


\subsection{High-level Schedule}
\label{sec:fdgen-slow-cryo-org-cs}

Table \ref{tab:fdgen-slow-cryo-schedule} shows key milestones on
the path to  commissioning of the first two DUNE detector modules.

\begin{dunetable}
[Key \dword{cisc} milestones]
{p{0.15\linewidth}p{0.60\linewidth}}
{tab:fdgen-slow-cryo-schedule}
{Key \dword{cisc} milestones leading towards commissioning of the first two DUNE detector modules.}   
Date & Milestone \\ \toprowrule
Aug.\ 2018 &	Validate instrumentation designs using data from ProtoDUNE  \\ \colhline
Jan.\ 2019 &	Complete architectural design for slow controls ready \\ \colhline
Feb.\ 2019 &	Full final designs of all cryogenic instrumentation devices ready \\ \colhline
Feb.\ 2023 &	Installation of Cryogenic Internal Piping for Cryostat 1 \\ \colhline
Apr.\ 2023 &	Installation of support structure for all instrumentation devices for Cryostat 1 \\ \colhline
Oct.\ 2023 &	All Instrumentation devices installed in Cryostat 1 \\ \colhline
Feb.\ 2024 &	All Slow Controls hardware and infrastructure installed for Cryostat 1  \\ \colhline
May 2024 &	Installation of Cryogenic Internal Piping for Cryostat 2 \\ \colhline
July 2024 &	Installation of support structure for all instrumentation devices for Cryostat 2 \\ \colhline
Jan.\ 2025 &	All Instrumentation devices installed in Cryostat 2 \\ \colhline
Apr.\ 2025 &	All Slow Controls hardware and infrastructure installed for Cryostat 2 \\ \colhline
July 2025 &	Full Slow controls systems commissioned and integrated into remote operations \\
\end{dunetable}


\cleardoublepage

\chapter{Technical Coordination}
\label{ch:fdsp-coord}

The \dword{tc} team is responsible for detector integration
and installation support. 
The \dword{dune} collaboration consists of a large number of
institutions distributed throughout the world. They are supported by a
large number of funding sources and collaborate with a large number of
commercial partners. Groups of institutes within \dword{dune} form
consortia that take complete responsibility for construction of their
system.  \dword{dune} has empowered several consortia (currently nine)
with the responsibility to secure funding and design, fabricate,
assemble, install, commission and operate their components of the
\dword{dune} \dword{fd}. There are three consortia focusing
exclusively on the \dword{spmod}: \dword{apa},
\single \dword{ce} and \single \dword{pds}. There are three
focusing exclusively on the \dword{dpmod}: 
\dword{crp}, \dual \dword{ce} and \dual \dword{pds}. There are
three joint consortia: \dword{hv}, \dword{daq} and \dword{cisc}. Other consortia may
be formed over time as concepts more fully emerge, such as a
\dword{fd} calibration system and various aspects of the \dword{nd}.
\dword{dune} \dword{tc}, under the direction of the
Technical Coordinator, has responsibility to monitor the
technical aspects of the detector construction, to integrate and
install the \dwords{detmodule} and to deliver the common projects. The
\dword{dune} \dword{tc} organization is shown in Figure~\ref{fig:TC_orgchart}.

\begin{dunefigure}[Organization of \dword{tc}]{fig:TC_orgchart}
  {Organization of \dword{tc}. This organization
 oversees the construction of the \dword{fd}, both \single and
 \dual, and the \dword{nd}.}
\includegraphics[width=0.85\textwidth,trim=20mm 90mm 30mm 90mm,clip]{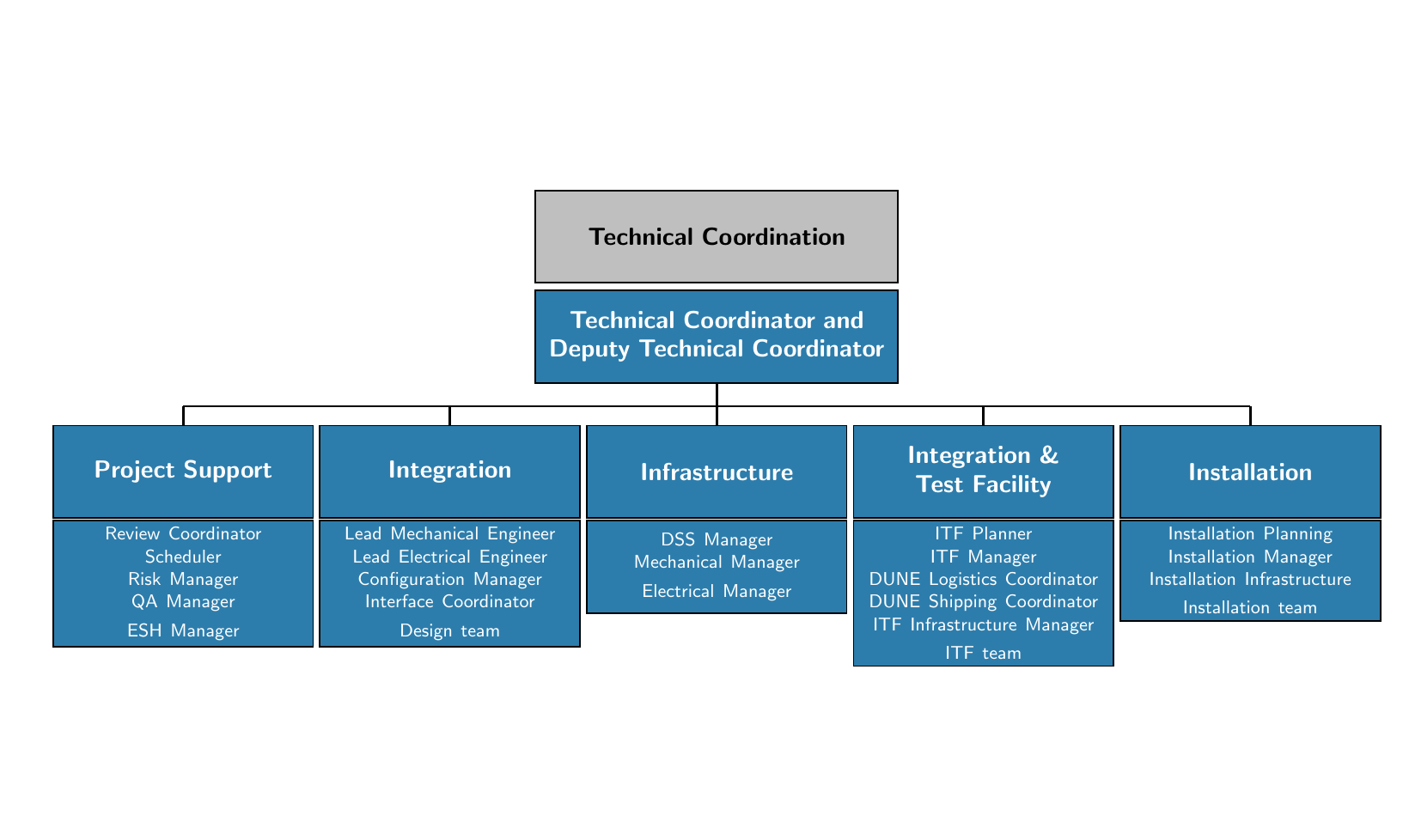}
\end{dunefigure}

The \dword{tc} organization staffing will grow over time as the project
advances. \Dword{tc} will provide staffing for teams underground at \surf, at
integration facilities, and at the near site at \fnal, in addition to
the core team distributed among collaborating institutions.

The \dword{dune} Project consists of a \dword{fd} and a
\dword{nd}. The \dword{nd} is at a pre-conceptual state; as the
conceptual design and organization emerges, it will become part of the
\dword{dune} Project. Currently the \dword{dune} Project consists of
the \dword{dune} \dword{fd} consortia and \dword{tc}.  The
\dword{dune} Project is moving towards a \dword{tdr} for
the \dword{fd}, both \single and \dual options, in 2019. It is
expected that a Conceptual Design Report for the \dword{nd} will be
prepared at the same time. 

The \dword{fd} components will be shipped
from the consortia construction sites to the \dword{itf}. 
\Dword{tc} will
evaluate and accept consortia components either at integration
facilities or the installation site and oversee the integration of
components as appropriate. The scope of the \dword{fd} integration
and installation effort is shown graphically in
Figure~\ref{fig:TC_flow}.

\begin{dunefigure}[Flow of components from the consortia to the \dword{fd}.]{fig:TC_flow}
  {Flow of components from the consortia to the \dword{fd}.}
 \includegraphics[width=\textwidth]{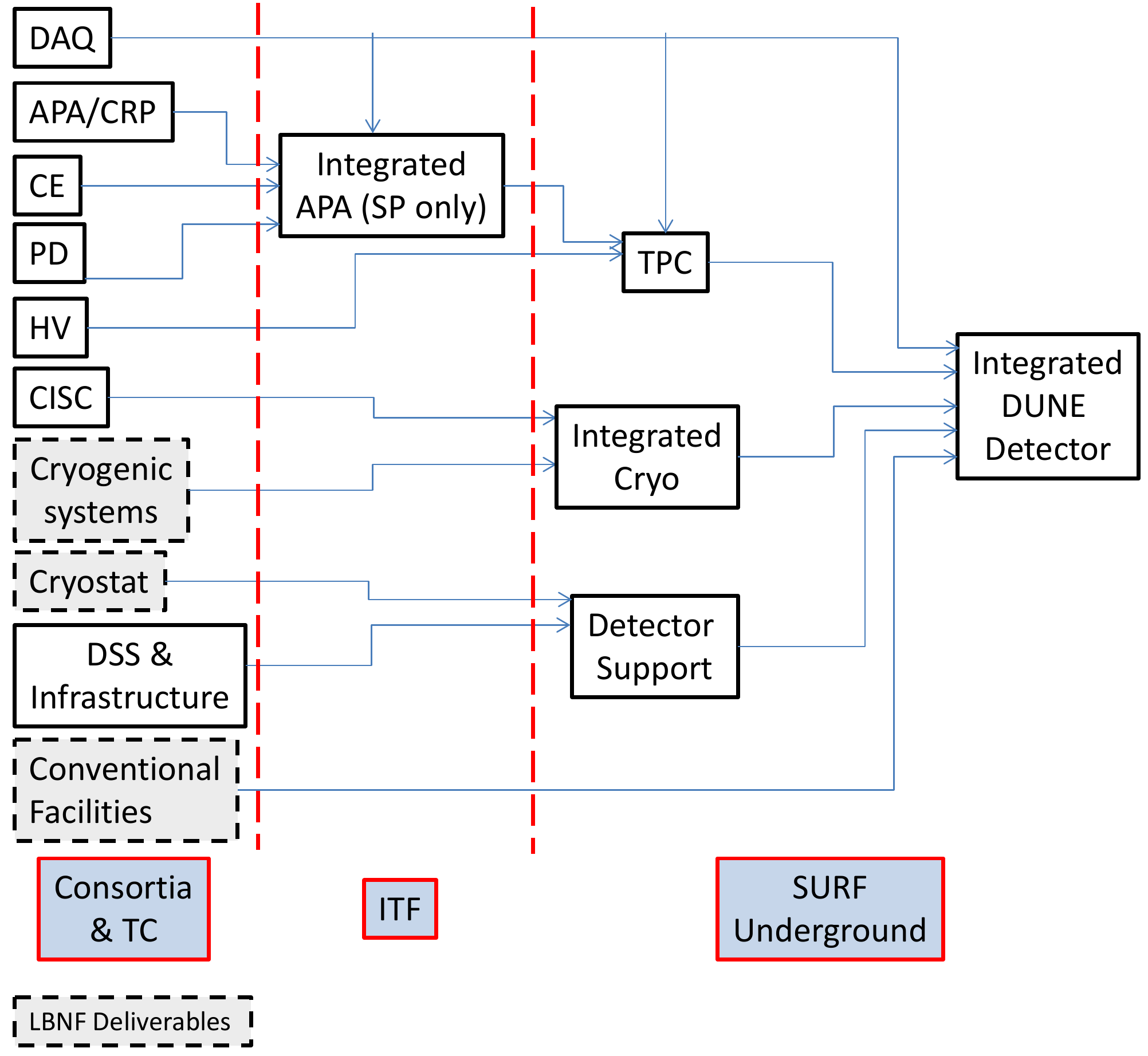}
\end{dunefigure}

\Dword{tc} interacts with the consortia via three main areas: project
coordination, integration, and installation.  Construction of the
\dword{dune} \dword{fd} requires careful technical coordination due to
its complexity.  Given the horizontal nature of the consortia
structure and the extensive interdependencies between the systems, a
significant engineering organization is required to deliver
\dword{dune} on schedule and within specifications and funding
constraints.

The responsibilities of \dword{tc} include:
\begin{itemize}
  \item management and delivery of all common projects;
  \item development and monitoring of the consortia interfaces;
  \item configuration control of all interface documents, drawings and envelopes;
  \item installation of detectors at the near and far sites;
  \item logistics for detector integration and installation at the near and far sites;
  \item survey of the detector;
  \item primary interface to \dword{lbnf} for conventional facilities, cryostat and cryogenics;
  \item primary interface to the host laboratory for infrastructure and operations support;
  \item development and tracking of project schedule and milestones;
  \item review of all aspects of the project;
  \item recording and approving all project engineering information, including: documents, drawings and models;
  \item project work breakdown schedules;
  \item project risk register;
  \item \dword{dune} engineering and safety standards, including grounding and shielding;
  \item monitoring of all consortia design and construction progress;
  \item \dword{qa} and all \dword{qa} related studies and documents;
  \item \dword{esh} organization and all safety related studies and documents.
\end {itemize}

\dword{dune} \dword{tc} interacts with \dword{lbnf} primarily through the
\dword{lbnf}/\dword{dune} systems engineering organization. \dword{tc}
provides the points of contact between the consortia and \dword{lbnf}.
\Dword{tc} will work with the \dword{lbnf}/\dword{dune} Systems Engineer to
implement the \dword{lbnf}/\dword{dune} Configuration Management Plan
to assure that all aspects of the overall \dword{lbnf}/\dword{dune}
project are well integrated. \Dword{tc} will work with \dword{lbnf} and the
host laboratory to ensure that adequate infrastructure and operations support
are provided during construction, integration, installation,
commissioning and operation of the detectors. The \dword{lbnf}/\dword{dune}  systems
engineering organization is shown in Figure~\ref{fig:DUNE_SE_org}.

\begin{dunefigure}[LBNF/DUNE systems engineering organizational structure.]{fig:DUNE_SE_org}
  {LBNF/DUNE systems engineering organizational structure.}
   \includegraphics[width=\textwidth,trim=20mm 105mm 30mm 110mm,clip]{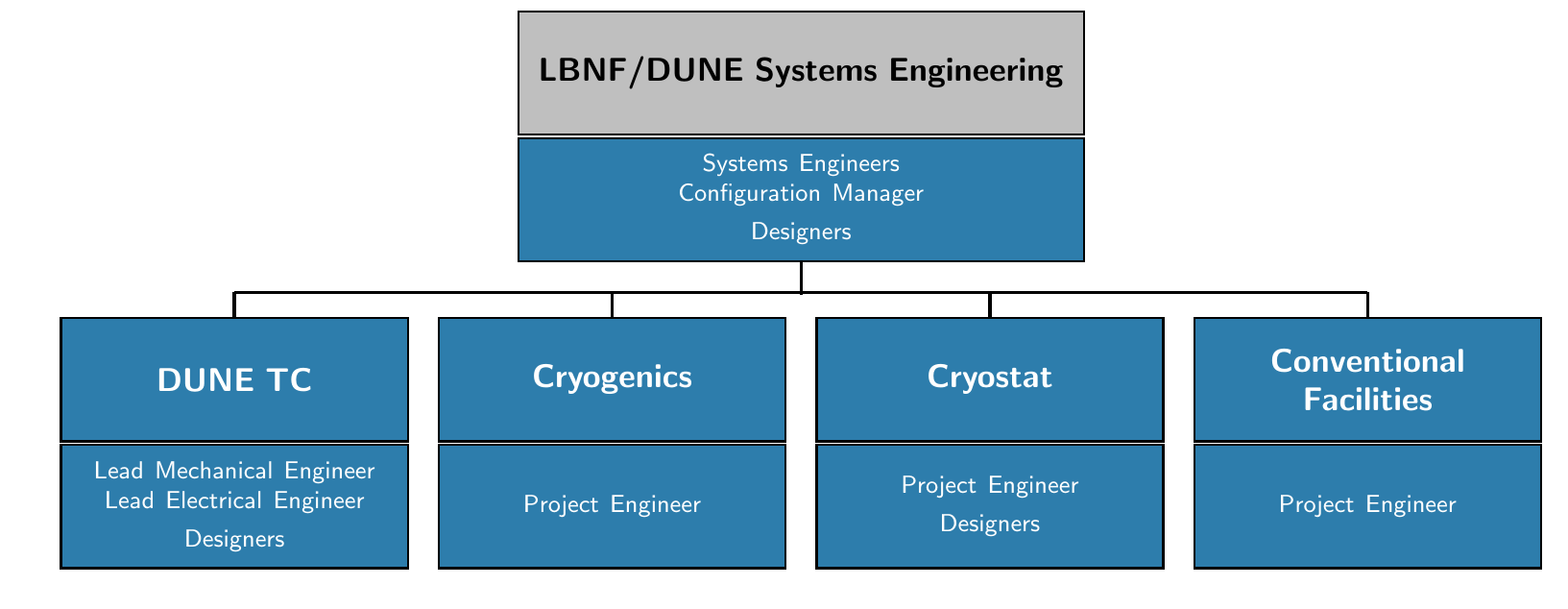}
\end{dunefigure}

Proper integration of the \dwords{fd} within the supporting
facilities and infrastructure at \surf is a major engineering task.
The \dword{lbnf}/\dword{dune} Systems Engineer is responsible for the
interfaces between the major \dword{lbnf} and \dword{dune} systems
(conventional facilities, cryostats, cryogenics systems and
\dwords{detmodule}). The \dword{lbnf}/\dword{dune} systems engineering team
includes several engineers and designers with responsibility for
maintaining computer aided design (CAD) models. \Dword{dune} \dword{tc}
supports an engineering team that works directly with the
\dword{lbnf}/\dword{dune} systems engineering team to ensure that the
detector is properly integrated into the overall system.

\dword{tc} has been working with the \dword{lbnf}/\dword{dune} systems engineering team to
define requirements from \dword{dune} for the conventional facilities final
design for the detector chambers, \dword{cuc}, drifts
and utilities. \Dword{tc} is representing the interests of the \dword{dune} detector
in the conventional facilities (CF) design. This includes refining the
detector installation plan to understand how much space is needed in
front of the \dwords{tco} of the cryostats
and therefore of the size of the chambers. \dword{tc} continues to refine the
detector needs for utilities in the detector caverns and the \dword{cuc} 
where the \dword{daq} will be housed.

Physics requirements on \dword{tc} include cleanliness in the cryostats,
survey and alignment tolerances, and grounding and shielding
requirements. The cleanliness requirement is for ISO 8 (class
100,000), which will keep rates from dust radioactivity below those of
the inherent $^{39}$Ar background. The alignment tolerances are driven
by physics requirements on reconstructing tracks. Grounding and
shielding are critical to enable this very sensitive, low-noise
detector to achieve the required \dword{s/n}. The physics
requirements for \dword{lbnf} and \dword{dune} are maintained in
DocDB-112.

\section{Project Support}
\label{sec:fdsp-coord-supp}

As defined in the \dword{dune} Management Plan (DMP), the \dword{dune}
Technical Board (TB) generates and recommends technical decisions to the 
collaboration executive board (EB) (see Figure~\ref{fig:TB_org}).
\begin{dunefigure}[DUNE Technical Board.]{fig:TB_org}
  {DUNE Technical Board.}
 \includegraphics[width=0.9\textwidth,trim=20mm 110mm 30mm 110mm,clip]{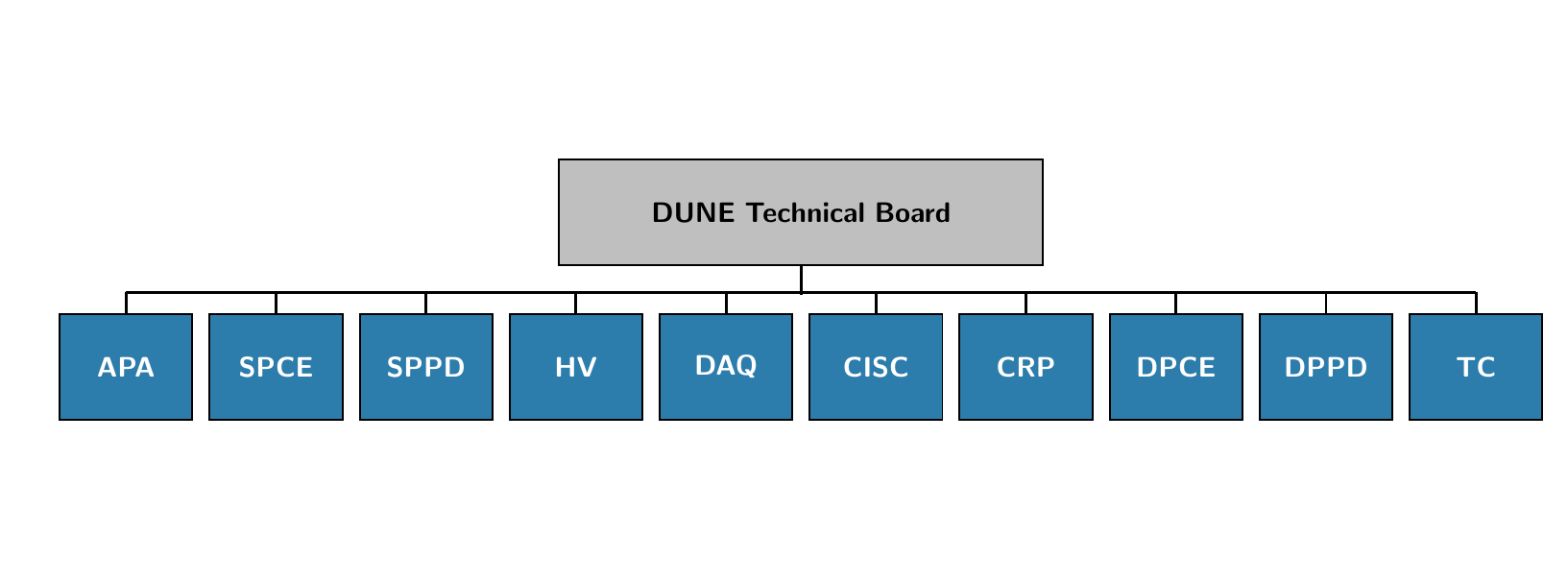}
\end{dunefigure}

It consists of all consortia scientific and technical leads. It meets
on a regular basis (approximately monthly) to review and resolve any
technical issues associated with the detector construction. It reports
through the EB to collaboration management. The \dword{dune} TB
is chaired by the technical coordinator. \dword{dune} collaboration
management, including the EB, is shown in Figure~\ref{fig:DUNE_org}. The
\dword{tc} engineering team also meets on a regular basis (approximately monthly)
to discuss more detailed technical issues. \Dword{tc} does not have
responsibility for financial issues; that will instead be referred to
the EB and Resource Coordinator (RC).

\begin{dunefigure}[DUNE management organizational structure.]{fig:DUNE_org}
  {DUNE management organizational structure.}
 \includegraphics[width=\textwidth]{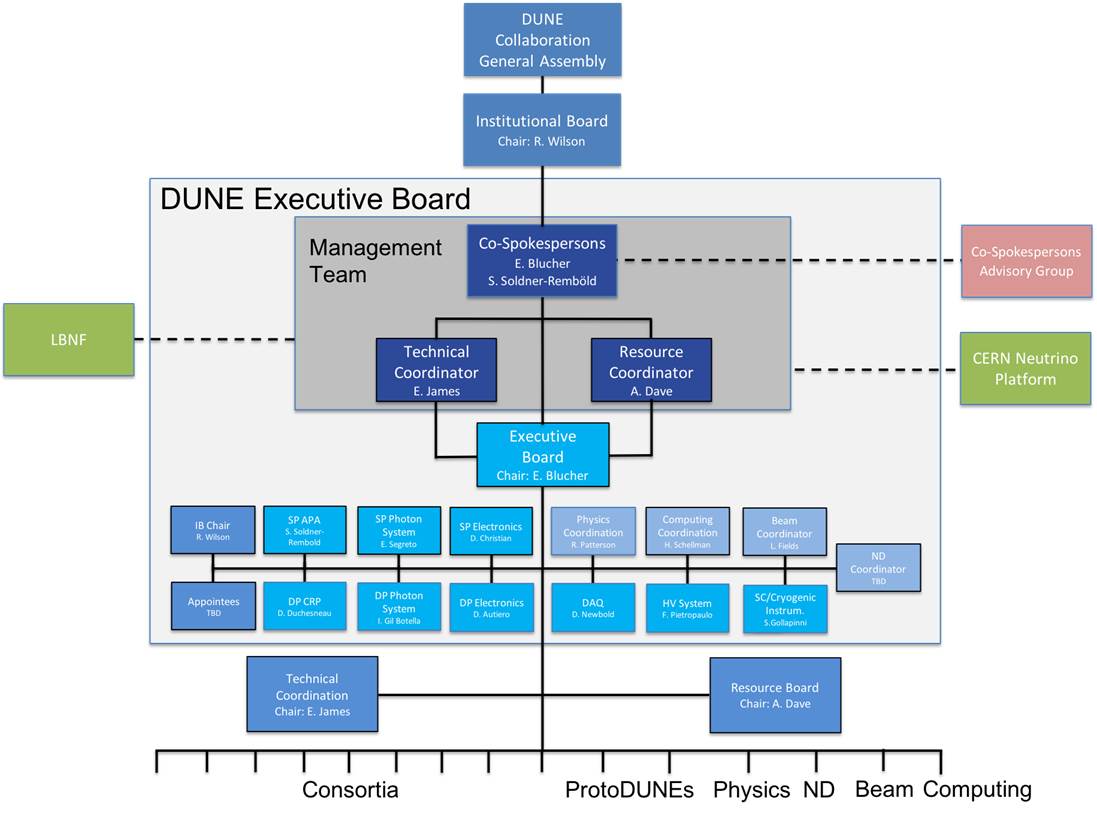}
\end{dunefigure}

\Dword{tc} has several major project support tasks that need to be accomplished:
\begin{itemize}
  \item Assure that each consortium has a well defined and complete
    scope, that the interfaces between the consortia are sufficiently
    well defined and that any remaining scope can be covered by \dword{tc}
    through \dword{comfund} or flagged as missing scope to the EB and RC. In
    other words, assure that the full detector scope is
    identified. Monitor the interfaces and consortia progress in
    delivering their scope.
  \item Develop an overall project \dlong{ims}
    that includes reasonable production schedules, testing plans and a
    well developed installation schedule from each consortium. Monitor
    the \dword{ims} as well as the individual consortium schedules.
  \item Ensure that appropriate engineering and safety standards are
    developed and agreed to by all key stakeholders and that these
    standards are conveyed to and understood by each
    consortium. Monitor the design and engineering work.
  \item Ensure that all \dword{dune} requirements on \dword{lbnf} for
    conventional facilities, cryostat and cryogenics have been clearly
    defined and understood by each consortium. Negotiate scope
    boundaries with \dword{lbnf}. Monitor \dword{lbnf} progress on
    final conventional facility design, cryostat design and cryogenics
    design.
  \item Ensure that all technical issues associated with scaling from
    \dword{protodune} have sufficient resources to converge on
    decisions that enable the detector to be fully integrated and
    installed.
  \item Ensure that the integration and \dword{qc} processes for each
    consortium are fully developed and reviewed and that the
    requirements on an \dword{itf} are well defined.
\end{itemize}

\Dword{tc} is responsible for technical quality and schedule and is not
responsible for consortia funding or budgets.  \Dword{tc} will try to help
resolve any issue that it can, but will likely have to push all
financial issues to the TB, EB and RC for resolution.

\Dword{tc} maintains a web page\footnote{\url{https://web.fnal.gov/collaboration/DUNE/DUNE\%20Project/\_layouts/15/start.aspx\#/}.}
with links to project documents. \Dword{tc} maintains repositories of project
documents and drawings. These include the \dword{wbs}, schedule, risk
register, requirements, milestones, strategy, detector models and
drawings that define the \dword{dune} detector.

\subsection{Schedule}
\label{sec:fdsp-coord-controls}

A series of tiered milestones are being developed for the \dword{dune}
project. The Tier-0 milestones are held by the spokespersons and host
laboratory director. Three have been defined and the current milestones and
target dates are:
\begin{enumerate}
\item Start main cavern excavation \hspace{2.1in} 2019
\item Start \dword{detmodule}~1 installation \hspace{2.1in} 2022
\item Start operations of \dword{detmodule}~1--2 with beam \hspace{1in} 2026
\end{enumerate}
These dates will be revisited at the time of the \dword{tdr} review.  Tier-1
milestones will be held by the technical coordinator and \dword{lbnf} Project
Manager and will be defined in advance of the \dword{tdr} review. Tier-2
milestones will be held by the consortia.

A high level version of the \dword{dune} milestones from the \dword{ims}
can be seen in Table~\ref{tab:DUNE_schedule}. 

\begin{dunetable}
[Overall \dword{dune} Project Tier-1 milestones.]
{p{0.84\linewidth}p{0.14\linewidth}}
{tab:DUNE_schedule}
{Overall \dword{dune} Project Tier-1 milestones.}
Milestone & Date   \\ \toprowrule
RRB Approval of Technical Design Review                       & 09/02/2019 \\ \colhline 
Beneficial Occupancy of Integration Test Facility             & 09/01/2021 \\ \colhline 
Construction of steel frame for Cryostat \#1 complete         & 12/17/2021 \\ \colhline 
Construction of Mezzanine for Cryostat \#1 complete           & 01/17/2022 \\ \colhline 
Begin integration/testing of Detector \#1 components at ITF   & 02/01/2022 \\ \colhline 
Beneficial Occupancy of Central Utility Cavern Counting room  & 04/16/2022 \\ \colhline 
Construction of steel frame for Cryostat \#2 complete         & 07/01/2022 \\ \colhline 
Construction of Mezzanine for Cryostat \#2 complete           & 08/01/2022 \\ \colhline 
\textbf{Beneficial occupancy of Cryostat \#1}                 & \textbf{12/23/2022} \\ \colhline 
Cryostat \#1 ready for TPC installation                       & 05/01/2023 \\ \colhline 
Begin integration/testing of Detector \#2 components at ITF   & 11/01/2023 \\ \colhline 
\textbf{Beneficial occupancy of Cryostat \#2}                 & \textbf{03/01/2024} \\ \colhline 
Begin closing Temporary Construction Opening for Cryostat \#1 & 05/01/2024 \\ \colhline 
Cryostat \#2 ready for TPC installation                       & 08/01/2024 \\ \colhline 
Cryostat \#1 ready for filling                                & 10/01/2024 \\ \colhline 
Begin closing Temporary Construction Opening for Cryostat \#2 & 07/18/2025 \\ \colhline 
\textbf{Detector \#1 ready for operations}                    & \textbf{10/01/2025} \\ \colhline 
Cryostat \#2 ready for filling                                & 12/05/2025 \\ \colhline 
\textbf{Detector \#2 ready for operations}                    & \textbf{12/18/2026} \\
\end{dunetable}

\Dword{tc} will maintain the \dword{ims} that links all consortium schedules
and contains appropriate milestones to monitor progress. The \dword{ims}
is envisioned to be maintained in MS-Project~\footnote{MicroSoft\texttrademark{} Project.} as it is expected
that many consortia will use this tool. It is currently envisioned as
three levels of control and notification milestones in addition to the
detailed consortium schedules. The highest level contains external
milestones, with the second level containing the key milestones for \dword{tc}
to monitor deliverables and installation progress, and the third level
containing the inter-consortium links. The schedules will go
under change control after agreement with each consortium on the
notification milestone dates and the \dword{tdr} is approved.

In addition to the overall \dword{ims} for construction and
installation, a schedule of key consortia activity in the period
2018--19 leading up to the \dword{tdr} has been developed.

To ensure that the \dword{dune} detector remains on schedule,
\dword{tc} will monitor schedule statusing from each consortium, will organize
reviews of schedules and risks as appropriate.  As schedule problems
arise \dword{tc} will work with the affected consortium to resolve the
problems. If problems cannot be solved, \dword{tc} will take the issue to the
TB and EB.

A monthly report with input from all consortia will be published by
\dword{tc}. This will include updates on consortium technical progress and
updates from \dword{tc} itself.

\subsection{Risk}
\label{sec:fdsp-coord-risk}

The successful operation of \dword{protodune} will retire a great many
potential risks to \dword{dune}. This includes most risks associated with the
technical design, production processes, \dword{qa}, integration
and installation. Residual risks remain relating to design and
production modifications associated with scaling to \dword{dune}, mitigations
to known installation and performance issues in \dword{protodune}, underground
installation at \surf and organizational growth.

The highest technical risks include: development of a system to
deliver \SI{600}{kV} to the \dual cathode; general delivery of the
required \dword{hv}; cathode and \dword{fc} discharge to the cryostat
membrane; noise levels, particularly for the \dword{ce}; 
number of dead channels; lifetime of components surpassing \dunelifetime{}; 
\dword{qc} of all components; verification of improved \dword{lem}
performance; verification of new cold  \dword{adc} and  \dword{coldata} performance;
argon purity; electron drift lifetime; \phel light yield;
incomplete calibration plan; and incomplete connection of design to
physics. Other major risks include insufficient funding, optimistic
production schedules, incomplete integration, testing and installation
plans.

Key risks for \dword{tc} to manage include the following:
\begin{enumerate}
    \item Too much scope is unaccounted for by the consortia and falls
      to \dword{tc} and \dword{comfund}.
    \item Insufficient organizational systems are put into place to
      ensure that this complex international mega-science project,
      including \dword{tc}, \fnal as host laboratory, \surf, DOE and all international
      partners continue to successfully work together to ensure
      appropriate rules and services are provided to enable success of
      the project.
  \item Inability of \dword{tc} to obtain sufficient personnel resources so as to
    ensure that \dword{tc} can oversee and coordinate all of its
    project tasks.  While the USA has a special responsibility towards
    \dword{tc} as host country, it is expected that personnel resources will
    be directed to \dword{tc} from each collaborating country. Related to this
    risk is the fact that consortium deliverables are not really
    stand-alone subsystems; they are all parts of a single \dword{detmodule}. This
    elevates the requirements on coordination between consortia.
\end{enumerate}

The consortia have provided preliminary versions of risk analyses that
have been collected on the \dword{tc} webpage. These are being developed into
an overall risk register that will be monitored and maintained by \dword{tc}
in coordination with the consortia.

\subsection{Reviews}
\label{sec:fdsp-coord-reviews}

\Dword{tc} is responsible for reviewing all stages of detector development
and works with each consortium to arrange reviews of the design
(\dword{pdr} and \dword{fdr}), production (\dword{prr} and
\dword{ppr}) and \dword{orr} of their system.  These
reviews provide input for the TB to evaluate technical decisions.
Review reports are tracked by \dword{tc} and provide guidance as to
key issues that will require engineering oversight by the \dword{tc}
engineering team. \Dword{tc} will maintain a calendar of \dword{dune}
reviews.

\Dword{tc} works with consortia leaders to review all detector designs,
with an expectation for a \dword{pdr}, followed by a \dword{fdr}.  All
major technology decisions will be reviewed prior to down-select.  \Dword{tc}
may form task forces as necessary for specific issues that need more
in-depth review.

Start of production of detector elements can commence only after
successful \dwords{prr}. Regular production progress
reviews will be held once production has commenced. The \dwords{prr}
will typically include review of the production of \textit{Module 0}, the
first such module produced at the facility. \Dword{tc} will work with
consortium leaders for all production reviews.

\Dword{tc} is responsible to coordinate technical documents for the LBNC
Technical Design Review.

\subsection{Quality Assurance}
\label{sec:fdsp-coord-qa}

The \dword{lbnf}/\dword{dune} \dword{qap} outlines the \dword{qa} requirements
for all \dword{dune} consortia and describes how the requirements
shall be met. The consortia will be responsible for implementing a
quality plan that meet the requirements of the
\dword{lbnf}/\dword{dune} \dword{qap}.  The consortia implement the
plan through the development of individual quality plans, procedures,
guides, \dword{qc} inspection and test requirements and travelers~\footnote{The
  traveler is a document that details the fabrication and inspection
  steps and ensures that all steps have been satisfactorily
  completed.} and test reports.  In lieu of a consortium-specific quality
plan, 
the consortia may work under the \dword{lbnf}/\dword{dune}
\dword{qap} and develop manufacturing and \dword{qc} plans, procedures and
documentation specific to their work scope.  The technical coordinator 
and consortia leaders are responsible for
providing the resources needed to conduct the Project successfully,
including those required to manage, perform and verify work that
affects quality.  The \dword{dune} consortia leaders are responsible
for identifying adequate resources to complete the work scope and to
ensure that their team members are adequately trained and qualified to
perform their assigned work.

The consortia work will be documented on travelers and applicable test
or inspection reports. Records of the fabrication, inspection and
testing will be maintained. When a component has been identified as
being in noncompliance to the design, the nonconforming condition
shall be documented, evaluated and dispositioned as one of the following: use-as-is (does
not meet design but can meet functionality as is), rework (bring into
compliance with design), repair (will be brought into meeting
functionality but will not meet design) or scrap. For nonconforming
equipment or material that is dispositioned as use-as-is or repair, a
technical justification shall be provided allowing for the use of the
material or equipment and approved by the design authority.

The \dword{lbnf}/\dword{dune} \dword{qam} reports
to the \dword{lbnf} Project Manager and \dword{dune} Technical
Coordinator and provides oversight and support to the consortium
leaders to ensure a consistent quality program.
\begin{enumerate}
  \item The \dword{qam} will plan reviews as independent assessments to assist
    the technical coordinator in identifying opportunities for
    quality and performance-based improvement and to ensure compliance
    with specified requirements.
  \item The \dword{qam} is responsible to work with the consortia in
    developing their \dword{qa} and \dword{qc} Plans.
  \item The \dword{qam} will review consortia \dword{qa} and \dword{qc} activity, including
    production site visits.
  \item The \dword{qam} will participate in consortia design reviews, conduct
    \dwords{prr} prior to the start of production,
    conduct \dwords{ppr} on a regular basis (as
    described in Section~\ref{sec:fdsp-coord-reviews}), and perform
    follow-up visits to consortium facilities prior to shipment of
    components to ensure all components and documentation are
    satisfactory.
\item The \dword{qam} is responsible for performing assessments at the
  \dword{itf}, the Far Site and the Near Site to
  ensure the activities performed at these locations are in accordance
  with the \dword{lbnf}/\dword{dune} \dword{qa} Program and applicable procedures,
  specifications and drawings.
\end{enumerate}

\subsubsection{Document Control}
\label{sec:fdsp-coord-document}

\Dword{tc} maintains repositories of project documents and drawings in two
document management systems.  DUNE Project documents will be stored in
the DUNE DocDB~\footnote{\url{https://docs.dunescience.org}}. DUNE drawings
will be stored in 
\dword{edms}~\footnote{\url{https://edms.cern.ch/ui/\#!master/navigator/project?P:1637280201:1637280201:subDocs}.}.
\Dword{tc} will maintain approved versions of \dword{qa}, \dword{qc} and testing plans,
installation plans, engineering and safety standards in the DUNE
DocDB.

Consortia have developed initial interface, risk, schedule and \dword{wbs}
documents that will be put under change control and managed by the \dword{tc}
engineering team along with the consortia involved. These
are currently in DocDB and will likely go under change control later
in 2018, although they will continue to be developed through the \dword{tdr}.

Thresholds for change control are described in the
\dword{lbnf}/\dword{dune} \dword{cmp}. The control process is further
described in Section~\ref{sec:fdsp-coord-integ-config}.

\subsection{ESH}
\label{sec:fdsp-coord-esh}

The \dword{dune} \dword{esh} program is described in the
\dword{lbnf}/\dword{dune} \dword{ieshp}. This plan is maintained by
the \dword{lbnf}/\dword{dune} \dword{esh} Manager, who reports to the
\dword{lbnf} Project Manager and the technical coordinator. The
\dword{esh} manager is responsible to work with the consortia in
reviewing their hazards and their \dword{esh} plans.  The \dword{esh}
Manager is responsible to review \dword{esh} at production sites,
integration sites and at \surf. It is expected that the \dword{esh}
reviews will be conducted as part of the \dword{prr} and \dword{ppr}
process described in Section~\ref{sec:fdsp-coord-reviews}.

A strong \dword{esh} program is essential to successful completion of
the \dword{lbnf}/\dword{dune} Project at \fnal, collaborating laboratories and
universities, the \dword{itf}, and \surf. \dword{dune} is committed to ensuring
a safe work environment for 
collaborators at
all institutions and to protecting the public from hazards associated
with construction and operation of \dword{lbnf}/\dword{dune}. In
addition, all work will be performed in a manner that preserves the
quality of the environment and prevents property damage. Accidents and
injuries are preventable and it is important that we work together to
establish an injury-free workplace.

To achieve the culture and safety performance required for this
project, it is essential that \dword{dune} ensure that procedures
are established to support the following \dword{esh} policy
statements:
\begin{itemize}
  \item Line managers are responsible for environmental stewardship
    and personal safety at the \dword{dune} work sites.
  \item Line managers, supported by the \dword{lbnf}/\dword{dune},
    \fnal, other collaborating laboratories, and \surf \dword{esh}
    organizations, will provide consistent guidance and enforcement of
    the \dword{esh} program that governs the activities of workers at each
    site where work is being performed.
  \item Incidents, whether they involve personal injuries or other
    losses, can be prevented through proper planning. All
    \dword{lbnf}/\dword{dune} work is planned.
  \item Workers are involved in the work planning process and
    continuous improvement, including the identification of hazards
    and controls.
  \item Working safely and in compliance with requirements is vital to
    a safe work environment. Line managers will enforce disciplinary
    policies for violations of safety rules.
  \item Each of us is responsible for  our own safety and for that of
    our co-workers. Together we create a safe work environment.
  \item A strong program of independent audits, self-assessments and
    surveillance will be employed to periodically evaluate the
    effectiveness of the \dword{esh} program.
  \item Any incidents that result or could have resulted in personal
    injury or illness, significant damage to buildings or equipment,
    or impact of the environment, will be investigated to determine
    corrective actions and lessons that can be applied to prevent
    recurrence. \dword{lbnf}/\dword{dune} encourages open reporting of
    errors and events.
\end{itemize}

To achieve the culture and safety performance required for this
project, it is essential that \dword{esh} be fully integrated into the
project and be managed as tightly as quality, cost, and schedule.

\section{Integration Engineering}
\label{sec:fdsp-coord-integ-sysengr}

The major aspects of detector integration focus on the mechanical and
electrical connections between each of the detector systems. This
includes verification that subassemblies and their interfaces are
correct (e.g., \dword{apa} and \single \dword{pds}). A second major area is in the support of
the 
\dwords{detmodule} and their interfaces to the cryostat and cryogenics. A
third major area is in assuring that the \dwords{detmodule} 
can be installed ---
that the integrated components can be moved into their final
configuration. A fourth major area is in the integration of the 
\dwords{detmodule} with the necessary services provided by the conventional
facilities.

\subsection{Configuration Management}
\label{sec:fdsp-coord-integ-config}

The \dword{tc} engineering team will maintain
configuration data in the appropriate format for the management of the
detector configuration. The consortia are responsible for providing
engineering data for their subsystems to \dword{tc}. The
\dword{tc} engineering team will work with the \dword{lbnf} project team to
integrate the full detector data into the global \dword{lbnf}
configuration files. Appropriate thresholds for tracking and for
change control will be established.

For mechanical design aspects, the \dword{tc} engineering team
will maintain full \threed CAD models of the \dwords{detmodule}. Appropriate level
of detail will be specified for each type of model. Each consortium
will be responsible for providing CAD models of their detector
components to be integrated into overall models. The \dword{tc} engineering
team will work with the \dword{lbnf} project team to integrate the
full \dword{detmodule} models into a global \dword{lbnf} CAD model which will
include cryostats, cryogenics systems, and the conventional
facilities. These will include models using varying software
packages. The \dword{tc} engineering team will work directly with the
consortium technical leads and their supporting engineering teams to
resolve any detector component interference or interface issues with
other detector systems, detector infrastructure, and facility
infrastructure.

For electrical design aspects, the \dword{tc} engineering team will maintain
high-level interface documents that describe all aspects of required
electrical infrastructure and electrical connections. All consortia
must document power requirements and rack space
requirements. Consortia are responsible for defining any cabling that
bridges the design efforts of two or more consortia. This agreed-upon
and signed-off interface documentation shall include cable
specification, connector specification, connector pinout and any data
format, signal levels and timing. All cables, connectors, printed
circuit board components, physical layout and construction will be
subject to project review. This is especially true of elements which
will be inaccessible during the project lifetime. Consortia shall
provide details on \lar temperature acceptance testing and
lifetime of components, boards, cables and connectors.

At the time of the release of the \dword{tdr}, the
\dword{tc} engineering team will work with the consortia to produce
formal engineering drawings for all detector components.  These
drawings are expected to be signed by the consortia technical leads,
project engineers, and Technical Coordinator.  Starting from that
point, the \dwords{detmodule} models and drawings will sit under formal change
control as discussed in Section~\ref{sec:fdsp-coord-qa}.  It is
anticipated that designs will undergo further revisions prior to the
start of detector construction, but any changes made after the release
of the \dword{tdr} will need to be agreed to by all of
the drawing signers and an updated, signed drawing produced. The major
areas of configuration management include:
\begin{enumerate}
  \item \threed model,
  \item Interface definitions,
  \item Envelope drawings for installation,
  \item Drawing management.
\end{enumerate}

\Dword{tc} will put into place processes for configuration
management.  Configuration management will provide \dword{tc}
and engineering staff the ability to define and control
the actual build of the detector at any point in time and to track any
changes which may occur over duration of the build as well as the
lifetime of the project as described in
Section~\ref{sec:fdsp-coord-qa}.

For detector elements within a cryostat, configuration management
will be frozen once the elements are permanently sealed within the
cryostat.  However, during the integration and installation process of
building the \dwords{detmodule} within the cryostat, changes may need to occur.
For detector elements outside the cryostat and accessible, all
repairs, replacements, hardware upgrades, system grounding changes,
firmware and software changes must be tracked.

Any change will require revision control, configuration
identification, change management and release management.

{\bf Revision Control}\\ Consortia will be responsible for providing
accurate and well documented revision control.  Revision control
will provide a method of tracking and maintaining the history of
changes to a detector element.  Each detector element must be clearly
identified with a document which includes a revision number and
revision history.  For mechanical elements, this will be reflected by
a drawing number with revision information.  For electrical elements,
schematics will be used to track revisions.  Consortia will be
responsible for identifying the revision status of each installed
detector element. Revisions are further controlled through maintenance
of the documents in DocDB and \dword{edms}.

{\bf Configuration Identification}\\
Consortia are responsible for providing unique identifiers or part
numbers for each detector element.  Plans will be developed on how
inventories will be maintained and tracked during the build.  Plans
will clearly identify any dynamic configuration modes which may be
unique to a specific detector element.  For example, a printed circuit
board may have firmware that affects its performance.

{\bf Change Management}\\
\dword{tc} will provide guidelines
for formal change management.  During the beginning phase of the
project, drawings and interface documents are expected to be signed by
the consortium technical leads, project engineers, and technical
coordinator.  Once this initial design phase is complete, the detector
models, drawings, schematics and interface documents will be under
formal change control.  It is anticipated that designs will undergo
further revisions prior to the start of detector construction, but any
changes made after the release of the \dword{tdr} will
need to be agreed to by all drawing signers and an updated signed
drawing produced.

{\bf Release Management}\\
Release management focuses on the delivery of the more dynamic aspects
of the project such as firmware and software.  Consortia with
deliverables that have the potential to affect performance of the
detector by changing firmware or software must provide plans on how
these revisions will be tracked, tested and released.  The
modification of any software or firmware after the initial release,
must be formally controlled, agreed upon and tracked.

\subsection{Engineering Process and Support}
\label{sec:fdsp-coord-integ-engr-proc}

The \dword{tc} organization will work with
the consortia through its \dword{tc} engineering team to ensure the proper
integration of all detector components.  The \dword{tc} engineering team will
document requirements on engineering standards and documentation that
the consortia will need to adhere to in the design process for the
detector components under their responsibility.  Similarly, the
project \dword{qa} and \dword{esh} managers will document \dword{qc} and safety
criteria that the consortia will be required to follow during the
construction, installation, and commissioning of their detector
components, as discussed in sections~\ref{sec:fdsp-coord-qa}
and~\ref{sec:fdsp-coord-esh}.

Consortia interfaces with the conventional facilities, cryostats, and
cryogenics are managed through the \dword{tc}
organization.  The \dword{tc} engineering team will work with the
consortia to understand their interfaces to the facilities and then
communicate these globally to the \dword{lbnf} Project team.  For conventional
facilities the types of interfaces to be considered are requirements
for bringing needed detector components down the shaft and through the
underground tunnels to the appropriate detector cavern, overall requirements for
power and cooling in the detector caverns, and the requirements on
cable connections from the underground area to the surface.
Interfaces to the cryostat include the number and sizes of the
penetrations on top of the cryostat, required mechanical structures
attaching to the cryostat walls for supporting cables and
instrumentation, and requirements on the global positioning of the
\dwords{detmodule} within the cryostat.  Cryogenics system interfaces include
requirements on the location of inlet and output ports, requirements on
the monitoring of the \lar both inside and outside the
cryostat, and grounding and shielding requirements on piping attached to
the cryostat.

\dword{lbnf} will be responsible for the design and construction of the
cryostats used to house the \dwords{detmodule}.  The consortia are required to
provide input on the location and sizes of the needed penetrations at
the top of the cryostats.  The consortia also need to specify any
mechanical structures to be attached to the cryostat walls for
supporting cables or instrumentation.  The \dword{tc} engineering
team will work with the \dword{lbnf} cryostat engineering team to understand
what attached fixturing is possible and iterate with the consortia as
necessary.  The consortia will also work with the \dword{tc} engineering
team through the development of the \threed CAD model to understand the
overall position of the \dwords{detmodule} within the cryostat and any issues
associated with the resulting locations of their detector components.

\dword{lbnf} will be responsible for the cryogenics systems used to purge,
cool, and fill the cryostats.  It will also be responsible for the
system that continually re-circulates \lar through filtering
systems to remove impurities.  Any detector requirements on the flow
of liquid within the cryostat 
will be developed by the consortia and
transmitted to \dword{lbnf} through the \dword{tc} engineering team.  Similarly,
any requirements on the rate of cool-down or maximum temperature
differential across the cryostat during the cool-down process will 
be specified by the consortia and transmitted to the \dword{lbnf} team.

The engineering design process is defined by a set of steps taken to
fulfill the requirements of the design.  By the time of the \dword{tdr}, 
all design requirements must be fully defined and
proposed designs must be shown to meet these requirements.  Based on
prior work, some detector elements may be quite advanced in the
engineering process, while others may be in earlier stages.  Each
design process shall closely follow the engineering steps described
below.

{\bf Development of specifications}\\ Each consortium is responsible
for the technical review and approval of the engineering
specifications.  The documented specifications for all major design
elements will 
include the scope of work, project milestones,
relevant codes and standards to be followed, acceptance criteria and
specify appropriate internal or external design reviews.
Specifications shall be treated as controlled documents and cannot be
altered without approval of the \dword{tc}
team.  The \dword{tc} engineering team will participate in and help facilitate
all major reviews as described in
Section~\ref{sec:fdsp-coord-reviews}.  Special TB reviews
will be scheduled for major detector elements.

{\bf Engineering Risk Analysis}\\ Each consortium is responsible for
identifying and defining the level of risk associated with their
deliverables as described in Section~\ref{sec:fdsp-coord-risk}.
\Dword{tc} will work with the consortia,
through its \dword{tc} engineering team, to document these risks in a risk
database and follow them throughout the project until they are
realized or can be retired.

{\bf Specification Review}\\
The \dword{dune} \dword{tc} organization and project engineers
shall review consortium specifications for overall compliance with the
project requirements.  Consortia must document all internal reviews
and provide that documentation to \dword{tc}.
Additional higher-level reviews may be scheduled by \dword{tc}
as described in Section~\ref{sec:fdsp-coord-reviews}.

{\bf System Design}\\
The system design process includes the production of mechanical
drawings, electrical schematics, calculations which verify compliance
to engineering standards, firmware, printed circuit board layout,
cabling and connector specification, software plans, and any other
aspects that lead to a fully documented functional design.  All
relevant documentation shall be recorded, with appropriate document
number, into the chosen engineering data management system and be
available for the review process.

{\bf Design Review}\\ The design review process is determined by the
complexity and risk associated with the design element.  For a simple
design element, a consortium may do an internal review.  For a more
complex or high risk element a formal review will be scheduled.
\dword{dune} \dword{tc} will facilitate the review,
bringing in outside experts as needed.  In all cases, the result of
any reviews must be well documented and captured in the engineering
data management system.  If recommendations are made, those
recommendations will be tracked in a database and the consortia will
be expected to provide responses. All results of these engineering
reviews will be made available for the subsystem design reviews
described in Section~\ref{sec:fdsp-coord-reviews}.

{\bf Procurement}\\ The procurement process will include the
documented technical specifications for all procured materials and
parts.  All procurement technical documents are reviewed for
compliance to engineering standards and \dword{esh} concerns.
\dword{dune} \dword{tc} will assist the consortia in working with
their procurement staff as needed.

{\bf Production and assembly}\\ During the implementation phase of the project,
the consortia shall provide the Technical Coordinator with updates on
schedule.  A test plan will be fully developed which will allow for
verification that the initial requirements have been met. This is part of
the \dword{qa} plan that will be documented and followed as described
in Section ~\ref{sec:fdsp-coord-qa}.

{\bf Testing and Validation\\} The testing plan documented in the
above step will be followed and results will be well documented in
consultation with the \dword{qam}.  The Technical Coordinator and \dword{tc}
engineering team will be informed as to the results and whether the
design meets the specifications.  If not, a plan will be formulated to
address any shortcomings and presented to the Technical Coordinator.

{\bf Final Documentation\\} Final reports will be generated that
describe the as built equipment, lessons learned, safety reports,
installation procedures, testing results and operations procedures.

\section{Detector Infrastructure}
\label{sec:fdsp-coord-infrastructure}

\dword{tc} is responsible for delivering the common infrastructure for
the \dword{fd}. This infrastructure is typically equipment that is used
by many groups. This may include: the electronics racks on top of the
cryostat with power, cooling, networking and safety systems, cable
trays, cryostat crossing tubes, and flanges, detector safety systems
and ground monitoring and isolation transformers.

\dword{tc} is responsible for the systems to support the
detectors. For the \dword{dpmod} this may consist of a cable
winch system similar to \dword{pddp}.  In the case of the
\dword{spmod} the installation group also provides the
\dword{dss}, which supports the detector and is needed to bring
equipment into the correct location in the cryostat.

\subsection{Detector Support System}

The \dword{dss} provides the structural support for the \dword{spmod}
inside the cryostat.  It also provides the necessary
infrastructure to move the detector elements into location during
assembly. As the \dword{dss} is a new design and is quite different
from the \dword{pdsp} \dword{dss} it is described in some detail in this
section. The detector elements supported by the \dword{dss} include
the \dwords{ewfc}, the \dwords{apa}, and the \dwords{cpa} with
top and bottom \dword{fc} panels.  The \dword{dss} is supported by the
cryostat outer steel structure through a series of \fdth{}s that
cross through the cryostat insulation and are anchored with flanges on
the cryostat roof. Inside the cryostat a series of stainless steel
I-beams are connected to the \fdth{}s and used to support the
detector. The \dword{dss} defines the location of the detector inside
the cryostat and it also defines how the detector elements
move and contract as the detector is brought to \dword{lar}
temperature. The design of the \dword{dss} encompasses the overall
structural design of the \dword{detmodule} as only after the elements are
mounted to the \dword{dss} and are connected together do they make a
unified mechanical structure. The requirements of the \dword{dss} are
as follows:
\begin{itemize}
 \setlength\itemsep{1mm}
\setlength{\parsep}{1mm}
\setlength{\itemsep}{-5mm}
\item Support the weight of the detector.
\item Accommodate cryostat roof movement during filling, testing, and operation.
\item Accommodate variation in \fdth locations and
  variation in the flange angles due to installation tolerances and
  loading on the warm structure.
\item Accommodate shrinkage of the detector and \dword{dss} from ambient
  temperature to \dword{lar} temperature.
\item Define the position of the detector components relative to each other. 
\item Provide electrical connection to the cryostat ground and remain electrically isolated from the detector.
\item Allow for the support penetrations to be purged with gaseous argon to prevent contaminants from diffusing back into the liquid. 
\item Ensure that the instrumentation cabling does not interfere with the \dword{dss}.
\item Consist entirely of components that can 
be installed through the \dword{tco}.
\item 
Meet AISC-360 or appropriate codes required at \surf.
\item 
Meet seismic requirements one mile underground at \surf.
\item Consist entirely of 
materials that are compatible for operation in ultrapure \dword{lar}.
\item Ensure that beams are completely submerged in \dword{lar}.
\item Ensure that detector components are not less than \SI{400}{mm} from the membrane flat surface.
\item Ensure that the supports do not interfere with the cryostat I-beam structures.
\item Ensure 
that the detector's lower \dword{gp} is above the cryogenic piping and the top of the \dword{dss} beams are submerged in \dword{lar} while leaving a \SI{4}{\%} ullage at the top of the cryostat.
\item Include the infrastructure necessary to move the \dword{apa} and
  \dword{cpa}-\dword{fc} assemblies from outside the cryostat through the
  \dword{tco} and to the correct position.
\end{itemize}

Figure~\ref{fig:DSS} (left) shows the \dword{dss} structure; there are
five rows of supports for the alternating rows of
\dword{apa}-\dword{cpa}-\dword{apa}-\dword{cpa}-\dword{apa}.  The
\dword{dss} is connected to the warm structure at a flange that is
mounted on the outside of the cryostat.  Figure~\ref{fig:DSS} (right)
shows the layout of these structural \fdth{}s.  The \dword{dss}
consists of pairs of \fdth{}s that support \SI{6.4}{m}-long
S8x18.4 stainless steel I-beam sections. The proposed design of the
\dword{dss} has \num{10} I-beam segments per row for a total of
\num{50} I-beam segments. Each I-beam is suspended on both ends by
rods from \fdth{}s that penetrate the roof.  
When cold, each beam shrinks, causing gaps to form between
\dwords{apa} that are adjacent but supported on separate beams.
\dwords{apa} that are supported on the same beam will not have gaps
develop because both the beam and \dwords{apa} are stainless steel so
they shrink together.  Each beam is supported by a nearly
\SI{2}{m} long rod that allows the beam support to move as the beam
contracts.

\begin{dunefigure}[\threed model of the \dword{dss} ]{fig:DSS}
  {\threed model of the \dword{dss} showing the entire
  structure on the left along with one row of \dword{apa} and
  \dword{cpa}-\dword{fc} at each end. The right panel is a zoomed image
  showing the connections between the vertical supports and the
  horizontal I-beams.}
\includegraphics[width=.49\textwidth]{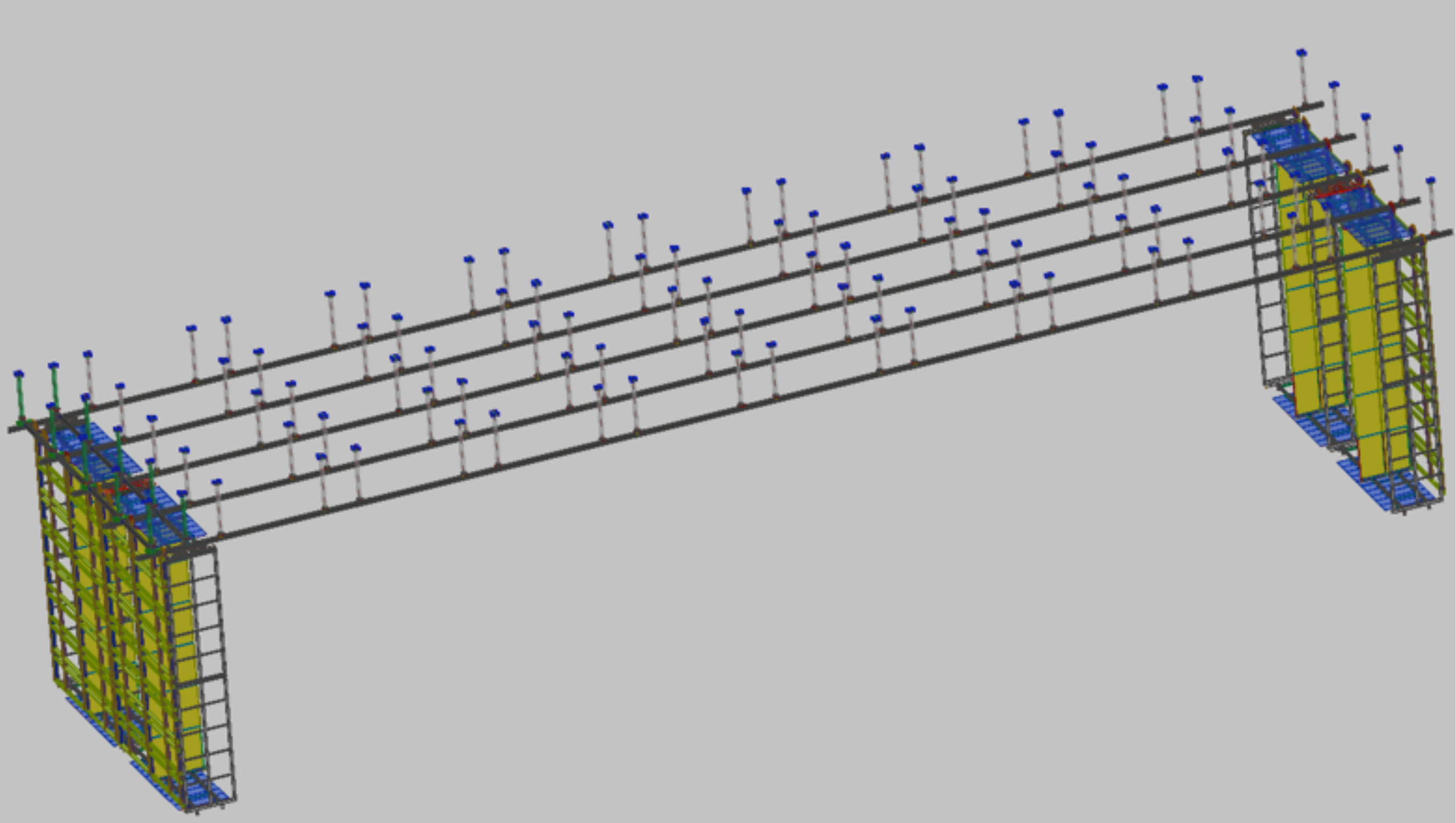}
 \includegraphics[width=.49\textwidth]{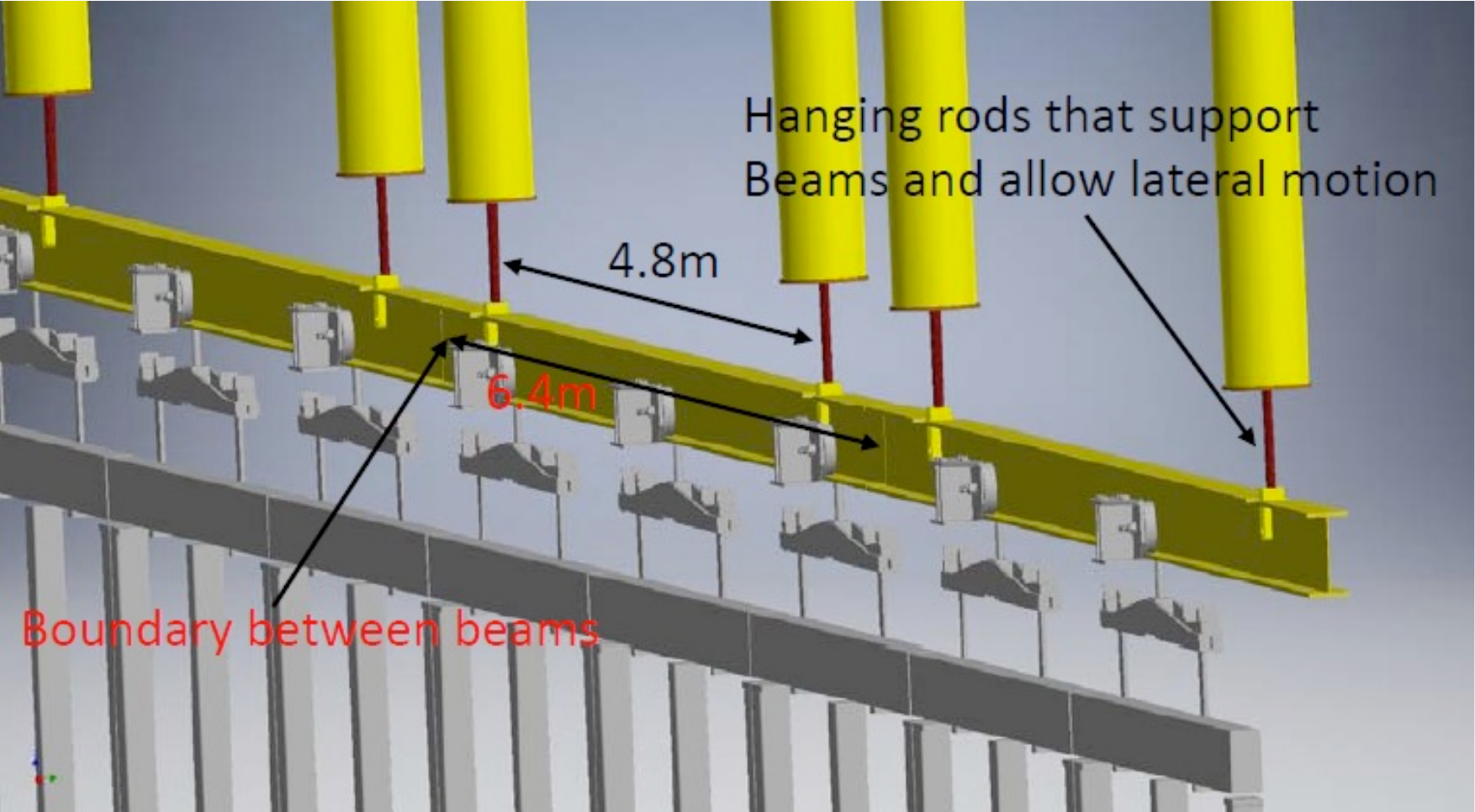}
\end{dunefigure}


Detector components are installed using a shuttle beam system
illustrated in Figure~\ref{fig:shuttle}.  The last two columns of
\fdth{}s (eastern-most) support temporary beams that run
north-south, perpendicular to the main \dword{dss} beams.  A shuttle
beam
has trolleys mounted to it and transverses 
north-south until aligned with the required row of \dword{dss} beam.  The last
\dword{apa} or \dword{cpa} in a row is supported by the shuttle beam which is bolted
directly to the \fdth{}s once it is in place.  As the last \dword{cpa} or
\dword{apa} in each row is installed, the north-south beams are removed.

A mechanical interlock system  prevents trolleys
from passing the end of the shuttle beam unless it is aligned with a
corresponding \dword{dss} beam.  The shuttle beam and each detector component are
moved using a motorized trolley.  A commercially available motorized
trolley will be modified as needed to meet the needs of the
installation.

\begin{dunefigure}[\threed models of the shuttle beam end of the \dword{dss}]{fig:shuttle}
  {\threed models of the shuttle beam end of the \dword{dss}. The figures show how an \dword{apa}
is translated into position using the north-south beams until it lines up with the correct
row of I-beams}
\includegraphics[width=.49\textwidth]{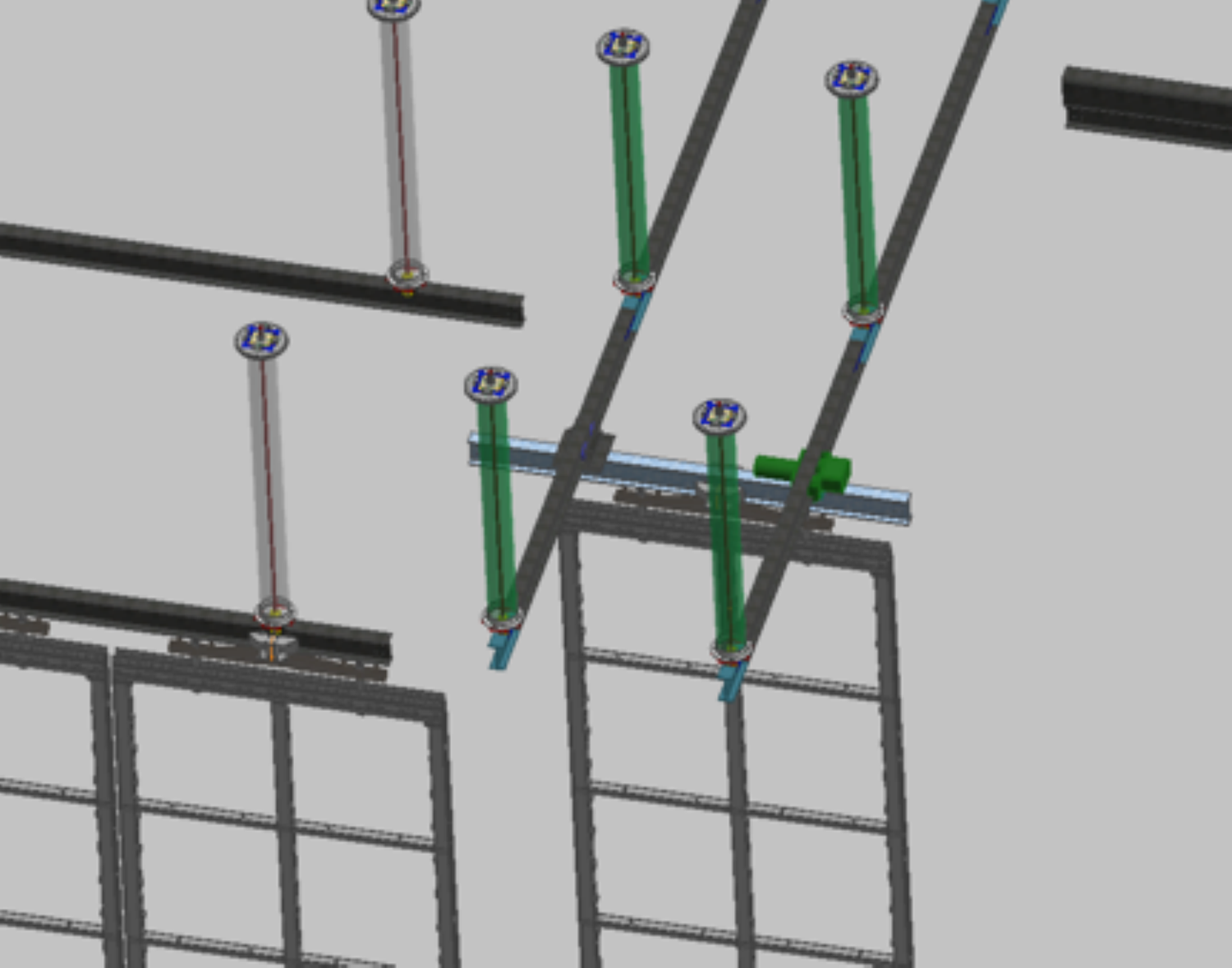}
 \includegraphics[width=.42\textwidth]{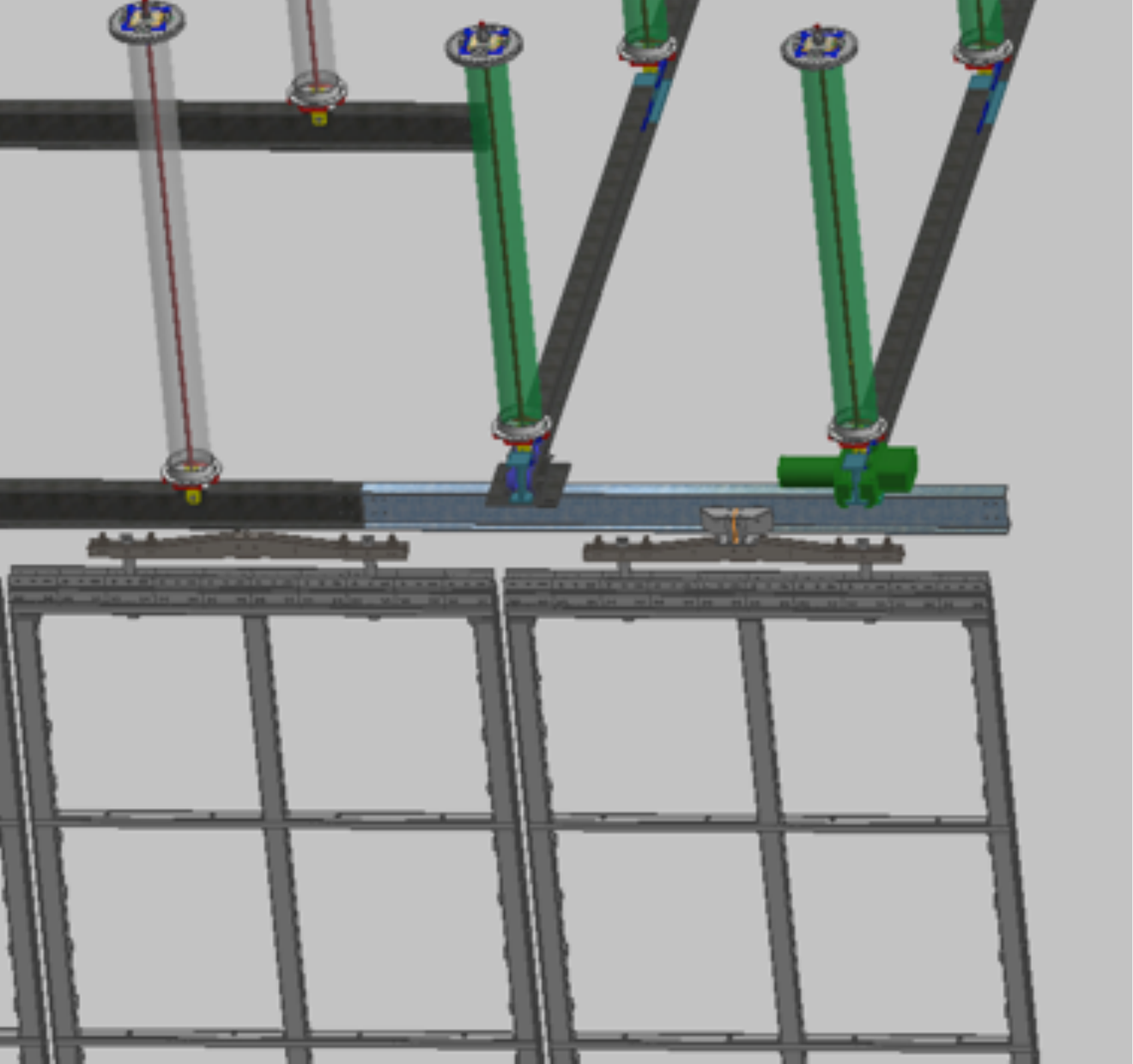}
\end{dunefigure}

The \dword{dss} installation begins with the placement and alignment of all
the \fdth{}s onto the flanges that are mounted to the warm vessel.
There are \num{25} \fdth{}s per row and five rows for a total of \num{125}
\fdth{}s.  A fixture with a tooling ball is attached to the
clevis of each \fdth.  The $xy$ position in the horizontal plane
and the vertical $z$ position of this tooling ball is defined, then a 
survey is performed to determine the location of each tooling
ball center and $xy$ and $z$ adjustments are made to get the tooling
ball centers to within $\pm$\SI{3}{mm}.  The \SI{6.4}{m} long I-beams are then 
raised and pinned to the clevis.  Each beam weighs roughly \SI{160}{kg} (\SI{350}{lbs}).
A lifting tripod is placed over each of the \fdth{}s
supporting a beam, and a \SI{0.64}{cm} \SI{0.25}{in}  
cable is fed through the top
flange of the \fdth down the \SI{14}{m} to the cryostat floor where it
is attached to the I-beam.  The winches on each tripod raise
the beam in unison in order to get it to the correct height to be
pinned to the \fdth clevis.  Once the beams are mounted, a final
survey of the beams takes place to ensure they are properly located and aligned
to each other.

A mock-up of the shuttle system is constructed to test the
mechanical interlock and drive systems for the shuttle beam
for each \dword{detmodule}.  Tests are conducted to evaluate the level of
misalignment between beams that can be tolerated and the amount of
positional control that can be achieved with the motorized trolley. It
is expected this will be finished prior to the \dword{tdr}. At the time of the
\dword{tdr} a larger prototype installation at Ash River will be under
construction. This prototype will use full scale elements and will be
used to develop the installation procedures and to test the
\dword{detmodule} installation process.

\section{The Integration and Test Facility}
\label{sec:fdsp-coord-integ-test}

\dword{dune} \dword{tc} is responsible for interfacing with the
\dword{lbnf} logistics team at \surf to coordinate transport of all
detector components into the underground areas.  Due to the limited
availability of surface areas at \surf for component storage, nearby
facilities will be required to receive, store and ship materials to
the Ross Shaft on an as-needed basis. A team within the \dword{tc}
organization will 
develop and execute the plan for
receiving detector components at a surface facility and transporting
them to the Ross Shaft in coordination with the on-site \dword{lbnf}
logistics team.  The surface facility will require warehouse space
with an associated inventory system, storage facilities, material
transport equipment and access to trucking.  Basic functions of this
facility will be to receive the detector components arriving from
different production sites around the world and prepare them for
transport into the underground areas, incorporating re-packaging and
testing as necessary. As a substantial facility will 
be
required, it can also serve as a location where some detector
components are integrated and undergo further testing prior to
installation.

The logistics associated with integrating and installing the
\dwords{fd} and their associated infrastructure present
a number of challenges. 
These include the size and
complexity of the detector itself, the number of sites around the
world that will be fabricating detector and infrastructure components,
the necessity for protecting components from dust, vibration and shock
during their journey to the deep underground laboratory, and the lack
of space on the surface near the Ross Shaft. To help
mitigate the associated risks, \dword{dune} plans to establish an
\dword{itf} somewhere in the vicinity of \surf. This facility
and its associated staff will need to provide certain functions and
services connected to the \dword{dune} \dword{fd} integration and
installation effort and will have other potential roles that are still under
consideration.  The areas to which the \dword{itf} will and could
contribute are the following:
\begin{itemize}
  \item {\bf Transport buffer:} The \dword{itf} needs to provide
    storage capacity for a minimum one-month buffer of detector
    components required for the detector installation process in the
    vicinity of \surf and be able to accept packaging materials
    returned from the underground laboratory.
  \item {\bf Longer-term storage:} The \dword{itf} needs to provide
    longer-term storage of detector components that need to be
    produced in advance of when they are to be installed and for which
    sufficient storage space does not exist at the production sites.
  \item {\bf Re-packaging} The \dword{itf} needs to have the capability
    to re-package components arriving from the various production
    sites into boxes that can be safely transported through the shaft
    into the underground areas.
  \item {\bf Component fabrication:} It could be convenient to fabricate some
  components 
  in the vicinity
    of \surf at the \dword{itf}, taking advantage of
    undergraduate science and engineering students from the South
    Dakota School of Mines and Technology (SDSMT).
  \item {\bf Component integration:} Integration of detector components
  received from different production sites 
  that must be done prior to transport underground,
    such as the installation of \dwords{pd} and 
    electronics on the \dwords{apa}, could be done
    prior to re-packaging at the \dword{itf}.    
  \item {\bf Inspection, testing and repair:} In cases where
    components are re-packaged for transport to the underground area,
    \dword{itf} support for performing tests on these components to
    ensure that no damage has occurred during shipping is likely
    desirable.  In addition, components integrated at the \dword{itf}
    would require additional testing prior to being re-packaged.  The
    \dword{itf} could provide facilities needed to repair some of the
    damaged components (others would likely need to be returned to
    their production sites).
  \item {\bf Collaborator support:} The host institution of the \dword{itf}
    will need to 
    provide support for a significant number
    of \dword{dune} collaborators involved in the above activities
    including services such as housing assistance, office space,
    computing access, and safety training.
  \item {\bf Outreach:} The host institution of the \dword{itf} would be
    ideally situated for supporting an outreach program to build upon
    the considerable public interest in the experiment that exists
    within South Dakota.
\end{itemize}

 The facilities project (\dword{lbnf}) will provide the cryostats that 
 house the \dwords{detmodule} and the cryogenics
systems that support them.  Additional large surface facilities in the vicinity
of \surf will be required to stage the components of these
infrastructure pieces prior to their installation in the underground
areas.  The requirements for these facilities, in contrast to the
\dword{itf}, are relatively straight-forward and can likely be met by
a commercial warehousing vendor, who would provide suitable storage
space, loading and unloading facilities, and a commercial inventory
management and control system.  It is currently envisioned that
operation of the \dword{lbnf} surface facilities and the \dword{dune}
\dword{itf} will be independent from one another.  However, the
inventory systems used at the different facilities will need to 
ensure proper coordination of all
deliveries being made to the \surf site.

A reasonable criterion for the locations of these surface staging
facilities is to be within roughly an hour's drive of \surf.
Population centers in South Dakota within this radius are Lead,
Deadwood, Sturgis, Spearfish, and Rapid City.  The \dword{lbnf}
surface facilities, which are not expected to have significant
auxiliary functionality, would logically be located as near to the
\surf site as possible.  In the case of the \dword{dune}
\dword{itf}, however, locating the facility in Rapid City to take
advantage of facilities and resources associated with the South Dakota
School of Mines \& Technologies (SDSMT) has a number of potential
advantages.  SDSMT and the local business community in Rapid City
have expressed interest in incorporating the \dword{dune} \dword{itf}
into an overall regional development program.

\subsection{Requirements}

The leadership teams associated with the \dword{dune} 
consortia taking responsibility for the different \dword{fd}
subsystems have provided input to \dword{tc} on which potential
\dword{itf} functions and services would be applicable to their
subsystems.  The consortia have also made preliminary assessments of
the facility infrastructure requirements that would be necessary in
order for these functions and services to be provided for their
subsystems at the \dword{itf}.  An attempt to capture preliminary,
global requirements for the \dword{itf} based on the information
received from the consortia results in the following:
\begin{itemize}
  \item Warehousing space on the order of \SI{2800}{m^2} (\num{30000} square feet); driven by
    potential need to store hundreds of the larger detector components
    needed to construct the TPCs.  The provided space will need to
    maintain some minimal cleanliness requirements (e.g.; no insects)
    and be climate-controlled within reasonable temperature and
    humidity ranges.
  \item Crane or forklift coverage throughout the warehousing
    space to access components as needed for further processing or
    transport to \surf.
  \item Docking area to load trucks with detector components being
    transported to \surf and to receive packaging materials
    returned from the site.
  \item Smaller clean room areas within the warehousing space to
    allow for the re-packaging of detector components for transport
    underground.  Re-packaging of larger components
    will require local crane coverage within the appropriate clean room
    spaces.
  \item Dedicated space for racks and cabinets available for dry-air
    storage and testing of electronics components.  Racks and cabinets
    must be properly connected to the building ground to avoid
    potential damage from electrostatic discharges.
  \item Climate-controlled dark room space for the handling and
    testing \dword{pd} components.
  \item Dedicated clean room on order of \SI{1000}{m^2} (\num{10000} square feet) to facilitate
    integration of electronics and \dwords{pd} on \dwords{apa}.
    The
    integrated \dwords{apa} are tested in cold boxes supported by cryogenics
    infrastructure.  Clean room lighting must be UV-filtered to avoid
    damaging the \dwords{pd}.  The height of the clean room must
    accommodate crane coverage needed for movement of the \dwords{apa} in and
    out of the cold boxes.  It will also be necessary to have
    platforms for installation crews to perform work at heights within
    different locations in the clean room.
  \item Access to machine and electronics shops for making simple
    repairs and fabricating unanticipated tooling.
  \item Access to shared office space for up to \num{30} collaborators
    contributing to the activities taking place at the \dword{itf}.
    Assistance for identifying temporary housing in the vicinity of
    the \dword{itf} for the visiting collaborators.
\end{itemize}

\subsection{Management}

Overall management of the \dword{itf} is envisioned to be
responsibility of one or more of the collaborating institutions on
\dword{dune}.  If the \dword{itf} is located in Rapid City, SDSMT
would be a natural choice due to its physical proximity, connections
to the local community and ability to provide access to resources and
facilities that would benefit planned \dword{itf} activities.  Initial
discussions with SDSMT representatives have taken place, in which
they have expressed a clear interest for hosting the \dword{itf}.
Additional discussions will be needed to understand the details of the
\dword{itf} management structure with in the context of a more
finalized set of requirements for the facility.

\subsection{Inventory System}

Effective inventory management will be essential for all aspects of
\dword{dune} detector development, construction, installation and
operation.  While its relevance and importance go beyond the
\dword{itf}, the \dword{itf} is the location at
which \dword{lbnf}, \dword{dune} project management, consortium
scientific personnel and \surf operations will interface.  We therefore
will develop standards and protocols for inventory management as part
of the \dword{itf} planning.  A critical requirement for the project
is that the inventory management system for procurement, construction
and installation be compatible with future \dword{qa}, calibration and
detector performance database systems.  Experience with past large
detector projects, notably \nova, has demonstrated that the capability
to track component-specific information is extremely valuable
throughout installation, testing, commissioning and routine operation.
Compatibility between separate inventory management and physics
information systems will be maintained for effective operation and
analysis of \dword{dune} data.

\subsection{ITF Infrastructure}

\dword{tc} is responsible for providing the common support
infrastructure at the \dword{itf}. This includes the cranes and
forklifts to move equipment in the \dword{itf}, 
any
cryostats and cryogenics systems to enable cold tests of consortium-provided 
components, 
the cleanrooms and cleanroom equipment
to enable work on or testing of consortia components, and UV-filtered
lighting as needed. This also includes racks and cable trays.

\section{Installation Coordination and Support}
\label{sec:fdsp-coord-install}


Installation Coordination and Support 
(also called simply \textit{Installation}) is
responsible for coordinating the detector installations, providing
detector installation support and providing installation-related
infrastructure. The installation group management responsibility is
shared by a scientific lead and a technical lead that report to the
Technical Coordinator. The 
group responsible for activities in the underground areas is referred to as
the \dword{uit}. The \dword{itf} group, which delivers equipment to the
Ross Shaft, and the \dword{uit}, which receives the equipment
underground, need to be in close communication and work closely
together.

Underground installation is in general responsible for coordinating
and supporting the installation of the \dwords{detmodule} and providing
necessary infrastructure for installation of the experiment. While the
individual consortia are responsible for the installation of their own
detector equipment, it is essential that the detector installation be
planned as a whole and that a single group coordinates the
installation and adapts the plans throughout the installation
process. The \dword{uit} has responsibility for overall coordination
of the installations. In conjunction with each consortium the
\dword{uit} makes the installation plan that describes how the
\dwords{detmodule} are to be installed. The installation plan is used to define
the spaces and infrastructure that will be needed to install them.
The installation plan will also be used to define the
interfaces between the Installation group and the consortium
deliverables.

\subsection{UIT Infrastructure}

The installation scope includes the infrastructure needed to install
the \dword{fd} such as the cleanroom, a small machine shop, special
cranes, scissor lifts, and access equipment.  Additional equipment
required for installation includes: rigging equipment, hand tools,
diagnostic equipment (including oscilloscopes, network analyzers, and
leak detectors), local storage with some critical supplies and some
personal protective equipment (PPE). The \dword{uit} will also provide
trained personnel to operate the installation infrastructure. The
consortia will provide the detector elements and custom tooling and
fixtures as required to install their detector components.

\subsection{Underground Detector Installation}
\label{sec:fdsp-coord-undergd}

For the \dwords{detmodule} to be installed in safe and efficient
manner, the efforts of the individual consortia must be coordinated
such that the installation is planned as a coherent process. The
interfaces between the individual components must be understood
and the spaces required for the installation process planned and
documented. The installation planning must take into account the
plans and scope of the \dword{lbnf} effort and the individual plans of
the nine consortia. By working with the \dword{lbnf} team and the
members of the consortia responsible for building and installing their
components, a joint installation plan and schedule, taking into account
all activities and needs of all stakeholders, can be developed. Although
the organization of the installation effort is still evolving, 
an installation coordinator will be the equivalent of a scientific lead for this effort.

One of the primary early responsibilities of the \dword{uit} is to
develop and maintain the \dword{dune} installation plan and the
installation schedule. This installation plan 
describes the installation process in sufficient detail to demonstrate
how all the individual consortium installation plans mesh and it 
gives an overview of the installation process. The installation plan
is used by the \dword{uit} to define the underground infrastructure
needed for detector installation and the interfaces it has with respect  to 
the consortia. The \dword{uit} is responsible for reviewing and
approving the consortium installation plans. Approved installation
plans, engineering design notes, signed final drawings, and safety
documentation and procedures are all prerequisites for the \dword{prr}. 
Approved procedures, safety approval, and
proper training are all required before the \dword{uit} performs
work. During the installation phase the installation leadership 
coordinates the \dword{dune} installation effort and adapts the schedule
as needed. The installation coordinator, together with management, will also
resolve issues when problems occur.

The installation infrastructure to be provided by the \dword{uit}
includes: the underground ISO 8 (or class \num{100000}) clean room
used for the installation; cranes and hoists (if they are not
delivered by \dword{lbnf}); and scissor lifts, aerial lifts, and the common
work platforms outside the cryostat. The \dword{uit} will have
responsibility for operating this equipment and assisting the
consortia with activities related to rigging, material transport, and
logistics. Each consortium is responsible for the installation of
their own equipment, so the responsibility of the installation group is
limited, but the material handling scope is substantial. To support
the installation process, an installation floor manager will lead a
trained crew with the main responsibility of transporting the
equipment to the necessary location and operating the cranes, hoists,
and other common equipment needed for the installation. It is expected
that the installation crew will work with the teams from the various
consortia but will mainly act in a supporting function. The
\dword{uit} floor manager will be responsible for supervising the
\dword{uit} crew, but the ultimate responsibility for all detector
components remains with the consortia even while the underground
team is rigging or transporting these components.  This will be
critical in the case where any parts are damaged during transport or installation,
as the consortia need to judge the necessary actions. 
\fixme{judge the situation and determine the necessary actions?}
For this reason,
a representative or point of contact (POC) from the consortia must be
present when any work is performed on their equipment. The consortium
is responsible for certifying that each installation step is properly
performed.

The \dword{uit} acts as the primary point of contact with
\dword{lbnf} and \surf from the time the components reach the Ross
headframe until the equipment reaches the experimental cavern. If
something goes wrong, \surf calls the \dword{uit} leader who then
contacts the responsible party. The consortia are responsible for
delivering to the \dword{uit} all approved procedures and specialized
tooling required for transport. The \dword{uit} leader acts as a point
of contact if the \dword{lbnf} or \surf team has questions or difficulties
with the underground transport.  The \dword{uit} receives the
materials from \dword{lbnf} and \surf at the entrance to the \dword{dune}
excavations. The \dword{uit} then delivers the equipment to the
required underground location.

In an effort to get an early estimate of the equipment required to
install the detectors the \dword{uit} has developed a preliminary
installation plan that outlines the installation process. At present
the installation plan consists of a \threed model of the cryostat in the
excavations. The \dword{spmod} elements are inserted in the
model and a proposal for how they are transported, assembled, and
inserted into the cryostat has been conceptually developed and expressed in a series of images
some of which are shown in Figures~\ref{fig:Install-seq} and~\ref{fig:cpa-fc-unpack-assy}. 
Conceptual designs of the infrastructure needed to support
the transport and assembly are also included in the model. See Figures~\ref{fig:Install-ISO-Top} and~\ref{fig:Install-TopView}. With this
as a tool, the proposed installation sequence can be iterated with the
consortia to converge on a baseline installation plan. A similar
process will be followed for the \dword{dpmod} once the base
configuration for the \dword{sp} installation is agreed upon. The
\dword{uit} has focused initially on the \dword{spmod} as the
\dword{sp} components are larger and the installation process more
complex. 

\begin{dunefigure}[APA and CPA installation steps]{fig:Install-seq}
  {Top row from left:  crated \dword{apa} rotating to vertical position;  crated vertical \dword{apa} placed in cart; \dword{apa} panels moved to fixture using the under-bridge crane. Bottom row: series showing \dword{cpa} panels uncrated and moved to fixture. }
\includegraphics[width=.9\textwidth]{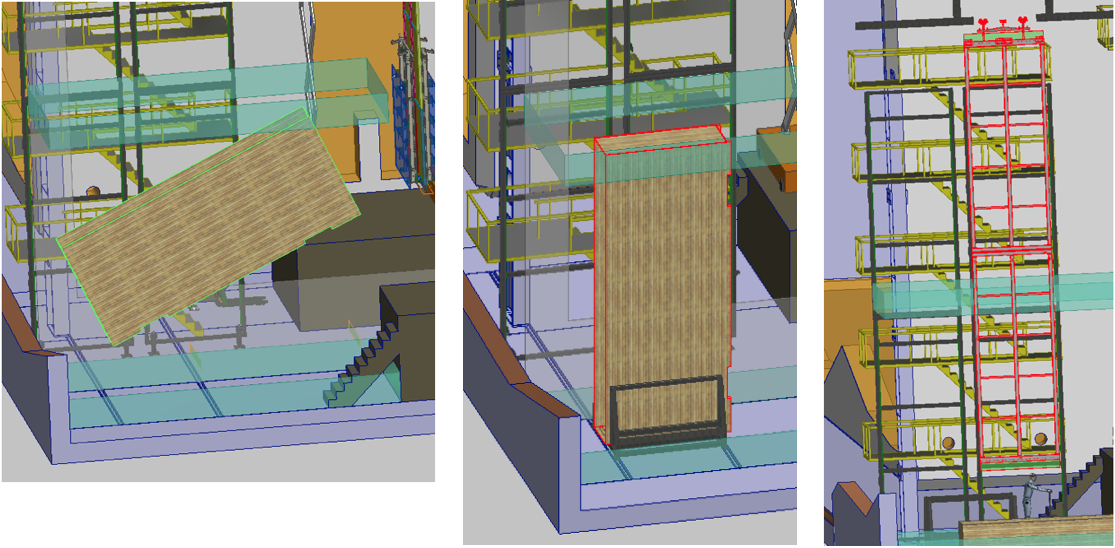}
\includegraphics[width=.9\textwidth]{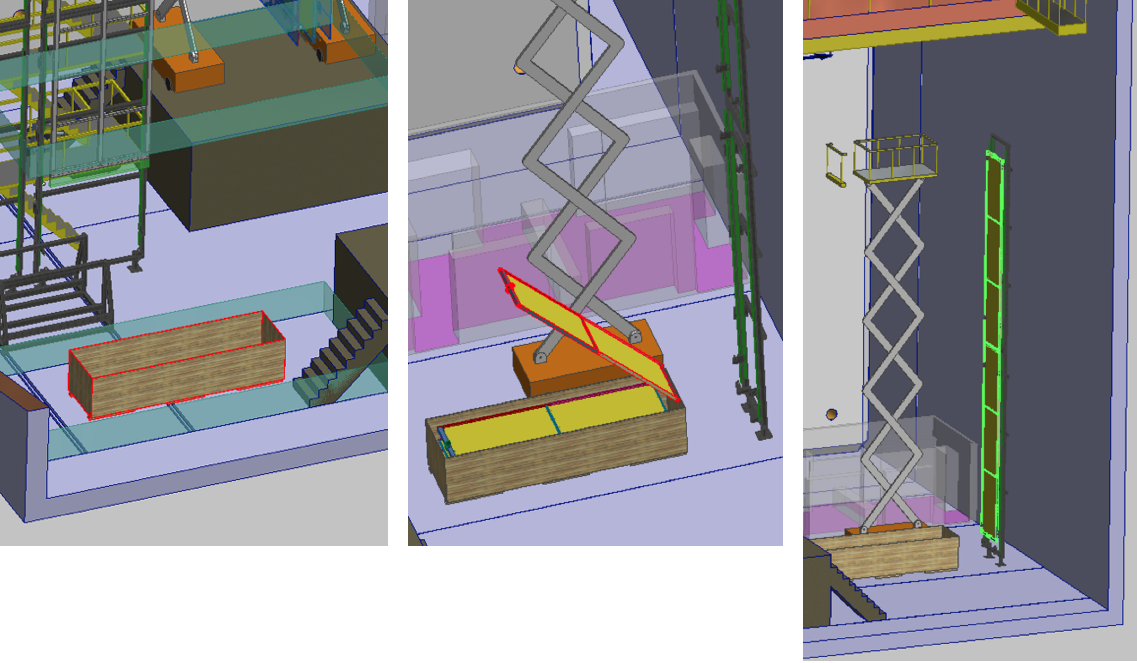}
\end{dunefigure}

\begin{dunefigure}[\dword{cpa} and \dword{fc} unpacking and assembly]{fig:cpa-fc-unpack-assy}
  {On the left, the assembled \dword{cpa} panel is placed onto the north \dword{tco} beam. On the right, the (green) \dword{fc} panels (already lowered into \dword{sas} and moved into the clean room) are installed as the \dword{cpa} array hangs under the \dword{tco} beam. }
\includegraphics[width=.9\textwidth]{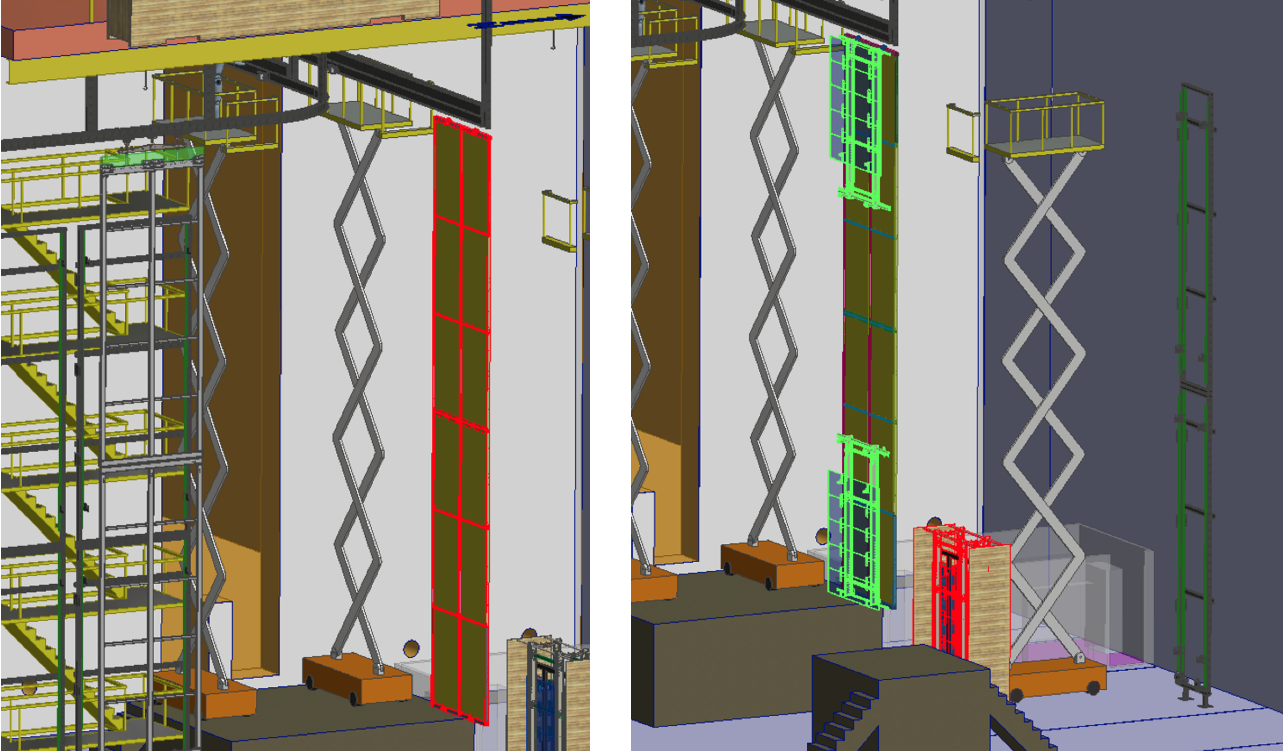}
\end{dunefigure}

\begin{dunefigure}[\threed model of underground area showing installation infrastructure]{fig:Install-ISO-Top}
  {\threed model of the underground area showing the infrastructure to install the \dword{spmod} in cryostat~1. The most significant features are presented including the \dword{apa} and \dword{cpa} assembly areas, the region around the \dword{tco} for materials entering the cryostat,  the changing room, the region for the materials air lock, (\dword{sas}), 
  and the means of egress.}
\includegraphics[width=.9\textwidth]{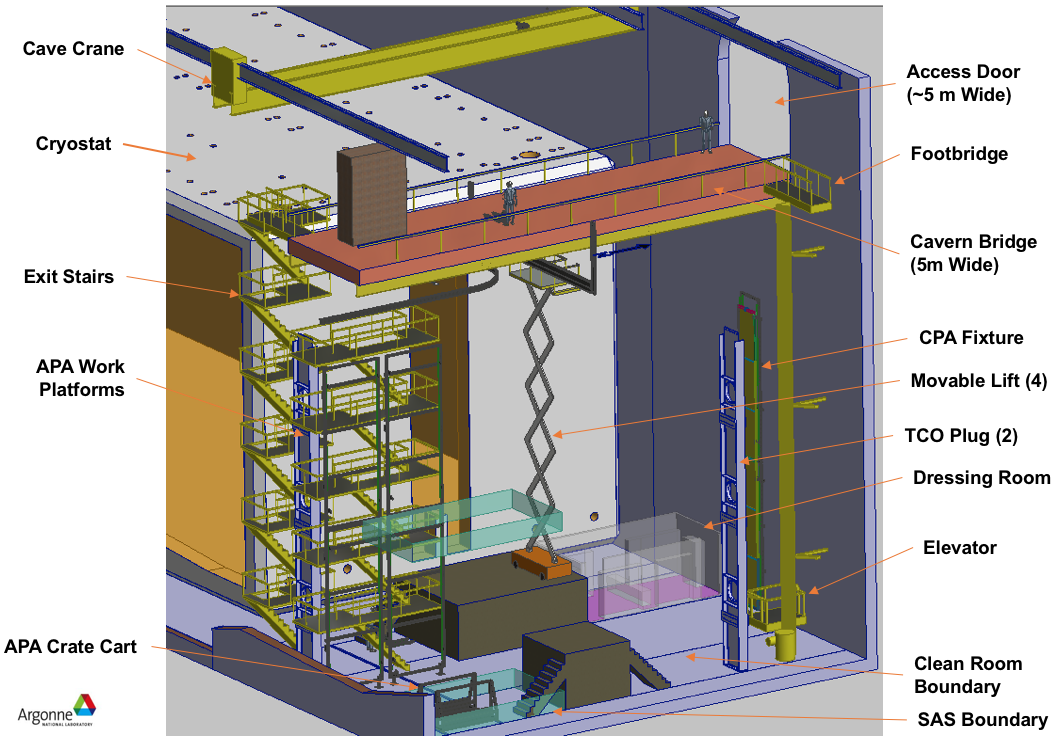}
\end{dunefigure}

\begin{dunefigure}[Section view of the \threed model showing layout]{fig:Install-TopView}
  {Section view of the \threed model showing layout, looking down on the installation area from below the bridge. Areas shown, left to right,  are the cryostat and \dword{tco}, the platform in front of the \dword{tco}, the dressing area, the \dword{apa} and \dword{cpa} assembly area (directly under the bridge), and the stairs and elevator. The lower right corner of the region is used as the materials air lock.}
\includegraphics[width=.9\textwidth]{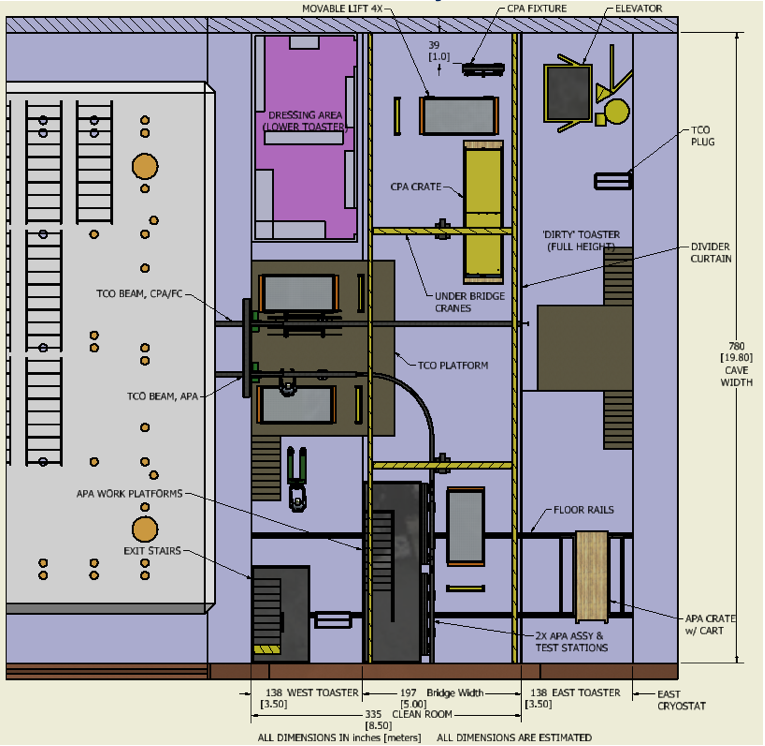}
\end{dunefigure}

In the current installation plan, \dword{dune} will take
ownership of the different underground areas at different times. The
surface data room and the underground room in the \dword{cuc} are available
significantly before the collaboration has access to the cryostats; 
the optical fibers between the surface and underground will be in
place even earlier. This will allow a \dword{daq} prototype to be developed
and tested early. The installation of the \dword{daq} hardware can also be
finished before the start of detector installation if desired, so the
\dword{daq} will not be on the critical path.  When the collaboration receives
access to Cryostat~1 the steel work for Cryostat~2 will be
finished and the work on installing the membrane will have
started. Excavation will be complete.  For planning purposes it is
assumed that the first \dword{detmodule} will be \dword{sp} and the second
\dword{dp}. The first step in the \dword{sp} installation is to
install the cryogenics piping and the \dword{dss}. As this piping will
require welding and grinding, it is a dirty process and must be
complete before the area can be used as a clean room. When this is
complete the cryostat can be cleaned and the false floor
re-installed. The clean infrastructure for installing the \dword{detmodule},
including the clean room, work platforms, scaffolding, the
fixturing to hold the detector elements during assembly, and all the
lifts need to be set up. Once the infrastructure is in place and the
area is clean, the installation of the main elements can start. The
general layout of the installation area showing the necessary space
and equipment is shown in Figure~\ref{fig:Install-seq}. 

The \dword{spmod}  is installed by first installing the west endwall or
endwall~1 (see Figure~\ref{fig:endwall}).

\begin{dunefigure}[End view of \dword{spmod} with \dword{ewfc} in
  place]{fig:endwall}
  {End view of \dword{spmod} with \dword{ewfc} in
  place, with one row of \dwords{apa} and \dwords{cpa}.}
\includegraphics[width=0.6\textwidth]{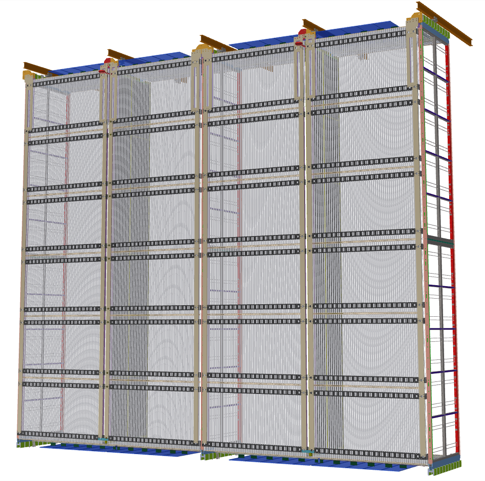}
\end{dunefigure}

The \dwords{apa} and \dwords{cpa} with top and bottom \dword{fc} panels are
installed next. The plan is to install six \dwords{apa} and four
\dwords{cpa} per week, which is enough to complete one of the \num{25}
rows every week. Additional time is built into the schedule to take
into account that the installation will be slower at the beginning and
some re-work may be needed. By building west-to-east, complete rows can
be finished and tested before moving to the next row. This reduces the
risk of finding a fault after final \dword{fc} deployment and cabling,
which would require dismantling part of the \dword{detmodule}. Some of the steps
needed to install the \dword{apa} and \dword{cpa} modules outside the
cryostat are also shown in Figure~\ref{fig:Install-seq}.  The middle three
panels show how the \dword{apa} needs to be handled in order to rotate
it and mount it to the assembly frame. After two \dwords{apa} are
mounted on top of each other, the cabling for the lower \dwords{apa}, and the
\dword{ce} and \dword{pd} cables can be installed. The
lower three panels show how the \SI{2}{m} \dword{cpa} sub-panels are
removed form the transport crates and assembled on a holding frame. Once
the \dword{cpa} module is assembled the \dword{fc} units can be
mounted. Finally, once the \dwords{apa} and \dwords{cpa} are installed,
the endwall~2 can be installed. A high-level summary of the schedule
is shown in Figure~\ref{fig:Install-Schedule}.

\begin{dunefigure}[High-level installation schedule]{fig:Install-Schedule}
  {High-level installation schedule.}
 \includegraphics[width=\textwidth]{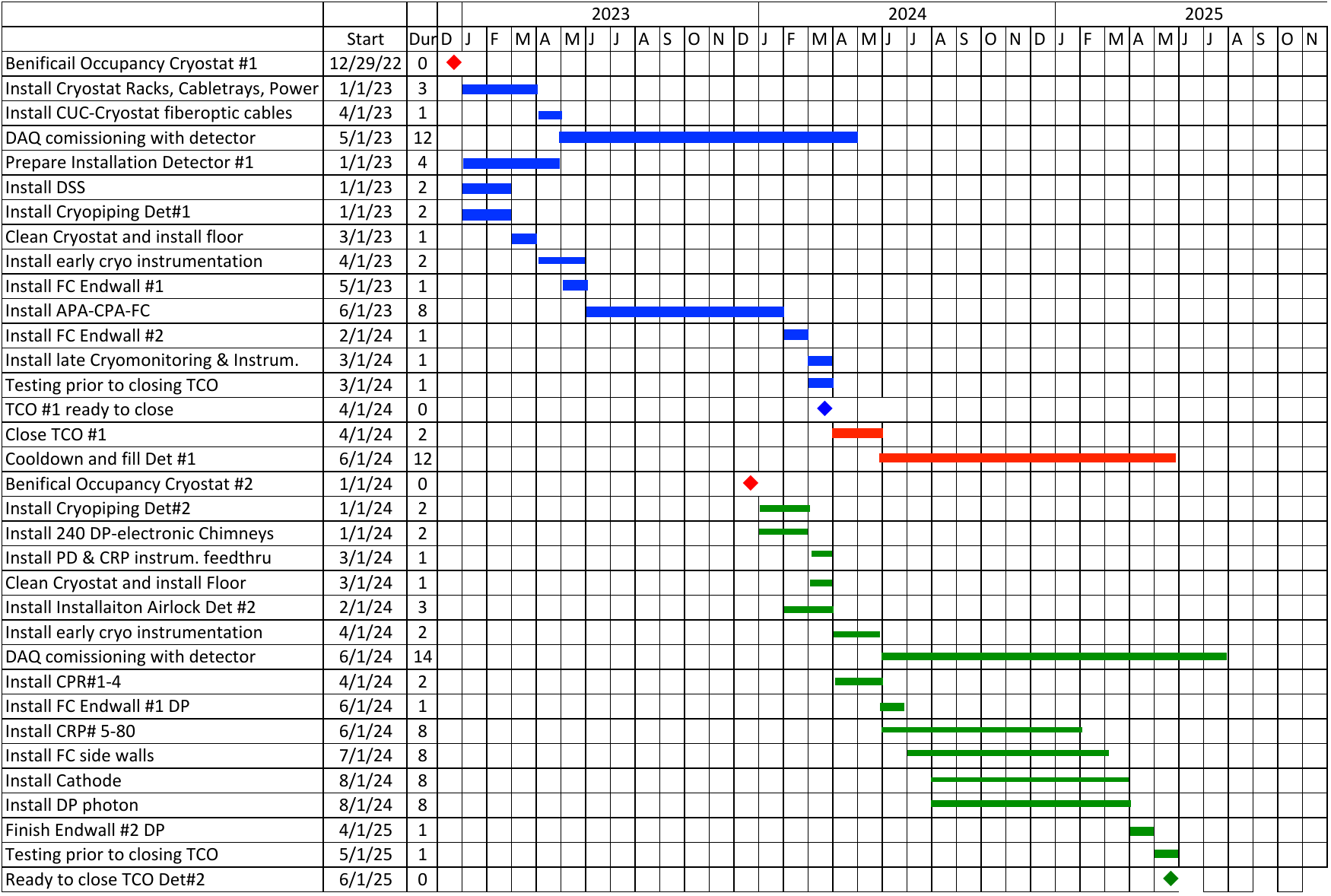}
\end{dunefigure}

As is seen in the installation schedule the second cryostat becomes
available four months before the first \dword{detmodule} installation is
complete. In this period, installation work for both \dwords{detmodule} 
proceeds in parallel. Like the \dword{spmod}, the first step is
the installation of the cryogenics piping, followed by a thorough cleaning
and installation of the false floor. While this piping is being
installed, the \dword{dp} chimneys for the electronics along with the
\dword{pds} and \dword{crp} instrumentation \fdth{}s can also be installed. Since the
chimneys are installed into the roof of the cryostat,  this work is
performed well away from the final installation work on the first
\dword{detmodule} so there should be no conflicts. Once the first \dword{detmodule} is
installed work on setting up the second \dword{detmodule}'s installation
infrastructure can begin. This work includes moving the cranes and
work platforms along with moving the walls of the clean room so that the
second cryostat is clean. The air filtration to the cryostat is also
moved to the second cryostat.  Since much of the work for the \dword{dp}
installation will be performed inside the cryostat, in principle, outside the cryostat a
clean room area smaller than that for the \dword{spmod}  would suffice. 
However, for
planning purposes, it will not be completely clear what type of \dword{detmodule}
will be installed in the second cryostat until fairly late. Therefore the \dword{uit}
will plan to provide a sufficiently large area outside the cryostat to
accommodate either detector technology.  
The much smaller clean room for a \dword{dpmod} 
could be installed just
outside the \dword{tco}. The installation process inside the \dword{dpmod} will
proceed east-to-west. At the start of the TPC installation the first
four \dwords{crp} -- comprising the first row -- will be installed. 
The left panel in Figure~\ref{fig:CRP-Install} shows two \dwords{crp}
being installed near the roof of the cryostat and the right panel
shows one of the \dwords{crp} in a transport box being moved into the cryostat.
Once the first \dword{crp} row is installed and tested, 
the first \dword{ewfc} can be installed.
In general rows of \dwords{crp} will be installed, followed by rows of \dword{fc} modules, followed by the cathode installation at the bottom of the \dword{detmodule}, followed by 
\dwords{pd} under the cathode plane. 
At the end of the installation, the second \dword{ewfc} is
installed and a final testing period for the full \dword{detmodule} is
foreseen. The \dword{dpmod} installation sequence is shown in green in
Figure~\ref{fig:Install-Schedule}.

\begin{dunefigure}[Image of the \dual CRPs being installed in
  the \dword{dpmod}]{fig:CRP-Install}
  {Left: Image of the \dword{dp} \dwords{crp} being installed in
  the \dword{dpmod}, showing the connection from the \dword{crp} to the
  electronics readout chimney. Right: Image of the \dword{crp} being
  inserted into the cryostat using a transport beam.  The \dual \dword{fc} modules will be inserted in a similar fashion.}
\includegraphics[width=0.45\textwidth]{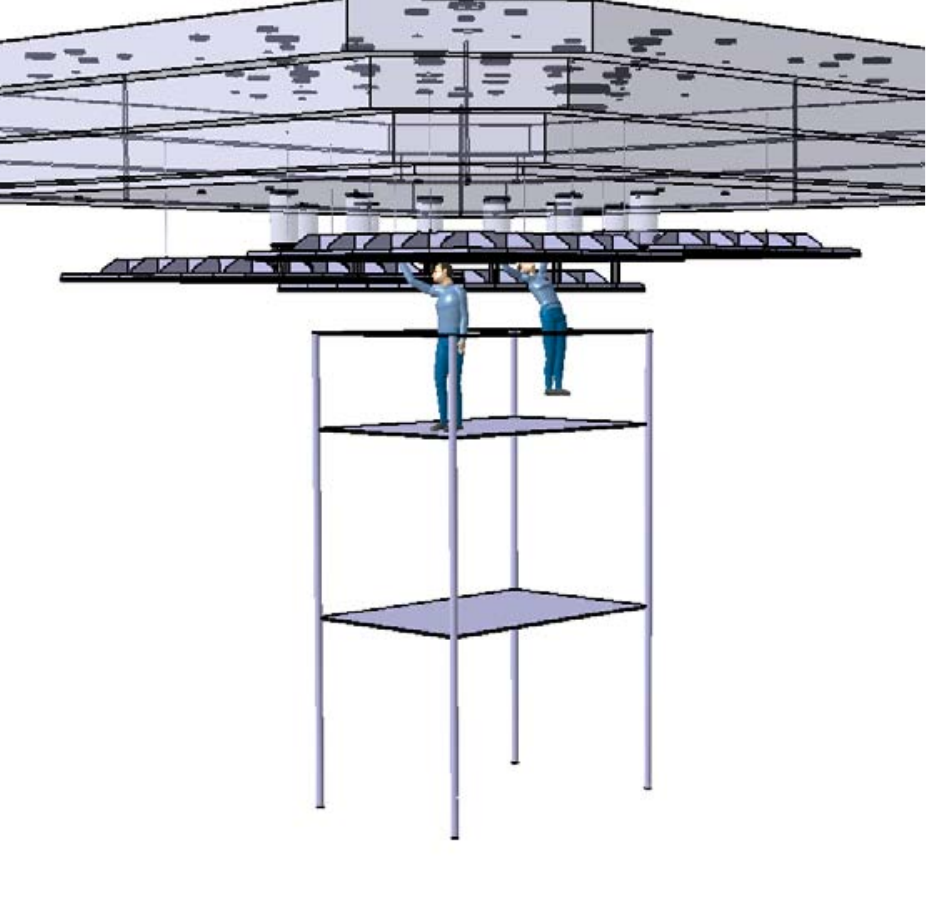}
\includegraphics[width=0.45\textwidth]{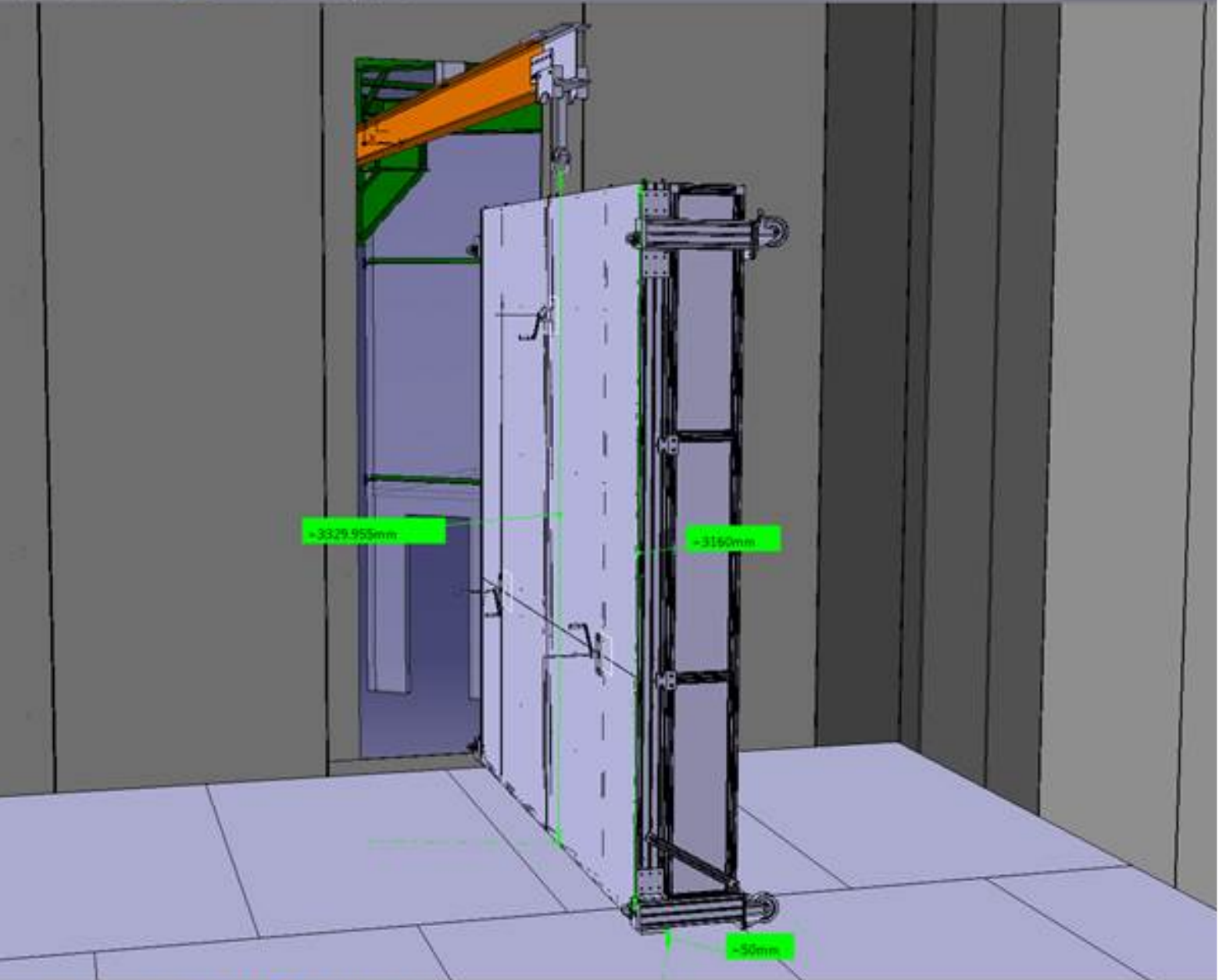}
\end{dunefigure}

Prior to the \dword{tdr}, mutually agreed upon installation plans must
be approved. These will set the schedule for the installation and will
determine the planning for staffing and budget. Having good estimates
for the time needed and having enough experience to ensure that the
interfaces are understood and the procedures are complete is important
for accurate planning. The experience at \dword{protodune} will be
very important as the \dword{protodune} installation establishes the
procedures for handling all the detector elements and in many cases
gives accurate estimates for the time needed. However, in the case of
the \dword{spmod}, many of these procedures need to revised or
newly developed. For example, the \dword{spmod} will be twice as high as
\dword{pdsp}, so two \dwords{apa} need to be assembled together
and a totally different cabling scheme is needed. Testing the
cabling must be done prior to the \dword{tdr} 
in order to
ensure the design is viable. The \dword{dp} will also need to develop
installation procedures as the \dword{dpmod} 
will have a significantly different \dword{fc} and cathode plane. 

By definition, the installation  is on the critical path, making it vital
that the work be performed efficiently and in a manner that has low
risk. In order to achieve this, a prototype of the installation
equipment for the \dword{spmod}  will be constructed at Ash
River (the \nova neutrino experiment \dword{fd} site in Ash River, Minnesota, USA), and the installation process tested with dummy detector
elements. It is expected that the setup will be available at the time
of the \dword{tdr}, but any lessons learned will need to be implemented and
tested after this. In the period just prior to the start of
installation, the Ash River setup will be used as a training ground for
the \dword{uit}.

\subsection{Preparation for Operations}

After the \dwords{detmodule} are installed in the cryostats there remains a lot
of work before they can be operated. First the \dword{tco}
must be closed. This requires bringing back the cryostat manufacturer. 
First the missing panel with the steel beams
and steel panel are installed to complete the cryostat's outer
structural hull. Then the remaining foam blocks and membrane panels
are installed from the inside using the roof access holes 
to enter the cryostat. 

In parallel, the \lar pumps are installed at
the ends of the cryostat and final connections are made to the
recirculation plant. Once the pumps are installed, the cryostat is
closed, and everything is leak tested, the cryogenics plant can be
brought into operation. First the air inside the cryostat is purged by
injecting pure argon gas at the bottom 
at a rate such
that the 
cryostat volume is filled uniformly but faster than the diffusion
rate. This produces a column of argon gas that rises through the volume 
and sweeps out the air. This process is referred to as the \textit{piston
purge}. When the piston purge is complete the cool-down of the \dword{detmodule}
can begin. Misting nozzles inject a liquid-gas mix into the cryostat
that cools the detector components at a controlled rate. 

Once the detector is
cold the filling process can begin. Gaseous argon stored at the surface 
at \surf is brought down the shaft and is re-condensed underground. The \lar then flows through filters to remove any H$_2$O or O$_2$ and
flows into the cryostat. Given the very large volume of the cryostats
and the limited cooling power for re-condensing, it is  
expected to take \num{12} months to fill the first \dword{detmodule} and \num{14} months to
fill the second. During this time the detector readout electronics
will be on monitoring the status of the detector. 
Once the
\dword{detmodule} is full, the drift high voltage can be carefully ramped up and
data taking can begin.

\cleardoublepage



\cleardoublepage
\printglossaries

\cleardoublepage
\cleardoublepage
\renewcommand{\bibname}{References}

\bibliographystyle{utphys} 
\bibliography{common/tdr-citedb}

\end{document}